TECHNISCHE
UNIVERSITÄT
DARMSTADT

# A New Approach to the Construction of Subdivision Algorithms

English Translation of the Dissertation:
Ein neuer Ansatz
zur Konstruktion von
Subdivisionsalgorithmen

by
M. Sc. Alexander Dietz

# Abstract

In this thesis, a new approach for constructing subdivision algorithms for generalized quadratic and cubic B-spline subdivision for subdivision surfaces and volumes is presented. First, a catalog of quality criteria for these subdivision algorithms is developed, serving as a guideline for the construction process.

The construction begins by generating the desired subdominant eigenvectors as the vertices of regular convex 2-polytopes for surfaces and convex 3-polytopes for volumes using circle packings. Subsequently, these polytopes are utilized to construct a Colin-de-Verdière-matrix for the generalized quadratic and a Colin-de-Verdière-like matrix for the generalized cubic B-spline subdivision. These matrices are then adjusted using the matrix exponential to obtain subdivision matrices with the desired properties.

All subdivision algorithms introduced in this paper empirically exhibit a subdominant eigenvalue of $1/2$ with the desired algebraic and geometric multiplicity. For the quadratic case, this property can even be formally proven. Moreover, the corresponding eigenvectors form a convex polytope in the central region for the generalized quadratic B-spline subdivision algorithms, while for the generalized cubic B-spline subdivision algorithms, they represent the refinement of a convex polytope. Additionally, the constructed subdivision algorithms fulfill various other quality criteria, such as affine invariance and convex hull preservation. Furthermore, they possess a dominant eigenvalue of $1$, respect all symmetries of the underlying structure, and exhibit appropriate support for the refinement rules.

The scientific contribution of this work lies in the elaboration of the quality criteria and the novel construction of the subdivision algorithms, including corresponding proofs and empirical analyses. Specifically, an alternative to the Catmull-Clark algorithm with a subdominant eigenvalue of $1/2$ is presented for generalized cubic B-spline subdivision surfaces. For generalized volumetric B-spline subdivision, subdivision algorithms with a suitable spectrum are introduced, making them promising candidates for applications such as simulations in isogeometric analysis.

Furthermore, it is demonstrated that the original Catmull-Clark algorithm [CC78] is not suitable for generalization to volumetric subdivision and that the established subdivision algorithms [Baj+02] and [JM99] do not exhibit a suitable spectrum for several combinatorial configurations. Additionally, research approaches for the volumetric case are proposed, aiming to generalize from hexahedral to arbitrary structures.

The construction of the subdivision matrices was also implemented algorithmically in Matlab. The corresponding software package, including functions for uniform refinement and for plotting subdivision volumes, can be found at: https://doi.org/10.48328/tudatalib-1841

# Disclaimer

This is the English translation of the corresponding German-language dissertation:

> Alexander Dietz. *Ein neuer Ansatz zur Konstruktion von Subdivisionsalgorithmen*
> Dissertation. Darmstadt: TU Darmstadt. 2025.
> DOI: https://doi.org/10.26083/tuprints-00030194

This version was translated using ChatGPT-4o, an artificial intelligence developed by OpenAI, in order to make the dissertation accessible to a broader audience in a low-threshold way. The translation has only been roughly checked for completeness, not for content or appropriate wording. In particular, it is explicitly possible that this version contains errors introduced during the translation by the artificial intelligence. If any parts of the content or formulation seem confusing, it is recommended to compare the relevant sections with the original German version. Care was taken during translation to ensure that the structure of paragraphs and formulas remains as close as possible to the German version, so that the corresponding sections can be easily located.

If any factual errors are found, they can be reported to the author via email:

alexander_dietz@t-online.de

They will be corrected in the next edition accordingly.

# Contents

# 1 Introduction

The discipline of subdivision originated in the 1970s and is thus about 50 years old today. The motivation for and applications of subdivision are very diverse. Besides the early works by de Rham [Rha47] and [Rha56], one of the first works on the topic is due to Chaikin [Cha74], whose motivation is already evident in the first sentence of his abstract [Cha74, p. 346]: "A fast algorithm for the generation of arbitrary curves is described." Obviously, he saw the main advantage of his construction in an algorithm that enables fast generation of arbitrarily shaped curves. Looking into the construction, it can be seen that the basis of the algorithm is a polygonal chain, i.e., a chain of points connected by lines. This polygonal chain is refined step by step by inserting new points, thus becoming smoother and converging to a curve with a certain smoothness. The proposed refinement rule of Chaikin can be equivalently represented by the matrix

$$S := \frac{1}{4}\begin{bmatrix} 3 & 1 \\ 1 & 3 \end{bmatrix} \tag{1.1}$$

Each pair of neighboring points in the polygonal chain is multiplied by this matrix, and the new control points are connected according to the neighborhood relation of the original control polygon.

The natural next question — to which curve the increasingly smoothed polygonal chain converges — was answered by Riesenfeld in [Rie75]. He showed that the increasingly refined polygonal chain converges to a quadratic B-spline curve. Thus, the polygonal chain can be interpreted as the control polygon of the corresponding B-spline. An illustration of the smoothing of a polygonal chain using the $[3/4, 1/4]$-rule from equation (1.1) and the convergence to the respective B-spline curve is shown in Figure 1.1. This insight was commented on by Riesenfeld in [Rie75, p. 309] as follows:

> Alas, one can feel disappointed to find that a new algorithm does not generate a new curve. On the other hand, this new scheme has clear merits of its own. It provides one more original way to view and to understand quadratic *B*-splines. As a hardware implementation scheme, it might have special appeal. But most importantly, it lends itself to natural generalizations that existing algorithms do not readily suggest.

It is therefore no reason for disappointment that Chaikin did not find a new type of curve, but rather an alternative way to represent quadratic B-spline curves. On the contrary, the representation of quadratic B-spline curves using Chaikin's subdivision algorithm offers three significant advantages that can be exploited in the application of subdivision.

The first major advantage is the simple representation of smooth objects in the context of computer-aided design (CAD). Instead of evaluating B-splines, it can suffice for certain applications to refine the control polygon sufficiently many times and to use it instead of evaluating the B-spline. For example, in the fourth image of Figure 1.1, it can be seen that after five refinement steps, the control polygon visually no longer differs from the B-spline curve.

The second important advantage is that Chaikin's subdivision algorithm generates a subdivision of the B-spline segments. More precisely, each segment is subdivided into two new segments, which can also be seen exemplarily in Figure 1.1. There, the frequency of the alternation of the black and gray segments of the B-spline curve doubles with each iteration. The refinement of the basis of quadratic B-spline curves thus comes "free of charge," which is useful for many approximation applications.

The third advantage is that the $[3/4, 1/4]$ rule is very easy to understand, implement, and apply. Since each pair of adjacent points in the polygonal chain is simply multiplied by the matrix from equation (1.1) and the new control points are connected according to the adjacency relation of the old control polygon, the application is not computationally expensive either in terms of implementation or runtime.

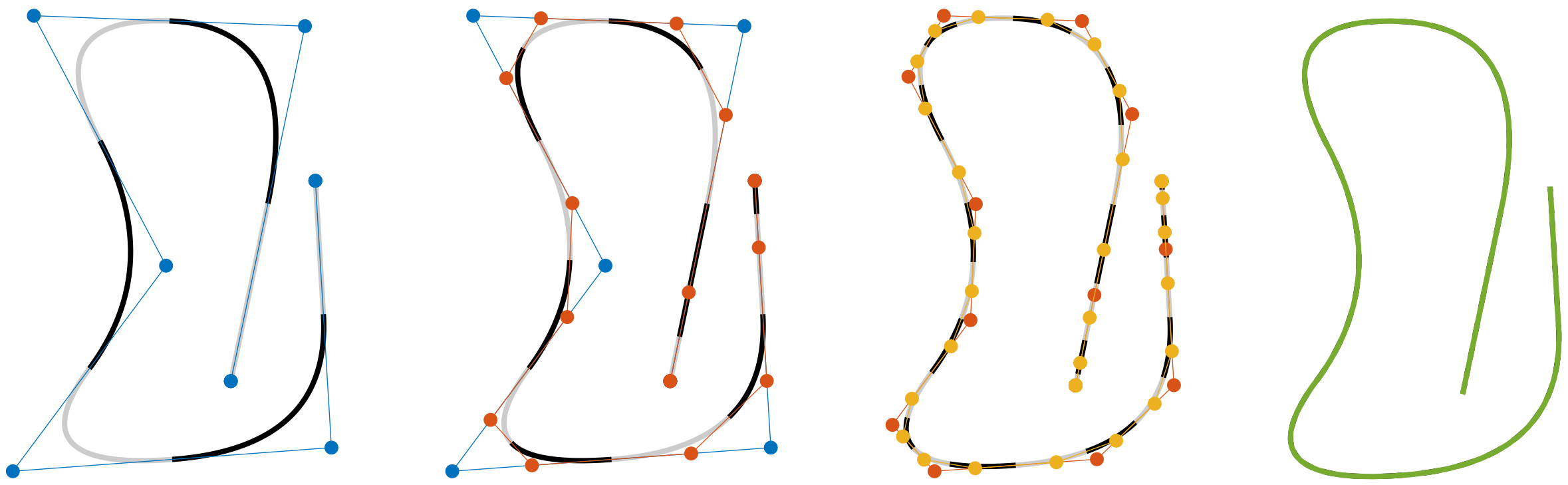

**Figure 1.1:** From left to right: Control polygon with doubled boundary knots in blue and the associated quadratic B-spline with gray-black shaded segments (1), the original in blue and the once refined control polygon in red with shaded segments of the B-spline relative to the refined control polygon (2), the counterpart to (2) with the control polygon refined once in red and twice in yellow (3), the control polygon refined five times in green (4).

These advantages enable the use of subdivision in various contexts for different purposes. First, the procedure can be generalized to surfaces. Doo and Sabin developed the surface counterpart to Chaikin's algorithm, i.e., a subdivision algorithm for quadratic B-spline surfaces, in [DS78]. Catmull and Clark developed a subdivision algorithm for cubic B-spline surfaces in [CC78]. Specifically, these two works developed matrices that refine a two-dimensional structure of control points.

Moreover, there is a mature theory concerning the smoothness of the generated surfaces, which can be found in [PR08]. An essential tool in this theory is the discrete Fourier transform. This allows concrete statements about eigenvalues and thus about the scaling behavior of subdivision matrices. In brief, the discrete Fourier transform exploits the rotational symmetry in subdivision matrices to reduce the information about eigenvalues and eigenvectors of the subdivision matrix to the information of the entries of the region rotated by the symmetry.

The largest industrial application area of subdivision surfaces generated by subdivision algorithms lies in the entertainment industry, especially in the context of animated films. Due to the three advantages described above, various objects and characters in animated films can be represented ideally. The probably best-known pioneering work in this context is the short film "Geri's Game" [Pin+97] from 1997, which won the Oscar for Best Animated Short Film the following year (see also [98]). Background information on the use of subdivision in "Geri's Game" can be found in [DKT98]. For this area, the theory of subdivision algorithms is largely mature.

Another application area is isogeometric analysis. However, here there remains considerable research potential. Isogeometric analysis is based on the idea of using the same basis for the description of the object (curve, surface, volume) and for the solution space. This approach is used for approximation and especially for the numerical solution of elliptic partial differential equations. In this context, the advantages of subdivision, especially the supplied refinement, can be fully exploited. However, in practice, objects are usually described by NURBS (non-uniform rational B-splines) in CAD, and classical methods like the finite element method (FEM) are mostly used for simulation.

Nonetheless, subdivision offers considerable potential for the numerical solution of partial differential equations in the framework of isogeometric analysis due to the advantages mentioned above. Besides the given refinement, the smoothness of the objects is a major advantage. Compared to triangle meshes of other established methods such as FEM, fewer basis functions are needed for approximation, leading to smaller systems of equations to solve the differential equations.

However, subdivision still leaves several questions open in this context. A list of challenges that should be solved for the application of subdivision in isogeometric analysis can be found in [Die+23].

One such challenge for subdivision surfaces is the presence of irregular points in the structure. In short, the generalization of the polygonal chain described above to surfaces consists of a grid made of quadrilaterals (or the dual grid). In this grid, arbitrarily many quadrilaterals can meet at a single vertex. If more or fewer than four quadrilaterals meet at such an inner vertex, this vertex is called an *irregular point*.

For the generalized cubic B-spline subdivision, smoothness is lost at these points according to [PR08]. Furthermore, for both the generalized quadratic and cubic B-spline subdivision, the representation in the region around the irregular points is only possible through a fragmented representation, concretely via a sequence of nested rings consisting of quadrilaterals. An illustrative example can be found in Figure 1.2.

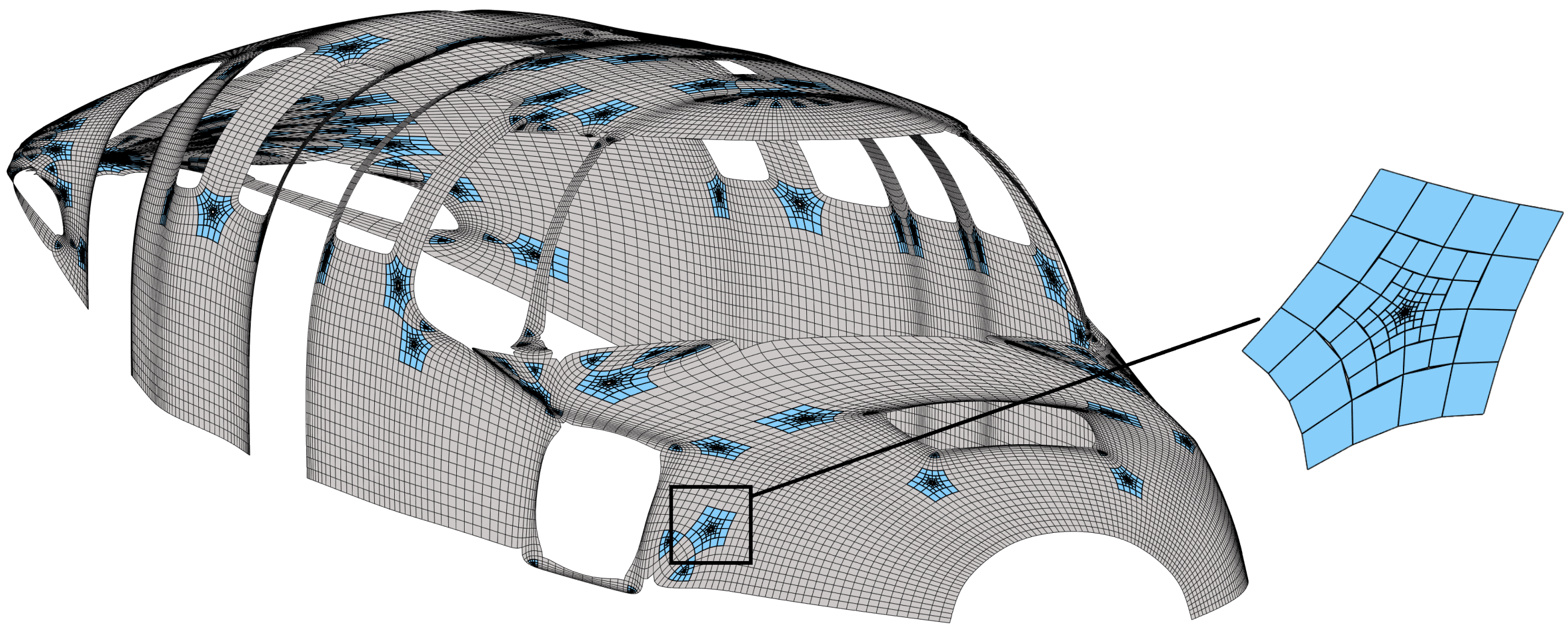

**Figure 1.2:** Example depiction of an airplane component as a subdivision surface with an enlarged view of the subdivision surface around an irregular point. It can be seen that the blue structure consists of infinitely many rings of quadrilaterals.

This sequence of rings leads to challenges in the integration of subdivision surfaces, which were already discussed in [Die18]. More relevant for the present work, however, is that for the generalized cubic B-spline subdivision, the scaling behavior of the ring sequence depends on the number of quadrilaterals around the irregular point, as will be shown in Section 2.4. This also affects the approximation of partial differential equations, since if the rings scale only slowly, more rings need to be generated and evaluated for the approximation.

In the volumetric case, the main interest lies in the area of isogeometric analysis, since for the visual representation of volumetric objects, it suffices to represent the boundary of the objects.

The volumetric case is largely unexplored so far. There is neither a regularity theory nor promising tools, such as the discrete Fourier transform for subdivision volumes, which would allow the eigenvalue information to be reduced to the matrix entries of a hexahedron. Due to the lack of theory, it is also not yet developed how the quality of a volumetric subdivision algorithm should be quantified.

The original goal of the present work was the simulation on surfaces and volumes generated by generalized B-spline subdivision. For a physical domain in $\mathbb{R}^2$, which is parametrized as a surface by generalized B-spline subdivision, first results have also been achieved. An exemplary solution of an elliptic differential equation on a physical domain in $\mathbb{R}^2$ is shown in Figure 1.3. Here, both adaptive refinement and techniques for efficient integration were used to produce the solution. Since the simulation was carried out in the context of isogeometric analysis, the graph of the solution can be represented as the trace of a generalized B-spline subdivision surface in $\mathbb{R}^3$.

With this first ”proof of concept,” the next step should have been the simulation of elliptic partial differential equations on generalized B-spline subdivision volumes. For generalized trivariate cubic B-spline subdivision, there are already two pioneering works in the literature [JM99] and [Baj+02] that can generate cubic B-spline subdivision volumes. However, both algorithms have significant weaknesses. The most serious weakness is that both algorithms produce matrices with up to three different subdominant eigenvalues, which in turn generate structures whose scaling behavior varies in different coordinate directions. This will be shown in Sections 2.4 and 2.5. They thus produce three-dimensional objects whose local asymptotic behavior initially converges towards something surface-like and, with further refinement, towards something line-like. The consequence in the context of isogeometric analysis is that entries of the Galerkin system matrix become very small, which in turn leads to matrices whose inverse can only be computed with a large potential for error.

In the further investigation of the two subdivision algorithms, it also turned out that for certain combinatorial arrangements, these two subdivision algorithms exhibit non-injective characteristic maps, assuming that one can even speak of a characteristic map with three distinct subdominant eigenvalues. This is discussed, among other places, in Section 2.5 and was previously unknown. Although there is no regularity theory yet for the volumetric case, since injectivity and regularity of the characteristic map in the surface case were equivalent to the generated surface being continuously differentiable at all points, the lack of injectivity of these two subdivision algorithms is at least disadvantageous. Moreover, this means that the asymptotic behavior is not only, as already described, surface- and later line-like. In the course of refinement, eigenvectors dominate the asymptotic behavior due to the unfavorable spectrum, which no longer have anything to do with the original volumetric structure.

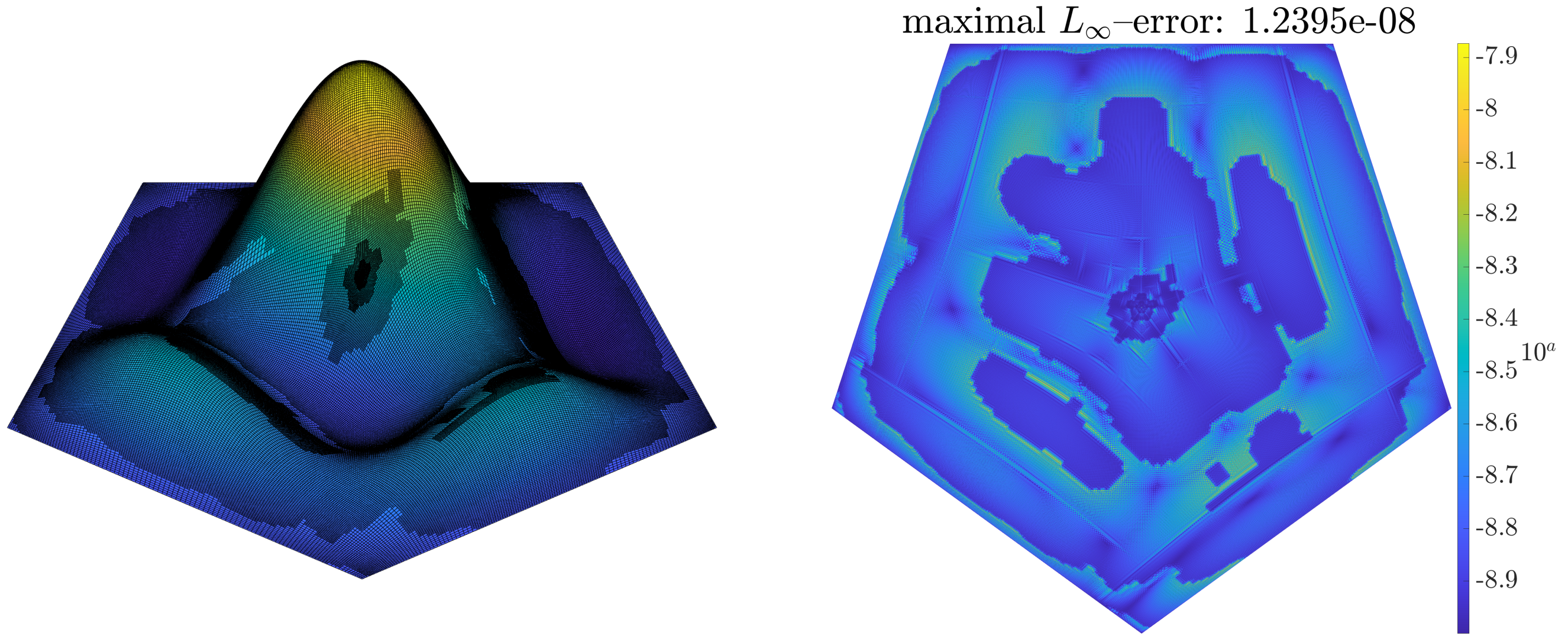

**Figure 1.3:** An exemplary illustration of a solution to an elliptic partial differential equation on a pentagonal physical domain in $\mathbb{R}^2$. The algorithm was intended to compute the differential equation to an accuracy of $10^{-7}$. The refinement of the cells was carried out adaptively. In the left image, the plotted graph of the solution is shown as the trace of a generalized B-spline subdivision surface in $\mathbb{R}^3$. In the right image, the pointwise error is shown. For this, the numerical solution was compared with the exact solution (which is known for this example).

Switching to generalized trivariate quadratic B-spline subdivision was also not possible, since, to our knowledge, no construction of the corresponding subdivision matrices is known yet. Quadratic B-spline subdivision volumes can therefore generally not be refined and thus not used so far.

Thus, the goal of the present work shifted significantly. To be able to use subdivision volumes in the context of isogeometric analysis as well as in other application areas, it is necessary to construct subdivision matrices that satisfy certain criteria, especially the criterion of having three identical subdominant eigenvalues. Therefore, the research goal of this work is:

> Create a list of quality criteria for subdivision matrices. Subsequently, construct subdivision algorithms for generalized quadratic and cubic B-spline subdivision surfaces and volumes that, if possible, meet these criteria. The subdivision matrices should, in particular, have a double (for the surface case) and a triple (for the volumetric case) subdominant eigenvalue $1/2$ with a suitable eigenstructure.

Besides the theoretical elaboration and proofs concerning the quality criteria to be developed, the focus is also on practical application. Since the construction of the subdivision matrices is constructive, it can be implemented algorithmically. Therefore, algorithms are always given in the corresponding chapters and sections in the form of pseudocode to present the information relevant for users in a condensed way. Furthermore, the implementation in Matlab of the algorithms described in this work will be made freely available under a CC-BY license in [Die25] without access restrictions.

## 1.1 Overview of Existing Approaches and Theory

In the previous section, some related works were already mentioned. In this section, they will be briefly recalled for classification purposes.

We begin by considering works on concrete subdivision algorithms. Besides the two works of de Rham [Rha47] and [Rha56], for the one-dimensional case, the work of Chaikin [Cha74] should be noted, who, according to Riesenfeld [Rie75], developed a subdivision algorithm for quadratic B-spline curves.

For the surface case, the two main works or algorithms are the Doo-Sabin algorithm from [DS78] for quadratic B-spline surfaces and the Catmull-Clark algorithm from [CC78] for cubic B-spline surfaces. Additionally, in [ZS01], a generalization of subdivision algorithms up to degree 9 was constructed, which also generates variants of the Doo-Sabin and Catmull-Clark algorithms.

There are also numerous other subdivision algorithms for the surface case. Among these, the Simplest Subdivision algorithm by Peters and Reif [PR97] for refining polytopes to closed $C^1$ surfaces is notable. The same refinement strategy was also used or developed in parallel in [HW99]. A subdivision algorithm for triangle meshes was introduced by Loop in [Loo87], and for structures consisting of both triangles and quadrilaterals, a method was developed in [SW05]. Furthermore, there are several other subdivision algorithms such as the Butterfly subdivision algorithm [DLG90] or the $\sqrt{3}$-subdivision algorithm [Kob96].

For the volumetric case, the two most important works are the subdivision algorithm by Joy and MacCracken [JM99] and the subdivision algorithm by Bajaj, Schaefer, Warren, and Xu [Baj+02]. Both algorithms construct cubic B-spline volumes from hexahedral meshes. However, the algorithm in [JM99] lacks a tensor product structure for the semi-irregular case, and neither algorithm has three identical subdominant eigenvalues. To our knowledge, no refinement rules are known so far for the volumetric case of generalized quadratic B-spline subdivision. A subdivision algorithm for tetrahedral meshes was presented in [SHW04].

Especially for the Catmull-Clark algorithm, i.e., the cubic surface case, there exist numerous variants that aim to improve it with respect to various aspects. This is also known as *tuning*. Regarding tuning subdivision matrices with respect to the ratio $\lambda^2/\mu$ between the subdominant and subsubdominant eigenvalues, the works [PU98], [ADS06], and [MM18] are noteworthy. The first two works provide the mask sizes of the Catmull-Clark algorithm, whereas the approach in [MM18] respects this structure.

Another form of subdivision adaptation for surfaces is the area of *guided subdivision*. One motivation is that classical subdivision algorithms like Catmull-Clark produce poor reflection lines near irregular points (see [KP18, p. 204]). To improve this, guided subdivision first creates a guide surface to direct the refinement. Using this guide surface, a sequence of rings is generated that can be closed after finitely many steps with a cap filling the remaining hole (cf. [KP18, p. 204]). Early works in this area are [KP05] and [KP07]. Alternatives related to Catmull-Clark are found in [KP18] and [KP19]. Variants with a scaling behavior of $1/2$ are also given in [KP17] and [KP24].

This work exclusively considers linear subdivision in the context of B-spline approximation, specifically generalized quadratic and cubic B-spline subdivision. This means that a mesh or polygonal chain of control points is refined linearly—that is, by applying a matrix—so that the increasingly refined mesh converges (in our case, to a B-spline element). Nonlinear subdivision also exists; here, Ewald's dissertation [Ewa20] on (non)linear subdivision of curves is notable. Subdivision is also used for interpolation. An overview for the surface case is given by Dyn in [Dyn02]. Moreover, the works [CMQ03] for tetrahedral meshes and [Xie+20] for hexahedral meshes present volumetric subdivision algorithms for interpolation.

The existing theoretical background on subdivision focuses on subdivision surfaces. The reason is that linear subdivision of B-spline curves requires little discussion: once constructed, the subdivision algorithms produce B-spline curves. There is a broad theory on general curves in the context of subdivision, but it is not relevant here. As already mentioned, no theory exists yet for the volumetric case, only the algorithms listed above. For the surface case, Peters and Reif's standard work [PR08] provides a broad and deep overview of subdivision algorithms for surfaces, their analysis, and regularity. In the following, some key milestones of this field are described, taken from the bibliography of [PR08].

The basis for the theory, especially the idea of the characteristic ring to describe local asymptotic behavior, originates from Reif's doctoral thesis [Rei93]. The condensed results were published in [Rei95]. Besides criteria for convergence, it also describes the role of subdominant eigenvalues in relation to the regularity of the characteristic map. The connection between regularity and injectivity of the characteristic map and $C^1$ smoothness of the subdivision surface is established in [PR98]. The effect of the ratio $\lambda^2/\mu$ on the curvature of the generated subdivision surfaces is described in [PR04], while the background of polar artifacts, i.e., nonuniform scaling behavior around irregular points, is discussed in [SB03]. A tuning for triangular meshes is also given in [BK04].

There remains substantial research potential regarding subdivision. A general overview is given in [RS19]. A more specialized overview of open research questions concerning subdivision in the context of isogeometric analysis can be found in [Die+23].

As mentioned, a central application area of subdivision is isogeometric analysis. The term was coined in [HCB05]. The idea is, as described, that the solution space of the differential equation is the same as the space used for representing the object. Subdivision has already been applied in this context, e.g., in [COS00] and [NKP14] for surfaces and [BHU10] and [Xie+20] for volumes. The latter, however, uses an interpolatory approach for volume generation.

Practical tools are also required for using subdivision elements in isogeometric analysis, such as element evaluation and integration over elements. Evaluation of subdivision surfaces, especially near irregular points, was introduced by Stam in [Sta98]. Integration of special integrals is described in [HR16]. Approaches for integrating functions over subdivision surfaces are found in [Die18] and for subdivision volumes in [Alt21].

Finally, alternatives for representing irregular regions of B-spline surfaces are considered. A good overview is found in [Pet19], especially regarding geometric continuity splines and singularly parameterized splines. The former were already described in [DeR90], and an overview is in [Pet02]. The latter was introduced in [Pet91] and treated in [Rei97] and [MR22]. Volumetric singular parameterizations were constructed in [Pet20] and investigated within isogeometric analysis in [YP22].

## 1.2 Structure of the Work and Reading Recommendation

As already described, the goal of this work is to construct subdivision algorithms for the four described cases of generalized quadratic and cubic B-spline subdivision of surfaces and volumes. For this, some preparations are needed first.

The first preparation is of a general nature. In the next section 1.3, the conventions and notation used in this work are presented. The aim of this section is to establish a clear notation and to improve readability.

Following the notation, it is recommended to briefly look into the appendix. Here, the foundations related to graphs and matrices that are prerequisites for this work are summarized. The section A on graph theory is subdivided into automorphisms (Section A.1), connectivity (Section A.2), planarity (Section A.3), faces (Section A.4), and two-dimensional duality (Section A.5). The section B on matrix fundamentals is subdivided into eigenvalues and eigenvectors (Section B.1), the matrix exponential (Section B.2), and concrete matrix exponentials (Section B.3).

Most of the appendix contents consist of basic definitions and can therefore be skipped; however, some special theorems can be found there. Especially Section B.3 contains statements and proofs on concrete matrix exponentials that are needed for this work but may be tedious to read. Thus, the appendix can be regarded as a reference work to be consulted at relevant points in the thesis.

After these preparations, the main part of the thesis begins with Chapter 2, which describes the fundamentals of subdivision. In Section 2.1, the structurally relevant terms are introduced first. This part is essential because these terms are developed specifically for this work to create a framework usable for all four initially described cases. Moreover, for the three-dimensional case, the combinatorial arrangements of structures are significantly more complex than in two dimensions, so a general structure is needed. In particular, the cells, the initial elements, and their interplay should be internalized to follow the further course of the work. The notion of regular initial elements is crucial, as it determines which parts of the structure can be evaluated as B-spline elements. Section 2.2 then deals with refining the described terms and explains the subdivision principle.

Section 2.3 considers structures that are invariantly mapped onto each other by refinement. Here, the motivation for defining initial elements becomes clear, since they are the smallest objects in the structure that can be invariantly mapped. Also, the classical case of subdivision, namely the sequence of rings around an irregularity, reappears here. For generalized cubic B-spline subdivision, it is important that not a sequence of rings but a sequence of double rings is used. This results from the desired freedom of choice of the subdivision matrix and is also used in [MM18]. Nonetheless, this description is uncommon in the literature and should therefore be noted even by experienced readers. Furthermore, this concept is extended to the three-dimensional case by generalizing rings and double rings to shells and double shells.

Section 2.4 marks the first substantial treatment of subdivision matrices. A catalog of 13 quality criteria is developed. Some criteria are mandatory; others are optional. For each criterion, motivation and the consequences of (non-)fulfillment are discussed.

Section 2.5 finally prepares the research goal of this work and particularly motivates the necessity of the triple subdominant eigenvalue for the three-dimensional case. Additionally, the previously unknown weaknesses of established volumetric subdivision algorithms [Baj+02] and [JM99] are shown, since for certain combinatorial arrangements the three largest eigenvalues less than 1 do not represent an injective shell. At the end of Chapter 2, both the necessary terms and structures as well as the goal in the form of quality criteria and the motivation and necessity of new subdivision algorithms are thus presented.

Chapters 3 through 6 deal with the actual construction of the subdivision algorithms. These are constructed for the initial elements, i.e., the smallest invariantly refinable elements. The refinement of larger structures then follows by combining the refinement of the initial elements. In this context, besides the subdivision matrices of the initial elements, the subdivision matrices of (double) rings and (double) shells are also examined.

The construction proceeds rather unconventionally compared to the usual methods in the literature. Typically, publications present constructions that end in explicit formulas for the entries of the subdivision matrix. Subsequently, certain quality criteria may be checked. The construction in this work proceeds completely differently, since the goal is to create a subdivision matrix with a triple subdominant eigenvalue $1/2$ and a suitable eigenvector structure.

Therefore, the construction strategy is exactly the reverse of the conventional approach. First, a suitable structure of the subdominant eigenvectors is generated. From this, a matrix is then constructed, and in the final step, this matrix is modified so that the spectrum matches the desired structure.

The construction of the eigenstructure is carried out in Chapter 3. The idea is that the eigenvectors of the subdominant eigenvalues for the generalized quadratic B-spline subdivision should represent convex $d$-polytopes and for the generalized cubic B-spline subdivision refinements of convex $d$-polytopes. This means that each row of the matrix formed by the $d$ eigenvectors represents a point in $\mathbb{R}^d$, and the convex hull of these points forms a convex polytope.

Thus, Chapter 3 deals with the construction of convex polytopes. The majority of the chapter focuses on three-dimensional polytopes. The central statement is Steinitz's theorem 3.8, which explains the connection between 3-connected planar graphs and convex 3-polytopes. Using this, the combinatorial structure in the form of a graph can be transformed into a convex polytope.

To this end, Section 3.1 first considers the basics of convex polytopes and algorithms for checking and generating the data structure, such as a 3-connectivity test, a planarity test, and the creation of the facet list of a 3-connected planar graph. Moreover, Section 3.1.1 describes the generation of random examples needed for empirical and numerical tests in this work.

Since convex polytopes are not unique for a given structure, Section 3.2 describes which shape the constructed 3-polytopes will have. The basis is the Koebe-Andreev-Thurston theorem 3.18. For this work, essentially only the final version Theorem 3.25 is relevant. Since this could not be directly taken from the literature in this strong formulation, the derivation of the final version is explained in Section 3.2.

The actual construction of 3-polytopes and their dual polytopes takes place in Section 3.3 and is based on the construction by Ziegler [Zie04]. The benefit of this section lies in the algorithmic implementation, which Ziegler did not provide. Therefore, the section is especially interesting for practitioners, as it describes numerical issues occurring in Ziegler's construction and offers practical solutions to generate the actual coordinates of the polytope vertices with high accuracy.

The chapter ends with the construction of 2-polytopes in Section 3.4, in which only regular $n$-gons centered at the origin are generated.

From these polytopes, matrices will then be generated. For this, the concept of Colin-de-Verdière-matrices is needed, introduced in Chapter 4 and specifically in Section 4.1. Simplified, a matrix can be generated from a $d$-polytope containing the origin whose kernel is spanned by the coordinate vectors of the polytope. For 3-polytopes, the construction by Lovász is used, described in Section 4.4. For 2-polytopes, a construction by Izmestiev exists but cannot be directly applied here. Thus, in Section 4.3, an explicit construction of a matrix for special 2-polytopes is developed and subsequently shown to agree with Izmestiev's theoretical results. For the generalized cubic B-spline subdivision, the eigenstructure is not a polytope but a refinement of a polytope. Hence, the results from Chapter 4 cannot be directly applied. For an alternative construction, Colin-de-Verdière-matrices of dimensions $0$ up to $d$ are required, so Section 4.2 additionally considers Colin-de-Verdière-matrices for 0- and 1-polytopes.

Based on the previous chapters, Chapters 5 and 6 treat the construction of generalized quadratic and cubic B-spline subdivision algorithms and form the core of this work. In Chapter 5, subdivision matrices for generalized quadratic B-spline subdivision are first presented. Section 5.1 elaborates the fundamentals and describes the Doo-Sabin algorithm. Refinement rules for regular initial elements are also explained.

Section 5.2 is dedicated to constructing new subdivision algorithms. For generalized quadratic B-spline subdivision, two variants are presented. The first variant directly uses the eigenstructure from Chapter 3 and produces a matrix with spectrum $1, 1/2, \dots, 1/2, 1/4, \dots, 1/4$. However, negative matrix entries arise in this construction.

The second variant first generates a Colin-de-Verdière-matrix from the eigenstructure in Chapter 3. The spectrum of this matrix is then shifted by means of the matrix exponential so that the final product has the desired spectrum. No negative matrix entries are produced here, but except for the dominant and subdominant eigenvalues, the spectrum cannot be exactly controlled.

Section 5.3 finally deals with the proofs regarding the quality criteria. For the different variants, it is investigated which quality criteria are (not) fulfilled. The random examples described in Section 3.1.1 are used to numerically verify whether the criteria are met.

Chapter 6 deals with generalized cubic B-spline subdivision and is structured analogously to Chapter 5, albeit considerably more complex. Section 6.1 first explains the fundamentals and the Catmull-Clark algorithm. The refinement of regular initial elements is discussed as well. Furthermore, it is explained why, for generalized cubic B-spline subdivision, the double ring structure for subdominant eigenvalues $1/2$ is necessary and how this relates to the evaluable region.

Section 6.2 then describes the construction. It begins with a $d$-polytope and refines it by inserting a point for each edge, each facet, and at the origin; these inserted points are connected with edges accordingly. Subsequently, Colin-de-Verdière-matrices for 0- to $d$-polytopes are generated for the individual areas of this structure and combined into a matrix. This matrix is decomposed into two triangular matrices, whose spectra are adjusted using the matrix exponential. The two thus modified matrices are finally multiplied to generate the subdivision matrix.

Section 6.3 again considers the quality criteria. As already evident from the title of the sixth chapter, the variant of generalized cubic B-spline subdivision is of experimental nature. The reason is that although the constructed matrices have a $d$-fold eigenvalue $1/2$, it could not be shown that the other eigenvalues different from $1$ have modulus smaller than $1/2$. Therefore, numerical experiments are particularly important for this variant, as they can at least provide a qualitative statement on whether the constructed algorithm fundamentally works.

The hexahedral structure of cells and the derived concepts used in this work can be extended for the volumetric case of hexahedral meshes to arbitrary meshes. The generalized quadratic B-spline subdivision algorithms can be generalized to this case without modification. This generalization is briefly considered but not detailed or rigorously treated in Chapter 7. Here, approaches for further research are presented, which are also discussed in the conclusion chapter 8.

## 1.3 Conventions and Notation

For reasons of clarity, some fundamental notations are used throughout this work. A complete overview of all notations can be found in the list of symbols; here only the basic structure of the notation is described, beginning with the number sets:

**Notation 1.1.** *Number sets are denoted by upright capital letters, for example* $\mathbb{N}$ *for the set of* natural numbers *(without zero),* $\mathbb{R}$ *for the set of* real numbers*, and* $\mathbb{C}$ *for the set of* complex numbers*. Specific variants are indicated by subscripts, for example* $\mathbb{N}_0$ *for the natural numbers including zero, or* $\mathbb{R}_{\geq 0}$ *for the set of non-negative real numbers. Furthermore, the* unit circle *is denoted by* $\mathbb{S}^1 \subset \mathbb{R}^2$*, and the* unit sphere *by* $\mathbb{S}^2 \subset \mathbb{R}^3$.

Next, the general symbolism is sketched:

**Notation 1.2.** *Scalar variables are denoted by lowercase letters, for example* $a$*, and matrices by uppercase letters, for example* $M$*. Vectors may be denoted either by lowercase or uppercase letters depending on convenience.*

*Functions (mappings) are denoted by small bold letters, such as* ***f****, and sets by large bold letters, such as* ***M****. Since a graph* ***G*** *can be interpreted as a tuple, i.e., a set consisting of a set of vertices* ***V*** *and a set of edges* ***E****, these three elements in particular are represented by bold letters.*

*Each object has one or more standard designations. For matrices, for example,* $M$ *is standard; for variables,* $a, b,$ *and* $c$*. Special objects have reserved letters, such as* $x$*,* $y$*, and* $z$ *for coordinate variables or* $E$ *for the identity matrix. Therefore, in unspecified cases, the standard designation is always used. Greek letters have no special function and merely extend the alphabet, but they can also be reserved for certain concepts. An overview of reserved letters can be found in the list of symbols.*

Next, the notation for matrices is described in more detail:

**Notation 1.3.** *As already mentioned, matrices are denoted by uppercase letters, for example* $M$ *for a general matrix. Entries of a matrix are described by subscript tuples enclosed in parentheses. Concretely, this means:*

- $M_{(i,:)}$ *denotes the* $i$*-th row of a matrix, with* $i \in \mathbb{N}$.
- $M_{(:,j)}$ *denotes the* $j$*-th column of a matrix, with* $j \in \mathbb{N}$.
- $M_{(i,j)}$ *denotes the* $(i, j)$*-th entry of a matrix, with* $i, j \in \mathbb{N}$.

*In contrast, indices of enumerated matrices are given without parentheses, for example* $\sum_{i=1}^{n} M_i$ *denotes the sum of* $n$ *matrices, not of* $n$ *entries of one matrix. When possible, indices are given as subscripts. However, when matrix entries and indices are used simultaneously, the indices appear as superscripts. To distinguish superscripts from powers, they are enclosed in parentheses.*

*For example, the expression* $\sum_{k=1}^{n} M_{(i,j)}^{(k)}$ *is the sum of all* $(i, j)$*-th entries of a sequence of* $n$ *matrices, whereas* $\sum_{k=1}^{n} M_{(i,j)}^{k}$ *is the sum of all* $(i, j)$*-th entries of powers of a matrix* $M$.

*The only other notation for matrices is* $M^T$*, denoting the* transpose *of a matrix. On the other hand, notations such as* $M'$*,* $\tilde{M}$ *or* $\overline{M}$ *denote specially defined matrices without uniform meaning. Hence,* $M'$ *does not explicitly mean the transpose of* $M$*, nor does* $\overline{M}$ *denote the adjoint of* $M$.

*Furthermore,* $E_n$ *denotes the identity matrix of size* $n \times n$*, and* $0_n$ *denotes the zero matrix of size* $n \times n$.

The notation for vectors is as follows:

**Notation 1.4.** *As already mentioned, vectors may be denoted by either lowercase or uppercase letters. Their dimension, unless explicitly defined, is inferred from the context. For example, if* $M \in \mathbb{R}^{3\times3}$*, then the vector* $a$ *in the equation*

$$Ma = \begin{bmatrix} 0 \\ 0 \\ 0 \end{bmatrix}$$

*is a column vector of dimension* $3 \times 1$*. The symbols* $0$ *and* $1$ *play a special role. They typically denote numbers but can also denote vectors. For instance, in the equation*

$$Ma = 0$$

*it is clear that* $0$ *here denotes a* $3 \times 1$ *zero vector. To emphasize this, the notation* $\vec{0}$ *is often, though not always, used. Thus, the equation could also be written as*

$$Ma = \vec{0}.$$

*The same applies analogously to* $1$ *and* $\vec{1}$*. For matrices, the notations* $0_n$ *and* $1_n$ *are used when a* $n \times n$ *matrix consisting of zeros or ones is intended.*

We also introduce the following notation with respect to the exponential function:

**Notation 1.5.** *Let $a \in \mathbb{R}$ and $M \in \mathbb{R}^{n \times n}$. Then the exponential* $\exp_2$ *is the notation for*

$$\exp_2(a) := \exp\big(\ln(2)a\big) = 2^a \quad \textit{and} \quad \exp_2(M) := \exp\big(\ln(2)M\big) = 2^M.$$

For graphs, the following convention applies, which is also embedded in Remark A.19 within the context of graph theory. Since this convention is important for reading this work, the relevant part is repeated here:

**Convention 1.6.** *Graphs in this work are initially described abstractly and can be interpreted abstractly. Treating the node set as an abstract set with $|\boldsymbol{V}|$ elements, a graph can be generated by the information of the adjacency matrix. Abstractly interpreted graphs thus initially encode only a structure between objects via their edges. Such a graph is also called an* abstract graph*.*

*Meaning can be assigned to the nodes by identifying them with concrete elements, or better, by defining the set* **V** *of nodes as these concrete elements. Thus, by the definition of the graph, the set of nodes can readily be defined as a subset of $\mathbb{R}^2$ or $\mathbb{R}^3$. Such a graph is also called an* embedded graph *or* realized graph*.*

*This work particularly deals in Chapter 3 with first generating an abstract graph from an adjacency matrix and then realizing it with nodes in $\mathbb{R}^2$ or $\mathbb{R}^3$. Since this realization is merely an isomorphism between abstract and concrete nodes, the abstract graph and its realization(s) are used synonymously. This enables a considerable simplification of the notation and makes the related concepts clearer.*

We also introduce the following notation regarding the eigenvalues of a matrix:

**Notation 1.7.** *Let $M \in \mathbb{C}^{n \times n}$ be a matrix.*

- *An unspecified eigenvalue of the matrix $M$ is denoted by $\lambda_i$, without further specifying $i$.*
- *A specific eigenvalue is denoted with distinct indices, for example $\lambda_0$, $\lambda_1$, $\lambda_t$, or $\lambda_n$.*
- *The subdominant eigenvalue of a matrix, as defined in Definition 2.34, is denoted by $\lambda$.*
- *The subsubdominant eigenvalue of a matrix, as defined in Definition 2.34, is denoted by $\mu$.*

This concludes the overview of the important notations for this work, so the content can begin with the next chapter.

# 2 Subdivision, Subdivision Matrices and their Quality Criteria

Since the 1970s, subdivision has been developed and researched as a branch of mathematics. To approach the concept for this work, we first formulate the following deliberately imprecise description:

**Description.** *A subdivision mapping is a mapping consisting of refinement rules that map control points with a certain structure in* $\mathbb{R}^d$*, with* $d \in \mathbb{N}$*, to control points with a certain structure in* $\mathbb{R}^d$ *and serve a specific purpose.*

It is hard to express the concept more vaguely. However, the above description enables us to describe the various areas and terms of subdivision.

We begin with the concept of a *mapping*. In this work, we restrict ourselves to linear subdivision, that is, mappings that can be represented by matrices. Of course, subdivision can also be realized with nonlinear mappings. For example, an analysis of one-dimensional subdivision algorithms, both linear and nonlinear, was conducted by Ewald in [Ewa20]. By restricting to linear subdivision, some freedom in designing the mapping is lost, but this restriction also offers many advantages. We thereby gain the advantage mentioned in the introduction of simplicity in representation and implementation. Furthermore, restricting to linear mappings allows us to analyze properties of the corresponding matrices. Thus, the theory becomes more specific and allows formulating results for the specific case.

The next striking phrase is *control points with a certain structure in* $\mathbb{R}^d$. Here three parameters must be specified: First, consider the *dimension* $d$. The dimension of the space in which the control points lie initially plays only a minor role in the theory. For example, in Figure 1.1 we see control points in $\mathbb{R}^2$ representing a planar curve. Chaikin's subdivision scheme [Cha74] can just as well be applied to a control polygon in $\mathbb{R}^3$, i.e., for a space curve. Similarly, surfaces and volumes in higher dimensions can be represented and used.

Particularly important, however, is the *structure* of the control points. Here one must first consider the dimension of the subdivision elements. Specifically, this work deals with subdivision curves, surfaces, and volumes, where subdivision curves are mentioned only for completeness. For the other two cases, i.e., subdivision surfaces and volumes, we will develop concrete subdivision algorithms. Moreover, the geometry of the elements and their relations are crucial. This will be discussed throughout this chapter. Subdivision can also be extended to higher-dimensional objects, which is beyond the scope of this work.

Also crucial is the *number* of control points. This work considers only subdivision matrices mapping $n$ control points to $n$ control points, with $n \in \mathbb{N}$. Thus, the subdivision matrix is always square. An example is the matrix from Equation (1.1). It always maps two control points to two new control points. Since this is done separately for each segment and segments overlap at one control point, the number of control points can increase through subdivision. The matrices are therefore applied locally. Thus, restricting to square matrices is not a true limitation, but it enables analysis of these matrices (e.g., with respect to the spectrum).

The last element to specify is the *purpose* of subdivision. Here the spectrum is very broad in the literature. We consider subdivision exclusively in the context of generalized quadratic and cubic B-spline subdivision, i.e., the (re)production of B-spline or subdivision elements.

These aspects will be concretized and elaborated in this chapter. We describe what subdivision is and should be in the context of this work. In Section 2.1 we first consider the fundamental concepts and structures we need. Then, in Section 2.2, the concept of subdivision is defined and we explain concretely how the concepts introduced at the beginning are refined. Invariant parts of the structure appear during subdivision, which we describe in Section 2.3. For the subdivision matrices of these invariant structures, we develop quality criteria in Section 2.4. There, we discuss the motivation behind each criterion as well as the effects and consequences if a subdivision matrix (does not) meet(s) a quality criterion.

Moreover, these quality criteria serve as guidelines for the subdivision matrices constructed in Chapters 5 and 6 and motivate why these particular constructions were chosen. The construction follows a common scheme for all newly developed subdivision algorithms in this work. This scheme is described in Section 2.5, after classifying the existing variants in the literature. In particular, this clarifies why Chapters 3 and 4 are needed and how they fit into the construction.

## 2.1 Definitions, Terms, and Structure

Based on the specifications just discussed, we can now define the terms needed for subdivision. We introduce these step by step and combine them into new concepts. As mentioned before, this work considers subdivision only in the context of generalized quadratic and cubic B-spline subdivision. Accordingly, the terms and definitions in this section are adapted to this application.

We start with objects in the domain and gradually add structure until we can describe the evaluation of elements (curve, surface, volume). The terms are illustrated with examples, which can only be described in detail once all terms are introduced. Nevertheless, it is recommended to look at Figures 2.2 and 2.3 for the different definitions, as this helps to gain an intuitive understanding of the terms.

Since the evaluated elements can obviously be different, we first define their type and degree.

**Definition 2.1.** *An* evaluated element *is, depending on its dimension, a* curve*, a* surface*, or a* volume*. We denote by $t \in \{1, 2, 3\}$ the* type *of the evaluated elements. Specifically, we set $t = 1$ for curves, $t = 2$ for surfaces, and $t = 3$ for volumes. Furthermore, we denote by $g \in \{2, 3\}$ the* degree *of the elements. Specifically, we use $g = 2$ for quadratic and $g = 3$ for cubic elements.*

This definition could be extended to other degrees and types. For example, both terms could be defined as natural numbers and thus generalized to quadratic elements or to elements of higher dimension. However, this work considers only the six combinations of degree and type. Moreover, the above definition is already given with respect to the purpose of subdivision: the (re)production of B-spline elements.

Using the type, we want to introduce the concept of the spline domain. This is based on the notion of a „spline domain“ from [PR08, p. 42–43] and generalizes it for arbitrary types. For this, we first need some terms and explanations which rely on [PR08, pp. 39–43]. We start with the definition of a *cell*:

**Definition 2.2.** *Let $t \in \{1, 2, 3\}$ be the type of the evaluated elements. We call*

$$s_i := [0, 1]^t \times \{i\} \quad \textit{with} \quad i \in \mathbb{N}$$

*a $t$-*variate cell*.*

An illustration of the cells (each without the index $i$) can be found in Figure 2.1. Each *cell* is itself a topological space (based on the open sets of $[0, 1]^t$). In the next step, we want to combine different cells into a topological space in a suitable way. For this, we define the following equivalence relation:

**Definition 2.3.** *Let $t \in \{1, 2, 3\}$ be the type of the evaluated elements and $[0, 1]^t \times \boldsymbol{M}$ with $\boldsymbol{M} \subset \mathbb{N}$ be a set of cells. Let $\boldsymbol{d}$ be an equivalence relation on $[0, 1]^t \times \boldsymbol{M}$ with the following properties:*

- *The equivalence relation $\boldsymbol{d}$ identifies only the $(t - 1)$-dimensional boundaries of cells pointwise. For $t = 1$ these boundaries are the two endpoints $0$ and $1$, for $t = 2$ the four outer edges, and for $t = 3$ the six outer faces.*
- *Only complete $(t - 1)$-dimensional boundaries are identified.*
- *No two $(t - 1)$-dimensional boundaries of the same cell are identified.*
- *At most two $(t - 1)$-dimensional boundaries of different cells are identified.*
- *Two cells share at most one identified $(t - 1)$-dimensional boundary.*

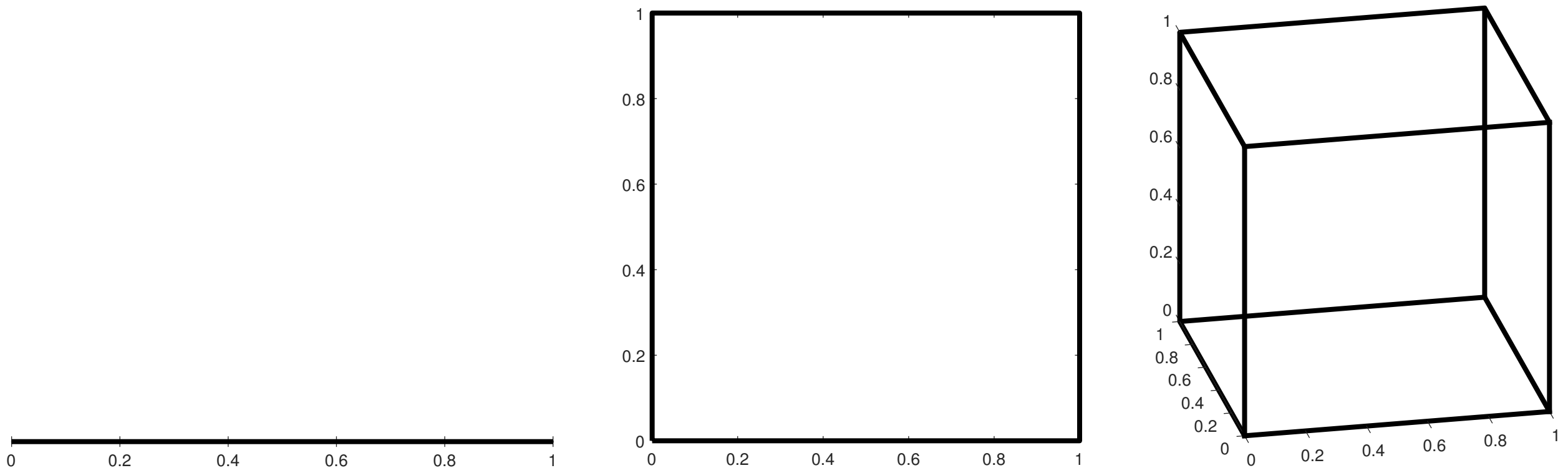

**Figure 2.1:** Illustration of the $t$-variate cells for $t = 1$ (left), $t = 2$ (middle) and $t = 3$ (right), each without the index $i$.

*Then* ***d*** *is called a* **D**-equivalence relation *on the set* $[0,1]^t \times \boldsymbol{M}$.

With these two concepts, it is possible to define the spline domain:

**Definition 2.4.** *Let $t \in \{1,2,3\}$ be the type of the evaluated elements. A set of cells $[0,1]^t \times \boldsymbol{M}$ with $\boldsymbol{M} \subset \mathbb{N}$ together with a* ***D****-equivalence relation* ***d*** *is called a* spline domain *and is denoted as a tuple by*

$$\boldsymbol{D} := \big([0,1]^t \times \boldsymbol{M}, \boldsymbol{d}\big) = \left(\{s_i \mid i \in \boldsymbol{M}\}, \boldsymbol{d}\right).$$

*If* ***M*** *is a finite set, we denote* ***M*** *without loss of generality by $\{1,\ldots,n\}$ (with $n \in \mathbb{N}$), and if* ***M*** *is an infinite set, we denote* ***M*** *without loss of generality by $\mathbb{N}$. Furthermore, we denote the number of cells in* ***D*** *by $|\boldsymbol{D}|$.*

Thus, **D** is again a topological space. A set in **D** is open if each part of every cell in the set is open.

The **D**-equivalence relation explains how the individual cells of the spline domain **D** are related to each other. Figuratively speaking, these cells are glued together at their boundaries with other cells. This principle of *gluing* is also described in [PR08, pp. 41–42]. Regarding the spline domain, the following remark should be noted:

**Remark 2.5.** *The topological background has only been briefly described here. Since this is only a marginal topic of the present work, we do not go into further detail. Background on topological spaces can be found, for example, in [Bro06, Chap. 2, pp. 19 ff.]. The glued spaces are known under the technical term* adjunction space *in the literature. Background information can also be found in [Bro06, Sec. 4.5, pp. 118 ff.].*

**Remark 2.6.** *The hexahedral structure of the cells in the volumetric case can be generalized to a trapezoidal structure. This is explained in Chapter 7.*

**Remark 2.7.** *Without loss of generality, we can assume that the spline domain is connected. That means one can reach from any cell to any other cell through the identification of points for $t = 1$, edges for $t = 2$, and facets for $t = 3$. If this is not the case, the different connected components can be considered separately as individual spline domains.*

As already mentioned, cells are glued together at their boundaries (vertices for $t = 1$, edges for $t = 2$, and facets for $t = 3$). This means there can be cells that are glued and thus adjacent to other cells on every boundary, and others that are not. For the following definitions, however, it is necessary that cells always have a neighbor at every glued boundary. Therefore, we introduce the following convention according to [PR08, p. 42]:

**Convention 2.8.** *Every definition in this chapter that involves a boundary can be embedded or extended in a suitable way into a scenario without boundaries but following the same rules. The elements at the boundary are then described by this embedding or extension. If additional elements are needed for a geometric description, we therefore assume in the following that this geometric description is given by an embedding of appropriate size. Hence, without loss of generality, we assume that all considered definitions and elements lie not on a boundary but inside.*

In the next step, we describe the realization or evaluation of the spline domain in $\mathbb{R}^d$. For this, we first consider the set of control points and then link them to the spline domain. Since the linking differs for degrees $g = 2$ and $g = 3$, we treat these two cases separately.

**Definition 2.9.** *Let* $\boldsymbol{D}$ *be a spline domain of type* $t$ *with* $|\boldsymbol{D}| = n$ *and let* $g = 2$ *be the degree. Let* $m, d \in \mathbb{N}$ *with* $m \geq n$. *The matrix* $P' \in \mathbb{R}^{m \times d}$ *is called the* control point matrix. *Each row in* $P'$ *represents a point in* $\mathbb{R}^d$ *and the entire matrix contains* $m$ *control points. Each control point is associated with a cell* $s_i \in \boldsymbol{D}$, *so the control points lie dual to the vertices of the cells. It holds that* $m \geq n$, *because for cells at the boundary control points must also be associated with cells of the respective embedding.*

**Definition 2.10.** *Let* $\boldsymbol{D}$ *be a spline domain of type* $t$ *and let* $g = 3$ *be the degree. Let* $n$ *be the number of all vertices of all cells of* $\boldsymbol{D}$ *(without double counting). Let* $m, d \in \mathbb{N}$ *with* $m \geq n$. *The matrix* $P' \in \mathbb{R}^{m \times d}$ *is called the* control point matrix. *Each row in* $P'$ *represents a point in* $\mathbb{R}^d$ *and the entire matrix contains* $m$ *control points. Each control point is associated with a vertex of a cell* $s \in \boldsymbol{D}$, *so the control points lie primal to the vertices of the cells. It holds that* $m \geq n$, *because for cells at the boundary control points must also be associated with vertices of cells of the respective embedding.*

The control points are thus objects in the image space and are meant to define the shape of the created elements (curve, surface, volume). To perform subdivision, i.e., refinement, we need a structure described by initial elements. Since these differ in their concrete form for degrees $g = 2$ and $g = 3$, we also define them separately.

**Definition 2.11.** *Let* $\boldsymbol{D}$ *be a spline domain of type* $t$ *and let* $g = 2$ *be the degree. Let* $d, m, n, k \in \mathbb{N}$ *with* $m \geq n$ *and let* $P' \in \mathbb{R}^{m \times d}$ *be a matrix of control points. Let the vertex with index* $k$ *be a vertex of* $n$ *cells in* $\boldsymbol{D}$. *For this vertex, we define an* initial element *as a quadruple*

$$\boldsymbol{I}_k(\boldsymbol{D}) = (\boldsymbol{G}_k, A_k, S_k, P_k),$$

*which is associated with the vertex of index* $k$ *in* $\boldsymbol{D}$. *The elements of the quadruple are defined as follows:*

- *The graph* $\boldsymbol{G}_k = (\boldsymbol{V}_k, \boldsymbol{E}_k)$ *consists of vertices* $\boldsymbol{V}_k$ *and edges* $\boldsymbol{E}_k$. *It has one vertex for each cell at which the vertex with index* $k$ *lies, so* $|\boldsymbol{V}_k| = n$. *Two vertices are connected by an edge if and only if the two associated cells (at a vertex, edge, or facet depending on* $t$*) are glued together. The graph is thus (like the control points for* $g = 2$*) dual to the vertices of the cells.*
- *The adjacency matrix* $A_k \in \{0, 1\}^{n \times n}$ *comes directly from the graph* $\boldsymbol{G}_k$. *Including it in the quadruple is therefore redundant, but the notation allows direct access to the graph's structure.*
- *The matrix* $P_k \in \mathbb{R}^{n \times d}$ *describes the matrix of control points of the initial element. These control points are those associated with the cells containing the vertex with index* $k$. *We thus get* $n$ *control points. We associate these with the vertices of the graph* $\boldsymbol{G}_k$. *The graph thus provides the control points of the initial element with a structure.*
- *The subdivision matrix* $S_k \in \mathbb{R}^{n \times n}$ *describes how the control points are refined. Its construction is described in Chapter 5. It depends completely on the structure of the cells containing the vertex with index* $k$, *i.e.,*

$$\{s_i \in \boldsymbol{D} \mid \text{the vertex with index } k \text{ is a vertex of } s_i\}.$$

**Definition 2.12.** *Let* $\boldsymbol{D}$ *be a spline domain of type* $t$ *and let* $g = 3$ *be the degree. Let* $d, m, n, k \in \mathbb{N}$ *with* $n \leq m$ *and let* $P' \in \mathbb{R}^{m \times d}$ *be a matrix of control points. Let the vertex with index* $k$ *be a vertex of* $j$ *cells in* $\boldsymbol{D}$ *and let* $n$ *be the number of vertices of these cells without double counting. For this vertex with index* $k$, *we define an* initial element *as a quadruple*

$$\boldsymbol{I}_k(\boldsymbol{D}) = (\boldsymbol{G}_k, A_k, S_k, P_k),$$

*which is associated with the vertex with index* $k$ *in* $\boldsymbol{D}$. *The elements of the quadruple are defined as follows:*

- *The graph* $\boldsymbol{G}_k = (\boldsymbol{V}_k, \boldsymbol{E}_k)$ *consists of vertices* $\boldsymbol{V}_k$ *and edges* $\boldsymbol{E}_k$. *It has one vertex for each vertex of a cell at which the vertex with index* $k$ *lies, so* $|\boldsymbol{V}_k| = n$. *Two vertices are connected by an edge if and only if the corresponding vertices of the cells are connected by an edge. The graph is thus (like the control points for* $g = 3$*) primal to the vertices of the cells.*
- *The adjacency matrix* $A_k \in \{0, 1\}^{n \times n}$ *comes directly from the graph* $\boldsymbol{G}_k$. *Including it in the quadruple is therefore redundant, but the notation allows direct access to the structure of the graph.*
- *The matrix* $P_k \in \mathbb{R}^{n \times d}$ *describes the matrix of control points of the initial element. These control points are those associated with the vertices of the cells containing the vertex with index* $k$. *We thus have* $n$ *control points. We associate these with the vertices of the graph* $\boldsymbol{G}_k$. *The graph thus provides the control points of the initial element with a structure.*

- *The subdivision matrix $S_k \in \mathbb{R}^{n \times n}$ describes how the control points are refined. Its construction is described in Chapter 6. It depends completely on the structure of the cells and their neighbors that contain the vertex with index $k$, i.e.,*

$$\{s_i \in \boldsymbol{D} \mid \text{the vertex with index } k \text{ is a vertex of } s_i \text{ or}$$
$$s_i \text{ shares a vertex with a cell whose vertex is the vertex with index } k\}.$$

We thus associate each vertex in the spline domain with an initial element and obtain a structure of size 2 for $t = 1$, $2 \times 2$ for $t = 2$, and $2 \times 2 \times 2$ for $t = 3$ initial elements per cell. This leads to the following relationship between the concepts:

$$\text{spline domain } \mathbf{D} \quad \leftrightarrow \quad \text{set of initial elements } \mathbf{I} \quad \leftrightarrow \quad \text{matrix of all control points } P',$$

and with respect to a single element:

$$\text{cell } s_i \quad \leftrightarrow \quad 2 \times \cdots \times 2 \text{ initial elements} \quad \leftrightarrow \quad \text{matrix of control points of the respective initial elements.}$$

This relationship is further illustrated in Figures 2.2 and 2.3. The connection, or more precisely the association, between cells and initial elements will be explained in more detail. Before doing so, we consider two examples to clarify the relation between these concepts.

**Example 2.13.** *This example aims to construct a quadratic B-spline curve in $\mathbb{R}^2$. For this, we first set $t = 1$ and $g = 2$. Furthermore, let*

$$\boldsymbol{D} := ([0,1] \times \{1,2\}, \boldsymbol{d}) = (\{s_1, s_2\}, \boldsymbol{d}),$$

*where **d** identifies the point $\{1\} \times \{1\}$ with $\{0\} \times \{2\}$. This gives the following structure:*

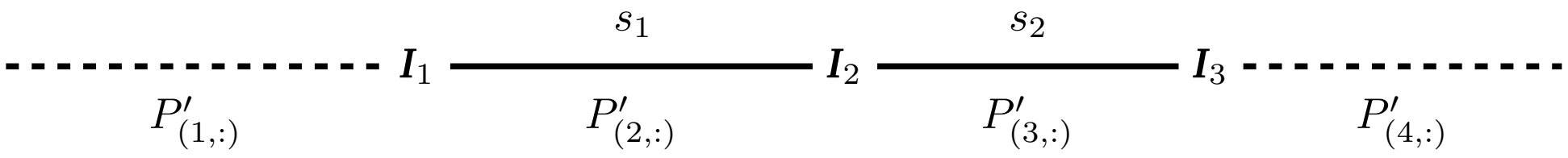

*Due to the embedding, the vertices of $s_1$ and $s_2$ belong to $n = 2$ cells. We define $m = 4$ and set*

$$P' := \begin{bmatrix} 1 & 1 \\ 2 & 2 \\ 4 & 2 \\ 5 & 1 \end{bmatrix}.$$

*Using these control points, we define three initial elements*

$$\boldsymbol{I}_1 := (\boldsymbol{G}_1, A, S, P'_{([1,2],:)}),$$
$$\boldsymbol{I}_2 := (\boldsymbol{G}_2, A, S, P'_{([2,3],:)}),$$
$$\boldsymbol{I}_3 := (\boldsymbol{G}_3, A, S, P'_{([3,4],:)}),$$

*with*

$$A := \begin{bmatrix} 0 & 1 \\ 1 & 0 \end{bmatrix} \quad \text{and} \quad S := \frac{1}{4}\begin{bmatrix} 3 & 1 \\ 1 & 3 \end{bmatrix}.$$

*The adjacency matrix $A$ arises directly from the structure of **D** and the control points identified with the elements of **D**. An illustration of this example can be found in Figure 2.2. It becomes clear here that neighboring initial elements share a control point. It should be noted that the identification of control points with the cells from **D** is arbitrary. It is also possible to add more cells or to change the identification (for example, swapping $P'_{(1,:)}$ with $P'_{(2,:)}$). Thus, the geometry is not created by the coordinates of the control points but by **D**.*

**Example 2.14.** *This example aims to construct a cubic B-spline curve in $\mathbb{R}^2$. For this, we first set $t = 1$ and $g = 3$. Furthermore, let*

$$\boldsymbol{D} := ([0,1] \times \{1,2\}, \boldsymbol{d}),$$

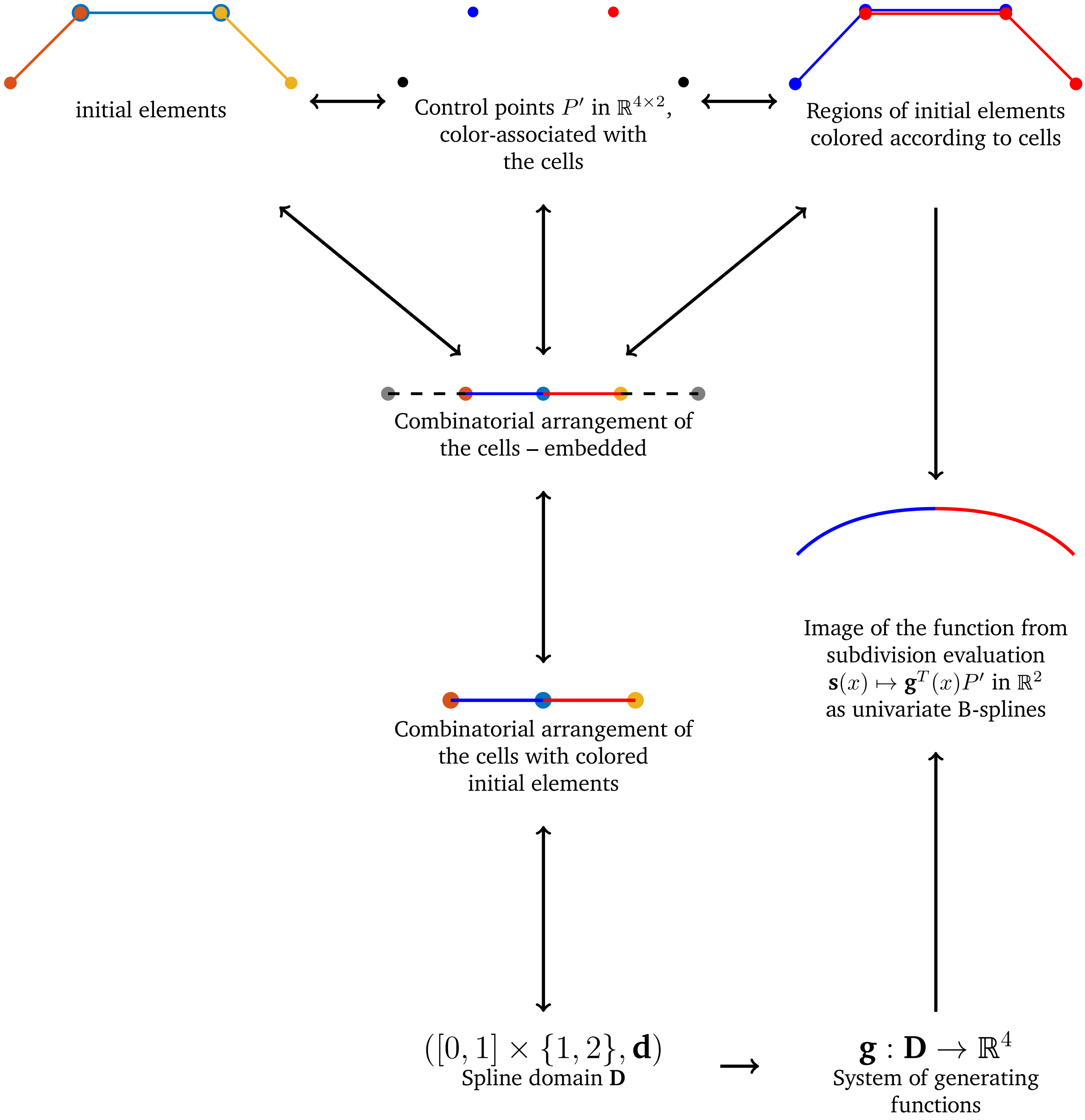

**Figure 2.2:** Illustration of the different concepts for the case $g = 2$ based on Example 2.13.

*where **d** identifies the point* $\{1\} \times \{1\}$ *with* $\{0\} \times \{2\}$. *This results in the following structure:*

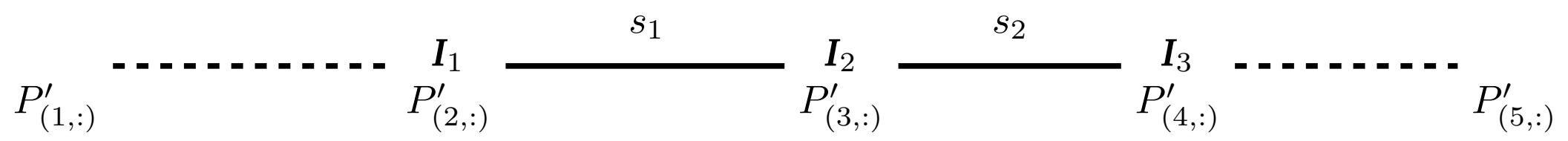

*Due to the embedding, the vertices of* $s_1$ *and* $s_2$ *each belong to* $j = 2$ *cells, which together have* $n = 3$ *distinct vertices without counting duplicates. We define* $m = 5$ *and set*

$$P' := \begin{bmatrix} 1 & 1 \\ 2 & 3 \\ 2 & 1 \\ 4 & 3 \\ 5 & 1 \end{bmatrix}.$$

*On these control points, we define three initial elements*

$$\begin{aligned} \boldsymbol{I}_1 &:= (\boldsymbol{G}_1, A, S, P'_{([1,2,3],:)}), \\ \boldsymbol{I}_2 &:= (\boldsymbol{G}_2, A, S, P'_{([2,3,4],:)}), \\ \boldsymbol{I}_3 &:= (\boldsymbol{G}_3, A, S, P'_{([3,4,5],:)}), \end{aligned}$$

*with*

$$A := \begin{bmatrix} 0 & 1 & 0 \\ 1 & 0 & 1 \\ 0 & 1 & 0 \end{bmatrix} \quad \text{and} \quad S := \frac{1}{8} \begin{bmatrix} 4 & 4 & 0 \\ 1 & 6 & 1 \\ 0 & 4 & 4 \end{bmatrix}.$$

*An illustration of this example is shown in Figure 2.3. In contrast to quadratic B-spline subdivision, it becomes clear that two neighboring initial elements overlap in two control points.*

As already described, the spline domain defines the structure of the initial elements. The graph **G** and therefore also the adjacency matrix $A$ of an initial element depend on the structure of the cells around the initial element. The same applies to the subdivision matrices of the initial elements. Since the control points $P$ are associated with the graph, the choice of control points also depends, as described, on the spline domain. The positions of the control points, that is, their concrete values in $\mathbb{R}^d$, are independent of this and determine the shape of the generated elements (curve, surface, volume).

Different initial elements can share control points. This is illustrated in Example 2.13. For instance, the blue initial element shares a control point with the red initial element. An illustration can be found in Figure 2.2. Similarly, in Example 2.14 and Figure 2.3, it can be seen that for $g = 3$, the overlap region increases to two control points.

For subdivision algorithms of higher type, the combinatorial arrangements or geometries of the initial elements can be correspondingly more complex. For example, if in a spline domain of type $t = 2$ and degree $g = 2$ five squares meet at a vertex of the cells, then the initial element associated with this point must correspondingly represent the geometry of the spline subdivision domain. In this example, this would mean that the initial element must be a pentagon. An illustration is given in Figure 2.4.

The next step is the evaluation of the structure described above into elements (curves, surfaces, volumes). For each combination of degree and type, we first define a regular initial element:

**Definition 2.15.** *Let* $t$ *be the type and* $g$ *the degree of the evaluated elements. Then every initial element*

$$\boldsymbol{I} = (\boldsymbol{G}, A, S, P) \quad \text{with } A := \begin{bmatrix} 0 & 1 \\ 1 & 0 \end{bmatrix} \quad \text{and} \quad S := \frac{1}{4} \begin{bmatrix} 3 & 1 \\ 1 & 3 \end{bmatrix}$$

*is a* regular initial element *for type* $t = 1$ *and degree* $g = 2$. *Furthermore, every initial element*

$$\boldsymbol{I} = (\boldsymbol{G}, A, S, P) \quad \text{with } A := \begin{bmatrix} 0 & 1 & 0 \\ 1 & 0 & 1 \\ 0 & 1 & 0 \end{bmatrix} \quad \text{and} \quad S := \frac{1}{8} \begin{bmatrix} 4 & 4 & 0 \\ 1 & 6 & 1 \\ 0 & 4 & 4 \end{bmatrix}$$

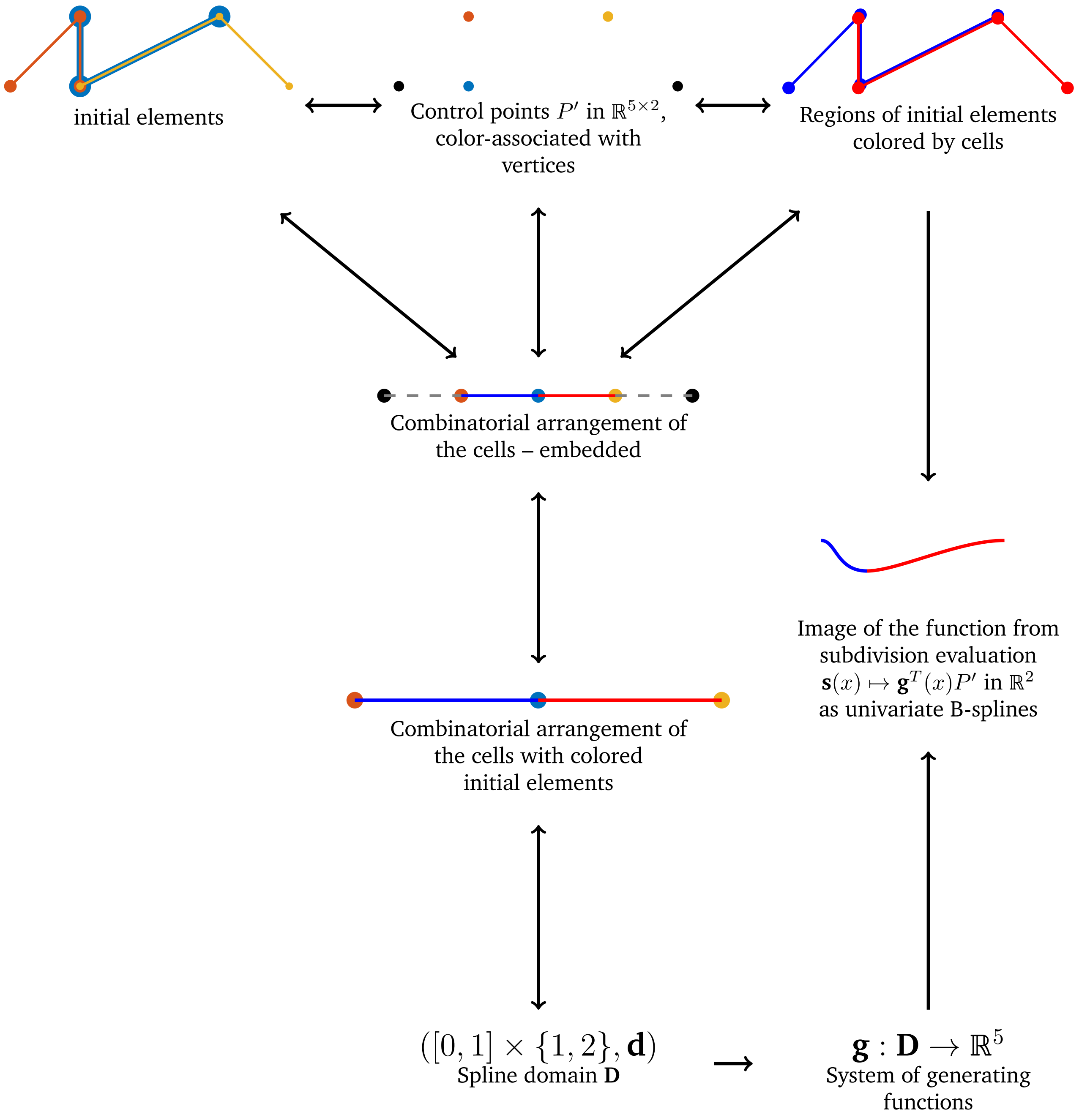

**Figure 2.3:** Illustration of the different concepts for the case $g = 3$ based on Example 2.14.

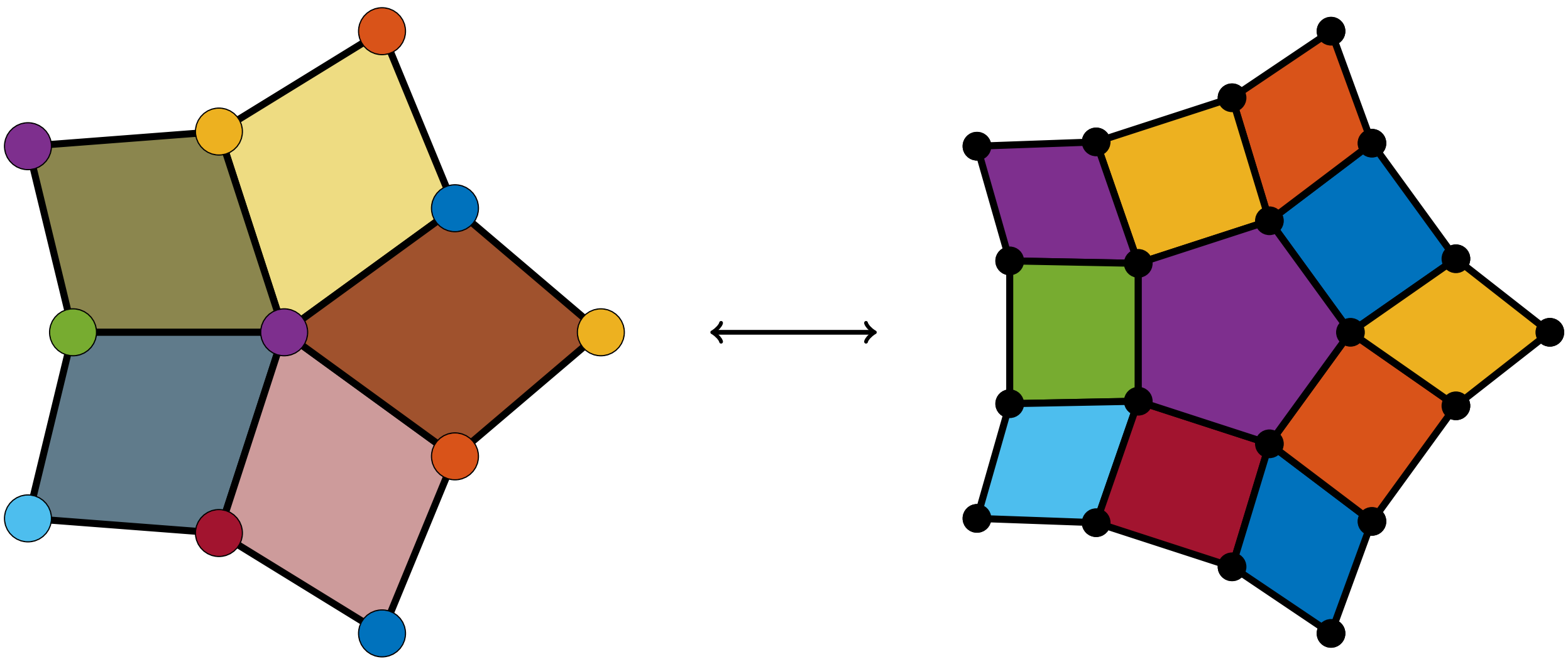

**Figure 2.4:** Example of a spline domain of type $t = 2$ and degree $g = 2$. On the left, five quadrilateral cells are shown, whose vertices are associated with initial elements. An illustration of the correspondingly colored initial elements and the associated control points in black can be seen on the right. Since the central point of the spline domain lies on five squares, the corresponding initial element must be a pentagon.

*is a regular initial element for type $t = 1$ and degree $g = 3$. The regular elements for types 2 and 3 are given by the $t$-fold tensor products of the above initial elements and are explicitly defined in Chapters 5 and 6.*

Whether an initial element is regular or not depends on its structure as defined above, but not on the specific control points. With the following definitions, we specify which initial elements must be regular initial elements.

**Definition 2.16.** *Let $\boldsymbol{D}$ be a spline domain of type $t$. An initial element has a* regular neighborhood *if it is a vertex of*

- *$2$ cells for $t = 1$,*
- *$2 \times 2$ cells for $t = 2$,*
- *$2 \times 2 \times 2$ cells for $t = 3$.*

*Such an arrangement of cells is called a* regular structure.

We use this for the next definition:

**Definition 2.17.** *We define the term* regular initial elements *separately for degrees $g = 2$ and $g = 3$:*

1. *For $g = 2$, all initial elements with a regular neighborhood are by definition* regular initial elements.
2. *Let $g = 3$. Let $\boldsymbol{I}_k$ be an initial element with a regular neighborhood, i.e., lying in a regular structure of cells. If all initial elements of these cells also have regular neighborhoods, then $\boldsymbol{I}_k$ is by definition a* regular initial element.

*All other initial elements may also be regular if they satisfy Definition 2.15, but unlike the above, they are not required to be regular.*

This automatically yields the following remark:

**Remark 2.18.** *For $t = 1$, it follows from the above definition that all initial elements are regular initial elements, since the spline domain always forms a chain of cells.*

For $g = 2$, this gives a simple description of regular initial elements. For $g = 3$, the description is somewhat more involved. Intuitively, this means only that the „regular“ region around each element must be larger by one element. We already refer to Example 2.26, where regular initial elements are illustrated.

The above Definition 2.15 of regular initial elements essentially anticipates Chapters 5 and 6. The adjacency matrix and the graph of an initial element arise directly from the structure of the cells containing the initial element. Hence, these two objects are uniquely determined for initial elements with regular neighborhoods as specified in the above definition. The value of this definition lies in fixing the subdivision matrices for certain constellations already at this stage. This is done here to enable an initial evaluation and to motivate subdivision in the next section. For the evaluation, we first describe the generating functions in the next definition:

**Definition 2.19.** *Let $\boldsymbol{D}$ be a spline domain related to a matrix of control points $P' \in \mathbb{R}^{m\times d}$. Then*

$$\boldsymbol{g} : \boldsymbol{D} \to \mathbb{R}^m, \quad \boldsymbol{g}(x) \mapsto \begin{bmatrix} \boldsymbol{g}_1(x) \\ \vdots \\ \boldsymbol{g}_m(x) \end{bmatrix}, \quad \textit{with} \quad \boldsymbol{g}_i : \boldsymbol{D} \to \mathbb{R} \quad \textit{and} \quad \sum_{i=1}^{m} \boldsymbol{g}_i(x) = 1 \quad \textit{for all} \quad x \in \boldsymbol{D},$$

*is called a* system of generating functions. *Here, $\boldsymbol{g}_i$ is the generating function associated to the control point $P'_{(i,:)}$. Furthermore,*

$$\boldsymbol{s} : \boldsymbol{D} \to \mathbb{R}^d, \quad \boldsymbol{s}(x) \mapsto \boldsymbol{g}^T(x)P'$$

*is called the* subdivision evaluation function.

It remains to specify how the generating functions and thus the subdivision evaluation function actually look. We formalize this in the next definition:

**Definition 2.20.** *Let $\boldsymbol{D}$ be a spline domain of type $t$ and $g$ the degree. A cell*

$$s_i = [0,1]^t \times \{i\} \in \boldsymbol{D},$$

*whose vertices are associated with regular initial elements, is called a* regular cell *and is evaluated as a $t$-variate B-spline of degree $g$. The corresponding generating functions $\boldsymbol{g}$ on the cell are uniform tensor-product B-spline basis functions of degree $g$, and the points $P_{s_i}$ of the initial elements of the cell $s_i$ (without double counting), together with their generating functions $\boldsymbol{g}_{s_i}$, form the control points of the B-spline arranged in a $(g+1) \times \cdots \times (g+1)$ structure by the initial elements.*

*Thus, the evaluation on the cell is fully determined by the above points and their generating function by*

$$\boldsymbol{s}\left([0,1]^t \times \{i\}\right) := \boldsymbol{g}_{s_i}^T\left([0,1]^t \times \{i\}\right) P_{s_i}.$$

*In particular, all other generating functions are identically zero on $[0,1]^t \times \{i\}$.*

This evaluation of Examples 2.13 and 2.14 is illustrated in the corresponding Figures 2.2 and 2.3. Since all initial elements are regular initial elements by definition in these examples, both cells consist exclusively of regular initial elements. Consequently, by Definition 2.20, these are evaluated as univariate uniform B-spline curves.

This raises the question of how to proceed with the other cells. For this purpose, we use the concept of subdivision, which is explained in the next section.

## 2.2 Subdivision

In the literal sense, subdivision means refinement or splitting that occurs on all levels of the structure. We begin by describing the subdivision of the spline domain:

**Definition 2.21.** *Let $\boldsymbol{D} = \boldsymbol{D}^{(1)} = ([0,1]^t \times \{1,\ldots,k\}, \boldsymbol{d}_1)$ be a spline domain of type $t$. Furthermore, let*

$$\boldsymbol{L} := \left[0, \tfrac{1}{2}\right] \quad \textit{and} \quad \boldsymbol{R} := \left[\tfrac{1}{2}, 1\right].$$

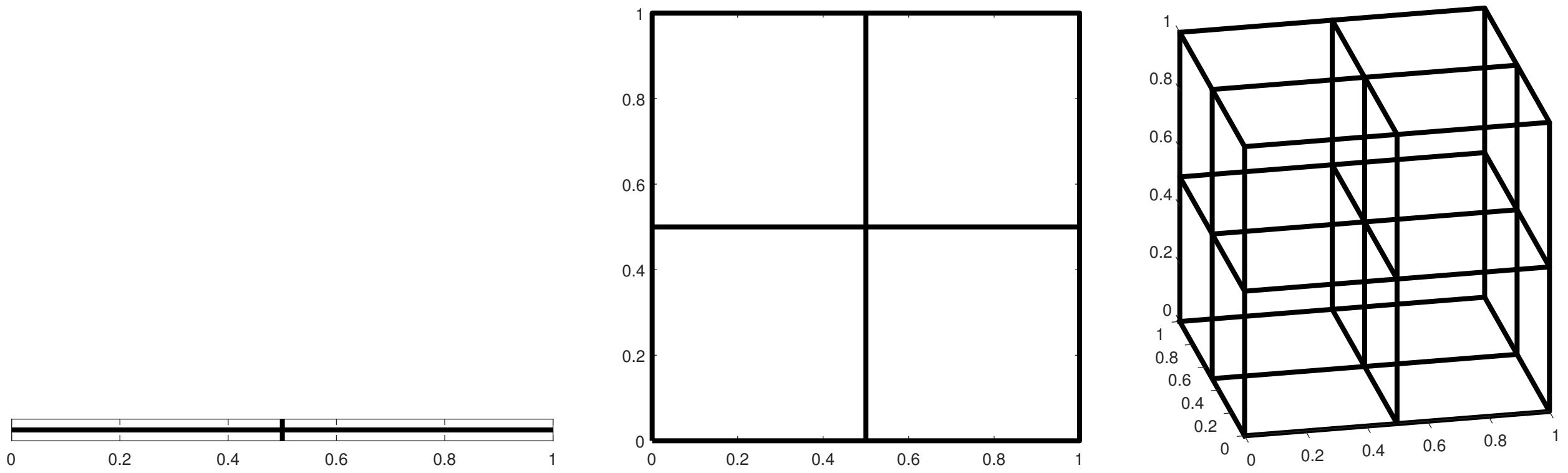

**Figure 2.5:** Illustration of the subdivision of a cell for $t = 1$ (left), $t = 2$ (middle) and $t = 3$ (right), each without the index of the cell.

*Then*

1. $\boldsymbol{D}^{(2)} := (\{\boldsymbol{L}, \boldsymbol{R}\} \times \{1, \ldots, k\}, \boldsymbol{d}_2)$ *for* $t = 1$,
2. $\boldsymbol{D}^{(2)} := (\{\boldsymbol{L} \times \boldsymbol{L}, \quad \boldsymbol{L} \times \boldsymbol{R}, \quad \boldsymbol{R} \times \boldsymbol{L}, \quad \boldsymbol{R} \times \boldsymbol{R}\} \times \{1, \ldots, k\}, \boldsymbol{d}_2)$ *for* $t = 2$, *and*
3. $\boldsymbol{D}^{(2)} := \big(\{\boldsymbol{L} \times \boldsymbol{L} \times \boldsymbol{L}, \quad \boldsymbol{L} \times \boldsymbol{L} \times \boldsymbol{R}, \quad \boldsymbol{L} \times \boldsymbol{R} \times \boldsymbol{L}, \quad \boldsymbol{L} \times \boldsymbol{R} \times \boldsymbol{R}, \quad \boldsymbol{R} \times \boldsymbol{L} \times \boldsymbol{L}, \quad \boldsymbol{R} \times \boldsymbol{L} \times \boldsymbol{R}, \quad \boldsymbol{R} \times \boldsymbol{R} \times \boldsymbol{L}, \quad \boldsymbol{R} \times \boldsymbol{R} \times \boldsymbol{R}\big\} \times \{1, \ldots, k\}, \boldsymbol{d}_2\big)$ *for* $t = 3$

*is called the* subdivision of the spline domain. *Thus,* $|\boldsymbol{D}^{(2)}| = 2^t \cdot k$. *The* $\boldsymbol{D}$*-equivalence relation* $\boldsymbol{d}_2$ *is induced by* $\boldsymbol{d}_1$. *Outer parts of cells are identified analogously to* $\boldsymbol{d}_1$*, and inner parts of cells are identified where they overlap.*

An illustration of the subdivision of a cell can be found in Figure 2.5. We define a reparametrization of the refined spline domain as follows:

**Definition 2.22.** *Let* $\boldsymbol{D} = \boldsymbol{D}^{(1)} = [0,1]^t \times \{1, \ldots, k\}$ *be a spline domain of type t. Further, let* $\boldsymbol{D}^{(2)}$ *be the subdivision of the spline domain* $\boldsymbol{D}$ *and let*

$$\boldsymbol{L} := \left[0, \tfrac{1}{2}\right], \quad \textit{and} \quad \boldsymbol{R} := \left[\tfrac{1}{2}, 1\right].$$

*We define the reparametrizations*

$$\boldsymbol{r_L} : \left[0, \tfrac{1}{2}\right] \to [0,1], \quad \boldsymbol{r_L}(a) := 2a,$$

*and*

$$\boldsymbol{r_R} : \left[\tfrac{1}{2}, 1\right] \to [0,1], \quad \boldsymbol{r_R}(a) := 2a - 1,$$

*and the index shifts*

$$\boldsymbol{i}^1 : \{\boldsymbol{L}, \boldsymbol{R}\} \to \{0,1\}, \qquad \boldsymbol{i}^1(\boldsymbol{M}) = \begin{cases} 0 & \textit{if } \boldsymbol{M} = \boldsymbol{L}, \\ 1 & \textit{if } \boldsymbol{M} = \boldsymbol{R}, \end{cases}$$

$$\boldsymbol{i}^2 : \{\boldsymbol{L}, \boldsymbol{R}\} \times \{\boldsymbol{L}, \boldsymbol{R}\} \to \{0, \ldots, 3\}, \qquad \boldsymbol{i}^2(\boldsymbol{M}_1, \boldsymbol{M}_2) = 2 \cdot \boldsymbol{i}^1(\boldsymbol{M}_1) + \boldsymbol{i}^1(\boldsymbol{M}_2),$$

$$\boldsymbol{i}^3 : \{\boldsymbol{L}, \boldsymbol{R}\} \times \{\boldsymbol{L}, \boldsymbol{R}\} \times \{\boldsymbol{L}, \boldsymbol{R}\} \to \{0, \ldots, 7\}, \qquad \boldsymbol{i}^3(\boldsymbol{M}_1, \boldsymbol{M}_2, \boldsymbol{M}_3) = 4 \cdot \boldsymbol{i}^1(\boldsymbol{M}_1) + 2 \cdot \boldsymbol{i}^1(\boldsymbol{M}_2) + \boldsymbol{i}^1(\boldsymbol{M}_3).$$

*Thus, we call*

$$\boldsymbol{r} : \boldsymbol{D}^{(2)} \to \left([0,1]^t \times \{1, \ldots, 2^t \cdot k\}, \tilde{\boldsymbol{d}}_2\right),$$

$$\boldsymbol{r}(\boldsymbol{M}_1 \times \cdots \times \boldsymbol{M}_t \times \{i\}, \boldsymbol{d}_2) \mapsto \left(\boldsymbol{r}_{\boldsymbol{M}_1}(\boldsymbol{M}_1) \times \cdots \times \boldsymbol{r}_{\boldsymbol{M}_t}(\boldsymbol{M}_t) \times \{2^t \cdot (i-1) + \boldsymbol{i}^t(\boldsymbol{M}_1, \ldots, \boldsymbol{M}_t) + 1\}, \tilde{\boldsymbol{d}}_2\right)$$

*the reparametrization mapping and define*

$$\tilde{\boldsymbol{D}}^{(2)} := \boldsymbol{r}\left(\boldsymbol{D}^{(2)}\right)$$

*as the reparametrization of* $\boldsymbol{D}^{(2)}$. *The* $\boldsymbol{D}$*-equivalence relation* $\tilde{\boldsymbol{d}}_2$ *identifies the images of already identified areas in* $\boldsymbol{D}^{(2)}$.

Intuitively, the reparametrization mapping transforms a refined spline domain into a form that itself is again a spline domain. Each refined element is mapped onto $[0,1]^t$ and the numbering is adjusted accordingly by the function $\mathbf{i}^t$. Thus, the reparametrization $\tilde{\mathbf{D}}^{(2)}$ satisfies the conditions of Definition 2.4 and is therefore itself a spline domain.

Thus, in order to transfer the subdivision to the image domain, the refinement of the control points is constructed as follows:

**Definition 2.23.** *Let $\boldsymbol{D}$ be a spline domain, $\boldsymbol{I} = \{\boldsymbol{I}_1, \dots, \boldsymbol{I}_k\}$ with $k \in \mathbb{N}$ the set of associated initial elements, and $P'$ the matrix of control points. These individual initial elements are denoted by*

$$\boldsymbol{I}_i = (\boldsymbol{G}_i, A_i, S_i, P_i).$$

*We call the matrix of all*

$$S_i P_i$$

*(without double counting) the* refined control points *and denote them by $P'^{(2)}$. The refined control points are assigned according to the refined spline domain. Concretely, this means:*

*Let $i$ be the index of a corner of cells in $\boldsymbol{D}$, which is associated with $\boldsymbol{I}_i$. Let $s_n$ be a cell in $\boldsymbol{D}$ that has the corner with index $i$ as a vertex.*

*Let $j$ be the index of a corner of cells in $\boldsymbol{D}^{(2)}$ that emerged by subdivision from the corner with index $i$, and let $s_m^{(2)}$ be the cell in $\boldsymbol{D}^{(2)}$ that originated from $s_n$ and has the corner with index $j$ as a vertex.*

- *For $g = 2$ and $P'_{(k,:)}$ the control point associated with $s_n$, let $P'^{(2)}_{(m,:)}$ be the control point that emerged from $P'_{(k,:)}$ through the refinement with $S_i$ of the initial element $\boldsymbol{I}_i$. Then we identify $P'^{(2)}_{(m,:)}$ with $s_m^{(2)}$.*
- *For $g = 3$ and $P'_{(1,:)}, \dots, P'_{(2^t,:)}$ the control points associated with the corners of $s_n$, let $P'^{(2)}_{(1,:)}, \dots, P'^{(2)}_{(2^t,:)}$ be the control points that emerged from $P'_{(1,:)}, \dots, P'_{(2^t,:)}$ through the refinement with $S_i$ of the initial element $\boldsymbol{I}_i$. Then we identify $P'^{(2)}_{(1,:)}, \dots, P'^{(2)}_{(2^t,:)}$ with the corners of $s_m^{(2)}$.*

The notation in the above definition is somewhat cumbersome. Pictorially speaking, the refinement of control points produces exactly the structure needed to associate the refined control points with the cells (for $g = 2$) or with the corners of the cells (for $g = 3$). An illustration can be found in Figure 2.6, where the structures from Examples 2.13 and 2.14 are refined.

It can be seen from the above definition that the refinement of the control points is determined by the initial elements. Or in other words: all control points generated by refinement arise from the initial elements. The structure is determined by the structure of the initial elements (the graph **G**), and the refinement is given by the subdivision matrix $S$, whose construction we still need to address. We add the following remark here:

**Remark 2.24.** *If initial elements overlap such that two initial elements produce the same refined control point, then the refinement rules for this control point must be identical to ensure that the refinement is well-defined. This aspect must be discussed in the concrete subdivision algorithms.*

At this point, it would be possible to also define the refined initial elements explicitly. However, initial elements are defined via the spline domain. Hence, the new initial elements arise directly by applying Definition 2.12 to $\tilde{\mathbf{D}}^{(2)}$. Finally, the question of evaluation arises, i.e., the refinement of the generating functions **g**. For this, we consider the following definition:

**Definition 2.25.** *Let $\boldsymbol{D}$ be a spline domain of type $t$, associated with a control point matrix $P' \in \mathbb{R}^{m \times d}$. Let **g** be a system of generating functions corresponding to $P'$.*
*Let $\boldsymbol{D}^{(2)}$ and $P'^{(2)}$ be the refinements of $\boldsymbol{D}$ and $P'$, respectively, with $|\boldsymbol{D}^{(2)}| = k$ and $m$ the number of rows, i.e., control points, of $P'^{(2)}$. Furthermore, let $\tilde{\boldsymbol{D}}^{(2)}$ be the reparametrization of $\boldsymbol{D}^{(2)}$. Then, for $(\tilde{\boldsymbol{D}}^{(2)}, P'^{(2)})$, using Definition 2.19, we define generating functions*

$$\mathbf{g}^{(2)} : \tilde{\boldsymbol{D}}^{(2)} \to \mathbb{R}^m, \quad \mathbf{g}^{(2)}(x) \mapsto \begin{bmatrix} \boldsymbol{g}_1^{(2)}(x) \\ \vdots \\ \boldsymbol{g}_m^{(2)}(x) \end{bmatrix}, \quad \textit{with} \quad \mathbf{g}_i^{(2)} : \tilde{\boldsymbol{D}}^{(2)} \to \mathbb{R} \quad \textit{and} \quad \sum_{i=1}^{m} \boldsymbol{g}_i^{(2)}(x) = 1 \quad \textit{for all} \quad x \in \boldsymbol{D}^{(2)}.$$

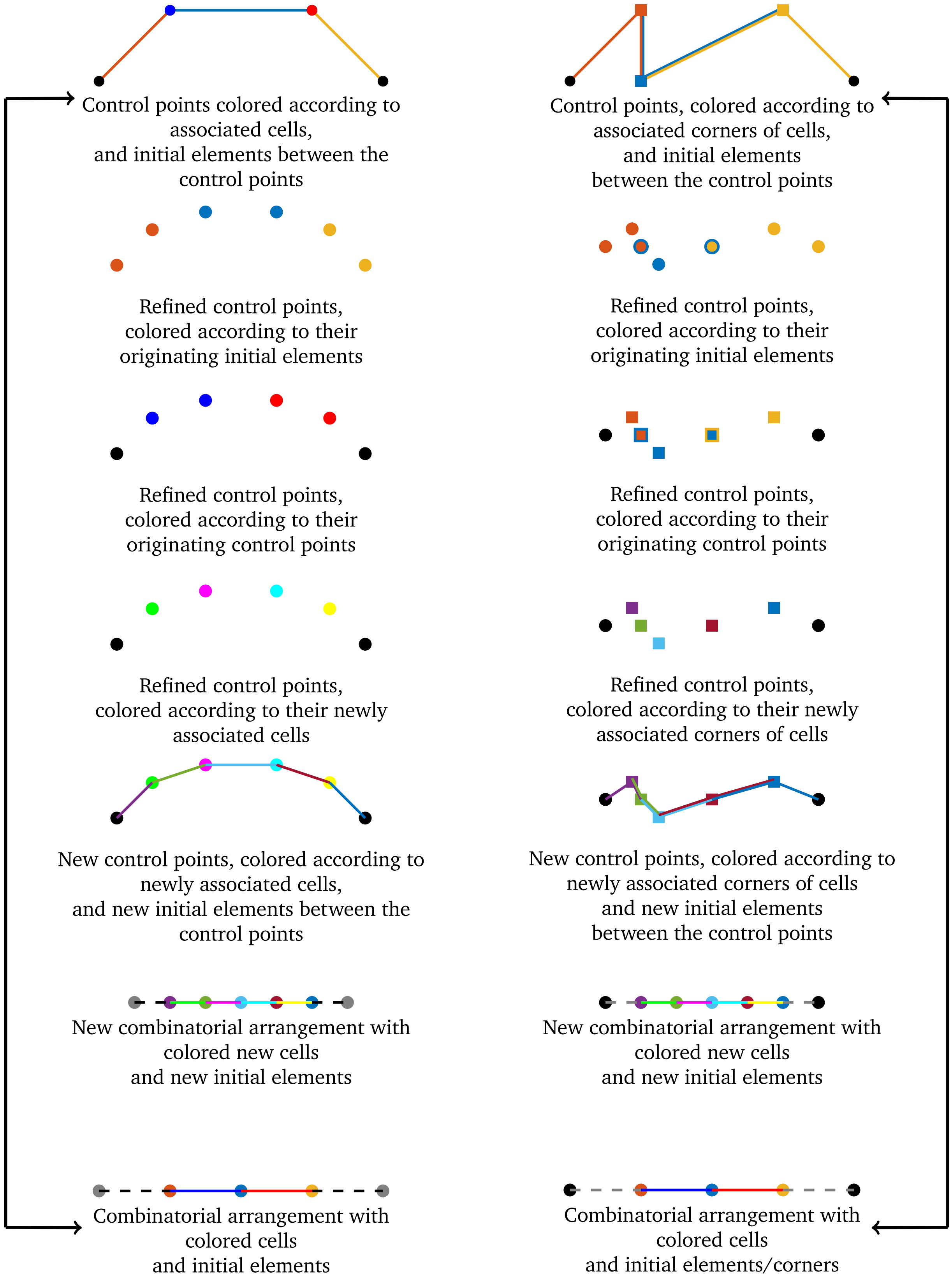

**Figure 2.6:** Refinement of cells and control points for Examples 2.13 with $g = 2$ (left column) and 2.14 with $g = 3$ (right column).

*With these generating functions, we recursively define*

$$\boldsymbol{s} : \boldsymbol{D} \to \mathbb{R}^d, \quad \boldsymbol{s}(x) \mapsto \left(\boldsymbol{g}^{(2)}(\boldsymbol{r}(x))\right)^T P'^{(2)} =: \boldsymbol{g}(x)^T P'.$$

*According to Definition 2.20, regular regions are evaluated as tensor product B-splines.*

Several aspects need to be discussed here. The first is well-definedness. As described in Definition 2.20, regular cells are evaluated to tensor product B-splines. For the above definition to be well-defined, the refined evaluation on regular cells must yield the same result as the coarse evaluation. In other words, evaluation of regular regions must be reproduced by refinement. The subdivision matrices in Definition 2.15 are chosen or defined exactly to satisfy this property.

The next aspect is refinability. The refined objects have exactly the form required as input for subdivision. Therefore, refinement can be applied again, repeatedly, as often as desired.

Another aspect is the benefit of refinement. This can be recognized on various levels. Regarding the generating functions, subdivision creates a finer representation of the same object. For simulation, this provides an object whose refinement is structurally embedded. It thus allows to easily produce the required resolution for simulation applications without having to worry about creating new geometries. This is a decisive advantage compared to simulation on triangle or tetrahedral meshes.

The main aspect to discuss, however, is the benefit of refinement regarding the evaluation or evaluability of irregular regions. For $t = 1$, there is none, since all areas are regular and can be evaluated directly. For $t = 2$ and $t = 3$, we consider the following general example:

**Example 2.26.** *Consider a cell of a spline domain* $\boldsymbol{D}$ *for* $t = 2$, *i.e.,* $[0, 1]^2 \times \{i\}$. *For this example, assume that nothing is known about the other cells in* $\boldsymbol{D}$. *In particular, it is unknown whether the initial elements associated with the vertices have a regular neighborhood. This cell is refined multiple times.*

*We use a square marker if the initial element has a regular neighborhood, and a round marker if this is not the case or is unknown. Furthermore, initial elements of the refined structure are colored yellow if it is unknown whether they are regular initial elements, and green if they definitely are regular initial elements. The same coloring is used for the cells, where green means that a cell has four regular initial elements as vertices and is therefore evaluable as a bivariate B-spline. Yellow means this is open or not the case. The result for* $g = 2$ *is shown in Figure 2.7 and for* $g = 3$ *in Figure 2.8.*

*Analogously, for* $t = 3$ *and a cell* $[0, 1]^3 \times \{i\}$, *the corresponding illustrations and results can be found in Figures 2.9 for* $g = 2$ *and 2.10 for* $g = 3$.

This example clearly shows the idea and benefit of subdivision. Regardless of the cell structure, the refinement process enables the evaluation of cells. With each refinement, the evaluable region around irregularities grows.

For $t = 2$, one can see that the evaluable region approaches the vertices of the cells. These vertices are the critical areas for $t = 2$. If the initial elements at the vertices are regular, then the yellow region around the vertex turns green. If they are not regular, refinement is needed to gradually describe the region around the initial element. We formalize this regularity in the next definition:

**Definition 2.27.** *Let* $t = 2$ *and let* $\boldsymbol{D}$ *be a spline domain. If a vertex of a cell lies on a regular structure of cells, we call this a* regular vertex. *Otherwise, we call it an* irregular vertex.

The regularity of a vertex is of course directly related to the regularity of the associated initial element. For $g = 2$ this is a one-to-one relation; for $g = 3$, the surrounding neighborhood must also be considered for the regularity of the initial element.

The truly crucial regions to consider in the context of subdivision are the irregular vertices. Section 2.3 treats this in detail. However, we can already note the following:

**Remark 2.28.** *For* $t = 2$, *it can be observed that after one refinement step, the cells have at least three regular vertices. Thus, after one refinement step, without loss of generality, each cell contains at most one irregular vertex.*

For $t = 3$, a similar but somewhat more complex picture emerges. Here, the critical regions occur both at vertices and edges of the cells. Analogous to the previous case, regularity for vertices occurs when they lie on a regular structure of cells, and regularity for edges occurs when $2 \times 2$ cells meet at an edge. We introduce the following definition:

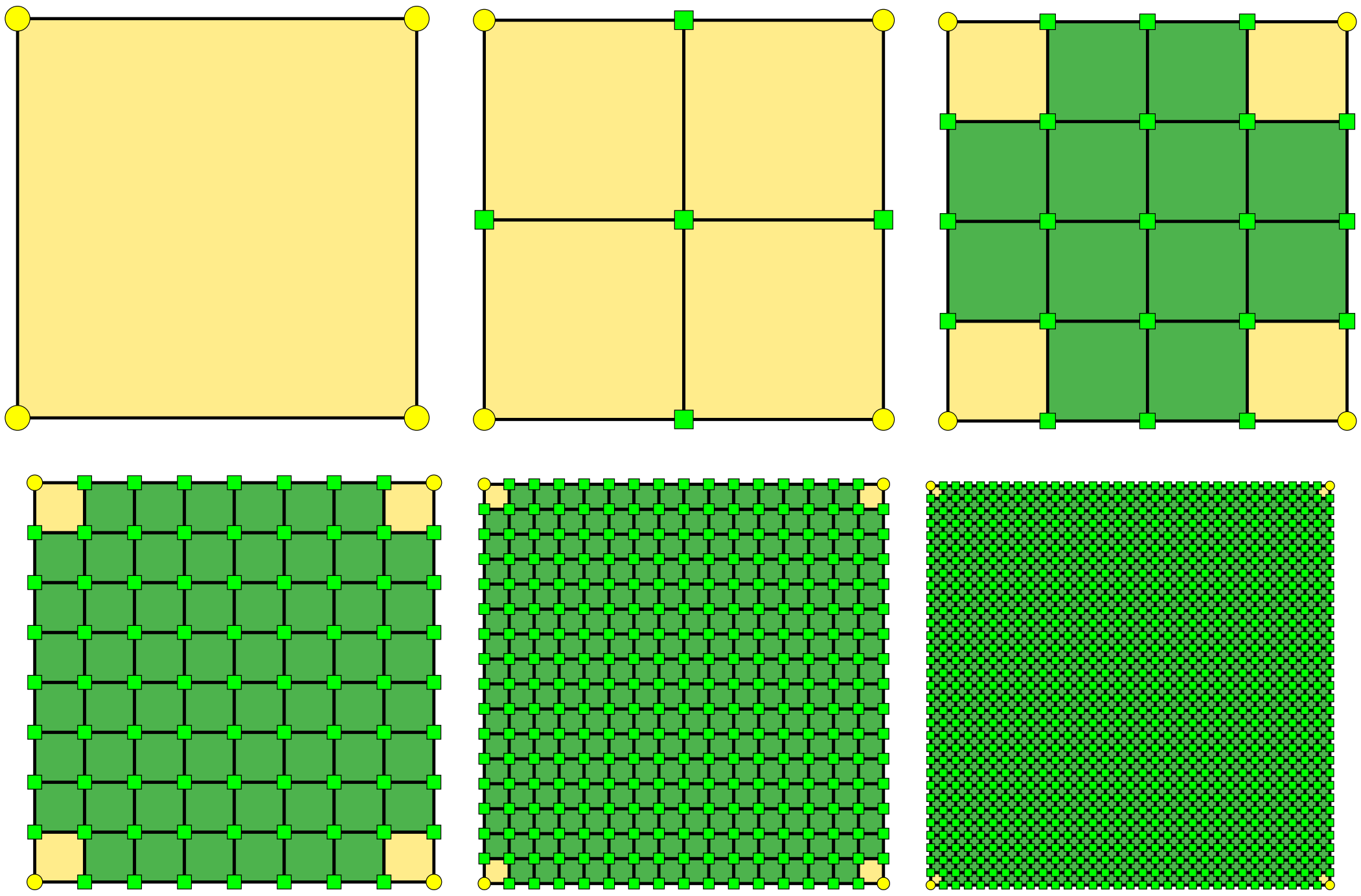

**Figure 2.7:** Initial elements with regular neighborhood (square markers) and still open neighborhood (round markers) of a cell in a spline domain with $g = 2$, uniformly refined five times. Regular initial elements and cells are shown in green; those with unknown regularity are shown in yellow.

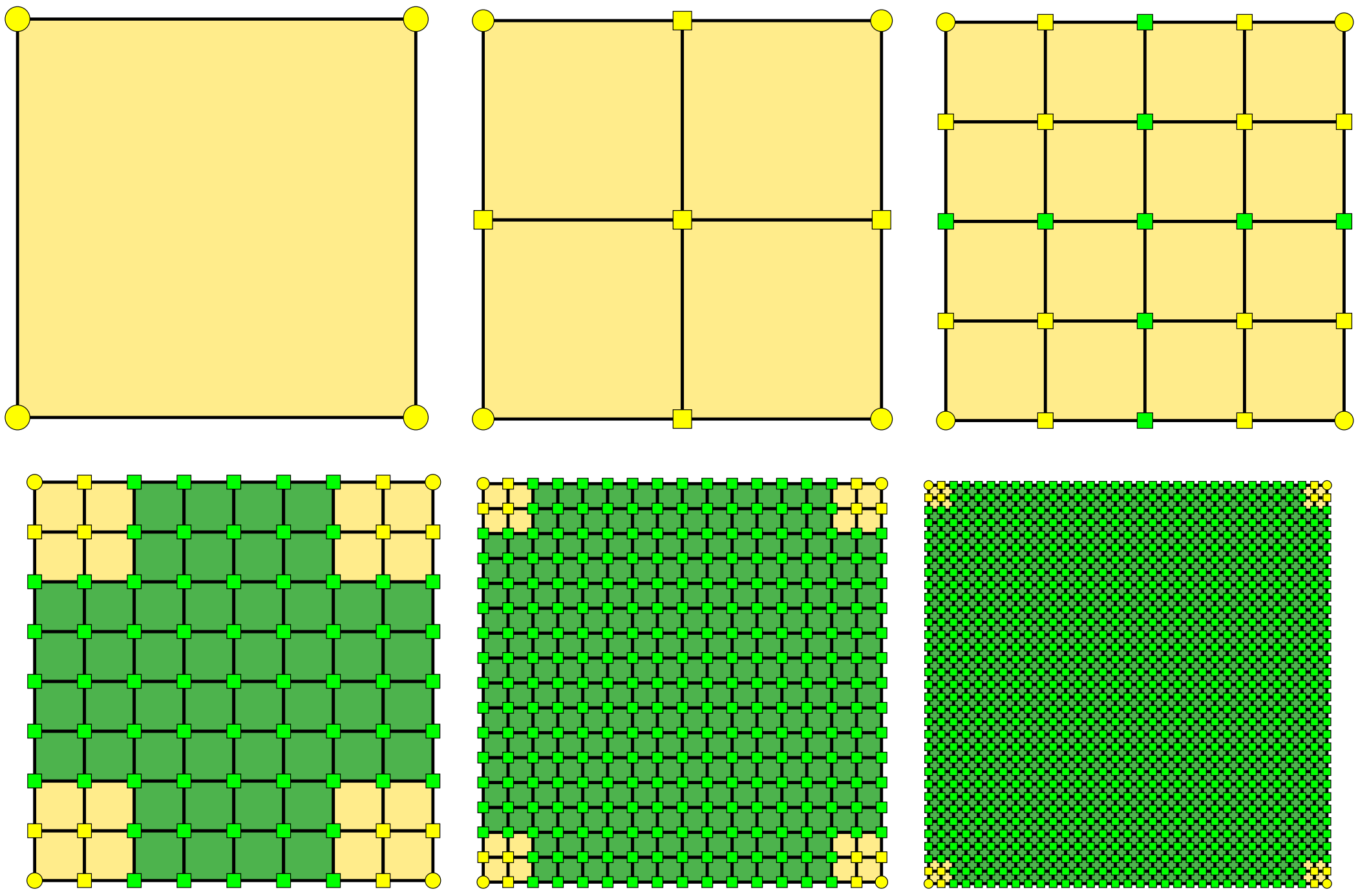

**Figure 2.8:** Analogous illustration to Figure 2.7 for $g = 3$.

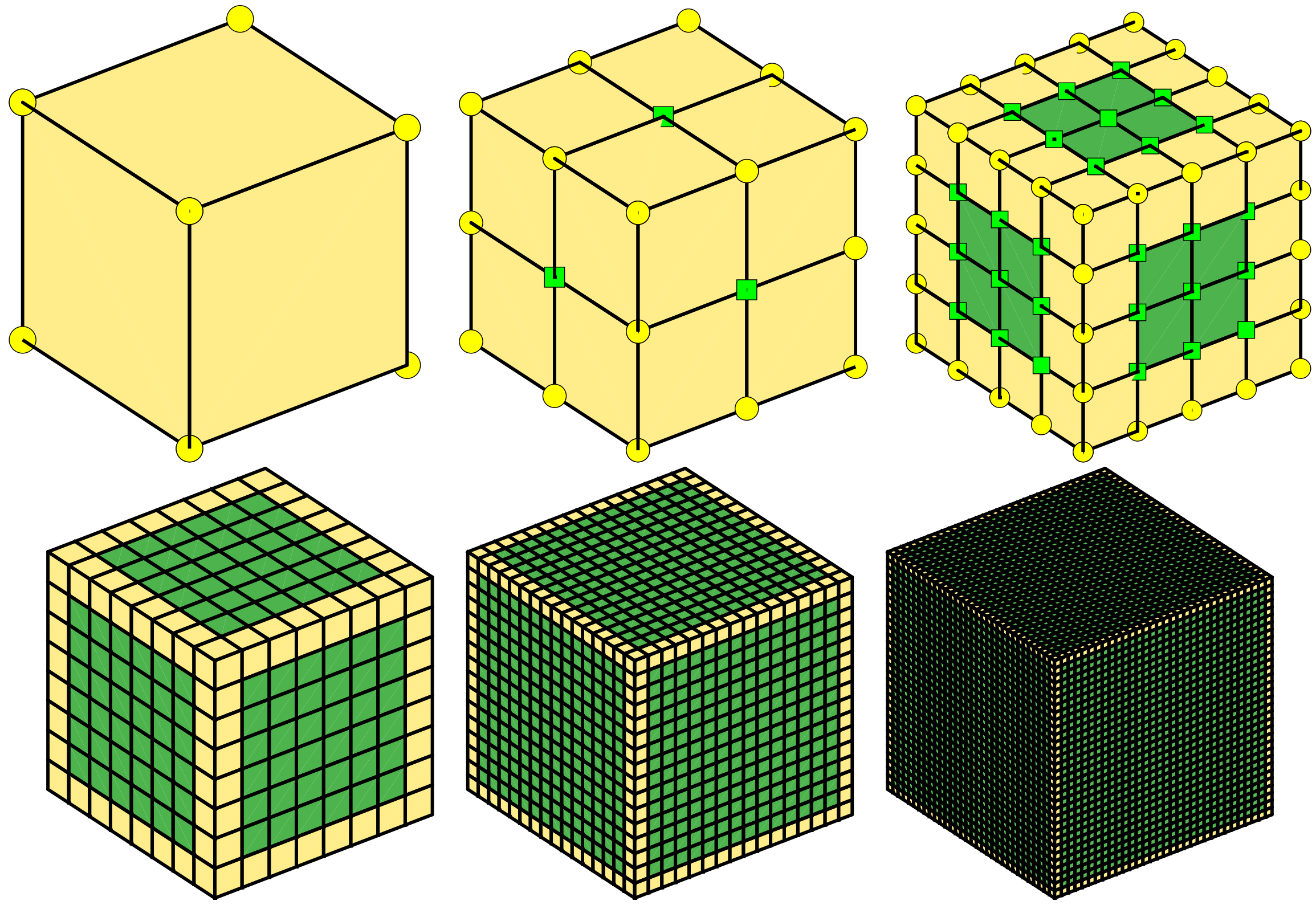

**Figure 2.9:** Analogous illustration to Figure 2.7 for $t = 3$ and $g = 2$.

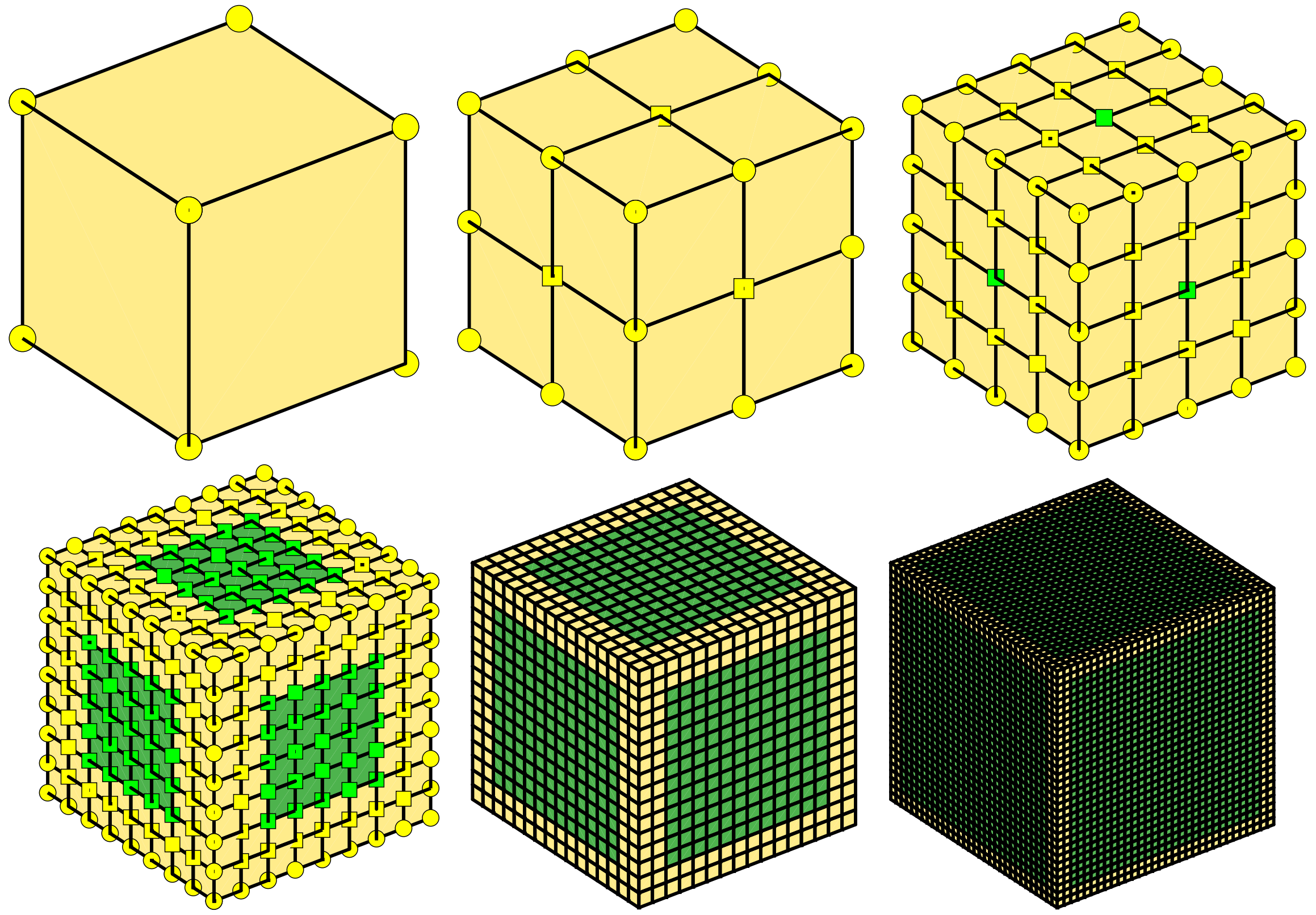

**Figure 2.10:** Analogous illustration to Figure 2.7 for $t = 3$ and $g = 3$.

**Definition 2.29.** *Let $t = 3$ and **D** be a spline domain. If a vertex of a cell lies on a regular structure of cells, we call it a* regular vertex. *If it lies on $n \times 2$ cells (with $n \in \mathbb{N}$ and $n \geq 3$), we call it a* semi-irregular vertex. *Otherwise, it is called an* irregular vertex.

*If an edge of a cell lies on $2 \times 2$ cells, we call it a* regular edge. *Otherwise, it is called an* irregular edge.

The following remark follows analogously to the $t = 2$ case:

**Remark 2.30.** *Let $t = 3$. From the above general example, it can be seen that after one refinement step, cells contain at most one irregular vertex and at most three semi-irregular vertices. All other vertices are regular by the structure. Furthermore, each cell contains at most three irregular edges; all other edges are regular. Hence, analogous to the $t = 2$ case, each cell can be assumed, without loss of generality, to contain at most one irregular vertex.*

Even in the case $t = 3$, it is clear that the evaluable region grows with each refinement. Moreover, for both $t = 2$ and $t = 3$, the non-evaluable region for $g = 3$ is exactly one refinement step larger than for $g = 2$. In other words, one more refinement step is needed for $g = 3$.

This leads to the question of evaluability in the limit, both for irregular edges and irregular vertices. We refer to the quality criteria Q1 and Q3 in Section 2.4. Also, the question of a direct and fast evaluation in the critical regions arises. Here, the work of Stam [Sta98] for the case $t = 2$ is noteworthy.

Through the above discussion, we have seen that subdivision allows cells to be successively evaluated as B-spline elements. For irregular regions, the cells must be refined further and further. In the next section, we focus on this refinement process and explain its asymptotic behavior based on invariant structures.

## 2.3 Invariant Structures

As already described, the evaluable regions of the spline domain depend on the structure of the cells. To describe the initially non-evaluable regions, it is helpful to consider invariant structures in the refinement process. This is discussed in [PR08] for the case $t = 2$. We extend these ideas here to the case $t = 3$. We begin with the smallest invariant elements, the initial elements.

Initial elements are identified with the vertices of cells. In the refinement of the cells, vertices are both reproduced and new vertices are introduced. For a reproduced vertex, the immediate neighborhood of the cells around that vertex does not change. In other words, the structure around a reproduced vertex is also reproduced by the refinement. This leads to the following lemma for initial elements:

**Lemma 2.31.** *Let **D** be a spline domain, $\boldsymbol{I} = \{\boldsymbol{I}_1, \ldots, \boldsymbol{I}_k\}$ with $k \in \mathbb{N}$ the set of associated initial elements, and $P'$ the matrix of control points. Let $\boldsymbol{D}^{(2)}$ and $P'^{(2)}$ be the refinements of **D** and $P'$. Let $k$ be the index of a vertex of a cell and*

$$\boldsymbol{I}_k = (\boldsymbol{G}_k, A_k, S_k, P_k)$$

*the associated initial element. For the initial element*

$$\boldsymbol{I}_k^{(2)} = \left(\boldsymbol{G}_k^{(2)}, A_k^{(2)}, S_k^{(2)}, P_k^{(2)}\right)$$

*of the vertex reproduced from the vertex with index $k$ it holds that*

$$\boldsymbol{G}_k^{(2)} = \boldsymbol{G}_k, \quad A_k^{(2)} = A_k, \quad \text{and} \quad P_k^{(2)} = S_k P_k.$$

*Furthermore, for $g = 2$ we have*

$$S_k^{(2)} = S_k.$$

*For $g = 3$, this also holds under the assumption that the vertex $k$ was already reproduced by a previous refinement.*

*Proof.* The equality of the graph structures follows directly from the reproduction of the surrounding structure of the initial elements. The equation for the control points follows directly from the definition of the refinement of control points given in Definition 2.23, which is defined precisely this way.

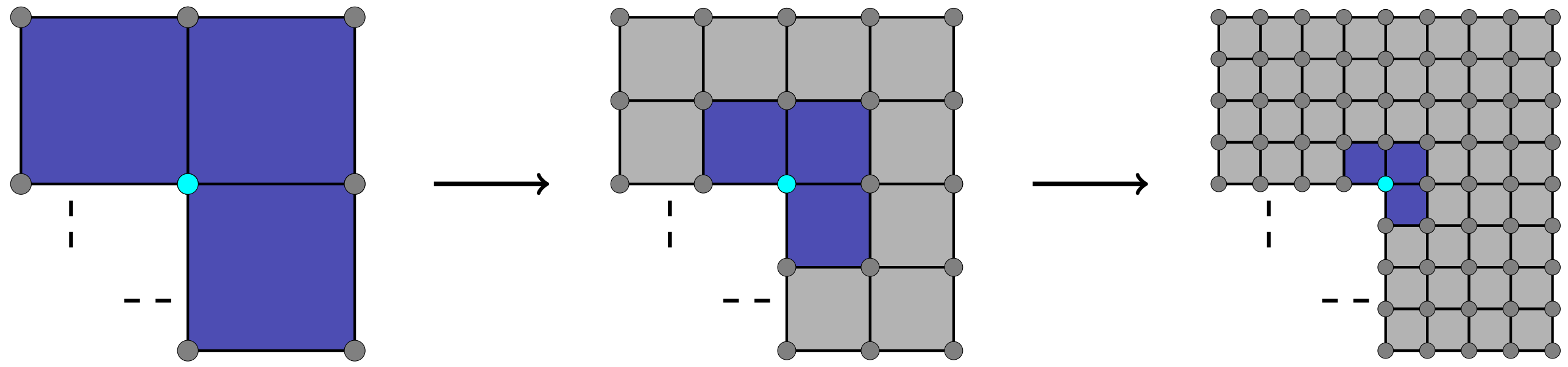

**Figure 2.11:** An initial element (cyan) and the region relevant for $S$ (dark blue) for $t = 2$ and $g = 2$. Through the refinement, the structure of the blue region and thus also $S$ does not change.

For $g = 2$, the subdivision matrix depends only on the surrounding cell structure. Hence the above equality holds for the subdivision matrices. For $g = 3$, the determination of $S$ also involves the next larger neighborhood of cells. This neighborhood is also reproduced after one refinement step, so the equality holds if the vertex with index $k$ was produced by refinement. □

For $t = 2$, the behavior described above is illustrated in Figure 2.11 for $g = 2$ and in Figure 2.12 for $g = 3$.

We observe here the same behavior as in the previous sections. The cases $g = 2$ and $g = 3$ behave analogously. For $g = 3$, however, an additional refinement step is necessary. To simplify the notation, we assume in the following that the corner with index $k$ has already emerged as a reproduction of a corner through refinement.

With the above lemma, we conclude that the asymptotic behavior of an initial element can be described by

$$\mathbf{I}_k^{(m)} = (\mathbf{G}_k, A_k, S_k, S_k^m P_k)\,.$$

Thus, the structure of the initial element remains the same, and the control points are generated by applying the subdivision matrix $S_k$ $m$ times to the initial control points. Hence, $S_k^m$ is the object to be analyzed with respect to the initial elements.

Another invariant structure can be observed on the level of the cells. We first consider the case $g = 2$ and $t = 2$. After sufficiently many refinement steps, by Remark 2.28, we have only cells with at most one irregular corner (and at least three regular corners). Refining such a cell produces three cells each with four regular corners, and a fourth cell with at most one irregular corner (and at least three regular corners).

This fourth cell thus has the same structure as the cell that was refined. Since the structure with respect to this fourth cell remains the same, the two cells can be mapped onto each other. For the four initial elements, the objects $\mathbf{G}, A, S$ reproduce one-to-one.

The description of the control points is more complex. Considering the refinement of all control points of all initial elements, the number of control points increases. However, since the structure of a reproduced cell remains the same, the number of control points needed for the new cell is the same as that of the original cell. Therefore, for the control points we can define a mapping as follows:

**Definition 2.32.** *Let $t = 2$, $g = 2$, let $\boldsymbol{D}$ be a spline definition domain, and let $s \in \boldsymbol{D}$ be a cell with at least three regular corners and a fourth corner $k$. Furthermore, let*

$$\boldsymbol{I}_i = (\boldsymbol{G}_i, A_i, S_i, P_i) \quad \textit{for} \quad i \in \{1, 2, 3, k\}$$

*be the four initial elements of the cell, and let $s^{(2)}$ be the refinement of $s$ that contains the reproduction of the corner $k$. Let*

$$\boldsymbol{I}_i^{(2)} = \left(\boldsymbol{G}_i, A_i, S_i, P_i^{(2)}\right) \quad \textit{for} \quad i \in \{1, 2, 3, k\}$$

*be the corresponding four new initial elements. We define $P$ as the matrix of control points of all initial elements of $s$, and $P^{(2)}$ as the matrix of control points of all initial elements of $s^{(2)}$ (both without double counting). Let $n$ be the number of*

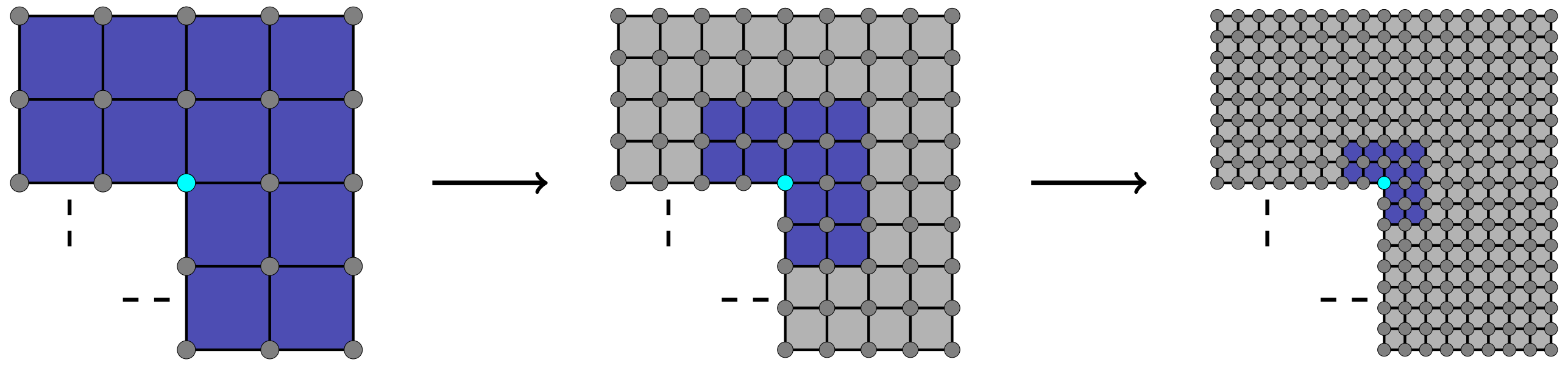

**Figure 2.12:** Analogous illustration to Figure 2.12 for $g = 3$. Here, the blue region is larger by one cell each time. Therefore, the cyan point must already have emerged through reproduction from a refinement step to guarantee the structure of the blue region. Afterwards, its reproduction proceeds analogously to the case $g = 2$.

*these control points in* $P$. *Furthermore, let*

$$\boldsymbol{G} := \bigcup_i \boldsymbol{G}_i$$

*be the associated graph that describes the combinatorial arrangement of* $P$. *Then we define*

$$S \in \mathbb{R}^{n\times n} \quad \textit{with} \quad SP = P^{(2)}$$

*as the subdivision matrix corresponding to* $\boldsymbol{G}$. *The rows of* $S$ *contain, at the appropriate entries, the corresponding rows of the subdivision matrices* $S_i$ *which belong to the respective new control points from* $P^{(2)}$, *and zero otherwise.*

Since we can use the refined cell again as the starting point for the above definition, the control points after $m$ refinements are given by

$$P^{(m)} = S^m P, \tag{2.1}$$

so the behavior is analogous to that of the initial elements. An illustration of the above definition can be found in Figure 2.13, where the yellow cell is mapped onto its refinement. From the above definition, it is clear that larger subdivision matrices can be constructed from the subdivision matrices of the initial elements, which correspondingly map larger structures invariantly onto their refinements.

In the following, we consider invariant structures and their prerequisites, to which the above definition can be transferred. Since the respective definitions are redundant with the above, we only explain which elements are mapped onto each other without describing a separate definition for each case. The subdivision matrix $S$ and the corresponding control points $P$ arise directly from the context of the cells and their initial elements. All cases are chosen so that equation (2.1) holds. The matrix $S$ is in particular always square.

For $t = 2$ and $g = 3$, we obtain the same definition under the condition that $s$ already comes from a refinement of a cell with at most one irregular corner. Then the structure is sufficiently refined so that the various elements reproduce. An illustration can be found in Figure 2.14. There, it can be seen that the yellow cell can be mapped onto its refinement. However, this is only possible after the cell with at most one irregular corner has already been refined once. Only then is the neighborhood of the initial elements of such shape that they reproduce themselves. The part relevant for determining the subdivision matrices of the initial elements is not fully shown in the figure. However, uniqueness follows here from the fact that the original cell has three regular corners. This can be understood clearly from Figure 2.8. In the second image, there are four cells, each with at most one irregular corner. Fixing one of these, the structure can be recognized from images 3–6.

This invariant structure can be transferred in exactly the same way to arbitrary refinements of cells. A refinement of a cell for $t = 2$ consists of four new cells. This structure of fours can be mapped onto the refinement of the cell with the corner $k$, which is also a structure of fours. The same can be done for structures of 16, 64, and so forth cells, for which an analogous but larger subdivision matrix $S$ can be formed each time.

The particularly interesting invariant structure is the one where evaluable cells are mapped onto evaluable cells again. Thus, invariant structures in the evaluated elements (for $t = 2$ surfaces) are created. In Figure 2.7 it can be seen that for $t = 2$ and $g = 2$, for a cell with three regular corners, three evaluable cells and one non-evaluable cell are formed after only one refinement. The three evaluable cells can then be evaluated to three B-spline surfaces arranged in an "L-shape" (see also [PR08, p. 60 ff.]). The four cells can be mapped by refinement onto the four cells arising from the refinement of the cell with the corner indexed by $k$. Since by the invariant structure three cells are regular again, the L-shaped piece is mapped onto a smaller L-shaped piece that fits exactly into the reentrant corner of the previous one. This process is also described in detail in [PR08]. The procedure is illustrated in Figure 2.13.

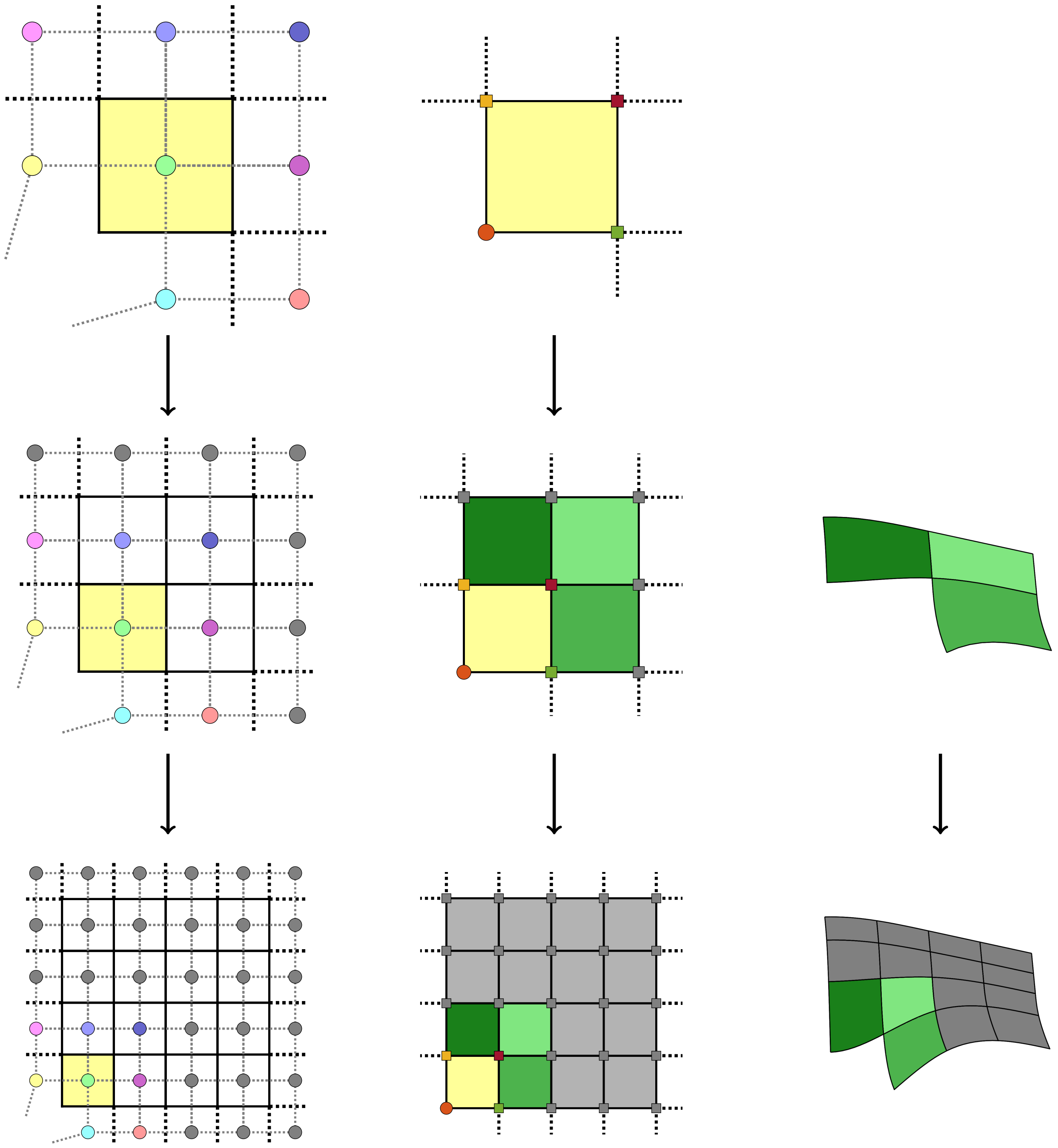

**Figure 2.13:** The first row shows, for $t = 2$ and $g = 2$, in the center a cell (yellow) with at least three regular corners. To the left, the control points belonging to the four initial elements are shown. In the second and third rows, the refinement of the yellow cell with the corresponding refinements of the control points can be seen. The green cells in the second row are evaluable cells. These have been exemplarily evaluated in the right graphic. The yellow and the three green cells form the refinement of the yellow cell in the first row and create a quadruple structure. This quadruple structure is invariantly mapped onto the corresponding quadruple structure in the third row. The invariant mapping of the evaluation is illustrated in the right image of the third row.

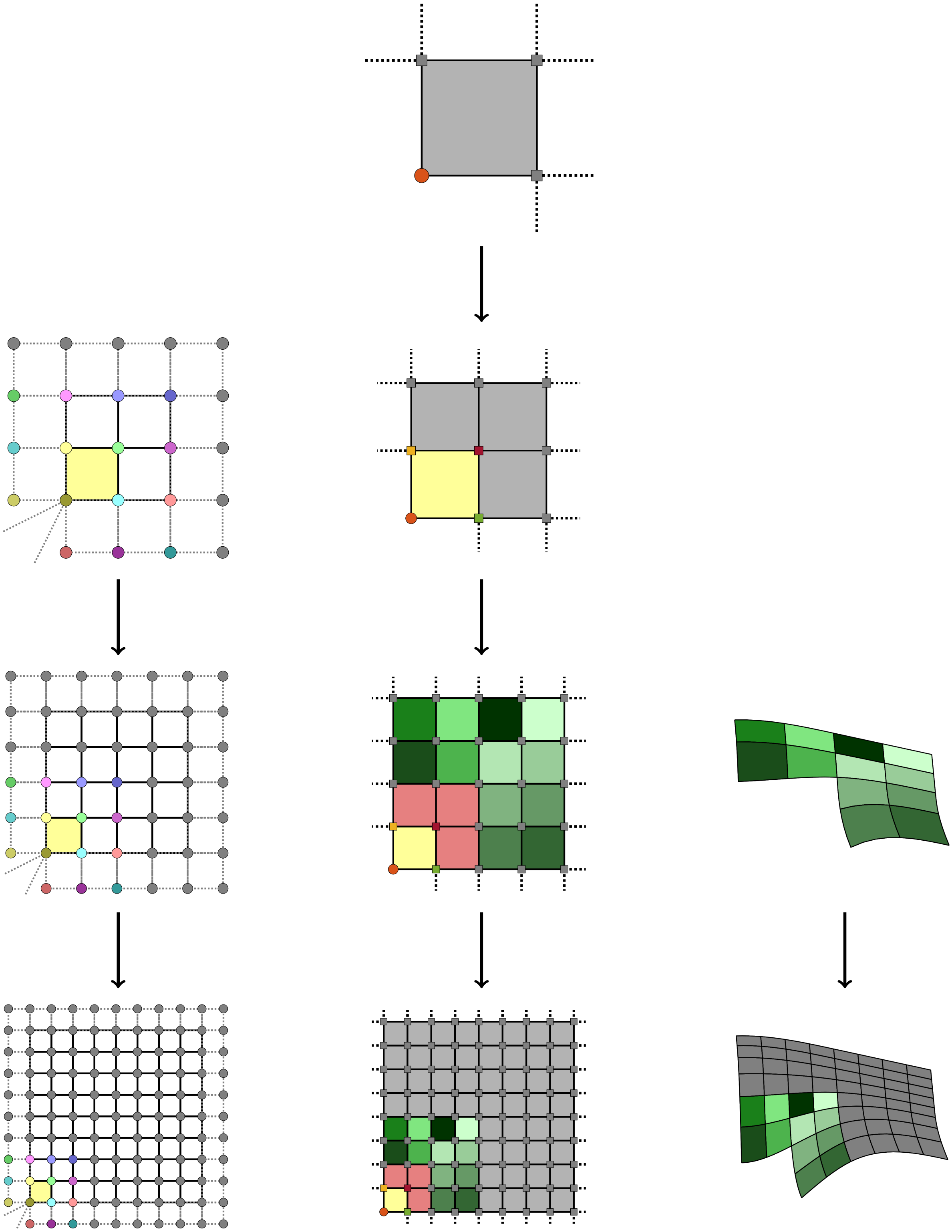

**Figure 2.14:** The illustration analogous to Figure 2.13 for $g = 3$. Here, an additional refinement step is necessary, which is why the yellow cell only appears in row 2. This cell is then invariantly mapped in rows 3 and 4. The control points in the left column lie primal to the corners of the cell, compared to the previous figure. In row 3, the yellow, red, and green cells form a structure of 16 cells, which is invariantly mapped onto the refinement in row 4. The evaluable cells form an L-shaped piece consisting of two layers.

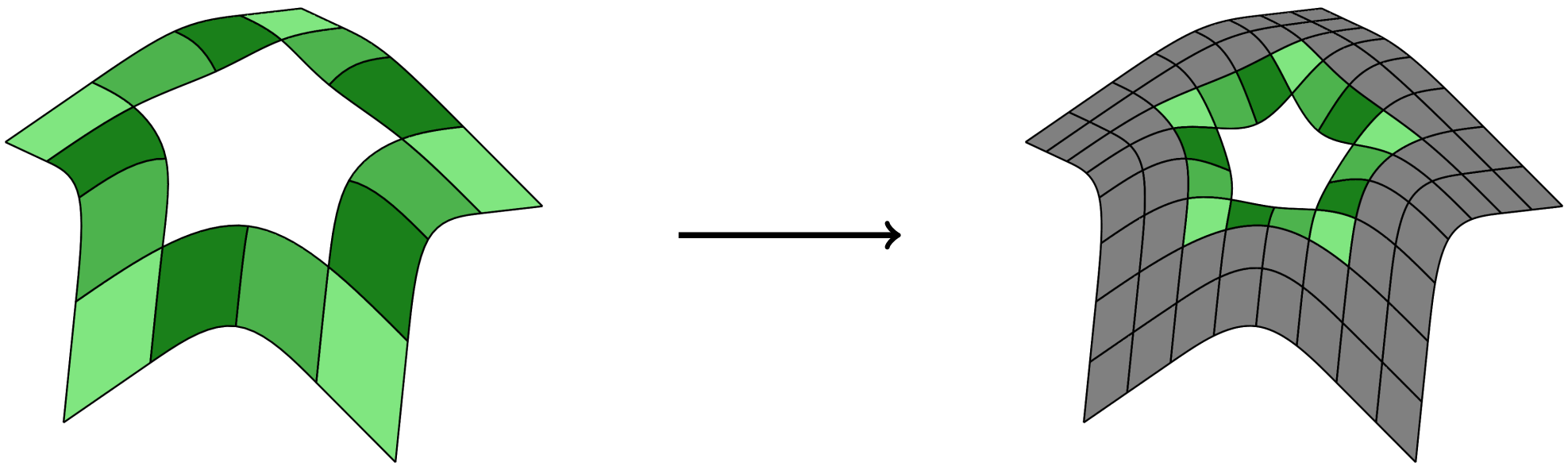

**Figure 2.15:** For $t = 2$ and $g = 2$, a ring consisting here of $3 \cdot 5$ elements is mapped onto a ring with $3 \cdot 5$ elements, which fits into the hole of the previous one.

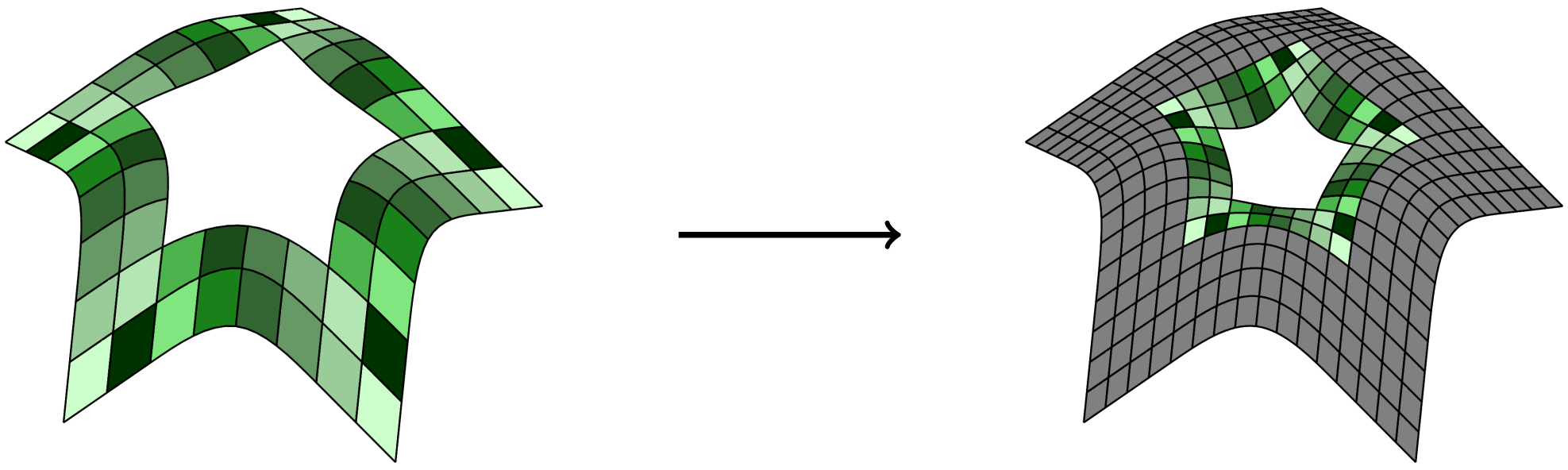

**Figure 2.16:** For $t = 2$ and $g = 3$, a double ring consisting of $12 \cdot 5$ elements is mapped onto a double ring with $12 \cdot 5$ elements, which fits into the hole of the previous one.

For $t = 2$ and $g = 3$, an evaluable region of a cell with at least three regular corners arises after two refinement steps. The initial element is refined by these two steps into $4 \times 4$ elements, of which twelve are evaluable. These $4 \times 4$ cells can then be invariantly mapped by refinement onto $4 \times 4$ cells. Hence, the twelve evaluated elements (which also form an L-shaped piece) are again mapped onto twelve evaluable elements. The new L-piece fits, as in the case $g = 2$, into the reentrant corner of the previous L-piece. This process is shown in Figure 2.14.

What applies to cells and their refinements can be generalized to groups of cells. In particular, structures of cells sharing an irregular corner are interesting, since their refinement enables the evaluation of the area around the irregular corner. For $t = 2$ and $g = 2$ we thus obtain the structure shown in Figure 2.13 for $n \geq 3$ cells. After one refinement step, three evaluable surfaces per cell arise, thus $3n$ evaluable surfaces for $n$ cells. The L-pieces fit together geometrically to form a *ring* (cf. [PR08, pp. 60 ff.]). By the invariant refinement structure, such a ring is mapped onto a ring. An illustration is given in Figure 2.15.

For $g = 3$, we also obtain a ring. Here, we get an L-shaped patch consisting of twelve surfaces, resulting in a ring of $12n$ elements. We refer to this as a *double ring structure*, since the $12n$ surfaces form two rings that are mapped onto two rings. These double ring structures have already been used in the literature, for example in [MM18], [KP17], [KP18], and [KP19]. The corresponding illustration can be found in Figure 2.16.

These two invariant structures, namely the mapping of (double) rings onto (double) rings, are the main subject of the study of subdivision surfaces. This structure is extensively treated in the monograph [PR08], since through the mapping of rings onto rings a surface with a hole is created, which is successively filled by subdivision. Under certain conditions (the quality criteria Q1 and Q3), the central point (i.e., the limit of the ring process) can also be evaluated.

This enables an analytical study of the smoothness of subdivision surfaces, which constitutes a major part of [PR08].

For the case $t = 2$, the investigation of two subdivision matrices is therefore of interest. On the one hand, those of the initial elements, since all other subdivision matrices are composed from them, and on the other hand, the subdivision matrices of the above-described (double) rings. For both, the study of the limit behavior in equation 2.1 is relevant, which is why the quality criteria in the next section focus on the local asymptotic behavior, i.e., on $S^m$ as $m \to \infty$.

Since Lemma 2.31 is independent of $t$, the initial elements behave for the case $t = 3$ similarly to those for $t = 2$. For $t = 3$ and $g = 2$, cells can be mapped onto each other that have at most one (semi-)irregular corner and additionally at most three semi-irregular corners. According to Remark 2.30, this holds for all cells after one refinement step.

For $g = 3$, we obtain an analogous invariant structure if the cell $s$ has already arisen from the refinement of such a cell (with at most one (semi-)irregular and additionally at most three semi-irregular corners) and contains the (semi-)irregular corner. This can be seen in the yellow element in Figure 2.17 (for $g = 2$) and Figure 2.18 (for $g = 3$).

For refinements of cells, we obtain results analogous to those for $t = 2$ under the aforementioned assumptions. This means that before refinement, a cell for $g = 2$ contains at most one (semi-)irregular corner and at most three semi-irregular corners. For $g = 3$, this means that the cell before refinement already originates from a refinement of a cell with at most one (semi-)irregular and at most three semi-irregular corners and has the same structure. Here, the interesting case is the one where evaluable elements map onto evaluable elements. For $g = 2$, these are combinations of $2 \times 2 \times 2$ elements, and for $g = 3$, combinations of $4 \times 4 \times 4$ elements. Illustrations can also be found in Figures 2.17 and 2.18. The shape of the evaluable elements here is a *tripod* (or, for $g = 3$, a *double tripod*). Under refinement, this is again mapped onto a (double) tripod. It can be observed that the evaluable area in the outer structure (under global refinement) also increases. Thus, the new tripod nests into the concave corner of the old one and, due to the enlargement of the outer region, surrounds it in all directions.

Considering the $n \geq 4$ cells around an irregular corner (under the same assumptions as in the previous part of the section), a structure emerges that deserves closer examination. The tripods assemble into a shell-like structure with holes. These holes appear around the irregular edges or the semi-irregular corners. If the subdivision matrices satisfy the quality criterion Q8 introduced in the next section, i.e., for the semi-regular case the three-dimensional structure is a tensor product of the irregular two-dimensional case and the one-dimensional case, these holes can be naturally filled. This filling is then again mapped by refinement onto the filling of the refined object.

This is best understood by considering the regular case. If the edges belonging to the holes are regular, the hole, consisting then of four regular cells, can be filled with four B-spline volumes. These B-spline volumes are then invariantly mapped onto their refinements. This can be seen in Figures 2.19 for $g = 2$ and 2.20 for $g = 3$ in the upper regions, respectively. At the same time, this explains why we spoke of at most three semi-irregular corners in the definitions above. These corners can also be regular, which simplifies the structure. The three blue regions in Figures 2.17 and 2.18 are green in this case. For the semi-irregular regions, the evaluation proceeds analogously, but the area associated with the cells must first be constructed through refinement. In the tensor product case, the evaluation can be reduced to the simpler surface case.

We thus obtain a structure which we call a *shell*. This shell (or for $g = 3$ also a *double shell*) is mapped by refinement onto a smaller shell, so that the interior is successively filled (see Figures 2.19 and 2.20). The central point, i.e., the limit result of the process, can again be explicitly given provided that the quality criteria Q1 and Q3 of the next section are satisfied. The shell can thus serve as the initial object for the analytical consideration.

Analogous to the case $t = 2$, for $t = 3$ the investigation of the subdivision matrices of the initial elements is of interest, since these determine the refinement rules. For analytical aspects, the subdivision matrices that map the mentioned shells to their refinements are also relevant. For both, the local asymptotic behavior, i.e., $S^m$, is important again.

We conclude this section with the concept of a subdivision algorithm. For $t = 2$, this is defined in [PR08, Def. 4.27, p. 80] as the combination of generating functions and a subdivision matrix of a ring around an (irregular) point, which must satisfy certain criteria. The idea is that with these two objects, the evaluation of the subdivision surface around an irregular point can be described. The subdivision algorithm is thus an algorithm for generating evaluated objects by refinement. We define this concept somewhat more generally:

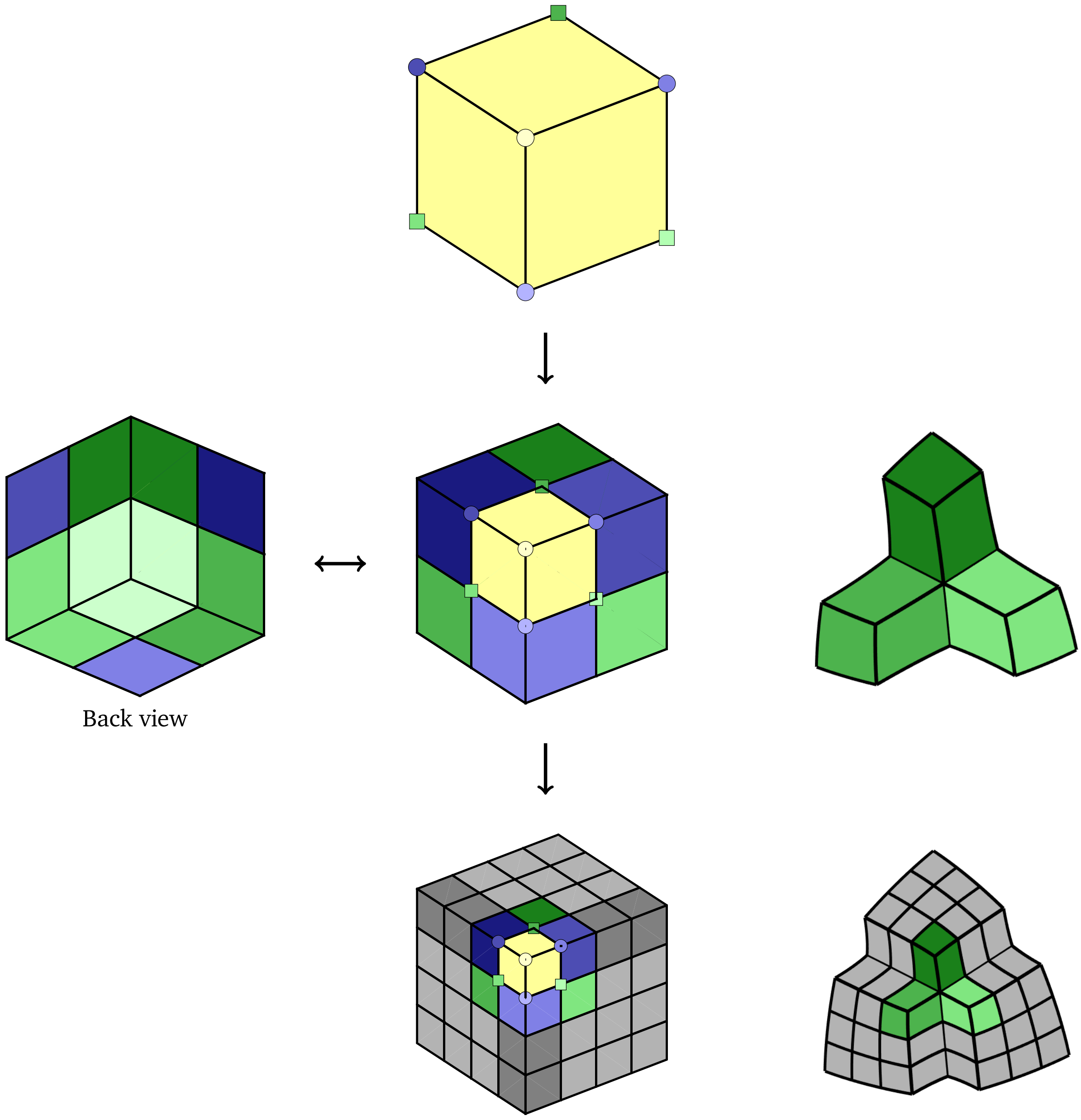

**Figure 2.17:** Refinement of a cell for $t = 3$ and $g = 2$. In the first row, a cell (yellow) is shown with at most one (semi-)irregular corner (yellow), at most three additional semi-irregular corners (blue), and four regular corners (green). Through the refinements in rows two and three, the structure of the yellow cell is reproduced. In row two, a structure of $2 \times 2 \times 2$ cells is visible, of which four are evaluable. The blue cells have (at most) one irregular edge and (at most) two semi-irregular points. This structure is reproduced through refinement, as visible in row three. The evaluation of this $2 \times 2 \times 2$ structure forms a tripod, visible in the right column. This shape is reproduced by refinement and fits into the now larger area of the previous tripod.

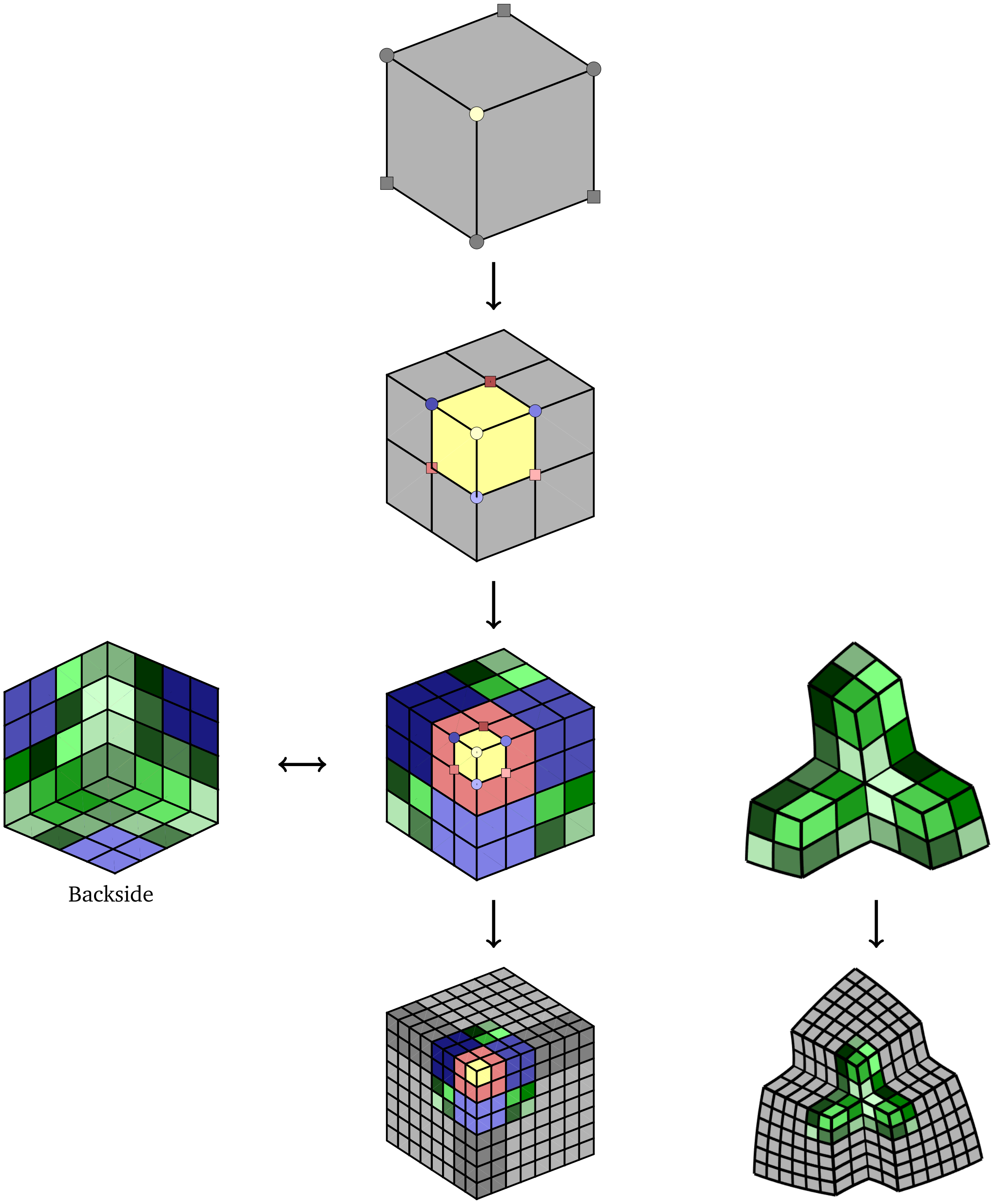

**Figure 2.18:** Analogous illustration to Figure 2.17 for $g = 3$. The yellow cell appears only in the second row, since it must emerge by refinement from a cell with at most one (semi-)irregular and at most three additional semi-irregular corners. In the third row, a structure of $4 \times 4 \times 4$ elements is shown, which is mapped invariantly to the finer structure in the fourth row. Unlike $g = 2$, the tripod consists of two layers. The invariant refinement is shown in the right column.

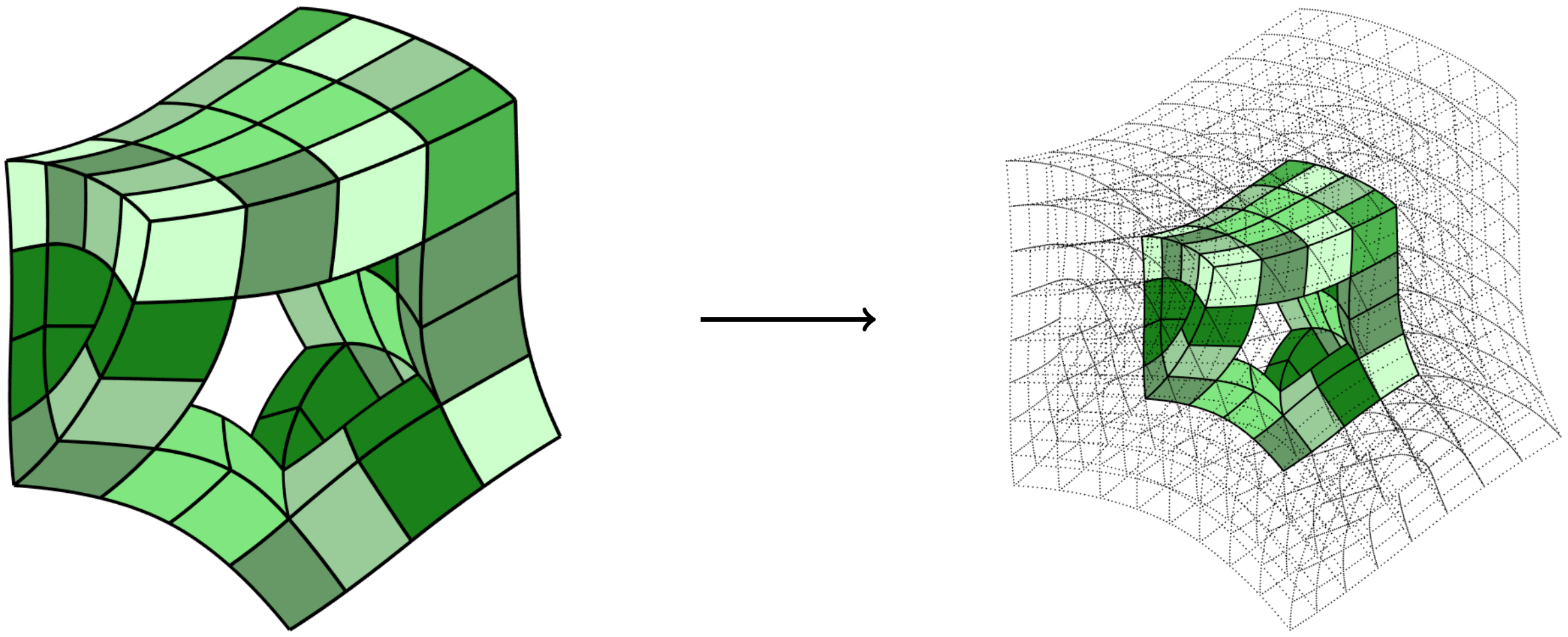

**Figure 2.19:** A shell for $t = 3$ and $g = 2$ around an irregular corner consisting of $8 \times 8$ cells and $8 \times 4 + 2 \times 4 = 40$ evaluated B-spline volumes (left) and its invariant refinement (right). The refinement of the outer shell is indicated only by black corner lines. It can be seen that the new shell nests inside the old one. Moreover, the geometry around the irregular corner includes two regular edges. Therefore, two holes in the shell (top and back right) are evaluable and filled. This evaluation is also invariantly mapped under refinement.

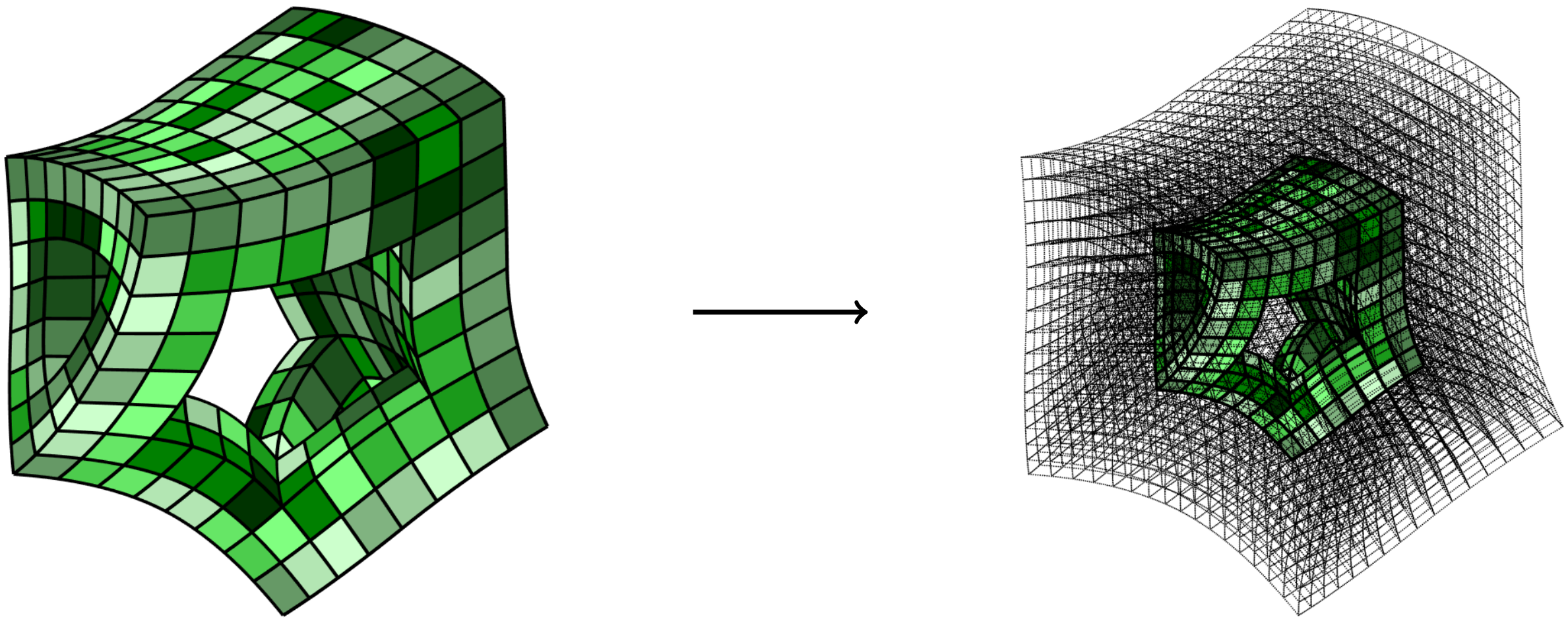

**Figure 2.20:** The illustration analogous to Figure 2.19 for $t = 3$ and $g = 3$. Here, the shell consists of two layers of B-spline volumes and thus represents the three-dimensional extension of Figure 2.16.

**Definition 2.33.** *Let **D** be a spline definition domain with an invariant structure as described in this section. Let* $P \in \mathbb{R}^{n\times d}$ *be the matrix of control points and* **g** *the associated generating functions. Furthermore, let* $P^{(2)} \in \mathbb{R}^{n\times d}$ *be the matrix of refined control points arising invariantly from* $P$ *and*

$$S \in \mathbb{R}^{n\times n} \quad \text{with} \quad SP = P^{(2)}$$

*the associated subdivision matrix that describes the invariant mapping. Then we call* $(\boldsymbol{g}, S)$ *a* subdivision algorithm.

It should first be noted that although the control points $P$ appear in the definition, they have no influence on the two elements **g** and $S$. Their mention is made partly because the generating functions are defined for the control points, and partly because this allows the subdivision matrix to be described clearly, without having to consider double counting of rows for different initial elements with identical control points.

This definition summarizes the content of this section well. The interesting elements with respect to refinement are the invariant structures which exhibit local irregularities. Through these invariant structures, the irregularity can be isolated and reduced, and the regular region can then be evaluated using the generating functions. Interpreting larger structures as collections of these invariant elements, the entire structure can be evaluated.

In the next section, we focus on the subdivision matrices associated with the invariant structures and describe the criteria they should meet. We especially explain the advantages of the criteria and the consequences of (non-)fulfillment.

## 2.4 Quality Criteria

In Definitions 2.11 and 2.12, the subdivision matrices of the initial elements were defined, but apart from the area of influence, no conditions were imposed on the matrices. In Section 2.3, larger subdivision matrices were additionally defined in Definition 2.32 and later in that section. However, these definitions also do not impose any conditions on the subdivision matrices. Thus, any real square matrix could be a subdivision matrix. Already in the previous section, we hinted that subdivision matrices should possess certain properties, for example, to allow evaluation of the central point of the refinement process.

In this section, we describe the desired properties. We also classify how central these properties are and what consequences occur if they are (not) fulfilled. For this, let

$$S \in \mathbb{R}^{n\times n}$$

be a subdivision matrix of arbitrary degree, type, and size (initial element, ring, shell, etc.). Before we begin, we need the following definition of special eigenvalues:

**Definition 2.34.** *Let* $S \in \mathbb{R}^{n\times n}$ *be a matrix with eigenvalues* $\lambda_0, \dots, \lambda_{n-1} \in \mathbb{C}$ *ordered such that* $|\lambda_0| \geq \dots \geq |\lambda_{n-1}|$.

1. *If* $|\lambda_0| > |\lambda_1|$*, then* $\lambda_0$ *is called the* dominant eigenvalue *of the matrix* $S$ *(cf. [PR08, p. 75]).*
2. *Furthermore, for* $k \in \{1, \dots, n-1\}$*, if the eigenvalue* $\lambda := \lambda_1 = \dots = \lambda_k$ *and (if* $k \neq n-1$*)*

$$|\lambda_0| > |\lambda_1| = \dots = |\lambda_k| > |\lambda_{k+1}|,$$

*and the geometric multiplicity of* $\lambda$ *is exactly* $k$*, then* $\lambda$ *is called the* $k$-fold subdominant eigenvalue *of the matrix* $S$ *(cf. [PR08, Def. 5.3, pp. 84–85]).*

3. *Furthermore, for* $m \in \{k+1, \dots, n-1\}$*, if* $\mu := \lambda_{k+1} = \dots = \lambda_m$*, and* $\mu \in \mathbb{R}_{>0}$*, and (if* $k \neq n-1$*)*

$$|\lambda_0| > |\lambda_1| = \dots = |\lambda_k| > |\lambda_{k+1}| = \dots = |\lambda_m| > |\lambda_{m+1}|,$$

*then* $\mu$ *is called the* subsubdominant eigenvalue *of the matrix* $S$ *(cf. [PR08, pp. 126–127]).*

With these prerequisites established, we can now begin to describe the quality criteria.

## Q1 Affine Invariance

**Definition 2.35.** *Let $S \in \mathbb{R}^{n\times n}$ be a subdivision matrix. The matrix $S$ satisfies quality criterion* Q1 affine invariance *if*

$$\sum_{j=1}^{n} S_{(i,j)} = 1 \quad \textit{for all} \quad i = 1, \dots, n,$$

*i.e., if each row sums up to* 1.

The motivation behind this criterion is geometric. When a row of the subdivision matrix is multiplied by the control points, each entry in the row can be interpreted as a weight. The subdivision matrix thus weights the old points to create new points. That the weights sum to 1 is therefore a natural and necessary choice. Affine invariance is a mandatory quality criterion that should be satisfied both theoretically and numerically with high precision. In [PR08, Def. 4.27, p. 80], this criterion is part of the definition of subdivision matrices for subdivision surfaces.

Affine invariance can also be equivalently expressed by the following lemma (cf. [PR08, p. 75]; also see [FHW79, Main Thm. 26, pp. 15–16] for stochastic matrices):

**Lemma 2.36.** *Let $S \in \mathbb{R}^{n\times n}$ be a matrix. The rows of the matrix $S$ sum to* 1 *if and only if the vector of all ones $\vec{1}$ is an eigenvector to the eigenvalue* 1.

*Proof.* It trivially follows that

$$S \cdot \vec{1} = \vec{1} \quad \Leftrightarrow \quad S_{(i,:)} \cdot \vec{1} = 1 \quad \text{for all} \quad i = 1, \dots, n.$$

And since

$$1 = S_{(i,:)} \cdot \vec{1} = \sum_{j=1}^{n} S_{(i,j)} \cdot 1 = \sum_{j=1}^{n} S_{(i,j)},$$

the above statement holds. □

## Q2 Convex Hull

**Definition 2.37.** *Let $S \in \mathbb{R}^{n\times n}$ be a subdivision matrix. The matrix $S$ satisfies quality criterion* Q2 convex hull *if it satisfies quality criterion Q1 and, additionally,*

$$S_{(i,j)} \geq 0 \quad \textit{for all} \quad i = 1, \dots, n, \quad \textit{and} \quad j = 1, \dots, n,$$

*that is, every entry of $S$ is non-negative.*

Although quality criterion Q1 is part of the definition of quality criterion Q2 to emphasize the property of affine invariance, we will always mention quality criterion Q1 in addition to quality criterion Q2 in the following.

Quality criterion Q2 is not a necessary condition for a subdivision matrix. However, the usefulness of the criterion and its connection to the concept of the convex hull becomes clear in the following lemma:

**Lemma 2.38.** *Let $S \in \mathbb{R}^{n\times n}$ be a matrix and $P \in \mathbb{R}^{n\times d}$ a matrix of control points. If quality criteria Q1 and Q2 are fulfilled for $S$, then each row of $S \cdot P$, and thus each new control point, lies in the convex hull (cf. Definition 3.1) of $\{P_{(1,:)}, \dots, P_{(n,:)}\}$, i.e., in the convex hull of the old control points.*

*Proof.* The convex hull of $\{P_{(1,:)}, \dots, P_{(n,:)}\}$ is by Lemma 3.2 exactly

$$\left\{ \varphi_1 P_{(1,:)} + \cdots + \varphi_n P_{(n,:)} \;:\; \varphi_j \in \mathbb{R}_{\geq 0} \text{ for all } j \in \{1, \dots, n\} \text{ and } \sum_{j=1}^{n} \varphi_j = 1 \right\}.$$

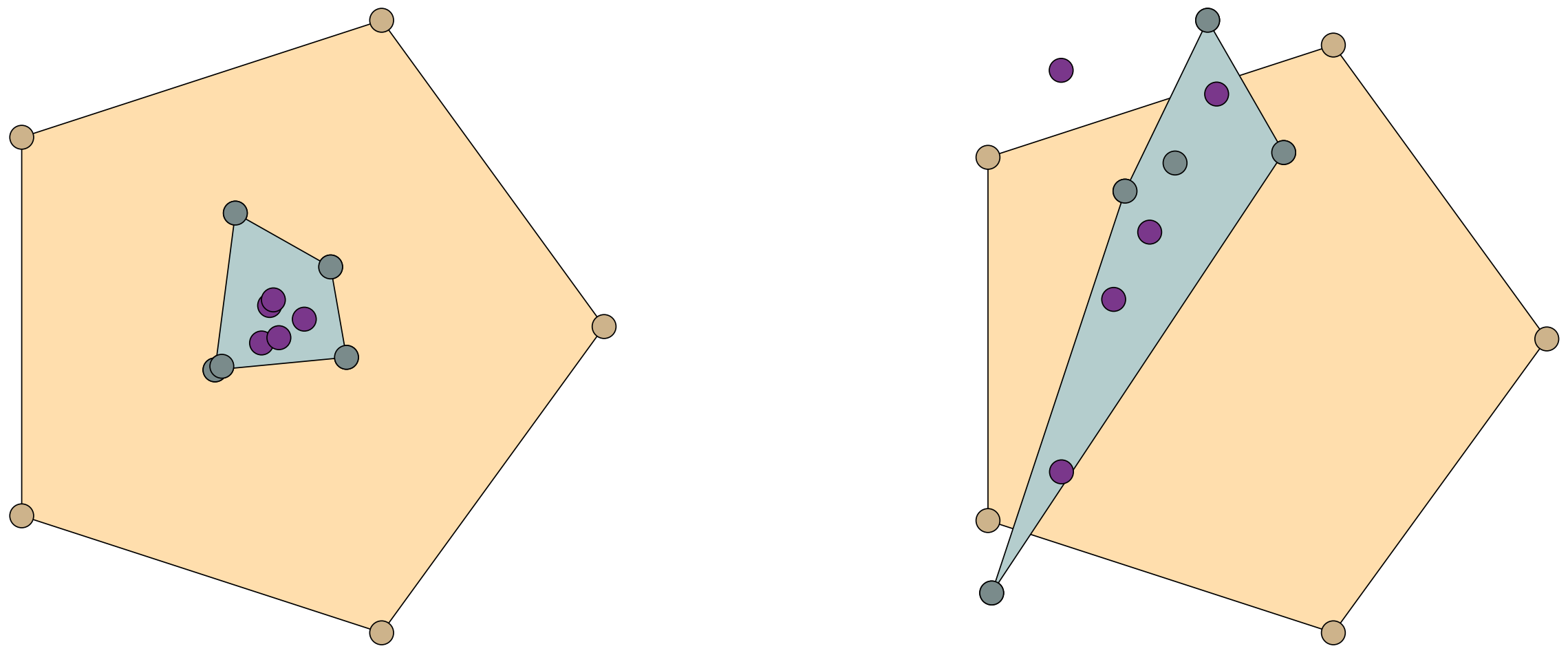

**Figure 2.21:** Illustration of Lemma 2.38. An initial configuration of five control points (yellow) is refined using a matrix that satisfies quality criterion Q1 but not quality criterion Q2 (right), and with a matrix that satisfies both quality criteria Q1 and Q2 (left). For the first (gray) and second (purple) refinement steps, it can be observed that on the left the new points lie within the convex hull of the old points, whereas on the right at least one point lies outside the convex hull in each refinement.

For $i \in \{1, \dots, n\}$ the $i$-th row of $S \cdot P$ is exactly $S_{(i,:)} \cdot P$. Hence,

$$S_{(i,:)} \cdot P = \sum_{j=1}^{n} S_{(i,j)} P_{(j,:)}.$$

Since by quality criterion Q1 we have $\sum_{j=1}^{n} S_{(i,j)} = 1$ for all $i$, and by quality criterion Q2 we have $S_{(i,j)} \geq 0$ for all $i, j$, we can set $\varphi_j := S_{(i,j)}$, which is a point in the convex hull of $\{P_{(1,:)}, \dots, P_{(n,:)}\}$. $\square$

With quality criteria Q1 and Q2 we thus gain control over the location of the new control points. Since the purpose of subdivision matrices is refinement and the evaluation of control points to elements with B-spline basis functions is performed, one expects this refinement to be local and the new control points to respect this locality. As mentioned before, this criterion is not necessary for this, but quality criterion Q1 is sufficient to guarantee local placement of the control points. An illustration of this behavior can be found in Figure 2.21.

It is also important to note that with quality criteria Q1 and Q2, the subdivision matrix $S$ is exactly a stochastic matrix according to Definition B.39. This allows us to apply theorems about stochastic matrices in the following, whenever these quality criteria are satisfied.

## Q3 Dominant Eigenvalue 1

**Definition 2.39.** *Let $S \in \mathbb{R}^{n \times n}$ be a subdivision matrix with eigenvalues $\lambda_0, \dots, \lambda_{n-1} \in \mathbb{C}$ ordered so that $|\lambda_0| \geq \dots \geq |\lambda_{n-1}|$. The matrix $S$ satisfies quality criterion* Q3 dominant eigenvalue 1 *if*

$$1 = \lambda_0 > |\lambda_i| \quad \textit{for all} \quad i \in \{1, \dots, n-1\},$$

*that is, if $1$ is a simple eigenvalue and the eigenvalue of largest magnitude.*

This is also a requirement for a subdivision algorithm according to [PR08, Def. 4.27, p. 80], thus an obligatory criterion for a subdivision matrix. To describe the usefulness of this quality criterion, we first need the concept of consistency (see [PR08, p. 75]):

**Definition 2.40.** *Let $S \in \mathbb{R}^{n\times n}$ be a matrix and $P \in \mathbb{R}^{n\times d}$ a matrix of control points. The matrix $S$ is called* consistent *if for every choice of control points $P$ and every vector $a \in \mathbb{R}^n$ with $\sum_{i=1}^{n} a_{(i)} = 1$ the limit*

$$\lim_{m\to\infty} a^T S^m P$$

*exists and is uniquely defined (depending on $P$).*

With this definition, the following lemma can be stated [PR08, Lem. 4.17, p. 75]:

**Lemma 2.41.** *Let $S \in \mathbb{R}^{n\times n}$ be a subdivision matrix satisfying quality criteria Q1 and Q3. Let $P \in \mathbb{R}^{n\times d}$ be a matrix of control points and $a \in \mathbb{R}^n$ with $\sum_{i=1}^{n} a_{(i)} = 1$. If $w_0 \in \mathbb{C}^n$ is the left eigenvector corresponding to the eigenvalue $1$, i.e., $w_0^T S = w_0^T$, then*

$$\lim_{m\to\infty} a^T S^m P = \frac{w_0^T}{\langle w_0^T, \vec{1}\rangle} P,$$

*and thus $S$ is consistent.*

*Proof.* The proof in a modified form can be found in [PR08, p. 75]. We nevertheless give a rough description here to show how the different concepts fit together. Decomposing the matrix $S$, it can be represented by Theorem B.29 (Jordan decomposition) as

$$S = VJV^{-1},$$

where

$$J = \begin{bmatrix} J_0 & 0 & 0 \\ 0 & \ddots & 0 \\ 0 & 0 & J_k \end{bmatrix}$$

is the block diagonal matrix of Jordan blocks of $S$. If quality criterion Q3 holds, i.e., $1$ is simple and the dominant eigenvalue of $S$, then

$$J = \begin{bmatrix} 1 & 0 & \dots & 0 \\ 0 & J_1 & \ddots & \vdots \\ \vdots & \ddots & \ddots & 0 \\ 0 & \dots & 0 & J_k \end{bmatrix}.$$

In [PR08, pp. 72–75] it is shown that for each $i \in \{1, \dots, k\}$, $J_i^m$ converges to the zero matrix of appropriate size as $m \to \infty$ when the corresponding eigenvalues satisfy $|\lambda| < 1$. Thus we get

$$\lim_{m\to\infty} a^T S^m P = \lim_{m\to\infty} a^T V J^m V^{-1} P = \lim_{m\to\infty} a^T V \begin{bmatrix} 1 & 0 & \dots & 0 \\ 0 & J_1^m & \dots & \vdots \\ \vdots & \ddots & \ddots & 0 \\ 0 & \dots & 0 & J_k^m \end{bmatrix} V^{-1} P = a^T V \begin{bmatrix} 1 & 0 & \dots & 0 \\ 0 & 0 & \dots & \vdots \\ \vdots & \ddots & \ddots & 0 \\ 0 & \dots & 0 & 0 \end{bmatrix} V^{-1} P.$$

Since $\frac{w_0^T}{\langle w_0^T, \vec{1}\rangle}$ is the first row of $V^{-1}$, we have

$$a^T V \begin{bmatrix} 1 & 0 & \dots & 0 \\ 0 & 0 & \dots & \vdots \\ \vdots & \ddots & \ddots & 0 \\ 0 & \dots & 0 & 0 \end{bmatrix} V^{-1} P = a^T V \begin{bmatrix} \frac{w_0^T}{\langle w_0^T, \vec{1}\rangle} \\ \vec{0}^T \\ \vdots \\ \vec{0}^T \end{bmatrix} P.$$

Since by quality criterion Q1 and Lemma 2.36 the vector $\vec{1}$ is the eigenvector for the eigenvalue $1$, and this eigenvalue

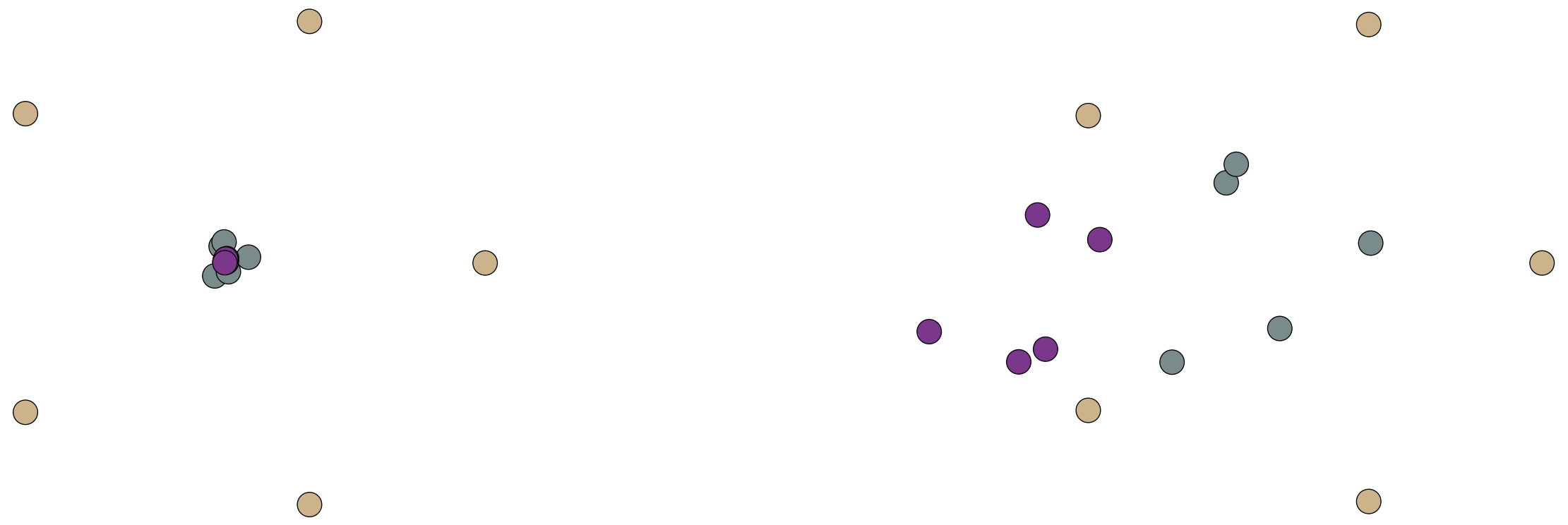

**Figure 2.22:** Illustration of Lemma 2.41. A starting configuration of five control points (yellow) is refined with a matrix that satisfies the quality criteria Q1 and Q3 (left), and with a matrix for which these quality criteria do not hold (right). For the refinement after two steps (gray) and after four steps (purple), it can be observed that on the left the new points converge, whereas on the right the new control points are arranged arbitrarily in space.

has algebraic multiplicity 1, it follows that

$$a^T V \begin{bmatrix} \frac{w_0^T}{\langle w_0^T, \vec{1} \rangle} \\ \vec{0}^T \\ \vdots \\ \vec{0}^T \end{bmatrix} P = a^T \begin{bmatrix} \frac{w_0^T}{\langle w_0^T, \vec{1} \rangle} \\ \vdots \\ \frac{w_0^T}{\langle w_0^T, \vec{1} \rangle} \end{bmatrix} P.$$

Since the entries of $a$ sum to 1, overall we get

$$\lim_{m\to\infty} a^T S^m P = a^T \begin{bmatrix} \frac{w_0^T}{\langle w_0^T, \vec{1} \rangle} \\ \vdots \\ \frac{w_0^T}{\langle w_0^T, \vec{1} \rangle} \end{bmatrix} P = \frac{w_0^T}{\langle w_0^T, \vec{1} \rangle} P.$$

□

We thus see that the quality criteria Q1 and Q3 are sufficient conditions for the convergence of a subdivision algorithm. With these two criteria, the central point—that is, the above limit—is evaluable. An illustration of this behavior can be found in Figure 2.22.

We also obtain the following relation:

**Lemma 2.42.** *Let $S \in \mathbb{R}^{n\times n}$ be an irreducible subdivision matrix that satisfies the quality criteria Q1 and Q2. Furthermore, assume that $\min_i S_{(i,i)} > 0$. Then the quality criterion Q3 also holds for the subdivision matrix $S$.*

*Proof.* This is merely a reformulation of Corollary B.41. □

## Q4 Subdominant eigenvalue with algebraic and geometric multiplicity $t$

**Definition 2.43.** *Let $S \in \mathbb{R}^{n\times n}$ be a subdivision matrix that satisfies the quality criteria Q1 and Q3, and let $\lambda_0, \dots, \lambda_{n-1} \in \mathbb{C}$ with $|\lambda_0| \geq \dots \geq |\lambda_{n-1}|$ be the eigenvalues of the matrix $S$. Further let $t \in \{1, 2, 3\}$ be the type of the evaluated elements. The matrix $S$ satisfies the quality criterion* Q4 subdominant eigenvalue with algebraic *and geometric multiplicity $t$, if*

$$\lambda_0 > |\lambda_1| = \cdots = |\lambda_t| > |\lambda_{t+1}|, \quad \lambda := \lambda_1 = \cdots = \lambda_t,$$

*and if the geometric multiplicity of $\lambda$ is exactly $t$, i.e., $\lambda$ is the $t$-fold subdominant eigenvalue of $S$.*

This criterion is motivated by [PR08, pp. 84 ff.] and together with quality criterion Q5 forms the main motivation for this work. We recall that the prerequisite for a subdominant eigenvalue according to Definition 2.34 is the existence of a dominant eigenvalue. Hence, for this criterion, we first require that $S$ possesses a dominant eigenvalue, and since we have already identified the quality criteria Q1 and Q3 as mandatory conditions for a subdivision matrix according to [PR08], we assume for this quality criterion that criteria Q1 and Q3 are satisfied.

Quality criterion Q4 is not necessary, but it is important for many practical subdivision applications. If quality criteria Q1 and Q3 are satisfied, the repeated application of the subdivision matrix on a control point matrix $P$ converges to a point, as we have seen in the previous criterion. Moreover, if the subdivision matrix has a $t$-fold subdominant eigenvalue, then in this process the shape of the control points for almost all configurations approaches a basis of the eigenspace of the subdominant eigenvalue.

How is this to be understood, and what does “almost all” mean? To answer these questions, we consider the expression

$$S^m P,$$

where $S \in \mathbb{R}^{n\times n}$ is a subdivision matrix of type $t$, and $P \in \mathbb{R}^{n\times d}$ is a matrix of control points. We assume that quality criteria Q1 and Q3 hold for $S$, that $S$ has a subdominant eigenvalue $\lambda_1 = \cdots = \lambda_j$ with $j \in \{1,\ldots,t\}$ of geometric multiplicity $j$, and that the eigenvalues and eigenvectors are ordered as

$$1 > |\lambda_1| \geq |\lambda_2| \geq \ldots$$

Thus, the matrix $S$ can be represented as

$$S = VJV^{-1} \quad \text{with} \quad J = \left[\begin{array}{cccc|cccc} 1 & 0 & \ldots & 0 & 0 & \ldots & \ldots & 0 \\ 0 & \lambda_1 & \ddots & \vdots & \vdots & & & \vdots \\ \vdots & \ddots & \ddots & 0 & \vdots & & & \vdots \\ 0 & \ldots & 0 & \lambda_j & 0 & \ldots & \ldots & 0 \\ \hline 0 & \ldots & \ldots & 0 & J_1 & 0 & \ldots & 0 \\ \vdots & & & \vdots & 0 & \ddots & \ddots & \vdots \\ \vdots & & & \vdots & \vdots & \ddots & \ddots & 0 \\ 0 & \ldots & \ldots & 0 & 0 & \ldots & 0 & J_k \end{array}\right]$$

where $J_1,\ldots,J_k$ are the other Jordan blocks of the matrix $S$. Accordingly, we have

$$S^m = VJ^mV^{-1} = V \left[\begin{array}{cccc|cccc} 1 & 0 & \ldots & 0 & 0 & \ldots & \ldots & 0 \\ 0 & \lambda_1^m & \ddots & \vdots & \vdots & & & \vdots \\ \vdots & \ddots & \ddots & 0 & \vdots & & & \vdots \\ 0 & \ldots & 0 & \lambda_j^m & 0 & \ldots & \ldots & 0 \\ \hline 0 & \ldots & \ldots & 0 & J_1^m & 0 & \ldots & 0 \\ \vdots & & & \vdots & 0 & \ddots & \ddots & \vdots \\ \vdots & & & \vdots & \vdots & \ddots & \ddots & 0 \\ 0 & \ldots & \ldots & 0 & 0 & \ldots & 0 & J_k^m \end{array}\right] V^{-1}.$$

Since $V$ is invertible, its columns form a basis of $\mathbb{R}^n$. Hence, there exists a matrix $P' \in \mathbb{R}^{n\times d}$ such that

$$VP' = P.$$

The columns of $P'$ are thus the coordinates of the columns of $P$ with respect to the basis $V$. They are also called *eigen-coefficients* in [PR08, Def. 4.16, p. 74]. We obtain

$$S^m P = VJ^m V^{-1} VP' = VJ^m P'.$$

Since the Jordan blocks $J_1^m, \dots, J_k^m$ decay faster than $1^m, \lambda_1^m, \dots, \lambda_j^m$ as $m$ grows (cf. [PR08, p. 72–75]), the above expression tends to

$$\left[\vec{1}\ \lambda^m V_{(:,2)}\ \dots\ \lambda^m V_{(:,j+1)}\ \vec{0}\ \dots\ \vec{0}\right] P' + \mathcal{O}(\lambda_{j+1}^m) = P'_{(1,:)} + \lambda^m \left(V_{(:,2)}P'_{(2,:)} + \dots + V_{(:,j+1)}P'_{(j+1,:)}\right) + \mathcal{O}(\lambda_{j+1}^m)$$

and, using the notation from Lemma 2.41, tends to

$$\frac{w_0^T}{\langle w_0^T, \vec{1}\rangle} P + \lambda^m \left(V_{(:,2)}P'_{(2,:)} + \dots + V_{(:,j+1)}P'_{(j+1,:)}\right) + \mathcal{O}(\lambda_{j+1}^m).$$

From this expression, several aspects can be read off. On the one hand, it confirms the statement from Lemma 2.41: For $m \to \infty$, only the summand

$$\frac{w_0^T}{\langle w_0^T, \vec{1}\rangle} P$$

remains. Furthermore, as $m \to \infty$, the term $\mathcal{O}(\lambda_{t+1}^m)$ tends to zero faster than

$$\lambda^m \left(V_{(:,2)}P'_{(2,:)} + \dots + V_{(:,j+1)}P'_{(j+1,:)}\right), \tag{2.2}$$

so that the local asymptotic behavior is governed by the expression (2.2). Let us examine this more closely. First,

$$\left(V_{(:,2)}P'_{(2,:)} + \dots + V_{(:,j+1)}P'_{(j+1,:)}\right) \in \mathbb{R}^{n\times j} \tag{2.3}$$

and thus represents $n$ control points in $\mathbb{R}^j$. If $j < t$, then the local asymptotic behavior corresponds to an object that does not have the same dimension as the object to be refined. The vectors $V_{(:,2)}, \dots, V_{(:,j+1)}$ span the eigenspace corresponding to the eigenvalue $\lambda$, and with the matrix $P'$ we obtain $j$ vectors from this eigenspace. If the dimension of the eigenspace is less than $t$, the local asymptotic behavior cannot have dimension $t$. For example, if $t = 3$ and $j = 1$, the initially three-dimensional object converges after rescaling to something one-dimensional. This can be seen in Figure 2.23.

Thus, we also obtain the justification for Quality Criterion Q4. The local asymptotic behavior of the refinement should have the same dimension as the original object, and this is only possible if the algebraic and geometric multiplicity of the subdominant eigenvalue is exactly $t$.

Quality Criterion Q4 is important for the application of subdivision in the context of simulation. When using the generated surface or volume as a physical domain for a differential equation, it is crucial that the volumes are indeed volumetric. Elements that approximate objects of lower dimension lead to nearly singular matrices in the simulation.

To examine this process more precisely, we consider the eigenstructure from a somewhat different perspective. Let $P' \in \mathbb{R}^{n\times t}$ with $VP' = P$ (thus, we consider the special case of surfaces in $\mathbb{R}^2$ or volumes in $\mathbb{R}^3$), and assume that Quality Criteria Q1 and Q3 are satisfied and that the eigenspaces of $\lambda_1, \dots, \lambda_t$ are full dimensional. Then $S$ has the eigenvalue 1 with eigenvector $\vec{1}$, and all other eigenvalues have modulus less than 1. We assume the eigenvalues and eigenvectors are ordered as

$$1 > |\lambda_1| \geq |\lambda_2| \geq \dots$$

and that the largest three eigenvalues (less than 1) each have maximal geometric multiplicity. Consider now the matrix

$$P'_{(2:t+1,:)}.$$

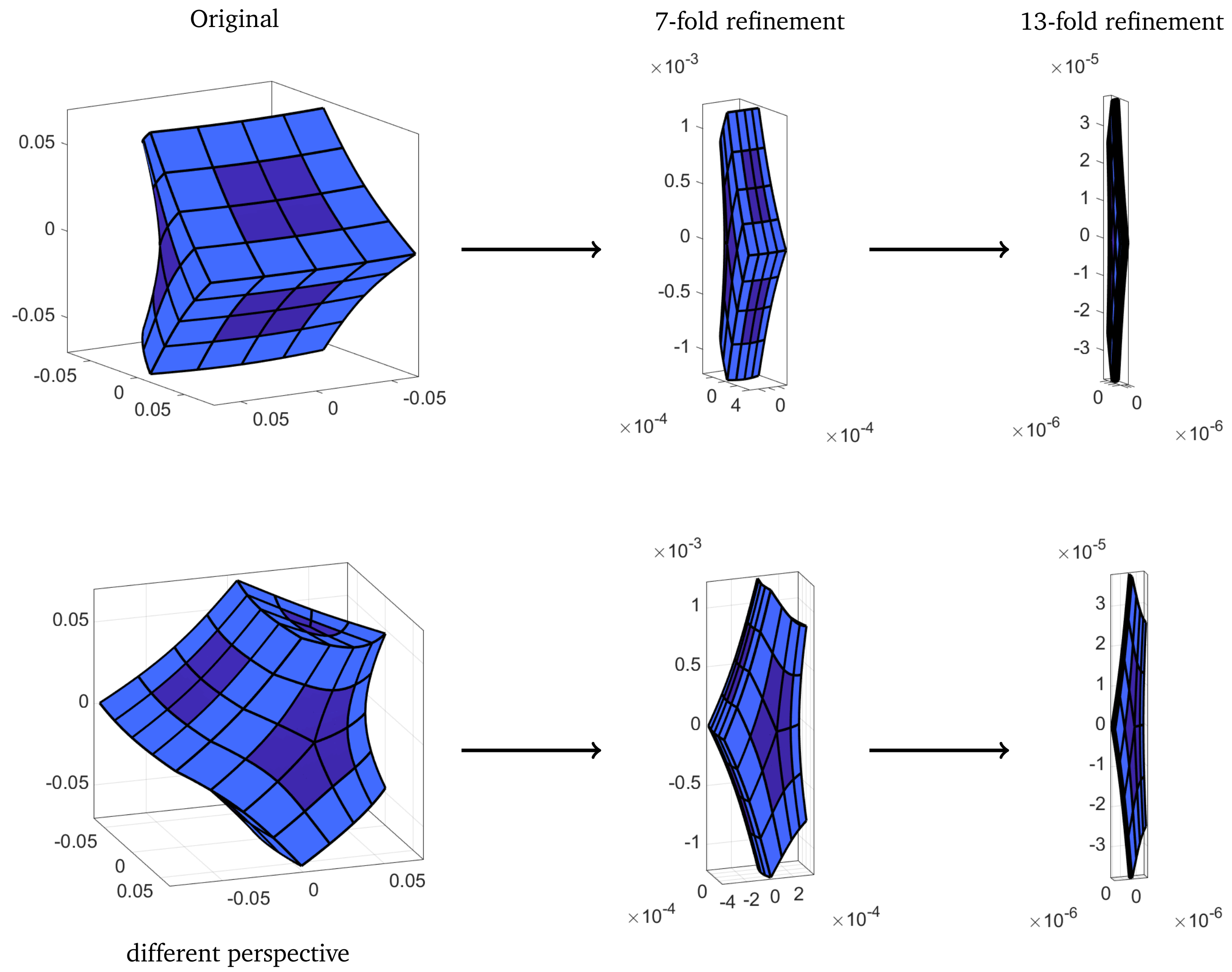

**Figure 2.23:** The combinatorial arrangement from Figure 2.20 with $t = 3$ and $g = 3$. It is refined using a subdivision matrix with the refinement rules described in [Baj+02]. The eigenvalues of the subdivision matrix are

$$[1, \quad 0.5604, \quad 0.492, \quad 0.4326, \quad \ldots] .$$

Hence, the subdivision matrix has a subdominant eigenvalue with algebraic multiplicity 1, which is why the volumetric structure looks planar after seven refinements and line-like after thirteen refinements.

The $(i, j)$-th entry of this matrix indicates how strongly the $i$-th eigenvector influences the $j$-th coordinate of the control points. If this matrix is invertible, then all $t$ eigenvectors influence the control points. In particular, $P'_{(2:t+1,:)}$ is a basis transformation matrix in $\mathbb{R}^t$, and we can transform the above expression to

$$S^m P = VJ^m P' = VJ^m P' P'^{-1}_{(2:t+1,:)} P'_{(2:t+1,:)} = VJ^m \begin{bmatrix} P'_{(1,:)} \cdot P'^{-1}_{(2:t+1,:)} \\ E_t \\ P'_{(t+2:n,:)} \cdot P'^{-1}_{(2:t+1,:)} \end{bmatrix} P'_{(2:t+1,:)}.$$

Regarding the new basis in $\mathbb{R}^t$, we obtain refinement coordinates

$$\begin{aligned} S^m P_{(:,1)} &= \big(P'_{(1,:)} \cdot P'^{-1}_{(2:t+1,1)} + \lambda_1^m V_{(:,2)} + \mathcal{O}(\lambda_{t+1}^m)\big) P'_{(2:t+1,:)} \\ &\vdots \\ S^m P_{(:,t)} &= \big(P'_{(1,:)} \cdot P'^{-1}_{(2:t+1,t)} + \lambda_t^m V_{(:,t+1)} + \mathcal{O}(\lambda_{t+1}^m)\big) P'_{(2:t+1,:)} \end{aligned} \quad . \tag{2.4}$$

If $P'_{(2:t+1,:)}$ is invertible and $|\lambda_t| > |\lambda_{t+1}|$, then the asymptotic behavior of the subdivision matrix is dominated by

$$S^m P = P'_{(1,:)} + \begin{bmatrix} \lambda_1^m & \dots & \lambda_t^m \end{bmatrix} V_{(:,2:t+1)} P'_{(2:t+1,:)} + \mathcal{O}(\lambda_{t+2}^m).$$

Each component here represents a function. $P'_{(1,:)}$ is the translation in space and simultaneously the convergence point of the subdivision. The eigenvalues $\lambda_1^m, \dots, \lambda_t^m$ scale the eigenvectors. The matrix $P'_{(2:t+1,:)}$ adapts the eigenvectors $V_{(:,2:t+1)}$ to the spatial structure of the object. Intuitively speaking, each configuration of control points $P$ has its own projection in the space spanned by $V_{(:,2:t+1)}$. This projection is thus an affine image of $V_{(:,2:t+1)}$. However, with $V_{(:,2:t+1)}$ we control the shape of the local asymptotic behavior.

Another interesting case is when $d > t$, that is, when the dimension of the space is larger than the type of the subdivision, for example, surfaces in $\mathbb{R}^3$ ($t = 2$, $d = 3$) or volumes in $\mathbb{R}^4$ ($t = 3$, $d = 4$). In this case, the construction proceeds analogously, except that the basis change matrix

$$P'_{(2:t+1,:)}$$

becomes

$$P'_{(2:d+1,:)}.$$

The rest of the construction proceeds similarly with $d + 1$ components. If it still holds that $|\lambda_t| > |\lambda_{t+1}|$, then the new components depend on $\lambda_{t+1}^m, \dots, \lambda_{d+1}^m$. Since these converge faster to zero, the local asymptotic behavior remains as described above. Thus, a surface in $\mathbb{R}^3$ after rescaling still converges to a two-dimensional structure, i.e., to a planar-like object. How this plane lies in space is then determined by $P'_{(2:d+1,:)}$.

As already mentioned, a subdominant eigenvalue with algebraic and geometric multiplicity $t$ allows for a local asymptotic behavior of the desired dimension for "almost all" configurations. Here, "almost all" means that $t$ columns of $P'_{(2:t+1,:)}$ are linearly independent. This is also required in [PR08, p. 84]. We consider two special cases to illustrate this.

The control points $P$ can be chosen so that $P'_{(2:t+1,:)} = 0$, for example, by choosing control points that are exactly eigenvectors of $S$ not associated with the first $t + 1$ largest eigenvalues. In this case, the local asymptotic behavior is determined by $\mathcal{O}(\lambda_{t+1}^m)$. This behavior is natural and cannot be avoided, since it is precisely the property of eigenvectors that

$$SV = \lambda_i V$$

holds. Furthermore, the control points can be chosen so that their columns are linearly dependent. In this case, the dimension of the original object is already smaller than $t$. The subdivision matrix thus does not, for example, turn an initially two-dimensional object into a three-dimensional one.

We conclude this criterion with several concepts that we will need later.

**Definition 2.44.** *Let $S \in \mathbb{R}^{n\times n}$ be a subdivision matrix of type $t$ with dominant eigenvalue $1$ corresponding to the eigenvector $\vec{1}$, and let the eigenvalues be ordered by magnitude, i.e.,*

$$1 = \lambda_0 = |\lambda_0| > |\lambda_1| \geq \cdots \geq |\lambda_{n-1}|.$$

*Assume furthermore that quality criteria Q1 and Q3 hold. Let $V$ be the matrix from the Jordan decomposition of $S$, so that $S = VJV^{-1}$.*

- *For $t = 2$, let $\boldsymbol{D}$ be a spline definition domain of a (double) ring, and $S$ a subdivision matrix that, as described in Section 2.3, maps a (double) ring onto a (double) ring. Let $\boldsymbol{g}$ be a system of generating functions and $(S, \boldsymbol{g})$ a subdivision algorithm. Replacing the control points in the subdivision evaluation function by $V_{(:,i)}$ for $i = 1, \ldots, n$, we call*

  $$\boldsymbol{s} : \boldsymbol{D} \to \mathbb{R}^d, \quad \boldsymbol{s}(x) \mapsto \boldsymbol{g}^T(x)V_{(:,i)}$$

  *an* eigenring *of the subdivision algorithm $(S, \boldsymbol{g})$ (cf. [PR08, Def. 4.16, p. 74]).*

- *For $t = 3$, let $\boldsymbol{D}$ be a spline definition domain of a (double) shell, and $S$ a subdivision matrix that, as described in Section 2.3, maps a (double) shell onto a (double) shell. Let $\boldsymbol{g}$ be a system of generating functions and $(S, \boldsymbol{g})$ a subdivision algorithm. Replacing the control points in the subdivision evaluation function by $V_{(:,i)}$ for $i = 1, \ldots, n$, we call*

  $$\boldsymbol{s} : \boldsymbol{D} \to \mathbb{R}^d, \quad \boldsymbol{s}(x) \mapsto \boldsymbol{g}^T(x)V_{(:,i)}$$

  *an* eigenshell *of the subdivision algorithm $(S, \boldsymbol{g})$.*

*If furthermore*

$$1 = \lambda_0 = |\lambda_0| > |\lambda_1| \geq |\lambda_t| > |\lambda_{t+1}|,$$

*then we define:*

- *The combination of eigenrings or eigenshells from $V_{(:,2:t+1)}$, that is specifically*

  $$\boldsymbol{s} : \boldsymbol{D} \to \mathbb{R}^d, \quad \boldsymbol{s}(x) \mapsto \boldsymbol{g}^T(x)V_{(:,2:t+1)}$$

  *is called the* characteristic ring *for $t = 2$, or the* characteristic shell *for $t = 3$ of the subdivision algorithm $(S, \boldsymbol{g})$ (cf. [PR08, Defs. 5.4, p. 85 and 5.10, pp. 92–93]). If $t$ is unspecified, the above is called the* characteristic map *of the subdivision algorithm $(S, \boldsymbol{g})$.*

- *Let $A$ be the union of all adjacency matrices of the initial elements contained in $S$, and let $\boldsymbol{G} = (\boldsymbol{V}, \boldsymbol{E})$ be the associated graph from $A$, where $\boldsymbol{V}$ is identified with the corresponding rows of $V_{(:,2:t+1)}$. Then we call $\boldsymbol{G}$ the* characteristic graph *of $S$. The realizations of the graph in $\mathbb{R}^t$, with nodes corresponding to the rows of $V_{(:,2:t+1)}$ and edges given by the shortest connections between these nodes, is called the* characteristic lattice *of $S$.*

## Q5 Subdominant Eigenvalue $1/2$ with Algebraic and Geometric Multiplicity $t$

**Definition 2.45.** *Let $S \in \mathbb{R}^{n\times n}$ be a subdivision matrix with eigenvalues $\lambda_0, \ldots, \lambda_{n-1} \in \mathbb{C}$ ordered such that $|\lambda_0| \geq \cdots \geq |\lambda_{n-1}|$. Furthermore, let $t \in \{1, 2, 3\}$ denote the type of the evaluated elements. The matrix $S$ satisfies quality criterion* Q5 subdominant eigenvalue $1/2$ with algebraic and geometric multiplicity $t$ *if it satisfies quality criterion Q4 and the subdominant eigenvalue $\lambda$ is exactly $1/2$.*

This criterion is an extension of quality criterion Q4. As explained before, the eigenspace of the subdominant eigenvalue describes the local asymptotic behavior of the refinement. The subdominant eigenvalue itself describes the scaling behavior. Intuitively, it determines how large the next invariant element is compared to the previous one. The value $1/2$ is the natural scaling factor. We explain this in the following:

A cell is refined into $2^t$ cells. New points are inserted at the midpoints of edges, faces, and (for $t = 3$) volumes. Likewise, the spline definition domain is subdivided into equally sized parts. Correspondingly, the scaling factor $1/2$ divides the structure of control points into equal parts.

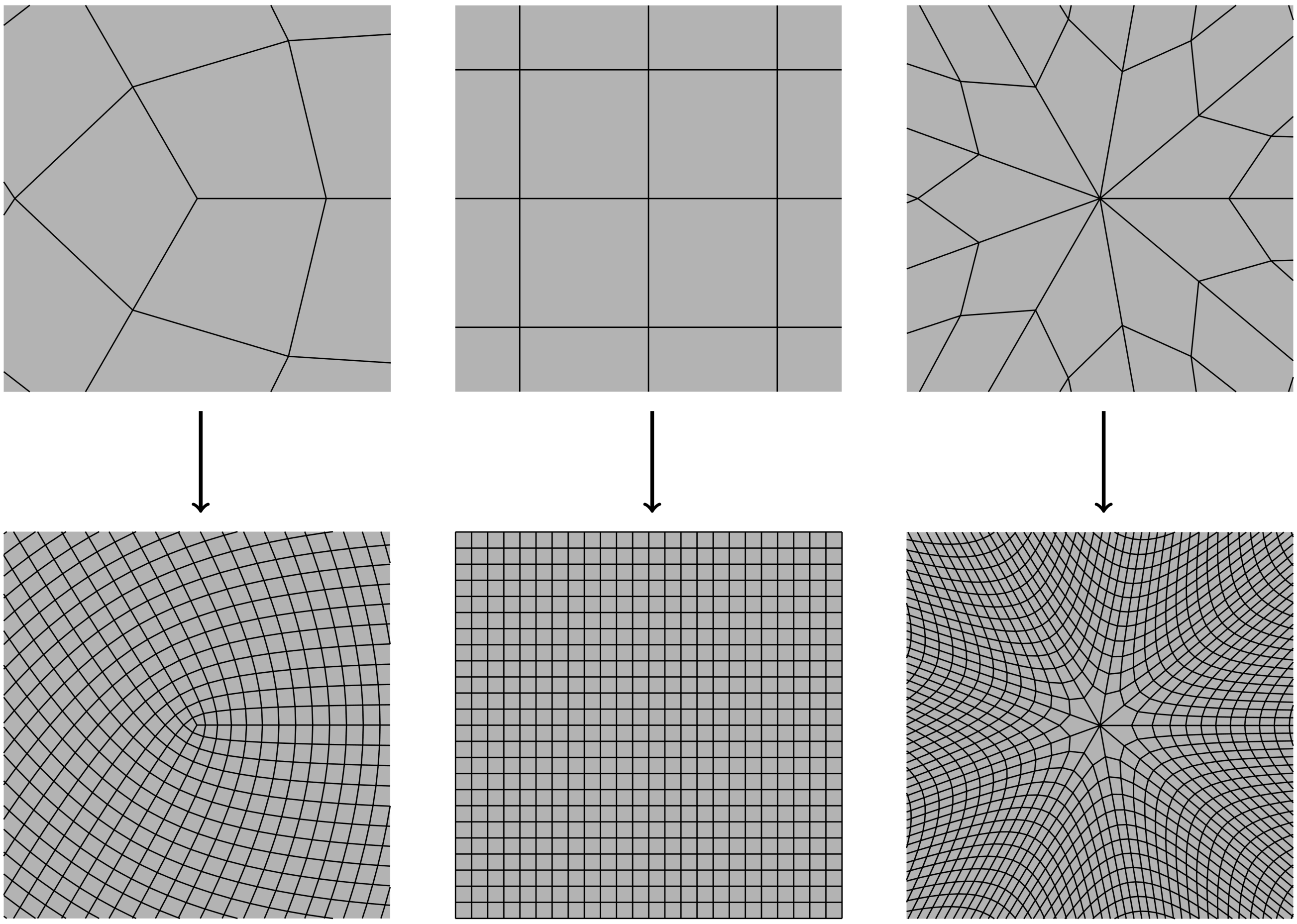

**Figure 2.24:** Three initial meshes for $t = 2$ and $g = 3$, each with one (ir)regular corner of valences 3, 4, and 9, refined using the Catmull-Clark algorithm [CC78]. The subdivision matrices have (double) subdominant eigenvalues $\lambda \approx 0.4101$ (left), $\lambda = 0.5$ (middle), and $\lambda \approx 0.6199$ (right). It can be seen that the left mesh contracts, the middle mesh refines evenly, and the right mesh expands.

The opposite effect can be seen in Figure 2.24. Here, three meshes are refined using the Catmull-Clark algorithm [CC78], which does not satisfy this quality criterion. It can be observed that the initial mesh is therefore not refined uniformly. For an irregularity involving three cells, the mesh contracts around the irregular corner, and for an irregularity involving nine cells, the mesh expands around the irregular corner.

## Q6 Subsubdominant eigenvalue $\mu$ with $\mu = \lambda^2$

**Definition 2.46.** *Let $S \in \mathbb{R}^{n \times n}$ be a subdivision matrix with eigenvalues $\lambda_0, \dots, \lambda_{n-1} \in \mathbb{C}$ ordered such that $|\lambda_0| \geq \dots \geq |\lambda_{n-1}|$. The matrix $S$ satisfies the quality criterion* Q6 subsubdominant eigenvalue $\mu$ with $\mu = \lambda^2$ *if it fulfills quality criterion Q4 with subdominant eigenvalue $\lambda$ and if $S$ has a subsubdominant eigenvalue $\mu$ such that $\mu = \lambda^2$.*

This quality criterion is rather peripheral in this work. For the case $t = 2$, it is a necessary condition for a subdivision matrix that generates a surface that is $C^2$ at irregular vertices (cf. [PR08, Thm. 7.15, p. 141]).

Our constructions in Chapters 5 and 6 will not satisfy this criterion. However, in general, a small subsubdominant eigenvalue is beneficial. The greater the gap to the subdominant eigenvalue, the faster the evaluated object converges to an element in the eigenspace of the subdominant eigenvalue. Intuitively, this means that the local asymptotic behaviour described in quality criterion Q4 sets in faster with a smaller subsubdominant eigenvalue because the powers of the other eigenvalues decay faster.

## Q7 Reproduction of the regular case

**Definition 2.47.** *Let $S \in \mathbb{R}^{n \times n}$ be a subdivision matrix of type $t$. The matrix $S$ satisfies the quality criterion* Q7 reproduction of the regular case *if every regular initial element included in $S$ is the $t$-fold tensor product of the respective univariate rule.*

This quality criterion is part of the definition of a subdivision matrix in this work. More precisely, it is described in Definitions 2.15 and 2.17. The background is, as already described in Section 2.1, that the tensor product structure in the regular areas allows evaluation as a tensor product B-spline. If this quality criterion is not fulfilled, the evaluated element whose cell consists of $2^t$ regular initial elements would not be reproduced under refinement. This is illustrated in Figure 2.25.

The criterion is therefore necessary for the well-definedness of the refinement. Here, two aspects must be discussed.

The first aspect is the question of which initial elements are regular initial elements. We have answered this question with Definition 2.20. However, for $g = 3$, the subdivision algorithms from [CC78], [JM99] and [Baj+02] produce regular initial elements and cells that, according to this work, would not necessarily have to be regular.

The backgrounds and the advantages and disadvantages of our definition of regular elements will be discussed in Chapter 6. However, it is already important to note that for each type of subdivision matrix it must be explained which areas are regular and can be evaluated.

The second aspect is the "natural reproduction" of the regular case. Subdivision rules and matrices are typically defined by a procedure applied to the respective combinatorial arrangement. If these procedures reproduce the regular subdivision rules for the regular combinatorial arrangement, they arise in a natural manner. However, this is not strictly necessary for the criterion. It is equally valid to specify concrete rules for each or certain combinatorial arrangements that differ from the procedure of others.

## Q8 Reproduction of the Semi-Regular Case

**Definition 2.48.** *Let $S^{(3)} \in \mathbb{R}^{n \times n}$ be a subdivision matrix of type $t = 3$ for a spline definition domain that is the tensor product of a two-dimensional spline definition domain and a one-dimensional spline definition domain. The matrix $S^{(3)}$ satisfies the quality criterion* Q8 reproduction of the semi-regular case *if there exists a permutation matrix $H \in \{0,1\}^{n \times n}$ such that $HS^{(3)}H^{-1}$ is the Kronecker product of the subdivision matrix $S^{(2)}$ corresponding to the*

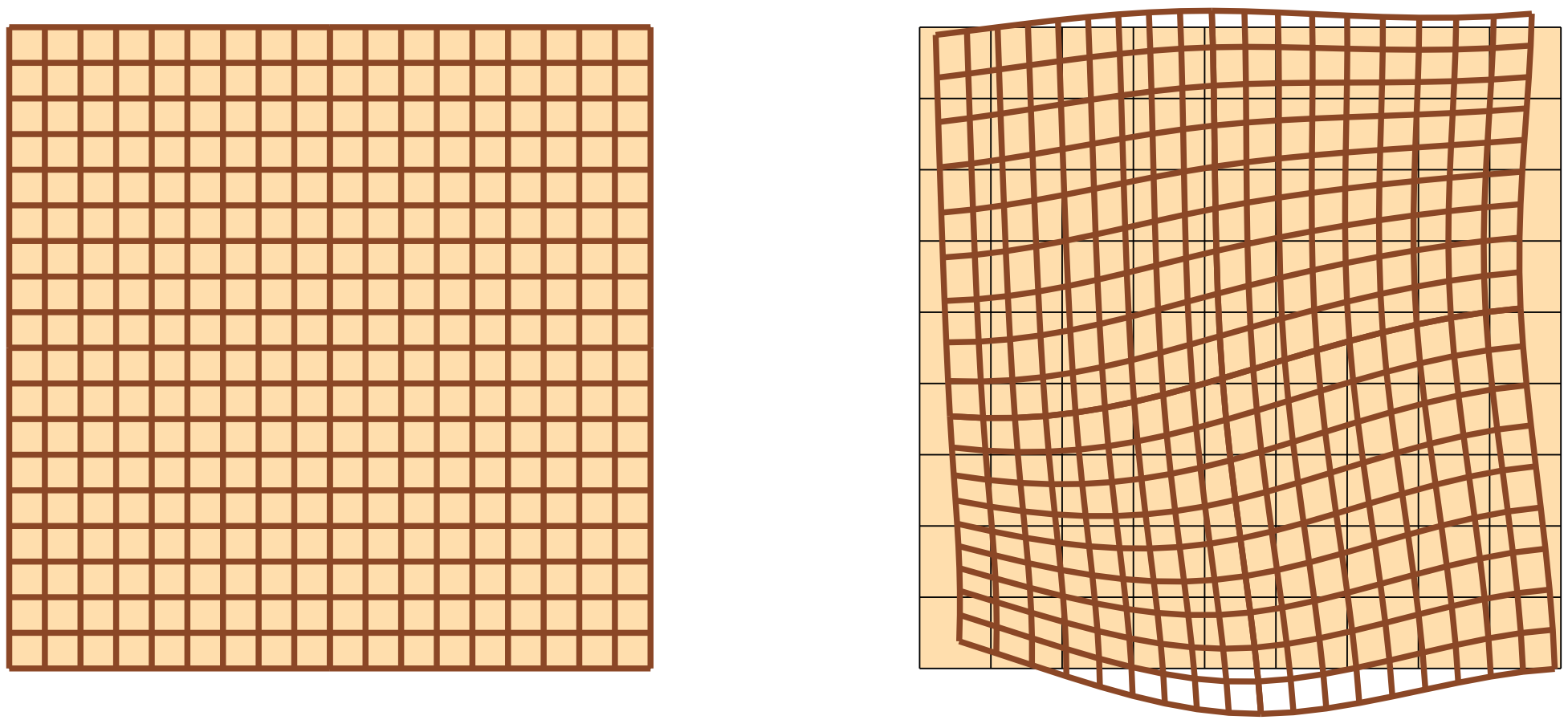

**Figure 2.25:** Evaluation of a regular cell for $t = 2$ and $g = 3$ (yellow) and the evaluation of its refinement as a mesh (brown). On the left, the regular case is reproduced, with the brown mesh lying exactly over the yellow surface. On the right, the regular case is not reproduced, and the brown mesh deviates from the yellow surface.

*two-dimensional structure and the subdivision matrix $S^{(1)}$ corresponding to the one-dimensional structure. Concretely, this means: The matrix $HS^{(3)}H^{-1}$ is the Kronecker product of $S^{(1)} \in \mathbb{R}^{n \times n}$ and $S^{(2)} \in \mathbb{R}^{m \times m}$, i.e.,*

$$HS^{(3)}H^{-1} = \text{kron}\left(S^{(1)}, S^{(2)}\right) = \begin{bmatrix} S^{(1)}_{(1,1)}S^{(2)} & \dots & S^{(1)}_{(1,n)}S^{(2)} \\ \vdots & & \vdots \\ S^{(1)}_{(n,1)}S^{(2)} & \dots & S^{(1)}_{(n,n)}S^{(2)} \end{bmatrix} \in \mathbb{R}^{n \cdot m \times n \cdot m}.$$

This somewhat cumbersome definition has a simple intuition. For $t = 2$, various subdivision matrices can be defined for different invariant regions. In Section 2.3, we have seen this for initial elements, cells around an irregular corner, rings, and double rings. These structures can be transferred to three dimensions by combinatorially stacking layers of these two-dimensional structures in the third dimension. For example, from $n$ two-dimensional cells sharing a corner, a second layer of $n$ hexahedra sharing an irregular edge arises, and a third layer of $2 \times n$ hexahedra, each sharing $n$ an irregular edge and all $2n$ a semi-irregular corner. Stacking these layers corresponds exactly to the tensor product of the two-dimensional structure with a one-dimensional structure. Illustrations can be found in Figure 2.26.
Combining two-dimensional spline definition domains with a one-dimensional spline definition domain, one can generate three-dimensional spline definition domains. These tensor product structures can also be parts of a larger three-dimensional structure. For example, consider the shells in Figures 2.19 and 2.20. Each evaluated (volumetric) facet of the shell is a combination of a two-dimensional and a one-dimensional structure. If the subdivision matrices and thus the refinement rules of these (sub)structures are tensor products of one-dimensional and two-dimensional subdivision matrices, then they satisfy Quality Criterion Q8. One can also say that the three-dimensional subdivision rules are generalizations of the corresponding two-dimensional ones. The permutation in the definition just described is necessary because the order of the rows and columns of the subdivision matrix is irrelevant for the tensor product structure. They only need to be arranged so that this structure becomes recognizable.

This quality criterion is not strictly necessary. For example, the subdivision matrices from [JM99] do not satisfy this criterion. This was already noted in [Baj+02, p. 353], whose subdivision matrices do satisfy the criterion. The advantage of subdivision matrices satisfying this criterion is, as already described in Section 2.3, that the tensor product structure can be found in all areas (graph, adjacency matrix, subdivision matrix). The description of the structures and especially the analysis can thus be represented as a tensor product of a two-dimensional and a one-dimensional structure. This provides access to theorems of one- and two-dimensional subdivision matrices for these regions. In particular, this allows the holes in the shells to be evaluated by two-dimensional methods.

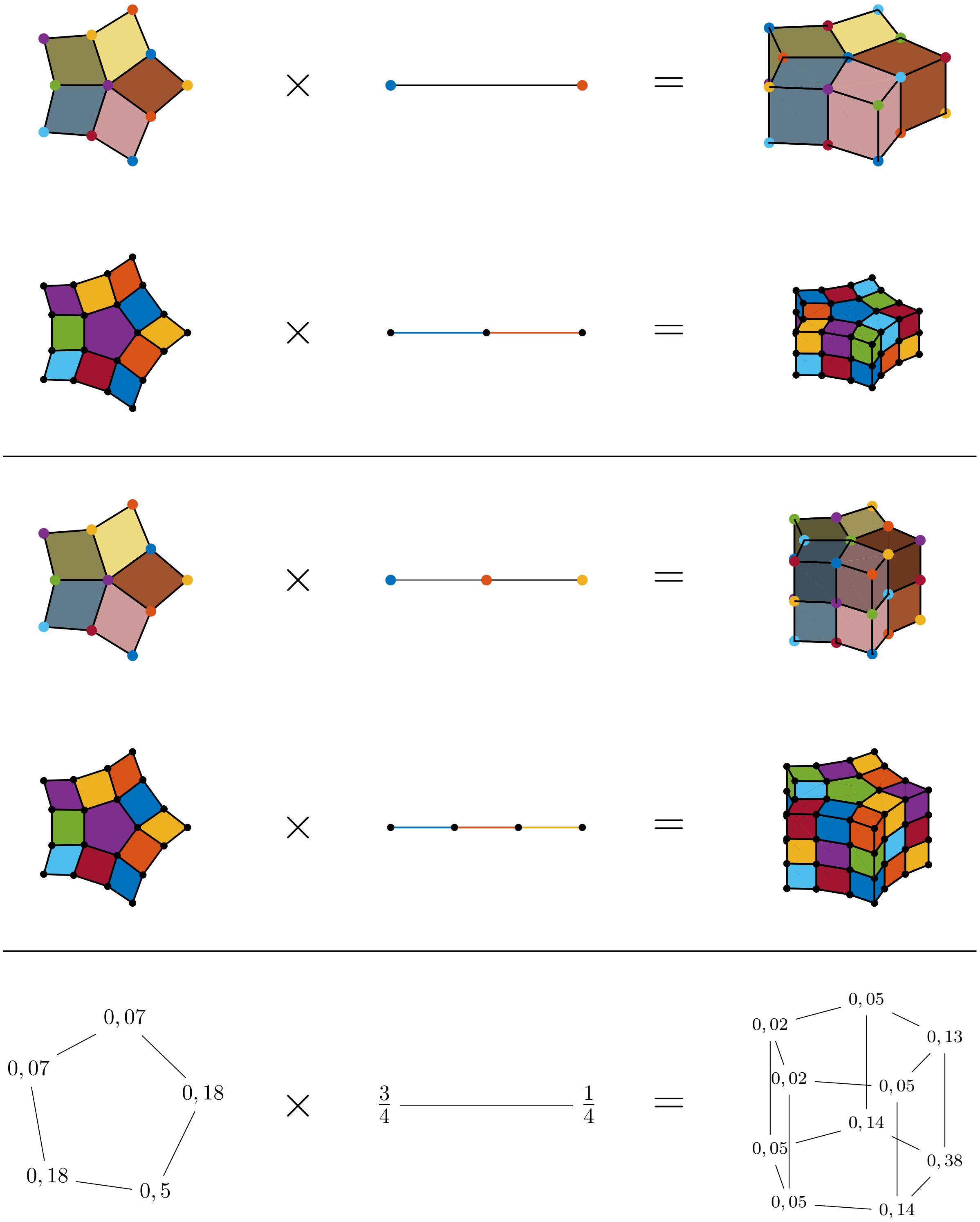

**Figure 2.26:** Three examples of the tensor product structure of a two-dimensional and a one-dimensional object. At the top, five two-dimensional cells (left) sharing an irregular corner are shown. Combined with a one-dimensional cell, this results in five three-dimensional cells sharing an irregular edge. In the row below, the corresponding initial elements are illustrated as an example. The middle example is analogous. Here, the combination of five two-dimensional and two one-dimensional cells yields $2 \cdot 5$ three-dimensional cells sharing an irregular corner. At the bottom, the combination of two concrete refinement rules is shown, each corresponding to a row of the respective subdivision matrices. Combined, they form a three-dimensional refinement rule that constitutes a row of the combined subdivision matrix.

It is worth noting that Quality Criterion Q7 is precisely a special case of Quality Criterion Q8. Translated, the former requires that the subdivision matrix of a structure composed of the tensor product of three one-dimensional structures is exactly the Kronecker product of the subdivision matrices of these three one-dimensional structures.

## Q9 For every combinatorial arrangement, a subdivision matrix can be generated

**Definition 2.49.** *Let **I** be an initial element of a spline definition domain **D**, which is a corner of $n \in \mathbb{N}$ cells. The algorithm for creating subdivision matrices satisfies Quality Criterion* Q9 for every combinatorial arrangement a subdivision matrix can be generated*, if for every* valid *combinatorial configuration of cells sharing a common corner, a subdivision matrix for the common initial element **I** can be generated.*

The term to discuss here is *valid*. To this end, we first define:

**Definition 2.50.** *Let $t = 2$ [$t = 3$] and **I** be an initial element of a spline definition domain **D**, which is a corner of $n \in \mathbb{N}$ cells. Then the* combinatorial graph *of **I** is the graph $\boldsymbol{G}_K = (\boldsymbol{V}_K, \boldsymbol{E}_K)$ with vertex set $\boldsymbol{V}_K := \{v_1, \ldots, v_n\}$ and edge set*

$$\boldsymbol{E}_K := \{\{v_i, v_j\} \mid \textit{cells with indices } i \textit{ and } j \textit{ share a common edge [facet] containing the corner}\}.$$

With this definition, we can describe the notion of validity:

**Definition 2.51.** *Let **I** be an initial element of a spline definition domain **D**, which is a corner of $n \in \mathbb{N}$ cells. The initial element, the combinatorial configuration, and the combinatorial graph $\boldsymbol{G}_K$ of **I** are* valid *if $\boldsymbol{G}_K$*

- *is a cycle (in the graph-theoretical sense according to Definition A.3) for $t = 2$,*
- *and is planar and 3-connected for $t = 3$.*

For $t = 2$, this definition is unsurprising. Due to Definition 2.4, no other combinatorial arrangements are possible. The definition for $t = 3$ is, however, at least unusual and deviates from definitions of subdivision matrices found in the literature. For example, in [JM99, p. 20] and [Baj+02, p. 353], all hexahedral meshes are valid, meaning all configurations in which the cells are hexahedra. For the combinatorial graph $\mathbf{G}_K$, this means that each node is incident to exactly three edges. Why should we deviate from this definition?

First, we note that planar 3-connected graphs are exactly the combinatorial structures of three-dimensional convex polytopes (see Theorem 3.8). To answer the question just posed, we compare the two definitions. On the one hand, not all combinatorial graphs of hexahedral initial elements (initial elements that lie exclusively on cells as required in Definition 2.4) are planar and 3-connected. On the other hand, not all planar 3-connected combinatorial structures correspond to hexahedral initial elements. Thus, the two sets have a non-empty intersection (for example, the graph corresponding to a hexahedron), but neither is a subset of the other. Therefore, we consider an example of a combinatorial graph where all nodes are incident to exactly three edges, but which is neither planar nor 3-connected.

**Example 2.52.** *We consider the graph $\boldsymbol{G}_K$ with associated adjacency matrix $A$. The graph $\boldsymbol{G}_K$ can also be described by the cycles encoded in the rows of $M$. Specifically, the two matrices*

$$\begin{bmatrix} 1 & 2 & 3 & 15 & 14 & 13 \\ 3 & 4 & 5 & 17 & 16 & 15 \\ 5 & 6 & 1 & 13 & 18 & 17 \\ 7 & 8 & 2 & 1 & 6 & 12 \\ 8 & 9 & 10 & 4 & 3 & 2 \\ 10 & 11 & 12 & 6 & 5 & 4 \\ 11 & 12 & 7 & 14 & 13 & 18 \\ 14 & 15 & 16 & 9 & 8 & 7 \\ 16 & 17 & 18 & 11 & 10 & 9 \end{bmatrix}$$

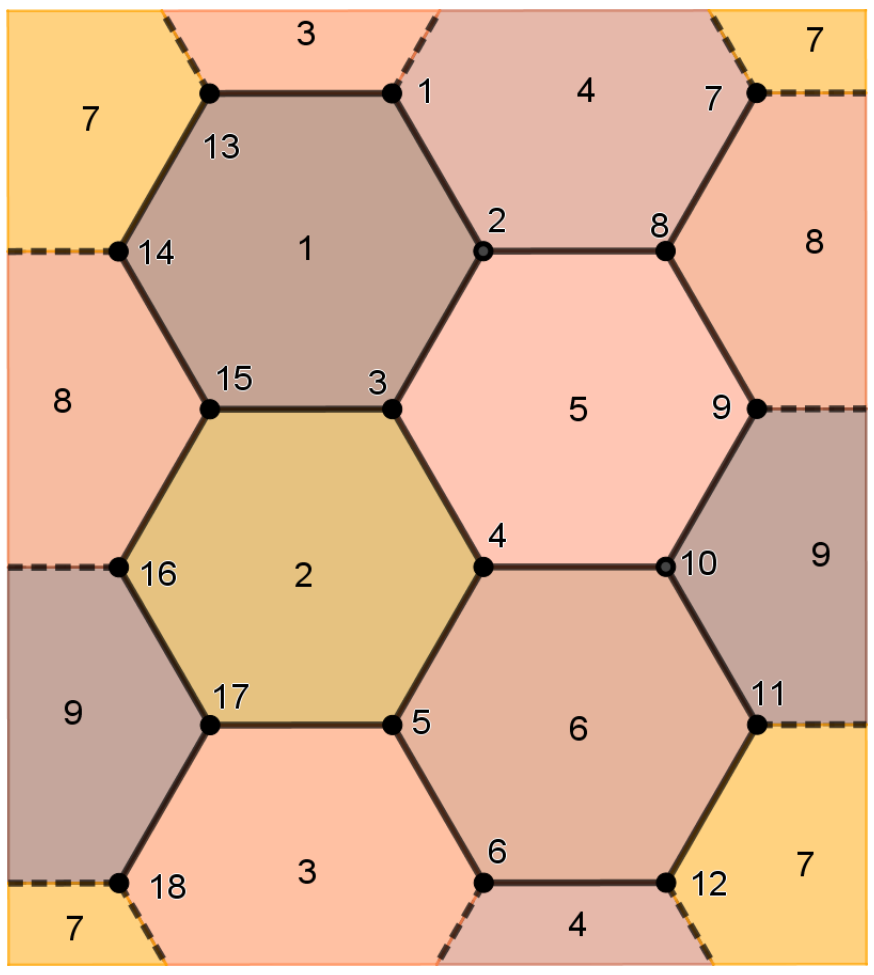

**Figure 2.27:** Illustration of the graph $\mathbf{G}_K$ from Example 2.52. If the figure is first glued together along the left and right edges (slightly shifted), and then along the top and bottom edges, a torus is formed.

*and*

$$A := \begin{bmatrix}
& 1 & & & & 1 & & & & & & & 1 \\
1 & & 1 & & & & & 1 \\
& 1 & & 1 & & & & & & & & & & & 1 \\
& & 1 & & 1 & & & & & 1 \\
& & & 1 & & 1 & & & & & & & & & & & 1 \\
1 & & & & 1 & & & & & & & 1 \\
& & & & & & & 1 & & & & 1 & & 1 \\
& 1 & & & & & 1 & & 1 \\
& & & & & & & 1 & & 1 & & & & & & 1 \\
& & & 1 & & & & & 1 & & 1 \\
& & & & & & & & & 1 & & 1 & & & & & & 1 \\
& & & & & 1 & 1 & & & & 1 \\
1 & & & & & & & & & & & & & 1 & & & & 1 \\
& & & & & & 1 & & & & & & 1 & & 1 \\
& & 1 & & & & & & & & & & & 1 & & 1 \\
& & & & & & & & 1 & & & & & & 1 & & 1 \\
& & & & 1 & & & & & & & & & & & 1 & & 1 \\
& & & & & & & & & & 1 & & 1 & & & & 1
\end{bmatrix}.$$

*The structure of $\mathbf{G}_K$ is shown in Figure 2.27 and can be interpreted as a torus consisting of 9 hexagons. In this example, each vertex is incident to exactly three edges, making it admissible for the algorithms in [JM99] and [Baj+02]. However, the graph is not planar and therefore not valid according to this work. Constructing the subdivision matrix for this example following [Baj+02] yields eigenvalues*

| *Index* | 0 | 1 | 2 | 3 | 4 | $\cdots$ | 13 | $\cdots$ |
|---|---|---|---|---|---|---|---|---|
| *Eigenvalue* | 1 | 0.5544 | 0.5544 | 0.5113 | 0.5113 | $\cdots$ | 0.25 | $\cdots$ , |

*where $0.25$ is the first simple eigenvalue less than 1. The local asymptotic behavior thus converges towards something surface-like. To get an idea of the shape of the local asymptotic behavior, we have plotted in Figure 2.28 volume evaluations of various combinations of eigenvectors (using the evaluable regions defined in [Baj+02], which, as mentioned, differ from the definitions in this work).*

*We display various combinations here because due to the structure, the subdivision matrix has no unique three largest eigenvalues less than 1. As can be seen, the local asymptotic behavior and thus the subdivision matrix do not represent a meaningful object. The eigen shells are not injective, and the structure intersects itself.*

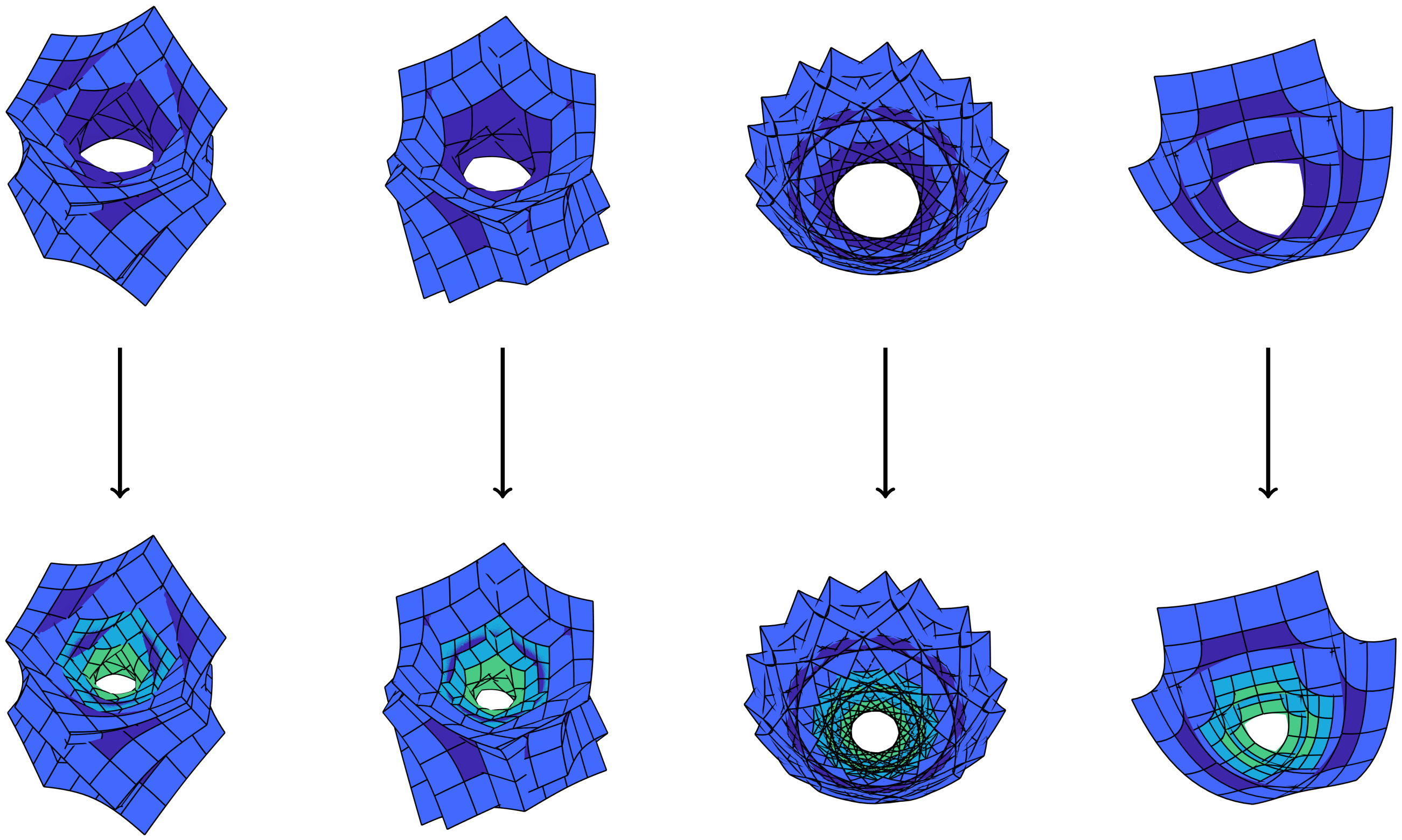

**Figure 2.28:** Illustrations of eigen shells for the eigenvectors with indices $[2, 3, 4]$, $[2, 3, 5]$, $[2, 3, 14]$ and $[4, 5, 14]$ (top) and their refinements (bottom).

*To avoid confusion, note here that although the combinatorial structure is a torus, the evaluated object cannot and should not be a torus since all cells share a single corner. However, it is certainly possible to generate a spline definition domain with cells whose evaluation represents a torus. For that torus, the refinement will not lie within its hole. The special behavior here results from the common corner of the cells.*

The above example is only a special case and not a proof that there is no meaningful example of hexahedral meshes whose combinatorial graph is not planar and 3-connected. However, it is geometrically intuitive that for practical applications, structures composed of hexahedra that do not relate to each other in a polytope-like structure are not meaningful. Therefore, the restriction from graphs with nodes of degree three to graphs that are planar, 3-connected, and with nodes of degree three is not restrictive.

The generalization to graphs that are planar and 3-connected but whose nodes do not necessarily have exactly three edges will be discussed in Chapter 7.

After clarifying which combinatorial configurations are valid, the question arises of how many valid combinatorial configurations there are. Since the set of prisms is valid, the answer is obviously infinite, but one might ask, for example, how many combinatorial configurations exist for a fixed number of nodes or edges. A first answer is given in [DF81]. This publication computes the number of combinatorial configurations with up to 22 edges. For exactly 22 edges, for instance, 485,704 different combinatorial configurations are obtained.

## Q10 The subdivision matrix respects all symmetries of the associated graph structure

**Definition 2.53.** *Let $S \in \mathbb{R}^{n \times n}$ be a subdivision matrix of an initial element or an invariant structure of a (double) ring or a (double) shell with associated graph $\mathbf{G}$. Let furthermore $\mathbf{G}_K$ be the combinatorial graph of the (central) initial element of the invariant structure. The matrix $S$ fulfills the quality criterion* Q10 all symmetries of the associated graph structure are respected *if for every automorphism $\mathbf{h}$ of $\mathbf{G}$ induced by $\mathbf{G}_K$ with corresponding permutation matrix $H$ it holds that*

$$H^{-1}SH = S.$$

This definition needs to be clarified first. Graphs are abstract structures consisting of nodes and edges, and automorphisms are bijective mappings from a graph to itself. Thus, automorphisms represent the symmetries of the underlying graph as described in Section A.1. If $P$ is a matrix of control points associated with the nodes of $\mathbf{G}$, then $P$ is a visualization of $\mathbf{G}$. If $\mathbf{h}$ is an automorphism of $\mathbf{G}$, then the geometric visualization of $P$ and $\mathbf{h}(P)$ is identical. The refinement of this realization should preserve this property. In other words, it should not matter whether one first permutes and then refines or first refines and then permutes. Concretely, this means that

$$S \cdot \mathbf{h}(P) = \mathbf{h}(SP)$$

and hence

$$S(HP) = H(SP).$$

These equations hold if and only if

$$SH = HS \quad \Leftrightarrow \quad H^{-1}SH = S.$$

Therefore, the subdivision matrix must satisfy the same commutation relation with automorphisms as the adjacency matrix does (cf. Characterization A.8).

The expression „induced by $\mathbf{G}_K$ automorphism“ here means that the symmetries represented by the automorphisms should be those of the combinatorial graph $\mathbf{G}_K$. For an initial element with $g = 2$, the graphs $\mathbf{G}$ and $\mathbf{G}_K$ coincide, so there is no restriction. For $g = 3$, the graph $\mathbf{G}$ consists of quadrilaterals or hexahedra. The restriction to the combinatorial graph means that only automorphisms of $\mathbf{G}$ are considered that map quadrilaterals to quadrilaterals and hexahedra to hexahedra. The same applies to the larger (double) ring and (double) shell structures.

If this criterion is not fulfilled, it means that the result of refinement for two combinatorially symmetric configurations depends on the assignment of control points to the nodes of the graph. This on the one hand allows more freedom in the design of subdivision matrices, but on the other hand makes their use significantly more complicated and can be undesirable with respect to application and simplicity. The effects are described in the following example:

**Example 2.54.** *We consider an initial element for $t = 2$ and $g = 2$ with four control points*

$$P = \begin{bmatrix} 1 & 1 \\ 4 & 0 \\ 4 & 6 \\ 0 & 4 \end{bmatrix}$$

*and a graph **G** with adjacency matrix*

$$A = \begin{bmatrix} 0 & 1 & 0 & 1 \\ 1 & 0 & 1 & 0 \\ 0 & 1 & 0 & 1 \\ 1 & 0 & 1 & 0 \end{bmatrix}.$$

*Furthermore, we consider the permutation*

$$\mathbf{h} = \begin{pmatrix} P_{(1,:)} & P_{(2,:)} & P_{(3,:)} & P_{(4,:)} \\ P_{(2,:)} & P_{(3,:)} & P_{(4,:)} & P_{(1,:)} \end{pmatrix} \quad \textit{with associated permutation matrix} \quad H = \begin{bmatrix} 0 & 1 & 0 & 0 \\ 0 & 0 & 1 & 0 \\ 0 & 0 & 0 & 1 \\ 1 & 0 & 0 & 0 \end{bmatrix}.$$

*Since $H^T AH = A$, **h** is an automorphism of **G**. For the initial element, we consider two different subdivision matrices*

$$S_1 = \frac{1}{16} \begin{bmatrix} 9 & 3 & 1 & 3 \\ 3 & 9 & 3 & 1 \\ 1 & 3 & 9 & 3 \\ 3 & 1 & 3 & 9 \end{bmatrix} \quad \textit{and} \quad S_2 = \frac{1}{16} \begin{bmatrix} 9 & 4 & 1 & 2 \\ 4 & 8 & 3 & 1 \\ 1 & 3 & 7 & 2 \\ 6 & 1 & 0 & 9 \end{bmatrix}.$$

*We find that*

$$H^T S_1 H = \frac{1}{16} \begin{bmatrix} 9 & 3 & 1 & 3 \\ 3 & 9 & 3 & 1 \\ 1 & 3 & 9 & 3 \\ 3 & 1 & 3 & 9 \end{bmatrix} = S_1 \quad \textit{but} \quad H^T S_2 H = \frac{1}{16} \begin{bmatrix} 9 & 6 & 1 & 0 \\ 2 & 9 & 4 & 1 \\ 1 & 4 & 8 & 3 \\ 2 & 1 & 3 & 7 \end{bmatrix} \neq S_2.$$

*Thus, $S_1$ satisfies the condition of quality criterion Q10 for the automorphism **h**. To prove the entire criterion, one would have to check all other automorphisms. The matrix $S_2$ does not satisfy quality criterion Q10. We obtain*

$$S_1(HP) = S_1 \cdot \begin{bmatrix} 4 & 0 \\ 4 & 6 \\ 0 & 4 \\ 1 & 1 \end{bmatrix} = \frac{1}{16} \begin{bmatrix} 51 & 25 \\ 49 & 67 \\ 19 & 57 \\ 25 & 27 \end{bmatrix} = H \cdot \frac{1}{16} \begin{bmatrix} 25 & 27 \\ 51 & 25 \\ 49 & 67 \\ 19 & 57 \end{bmatrix} = H(S_1 P)$$

*and*

$$S_2(HP) = S_2 \cdot \begin{bmatrix} 4 & 0 \\ 4 & 6 \\ 0 & 4 \\ 1 & 1 \end{bmatrix} = \frac{1}{16} \begin{bmatrix} 54 & 30 \\ 49 & 61 \\ 18 & 48 \\ 37 & 15 \end{bmatrix} \neq \frac{1}{16} \begin{bmatrix} 48 & 26 \\ 41 & 51 \\ 10 & 42 \\ 29 & 23 \end{bmatrix} = H \cdot \frac{1}{16} \begin{bmatrix} 29 & 23 \\ 48 & 26 \\ 41 & 51 \\ 10 & 42 \end{bmatrix} = H(S_2 P).$$

*The consequences are shown in Figure 2.29. Since the symmetry is not respected for $S_2$, the shape of the refinements of $P$ and $HP$ differ. Hence, despite the symmetry in the graph, the assignment of nodes matters.*

*It should be noted that the initial element described here is a regular initial element, so its refinement matrix according to Definition 2.15 must be exactly $S_1$. The matrix $S_2$ is used here only for illustration, to clarify the system with a simple example.*

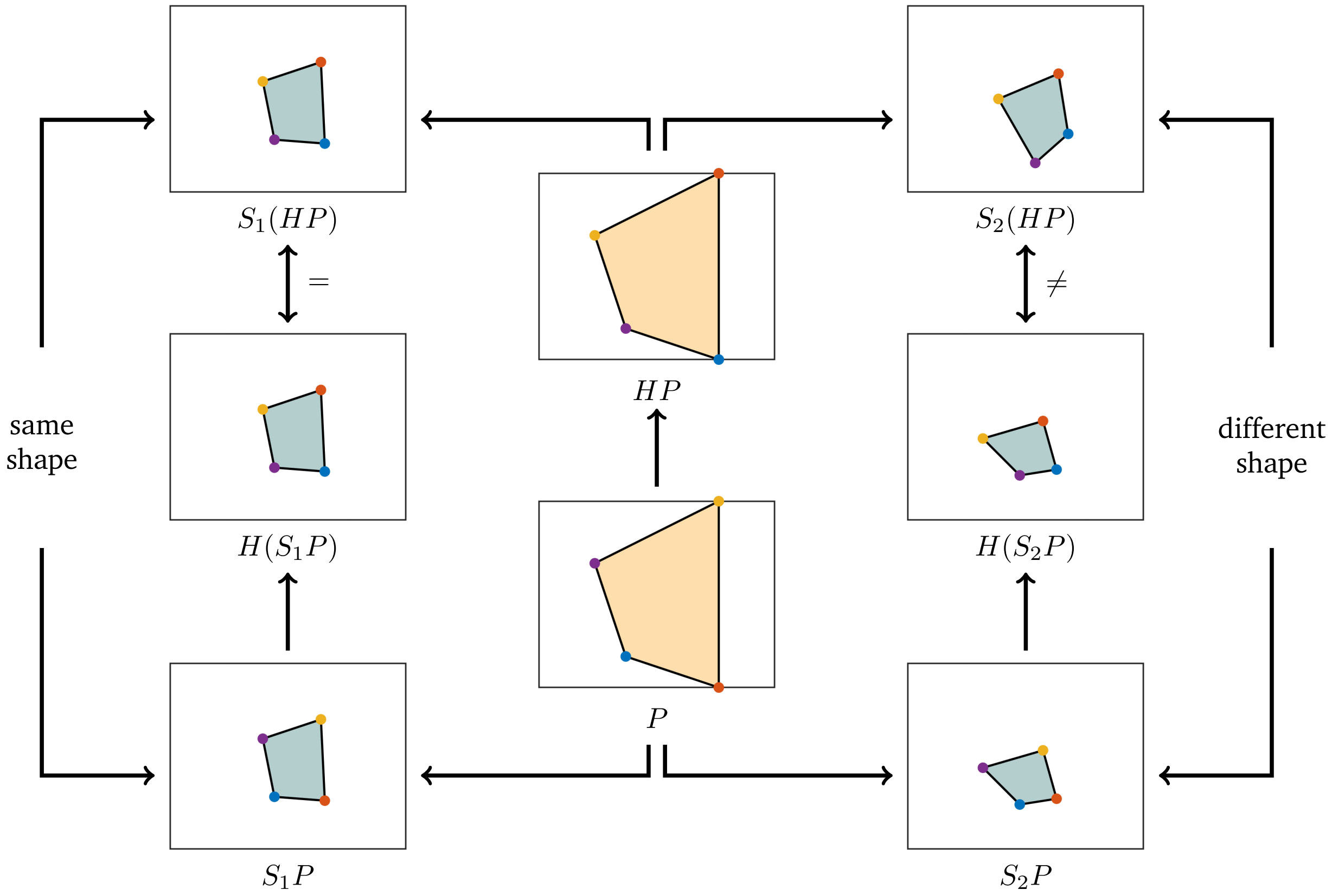

**Figure 2.29:** Illustration of Example 2.54. The initial configuration in the center is refined with $S_1$ on the left and with $S_2$ on the right. Since $S_1$ respects the symmetry, the same shape or object is obtained on the left. On the right, however, the symmetry is not respected, leading to different shapes and objects.

## Q11 Proper Support of the Refinement Rules

**Definition 2.55.** *Let **D** be a spline definition domain with control points $P'$. Let furthermore $\mathbf{D}^{(2)}$ and $P'^{(2)}$ denote their respective refinements. The algorithm for constructing the subdivision matrices fulfills quality criterion* Q11 proper support of the refinement rules*, if the refinement rule for every control point in $P'^{(2)}$, which arises from the refinement of several initial elements, consists exclusively of control points that are common to all these initial elements.*

At first glance, this quality criterion may seem self-evident and superfluous. For $g = 2$, it is always fulfilled since no refined control point originates from more than one initial element. Also, for the case $g = 3$, we have already established in Remark 2.24 that the refinement rules for overlapping initial elements must be identical to ensure well-defined refinement.

Although the definition of initial elements in this work is chosen sensibly, it could be defined differently, since control points are merely mapped linearly to new control points. If one regards the refinement of all control points as a large matrix, the entries of this matrix can theoretically be chosen freely. Thus, the refinement can be considered a global object and thereby represented as a matrix.

In this context, the refinement rules for initial elements at irregular corners for $g = 3$ are of particular interest. As an example, consider the case $t = 2$ and an irregular corner of valence 5. This yields an initial element with eleven control points. Exemplarily, we embed this initial element in a spline definition domain with an irregular corner, whose evaluable area corresponds to a double ring. The structure and control points are shown in Figure 2.30.

The described initial element has three different types of control points. First, points 1 to 5, which are not connected to point 11. Second, points 6 to 10, which are connected to point 11, and third, the central point 11. We now consider the refinements of these eleven control points with respect to the irregular initial element. The explanations in the following paragraphs are illustrated in Figure 2.30.

The refinement of point 1 arises from the initial elements of the corners associated with control points 1, 6, 10, and 11. These share the control points 1, 6, 10, and 11. According to quality criterion Q11, the refinement rule may therefore only consist of these four points. Points 2 to 5 behave analogously.

The refinement of point 10 arises from the initial elements of the corners associated with control points 10 and 11. These share the control points 1, 5, 6, 9, 10, and 11. According to quality criterion Q11, the refinement rule may therefore only consist of these six points. Points 6 to 9 behave analogously.

The refinement of point 11 arises only from the initial element of the corner associated with control point 11. Therefore, according to quality criterion Q11, the refinement rule may consist of all eleven points. This yields a subdivision matrix with a ”sparse” structure, as illustrated in Figure 2.30.

For various reasons, for example because more freedom in designing the subdivision matrix is desired or because the spectrum should be better controlled, it may be desired that the matrix described above is fully populated. This contradicts, as already mentioned, the structure described in this work.

The actual disadvantage is that the support of the generating functions belonging to the control points increases. This is intuitive. If a control point influences additional refinement rules, it also influences additional refined control points and thereby the evaluations of the cells that have these refined control points in their initial elements. Thus, adjusting the subdivision matrix of the central initial element leads to the support of the associated generating functions no longer being a $n$-gon but having indentations. The phenomenon is shown in Figure 2.31 and applies correspondingly to other configurations in two dimensions and to irregular corners in three dimensions.

The effects of such generating functions are not considered in this work. However, this behavior shows that the definitions are chosen meaningfully and that the quality criterion is justified.

## Q12 A Self-Penetration Free Eigengrid

**Definition 2.56.** *Let $S \in \mathbb{R}^{n \times n}$ be a subdivision matrix of an initial element or an invariant structure analogous to Definition 2.32 with associated graph **G**, for which Quality Criterion Q3 holds. Let $t$ be the type of the evaluated elements.*

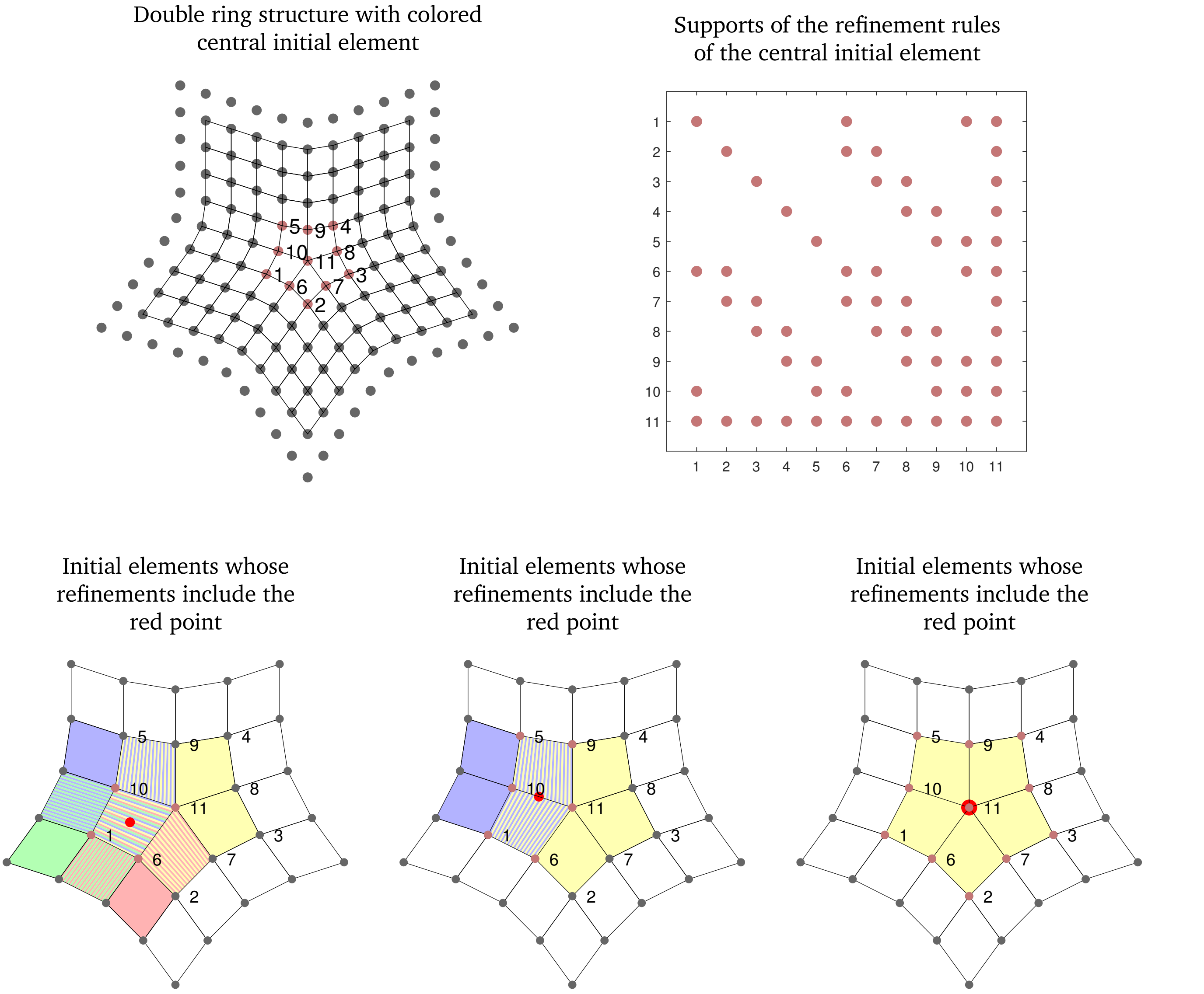

**Figure 2.30:** A double ring structure for $t = 2$ and $g = 3$ with a colored and numbered central initial element is shown exemplarily in row 1 (left). The inevitable supports of the refinement rules are depicted as a matrix in row 1 (right). The reasoning for the supports is shown in row 2. Here, for the red points, the initial elements whose refinements include the red points are displayed. For a facet point, these are four initial elements; for an edge point, two initial elements; and for the central point, one initial element. Accordingly, the refinement rules may only consist of the intersection of the points of these initial elements. The points of the intersection are colored light red. In the left figure these are four points, in the middle six points, and in the right eleven points.

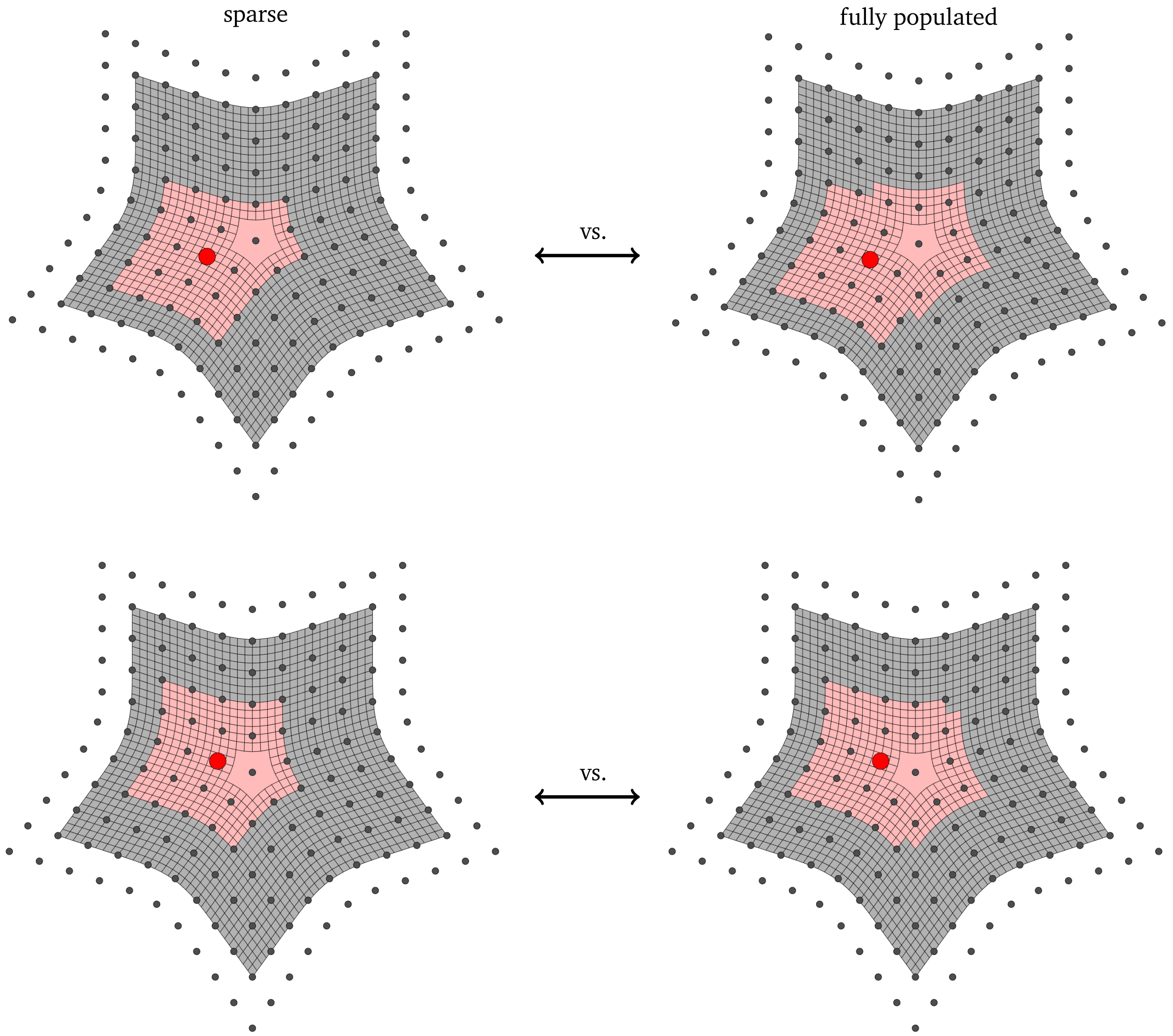

**Figure 2.31:** Support of the generating function corresponding to the red control point (light red). The spline domain was refined sufficiently many times to determine the support precisely. On the left, the support for the sparse version of the subdivision matrix shown in Figure 2.30 is displayed. On the right, the support for a fully populated variant is shown. It is visible that in the fully populated case, the support of the generating function develops inward indentations.

*Furthermore, assume the eigenvalues of $S$ satisfy*

$$1 = \lambda_0 = |\lambda_0| > |\lambda_1| \geq \cdots \geq |\lambda_t| > |\lambda_{t+1}| \geq \cdots \geq |\lambda_{n-1}|.$$

*The matrix $S$ satisfies the quality criterion* Q12 a self-penetration free eigen grid *if the eigenspace of the subdominant eigenvalue has a basis whose characteristic grid*

- *for $t = 2$ satisfies that two lines intersect only at endpoints,*
- *for $t = 3$ satisfies that the convex hulls (cf. Definition 3.2) of two facets intersect only at boundary edges or vertices,*

*and different control points have different coordinates.*

For $t = 3$ and $g = 2$, the term facet in the above definition refers to the facets of the polytopes of all initial elements of the underlying structure. For $t = 3$ and $g = 3$, the term facet refers to all hexahedra of the underlying structure.

This quality criterion is quite restrictive, but cannot be formulated otherwise. The idea is that the grid structure of the eigenspace is free of self-penetrations, i.e., it is not interwoven or knotted. For $t = 2$, this can be checked by verifying that two lines do not cross. The representation of the graph by the eigenspace must thus be planar. For $t = 3$ this is more complicated. Here one should imagine a structure where facets are not penetrated by other elements such as edges, control points, or other facets. Since facets do not necessarily lie in a single plane, we consider their convex hulls. Examples can be found in Figure 2.32.

This criterion serves as a preparation for the next criterion. It should be noted that a self-penetration free grid does not imply an injective mapping, and an injective mapping does not imply a self-penetration free grid. One can construct examples for which one condition holds but not the other. The hope here is that a self-penetration free structure also leads to an injective mapping. Conversely, there is little hope that a knotted structure will result in something injective.

## Q13 An Injective Regular Characteristic Mapping

**Definition 2.57.** *Let $S \in \mathbb{R}^{n \times n}$ be a subdivision matrix of a (double) ring or a (double) shell for which Quality Criterion Q3 holds. Let $t$ be the type of the evaluated elements,* **D** *the spline definition domain for the (double) ring or (double) shell, and* **g** *the system of generating functions. Moreover, let the eigenvalues of $S$ satisfy*

$$1 = \lambda_0 = |\lambda_0| > |\lambda_1| \geq \cdots \geq |\lambda_t| > |\lambda_{t+1}| \geq \cdots \geq |\lambda_{n-1}|.$$

*The subdivision algorithm $(S, \mathbf{g})$ fulfills Quality Criterion* Q13 an injective regular characteristic mapping *if its characteristic mapping is regular and injective.*

This quality criterion is to be considered a long-term goal of subdivision theory. The injectivity of the characteristic mapping has many advantages. First, it determines the local asymptotic behavior of the refinement and thus the visual and geometric structure of the surface or volume. If injectivity is not given here, the surface (or volume) overlaps with itself in the long run, making it unusable for aesthetic and simulation purposes. An exemplary illustration is shown in Figure 2.33.

On the other hand, injectivity and regularity are crucial for analysis. Peters and Reif have shown that for $t = 2$, a regular injective characteristic map is equivalent to the surface being exactly $C^1$ smooth at the evaluation of the central irregular points and $C^k$ smooth elsewhere, where $k$ is the smoothness of the regular B-spline functions (see [PR08, Thm. 5.8, p. 88]). For the case $t = 3$, the analysis of regularity is still completely open.

In this work, it will not be possible to make statements about injectivity and regularity. However, quality criterion Q12 can serve as a preparation for injectivity, to at least say something about the eigenstructure.

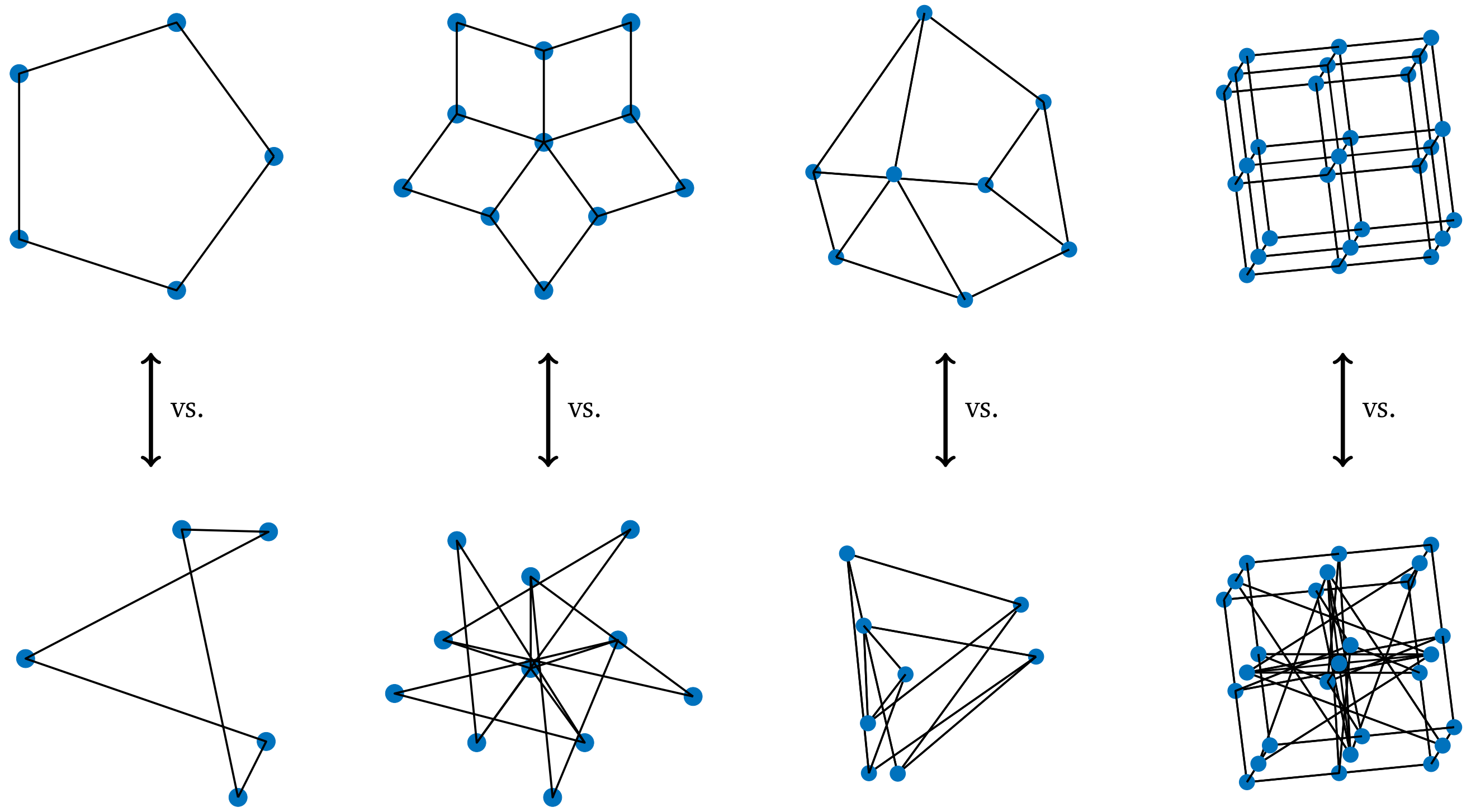

**Figure 2.32:** Illustrations of Quality Criterion Q12 for the subdominant eigenvector structure of initial elements. Above are examples where the criterion is fulfilled, below are examples where the criterion is not fulfilled. The initial elements are constructed from left to right for the cases $(t = 2, g = 2)$, $(t = 2, g = 3)$, $(t = 3, g = 2)$, and $(t = 3, g = 3)$.

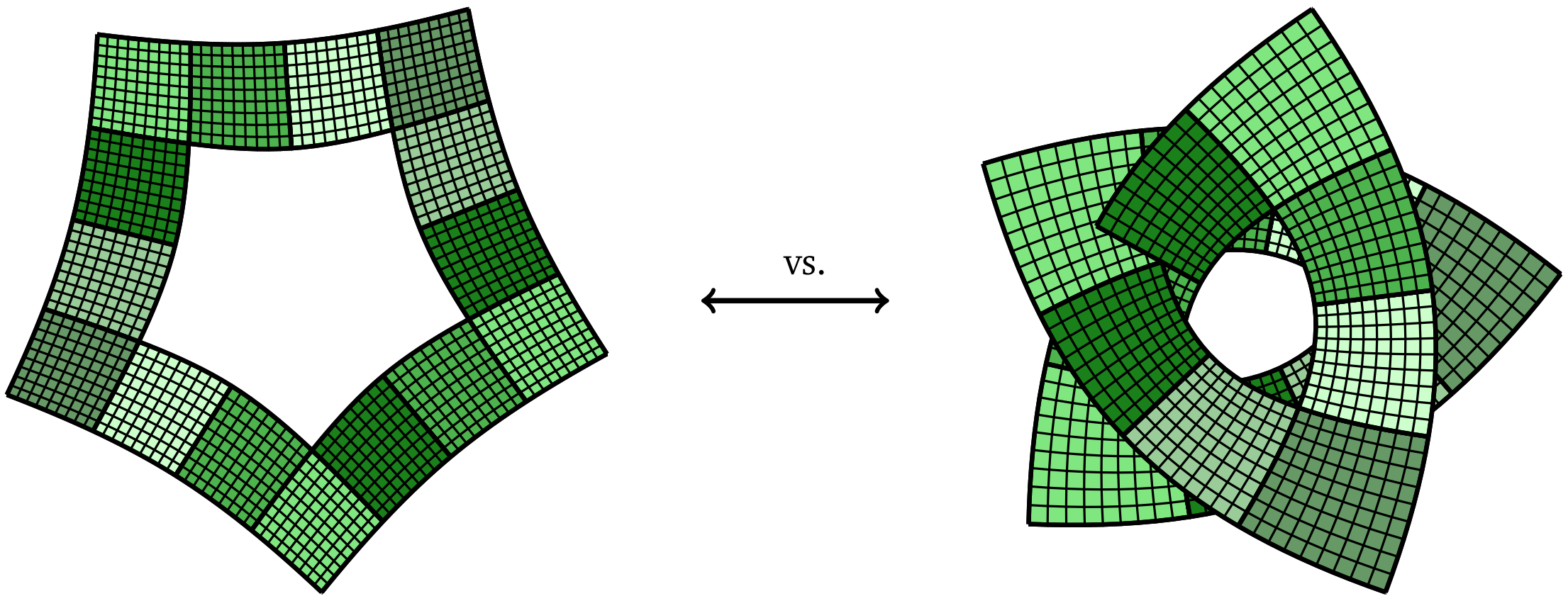

**Figure 2.33:** An injective regular characteristic mapping (left) and a non-injective characteristic mapping (right) for $t = 2$ and $g = 2$.

In this section, a total of 13 quality criteria have been presented. The criteria Q1 affine invariance, Q3 dominant eigenvalue, and Q7 regular case are mandatory. Without Q1 and Q3 the evaluated elements might not converge, and without Q7 the reproduction of the generated evaluation does not occur.

The fact that, as described in quality criterion Q9, a subdivision matrix can be generated for every combinatorial configuration is also crucial, because otherwise the structure described in this chapter would have to be further restricted for applications.

Regarding eigenvalues, we have also seen with quality criterion Q4 that at least a $t$-fold subdominant eigenvalue is essential for the local asymptotic behaviour. This eigenvalue is, in the best case by quality criterion Q5, $1/2$, and optionally the subsubdominant eigenvalue is exactly $1/4$ by quality criterion Q6.

Concerning the matrix structure, we have considered quality criteria Q2 convex hull, Q10 symmetry, and Q8 semi-regular case. These are optional but lead to advantageous properties of the subdivision matrix.

Regarding the generating functions, quality criterion Q11 proper support was described, though its effects could not be estimated. Regarding the structure, we discussed quality criteria Q12 self-penetration-free eigengrid and Q13 injective and regular characteristic map. Whether the former can be used for the latter is still unclear. The latter criterion is a long-term goal for the volumetric case, for which no theory yet exists.

## 2.5 Classification, Goals, and General Construction

The best-known algorithms for creating subdivision matrices in the context of generalized B-spline subdivision are the Doo-Sabin algorithm [DS78] (for $t = 2$ and $g = 2$), the Catmull-Clark algorithm [CC78] (for $t = 2$ and $g = 3$), and the algorithms from the works [JM99] and [Baj+02] (for $t = 3$ and $g = 3$). For the case $t = 3$ and $g = 2$, to our knowledge, no algorithm exists yet.

The case $t = 2$ is already extensively researched. The analysis of the algorithms is mature and very well summarized in [PR08]. In particular, for [DS78] and [CC78], the probably most important quality criterion Q13 of a regular and injective characteristic map holds, which guarantees the smoothness of the subdivision surface. The only weak point of the Catmull-Clark algorithm is that the subdominant eigenvalue is not $1/2$. Therefore, the current research task (as described in [Die+23, Task 2.1, p. 146]) is:

> Study and employ variants on Catmull-Clark [...] with small polar artifact $\left|\lambda - \frac{1}{2}\right|$ and smallest possible $\mu$.

Here, $\lambda$ is the subdominant and $\mu$ the subsubdominant eigenvalue.

Moreover, the two-dimensional Catmull-Clark algorithm cannot be extended to three dimensions without losing essential quality criteria. This will be explained below.

We start with a subdivision matrix $S \in \mathbb{R}^{n\times n}$ for $t = 2$ and $g = 3$ constructed according to Catmull-Clark [CC78] for an evaluable ring structure. This matrix has a dominant eigenvalue $1$ with eigenvector $\vec{1}$ and a double subdominant eigenvalue.

An extension of the Catmull-Clark rules for the case $t = 3$ in the semi-regular case would mean that the three-dimensional variant in the tensor product case, i.e. for quality criterion Q8, can be represented as

$$S^{(3)} = \operatorname{kron}(M, S), \quad \text{with} \quad M = \frac{1}{8}\begin{bmatrix} 0 & 4 & 4 & & & & \\ & 1 & 6 & 1 & & & \\ & & 4 & 4 & & & \\ & & 1 & 6 & 1 & & \\ & & & 4 & 4 & & \\ & & & 1 & 6 & 1 & \\ & & & & 4 & 4 & 0 \end{bmatrix}.$$

Note that we use not a double-ring structure but the evaluation given by [CC78]. Thus, we obtain seven layers of $S$, which correspond to seven layers of control points that yield an evaluable shell as in [CC78]. [Baj+02] and [JM99] use the same evaluation domain. The construction can also be described on an initial element, then $M$ would be accordingly

$$\frac{1}{8}\begin{bmatrix}4 & 4 & 0\\ 1 & 6 & 1\\ 0 & 4 & 4\end{bmatrix},$$

and the rest of the construction proceeds analogously. The same holds for $g = 2$. Here, the matrices

$$M = \frac{1}{4}\begin{bmatrix}3 & 1\\ 1 & 3\end{bmatrix} \quad \text{and} \quad M = \frac{1}{4}\begin{bmatrix}0 & 3 & 1 & & & \\ & 1 & 3 & & & \\ & & 3 & 1 & & \\ & & 1 & 3 & & \\ & & & 3 & 1 & \\ & & & 1 & 3 & 0\end{bmatrix}$$

are obtained. For the original scenario, both matrices can be decomposed into their Jordan form. We get accordingly

$$S := VJV^{-1} \quad \text{and} \quad M := V'J'V'^{-1} = \begin{bmatrix}1 & 3 & \\ 1 & 2 & \\ 1 & 1 & \\ 1 & 0 & V'_{(:,3:7)}\\ 1 & -1 & \\ 1 & -2 & \\ 1 & -3 & \end{bmatrix}\begin{bmatrix}1 & & & & & & \\ & \frac{1}{2} & & & & & \\ & & \frac{1}{4} & & & & \\ & & & \frac{1}{8} & & & \\ & & & & \frac{1}{8} & & \\ & & & & & 0 & \\ & & & & & & 0\end{bmatrix}V'^{-1}.$$

Using Lemma B.9 we obtain

$$S^{(3)} = \text{kron}\left(V'J'V'^{-1}, VJV^{-1}\right) = \text{kron}\left(V', V\right)\text{kron}\left(J'V'^{-1}, JV^{-1}\right) = \text{kron}\left(V', V\right)\text{kron}\left(J', J\right)\text{kron}\left(V'^{-1}, V^{-1}\right).$$

Since $J'$ is a diagonal matrix, we have

$$\text{kron}\left(J', J\right) = \begin{bmatrix}J & & & & & & \\ & \frac{1}{2}J & & & & & \\ & & \frac{1}{4}J & & & & \\ & & & \frac{1}{8}J & & & \\ & & & & \frac{1}{8}J & & \\ & & & & & 0J & \\ & & & & & & 0J\end{bmatrix}$$

which is a Jordan matrix for $S^{(3)}$. By Lemma B.10 we also get

$$\text{kron}\left(V', V\right)^{-1} = \text{kron}\left(V'^{-1}, V^{-1}\right).$$

Therefore,

$$S^{(3)} = \text{kron}\left(V', V\right)\text{kron}\left(J', J\right)\text{kron}\left(V', V\right)^{-1}$$

is the Jordan decomposition of $S^{(3)}$. This is exactly the statement of Theorem B.32. The eigenvalues of $S^{(3)}$ are thus the eigenvalues of $S$ multiplied by the eigenvalues of $M$.

Next, we look at some special eigenvalues and their eigenvectors. We start with the eigenvalue 1. This is simple again since 1 is the dominant eigenvalue of $S$ and $M$. This eigenvalue has the eigenvector $\vec{1}$ because the corresponding row in $\text{kron}\left(V', V\right)$ is exactly $\text{kron}(\vec{1}, \vec{1})$.

The second interesting eigenvalue is the double subdominant eigenvalue $\lambda$ of $S$, combined with the eigenvalue $1$ from $M$. In $S^{(3)}$ this is again a double eigenvalue with eigenvectors

$$\text{kron}\left(\vec{1}, V_{(:,2:3)}\right) = \begin{bmatrix} V^T_{(:,2:3)} & V^T_{(:,2:3)} & V^T_{(:,2:3)} & V^T_{(:,2:3)} & V^T_{(:,2:3)} & V^T_{(:,2:3)} & V^T_{(:,2:3)} \end{bmatrix}^T .$$

For each of the seven layers, we get the same values. If $|\lambda| > 1/2$, then $S^{(3)}$ again has a double subdominant eigenvalue $\lambda$. If $|\lambda| < 1/2$, then $S^{(3)}$ has a simple subdominant eigenvalue $1/2$. And if $|\lambda| = 1/2$, then $S^{(3)}$ has a triple subdominant eigenvalue $1/2$.

The (third) eigenvalue $1/2$ arises from the combination of the eigenvalue $1$ of $S$ and the eigenvalue $1/2$ of $M$. The corresponding eigenvector is

$$\text{kron}\left(V'_{(:,2)}, \vec{1}\right) = \begin{bmatrix} 3 \cdots 3 & 2 \cdots 2 & \cdots & -3 \cdots -3 \end{bmatrix}^T .$$

Each layer thus has a constant value, but the layers differ. If the local asymptotic behavior depended on these three eigenvectors, one would obtain a volumetric structure that approaches a line for $|\lambda| < 1/2$, a volume for $|\lambda| = 1/2$, and a surface for $|\lambda| > 1/2$. This is illustrated in Figure 2.34. We thus obtain the first theorem of this work:

**Theorem 2.58.** *Let $S^{(3)}$ for $t = 3$ be a three-dimensional extension of a two-dimensional subdivision matrix $S$ of an initial element or a (double) ring. Assume quality criterion Q8 holds with*

$$S^{(3)} = \text{kron}\left(S^{(1)}, S\right) .$$

*Furthermore, let $S^{(1)}$ be, according to Definition 2.48, exactly*

$$S^{(1)} := M = \frac{1}{4}\begin{bmatrix} 3 & 1 \\ 1 & 3 \end{bmatrix} \qquad \textit{for } g = 2 \textit{ and } S \textit{ belongs to an initial element},$$

$$S^{(1)} := M = \frac{1}{4}\begin{bmatrix} 0 & 3 & 1 & & & \\ & 1 & 3 & & & \\ & & 3 & 1 & & \\ & & 1 & 3 & & \\ & & & 3 & 1 & \\ & & & 1 & 3 & 0 \end{bmatrix} \qquad \textit{for } g = 2 \textit{ and } S \textit{ belongs to a ring},$$

$$S^{(1)} := M = \frac{1}{8}\begin{bmatrix} 4 & 4 & 0 \\ 1 & 6 & 1 \\ 0 & 4 & 4 \end{bmatrix} \qquad \textit{for } g = 3 \textit{ and } S \textit{ belongs to an initial element},$$

$$S^{(1)} := M = \frac{1}{8}\begin{bmatrix} 0 & 0 & 4 & 4 & & & & & & & \\ & & 1 & 6 & 1 & & & & & & \\ & & & 4 & 4 & & & & & & \\ & & & 1 & 6 & 1 & & & & & \\ & & & & 4 & 4 & & & & & \\ & & & & 1 & 6 & 1 & & & & \\ & & & & & 4 & 4 & & & & \\ & & & & & 1 & 6 & 1 & & & \\ & & & & & & 4 & 4 & & & \\ & & & & & & 1 & 6 & 1 & & \\ & & & & & & & 4 & 4 & 0 & 0 \end{bmatrix} \qquad \textit{for } g = 3 \textit{ and } S \textit{ belongs to a double ring}.$$

*Then it holds that*

$$S \textit{ satisfies Q3} \quad \Leftrightarrow \quad S^{(3)} \textit{ satisfies Q3}$$

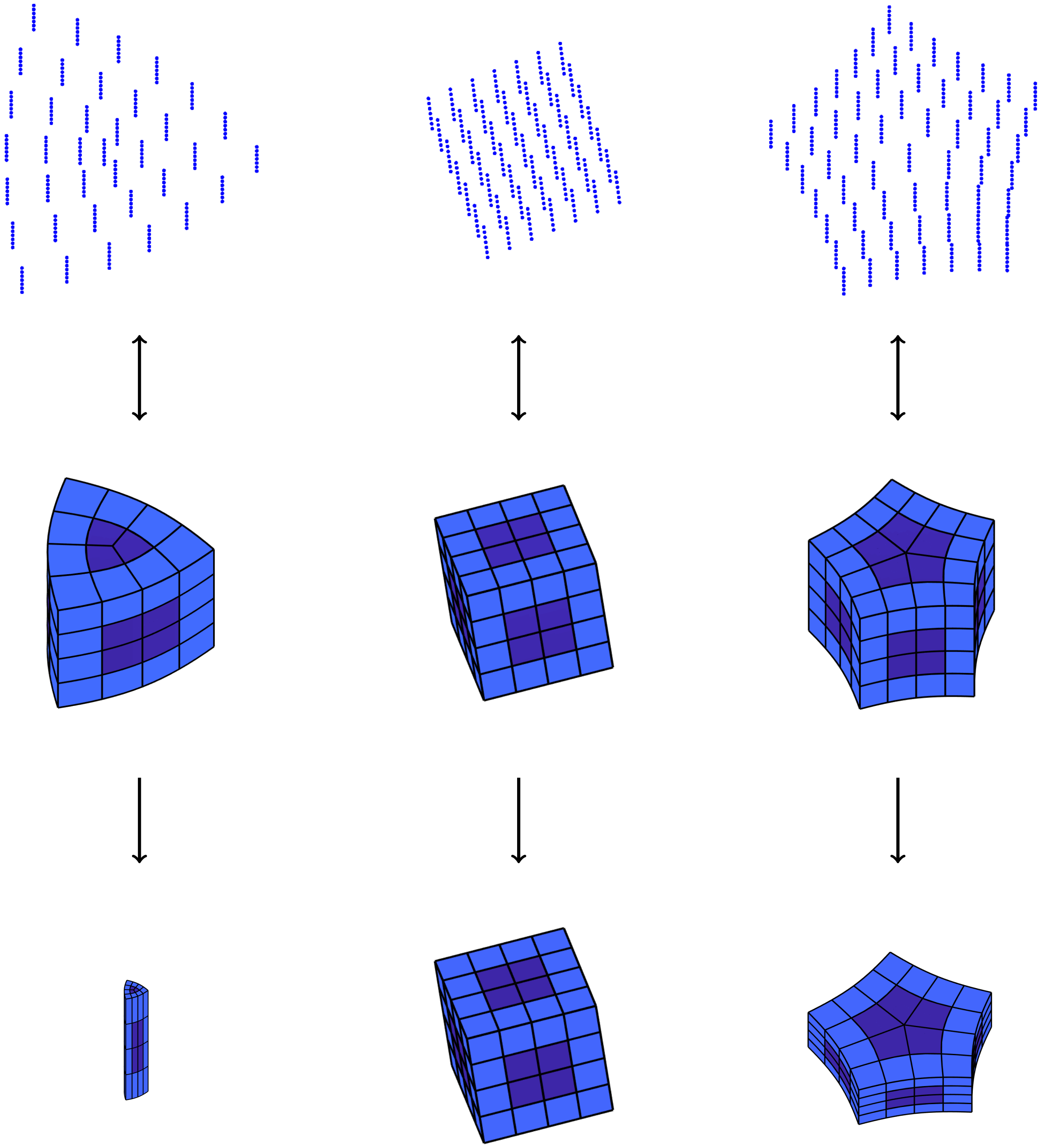

**Figure 2.34:** Tensor product extension of the Catmull-Clark algorithm for valences 3 (left), 4 (middle), and 5 (right). The first row shows the control points of the three largest eigenvectors smaller than 1, the second row their evaluation, and the third row the evaluation after ten refinements (each rescaled).

*and if quality criterion Q3 is satisfied for $S$ or $S^{(3)}$, then*

$$S \text{ satisfies Q5} \quad \Leftrightarrow \quad S^{(3)} \text{ satisfies Q5}$$

*and*

$$S \text{ satisfies Q5 or has a triple subdominant eigenvalue } > \tfrac{1}{2} \quad \Leftrightarrow \quad S^{(3)} \text{ satisfies Q4.}$$

*Proof.* For this proof, we briefly recall some statements. First, quality criterion Q3 concerns the dominant eigenvalue 1 and quality criterion Q5 concerns the subdominant eigenvalue $1/2$. We will also use Theorem B.32 multiple times without explicitly citing it each time. This theorem states that the eigenvalues of the Kronecker product of two square matrices are the Kronecker product of the eigenvalues of the two matrices. Thus, the eigenvalues of $S^{(3)}$ are the Kronecker product of the eigenvalues of $S^{(1)}$ and $S$. First, the eigenvalues of the above matrices $M$ are exactly

$$\left(1, \tfrac{1}{2}\right) \quad \text{for} \quad M = \frac{1}{4}\begin{bmatrix} 3 & 1 \\ 1 & 3 \end{bmatrix}, \qquad \left(1, \tfrac{1}{2}, \tfrac{1}{4}, \tfrac{1}{8}, \tfrac{1}{8}, 0 \ldots 0\right) \quad \text{for} \quad M = \frac{1}{8}\begin{bmatrix} 0 & 0 & 4 & 4 & & & & & & & \\ & & 1 & 6 & 1 & & & & & & \\ & & & 4 & 4 & & & & & & \\ & & & 1 & 6 & 1 & & & & & \\ & & & & 4 & 4 & & & & & \\ & & & & 1 & 6 & 1 & & & & \\ & & & & & 4 & 4 & & & & \\ & & & & & 1 & 6 & 1 & & & \\ & & & & & & 4 & 4 & & & \\ & & & & & & 1 & 6 & 1 & & \\ & & & & & & & 4 & 4 & 0 & 0 \end{bmatrix}$$

and

$$\left(1, \tfrac{1}{2}, \tfrac{1}{4}\right) \quad \text{for} \quad M = \frac{1}{8}\begin{bmatrix} 4 & 4 & 0 \\ 1 & 6 & 1 \\ 0 & 4 & 4 \end{bmatrix}, \qquad \left(1, \tfrac{1}{2}, \tfrac{1}{4}, \tfrac{1}{4}, 0, 0\right) \quad \text{for} \quad M = \frac{1}{4}\begin{bmatrix} 0 & 3 & 1 & & & \\ & 1 & 3 & & & \\ & & 3 & 1 & & \\ & & 1 & 3 & & \\ & & & 3 & 1 & \\ & & & 1 & 3 & 0 \end{bmatrix}$$

All have full geometric multiplicity. We prove the implications step by step:

- $S$ satisfies Q3 $\Rightarrow$ $S^{(3)}$ satisfies Q3:

  Since 1 is the simple and largest eigenvalue of $S$, and also 1 is the largest and simple eigenvalue of each variant of $M$, the matrix $S^{(3)}$ has a simple eigenvalue 1, which is the largest in magnitude because all other eigenvalues of $S^{(3)}$ are products of eigenvalues with magnitude less or equal to 1 and strictly less than 1.

- $S$ satisfies Q3 $\Leftarrow$ $S^{(3)}$ satisfies Q3:

  Since 1 is the simple and largest eigenvalue of $S^{(3)}$, assume for contradiction that $S$ has an eigenvalue greater than 1. Then, by the Kronecker product with $M$, this eigenvalue would also be an eigenvalue of $S^{(3)}$, contradicting the assumption.

  Furthermore, $S$ must have an eigenvalue 1 because the eigenvalues of $S^{(3)}$ have magnitude less or equal to 1 and 1 must arise as the product of eigenvalues equal to 1 from $S$ and $M$.

  Also, $S$ cannot have multiple eigenvalues of magnitude 1 because those would also appear in $S^{(3)}$, contradicting the assumption. Thus, $S$ has a simple dominant eigenvalue 1.

- $S$ satisfies Q5 $\Rightarrow$ $S^{(3)}$ satisfies Q5:

  If $S$ has a double subdominant eigenvalue $\frac{1}{2}$, then combined with the eigenvalue 1 from $M$ in the Kronecker product, this eigenvalue $\frac{1}{2}$ appears doubly in $S^{(3)}$. The third subdominant eigenvalue arises from the combination of eigenvalue 1 of $S$ and eigenvalue $\frac{1}{2}$ of $M$. Since neither $S$ nor $M$ have eigenvalues with magnitude strictly between 1 and $\frac{1}{2}$, all other eigenvalues of $S^{(3)}$ have magnitude less than $\frac{1}{2}$.

- $S$ satisfies Q5 $\Leftarrow S^{(3)}$ satisfies Q5:
  Suppose $S^{(3)}$ has a triple subdominant eigenvalue $\frac{1}{2}$. Since the eigenvalues of $S^{(3)}$ are the Kronecker product of eigenvalues of $S$ and $M$, one of the three eigenvalues $\frac{1}{2}$ in $S^{(3)}$ must come from eigenvalue $1$ of $S$ and eigenvalue $\frac{1}{2}$ of $M$.
  Assume one of the other eigenvalues $\frac{1}{2}$ in $S^{(3)}$ does not come from eigenvalue $1$ of $M$ and eigenvalue $\frac{1}{2}$ of $S$. Then $S$ would have an eigenvalue larger than $\frac{1}{2}$, which would also appear in $S^{(3)}$ (contradiction). Hence, the other two eigenvalues $\frac{1}{2}$ in $S^{(3)}$ come from eigenvalue $1$ of $M$ and eigenvalue $\frac{1}{2}$ of $S$.
  By the same reasoning, $S$ cannot have eigenvalues between $\frac{1}{2}$ and $1$ nor a third eigenvalue $\frac{1}{2}$. Hence, $S$ satisfies quality criterion Q5.
- $S$ satisfies Q5 or has a triple subdominant eigenvalue $> 1/2 \Leftarrow S^{(3)}$ satisfies Q4:
  If the subdominant eigenvalue of $S^{(3)}$ is exactly $\frac{1}{2}$, then the above implication

  $$S \text{ satisfies Q5 } \Leftarrow S^{(3)} \text{ satisfies Q5}$$

  applies and $S$ has a double subdominant eigenvalue $\frac{1}{2}$. Otherwise, $S^{(3)}$ has a triple subdominant eigenvalue greater than $\frac{1}{2}$, which must arise from the combination of eigenvalue $1$ of $S$ and eigenvalue $\frac{1}{2}$ of $M$. Since neither $S$ nor $M$ have eigenvalues greater than 1, this implies $S$ has a triple subdominant eigenvalue.
- $S$ satisfies Q5 or has a triple subdominant eigenvalue $> 1/2 \Rightarrow S^{(3)}$ satisfies Q4:
  If $S$ has a double subdominant eigenvalue $\frac{1}{2}$, then the above implication

  $$S \text{ satisfies Q5 } \Rightarrow S^{(3)} \text{ satisfies Q5}$$

  applies and $S^{(3)}$ satisfies the corresponding criterion. If $S$ has a triple subdominant eigenvalue greater than $\frac{1}{2}$, then this eigenvalue also appears triply in $S^{(3)}$. Since $M$ has no eigenvalues between $\frac{1}{2}$ and $1$, this is the triple subdominant eigenvalue of $S^{(3)}$ as well.

□

With this theorem, we see that the Catmull-Clark algorithm is not a good starting point for a three-dimensional generalization. According to [PR08, Thm. 6.1, p. 114], the (double) subdominant eigenvalue of the Catmull-Clark algorithm is exactly

$$\lambda = \frac{\cos\left(\frac{2\pi}{n}\right) + 5 + \sqrt{\left(\cos\left(\frac{2\pi}{n}\right)+9\right)\left(\cos\left(\frac{2\pi}{n}\right)+1\right)}}{16},$$

where $n$ is the number of cells at an initial element. For $n > 4$, we have $\cos\left(\frac{2\pi}{n}\right) > 0$. Hence,

$$\lambda > \frac{0+5+\sqrt{(0+9)(0+1)}}{16} = \frac{5+3}{16} = \frac{1}{2}.$$

Since all two-dimensional Catmull-Clark subdivision matrices have a double subdominant eigenvalue, it follows from Theorem 2.58 that the tensor product variant for $n > 4$ cannot satisfy quality criterion Q4.

The initial situation for the Catmull-Clark algorithm is actually even less suitable than shown so far. For $n > 4$, according to [PR08, p. 112], the Catmull-Clark algorithm has eigenvalues of the form

$$\lambda_1' = \frac{\cos\left(\frac{4\pi}{n}\right) + 5 + \sqrt{\left(\cos\left(\frac{4\pi}{n}\right)+9\right)\left(\cos\left(\frac{4\pi}{n}\right)+1\right)}}{16}$$

and

$$\lambda_2' = \frac{\cos\left(\frac{2(n-2)\pi}{n}\right) + 5 + \sqrt{\left(\cos\left(\frac{2(n-2)\pi}{n}\right)+9\right)\left(\cos\left(\frac{2(n-2)\pi}{n}\right)+1\right)}}{16}.$$

Because

$$\cos\left(\frac{4\pi}{n}\right) = \cos\left(2\pi - \frac{4\pi}{n}\right) = \cos\left(\frac{2\pi n - 4\pi}{n}\right) = \cos\left(\frac{2\pi(n-2)}{n}\right),$$

we have $\lambda' := \lambda'_1 = \lambda'_2$. For $n > 8$ it holds that $\cos\left(\frac{4\pi}{n}\right) > 0$ and thus, analogously to the subdominant eigenvalue,

$$\lambda' > \frac{1}{2} \quad \text{for} \quad n > 8.$$

The only eigenvalue according to [PR08, p. 112] that could lie between $\lambda'$ and $\lambda$ (for $n > 8$) is

$$\lambda_{a,b}^{1,2} = \frac{4a - 1 \pm \sqrt{(4a-1)^2 + 8b - 4}}{8}.$$

For the Catmull-Clark algorithm, $a = 1 - \frac{7}{4n}$ and $b = \frac{3}{2n}$. Thus, for $n > 8$ there are two possibilities:

Either one of the $\lambda_{a,b}^{1,2}$ is the subsubdominant eigenvalue, or $\lambda'$ is the subsubdominant eigenvalue. In both cases, the subsubdominant eigenvalue is greater than $\frac{1}{2}$ and hence the subsubdominant eigenvalue of $S^{(3)}$, being the product of this and 1 from $M$, is greater than $\frac{1}{2}$. The eigenvector(s) corresponding to this eigenvalue are combinations of the respective eigenvectors of $S$ and the eigenvector $\vec{1}$. Thus, all control points that are combinations from the respective eigenspaces of $\lambda$ and $\mu$ of $S^{(3)}$, arising from one control point of $S$, have the same coordinates. The seven layers described above therefore collapse with respect to $\lambda$ and $\mu$ into a single layer. Hence, we obtain the second theorem of this work:

**Theorem 2.59.** *Let $t = 3$ and $S^{(3)}$ be a subdivision matrix that is a three-dimensional extension of a two-dimensional subdivision matrix $S$ of an initial element or a ring. Hence, quality criterion Q8 holds. Further, quality criteria Q3 and Q4 hold for $S$. If $S^{(1)}$ from Definition 2.48 is as in Theorem 2.58, then:*

*If the subsubdominant eigenvalue of $S$ is greater than $\frac{1}{2}$, then the evaluation of any linearly independent selection of eigenvectors from the eigenspaces of the two largest eigenvalues different from 1 forms a surface in $\mathbb{R}^3$ and is therefore not injective. If $S^{(3)}$ has a characteristic map, then this is also not injective.*

*Proof.* Let the geometric multiplicity of the subsubdominant eigenvalue of $S$ be denoted by $i \in \mathbb{N}$. As described before, the two largest eigenvalues different from 1 are greater than $1/2$. The vectors spanning the corresponding eigenspaces are thus

$$\text{kron}\left(\vec{1}, V_{(:,2:3+i)}\right).$$

Therefore, each layer of a point from $S$, generated by $M$, has the same coordinates, and the evaluation forms a surface in $\mathbb{R}^3$. Since the evaluation is volumetric but maps onto a surface, it is in particular not injective. □

If one of the $\lambda_{a,b}^{1,2}$ is the subsubdominant eigenvalue, then the eigenspaces of the three largest eigenvalues different from 1 are uniquely determined. From this, it follows directly that the characteristic map forms a surface in $\mathbb{R}^3$ and is therefore not injective.

If, however, $\lambda'$ is the subsubdominant eigenvalue, then the eigenspaces together are no longer $t$-dimensional. The evaluation of the vectors of the local asymptotic behavior is thus not only a surface and therefore not injective, but their shape also depends on the control points, and control over the local asymptotic behavior is lost.

It is important to mention that this behavior is independent of the choice of the refinement rule at the irregular point for the two-dimensional Catmull-Clark. The rule can only influence whether the third eigenvector is unique or depends on the choice of control points.

This behavior has several consequences. On the one hand, it proves that the Catmull-Clark algorithm is not a good starting point for a three-dimensional subdivision algorithm. The regular rules around the irregular point cause the spectrum to be uncontrollable. Thus, sticking to these rules must inevitably fail. This supports the argument of this work to broaden the irregular region around an element and to understand the generalized cubic B-spline subdivision around irregular points as an invariant mapping from double rings to double rings or double shells to double shells, rather than as a mapping from rings to rings. To avoid this behavior, we concretize the above task as follows:

> Study and employ variants on Catmull-Clark with $\lambda = \frac{1}{2}$ and smallest possible $\mu$.

The consequences of this behavior also extend to the subdivision algorithms found in the literature. The algorithm by [Baj+02] must necessarily exhibit the same behavior as the above-described variant of the Catmull-Clark algorithm in the tensor product case. The variant from [Baj+02] contains the same rules as the Catmull-Clark algorithm for the tensor product case; only the rules for the central point differ. Thus, for $n > 8$, the same behavior arises as described for the extension of the Catmull-Clark algorithm. Although the algorithm by [JM99] does not have a tensor product structure, it exhibits the same behavior as the other two algorithms. This is shown in Figure 2.35.

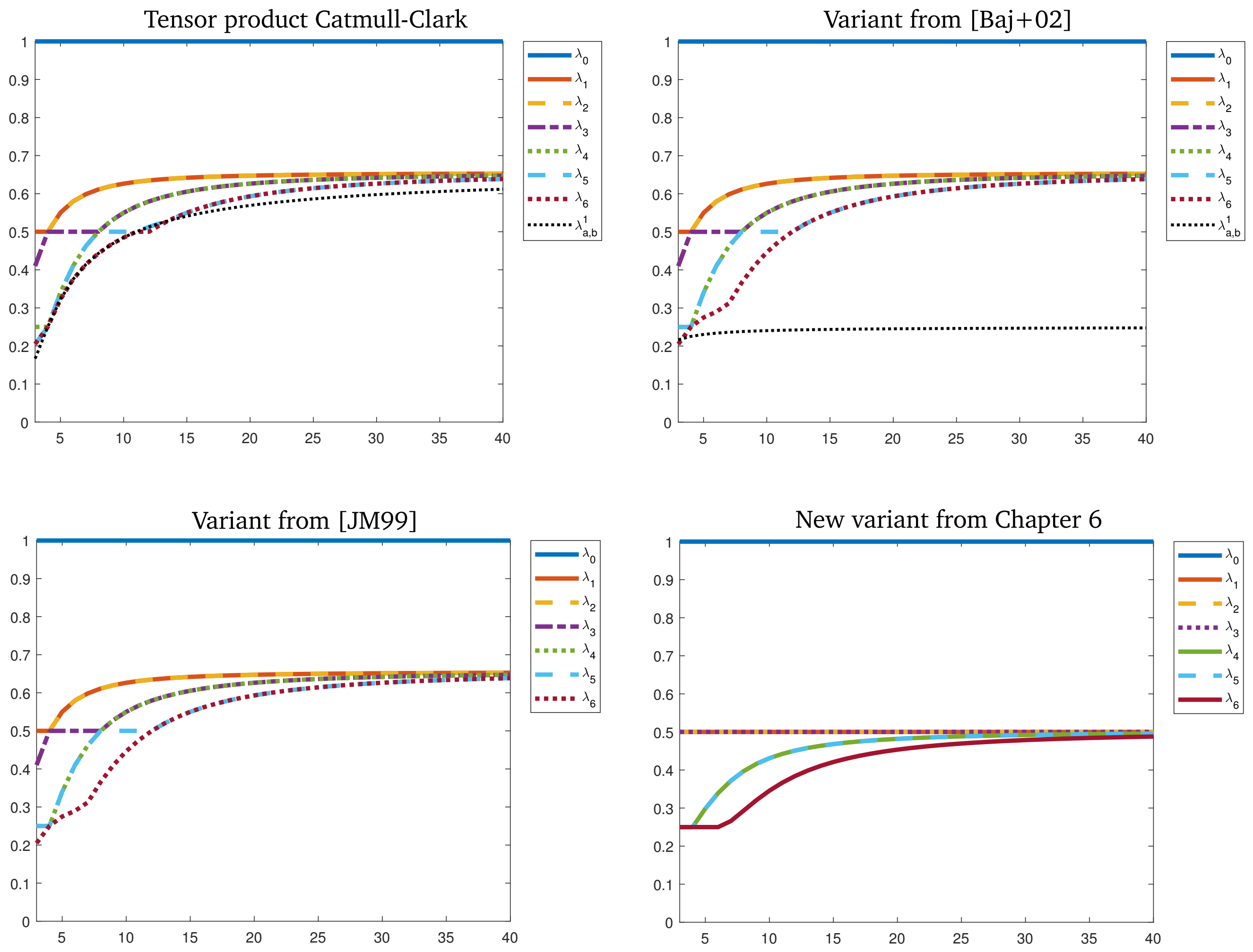

**Figure 2.35:** The seven largest eigenvalues of the subdivision matrices for $t = 3$ and $g = 3$ in the semi-regular case depending on the valence of the central vertex, shown for the three discussed variants and for the corresponding subdivision matrices constructed in Section 6.

The variant by [JM99] also has an eigenvalue structure of the form

$$1 > |\lambda_1| = |\lambda_2| > |\lambda_3| = |\lambda_4|.$$

Tests further show that the selection of three linearly independent eigenshells corresponding to the two largest eigenvalues different from $1$ with [JM99] for the tensor product cases are likewise not injective. See, for example, Figure 2.36.

However, these are not the only cases where the characteristic maps of [Baj+02] and [JM99] fail to be injective. Experiments show that both algorithms seem to work well when the distances between the corners of the dual structure are as evenly spread as possible. When these distances are distributed as evenly as possible, the characteristic map appears to be injective. But as soon as the distances become unevenly distributed, the characteristic map seems to lose injectivity. Loosely speaking, the soccer ball works well, while the pie divided into $n$ slices does not. Another example is shown in Figure 2.37. For this example, the largest eigenvalues in magnitude of the algorithms are:

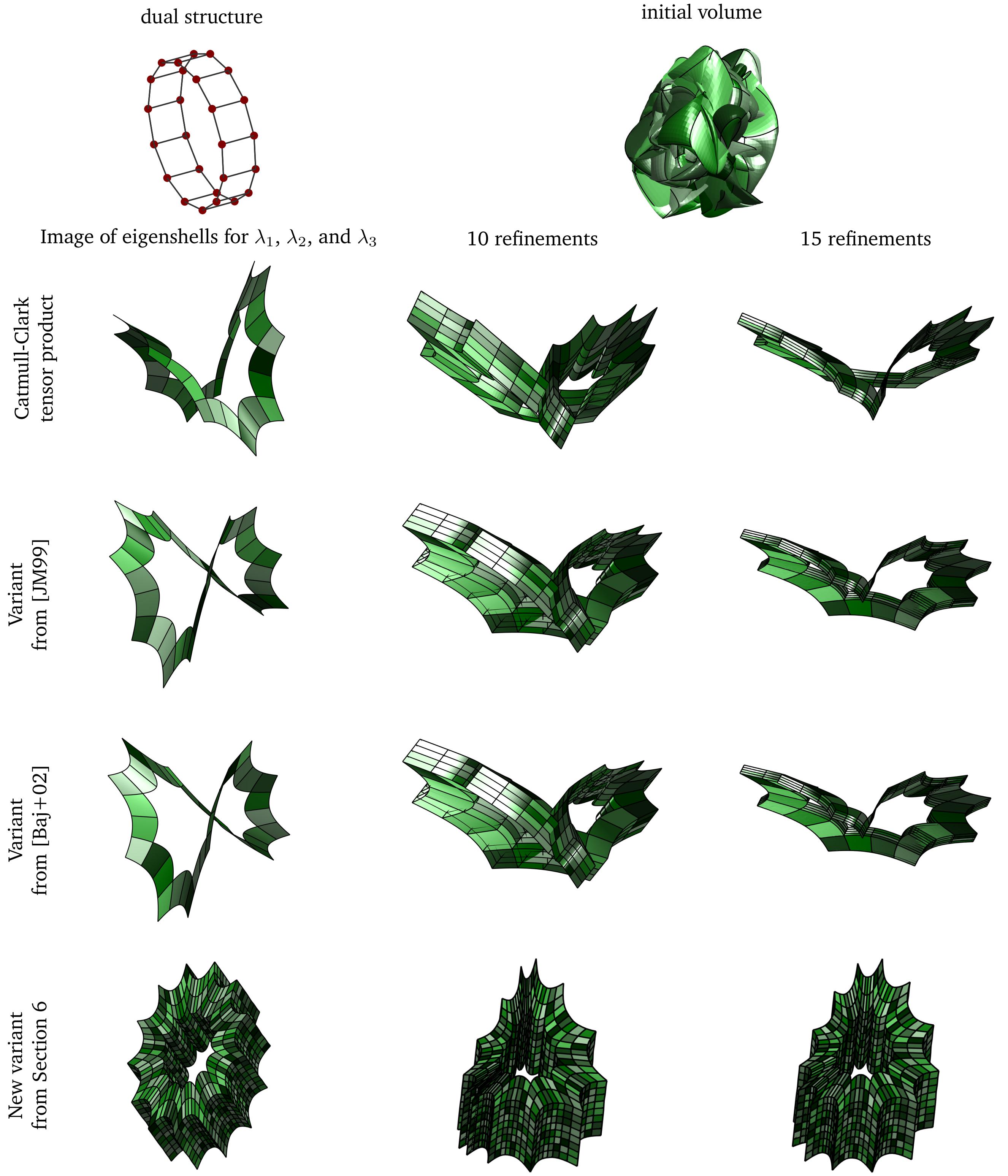

**Figure 2.36:** The tensor product structure for $t = 3$, arising from a two-dimensional ring with 13 cells. In the left column one sees the dual structure and the image of the characteristic map or the image of the eigenshells for $\lambda_1$, $\lambda_2$, and $\lambda_3$ (for Catmull-Clark tensor product, [Baj+02], and [JM99] holds $\lambda_3 = \lambda_4$) of the corresponding subdivision matrices. This was applied 10 and 15 times to randomly chosen control points. The initial volume is shown in the first row, with refinements of the respective subdivision algorithms below.

| | $\lambda_0$ | $\lambda_1$ | $\lambda_2$ | $\lambda_3$ | $\lambda_4$ |
|---|---|---|---|---|---|
| [Baj+02] | 1 | 0,6957 | 0,6951 | 0,6347 | 0,6312 |
| [JM99] | 1 | 0,6957 | 0,6951 | 0,6347 | 0,6312 |
| Variant Chapter 6 | 1 | 0,5 | 0,5 | 0,5 | 0,4713 |

The two established algorithms are therefore generally not recommended. Accordingly, the current tasks (following [Die+23, Tasks 2.2 and 2.3, p. 147]) consist of:

> Construct a trivariate generalization of Catmull-Clark [Doo-Sabin] with triple subdominant eigenvalue $\lambda = \frac{1}{2}$ and smallest possible $\mu$.

The core of this work is dedicated to solving or answering these three tasks. However, the approach to construction deviates from the usual procedure. Instead of prescribing rules or formulas for individual matrix entries, we start with the construction of a convex polytope. This is done in Chapter 3. This polytope, or for $g = 3$ a variant of it, is intended to serve as the basis for the eigenspace of the subdominant eigenvalue. In a second step, a matrix is constructed using this polytope, which has a $t$-fold eigenvalue with an eigenspace having the polytope as basis. There are various ways to do this, but we will prioritize using the Colin-de-Verdière-matrices from Chapter 4 to construct this matrix. Finally, the constructed matrices are adjusted if necessary to obtain the correct eigenvalue structure. This is done in Chapters 5 and 6.

This chapter has laid the foundation for the new approaches to constructing subdivision algorithms presented in this work. First, in Section 2.1 we introduced all relevant concepts regarding subdivision and developed a structure applicable to all dimensions of objects. Next, in Section 2.2 we described the subdivision itself, and in Section 2.3 we identified the invariant structures relevant to us. In Section 2.4 we detailed quality criteria that subdivision matrices should satisfy. We particularly emphasized the advantages of subdivision algorithms that meet these criteria and also discussed the consequences of (not) meeting them. Finally, in Section 2.5 we investigated the problems of subdivision algorithms found in the literature and identified necessary improvements for their use.

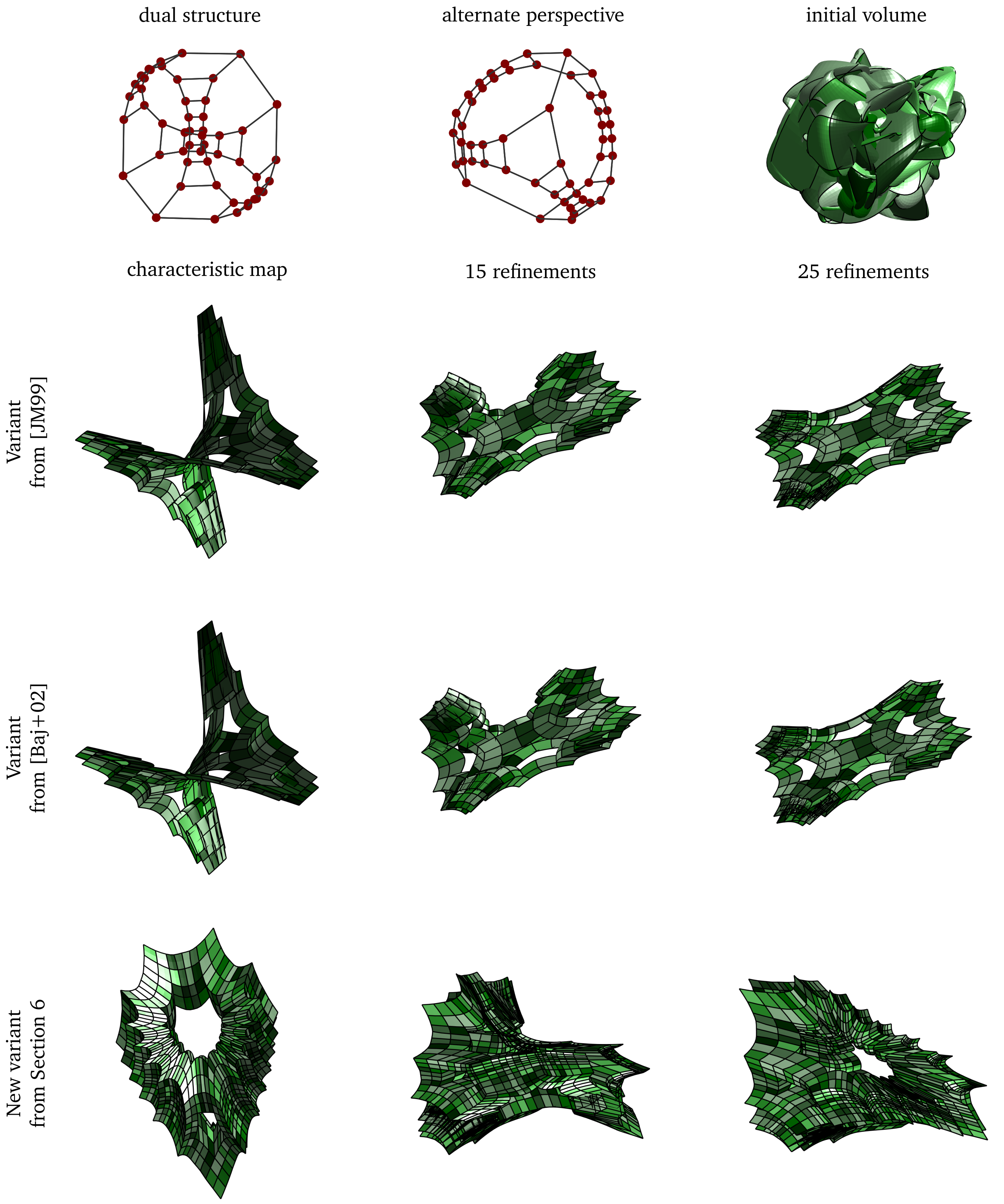

**Figure 2.37:** Analogous illustration to Figure 2.36. The example stems from a 28-gon where an additional line was introduced in the dual prism structure at the top and bottom. The two new lines are not parallel. It can be seen that both the eigenshell and after several refinement steps the variants of [Baj+02] and [JM99] are not injective.

# 3 Construction of Convex Polytopes

In this chapter, we describe the construction of convex polytopes. Specifically, we compute the vertices of the associated polytope in the corresponding space, $\mathbb{R}^2$ or $\mathbb{R}^3$, starting from a combinatorial structure. This chapter focuses on three-dimensional polytopes. Only at the end do we briefly address the construction of two-dimensional polytopes. The coordinate vectors of the vertices will later serve as the basis of the subdominant eigenspaces of the subdivision matrices.

The construction method is based on the work of [Zie04]. There, a constructive proof of the Koebe-Andreev-Thurston Theorem is given (Theorem 3.18 in the version by Ziegler and Theorem 3.25 in the version we use), which serves as a blueprint for constructing convex polytopes. The main result used by Ziegler in this construction is Theorem 3.35, which uses the functional from Bobenko and Springborn in [BS04]. These two works form the core theory of this chapter. Additionally, we examine further properties of the constructed polytopes that are needed for the proofs of the quality criteria of the subdivision matrices in Chapters 5 and 6.

As already mentioned, the construction of polytopes in [Zie04] is well developed. However, it remains theoretical. The algorithmic implementation is symbolic and not feasible with realistic runtime, which is why our implementation includes numerical methods. Since the basis of the eigenspace should represent the structure and shape of the polytope as precisely as possible, it is necessary to compute with high accuracy during the construction. Therefore, this chapter focuses on the concrete practical implementation. We will particularly address the numerically challenging construction steps and describe the algorithmic implementation. This is also the main contribution of this chapter.

It is recommended to read this chapter from a user's perspective. The corresponding implementation in Matlab can be found in [Die25], and this chapter can serve as a guide for it.

We begin in Section 3.1 by introducing the basics of convex polytopes. Since we need an algorithm for constructing convex polytopes, we also discuss in this section the requirements and input types of the algorithm and explain how we can check them for validity. Section 3.2 is more theoretical. Since convex polytopes are not uniquely determined by a given abstract structure, we discuss in this section which polytope we want to construct and what properties it should have. In Section 3.3, we then describe the actual construction and focus especially on the algorithmic implementation and numerical challenges. Section 3.4 concludes the chapter with the two-dimensional analogue of convex polytopes.

## 3.1 Fundamentals, Input Types, and Validity Checks

In this section, we discuss all the requirements needed for the construction of 3-polytopes. To do this, we first introduce terms used in the context of polytopes. We also describe which data structure can be used for 3-polytopes and the graphs that represent them. Since polytopes can be associated with certain graphs, we then look at how the properties of these graphs can be checked algorithmically. Finally, we examine a generator for random combinatorial configurations of 3-polytopes.

Before we can describe the construction of convex polytopes, we must first explain what a convex polytope is. Therefore, we begin with the definition of convexity according to [Zie94, pp. 3–4]:

**Definition 3.1.** *Let* $\boldsymbol{M} \subseteq \mathbb{R}^d$ *be a set.*

1. *The set* $\boldsymbol{M}$ *is called* convex *if for any two points* $a, b \in \boldsymbol{M}$*, the connecting line*

$$\{\varphi a + (1 - \varphi) b \mid 0 \leq \varphi \leq 1\}$$

   *is also contained in* $\boldsymbol{M}$*.*

2. *The* convex hull *of* $\boldsymbol{M}$ *is the intersection of all convex sets that contain* $\boldsymbol{M}$*, that is,*

$$\operatorname{conv}(\boldsymbol{M}) := \bigcap \left\{ \boldsymbol{M}' \subseteq \mathbb{R}^d \mid \boldsymbol{M} \subseteq \boldsymbol{M}', \boldsymbol{M}' \text{ convex} \right\}.$$

If the set **M** consists of finitely many points, then the convex hull of **M** can be described as follows, according to [Zie94, p. 4]:

**Lemma 3.2.** *Let* $\boldsymbol{M} = \{a_1, \ldots, a_n\} \subset \mathbb{R}^d$ *be a finite set with* $n \in \mathbb{N}$ *and let* $M := \left[a_1^T, \ldots, a_n^T\right]^T \in \mathbb{R}^{n \times d}$ *be the corresponding matrix whose rows represent the points from* $\boldsymbol{M}$*. Then the convex hull of* $\boldsymbol{M}$ *is given by*

$$\operatorname{conv}(M) := \operatorname{conv}(\boldsymbol{M}) = \left\{ \varphi_1 a_1 + \cdots + \varphi_n a_n \mid \varphi_i \in \mathbb{R}_{\geq 0} \text{ for all } i \in \{1, \ldots, n\} \text{ and } \sum_{i=1}^{n} \varphi_i = 1 \right\}.$$

We also obtain the following lemma according to [HL01, prop. 1.2.1, p. 22]:

**Lemma 3.3.** *Let* $\boldsymbol{M}_1$ *and* $\boldsymbol{M}_2$ *be two convex sets. Then* $\boldsymbol{M}_1 \cap \boldsymbol{M}_2$ *is also convex.*

Using these descriptions of convex sets, the concept of a polytope can be defined as follows, according to [Zie94, Def. 0.1, pp. 4–5]:

**Definition 3.4.** *We define:*

- *A* V-polytope *is the convex hull of a finite set of points in* $\mathbb{R}^d$*.*
- *An* H-polytope *is the intersection of finitely many closed half-spaces in* $\mathbb{R}^d$*, which is bounded in the sense that it contains no ray of the form*

$$\left\{ a + \varphi b \mid \varphi \geq 0, a, b \in \mathbb{R}^d \right\}.$$

- *A* polytope $\boldsymbol{P}$ *is a subset of* $\mathbb{R}^d$ *that can be represented either as a V-polytope or an H-polytope.*
- *A* $d$-polytope *is a* $d$*-dimensional polytope in* $\mathbb{R}^{d'}$ *with* $d \leq d'$*.*

This definition contains two representations of polytopes: the V-polytope, which is described by its vertices, and the H-polytope, which is described by half-spaces or facets. The following theorem from [Zie94, Thm. 1.1, p. 29] shows the equivalence of these two definitions:

**Theorem 3.5.** *A subset* $\boldsymbol{M} \subseteq \mathbb{R}^d$ *is the convex hull of finitely many points (V-polytope)*

$$\boldsymbol{M} = \operatorname{conv}(P) \quad \text{with} \quad P \in \mathbb{R}^{n \times d}$$

*if and only if it is a bounded intersection of finitely many closed half-spaces (H-polytope)*

$$\boldsymbol{M} = \left\{ x \in \mathbb{R}^d \mid Ax \leq b \right\} \quad \text{with} \quad A \in \mathbb{R}^{m \times d} \quad \text{and} \quad b \in \mathbb{R}^m.$$

As described in [Zie94], this theorem allows us to represent polytopes in two different ways. Depending on the application, one or the other representation may be more advantageous. Also, this theorem ensures that all polytopes are convex. When we refer to polytopes in the following, we always mean convex polytopes.

For a convex polytope, a corresponding dual polytope can also be defined. Since there are different types of duality, we define here the polar duality according to [Lov01, p. 224] and [Zie94, Def. 2.10, p. 61]:

**Definition 3.6.** *Let* $\boldsymbol{P}$ *be a convex* $d$*-polytope in* $\mathbb{R}^d$ *that contains the origin. Then the* polar dual polytope *is defined as*

$$\boldsymbol{P}^* = \{x \in \mathbb{R}^d \mid a^T x \leq 1 \textit{ for all } a \in \boldsymbol{P}\} \tag{3.1}$$

In the following, we will simply speak of dual polytopes, always meaning polar dual polytopes. The following properties hold for them [Zie94, Thm. 2.11, p. 62]:

**Theorem 3.7.** *Let* $\boldsymbol{P}$ *be a* $d$*-polytope that contains the origin, and let* $\boldsymbol{P}^*$ *be its dual polytope. Then:*

1. $\boldsymbol{P}^*$ *is convex.*
2. $0 \in \boldsymbol{P}^*$.
3. $(\boldsymbol{P}^*)^* = \boldsymbol{P}$.
4. *If a polytope is defined as* conv$(P)$ *with* $P \in \mathbb{R}^{n \times d}$*, then the dual polytope is*

$$\boldsymbol{P}^* = \{x \in \mathbb{R}^d \mid P_{(i,:)} x \leq 1, \textit{ for } i = 1 \ldots, n\}.$$

The dual polytope is thus convex, contains the origin, and its facets can be described using the vertices of the primal polytope. Since the dual of the dual polytope is again the primal polytope, the facets of the primal polytope can also be described using the vertices of the dual polytope.

In the next step, we connect the concept of polytopes with graph theory. For this, we use the works of Steinitz.[1] He formulated the following theorem [SR76, p. 192]:

> In the previous section, we examined two classes of regular polyhedral complexes that are characterized by particularly simple properties: the Eulerian complexes, and the polyhedra without overlapping elements. From this, one might already guess that those polyhedra which belong to both classes at once – that is, Eulerian polyhedra without overlapping elements – must play a particularly important role. We refer to them as $K$-polyhedra, with the $K$ indicating their close connection to convex polyhedra. That every convex polyhedron satisfies Euler's formula and contains no overlapping elements, and thus represents a $K$-polyhedron, has already been noted in Section 1, p. 91. But it also holds true the other way around: every $K$-polyhedron can be realized as a convex polyhedron. For this „Fundamental Theorem of Convex Types“, we will give several proofs in the next section.

In [Ste22, p. 77], the formulation is more precise:

> *Every* $K$*-polyhedron can be realized as a convex polyhedron.*
> Conversely, it is evident that every convex polyhedron represents a $K$-polyhedron.

Since the phrasing is rather awkward, it can be reformulated as follows, according to [Zie94, Thm. 4.1, p. 103]:

**Theorem 3.8** (Steinitz's Theorem)**.** *The graph* $\boldsymbol{G} = (\boldsymbol{V}, \boldsymbol{E})$ *is the graph of a* 3-polytope *if and only if it is simple, planar, and 3-connected.*

We therefore define:

**Definition 3.9.** *A simple planar 3-connected graph is called the* skeleton *or* structure graph *of a 3-polytope.*

The combinatorial structure of a 3-polytope can thus be described by a simple planar 3-connected graph. Intuitively, each node of a structure graph corresponds to a vertex of a polytope, and each edge of the structure graph corresponds to an edge of a polytope. With this connection, we can fully utilize the power of graph theory in the context of polytopes.

To construct a 3-polytope, we now only need the information from its structure graph. Therefore, we define the input to the construction algorithm as follows:

[1] Steinitz's work on polytopes first appeared in [Ste22] and later again in [Ste34]. For the reference [Ste34], the reprint version from [SR76] was used.

**Definition 3.10.** *Let $\mathbf{G} = (\mathbf{V}, \mathbf{E})$ be the structure graph of a 3-polytope. A* valid input *for the algorithm to construct a 3-polytope is one of the following three types:*

- *The adjacency matrix*
$$A \in \{0,1\}^{|\mathbf{V}|\times|\mathbf{V}|}$$
*of **G**.*

- *A list of edges*
$$E \in \{1,\dots,|\mathbf{V}|\}^{|\mathbf{E}|\times 2}.$$
*Each row in $E$ represents an edge of **G**, and the two entries in a row represent the two nodes of that edge.*

- *A list of facets*
$$F \in \{0,1,\dots,|\mathbf{V}|\}^{(2-|\mathbf{V}|+|\mathbf{E}|)\times m},$$
*where $m$ is the maximum size or number of nodes in any facet of **G**. Each row in $F$ represents a facet of **G** (see also Euler's formula A.29). Each entry in a row represents a node. Two nodes are connected by an edge if they are adjacent in the row. Additionally, the last and the first node are connected by an edge. Since facet sizes can vary, the number of columns in $F$ depends on the largest facet. The first entry in each row contains a node, and any extra entries are padded with zeros.*

All three forms are sufficient to describe the graph **G**. The adjacency matrix is a matrix representation of the graph. A list of edges can be used to construct the adjacency matrix, and since each edge in a 3-connected planar graph is part of exactly two facets (cf. Theorem A.30), the list of facets includes every edge of **G**. We allow different input formats since, depending on context, certain storage types may be more convenient or clearer than others.

From a practical standpoint, it is important to note that the nodes must be numbered uniformly from 1 to $|\mathbf{V}|$. Adjacency matrices with empty rows or columns, or edge or facet lists with missing nodes, will cause errors, as they would be interpreted as graphs with isolated nodes.

All three input formats are converted to an adjacency matrix in the next step. For this, we create a zero matrix of the appropriate size. A list of facets is first converted into a list of edges by creating one edge for each pair of adjacent nodes. A list of edges is then converted into an adjacency matrix by setting both $A_{(i,j)}$ and $A_{(j,i)}$ to 1 for each pair $(i, j)$.

Once the input is defined, we must check whether it is valid. This means checking whether the graph described by the input is simple, planar, and 3-connected. We describe the approach for this in the next three paragraphs:

**Simple** To check whether the graph is simple, we can proceed as follows. To test for loops, we check the diagonal of the adjacency matrix. If any diagonal entry is nonzero, the input is invalid. Multiple edges are ruled out from the start, since the adjacency matrix only allows entries in $\{0, 1\}$. Duplicate entries in the edge or facet list are ignored when the adjacency matrix is constructed. To check for directed edges, we test whether

$$A^T - A = 0$$

holds. If any entry is nonzero, the input is invalid. If all criteria are met, the graph is simple.

**3-connected** To test whether a graph $\mathbf{G} = (\mathbf{V}, \mathbf{E})$ is 3-connected, we use the algorithm from [Eve11, Sec. 6.2, pp. 121–129], which in turn relies on Menger's Theorem A.15 and the algorithm by George Bernard Dantzig and Delbert Ray Fulkerson [DF57]. The algorithm works as follows:

Let $v_i, v_j \in \mathbf{V}$ be two nodes, and let $\mathbf{n}(v_i, v_j)$ be the minimum number of nodes from **V** that must be removed so that $v_i$ and $v_j$ are no longer connected by a path. If $v_i$ and $v_j$ are adjacent, we set

$$\mathbf{n}(v_i, v_j) := |\mathbf{V}| - 1,$$

because all others and one of the two must be removed to disconnect them.

Otherwise, $\mathbf{n}(v_i, v_j)$ gives the number of disjoint paths between $v_i$ and $v_j$. The algorithm from [Eve11] computes $\mathbf{n}(v_i, v_j)$ for all combinations $v_i, v_j \in \mathbf{V}$ and returns

$$\min_{v_i, v_j \in \mathbf{V}} \mathbf{n}(v_i, v_j)$$

We describe our implementation based on the proof of [Eve11, Thm. 6.4, pp. 122–124] in Algorithm 1:

**Algorithm 1:** 3ConnectedTest

**Data:** Adjacency matrix $A$
**Result:** true or false

**begin**
  $\mathbf{G} = (\mathbf{V}, \mathbf{E}) = \text{graph}(A)$;
  $n = |V|$;
  $\mathbf{n} = \infty$;
  Replace each undirected edge in **G** with two directed edges;
  Replace each node $v$ with two nodes $v'$ and $v''$. Connect $v'$ and $v''$ with a directed edge $(v', v'')$ of weight 1;
  Redirect each edge pointing to $v$ to $v'$. Set its weight to $\infty$;
  Redirect each edge with $v$ as the starting node to start at $v''$. Set its weight to $\infty$;
  **for** $i = 1 : n$ **do**
    **for** $j = 1 : n$ **do**
      **if** $i \neq j$ **then**
        **if** $(v_i'', v_j') \in \boldsymbol{E} \,||\, (v_j'', v_i') \in \boldsymbol{E}$ **then**
          $\mathbf{n}' = |\mathbf{V}| - 1$;
        **else**
          $\mathbf{n}' =$ Compute the max flow from source $v_i''$ to sink $v_j'$;
        $\mathbf{n} = \min(\mathbf{n}, \mathbf{n}')$;
  **if** $\boldsymbol{n} \geq 3$ **then**
    **return** *true*;
  **else**
    **return** *false*;

First, the graph is transformed in such a way that the number of paths can be computed using flow. For this, a construction is created for each node, consisting of $v'$ and $v''$, which can pass a value of at most 1. In a figurative sense, this ensures that each node in the flow can be visited only once.

If $v_i$ and $v_j$ are adjacent, then, as already mentioned, we set $\mathbf{n}(v_i, v_j) = |\mathbf{V}| - 1$. Otherwise, we compute the maximum flow. The minimum among all such computed values is stored in the variable **n**. If, in the end, this value is $\geq 3$, the graph is 3-connected and the algorithm returns *true*. Otherwise, the graph is not valid and the result is *false*.

**Planarity** To test whether the graph $\mathbf{G} = (\mathbf{V}, \mathbf{E})$ is planar, we use the method of Fraysseix and Rosenstiehl [FR85] as implemented by Brandes [Bra09]. First, we note the following:

**Remark 3.11.** *There is a wide variety of algorithms in the literature for testing planarity. It is possible to test planarity in runtime $\mathcal{O}(|\boldsymbol{E}|)$. The algorithms described in the literature use clever data structures to achieve this runtime. However, this makes both the programming and the explanation of the algorithms very complex, which is why we have chosen the version by Ulrik Brandes [Bra09]. Although it has a runtime of only $\mathcal{O}(|\boldsymbol{E}|^2)$, it is easier to understand and implement. Since the overall algorithm does not have linear runtime anyway, the mentioned advantages outweigh the disadvantages.*

In Algorithm 2, we only explain the idea behind Brandes' algorithm and refer to [Bra09] for details.

**Algorithm 2:** PlanarityTest

**Data:** Adjacency matrix $A$
**Result:** true or false
**begin**
  $\mathbf{G} = (\mathbf{V}, \mathbf{E}) = \text{graph}(A)$;
  **if** $|\boldsymbol{E}| > 3|\boldsymbol{V}| - 6$ **then**
    **return** *false*;
  Select a random node and construct the oriented DFS (depth-first search) graph;
  Add all remaining edges in the opposite direction;
  Assign each node its height and base value;
  Compute the dependency matrix of the backward edges;
  **if** *the dependency matrix is balanced* **then**
    **return** *true*;
  **else**
    **return** *false*;

**Initialization** First, the graph is constructed from the adjacency matrix and the necessary condition for planar graphs $|\mathbf{E}| \leq 3|\mathbf{V}| - 6$ is checked. Verifying this condition takes almost no time and can rule out planarity in many cases. If this condition is met, the actual algorithm can begin.

**Oriented DFS Tree and Backward Edges** DFS stands for „depth first search“ and describes a *depth-first search* starting from a root node. The result is a tree in the graph-theoretic sense with directed edges. The height of each node is its position in the order in which the DFS algorithm encounters it. The edges point from a node with lower height to one with higher height, i.e., they point away from the source.

To the DFS tree, we add all remaining edges from the original graph. These also receive a direction, from higher to lower height. They thus point in the opposite direction to the tree edges and are therefore called *backward edges*. The base value of a node $v_i$ is the smallest height of any node that can be reached from $v_i$ via arbitrary tree edges and at most one backward edge. If no such node exists, the base value of $v_i$ is its own height.

**Dependency Matrix of the Backward Edges** The idea of the planarity test begins with the DFS tree. This can be drawn planarly. Backward edges are then added step by step to complete the graph. When drawing a backward edge, a decision must be made as to whether it will be placed to the left or right of the two nodes. This is not initially clear, and depending on the graph, it might even be possible to place an edge on either side.

However, there are edges that must be placed on opposite sides, because otherwise they would necessarily cross. These edges are called *T-opposite*. Similarly, there are edges that must be placed on the same side because both are T-opposite to the same edge. These are called *T-alike*. This information is encoded in the dependency matrix. If there are two edges that cannot be on the same side or on opposite sides, the graph is not planar. For details, see [Bra09].

**Balanced Dependency Matrix** This means checking whether all conditions encoded in the dependency matrix can be fulfilled simultaneously. An algorithm with a detailed description is provided in [HK80].

In this section, we already explained how we construct the adjacency matrix from the different input options. The algorithm for constructing the polytope also requires the list of facets of the graph. With Section A.4, we already have all the required results to create such a list. First, we note that facets are defined only for planar graphs according to Definition A.22, and many results from Section A.4 refer to planar 3-connected graphs. If the planarity and 3-connectivity tests described above are successful, a list of facets $F$ can be constructed using Algorithm 3. We now describe the individual steps of the algorithm:

**Algorithm 3:** CreateFacetList

**Data:** Adjacency matrix $A$ of a planar 3-connected graph

**Result:** Facet list $F$

**begin**
- FacetSize=3;
- FacetList=[];
- $\mathbf{G} = (\mathbf{V}, \mathbf{E}) = \text{graph}(A)$;
- **while** $|FacetList| < 2 - |\boldsymbol{E}| + |\boldsymbol{V}|$ **do**
  - Find all cycles of size *FacetSize* in **G**;
  - Check each found cycle to see if it is an induced non-separating cycle. If so, add it to the facet list;
  - Check each pair of found facets for a shared edge. If found, delete the edge from the graph **G**;
  - FacetSize = FacetSize + 1;

With Theorem A.27, Corollary A.28, and Theorem A.29, we know that the number of facets in any representation of the graph is $2 - |\mathbf{E}| + |\mathbf{V}|$.

A facet is, according to Theorem A.25, an induced non-separating cycle. The induced property follows directly from the adjacency matrix. If two non-adjacent nodes in a cycle share an edge, the cycle is not induced. The non-separating property can be checked by removing the found cycle from the graph and testing the connectivity using the Matlab function `conncomp` (see [Mata]).

By Theorem A.30, each edge in a 3-connected planar graph lies on exactly two facets. Once these two are found, the edge can be deleted from the graph to improve the algorithm's runtime.

With this, all preparations are complete. Before we describe the properties of the generated 3-polytopes in the next section, we explain the generation of random examples.

### 3.1.1 Generation of Random Examples

To empirically test the generation of the subdivision matrices in Chapters 5 and 6, we need a large number of examples. Since both algorithms require the same input as defined in Definition 3.10, we describe here how we construct random adjacency matrices of 3-connected planar graphs. For this, we use Algorithm 4, which we explain in the following:

**Dual Polytope** We first create a dual polytope by generating $n$ points on the unit circle. Choosing between $4$ and $k$ ensures that we have the minimum number (4) of points for a dual polytope, and the maximum (k) ensures that the examples do not become too large. For the examples used in this work, we arbitrarily set this maximum to 100. For these random points, we generate the facets of the corresponding convex hull. If (by chance) more than three points lie in a facet, this facet is triangulated at random.

Using the facet information, we can then generate an adjacency matrix $A^*$ of the points. It is important to note that the polytope itself is no longer relevant at this point; only the structure generated in $A^*$ matters.

**Dual Graph and Edge Deletion** The adjacency matrix $A^*$ is the adjacency matrix of a dual graph. Since the facets in $F^*$ are all triangles, each point of a potential primal graph has three neighbors. This is our preferred type of graph, as it corresponds to a hexahedral structure defined in Chapter 2 (see Definition 2.51 and Quality Criterion Q9). However, since we also want to test other 3-connected planar graphs, we randomly delete some edges in certain examples so that nodes in the primal graph have more than three neighbors.

**Primal Graph** If the dual graph $\mathbf{G}^*$ created from $A^*$ is planar and 3-connected, we construct the corresponding primal graph **G**. For safety, we then check again whether it is also planar and 3-connected. If this is the case and **G** has at most $k$ nodes, then the example is valid and the adjacency matrix is returned. Since we are considering many

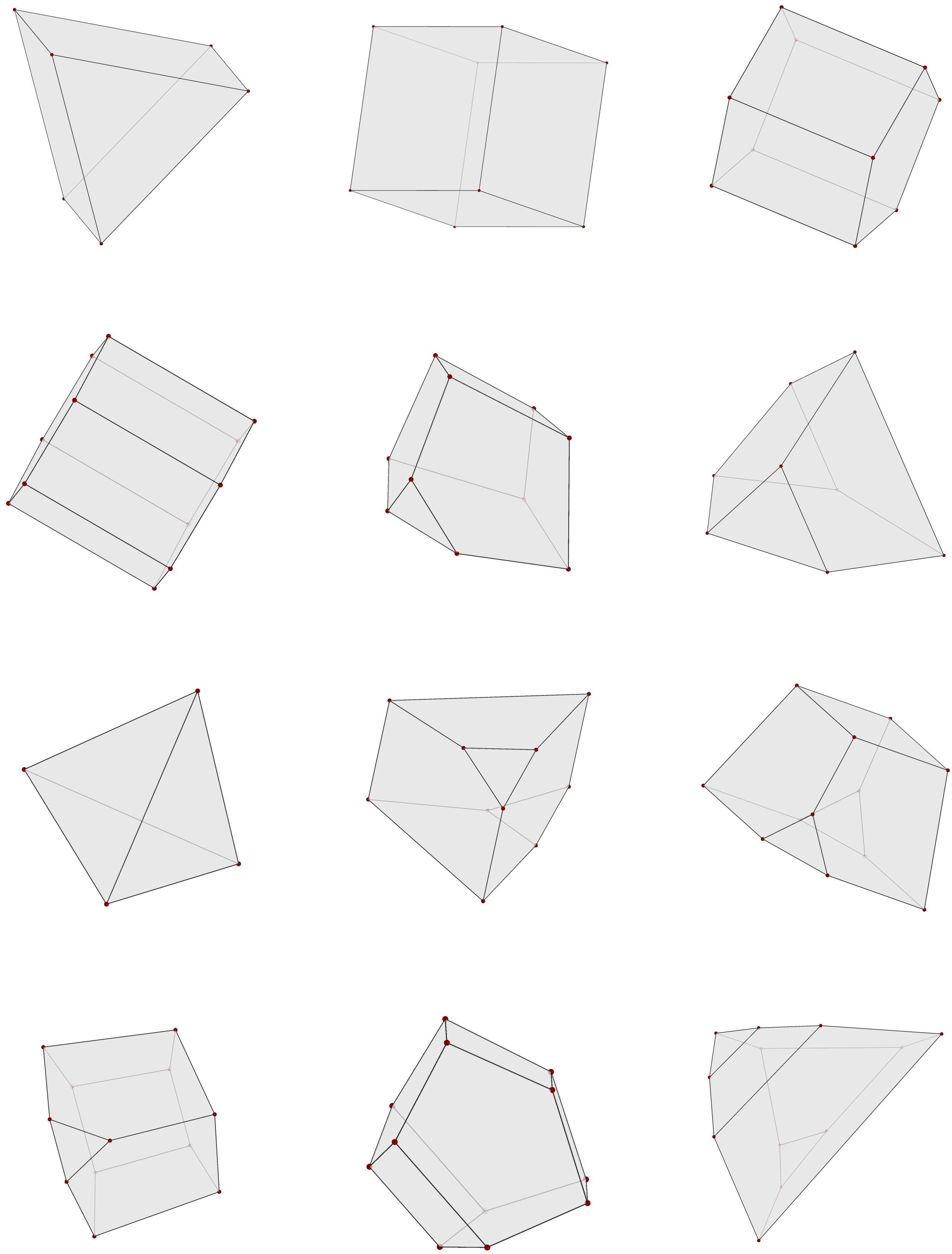

**Figure 3.1:** Gallery of random 3-polytopes whose graph structure was generated using Algorithm 4.

**Algorithm 4:** GenerateRandomGraph

**Result:** Adjacency matrix $A$ of a 3-connected planar graph
**begin**
  EdgeDeletion = random value from $0$ or $1$;
  **while** *true* **do**
    $k = 100$;
    $n =$ random number from $\{4, \dots, k\}$;
    $P^* =$ random matrix from $[-0.5, 0.5]^{(n \times 3)}$, where each row is non-zero;
    Normalize each row of $P^*$ so that each point lies on the unit circle;
    $F^* =$ create a triangulated facet list of the convex hull of $P^*$;
    $A^* =$ adjacency matrix corresponding to $F^*$;
    **if** *EdgeDeletion* **then**
      Set random pairs $A^*_{(i,j)}$ and $A^*_{(j,i)}$ to 0 if they were 1;
    Delete all empty rows and columns from $A^*$;
    $\mathbf{G}^* =$ graph corresponding to $A^*$;
    **if** $\mathbf{G}^*$ *is planar and 3-connected* **then**
      Construct the primal graph **G** from $\mathbf{G}^*$ with adjacency matrix $A \in \{0, 1\}^{m \times m}$;
      **if** **G** *is planar and 3-connected and* $m \leq k$ **then**
        **return** $A$

examples, we need to store the generated adjacency matrices accordingly. For storage, all adjacency matrices are converted to the same format, for which we fix the size to $k \times k$ matrix entries.

**Dataset** For this work, a total of 305,095 examples were constructed, each with its corresponding adjacency matrix saved. Duplicates, i.e. exactly identical adjacency matrices, were removed before saving. However, the dataset does contain isomorphic graphs. If a graph is isomorphic to another, this information is also stored, in order to later test the corresponding subdivision matrices for isomorphism. The dataset contains a total of 36,336 isomorphic graphs and thus 268,759 unique examples.[2] Additionally, 186,506 of the examples are graphs in which each node has exactly three neighbors, of which 158,527 are unique examples.

## 3.2 Selection and Properties of Convex Polytopes

Before we can describe the construction of $3$-*polytopes* in Section 3.3, we examine some theoretical results, since we will need polytopes with specific properties for the remainder of this work. The constructed polytopes should consist of edges that are tangent to the unit sphere. Furthermore, the tangency points should have their centroid at the origin, and all symmetries of the underlying structure should be reflected by the polytope. The same should apply to the corresponding dual polytopes. The 3-polytopes in this work are generated using the special technique of *circle packing*, from which the desired properties follow. The core of this theory is the Koebe-Andreev-Thurston Theorem. The specific version from Theorem 3.25 at the end of this section is the one relevant to us. This section thus serves as a (historical) lead-up to this final version.

The first question that arises when constructing convex polytopes from graphs is whether a convex polytope has a unique representation. The trivial answer is of course no. Every convex hull of four points in $\mathbb{R}^3$ that do not lie in a single plane is a tetrahedron and therefore a convex 3-polytope. If different realizations of a polytope exist for a given graph structure, the question becomes: What is the right choice for constructing and realizing a convex 3-polytope for our purposes?

[2] If multiple graphs are isomorphic to each other, one is counted as a unique example and the others as isomorphisms.

To answer this, we require the Koebe-Andreev-Thurston Theorem from the field of *circle packings*. This theorem guarantees the existence of a polytope with the properties we desire, as summarized in Theorem 3.25. The history of this theorem is quite extensive and there are multiple formulations and variants. Therefore, we will go through these step by step, gradually adding more expressive power to the theorem until we arrive at our final version in Theorem 3.25. Much of the background information and historical context provided here comes from [Zie94, pp. 115–119].

The variants of the theorem introduce terminology that is not explained in detail here. The goal of this section is to provide an overview of the different versions and to extract the relevant aspects from each. All terms important for this work will be defined precisely during the polytope construction in Section 3.3.

The first version of the theorem is the formulation by Koebe [Koe36, p. 162].

**Theorem 3.12** (Koebe). *The task of placing circular disks, with unknown sizes, side by side on the surface of the sphere $n$ without overlapping, so that they fulfill a contact scheme prescribed by an arbitrary common triangulation pattern (closure problem), always permits one and, except for a circle transformation, only one solution.*

A more modern formulation can be found in the dissertation by Springborn [Spr03, Thm. 1.1, p. 1]:

**Theorem 3.13** (Koebe as cited by Springborn). *For every abstract triangulation of the sphere there is a circle packing whose adjacency graph is a geometric realization of the triangulation. The circle packing corresponding to a triangulation is unique up to Möbius transformations of the sphere.*

Koebe's theorem was rediscovered by Thurston, who traced its proof back to a theorem by Andreev. In this context, Ziegler refers to [And70], where the following theorem can be found, among others [And70, p. 431]:

**Theorem 3.14** (Andreev). *Let $m$ be a convex abstract three-dimensional polyhedron with vertices of simplicial type, but not a simplex. The conditions $m_0 - m_4$ are necessary and sufficient for the existence of a convex bounded polyhedron $M$ in three-dimensional Lobachevsky space $\lambda^3$ with dihedral angles not greater than $\pi/2$, such that $a_{ij}(M) = a_{ij}$.*

The original source regarding Thurston is his lecture notes [Thu77]. These were also published decades later in [Thu22], and so we use this version as our reference. Thurston uses the following formulation of Andreev [Thu22, Thm. 13.6.1, p. 290]:

**Theorem 3.15** (Andreev as cited by Thurston). *Let $O$ be a Haken orbifold with*

$$X_O = D^3, \quad \sum_O = \partial D^3.$$

*Then $O$ has a hyperbolic structure if and only if $=$ has no incompressible Euclidean suborbifolds. If $O$ is a Haken orbifold with $X_O = D^3 -$ (finitely many points) and $\sum_O = \partial D^3$, and if a neighborhood of each deleted point is the product of a Euclidean orbifold with an open interval, (but $O$ itself is not such a product) then $O$ has a complete hyperbolic structure with finite volume if and only if each incompressible Euclidean suborbifold can be isotoped into one of the product neighborhoods.*

This wording does not appear in [And70], but it is nevertheless cited in the literature as the corresponding source.

Independently, based on the theorem just described, Thurston formulated the following two theorems (Thurston I: [Thu22, Cor. 13.6.2, p. 290] and Thurston II: [Thu22, Cor. 13.6.3, p. 293]):

**Theorem 3.16** (Thurston I). *Let **G** be any graph in $\mathbb{R}^2$ such that each edge has distinct ends and no two vertices are joined by more than one edge. Then there is a packing of circles in $\mathbb{R}^2$ whose nerve is isotopic to **G**. If **G** is the one-skeleton of a triangulation of $\mathbb{S}^2$, then this circle packing is unique up to Möbius transformation.*

**Theorem 3.17** (Thurston II). *Let $T$ be any triangulation of $\mathbb{S}^2$. Then there is a convex polyhedron in $\mathbb{R}^3$3, combinatorially equivalent to $T$ whose one-skeleton is circumscribed about the unit sphere (i.e., each edge of $T$ is tangent to the unit sphere). Furthermore, this polyhedron is unique up to a projective transformation of $\mathbb{R}^3$ which preserves the unit sphere.*

The work of the three authors Koebe, Andreev, and Thurston led to the so-called „Koebe-Andreev-Thurston Circle Packing Theorem“. As already mentioned, there are numerous variants of the theorem, some of which we will consider in the following. We begin with the version by Ziegler [Zie04, Thm. 1.3, p. 10]:

**Theorem 3.18** (Koebe-Andreev-Thurston Theorem as cited by Ziegler)**.** *Each 3-connected planar graph can be realized by a [convex] 3-polytope which has all edges tangent to the unit sphere. Moreover, this realization is unique up to M¨obius transformations (projective transformations that fix the sphere). The edge-tangent realization for which the barycenter of the tangency points is the center of the sphere is unique up to orthogonal transformations.*

It should first be noted that Ziegler does not explicitly speak of convex polytopes in his formulation. However, he states in [Zie04, p. 3] that all polytopes treated in these lecture notes are convex, so we can include that here.

This theorem provides us with an initial result that describes a unique polytope (up to rotations and reflections about the origin) and offers a first indication of the choice of polytope for this work. However, the above formulation does not yet include all the desired properties of the polytope, which is why we need to discuss further variants of the theorem.

The added value of [Zie04] is that Ziegler presents a constructive proof of the theorem using the functional by Bobenko and Springborn [BS04]. The polytope mentioned above can thus be constructed explicitly using Ziegler's instructions, and we use this construction in Sections 3.3.1 and 3.3.2.

Ziegler's construction, however, also has some disadvantages with regard to implementation. The proof itself ends after the construction of the special quad graph (see Section 3.3.1) with the statement [Zie04, p. 24] that, starting from a correct rectangular circle packing, the spherical circle packing can easily be reconstructed (using an inverse stereographic projection).

However, the choice of the correct M"obius transformation that ensures that the centroid of the tangency points lies at the origin after stereographic projection is not described. While Ziegler's proof establishes the uniqueness of the polytope up to M"obius transformations, it does not provide statements about the existence of a specific M"obius transformation with the desired properties. For this reason, we consider a second formulation of the theorem from [Zie94], which represents a generalization of the above theorem [Zie94, Thm. 4.13, p. 118]:

**Theorem 3.19.** *For every planar 3-connected graph, there is a representation as the graph of a 3-polytope whose edges are all tangent to the unit sphere $\mathbb{S}^2 \subseteq \mathbb{R}^3$, and such that $0$ is the barycenter of the contact points. This representation is unique up to rotations and reflections of the polytope in $\mathbb{R}^3$. In particular, in this representation every combinatorial symmetry of the graph is realized by a symmetry of the polytope.*

Convexity can again be added here, since Ziegler's definition of 3-polytopes agrees with Definition 3.4. Ziegler refers here among others to the work of Schramm [Sch92], who generalized the Koebe-Andreev-Thurston Theorem from the sphere to general convex bodies.

The above theorem focuses on the realization of the 3-polytope with centroid at $0$ and adds a statement about the symmetry of the polytope. In Ziegler's book [Zie94], there is no detailed proof of the theorem, but he refers to his lecture notes [Zie04], from which we already took Theorem 3.18. These notes also include the reference for the existence and uniqueness of the 3-polytope with centroid at 0. The corresponding argument and proof are found in [Spr05], from which we cite the following theorem [Spr05, Thm. 2, p. 514] and lemma [Spr05, Lem. 1, p. 514]:

**Theorem 3.20.** *For every combinatorial type of convex 3-dimensional polyhedra there is a unique polyhedron (up to isometry) with edges tangent to the unit sphere $\mathbb{S}^2 \subset \mathbb{R}^3$, such that the origin $0 \in \mathbb{R}^3$ is the barycenter of the points where the edges touch the sphere.*

**Lemma 3.21.** *Let $v_1, \ldots, v_n$ with $n \geq 3$ be distinct points on the $d$-dimensional unit sphere $\mathbb{S}^d \subset \mathbb{R}^{d+1}$. There exits a M¨obius transformation $T$ of $\mathbb{S}^d$ such that*

$$\sum_{j=1}^{n} T v_j = 0.$$

*If $\tilde{T}$ is another such Möbius transformation, then $\tilde{T} = RT$, where $R$ is an isometry of $\mathbb{S}^d$.*

This establishes the existence and uniqueness of the polytope with centroid of tangency points at the origin, up to isometries that preserve the sphere. What remains to be shown for Theorem 3.19 is the argument about symmetry. For this, we need the work of [BE01], which uses „quasiconvex programming" to find sets of circles on the unit sphere subject to specific conditions [BE01, pp. 14–15]:

> Spherical graph drawing [Koe36]. Any embedded planar graph can be represented as a collection of tangent circles on a sphere $\mathbb{S}^2$; this representation is unique for maximal planar graphs, up to M¨obius transformation. Our algorithms can find a canonical spherical realization of any planar graph that optimizes the minimum circle radius or the minimum separation between two vertices, and that realizes any symmetries implicit in the given embedding.

This is further explained in [BE01, p. 20] as follows:

> Due to the fact that a quasiconvex program only has a single global optimum, the transformed coin graph will display any symmetries present in the initial graph embedding. That is, any homeomorphism of the sphere that transforms the initial embedded graph into itself becomes simply a rotation or reflection of the sphere in the optimal embedding. If the graph has a unique embedding then any isomorphism of the graph becomes a rotation or reflection.

Springborn describes the connection between his work and the previously cited [BE01] as follows [Spr05, pp. 514–515]:

> A similar interplay of geometries [as in Theorem 3.20 and Lemma 3.21] leads Bern and Eppstein, to another choice of a unique representative for each polyhedral type. Given $n$ spheres in $\mathbb{S}^d$ , Bern and Eppstein apply that Möbius transformation which makes the smallest sphere as large as possible [BE01]. It is not difficult to see that this Möbius transformation is unique up to post-composition with a rotation if $n \geq 3$. Since edge-tangent polyhedra correspond to circle packings, this leads to another choice of unique representative for each polyhedral type [Epp03].
>
> For symmetric polyhedral types (more precisely, for those polyhedral types with a symmetry group of orientation preserving isomorphisms which is not just a cyclic group) the unique representative of Bern and Eppstein coincides with ours.

This provides us with the symmetry property, which will play an essential role for the quality criterion Q10.

The next step is a further generalization of Theorem 3.19: the existence and construction of a dual polytope. This will be relevant in the construction of the Colin-de-Verdière-matrix in Section 4.4 as well as in the construction of the subdivision matrices for $t = 3$ and $g = 3$. The first result concerning duality mentioned in [Zie94] is the following theorem by Brightwell and Scheinerman [BS93, Thm. 6, pp. 217–218]:

**Theorem 3.22** (Brightwell and Scheinerman I)**.** *Let* ***G*** *be a 3-connected planar graph, with a designated outside facet. There is a collection of circles in the plane, one circle representing each vertex and each facet of* ***G****, satisfying properties (1)–(5). Furthermore, this collection is unique up to linear fractional transformations and reflections of the plane.*

*(1) No two vertex-circles cross, and no two facet-circles cross.*

*(2) Corresponding to every edge $e$ of* ***G****, there is a point in the plane where four circles meet, namely, those corresponding to the two endpoints of $e$ and the two faces bounded by $e$. This point will be called an edge-point and is to be thought of as representing $e$.*

*(3) A facet-circle and a vertex-circle intersect only when the corresponding vertex is on the boundary of the corresponding facet.*

*(4) The region bounded by the circle corresponding to the outside facet contains all other facet-circles. With this exception, none of the disks bounded by one of the circles contains another of the circles.*

*(5) At each edge-point, the two vertex-circles cross the two facet-circles at right angles.*

We already refer here to the quad graph from Definition 3.27, which matches this description. Based on this theorem, Brightwell and Scheinerman prove the following result [BS93, Thm. 8, p. 218]:

**Theorem 3.23** (Brightwell and Scheinerman II)**.** *Let* ***G*** *be a 3-connected planar graph and* $\boldsymbol{G}^*$ *its planar dual. It is possible to draw* ***G*** *and* $\boldsymbol{G}^*$ *simultaneously in the plane with straight-line edges so that the edges of* ***G*** *cross the edges of* $\boldsymbol{G}^*$ *at right angles.*

Building on this, Grünbaum formulates the following generalization of the Koebe-Andreev-Thurston Theorem [Grü07, Thm. 3.1, p. 449]:

**Theorem 3.24** (Grünbaum)**.** *If* ***G*** *is a planar 3-connected graph, then* ***G*** *can be realized by a convex polyhedron* ***P*** *with the following properties:*

- *All edges of* ***P*** *are tangent to a sphere* ***K****, and the centroid of the points of tangency is the center of* ***K****.*
- *The graph* $\boldsymbol{G}^*$ *dual to* ***G*** *is realized by a convex polyhedron* $\boldsymbol{P}^*$ *dual to* ***P****, with all edges of* $\boldsymbol{P}^*$ *tangent to* ***K****.*
- *the edges of* ***P*** *and* $\boldsymbol{P}^*$ *that correspond to each other under the duality of* ***G*** *and* $\boldsymbol{G}^*$ *(and of* ***P*** *and* $\boldsymbol{P}^*$*) are mutually perpendicular.*

The combination of Theorems 3.19 and 3.24 leads us to our final formulation of the Koebe-Andreev-Thurston Theorem:

**Theorem 3.25** (Final Version of the Koebe-Andreev-Thurston Theorem)**.** *For every simple, planar, 3-connected graph* ***G****, there exists a representation as the graph of a convex 3-polytope* ***P****, whose edges all touch the unit sphere* $\mathbb{S}^2 \subseteq \mathbb{R}^3$*, such that the origin* $0$ *is the centroid of the tangency points.*

*Moreover, there exists a 3-connected planar graph* $\boldsymbol{G}^*$ *that is dual to* ***G****, and a convex dual 3-polytope* $\boldsymbol{P}^*$*, which contains one vertex for each facet, one edge for each edge, and one facet for each vertex of the primal polytope. All edges of this polytope also touch the unit sphere* $\mathbb{S}^2 \subseteq \mathbb{R}^3$*, and the centroid of the tangency points of the dual polytope* $\boldsymbol{P}^*$ *is also 0. Each pair of corresponding edges of the primal and dual polytopes intersect at their tangency point and are orthogonal to each other.*

*Both representations are unique up to rotations and reflections of the corresponding polytope in* $\mathbb{R}^3$*. In particular, in this representation, every combinatorial symmetry of the (primal and dual) graph is realized as a symmetry of the (primal and dual) polytope.*

It is worth noting that there is a similar formulation in [Spr03, Thm. 1.3, pp. 2–3] with the same strength and generality, including a proof. The difference is that Springborn expresses the theorem using the terminology of cell decompositions of the sphere rather than graph theory.

This theorem answers the initial question regarding the choice of realization for the convex polytope. The polytope described in Theorem 3.25 can be constructed uniquely, and a corresponding dual polytope exists uniquely for Chapter 4, with both fulfilling all symmetries of the underlying graph. This property will be a central part of the proof of the quality criteria Q7 (regular case), Q8 (semi-regular case), and Q10 (symmetries). Although the description of the various theorems in this section was extensive, such thoroughness is justified here, as this constitutes one of the key foundations for constructing the polytopes and proving the quality criteria. In the next section, we use the construction by [Zie04] to generate the polytope and its dual as described in Theorem 3.25.

## 3.3 Construction of Three-Dimensional Polytopes

In this section, we describe the concrete construction of the 3-polytopes in $\mathbb{R}^3$. The ideas presented here are based on the constructive proof by Ziegler [Zie04]. Since his work is of a theoretical nature, we focus here on the concrete implementation. In particular, we address numerical aspects and describe challenges and solutions related to practical realization.

To better understand the construction process and to get an idea of what the construction steps might look like, we begin by briefly describing the simpler reverse direction—namely, how to obtain a planar representation in $\mathbb{R}^2$ of the associated graph from a polytope whose edges are tangent to the unit sphere. This reverse direction is also outlined in [Zie04, p. 11] and serves here only for orientation and classification. A detailed description is given later in Section 3.3.1, which will serve as the basis for deriving the individual steps of the forward construction.

First, a circle can be created for each primal and dual vertex of the polytopes. Each such circle passes through all the tangency points of edges on the unit sphere at which the respective point lies. The dual circles lie on the facets of the primal polytope and vice versa. These circles can then be mapped to $\mathbb{R}^2$ via stereographic projection. Before doing so, the polytope is rotated so that one of the tangency points lies at the north pole of the projection. The four circles passing through the rotated point $[0, 0, 1]$ are then mapped to straight lines, while all others are mapped to circles. From the centers and intersection points of the projected circles, a graph can be constructed, which we will refer to as the restricted quad graph. The centers of primal circles correspond to primal points, the centers of dual

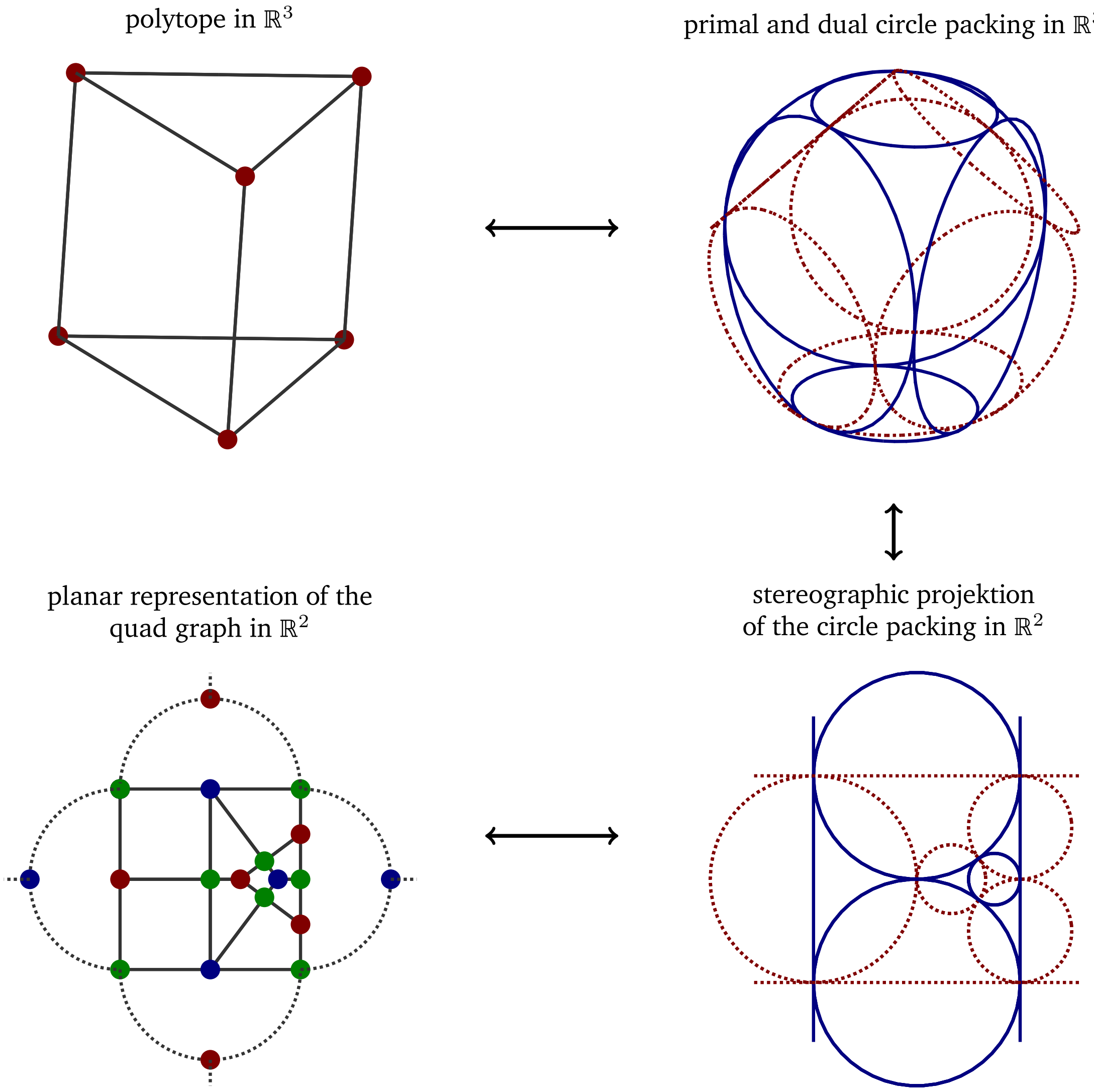

**Figure 3.2:** General procedure for converting a polytope into a quad graph and vice versa. The polytope is first transformed into a circle packing, which is then mapped to $\mathbb{R}^2$ via stereographic projection and subsequently transformed into a quad graph. The construction of the polytope in this chapter proceeds in the reverse order.

**Algorithm 5:** Construction3Polytope

**Data:** Adjacency matrix $A$, or edge list $E$, or facet list $F$ of a 3-connected planar graph **G** as defined in Definition 3.10

**Result:** Vertex matrix $P \in \mathbb{R}^{n \times 3}$ of the primal 3-polytope and vertex matrix $P^\star \in \mathbb{R}^{n \times 3}$ of the dual 3-polytope, with the properties from Theorem 3.25

**begin**
- **if** *Input is $E$ or $F$* **then**
  - Generate adjacency matrix $A$ from $E$ or $F$;
- Test with $\max|\mathtt{diag}(A)| = 0$ whether **G** contains no loops;
- Test with $A^T = A$ whether **G** is undirected;
- Test with Algorithm 1 whether **G** is 3-connected;
- Test with Algorithm 2 whether **G** is planar;
- **if** *any of the tests failed* **then**
  - **return** *error('Input was not valid')*;
- **if** *Input is not $F$* **then**
  - Generate facet list $F$ using Algorithm 3;
- Construct from **G** the quad graph **Q**(**G**) from Definition 3.27 as described in Section 3.3.1;
- Construct from **Q**(**G**) the restricted quad graph **Q**′(**G**) from Definition 3.32 as described in Section 3.3.1;
- Create a list $K$ of all (quadrilateral) faces from **Q**(**G**) that consist only of vertices from **Q**′(**G**). These are the combinatorial representations of the right-angled kites we will construct in the next steps. Each quadrilateral consists of a primal point $p$, two edge points $e_1$ and $e_2$, and a dual point $f$;
- Assign a radius $r$ to each primal point $p \in \mathbf{V}$ and each dual point $f \in \mathbf{V}^\star$. Compute these radii as described in Section 3.3.1 using Algorithm 6;
- Realize the restricted quad graph in $\mathbb{R}^2$ by drawing each quadrilateral from $K$ as a right-angled kite using the radii $r$ in Algorithm 7;
- Compute the degrees of freedom for the inverse stereographic projection $a_x$, $a_y$, and $a_z$ using Algorithm 8;
- Translate the realization of the restricted quad graph by $(a_x, a_y)$ and then scale it by $a_z$;
- Project the tangency points $e$ of the translated and scaled realization of the restricted quad graph onto the unit sphere using the inverse stereographic projection;
- Determine the vertices $P$ of the primal polytope **P** and the vertices $P^\star$ of the dual polytope $\mathbf{P}^\star$ by solving a system of equations derived from the tangent planes at the tangency points $e$ neighboring the vertices $p$ and $f$ on the sphere;
- **return** *Vertices $P$ of the primal polytope and vertices $P^\star$ of the dual polytope*;

circles to dual points, and the intersection points of the circles correspond to the tangency points. An illustration is provided in Figure 3.2.

The construction of the polytope reverses these steps and is divided into two sections. In the first Section 3.3.1, we describe the construction of the restricted quad graph, including the determination of the two-dimensional coordinates for the realization of the vertices in $\mathbb{R}^2$. The tangency points are of particular importance in this context. In the second Section 3.3.2, we explain how these tangency points are mapped back to $\mathbb{R}^3$ or to the unit sphere $\mathbb{S}^2$ using inverse stereographic projection, and how the primal and dual polytopes are constructed from them. An overview of all construction steps described below is given in Algorithm 5.

Additionally, we introduce the following notation for this section:

**Notation 3.26.** *In this section, we construct a 3-polytope from a planar 3-connected graph. For this purpose, we require three categories of points:*

- *The first category of points refers to the vertices of the primal polytope. These are associated with the nodes of* **G**. *We denote all related concepts (node, realization in $\mathbb{R}^2$, and realization in $\mathbb{R}^3$) of these vertices with $p$. In illustrations, these points are always shown in red.*

- *The second category of points refers to the tangential edge points, which will lie on the unit sphere. These will also be briefly called tangency points. They are associated with the edges of* **G**. *We denote all related concepts (realization in* $\mathbb{R}^2$ *and realization in* $\mathbb{R}^3$*) of these tangential edge points with* $e$. *In illustrations, these points are always shown in green.*
- *The third category of points refers to the vertices of the dual graph. These are associated with the faces of* **G**. *We denote all related concepts (realization in* $\mathbb{R}^2$ *and realization in* $\mathbb{R}^3$*) of these dual facet points with* $f$. *In illustrations, these points are always shown in blue.*

*Furthermore, throughout this section, we use the respective graphs and their realizations in* $\mathbb{R}^2$ *and* $\mathbb{R}^3$ *interchangeably. The same applies to the terms node and point.*

### 3.3.1 Construction of the Quad Graph

As described at the beginning, the first step in the construction is to find a representation of the primal and dual graphs in $\mathbb{R}^2$, which can then be transformed into a polytope in the next section. This construction forms the core of Ziegler's proof of the Koebe-Andreev-Thurston Theorem. Accordingly, this section follows the structure of [Zie04]. We represent the primal and dual graphs in a common graph, which we call the *quad graph* (cf. [Zie04, p. 11–15]):

**Definition 3.27.** *Let* $\boldsymbol{G} = (\boldsymbol{V}, \boldsymbol{E})$ *be a planar 3-connected graph and* $\boldsymbol{G}^* = (\boldsymbol{V}^*, \boldsymbol{E}^*)$ *its dual graph. Then the* quad graph $\boldsymbol{Q}(\boldsymbol{G}) = (\boldsymbol{V_Q}, \boldsymbol{E_Q})$ *is a graph with the following properties:*

- *Each node in* $\boldsymbol{V_Q}$ *represents either a node of* **G***, an edge of* **G***, or a facet of* **G** *(and thus a node of* $\boldsymbol{G}^*$*). Let* $\boldsymbol{V_E}$ *be the nodes that represent the edges of* **G**. *Then we define*

  $$\boldsymbol{V_Q} := \boldsymbol{V} \cup \boldsymbol{V_E} \cup \boldsymbol{V}^*.$$

  *The ordering of the nodes in* $\boldsymbol{Q}(\boldsymbol{G})$ *in the adjacency matrix and in any other object in this chapter is* $[\boldsymbol{V}, \boldsymbol{V_E}, \boldsymbol{V}^*]$.
- *There are two types of edges in* $\boldsymbol{Q}(\boldsymbol{G})$. *A node from* **V** *is connected to a node from* $\boldsymbol{V_E}$ *by an edge if the corresponding node from* **V** *is part of the corresponding edge from* **E**.
  *A node from* $\boldsymbol{V}^*$ *is connected to a node from* $\boldsymbol{V_E}$ *by an edge if the corresponding edge from* **E** *is part of the corresponding facet that is dual to the node from* $\boldsymbol{V}^*$. *Thus, edge points are connected to the corresponding node and facet points on which they lie, and these form all the edges of* $\boldsymbol{Q}(\boldsymbol{G})$.

The structure of the quad graph of **G** can be generated directly from the adjacency matrix $A$ of **G** and the facet list $F$ of **G**. The creation of these two objects has already been explained in Section 3.1.

For this quad graph, we want to find a representation in $\mathbb{R}^2$ that satisfies certain conditions. To work out these conditions, we return to the reverse construction described at the beginning of Section 3.3 and elaborate on it. If we understand how the realization of the quad graph in $\mathbb{R}^2$ is formed from the 3-polytope, we can determine how the quad graph must be drawn so that we can build a 3-polytope from it. For the reverse construction, we refer to Figure 3.2, which illustrates the individual steps.

#### Reverse Construction

We begin with a convex polytope whose edges are tangent to the unit sphere, along with its dual polytope, which also has edges tangent to the unit sphere. Each pair of edges from the primal and dual polytope shares the same tangency point, and the two edges are orthogonal to each other (see Theorem 3.25).

Since each facet (primal and dual) lies in a plane that intersects the sphere, all tangency points associated with a given facet lie on a corresponding primal or dual circle. Two primal (or dual) circles sharing a common tangency point are tangent to each other, and the tangents at that point are identical. Two tangents—one from a primal circle and one from a dual circle—that meet at the same tangency point are orthogonal to each other. We identify the circle of a facet of the primal polytope with the corresponding vertex of the dual polytope, and vice versa. Specifically, this means that a primal circle is associated with a primal point and lies in a dual facet, and a dual circle is associated with a dual point and lies in a primal facet. An illustration is shown in Figure 3.3. The collection of these circles is referred to as a *circle packing*.

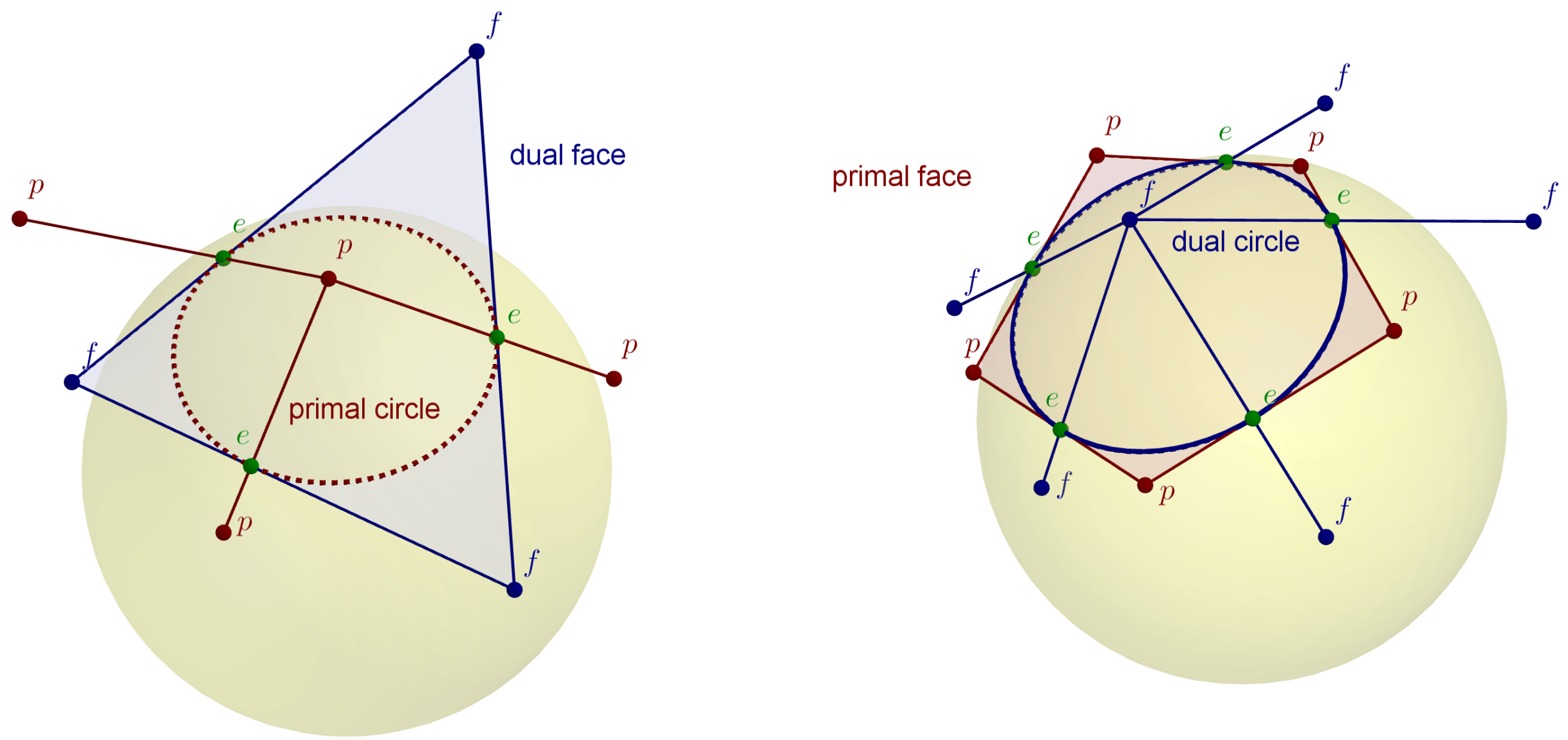

**Figure 3.3:** Illustrative example of a primal circle associated with a primal point and lying in a dual facet (left), and a dual circle associated with a dual point and lying in a primal facet (right).

We aim to map such a circle packing to $\mathbb{R}^2$ using stereographic projection. For this purpose, we define the projection following [Lee20, Def. 3.1 and Def. 3.3, pp. 47, 49]:

**Definition 3.28.** *For every point $a \in \mathbb{S}^2 \setminus \{[0,0,1]\}$, there exists a unique point $b = [b_x, b_y]$ such that the line from $[0,0,1]$ to $[b_x, b_y, 0]$ passes through $a$. The bijective mapping*

$$\phi : \mathbb{S}^2 \to \mathbb{R}^2 \cup \{\infty\}, \quad \textit{with} \quad \phi(a) = \begin{cases} \infty & \textit{for } a = [0,0,1] \\ b & \textit{otherwise} \end{cases}$$

*is called the* stereographic projection. *Its inverse*

$$\phi^{-1} : \mathbb{R}^2 \cup \{\infty\} \to \mathbb{S}^2 \quad \textit{with} \quad \phi^{-1}(b) = a$$

*is called the* inverse stereographic projection.

This mapping can be made explicit using the following theorem [Lee20, Theorem 3.2, p. 48]:

**Theorem 3.29.** *We obtain*

$$\phi(x,y,z) = \left(\frac{x}{z-1}, \frac{y}{z-1}\right) =: (\overline{x}, \overline{y})$$

*and*

$$\phi^{-1}(\overline{x}, \overline{y}) = \left(\frac{2\overline{x}}{\overline{x}^2 + \overline{y}^2 + 1}, \frac{2\overline{y}}{\overline{x}^2 + \overline{y}^2 + 1}, \frac{\overline{x}^2 + \overline{y}^2 - 1}{\overline{x}^2 + \overline{y}^2 + 1}\right) = (x,y,z).$$

An illustration of the stereographic projection can be found in Figure 3.4. We also obtain the following statement about angles, which we will need later on [Lee20, Thm. 3.5, p. 53]:

**Theorem 3.30.** *The stereographic projection preserves angles.*

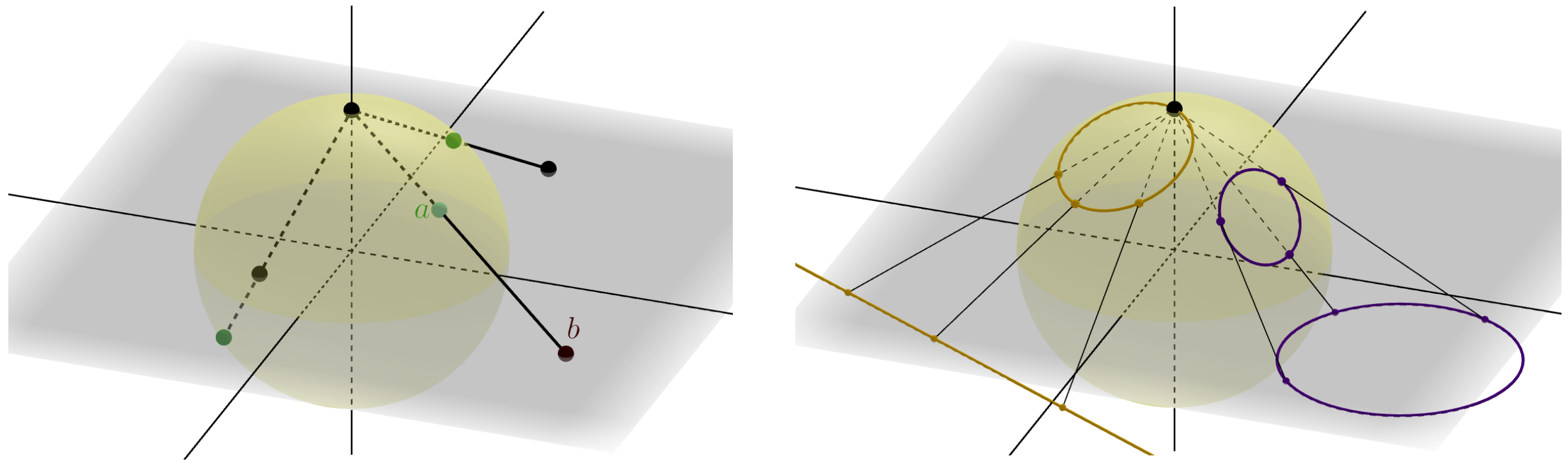

**Figure 3.4:** Illustrative example of the stereographic projection for three points (left) and illustration of Theorem 3.31 (right).

Furthermore, we obtain the following result about the mapping of circles under stereographic projection from [Lee20, Thm. 3.4, p. 49]:

**Theorem 3.31.** *The stereographic projection maps circles on the unit sphere that contain the point* $[0, 0, 1]$ *to straight lines in* $\mathbb{R}^2$*, and maps circles on the unit sphere that do not contain the point* $[0, 0, 1]$ *to circles in* $\mathbb{R}^2$.

Since the stereographic projection is bijective, the inverse also holds accordingly: lines in $\mathbb{R}^2 \cup \{\infty\}$ are mapped to circles on the sphere that contain the point $[0, 0, 1]$, and circles in $\mathbb{R}^2 \cup \{\infty\}$ are mapped to circles on the sphere that do not contain the point $[0, 0, 1]$.

We use these theorems in the construction of the quad graph. We start with the polytope and rotate it so that one of the tangency points becomes $[0, 0, 1]$. This can be achieved by rotating the polytope on the unit sphere. The resulting circle packing contains exactly four circles that include the point $[0, 0, 1]$. Under stereographic projection, these are mapped to straight lines in $\mathbb{R}^2 \cup \{\infty\}$. Since the stereographic projection preserves angles, the images of the two primal circles and the images of the two dual circles under $\phi$ are parallel, and the combination of primal and dual lines is orthogonal. Furthermore, the polytope can be rotated such that these lines are aligned with the $x$- and $y$-axes (see Figure 3.2). All other circles of the circle packing are mapped to circles by $\phi$.

This two-dimensional circle packing can be converted into a quad graph, or more precisely, into a realization of the quad graph in $\mathbb{R}^2$, by proceeding as follows:

- For the center of each primal circle, a primal point $p$ is placed. This point is the projection of the corresponding vertex of the primal polytope.
- For the center of each dual circle, a dual point $f$ is placed. This point is the projection of the corresponding vertex of the dual polytope.
- For each intersection point of a primal and a dual circle, a tangency point $e$ is placed. This point is the projection of the corresponding tangency point lying on the sphere.
- For each line, a primal or dual point is placed outside the rectangle formed by these lines.
- For the tangency point at $[0, 0, 1]$, a tangency node is placed outside the rectangle formed by the lines.
- Each primal point is connected by a line segment to the tangency points lying on the corresponding primal circle.
- Each dual point is connected by a line segment to the tangency points lying on the corresponding dual circle.

This is also illustrated in Figure 3.2. With this, we have explained the construction of the quad graph from a polytope and can now formulate the conditions required for the realization of the quad graph.

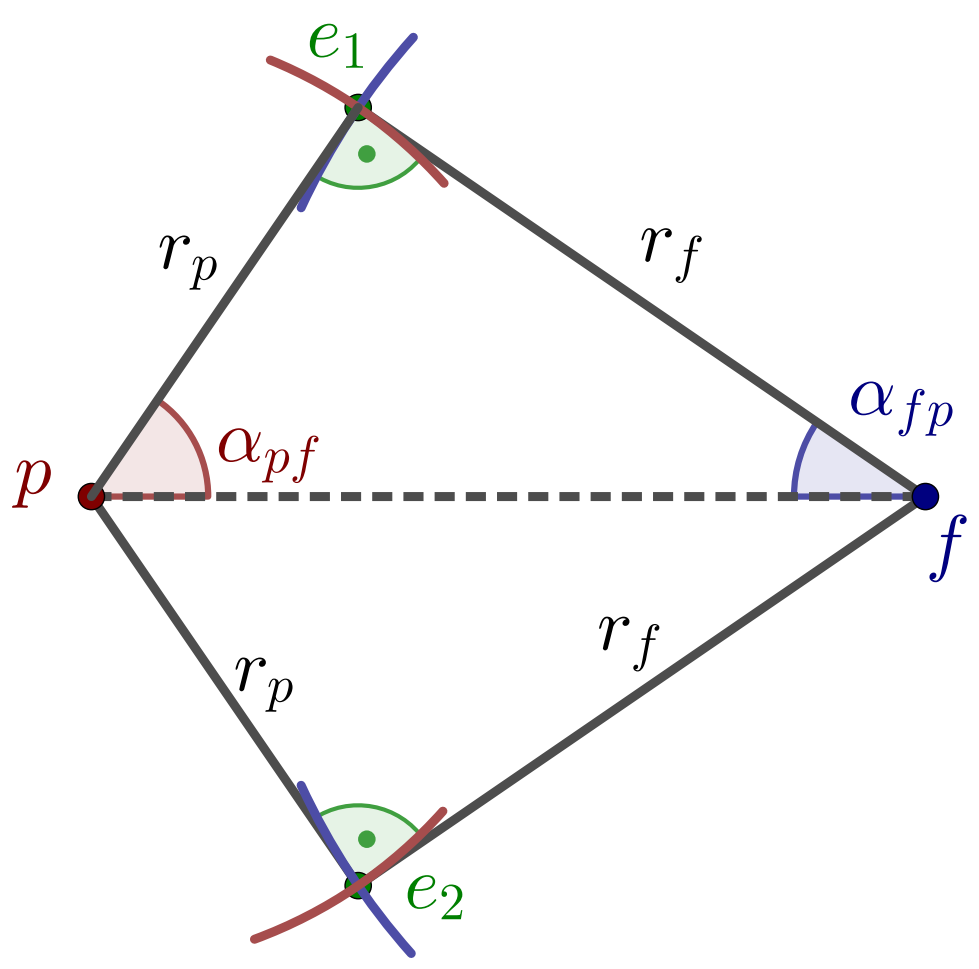

**Figure 3.5:** Illustrative example of a right-angled kite including labeled components, based on [Zie04, Fig. 1.13, p. 15].

### Conditions for the Realization of the (Restricted) quad graph

In constructing the realization of the quad graph, we focus only on the part of the graph that is enclosed by the four lines (including the points that lie on the lines themselves). We define this part more precisely as follows:

**Definition 3.32.** *Let* $\boldsymbol{G}$ *be a 3-connected planar graph and let* $\boldsymbol{Q}(\boldsymbol{G})$ *be the corresponding quad graph. Let* $e$ *be a distinguished edge point from* $\boldsymbol{V_E}$*. The graph* $\boldsymbol{Q}'(\boldsymbol{G}) = (\boldsymbol{V}'_{\boldsymbol{Q}}, \boldsymbol{E}'_{\boldsymbol{Q}})$ *with*

$$\boldsymbol{V}'_{\boldsymbol{Q}} = \{v \in \boldsymbol{V_Q} \mid v \text{ is not } e \text{ and not adjacent to } e\} \quad \text{and} \quad \boldsymbol{E}'_{\boldsymbol{Q}} = \left\{\{v_1, v_2\} \in \boldsymbol{E_Q} \mid v_1, v_2 \in \boldsymbol{V}'_{\boldsymbol{Q}}\right\}$$

*is called the* restricted quad graph of **G**.

This is sufficient to reconstruct the polytope. The realization of each node of the restricted quad graph should lie within the rectangle defined by the four lines (including the boundary). Since each edge of the polytope is incident to two vertices and two faces, every tangency point in the quad graph has four neighbors. Because edges in the quad graph only connect primal points to tangency points and tangency points to dual points, every two-dimensional region bounded by edges in the planar realization of the quad graph—i.e., every facet of the quad graph—is a quadrilateral of the form:

$$\text{primal node } p \text{ – tangency node } e_1 \text{ – dual node } f \text{ – tangency node } e_2.$$

The realization of these quadrilaterals in $\mathbb{R}^2$, as described in the reverse construction, has a special geometric shape which we call a right-angled kite, defined as follows (cf. [Zie04, p. 14]):

**Definition 3.33.** *A* right-angled kite *is a quadrilateral in* $\mathbb{R}^2$*, consisting of four corners and four edges with the following properties:*

1. *There are two opposite points* $p$ *and* $f$*, such that the two edges incident to* $p$ *have equal length, and the two edges incident to* $f$ *have equal length.*
2. *The two other points are denoted* $e_1$ *and* $e_2$*. The interior angles at these points are 90°.*

An illustrative example can be found in Figure 3.5. A right-angled kite is therefore the union of a right triangle and its reflection along the hypotenuse.

Each quadrilateral in the realization of the restricted quad graph should be a right-angled kite. The edge lengths of the kite can be derived from the radii of the primal and dual circles whose centers are exactly the points $p$ and $f$. The 90° angles at the other two points follow from the angle-preserving property of the stereographic projection, stated in Theorem 3.30. For the kites in the restricted quad graph, we introduce the following notation:

**Definition/Notation 3.34.** *Let $\boldsymbol{G} = (\boldsymbol{V}, \boldsymbol{E})$ be a planar 3-connected graph and $\boldsymbol{G}^* = (\boldsymbol{V}^*, \boldsymbol{E}^*)$ its dual graph. Let $\boldsymbol{Q}'(\boldsymbol{G})$ be the restricted quad graph corresponding to $\boldsymbol{G}$. Then we define*

$$\boldsymbol{K} := \big\{\{p,f\} \,|\, p \in \boldsymbol{V}, f \in \boldsymbol{V}^* \text{ and } p \text{ and } f \text{ share a quadrilateral in the realization of } \boldsymbol{Q}'(\boldsymbol{G})\big\}$$

*as the set of all primal and dual nodes that share a quadrilateral. Furthermore, for $p \in \boldsymbol{V}$ and $f \in \boldsymbol{V}^*$, we define*

$$\boldsymbol{K}(f) := \{p \in \boldsymbol{V} \,|\, \{p,f\} \in \boldsymbol{K}\} \quad \text{and} \quad \boldsymbol{K}(p) := \{f \in \boldsymbol{V}^* \,|\, \{p,f\} \in \boldsymbol{K}\}$$

*as the set of all primal points sharing a quadrilateral with $f$, and the set of all dual points sharing a quadrilateral with $p$, respectively.*

With this notation and under the given conditions for the realization of the restricted quad graph (the outer edges form a rectangle and every facet is a right-angled kite), we can now begin the actual construction. In the first step, we compute the radii of the primal and dual circles, which correspond to the edge lengths of the right-angled kites in the restricted quad graph.

### Computation of Radii

The challenge in constructing the realization of the quad graph in $\mathbb{R}^2$ lies in the fact that while the structure is known, the specific layout is not. So how can the nodes of the graph be arranged in $\mathbb{R}^2$ so that the resulting restricted quad graph consists entirely of right-angled kites? The first step is to compute the side lengths of these right-angled kites, which correspond to the radii of the associated primal and dual circles. This step is the core of the construction described in [Zie04].

First, we draw a line between the corners $p$ and $f$ within each right-angled kite. This divides the kite into two congruent right-angled triangles. The two remaining angles of these triangles can be expressed as

$$\alpha_{pf} = \arctan\left(\frac{r_f}{r_p}\right) \quad \text{and} \quad \alpha_{fp} = \arctan\left(\frac{r_p}{r_f}\right),$$

where $r_p$ denotes the length of the edges at $p$ and $r_f$ the length of the edges at $f$ (see Figure 3.5).

Fixing a dual point $f$, we can sum the angles at point $f$ for all right-angled kites that share this point. The total is $2\pi$ if $f$ lies in the interior of the restricted quad graph, and $\pi$ if $f$ lies on the boundary. The value $\pi/2$ is not possible, since the corners of the outer rectangle are always tangent points. This gives us the equation

$$\sum_{p\in\mathbf{K}(f)} 2\alpha_{fp} = \sum_{p\in\mathbf{K}(f)} 2\arctan\left(\frac{r_p}{r_f}\right) =: \alpha_f := \begin{cases} \pi & \text{if } f \text{ lies on the boundary} \\ 2\pi & \text{if } f \text{ lies in the interior} \end{cases} \tag{3.2}$$

and analogously, for a fixed primal point $p$:

$$\sum_{f\in\mathbf{K}(p)} 2\alpha_{pf} = \sum_{f\in\mathbf{K}(p)} 2\arctan\left(\frac{r_f}{r_p}\right) =: \alpha_p := \begin{cases} \pi & \text{if } p \text{ lies on the boundary} \\ 2\pi & \text{if } p \text{ lies in the interior} \end{cases}, \tag{3.3}$$

To determine whether a point $f$ or $p$ lies on the boundary, we check if it shares a quadrilateral with one of the points that shares a quadrilateral with the point projected to $[0,0,1]$. By formulating the equation (3.2) for each dual point $f$ and equation (3.3) for each primal point $p$ (only for points in the restricted quad graph), we obtain a nonlinear system of equations:

$$\mathbf{q}(r) := \left[\sum_{p\in\mathbf{K}(f)} 2\arctan\left(\frac{r_p}{r_f}\right) - \alpha_f, \quad \dots \quad \sum_{f\in\mathbf{K}(p)} 2\arctan\left(\frac{r_f}{r_p}\right) - \alpha_p, \quad \dots \quad \right]^T = \vec{0}.$$

The vector $\mathbf{q}$ contains one equation for each primal and dual point in the restricted quad graph, resulting in a total of $|\mathbf{V}| + |\mathbf{F}| - 4$ equations, where $\mathbf{V}$ is the set of vertices and $\mathbf{F}$ the set of faces of the graph $\mathbf{G}$ and the associated polytope.

To solve this system, we first replace each radius with its logarithm:

$$r'_f := \ln(r_f) \quad \text{and} \quad r'_p := \ln(r_p).$$

Since the realization of the restricted quad graph can be scaled arbitrarily, we normalize the solution by imposing:

$$\prod_f r_f \cdot \prod_p r_p = 1 \quad \Rightarrow \quad \sum_f r'_f + \sum_p r'_p = 0.$$

Next, we define the auxiliary function:

$$\boldsymbol{f}(a) := \arctan\left(\exp(a)\right).$$

From this, we get:

$$\boldsymbol{f}\left(r'_p - r'_f\right) = \arctan\left(\exp\left(r'_p - r'_f\right)\right) = \arctan\left(\frac{\exp\left(r'_p\right)}{\exp\left(r'_f\right)}\right) = \arctan\left(\frac{r_p}{r_f}\right).$$

Thus, we can rewrite equations (3.2) and (3.3) as:

$$\sum_{p \in \mathbf{K}(f)} 2\boldsymbol{f}\left(r'_p - r'_f\right) = \alpha_f \quad \text{and} \quad \sum_{f \in \mathbf{K}(p)} 2\boldsymbol{f}\left(r'_f - r'_p\right) = \alpha_p,$$

which leads us to the following system of equations:

$$\mathbf{q}'\left(r'\right) := \left[\sum_{p \in \mathbf{K}(f)} 2\boldsymbol{f}\left(r'_p - r'_f\right) - \alpha_f, \quad \ldots \quad \sum_{f \in \mathbf{K}(p)} 2\boldsymbol{f}\left(r'_f - r'_p\right) - \alpha_p, \quad \ldots \right]^T = \vec{0}. \tag{3.4}$$

With

$$\overline{\boldsymbol{f}}(a) := \int_{-\infty}^{a} \boldsymbol{f}(b)\,\mathrm{d}b$$

we obtain the following functional developed by Bobenko and Springborn [BS04]:

$$\mathrm{bs}\left(r'\right) := \sum_{\{p,f\} \in \mathbf{K}} \left(\overline{\boldsymbol{f}}\left(r'_f - r'_p\right) + \overline{\boldsymbol{f}}\left(r'_p - r'_f\right) - \frac{\pi}{2}\left(r'_f + r'_p\right)\right) + \sum_f \alpha_f r'_f + \sum_p \alpha_p r'_p.$$

The first sum runs over all unordered pairs of $p$ and $f$, meaning there is one summand for each right-angled kite. This functional has the following properties [Zie04, p. 16]:

**Theorem 3.35.** *For the above functional* $\mathrm{bs}(r')$*, the following statements hold:*

1. *The critical points of* $\mathrm{bs}(r')$ *are exactly the solutions of the system of equations* (3.4).
2. *The functional is convex: If we add the normalization constraint* $\sum_f r'_f + \sum_p r'_p = 0$*, then the functional is positive definite, and its critical point is unique if it exists.*
3. *The functional increases when any difference* $r'_i - r'_j$ *increases. Therefore, the functional must have a critical point.*

*As a result, there exists a unique critical point. This point is the desired solution of the system of equations* (3.4).

*Proof.* The proof can be found in [Zie04, p. 17–23] and is the core of that work. Here we only show that the gradient of the functional matches the left-hand side of the system (3.4).

If we differentiate $\text{bs}(r')$ with respect to $r'_p$, we get:

$$\frac{\text{bs}}{\partial r'_p}(r') = \sum_{f \in \mathbf{K}(p)} \left( -\boldsymbol{f}(r'_f - r'_p) + \boldsymbol{f}(r'_p - r'_f) - \frac{\pi}{2} \right) + \alpha_p$$
$$= \sum_{f \in \mathbf{K}(p)} \left( -\arctan\left(\frac{r_f}{r_p}\right) + \arctan\left(\frac{r_p}{r_f}\right) - \frac{\pi}{2} \right) + \alpha_p.$$

With

$$\arctan(a) = \frac{\pi}{2}\,\text{sgn}(a) - \arctan\left(\frac{1}{a}\right) \quad \text{and} \quad \frac{r_p}{r_f} > 0$$

we obtain:

$$\frac{\text{bs}}{\partial r'_p}(r') = \sum_{f \in \mathbf{K}(p)} \left( -\arctan\left(\frac{r_f}{r_p}\right) + \arctan\left(\frac{r_p}{r_f}\right) - \frac{\pi}{2} \right) + \alpha_p$$
$$= \sum_{f \in \mathbf{K}(p)} \left( -\arctan\left(\frac{r_f}{r_p}\right) + \frac{\pi}{2} - \arctan\left(\frac{r_f}{r_p}\right) - \frac{\pi}{2} \right) + \alpha_p$$
$$= \sum_{f \in \mathbf{K}(p)} \left( -2\arctan\left(\frac{r_f}{r_p}\right) \right) + \alpha_p$$
$$= \sum_{f \in \mathbf{K}(p)} \left( -2\boldsymbol{f}(r'_f - r'_p) \right) + \alpha_p$$
$$= -\left( \sum_{f \in \mathbf{K}(p)} \left( 2\boldsymbol{f}(r'_f - r'_p) \right) - \alpha_p \right).$$

If we set this expression equal to 0, we obtain the corresponding equation from the system. The same argument applies for the derivative with respect to $r'_f$. □

At this point, we note the following:

**Remark 3.36.** *Ziegler's construction is theoretical and focuses in the next part of his work on proving Theorem 3.35. For his purpose, the construction is essentially complete at this point. After computing $r'$, we can determine $r$ from it. This makes it possible to construct the quad graph, and using the inverse stereographic projection, we obtain the polytope. Since our focus is on the practical construction, we will now work out these next steps in detail. From a theoretical viewpoint, they may seem trivial. However, in practice, these steps involve many interesting numerical challenges. It is precisely these challenges that we now address, and we describe our concrete algorithmic implementation.*

In the next step, we aim to find a solution to $\mathbf{q}'(r')$. For this, we use the Newton method (background and descriptions can be found, for example, in [Kel03]). To apply it, we first need to compute the Hessian matrix of $\text{bs}(r')$, which is the derivative of $\mathbf{q}'(r')$.

For the notation, we use the index variables $i$ and $j$. The variable $i$ refers to the $i$-th row of $\mathbf{q}'(r')$, and $j$ refers to the $j$-th variable in $r'$. Both variables can take values from the set $\{1, \dots, |\mathbf{V}| + |\mathbf{F}| - 4\}$, meaning they can refer to either primal or dual nodes. We also introduce the index variable $k \in \{1, \dots, |\mathbf{V}| + |\mathbf{F}| - 4\}$ to indicate summation.

With this notation, the entries of the Hessian matrix $M$ are given by:

- Case $i = j$:

$$M_{(i,i)} := \frac{\partial}{\partial r'_i}\mathbf{q}'_i(r') = \frac{\partial}{\partial r'_i}\left( \sum_{k \in \mathbf{K}(i)} 2\boldsymbol{f}(r'_k - r'_i) - \alpha_i \right)$$

And further:

$$M_{(i,i)} := \frac{\partial}{\partial r'_i}\mathbf{q}'_i(r') = 2\frac{\partial}{\partial r'_i}\left(\sum_{k\in\mathbf{K}(i)} \boldsymbol{f}\left(r'_k - r'_i\right)\right)$$
$$= 2\frac{\partial}{\partial r'_i}\left(\sum_{k\in\mathbf{K}(i)} \arctan\left(\exp\left(r'_k - r'_i\right)\right)\right)$$
$$= -2\left(\sum_{k\in\mathbf{K}(i)} \frac{\exp\left(r'_k + r'_i\right)}{\exp\left(2r'_k\right) + \exp\left(2r'_i\right)}\right)$$

- Case $i \neq j$ and nodes $i$ and $j$ share a common right-angled kite:

$$M_{(i,j)} := \frac{\partial}{\partial r'_j}\mathbf{q}'_i(r') = \frac{\partial}{\partial r'_j}\left(\sum_{k\in\mathbf{K}(i)} 2\boldsymbol{f}\left(r'_k - r'_i\right) - \alpha_i\right)$$
$$= 2\frac{\partial}{\partial r'_j}\boldsymbol{f}\left(r'_j - r'_i\right)$$
$$= 2\frac{\partial}{\partial r'_j}\arctan\left(\exp\left(r'_j - r'_i\right)\right)$$
$$= 2\frac{\exp\left(r'_j + r'_i\right)}{\exp\left(2r'_j\right) + \exp\left(2r'_i\right)}$$

- Case $i \neq j$ and nodes $i$ and $j$ do not share a kite:

$$M_{(i,j)} := \frac{\partial}{\partial r'_j}\mathbf{q}'_i(r') = 0$$

This Hessian matrix does not have full rank. This can be seen directly by summing all columns:

$$\sum_j M_{(i,j)} = 2\left(\sum_{j\in\mathbf{K}(i)} \frac{\exp(r'_j + r'_i)}{\exp(2r'_j) + \exp(2r'_i)}\right) - 2\left(\sum_{k\in\mathbf{K}(i)} \frac{\exp(r'_k + r'_i)}{\exp(2r'_k) + \exp(2r'_i)}\right) = 0$$

This behavior also has a geometric explanation. In the equation system above, any solution remains valid if we add a constant to every variable. This is due to the properties of the function $\boldsymbol{f}$. If we add a constant $c$ to each logarithmic radius $r'$, we get:

$$\boldsymbol{f}\left((r'_j + c) - (r'_i + c)\right) = \boldsymbol{f}(r'_j + c - r'_i - c) = \boldsymbol{f}(r'_j - r'_i).$$

Since the radii are just the exponentials of the values $r'$, adding a constant to each $r'$ is the same as multiplying every radius $r$ by a positive constant:

$$r = \exp\left(r'\right) \quad \text{and} \quad \tilde{r} = \exp\left(r' + c\right) = \exp\left(r'\right)\exp\left(c\right) = r\exp\left(c\right)$$

We have already observed this behavior: the quad graph can be scaled freely, which means the equation system $\mathbf{q}'(r')$ cannot have a unique solution.

From the point of view of the Hessian matrix, this is problematic, because we need its inverse to apply Newton's method. Therefore, we will modify the system step by step, starting with the following lemma:

**Lemma 3.37.** *All rows of $\mathbf{q}'(r')$ sum to zero.*

*Proof.* Let $n$ be the number of rows of $\mathbf{q}'$. If we sum all rows, we get:

$$\sum_{i=1}^{n} \mathbf{q}'_i(r') = \sum_{f} \sum_{p\in\mathbf{K}(f)} 2\boldsymbol{f}(r'_p - r'_f) + \sum_{p} \sum_{f\in\mathbf{K}(p)} 2\boldsymbol{f}(r'_f - r'_p) - \sum_{f} \alpha_f - \sum_{p} \alpha_p$$

The first two sums each run over all right-angled kites. Since each kite appears exactly once in both sums, we can write:

$$\sum_{i=1}^{n} \mathbf{q}'_i(r') = \sum_{\{p,f\}\in\mathbf{K}} \left(2\boldsymbol{f}(r'_p - r'_f) + 2\boldsymbol{f}(r'_f - r'_p)\right) - \sum_{f} \alpha_f - \sum_{p} \alpha_p$$

Using the identity

$$\frac{\pi}{2}\,\mathrm{sgn}(a) = \arctan(a) + \arctan\left(\frac{1}{a}\right), \quad \text{and since } \frac{r_p}{r_f} > 0,$$

we find:

$$\sum_{i=1}^{n} \mathbf{q}'_i(r') = \sum_{\{p,f\}\in\mathbf{K}} \pi - \sum_{f} \alpha_f - \sum_{p} \alpha_p \tag{3.5}$$

To continue, we introduce some notation from graph theory:

- $|\mathbf{V}|$ is the number of nodes
- $|\mathbf{E}|$ is the number of edges or edge-points
- $|\mathbf{F}|$ is the number of faces or facet-points
- $|\mathbf{Q}|$ is the number of faces in the full quad graph $\mathbf{Q}(\mathbf{G})$
- $|\mathbf{Q}'|$ is the number of right-angled kites in the restricted quad graph

Each edge point in $\mathbf{Q}(\mathbf{G})$ belongs to exactly four faces, and each facet contains two edge points. Since the number of edge points equals the number of edges in $\mathbf{G}$, we have:

$$2|\mathbf{E}| = |\mathbf{Q}|$$

The restricted quad graph includes all nodes of $\mathbf{Q}(\mathbf{G})$, except for one edge point $e$, two primal nodes $p_1$ and $p_2$, and two dual nodes $f_1$ and $f_2$. Let $\mathbf{v}(p)$ and $\mathbf{v}(f)$ be the number of faces that a point $p$ or $f$ belongs to in $\mathbf{Q}(\mathbf{G})$.

Then, the number of right-angled kites in the restricted graph is:

$$|\mathbf{Q}'| = |\mathbf{Q}| - \mathbf{v}(p_1) - \mathbf{v}(p_2) - \mathbf{v}(f_1) - \mathbf{v}(f_2) + 4 = 2|\mathbf{E}| - \mathbf{v}(p_1) - \mathbf{v}(p_2) - \mathbf{v}(f_1) - \mathbf{v}(f_2) + 4$$

Therefore, the first term in equation (3.5) becomes:

$$\sum_{\{p,f\}\in\mathbf{K}} \pi = \left(2|\mathbf{E}| - \mathbf{v}(p_1) - \mathbf{v}(p_2) - \mathbf{v}(f_1) - \mathbf{v}(f_2) + 4\right)\pi$$

The two remaining sums $\sum_f \alpha_f$ and $\sum_p \alpha_p$ each contain $|\mathbf{F}| - 2$ and $|\mathbf{V}| - 2$ terms, respectively. The terms are either $2\pi$ or $\pi$, depending on whether the corresponding point lies in the interior or on the boundary. By checking which faces are shared with the excluded points, we find:

$$\sum_{f} \alpha_f = 2\pi(|\mathbf{F}| - 2) - \pi(\mathbf{v}(p_1) - 2 + \mathbf{v}(p_2) - 2)$$

$$\sum_{p} \alpha_p = 2\pi(|\mathbf{V}| - 2) - \pi(\mathbf{v}(f_1) - 2 + \mathbf{v}(f_2) - 2)$$

Substituting into equation (3.5), we get:

$$
\begin{aligned}
\sum_{i=1}^{n} \mathbf{q}'_i(r') &= (2|\mathbf{E}| - \mathbf{v}(p_1) - \mathbf{v}(p_2) - \mathbf{v}(f_1) - \mathbf{v}(f_2) + 4)\,\pi \\
&\quad - (2\pi(|\mathbf{F}| - 2) - \pi(\mathbf{v}(p_1) - 2 + \mathbf{v}(p_2) - 2)) \\
&\quad - (2\pi(|\mathbf{V}| - 2) - \pi(\mathbf{v}(f_1) - 2 + \mathbf{v}(f_2) - 2)) \\
&= \pi\,(2|\mathbf{E}| - 2|\mathbf{F}| - 2|\mathbf{V}| + 4)
\end{aligned}
$$

Using Euler's formula from Theorem A.27, we know that

$$2|\mathbf{E}| - 2|\mathbf{F}| - 2|\mathbf{V}| + 4 = 0$$

Thus, we finally conclude:

$$\sum_{i=1}^{n} \mathbf{q}'_i(r') = 0$$

So all rows of $\mathbf{q}'(r')$ sum to zero. □

This formulation is not found in [And70], but is cited in the literature as the corresponding source.
Using this lemma, we can remove the last row from the equation system $\mathbf{q}'(\mathbf{r}')$ without loss of generality. Additionally, we introduce the constraint $\sum_i r'_i = 0$ into the system. With $n$ denoting the number of rows of $\mathbf{q}'$, we get the extended system:

$$\overline{\mathbf{q}'}(r') := \begin{bmatrix} \mathbf{q}'(r')_{(1:n-1)} \\ \sum_i r'_i \end{bmatrix} = \vec{0}.$$

This constraint guarantees, according to Theorem 3.35, that the solution to the system is unique. The corresponding Hessian matrix is then

$$\overline{M} := \begin{bmatrix} & M_{(1:n-1,:)} & \\ 1 & \cdots & 1 \end{bmatrix}$$

which is now square. We apply Newton's method using the steps described in Algorithm 6, which we explain below:

**Algorithm 6:** ComputeRadii

**Data:** Equation system $\overline{\mathbf{q}'}(r')$
**Result:** Radii $r$

**begin**
- Set $r' = 0$;
- Set counter $c = 0$;
- Set maximum iterations $m = 100$;
- **while** $\left(\max\left(\left|\overline{\mathbf{q}'}(r')\right|\right) > 10^{-13}\right)$ && $(c < m)$ **do**
  - $a = \overline{M}(r') \backslash \overline{\mathbf{q}'}(r')$;
  - Update $r' = r' - a$;
  - Increment counter $c = c + 1$;
- **if** $c == m$ **then**
  - Use `fsolve` to solve $\left(\overline{\mathbf{q}'}(r') = 0\right)$;
- Compute $r = \exp(r')$;

We start the Newton method with an initial value of $r' = 0$ and apply it until the accuracy of the solution in each individual equation is better than $10^{-13}$. This tolerance is chosen arbitrarily. Since machine precision is usually around $10^{-15}$, a threshold of $10^{-13}$ is suitable. This still allows the individual terms in the equation system to be represented accurately enough. In the actual implementation, the tolerance can also be set manually.

In each iteration, the equation system and its Hessian matrix are computed for the current values of $r'$. The line

$$a = \overline{M}(r') \backslash \overline{\mathbf{q}'}(r') \quad \Leftrightarrow \quad \overline{M}(r') \cdot a = \overline{\mathbf{q}'}(r')$$

determines the Newton direction. The *backslash* operator \ in Matlab (see [Matd]) is used to compute the vector $a$. This avoids the computationally expensive inversion of the matrix $\overline{M}(r')$. Then the new value for $r'$ is calculated using the direction $a$.
The variable $c$ is a counter and serves as a safeguard. To prevent an infinite loop, the Newton method stops after at most $m$ iterations. We chose $m = 100$ arbitrarily. If no solution is found by then, we fall back on Matlab's built-in solver `fsolve` (see [Matc]). In practice, however, the Newton method usually produces reliable results and is significantly faster than the standard solver.

Finally, we compute the actual radii $r$ as the exponentials of $r'$. With these radii, we can construct the quad graph in the next step.

### Computing the coordinates of the restricted quad graph

From a theoretical point of view, the radii contain all information necessary to build the restricted quad graph. However, computing the exact coordinates is non-trivial and is carried out using Algorithm 7. We explain each step of the algorithm below. An illustrative example is shown in Figure 3.6.

**Algorithm 7:** DrawRestrictedQuadGraph

**Data:** Restricted quad graph $\mathbf{Q} = (\mathbf{V_Q}, \mathbf{E_Q})$,
List of right-angled kites $M \in \{1, \ldots, |\mathbf{V}|\}^{n \times 4}$, vector of radii $r$
**Result:** Coordinates of each vertex $P \in \mathbb{R}^{|\mathbf{V_Q}| \times 2}$

**begin**
- KiteVisited = Zero vector of dimension $|n| \times 1$;
- VertexVisited = Zero vector of dimension $|\mathbf{V_Q}| \times 1$;
- EdgePriority = Vector with entries $\infty$ of dimension $|\mathbf{E_Q}| \times 1$;
- $e := v_e$ = Find one of the four corner points of the restricted quad graph;
- $v := v_v$ = Find the primal node that shares an edge with $e$;
- $P_{(e,:)} = [0, 0]$;
- $P_{(v,:)} = [0, -r_v]$;
- EdgePriority($[v, e]$) = 1;
- Compute all kite constraints;
- **while** $\sum$ *KiteVisited* $< n$ **do**
  - MinPriority = min(EdgePriority);
  - CurrentKite = Find an unvisited right-angled kite with an edge of minimal priority;
  - Determine the coordinates of the four corners of the current kite and mark their VertexVisited values as 1;
  - Set the EdgePriority of each edge in the kite to MinPriority + 1;
  - **if** *any kite constraint is violated* **then**
    - Recompute the coordinates of the previously placed nodes using Newton's method;

**Initialization** We begin the construction of the restricted quad graph at one of its corner points. To do so, we first look for one of the four right-angled kites that is located at a corner. Such a kite includes both a primal and a dual point on the boundary, which means that the corresponding angles $\alpha_f$ and $\alpha_p$ take the value $\pi$.

The specific choice of the starting kite is arbitrary but reproducible, since it depends on the internal numbering of the nodes and edges. Furthermore, this choice has no effect on the final polytope (except for rotation or reflection on the unit sphere).

Next, we set the corner node to the coordinate $[0, 0]$ and the associated primal node to $[0, -r_p]$. This satisfies the radius of the primal node and defines a direction in which the graph will be constructed. This direction is again arbitrary but does not affect the outcome (except for a possible rotation or reflection).

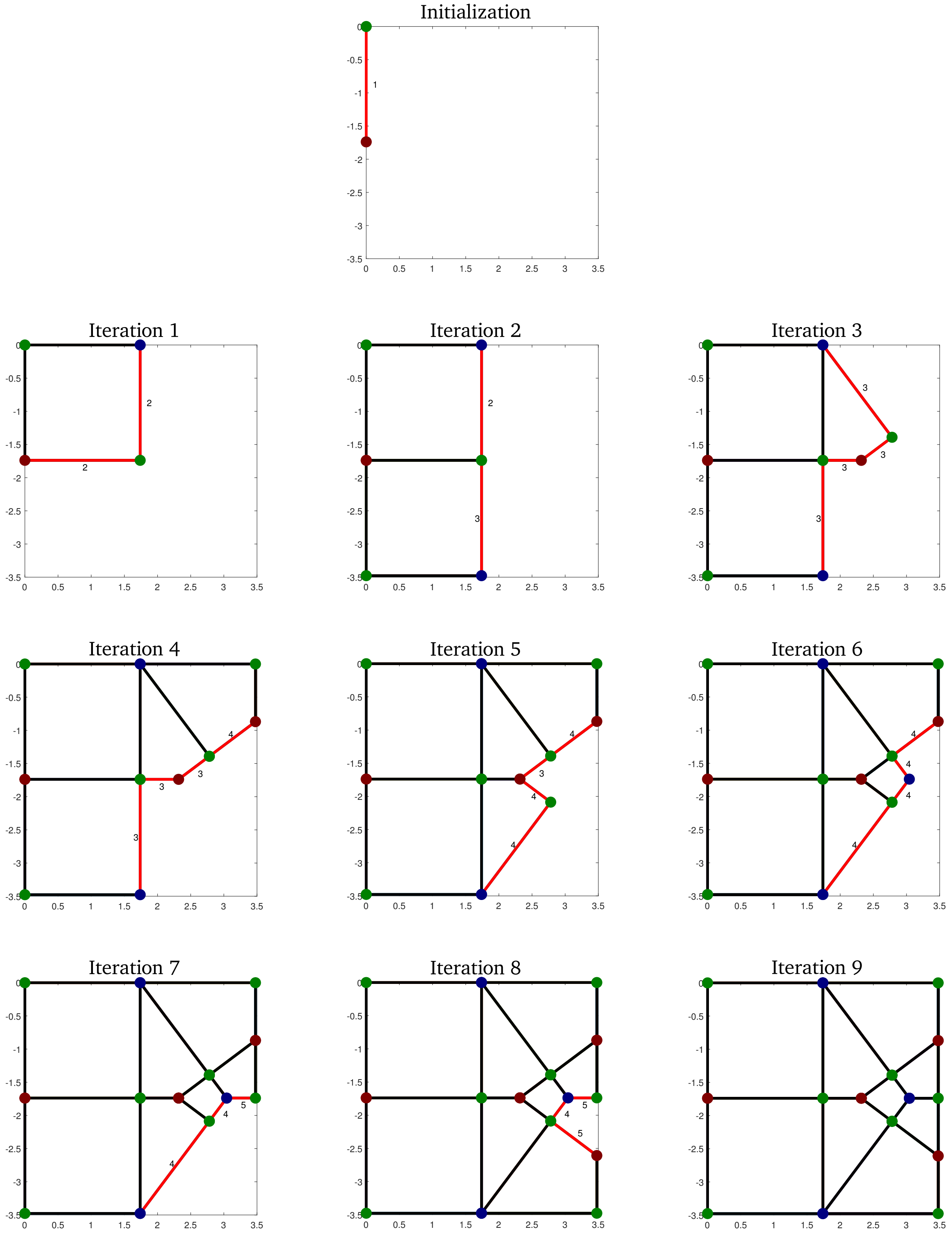

**Figure 3.6:** Example illustration of the construction of the quad graph using Algorithm 7. The first image shows the initialization. Each subsequent image represents one full iteration of the while-loop. Edges connected to unseen kites are marked in red, and the edge priority is displayed.

With these two points fixed, and the direction of the associated dual point set to $[a, 0]$ for some $a > 0$, all remaining points of the restricted quad graph are uniquely determined and computed in the next steps.

**General Procedure – BFS** The algorithm constructs each right-angled kite one after another. To determine which kite should be drawn next, we assign a priority number to each edge. Initially, all edges are assigned an infinite priority. As soon as an edge is drawn, its priority is updated.

In the initialization step, the edge $[p, e]$ is already drawn and assigned priority $1$. In each iteration, the algorithm finds the edge with the smallest priority number and draws the right-angled kite that shares this edge and has not yet been drawn. The other edges of the newly drawn kite then receive the same priority value plus one.

This means the priority numbers represent the increased distance (in number of already drawn kites) from the first kite. This strategy is known as *breadth-first search*.

Since the radii contain numerical errors, the breadth-first approach is well suited here. Because the construction stays close to the starting kite, numerical errors do not accumulate as much throughout the process.

**Constructing a Right-Angled Kite** Due to the initialization and the structure of the algorithm, there is always one known edge when constructing a new kite. This edge consists of either a primal or a dual point and an edge point. We denote this edge point by $e_1$, so the edge is of the form $\{p, e_1\}$ or $\{e_1, f\}$. To simplify notation, we assume that the primal point $p$ is already known. The case where a dual point $f$ is known instead is handled analogously.

With this information, the dual point can be directly constructed using the right angle at $e_1$:

$$f = e_1 \pm r_f \frac{M \cdot (p - e_1)}{|p - e|} \quad \text{with} \quad M = \begin{bmatrix} 0 & -1 \\ 1 & 0 \end{bmatrix}$$

The sign depends on which side of the segment $(p, e_1)$ the previously drawn kite lies. For the initial edge, we define the direction as $[r_f, 0]$. Internally, we assign each edge a direction, which helps us determine on which side the new right-angled kite should be constructed.

The fourth point of the kite is calculated by first projecting $e_1$ perpendicularly onto the segment $(p, f)$. We call this projection point $a$. The fourth point is then given by

$$e_2 = e_1 + 2(a - e_1).$$

If either of the two new points $f$ or $e_2$ has already been drawn (as seen, for example, in Iterations 5, 6, 7, and 9 in Figure 3.6), we set the point as the arithmetic mean of the previously stored and the newly computed position.

A visual illustration of each construction step can be found in Figure 3.7.

**Kite Conditions** From a theoretical point of view, the following conditions are not necessary because the shape of the restricted quad graph is fully determined by the radii and the initial placement of the first right-angled kite. However, since the radii include numerical errors, inaccuracies can arise in the construction of the restricted quad graph, and these can accumulate in the worst case. To counteract this, we already use a breadth-first search to determine the order of drawing the kites. In addition, to ensure numerical stability especially in large graphs, we now define a number of conditions that are checked after each step of the construction. The first three conditions refer to the two triangles that make up a right-angled kite. The next three conditions ensure that kites do not overlap in the drawing.

1. In every right-angled kite, the distance between $p$ and $e_1$ and the distance between $p$ and $e_2$ must equal the radius $r_p$:
$$|p - e_1| = |p - e_2| = r_p.$$
2. In every right-angled kite, the distance between $f$ and $e_1$ and the distance between $f$ and $e_2$ must equal the radius $r_f$:
$$|f - e_1| = |f - e_2| = r_f.$$
3. The angles at $e_1$ and $e_2$ must be right angles. According to the Pythagorean theorem, this is the case if
$$|p - f| = \sqrt{r_p^2 + r_f^2}.$$

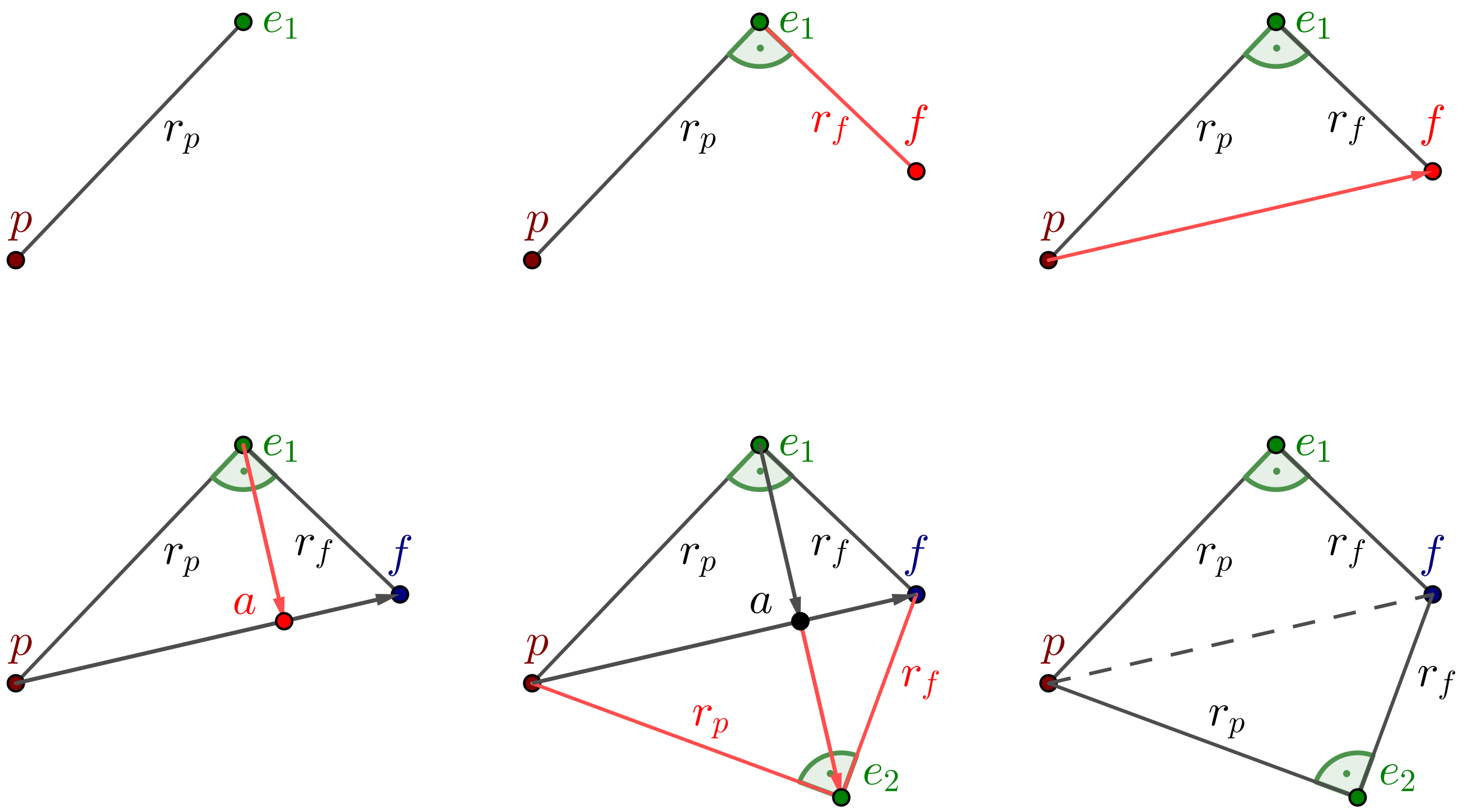

**Figure 3.7:** Illustrative example of constructing a right-angled kite.

4. The two triangles that form the kite must not overlap. This means that the distance between $e_1$ and $e_2$ must be strictly positive. This distance can be calculated using Thales' theorem (see [Euc86, Book 3, Proposition 31, p. 66] and [KK07, p. 144]) and the sine rule (see [CG67, Thm. 1.11, p. 2]):

$$\frac{r_p r_f \sqrt{r_p^2 + r_f^2}}{2 \cdot \left(\frac{1}{2}\sqrt{r_p^2 + r_f^2} \cdot \bar{a}\right)} = 2 \cdot \frac{1}{2}\sqrt{r_p^2 + r_f^2} \quad \Leftrightarrow \quad \frac{r_p r_f}{\bar{a}} = \sqrt{r_p^2 + r_f^2} \quad \Leftrightarrow \quad \bar{a} = \frac{r_p r_f}{\sqrt{r_p^2 + r_f^2}}.$$

Here, $\bar{a}$ is the length of the segment between $e_1$ and the projection point $a$ shown in Figure 3.7. The value $\sqrt{r_p^2 + r_f^2}$ is the distance between $p$ and $f$, $\frac{1}{2}\sqrt{r_p^2 + r_f^2} \cdot \bar{a}$ is the area of the triangle, and $\frac{1}{2}\sqrt{r_p^2 + r_f^2}$ is the circumradius of the right triangle. The distance between $e_1$ and $e_2$ is then:

$$|e_1 - e_2| = 2 \cdot \bar{a} = \frac{2 r_p r_f}{\sqrt{r_p^2 + r_f^2}}.$$

5. Two right-angled kites sharing an edge that contains a dual point must not overlap. Therefore, the distance between the two primal points $p$ and $p_2$ must satisfy:

$$|p - p_2| = r_p + r_{p_2}.$$

6. Two right-angled kites sharing an edge that contains a primal point must not overlap. Therefore, the distance between the two dual points $f$ and $f_2$ must satisfy:

$$|f - f_2| = r_f + r_{f_2}.$$

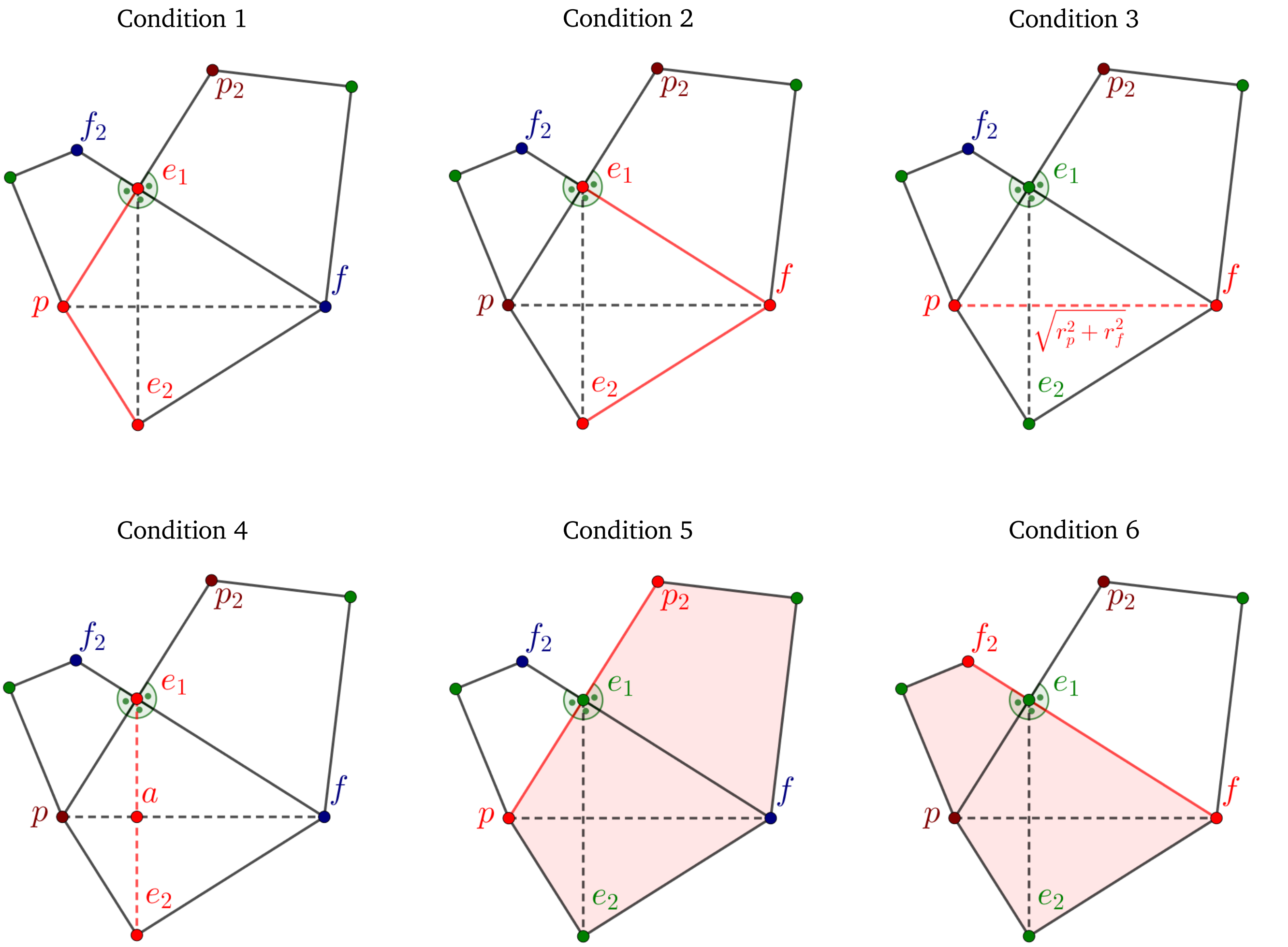

**Figure 3.8:** Illustration of the six kite conditions.

The distances described in the six kite conditions are illustrated in Figure 3.8. These conditions can be expressed as a system of nonlinear equations for all already constructed right-angled kites. The variables are the $x$- and $y$-coordinates of the vertices. We explain one of these equations using condition 3 as an example; the other equations follow analogously. The distance between $p$ and $f$ is given by

$$\sqrt{(p_x - f_x)^2 + (p_y - f_y)^2}$$

and the corresponding equation becomes

$$\mathbf{q}(p, f) := \sqrt{(p_x - f_x)^2 + (p_y - f_y)^2} - \sqrt{r_p^2 + r_f^2} = 0,$$

where $r_p$ and $r_f$ are constants in the context of the equation system.

If, after drawing a new right-angled kite, one of these conditions is violated by the current geometry, we apply the Newton method, as described in Algorithm 6, to the system of all equations involving already constructed kites. We only consider those equations for which all required points have already been computed.

The entries of the Jacobian matrix (in this context also referred to as the Hessian) are given by:

$$\frac{\partial}{\partial p_x}\mathbf{q}(p, f) = \frac{p_x - f_x}{\sqrt{(p_x - f_x)^2 + (p_y - f_y)^2}} \qquad \frac{\partial}{\partial f_x}\mathbf{q}(p, f) = \frac{f_x - p_x}{\sqrt{(p_x - f_x)^2 + (p_y - f_y)^2}}$$
$$\frac{\partial}{\partial p_y}\mathbf{q}(p, f) = \frac{p_y - f_y}{\sqrt{(p_x - f_x)^2 + (p_y - f_y)^2}} \qquad \frac{\partial}{\partial f_y}\mathbf{q}(p, f) = \frac{f_y - p_y}{\sqrt{(p_x - f_x)^2 + (p_y - f_y)^2}}.$$

As a starting point, we use the coordinates already computed in the previous steps of the restricted quad graph construction, since they are already very close to a valid solution.

With this step, the construction of the restricted quad graph is complete, and we can proceed in the next section to generate the polytope using the inverse stereographic projection.

### 3.3.2 Inverse Stereographic Projection and Construction of the Primal and Dual Polytope

In this step, we want to project the tangent points $e$ of the quad graph constructed in the previous section onto the unit sphere using the inverse stereographic projection. The center of mass of the projected tangent points should lie at the origin, as described in Theorem 3.25. Ziegler does not discuss this detail in [Zie04], since his work focuses on proving Theorem 3.35. We therefore describe below how to construct the polytope such that the centroid of its edge directions is at $0$:

During the construction of the realization of the restricted quad graph, we made two important assumptions. The first was a normalization of the radii $r$, and the second was fixing the outer corner of the initial right kite at the point $[0, 0]$. Both assumptions were made arbitrarily and can be adjusted. This leads to the following three degrees of freedom for the already constructed quad graph:

1. Shift the restricted quad graph in $x$-direction by a constant $a_x \in \mathbb{R}$.
2. Shift the restricted quad graph in $y$-direction by a constant $a_y \in \mathbb{R}$.
3. Scale the restricted quad graph by a factor $a_z \in \mathbb{R}$.

Applying these degrees of freedom to the coordinates of the tangent points of the restricted quad graph, the inverse stereographic projection becomes:

$$\boldsymbol{\phi}^{-1}((\overline{x} + a_x)a_z, (\overline{y} + a_y)a_z) := \boldsymbol{\phi}^{-1}_{a_x,a_y,a_z}(\overline{x}, \overline{y}) = \frac{\left[2(\overline{x} + a_x)a_z, \quad 2(\overline{y} + a_y)a_z, \quad ((\overline{x} + a_x)a_z)^2 + ((\overline{y} + a_y)a_z)^2 - 1\right]}{((\overline{x} + a_x)a_z)^2 + ((\overline{y} + a_y)a_z)^2 + 1}. \tag{3.6}$$

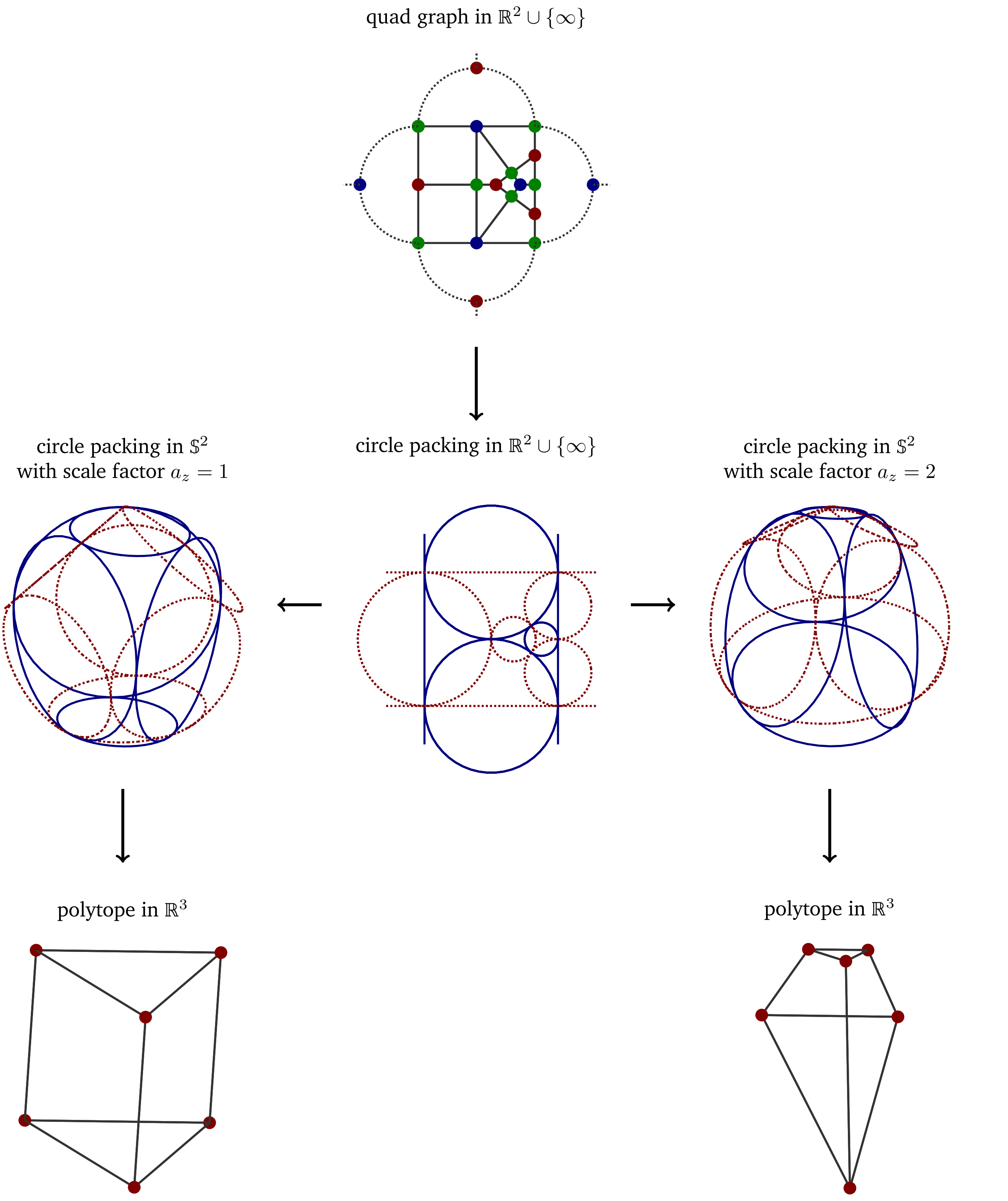

**Figure 3.9:** The inverse stereographic projection of the circle packing results in different polytopes depending on the scaling. Illustrated here with two different scalings as examples.

With these three degrees of freedom, we can influence the projection onto the unit sphere. This is shown in Figure 3.9, where the right column illustrates a quad graph that has been scaled by a factor of 2 before the projection.

To determine the values for $a_x$, $a_y$, and $a_z$, we use the following system of equations:

$$\begin{aligned} \frac{1}{m}\sum_{i=1}^{m-1} \frac{2(\overline{x}_i + a_x)a_z}{((\overline{x}_i + a_x)a_z)^2 + ((\overline{y}_i + a_y)a_z)^2 + 1} &= 0 \\ \frac{1}{m}\sum_{i=1}^{m-1} \frac{2(\overline{y}_i + a_y)a_z}{((\overline{x}_i + a_x)a_z)^2 + ((\overline{y}_i + a_y)a_z)^2 + 1} &= 0 \\ \frac{1}{m} + \frac{1}{m}\sum_{i=1}^{m-1} \frac{((\overline{x}_i + a_x)a_z)^2 + ((\overline{y}_i + a_y)a_z)^2 - 1}{((\overline{x}_i + a_x)a_z)^2 + ((\overline{y}_i + a_y)a_z)^2 + 1} &= 0. \end{aligned} \tag{3.7}$$

This system arises from requiring that each coordinate of the centroid of the tangent points must equal zero. The coordinates $\overline{x}_i$ and $\overline{y}_i$ with $i \in \{1, \ldots, m-1\}$ are the coordinates of the tangent points of the restricted quad graph. Together with the tangent point $[0, 0, 1]$, which is part of the full quad graph but not of the restricted one, and which is independent of the values of $a_x$, $a_y$, and $a_z$, the centroid is determined by the above system. This gives us three equations for three unknowns or degrees of freedom.

According to the version of the Koebe-Andreev-Thurston Theorem stated in Theorem 3.25, this system is uniquely solvable, since the theorem guarantees the existence of a 3-polytope with edge centroid at the origin. This also holds in the case where one of the tangent points of the 3-polytope lies at $(0, 0, 1)$ and where the circles corresponding to that point are mapped onto the rectangle used in the packing. This fixed setup ensures uniqueness with respect to rotations and reflections, assuming that it is specified which circle maps to which line. Because the stereographic projection is a bijection between $\mathbb{S}^2$ and $\mathbb{R}^2 \cup \{\infty\}$, the tangent points of the 3-polytope can be mapped to $\mathbb{R}^2 \cup \{\infty\}$.

The reverse direction holds as well, due to the bijectivity of the stereographic projection. However, to ensure the uniqueness of the solution to the above system, the sign of $a_z$ would also need to be fixed. This sign determines whether the $x$ and $y$ coordinates of the edge points of the restricted quad graph are reflected, and must be specified to fix the position of the lines. For our purposes, it is sufficient to find a 3-polytope whose edge centroid lies at zero, so uniqueness is not critical here. The key point is the existence of a solution.

We will also solve the system of equations (3.7) using Newton's method, analogous to Algorithm 6. For this, we first define

$$z_z := ((\overline{x}_i + a_x)a_z)^2 + ((\overline{y}_i + a_y)a_z)^2 - 1 \quad \text{and} \quad z_n := ((\overline{x}_i + a_x)a_z)^2 + ((\overline{y}_i + a_y)a_z)^2 + 1,$$

and consider the following partial derivatives:

$$\begin{aligned} &\frac{\partial}{\partial a_x} 2(\overline{x}_i + a_x)a_z = 2a_z, && \frac{\partial}{\partial a_y} 2(\overline{x}_i + a_x)a_z = 0, && \frac{\partial}{\partial a_z} 2(\overline{x}_i + a_x)a_z = 2(\overline{x}_i + a_x), \\ &\frac{\partial}{\partial a_x} 2(\overline{y}_i + a_y)a_z = 0, && \frac{\partial}{\partial a_y} 2(\overline{y}_i + a_y)a_z = 2a_z, && \frac{\partial}{\partial a_z} 2(\overline{y}_i + a_y)a_z = 2(\overline{y}_i + a_y), \\ &\frac{\partial}{\partial a_x} z_z = 2a_z^2(\overline{x}_i + a_x), && \frac{\partial}{\partial a_y} z_z = 2a_z^2(\overline{y}_i + a_y), && \frac{\partial}{\partial a_z} z_z = 2a_z\left((\overline{x}_i + a_x)^2 + (\overline{y}_i + a_y)^2\right), \\ &\frac{\partial}{\partial a_x} z_n = 2a_z^2(\overline{x}_i + a_x), && \frac{\partial}{\partial a_y} z_n = 2a_z^2(\overline{y}_i + a_y), && \frac{\partial}{\partial a_z} z_n = 2a_z\left((\overline{x}_i + a_x)^2 + (\overline{y}_i + a_y)^2\right). \end{aligned}$$

The entries of the Hessian matrix $M$ for the system (3.7) result from combining the above derivatives using the quotient rule and summing over the $m-1$ points $(\overline{x}_i, \overline{y}_i)$:

$$M = \frac{1}{m}\sum_{i=1}^{m-1} \frac{1}{\left(a_z^2\overline{\overline{x}}_i^2 + a_z^2\overline{\overline{y}}_i^2 + 1\right)^2} \begin{bmatrix} 2a_z\left(-a_z^2\overline{\overline{x}}_i^2 + a_z^2\overline{\overline{y}}_i^2 + 1\right) & 4a_z^3\overline{\overline{x}}_i\overline{\overline{y}}_i & -2\overline{\overline{x}}_i\left(a_z^2\overline{\overline{x}}_i^2 + a_z^2\overline{\overline{y}}_i^2 - 1\right) \\ 4a_z^3\overline{\overline{x}}_i\overline{\overline{y}}_i & 2a_z\left(a_z^2\overline{\overline{x}}_i^2 - a_z^2\overline{\overline{y}}_i^2 + 1\right) & -2\overline{\overline{y}}_i\left(a_z^2\overline{\overline{x}}_i^2 + a_z^2\overline{\overline{y}}_i^2 - 1\right) \\ 4a_z^2\overline{\overline{x}}_i & 4a_z^2\overline{\overline{y}}_i & 4a_z\left(\overline{\overline{x}}_i^2 + \overline{\overline{y}}_i^2\right) \end{bmatrix} \tag{3.8}$$

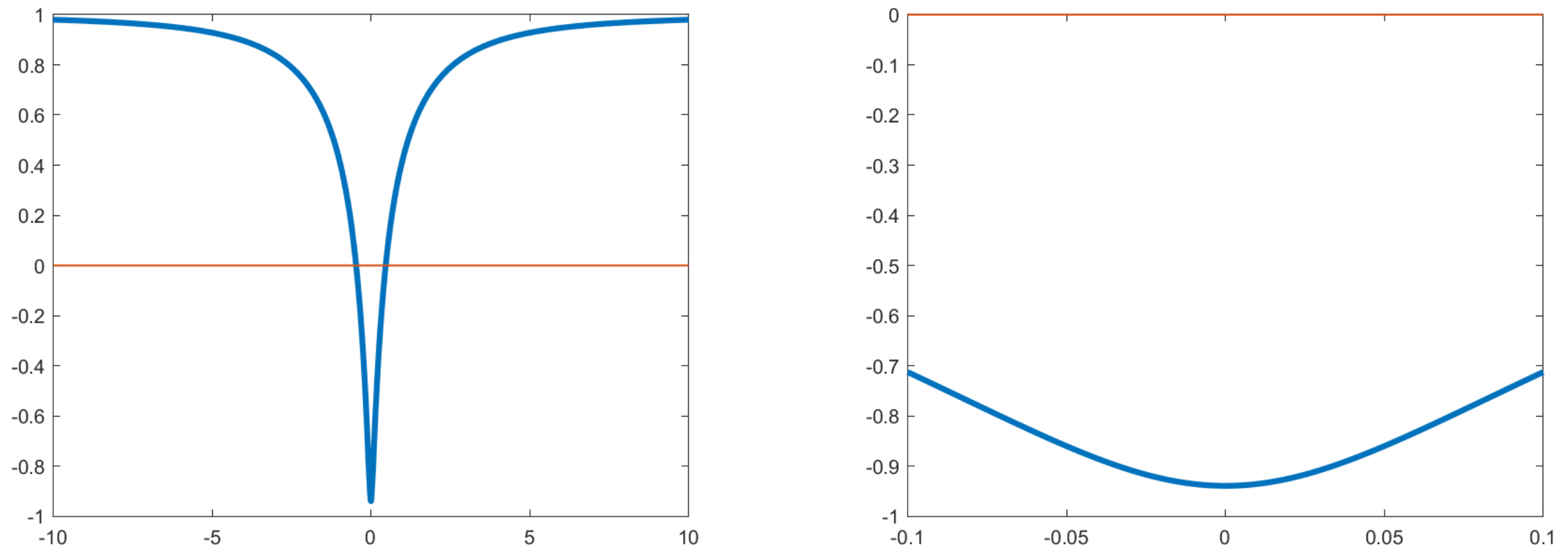

**Figure 3.10:** The $z$-value of the tangent centroid as a function of $a_z$ for an example with already exactly determined $a_x$ and $a_y$ (left), and a zoomed-in view of the same curve around the zero point of the $x$-axis. The equation system is solved by those values of $a_z$ for which the blue curve intersects the red zero-line. Since Newton's method relies on the tangent of the blue curve, the method can diverge if the starting point is poorly chosen.

with

$$\overline{\overline{x}}_i = \overline{x}_i + a_x \quad \text{and} \quad \overline{\overline{y}}_i = \overline{y}_i + a_y.$$

Since the point $[0, 0, 1]$ is not dependent on the three variables, it has no effect on the Hessian matrix. With this, all components necessary for the Newton method are now defined.

Determining the degrees of freedom $a_x$, $a_y$, and $a_z$ using Newton's method is, however, highly unstable. This can be illustrated with the following example: let us assume we have already found the exact values for $a_x$ and $a_y$, and we only need to determine the value of $a_z$. In that case, the third equation, as a function of $a_z$, would take the shape shown in Figure 3.10.
This shape has been reproduced in similar form across various tests, indicating that the equation system consistently exhibits this characteristic. If the starting point for Newton's method is not chosen close enough to the actual solution, the method may diverge.

To counteract this and find a good initial guess for Newton's method, we use a heuristic approach. The stereographic projection maps points from the lower hemisphere of the sphere to the interior of the unit circle in $\mathbb{R}^2 \cup \{\infty\}$, while points from the upper hemisphere are mapped outside the unit circle. Assuming that the tangent points are roughly evenly distributed between the northern and southern hemispheres of the sphere, the point density in the planar quad graph is higher for points corresponding to the southern hemisphere than for the northern side. Furthermore, the deeper a point lies towards the south pole of the sphere, the denser its corresponding region in the planar graph becomes.

Therefore, the node of the planar quad graph with the smallest average distance to its neighbors serves as a good reference point for the $x$- and $y$-shift of the graph. The graph is thus translated so that this node is located at $[0, 0]$. Using this heuristic and Algorithm 8, we can determine the degrees of freedom for the inverse stereographic projection. The individual steps of the algorithm are explained in the following section.

**Initialization and Point Distance** First, each drawn point of the restricted quad graph is assigned a value called the *point distance*. For the primal points $p$ and the dual points $f$, this distance is given by the radius of the corresponding circles. For the tangency points, the distance is defined as the average distance to the four neighboring points. We assume that a good starting point for Newton's method lies in the interior of the restricted quad graph. Therefore, the point distance for points on the boundary of the graph is set to $\infty$.

After computing the point distances, the points are sorted in ascending order based on these values. As a result, the point with the smallest point distance will be considered first in the subsequent part of the algorithm.

**Algorithm 8:** ComputeDegreesOfFreedomOfInverseStereographicProjection

**Data:** Coordinates of the nodes of the restricted quad graph $P = (P_x, P_y)$

**Result:** Degrees of freedom of the inverse stereographic projection $a_x, a_y, a_z$

1 **begin**
2   $P_z$ = zero vector of dimension $|P| \times 1$;
3   PointDistance = zero vector of dimension $|P| \times 1$;
4   SolutionFound = false;
5   **foreach** $P_{(i,:)}$ *in* $P$ **do**
6     PointDistance = Calculate the average distance from $P_i$ to its neighbors;
7     **if** $P_i$ *lies on the boundary of the restricted quad graph* **then**
8       PointDistance = $\infty$;
9   Sort points ascendingly by their PointDistance;
10   $i = 1$;
11   **while** *SolutionFound == false* && $i \leq |P|$ **do**
12     Set $b_x = -(P_x)_{(i)}$ and $b_y = -(P_y)_{(i)}$;
13     Find $b_z$ satisfying $\sum_j \frac{(z_z)_j}{(z_n)_j} = 0$ for $a_z = b_z$, $a_x = b_x$, and $a_y = b_y$ using the bisection method;
14     Set $(P_z)_{(i)} := b_z$;
15     Use the Newton method with start values $b_x, b_y, b_z$ to determine $a_x, a_y, a_z$;
16     **if** *Newton method succeeded* **then**
17       SolutionFound = true;
18     $i = i + 1$;
19   $i = 1$;
20   **while** *SolutionFound == false* && $i \leq |P|$ **do**
21     Set $b_x = -(P_x)_{(i)}$, $b_y = -(P_y)_{(i)}$ and $b_z = -(P_z)_{(i)}$;
22     Use `fsolve` with start values $b_x, b_y, b_z$ to determine $a_x, a_y, a_z$;
23     **if** `fsolve` *succeeded* **then**
24       SolutionFound = true;
25     $i = i + 1$;
26   **if** *SolutionFound == false* **then**
27     error('Degrees of freedom of the inverse stereographic projection could not be computed');
28   **else**
29     **return** $a_x, a_y, a_z$;

The variable $P_z$ is used to store the starting values for $a_z$ so that they do not need to be recomputed in the second part of the algorithm. The variable *SolutionFound* ensures that the algorithm terminates once a solution has been found, and does not continue searching with other starting values.

**General Structure** After the initialization, the algorithm attempts to find the values $a_x$, $a_y$, and $a_z$. For this, we use two methods: Newton's method (see the first part of Algorithm 6) and the `fsolve` command from Matlab [Matc]. The `fsolve` command is only used if Newton's method fails to find a solution. Both strategies require starting values $b_x$, $b_y$, and $b_z$, which are chosen based on the coordinates of the nodes of the restricted quad graph. Using these starting values, the algorithm tries to solve the above system. If a solution is found, the algorithm stops and returns it. If not, it tries again with the starting values from a different coordinate point.

If Newton's method fails for all starting values, the algorithm tries to find a solution using Matlab's standard solver with the same starting values. If this also fails, the algorithm returns an error.

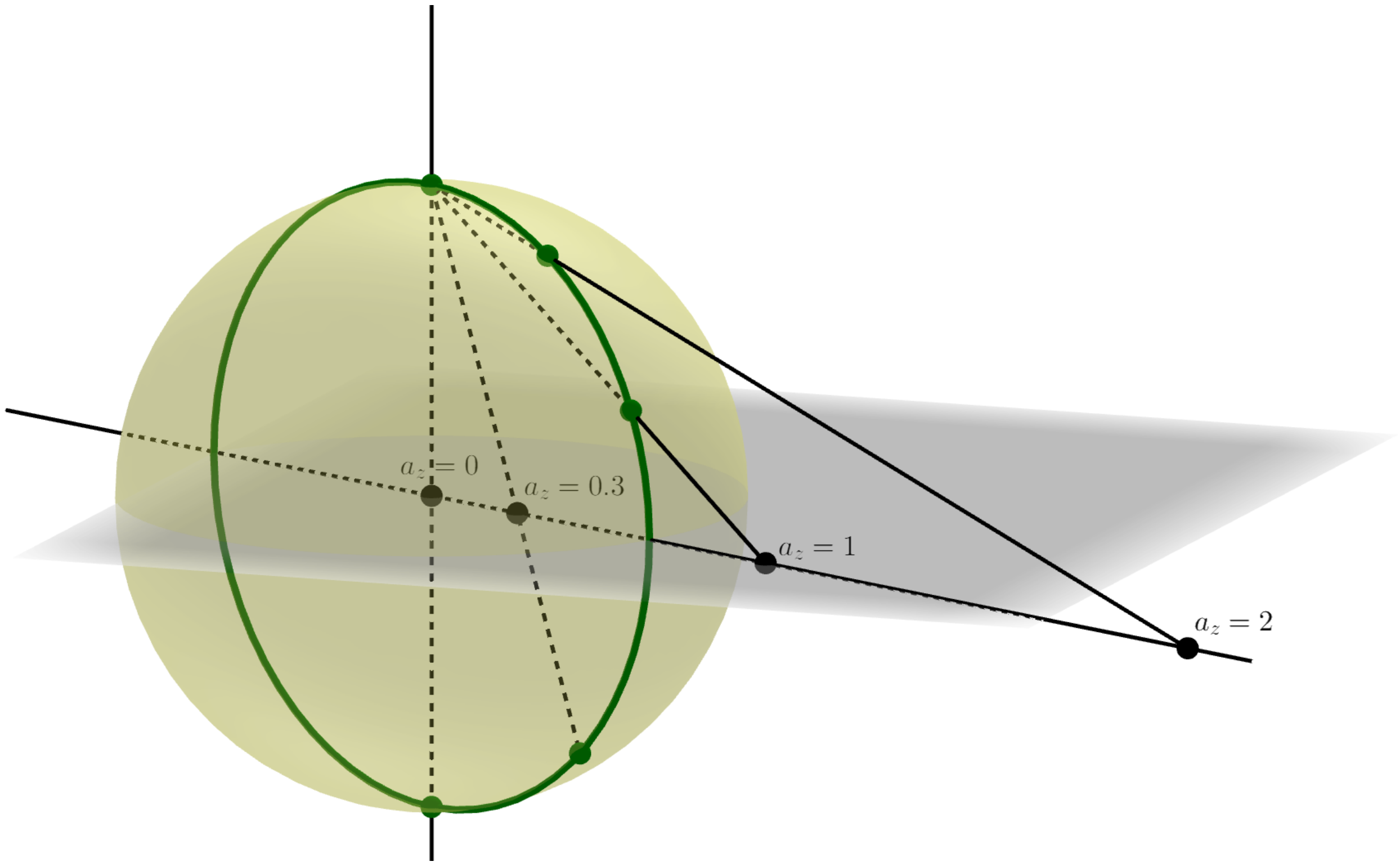

**Figure 3.11:** A point in the $\mathbb{R}^2 \cup \{\infty\}$ plane is stretched by the factor $a_z$. The points projected onto the sphere are shifted along a meridian.

**Initial Value Calculation for Newton's Method** As already mentioned, we use the negative coordinate values of the point with the smallest point distance as starting values for $a_x$ and $a_y$. This shifts the realization of the restricted quad graph so that the selected point lies at the origin. We only need a suitable starting value for $a_z$. To find this, we study the behavior of the third coordinate of the inverse stereographic projection as we vary $a_z$.

According to Equation (3.6), this coordinate is

$$\frac{((\overline{x}+a_x)a_z)^2 + ((\overline{y}+a_y)a_z)^2 - 1}{((\overline{x}+a_x)a_z)^2 + ((\overline{y}+a_y)a_z)^2 + 1} =: \frac{ca_z^2 - 1}{ca_z^2 + 1} \quad \text{with} \quad c := (\overline{x}+a_x)^2 + (\overline{y}+a_y)^2 \geq 0. \tag{3.9}$$

If $c = 0$, this expression is constantly $-1$. Assume now that $c > 0$. Taking the derivative of expression (3.9) with respect to $a_z$ gives

$$\frac{\partial}{\partial a_z}\left(\frac{ca_z^2 - 1}{ca_z^2 + 1}\right) = \frac{4ca_z}{(ca_z^2 + 1)^2}. \tag{3.10}$$

Since $c > 0$, the denominator in (3.10) is strictly greater than zero for all $a_z$. The numerator can only be zero if $a_z = 0$. By inserting values and using continuity, we conclude that expression (3.9) has a global minimum at $a_z = 0$ when $c > 0$. The derivative (3.10) is also positive for $a_z > 0$ and negative for $a_z < 0$.

This leads to the following conclusions: For $c = 0$, the third coordinate of the inverse stereographic projection of the corresponding point is $-1$ regardless of $a_z$. If $c > 0$ and $a_z = 0$, then the third coordinate is also $-1$.

For $a_z > 0$, the third coordinate of the inverse stereographic projection increases strictly monotonically with increasing $a_z$. Similarly, for $a_z < 0$, it increases strictly monotonically with decreasing $a_z$. Since the limit of (3.9) approaches 1 as $|a_z| \to \infty$, the third coordinate of the inverse stereographic projection increases as $|a_z|$ increases.

Geometrically, varying $a_z$ moves the projection of a point along its corresponding meridian on the unit sphere. Increasing $|a_z|$ moves the point toward $[0, 0, 1]$ on the sphere. This behavior is illustrated in Figure 3.11.
This behavior also transfers to the centroid of the tangent points. We can see this clearly from the equation (3.7) for the third coordinate of the tangent centroid. For $a_z > 0$ we get

$$\frac{1}{m}+\frac{1}{m}\sum_{i=1}^{m-1}\frac{((\overline{x}_i+a_x)a_z)^2+((\overline{y}_i+a_y)a_z)^2-1}{((\overline{x}_i+a_x)a_z)^2+((\overline{y}_i+a_y)a_z)^2+1}=\frac{1}{m}+\frac{1}{m}\sum_{i=1}^{m-1}\frac{(\overline{x}_i+a_x)^2+(\overline{y}_i+a_y)^2-\frac{1}{a_z^2}}{(\overline{x}_i+a_x)^2+(\overline{y}_i+a_y)^2+\frac{1}{a_z^2}}=:\frac{1}{m}+\frac{1}{m}\sum_{i=1}^{m-1}\frac{c_i-\frac{1}{a_z^2}}{c_i+\frac{1}{a_z^2}}$$

with $c_i \geq 0$. Let $a'_z > a_z$. Then we have

$$\frac{1}{m}+\frac{1}{m}\sum_{i=1}^{m-1}\frac{c_i-\frac{1}{a_z^2}}{c_i+\frac{1}{a_z^2}}<\frac{1}{m}+\frac{1}{m}\sum_{i=1}^{m-1}\frac{c_i-\frac{1}{a_z^2}}{c_i+\frac{1}{a_z'^2}}<\frac{1}{m}+\frac{1}{m}\sum_{i=1}^{m-1}\frac{c_i-\frac{1}{a_z'^2}}{c_i+\frac{1}{a_z'^2}}.$$

The third coordinate of the centroid of the tangent points is thus strictly monotonically increasing for $a_z > 0$ with increasing $a_z$.

Since we want to find a starting value $b_z$ for $a_z$ with given $b_x$ and $b_y$, which is as close as possible to a solution, we choose $b_z$ such that the third coordinate of the tangent centroid is zero, i.e., such that

$$\frac{1}{m}+\frac{1}{m}\sum_{i=1}^{m-1}\frac{((\overline{x}_i+b_x)b_z)^2+((\overline{y}_i+b_y)b_z)^2-1}{((\overline{x}_i+b_x)b_z)^2+((\overline{y}_i+b_y)b_z)^2+1}=0$$

is satisfied. For $b_z = 0$, the third coordinate of the tangent centroid is negative. Due to monotonicity, there is thus exactly one zero for $b_z > 0$, which we determine by the bisection method. This method is described in Algorithm 9 and further details can be found, for example, in [Kön04, Section 2.3 I, pp. 11–13].

In the bisection method, we first search for bounds where the third coordinate of the tangent centroid is less than ($b_l = 0$) and greater than ($b_r$) zero. Then, we compute the third coordinate of the tangent centroid at the midpoint of the interval $(b_l + b_r)/2$. If this value is less than zero, the left bound ($b_l$) is adjusted. If it is greater than zero, the right bound ($b_r$) is adjusted accordingly. This process continues until the third coordinate of the tangent centroid is sufficiently close to the target value (zero), in our case better than $10^{-13}$. The resulting value is then used as the start value $b_z$.

**Determine $\mathbf{a_x}$, $\mathbf{a_y}$ and $\mathbf{a_z}$** After determining the start values $b_x$, $b_y$ and $b_z$, we use the Newton method as described before. We have already constructed the gradient in equation (3.7) and the Hessian matrix in equation (3.8). In Figures 3.13 and 3.14 there is a graphic each, where the regions for which the Newton method succeeded are shaded. We see that for these examples the heuristic was able to compute the inverse stereographic projection. The green admissible area is also located around where the graph nodes are most densely packed. In total, over 300,000 different polytopes were constructed for this work (see Section 3.1.1). The construction of the inverse stereographic projection was possible for every example, indicating that the heuristic is well chosen.

As a backup, we nevertheless included Matlab's standard solver `fsolve`. This also works here with an acceptable runtime, since it only has to solve a system of three equations with three unknowns.

With the inverse stereographic projection, it is now possible to construct the polytope. For this, all tangent points $e$ of the restricted quad graph are first projected directly onto the sphere. The tangent point at $\infty$ is additionally mapped to the point $[0, 0, 1]$. We describe the respective tangent planes to the unit sphere at the projected tangent points by the normal form

$$(x - e)\vec{n} = 0.$$

Here, $x$ is a point on the plane, $e$ the tangent point, and $\vec{n}$ a normal vector of the plane. In this special case it holds that $\vec{n} = e$, and we obtain

$$(x - e)e = 0 \quad \Leftrightarrow \quad \langle x, e\rangle = \langle e, e\rangle = 1.$$

The connecting line between two primal (dual) points should, by construction, be tangent to the sphere. Hence, all primal (dual) points of the polytope lie in the tangent planes of the corresponding tangent points. This is illustrated in Figure 3.12. We can therefore form from the above equation a system of equations of the form

**Algorithm 9:** BisectionMethod

**Data:** Coordinates of the tangent points of the restricted quad graph $P = (P_x, P_y)$, start values $b_x$ and $b_y$ for $a_x$ and $a_y$

**Result:** Start value $b_z$ for $a_z$

**begin**
- $b_l = 0;$
- $b_r = 1;$
- $m = |P| + 1;$
- **for** $i = 1 : |P|$ **do**
  - $c_i = ((P_x)_i + b_x)^2 + \left((P_y)_i + b_y\right)^2;$
- LimitFound = false;
- **while** *LimitFound == false* **do**
  - $z = \frac{1}{m} + \frac{1}{m}\sum_i \frac{c_i - \frac{1}{b_r^2}}{c_i + \frac{1}{b_r^2}};$
  - **if** $z > 0$ **then**
    - LimitFound = true;
  - **else**
    - $b_r = 10 \cdot b_r;$
- $b_z = \frac{b_l + b_r}{2};$
- $z = \frac{1}{m} + \frac{1}{m}\sum_i \frac{c_i - \frac{1}{b_z^2}}{c_i + \frac{1}{b_z^2}};$
- **while** $|z| > 10^{-13}$ **do**
  - **if** $z < 0$ **then**
    - $b_l = b_z;$
  - **else**
    - $b_r = b_z;$
  - $b_z = \frac{b_l + b_r}{2};$
  - $z = \frac{1}{m} + \frac{1}{m}\sum_i \frac{c_i - \frac{1}{b_z^2}}{c_i + \frac{1}{b_z^2}};$
- **return** $b_z$;

$$\begin{bmatrix} e_1^T \\ \vdots \\ e_n^T \end{bmatrix} p = \vec{1} \quad \text{respectively} \quad \begin{bmatrix} e_1^T \\ \vdots \\ e_n^T \end{bmatrix} f = \vec{1}$$

for each primal and dual point. Here, $e_1, \ldots, e_n$ are the tangent points connected to the respective primal or dual point in the quad graph. Since both the primal and dual graphs are 3-connected, each point has at least three neighbors, so the system is uniquely solvable for every point. If a point has more than three neighbors, we solve the system using the linear least squares problem

$$\begin{bmatrix} e_1 & \cdots & e_n \end{bmatrix} \begin{bmatrix} e_1^T \\ \vdots \\ e_n^T \end{bmatrix} p = \begin{bmatrix} e_1 & \cdots & e_n \end{bmatrix} \vec{1} \quad \text{respectively} \quad \begin{bmatrix} e_1 & \cdots & e_n \end{bmatrix} \begin{bmatrix} e_1^T \\ \vdots \\ e_n^T \end{bmatrix} f = \begin{bmatrix} e_1 & \cdots & e_n \end{bmatrix} \vec{1}. \tag{3.11}$$

It would also be possible in this case to select only three equations at random. However, since the coordinates of the tangent points are subject to numerical errors, the linear least squares problem is more appropriate here. After performing this for each primal and dual vertex, the polytope is completely constructed. Figures 3.13 and 3.14 illustrate the individual steps of the entire chapter on the construction of polytopes using two somewhat more complex examples.

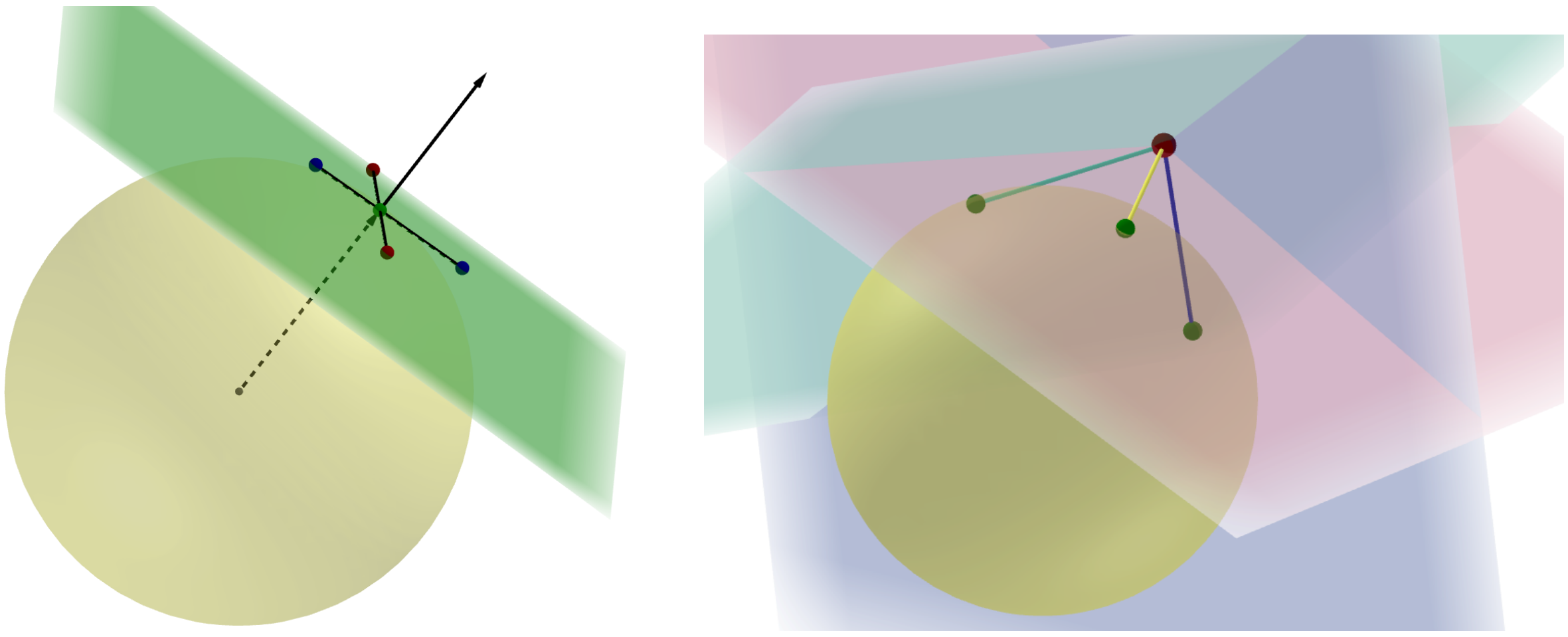

**Figure 3.12:** Tangent plane at a tangent point with the two associated primal and dual points lying on the plane (left) and three tangent planes at three tangent points whose intersection is a primal point (right).

We conclude this section with four statements about the constructed convex polytopes. We begin with the following lemma, which—slightly differently formulated—can be found in [Cas89, Num. 2, p. 2]:

**Lemma 3.38.** *Let $\mathbb{S}^2$ be the unit sphere and **M** a plane intersecting $\mathbb{S}^2$. Assume that **M** is not tangent to $\mathbb{S}^2$ and is not a plane through the origin. Furthermore, let $\vec{n}$ be the unit normal vector of the plane **M** pointing from the origin towards the plane. Then the intersection of **M** and $\mathbb{S}^2$ is a circle with center $a$. Moreover, the position vector of $a$ has the same direction as the normal vector $\vec{n}$.*

*Proof.* Let $\vec{n}$ be the normal vector of the plane **M**, and let $a$ be the intersection of the plane **M** with the line through the origin spanned by $\vec{n}$. Let $c$ be any point in the intersection of **M** and $\mathbb{S}^2$. Then the triangle $(0, a, c)$ has a right angle at $a$, since the normal vector is perpendicular to the plane. The distance between $a$ and $c$ can be calculated using the Pythagorean theorem as

$$|a - c| = \sqrt{|c|^2 - |a|^2} = \sqrt{1 - |a|^2},$$

since the length of the position vector $c$ is 1 because $c$ lies on the sphere. Hence, the distance from $a$ to every point on the intersection of **M** and $\mathbb{S}^2$ depends only on $a$ and is thus constant. Therefore, the intersection between **M** and $\mathbb{S}^2$ is a circle with radius $\sqrt{1 - |a|^2}$ and center $a$. Because $a$ is defined as the intersection of the plane **M** and the line through the origin in the direction of the normal vector, the position vector of the circle center points in the same direction as the normal vector of the plane **M**. An illustration of the proof can be found in Figure 3.15. □

The following lemma is based on the above and describes an alternative method for computing the primal and dual points of the polytope:

**Lemma 3.39.** *Let $\mathbb{S}^2$ be the unit sphere and **M** a plane intersecting $\mathbb{S}^2$. Assume that **M** is not tangent to $\mathbb{S}^2$ and is not a plane through the origin. Furthermore, let $\vec{n}$ be the unit normal vector of the plane **M** pointing from the origin towards the plane, and let $a$ be the center of the circle formed by the intersection of **M** and $\mathbb{S}^2$, and $c$ be any point in this intersection. The point*

$$b := \frac{1}{|a|}\vec{n} = \frac{1}{|a|^2}a$$

*lies in the tangent plane of $\mathbb{S}^2$ at the point $c$.*

*Proof.* As already mentioned, the tangent plane to the sphere at the point $c$ can be expressed using the normal form as

$$\langle x, c\rangle = \langle c, c\rangle = 1.$$

Consider $\langle b, c\rangle$, then we have

$$\langle b, c\rangle = \left\langle \frac{1}{|a|}\vec{n}, c \right\rangle = \frac{1}{|a|}\langle \vec{n}, c\rangle.$$

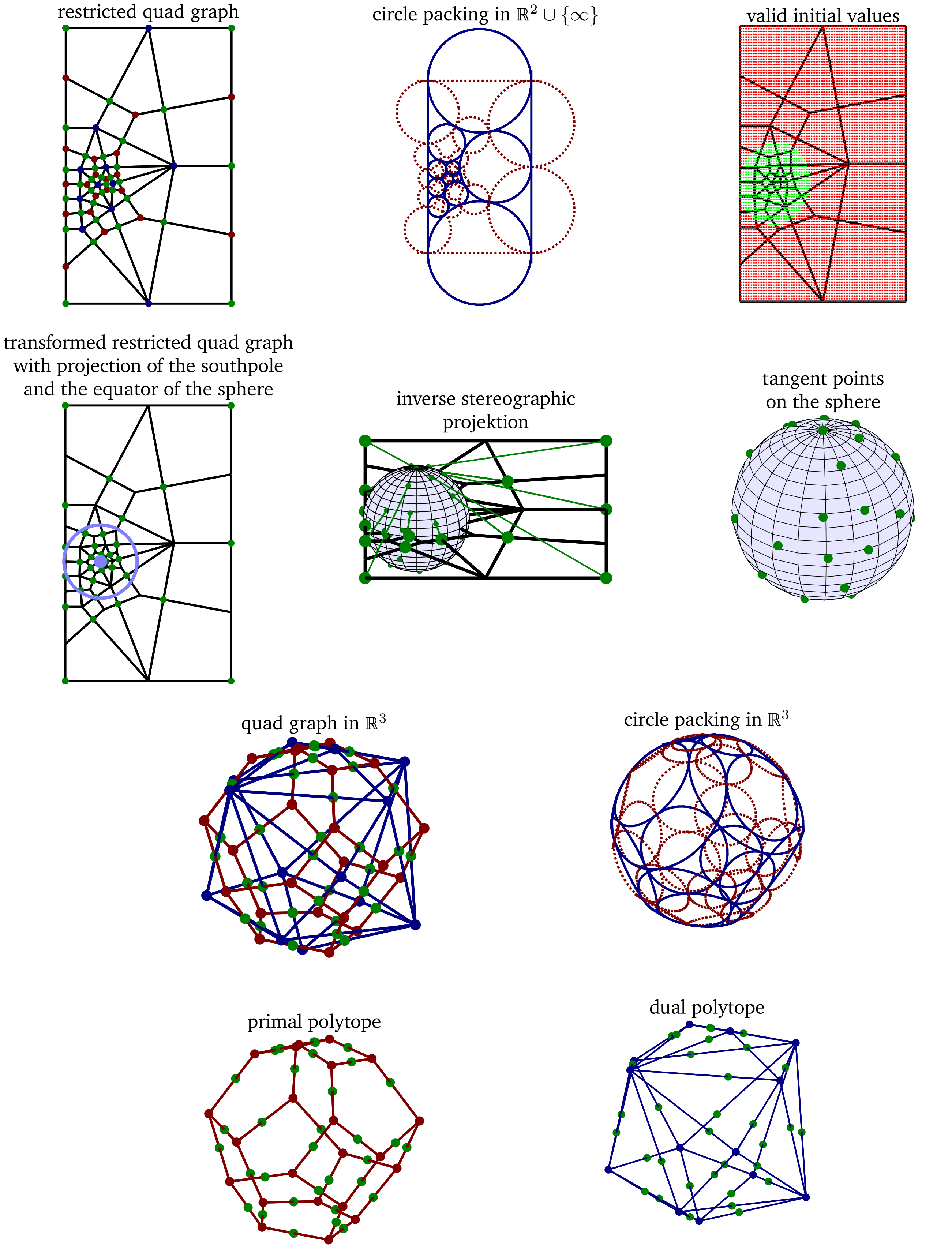

**Figure 3.13:** The individual steps of constructing a polytope, exemplified here for an asymmetric polytope.

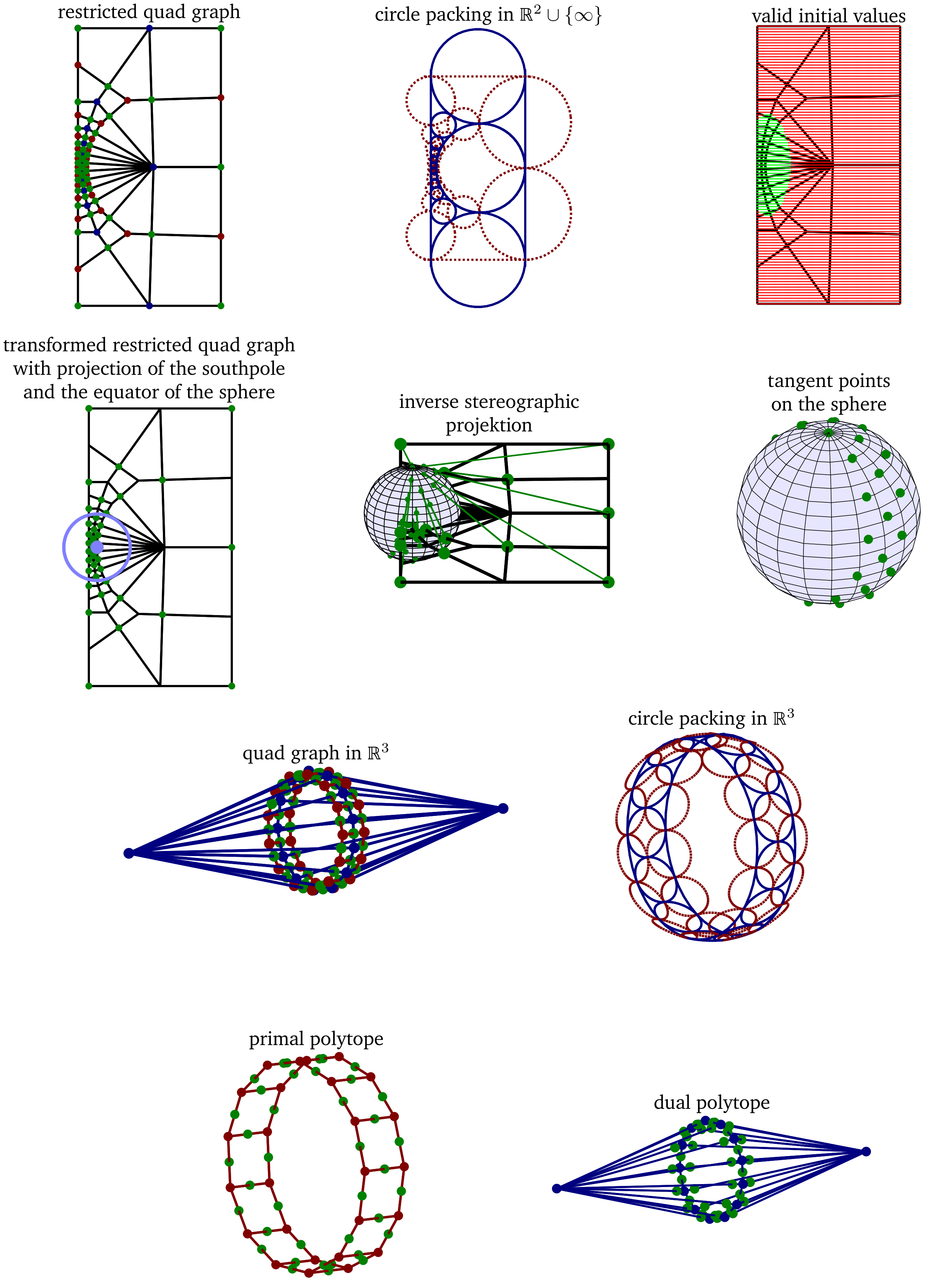

**Figure 3.14:** The individual steps of constructing a polytope, here exemplified for a 13-sided prism.

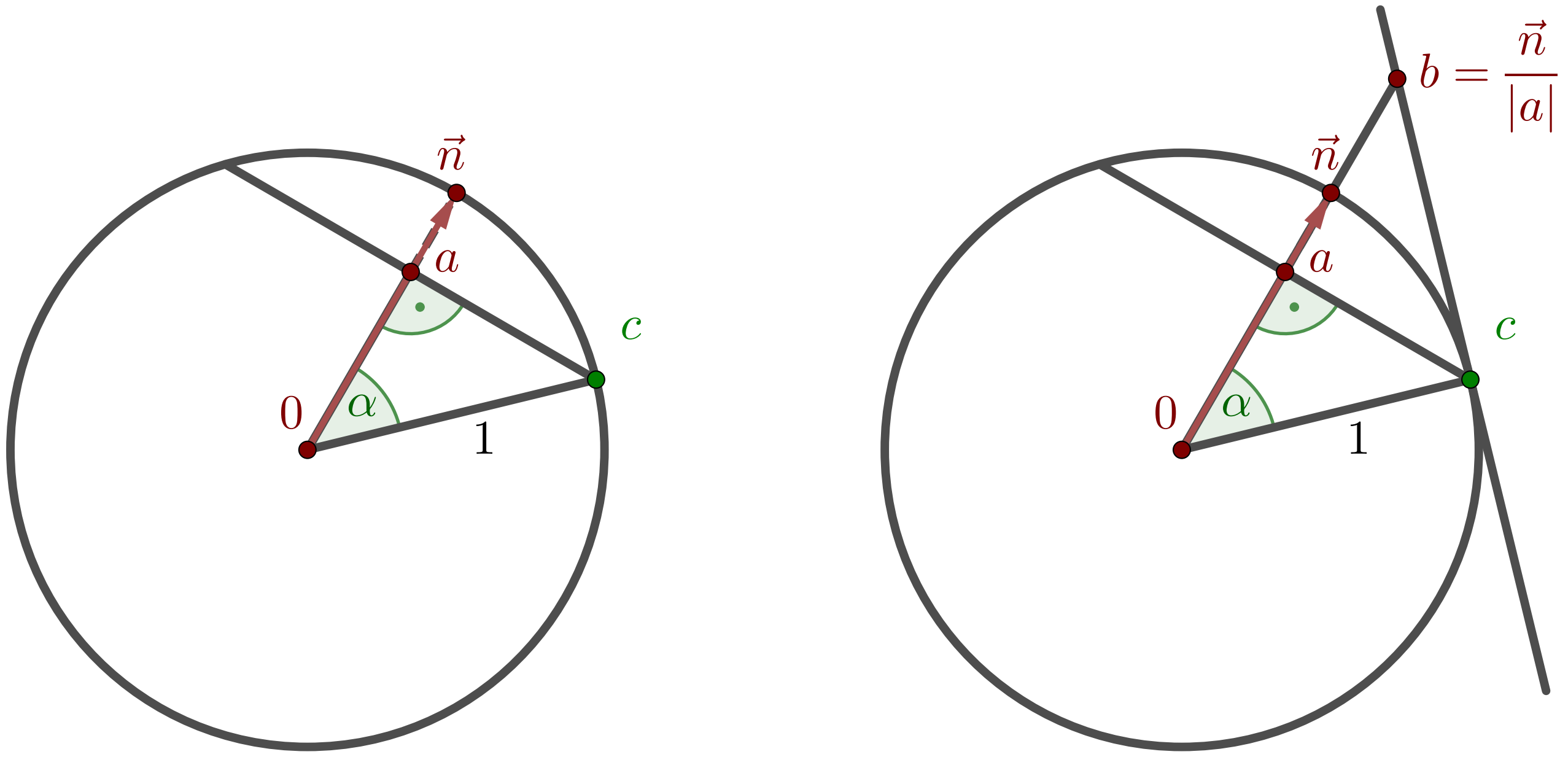

**Figure 3.15:** Sketches of the proofs of Lemma 3.38 (left) and Lemma 3.39 (right). Since all points lie in a plane, only the respective plane is drawn.

Let $\alpha$ be the angle between $\vec{n}$ and the position vector of $c$. Then by definition of the scalar product,

$$\langle b, c\rangle = \frac{1}{|a|}\langle \vec{n}, c\rangle = \frac{1}{|a|}|c||\vec{n}|\cos(\alpha) = \frac{1}{|a|}\cos(\alpha).$$

Since the position vectors $a$ and $\vec{n}$ point in the same direction, the angle between $a$ and $c$ is exactly $\alpha$. Because the triangle $(0, a, c)$ is right-angled, we can rewrite $\cos(\alpha)$ as

$$\cos(\alpha) = \frac{|a|}{|c|} = |a|.$$

Inserting this back into the above equation yields

$$\langle b, c\rangle = \frac{1}{|a|}\cos(\alpha) = \frac{1}{|a|}|a| = 1,$$

thus the point $b$ lies on the tangent plane at the point $c$. An illustration of the proof can be found in Figure 3.15. □

With this lemma, we can directly formulate the following corollary:

**Corollary 3.40.** *Let $\mathbb{S}^2$ be the unit sphere and **M** a plane intersecting $\mathbb{S}^2$. Assume that **M** is not tangent to $\mathbb{S}^2$ and is not a plane through the origin. Let the intersection of $\mathbb{S}^2$ and **M** be a primal or dual circle of the three-dimensional circle packing with center $a$. Then the point*

$$b := \frac{1}{|a|^2}a$$

*is a vertex of the primal or dual polytope corresponding to the respective circle of the circle packing.*

*Proof.* Since the point $b$ lies in every tangent plane of every point on the circle, it also lies in the intersection of all tangent planes. Because the primal or dual point is defined exactly as this intersection of finitely many tangent planes, $b$ is precisely this point. □

Our construction uses the described linear system (3.11) to compute the primal and dual points. Nevertheless, these three statements are theoretically useful. In particular, they allow describing the projection of the polytope's vertices onto the cutting plane. Since

$$|p| = \frac{1}{|a|} \Leftrightarrow |a| = \frac{1}{|p|} \quad \text{respectively} \quad |f| = \frac{1}{|a|} \Leftrightarrow |a| = \frac{1}{|f|}$$

we obtain

$$a = \frac{|a|}{|p|}p = |a|^2 p = \frac{1}{|p|^2}p \quad \text{respectively} \quad a = \frac{|a|}{|f|}f = |a|^2 f = \frac{1}{|f|^2}f. \tag{3.12}$$

We will need this equation, among other places, in Chapter 6.

We use the previously proven statements for the next lemma, which concerns the dual polytope. We want to show that the constructed dual polytope is the polar dual polytope as defined in Definition 3.6.

**Lemma 3.41.** *The dual polytope* $\mathbf{P}^*$ *constructed in this section from* $\mathbf{P}$ *is the polar dual polytope to* $\mathbf{P}$.

*Proof.* Since the edges of the polytope are tangent to the sphere, the constructed polytope contains the origin. Let $p$ be a vertex of the primal polytope. We need to show that the corresponding facet of the dual polytope can be described as a subset of

$$\{x \in \mathbb{R}^3 \mid p \cdot x = 1\}.$$

If this holds, then by Theorem 3.7 the constructed polytope is exactly the polar dual polytope.

If $p$ is the primal point, then by Corollary 3.40 the point

$$a := \frac{1}{|p|^2}p \tag{3.13}$$

is the center of the primal circle. Since the tangent points of the primal and dual polytope coincide, the corresponding tangent points of the dual polytope lie on the primal circle, and the primal circle lies in a facet of the dual polytope. By Lemma 3.39, the position vector of the circle center and the normal vector of the corresponding plane point in the same direction, so the plane can be described via the normal form as

$$(x-a)\cdot \vec{n} = 0 \quad \Leftrightarrow \quad (x-a)\cdot \frac{a}{|a|} = 0 \quad \Leftrightarrow \quad \left\langle x, \frac{a}{|a|} \right\rangle - \left\langle a, \frac{a}{|a|} \right\rangle = 0.$$

Since the length of $a$ is exactly $1/|p|$, using the definition of $a$ from equation (3.13) yields

$$\left\langle x, \frac{a}{|a|} \right\rangle - \left\langle a, \frac{a}{|a|} \right\rangle = 0 \quad \Leftrightarrow \quad \left\langle x, \frac{\frac{1}{|p|^2}p}{\frac{1}{|p|}} \right\rangle - \left\langle \frac{1}{|p|^2}p, \frac{\frac{1}{|p|^2}p}{\frac{1}{|p|}} \right\rangle = 0 \quad \Leftrightarrow \quad \left\langle x, \frac{p}{|p|} \right\rangle - \left\langle \frac{p}{|p|^2}, \frac{p}{|p|} \right\rangle = 0.$$

This expression can be simplified further to

$$\left\langle x, \frac{p}{|p|} \right\rangle - \left\langle \frac{p}{|p|^2}, \frac{p}{|p|} \right\rangle = 0 \quad \Leftrightarrow \quad \frac{1}{|p|}\langle x, p\rangle - \frac{1}{|p|} = 0 \quad \Leftrightarrow \quad \langle x, p\rangle = 1.$$

Thus, the facets can be represented by $p \cdot x = 1$. Defining the dual polytope by its facets as

$$\{x \in \mathbb{R}^d \mid p_i \cdot x \leq 1, \text{ for } i = 1, \ldots, n\},$$

we obtain exactly the representation of the polar dual polytope from Theorem 3.7. Each inequality is indeed $\leq 1$ because $p_i \cdot 0 < 1$ holds and the constructed dual polytope contains the origin. Therefore, the dual polytope constructed from **P** is the polar dual polytope $\mathbf{P}^*$ of **P**. □

Thus, we have completed the construction of primal and dual polytopes. In the next section, we will briefly consider the construction of two-dimensional polytopes.

## 3.4 Construction of Two-Dimensional Polytopes

In this section, we describe the construction of convex 2-polytopes that are to have analogous properties to those in Section 3.2. Our goal is to construct a convex polytope with edges tangent to the unit circle and tangent centroid at zero. Although the construction of these 2-polytopes seems trivial, we require the explicit coordinates of the vertices for the continuation of this work, specifically for the two-dimensional variant of the generalized quadratic B-spline subdivision ($t = 2$, $g = 2$) and for the two- and three-dimensional variants of the generalized cubic B-spline subdivision ($t = 2$, $g = 3$ and $t = 3$, $g = 3$). Additionally, the vertex coordinates of the corresponding dual polytopes are needed.

For the construction, we start with the following definition:

**Definition 3.42.** *Let* $\mathbf{G} = (\mathbf{V}, \mathbf{E})$ *be a graph with* $n$ *vertices and* $n \geq 3$ *edges, which consists only of a cycle. Then* **G** *is the* graph of a 2-polytope ***P****, and* ***P*** *is a* realization *of* **G**.

Note that the graph **G** is obviously outerplanar.

Let $n \geq 3$ be the number of vertices of the polytope. Then the polytope also has $n$ edges. We define the tangent points of the $n$ edges as

$$e_j = \left[\cos\left(\frac{2\pi j}{n}\right), \sin\left(\frac{2\pi j}{n}\right)\right], \quad \text{for} \quad j \in \{0, \dots, n-1\}. \tag{3.14}$$

To determine the centroid of the tangent points, we first state the following lemma following [Kna09, p. 371]:

**Lemma 3.43.** *Let* $a, b \in \mathbb{R}$ *with* $b \neq 0$*. Then for* $n \in \mathbb{N}$ *it holds that*

$$\sum_{j=0}^{n-1} \sin(a + jb) = \frac{\sin\left(\frac{nb}{2}\right)}{\sin\left(\frac{b}{2}\right)} \sin\left(a + \frac{(n-1)b}{2}\right)$$

*and*

$$\sum_{j=0}^{n-1} \cos(a + jb) = \frac{\sin\left(\frac{nb}{2}\right)}{\sin\left(\frac{b}{2}\right)} \cos\left(a + \frac{(n-1)b}{2}\right).$$

From this lemma, we directly obtain the following corollary:

**Corollary 3.44.** *Let* $e_j$ *for* $j \in \{0, \dots, n-1\}$ *be defined as in Equation* (3.14)*. Then it holds that*

$$\sum_{j=0}^{n-1} e_j = \sum_{j=0}^{n-1} \left[\cos\left(\frac{2\pi j}{n}\right), \sin\left(\frac{2\pi j}{n}\right)\right] = [0, 0].$$

*Proof.* Substituting $a = 0$ and $b = 2\pi/n$ into Lemma 3.43 yields

$$\sum_{j=0}^{n-1} \sin\left(0 + j\frac{2\pi}{n}\right) = \frac{\sin\left(\frac{n\frac{2\pi}{n}}{2}\right)}{\sin\left(\frac{\frac{2\pi}{n}}{2}\right)} \sin\left(0 + \frac{(n-1)\frac{2\pi}{n}}{2}\right) = \frac{\sin(\pi)}{\sin\left(\frac{\frac{2\pi}{n}}{2}\right)} \sin\left(\frac{(n-1)\pi}{n}\right) = 0,$$

and

$$\sum_{j=0}^{n-1} \cos\left(0 + j\frac{2\pi}{n}\right) = \frac{\sin\left(\frac{n\frac{2\pi}{n}}{2}\right)}{\sin\left(\frac{\frac{2\pi}{n}}{2}\right)} \cos\left(0 + \frac{(n-1)\frac{2\pi}{n}}{2}\right) = \frac{\sin(\pi)}{\sin\left(\frac{\frac{2\pi}{n}}{2}\right)} \cos\left(\frac{(n-1)\pi}{n}\right) = 0.$$

□

Thus, the centroid of the tangent points is exactly zero. From the tangent points, the primal points of the polytope can be determined. Each primal point lies on the tangents of two adjacent tangent points. The tangents can be described similarly to the tangent planes in Section 3.3 using the normal form

$$\langle x, e\rangle = \langle e, e\rangle = 1. \tag{3.15}$$

Hence, for each point $p_j$, the linear system

$$\begin{bmatrix} \cos\left(\frac{2\pi j}{n}\right) & \sin\left(\frac{2\pi j}{n}\right) \\ \cos\left(\frac{2\pi(j+1)}{n}\right) & \sin\left(\frac{2\pi(j+1)}{n}\right) \end{bmatrix} \begin{bmatrix} (p_j)_x \\ (p_j)_y \end{bmatrix} = \begin{bmatrix} 1 \\ 1 \end{bmatrix}$$

follows. This yields

$$p_j = \begin{bmatrix} -\frac{\sin\left(\frac{2\pi j}{n}\right) - \sin\left(\frac{2\pi(j+1)}{n}\right)}{\sin\left(\frac{2\pi}{n}\right)} \\ \frac{\cos\left(\frac{2\pi j}{n}\right) - \cos\left(\frac{2\pi(j+1)}{n}\right)}{\sin\left(\frac{2\pi}{n}\right)} \end{bmatrix}. \tag{3.16}$$

An exemplary illustration for $n = 5$ is shown in Figure 3.16. Since the tangent points are equidistantly distributed on the unit circle, the length of each edge of the polytope as well as the angle at every vertex are equal. Thus, the 2-polytope is a regular $n$-gon exhibiting all symmetries of the underlying structure.

For convexity, we cite the following lemma from [Lee13, Thm. 14.31, p. 266]:

**Lemma 3.45.** *Every 2-polytope whose edges are all tangent to a circle and which contains that circle is convex.*

Thus, all 2-polytopes generated in this section are convex, and dual polytopes can also be constructed for them. By Theorem 3.7, the edges of such a dual polytope can be described by the equation

$$\begin{bmatrix} -\frac{\sin\left(\frac{2\pi j}{n}\right) - \sin\left(\frac{2\pi(j+1)}{n}\right)}{\sin\left(\frac{2\pi}{n}\right)} & \frac{\cos\left(\frac{2\pi j}{n}\right) - \cos\left(\frac{2\pi(j+1)}{n}\right)}{\sin\left(\frac{2\pi}{n}\right)} \end{bmatrix} x = 1.$$

We obtain for the vertices of the 2-polytope the linear system

$$\begin{bmatrix} -\frac{\sin\left(\frac{2\pi j}{n}\right) - \sin\left(\frac{2\pi(j+1)}{n}\right)}{\sin\left(\frac{2\pi}{n}\right)} & \frac{\cos\left(\frac{2\pi j}{n}\right) - \cos\left(\frac{2\pi(j+1)}{n}\right)}{\sin\left(\frac{2\pi}{n}\right)} \\ -\frac{\sin\left(\frac{2\pi(j+1)}{n}\right) - \sin\left(\frac{2\pi(j+2)}{n}\right)}{\sin\left(\frac{2\pi}{n}\right)} & \frac{\cos\left(\frac{2\pi(j+1)}{n}\right) - \cos\left(\frac{2\pi(j+2)}{n}\right)}{\sin\left(\frac{2\pi}{n}\right)} \end{bmatrix} \begin{bmatrix} (f_j)_x \\ (f_j)_y \end{bmatrix} = \begin{bmatrix} 1 \\ 1 \end{bmatrix}.$$

Using various addition theorems, the solution is

$$f_j = \begin{bmatrix} \cos\left(\frac{2\pi(j+1)}{n}\right) \\ \sin\left(\frac{2\pi(j+1)}{n}\right) \end{bmatrix}.$$

That the dual points coincide with the tangent points is also clear geometrically. The equations for the primal points arose from

$$\langle x, e\rangle = \langle e, e\rangle = 1.$$

By Theorem 3.7, these are exactly the equations of the edges of the dual polytope corresponding to the tangent points. Since the dual of the dual polytope, which contains the origin, is again the primal polytope by Theorem 3.7, this is also geometrically intuitive. An exemplary illustration of a primal and dual polytope can be found in Figure 3.16.

With this, we have discussed all statements about polytopes needed in the following chapters. Specifically, we have explained the theory of convex polytopes and described, using the Koebe-Andreev-Thurston theorem, which representation of 3-polytopes we choose. We then constructively produced this representation following the proof in [Zie04]. Since the original description was rather theoretical, we worked out the individual steps with a focus on practical use and especially employed the inverse stereographic projection to generate a polytope with tangent centroid zero. Finally, we proved some statements about convex polytopes and described the construction of 2-polytopes.

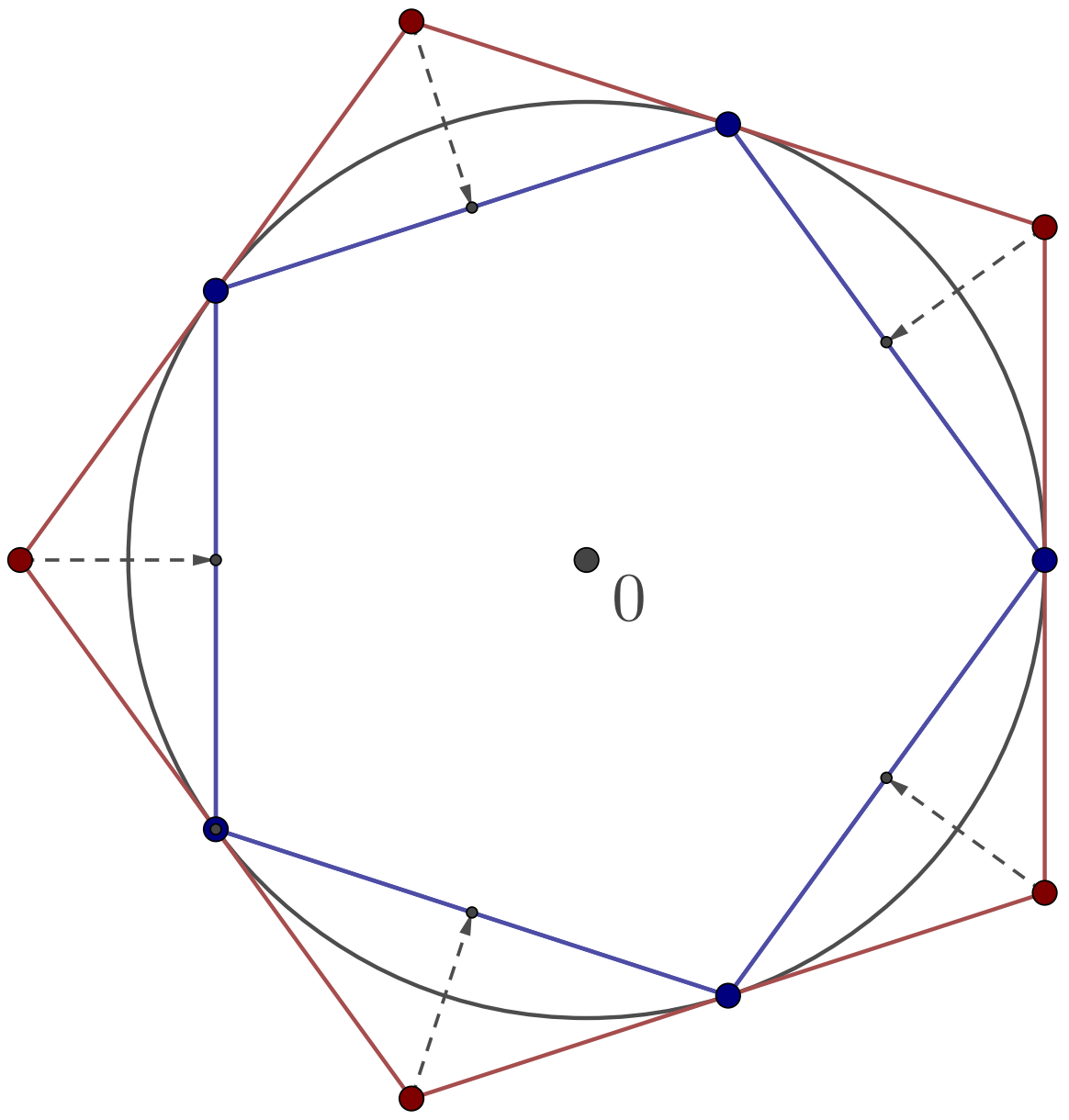

**Figure 3.16:** Primal (red) and dual (blue) $2$-polytope for $n = 5$.

The polytopes form the basis for Chapters 5 and 6, since the eigenspace of the subdominant eigenvalue of the subdivision matrices is constructed from these polytopes. On one hand, the theoretical aspects such as the implementation of symmetry and the tangential edges are important, as they are needed for the proofs of quality criteria for the subdivision matrices. On the other hand, the practical aspects such as implementability itself and minimizing numerical errors are crucial, since they play a central role for feasibility and empirical results.

In the next chapter, we will use the polytopes constructed in this chapter to generate Colin-de-Verdière-matrices, which serve as a preliminary stage for the subdivision matrices of Chapters 5 and 6.

# 4 Construction of Colin-de-Verdière-Matrices

In the previous chapter, we discussed the construction of convex polytopes, and in this chapter, we focus on the concrete construction of Colin-de-Verdière-matrices. These matrices will have eigenspaces generated by the polytopes treated in the previous chapter. The $n$ vertices of convex 2- and 3-polytopes can be interpreted as a matrix in $\mathbb{R}^{n\times 2}$ respectively $\mathbb{R}^{n\times 3}$, and the columns of this matrix shall be eigenvectors of the Colin-de-Verdière-matrices constructed in this chapter.

This serves as a preliminary step towards the construction of the subdivision matrices. The Colin-de-Verdière-matrices have special properties that we will need for the proofs of the quality criteria. How these interact will be described in Chapters 5 and 6.

In Section 4.1, we start with the theoretical aspects and fundamentals of Colin-de-Verdière-matrices. In the following sections, we construct Colin-de-Verdière-matrices for polytopes of dimensions $0$ to $3$. In Section 4.2 we first consider the trivial cases of Colin-de-Verdière-matrices for polytopes of dimensions 0 and 1. Afterwards, in Section 4.3 we describe the construction of Colin-de-Verdière-matrices for 2-polytopes, and in Section 4.4 for 3-polytopes. For the generalized quadratic B-spline subdivision algorithms, we will only need the Colin-de-Verdière-matrices for 2-polytopes (for $t = 2$) and 3-polytopes (for $t = 3$). Since the relevant eigenspaces of the generalized cubic B-spline subdivision matrices arise from decompositions of polytopes, for their construction we will need Colin-de-Verdière-matrices of all the dimensions described here. Therefore, Sections 4.2 and 4.3 are not mentioned here only for completeness but are applied in the construction of subdivision matrices in Chapter 6.

## 4.1 Definition and Properties of Colin-de-Verdière-Matrices

The initial definition of Colin-de-Verdière-matrices originates from the work of the mathematician Colin de Verdière [Col90]. It was translated from French to English in 1993 in [Col93], and these two works thus form the origin of the term. A fairly detailed overview of Colin-de-Verdière-matrices and their applications can be found in [HLS99], from which we also took the definition of Colin-de-Verdière-matrices [HLS99, Def. 1.1, p. 31]:

**Definition 4.1.** *Let $\boldsymbol{G} = (\boldsymbol{V}, \boldsymbol{E})$ be a graph with $n$ vertices, i.e. $|\boldsymbol{V}| = n$. Let $C \in \mathbb{R}^{n\times n}$ be a symmetric matrix with the following properties:*

*C1 For all $i, j \in \{1, \dots, n\}$ with $i \neq j$ holds*

$$C_{(i,j)} \begin{cases} < 0 & \text{if } \{v_i, v_j\} \in \boldsymbol{E}, \\ = 0 & \text{if } \{v_i, v_j\} \notin \boldsymbol{E}. \end{cases}$$

*C2 The matrix $C$ has exactly one negative eigenvalue with algebraic multiplicity 1.*

*C3 There is no nonzero matrix $M \in \mathbb{R}^{n\times n}$ with $M_{(i,j)} = 0$ if $i = j$ and $M_{(i,j)} = 0$ if $C_{(i,j)} \neq 0$, such that $CM = 0$.*

*Then $C$ is called a* Colin-de-Verdière-matrix.

We now examine this definition more closely. Property C1 determines the shape of $C$. With this condition, $C$ resembles an adjacency matrix, since apart from the diagonal, both matrices share the same pattern of nonzero entries. However, the diagonal entries of $C$ can take arbitrary values, including positive ones.

Property C2 is well-defined since symmetric matrices, by Theorem B.44, have only real eigenvalues. Hence, $C$ has exactly one negative eigenvalue, and all other eigenvalues are $\geq 0$.
Property C3 is called the *strong Arnold property*, originally introduced in [Arn72].

With this definition, it is possible to define an invariant quantity for a graph. This is also introduced in [Col93]. Here, we use the definition from [Izm10, Def. 1.1, pp. 427–428]:

**Definition 4.2.** *Let $\boldsymbol{G}$ be a graph and $\boldsymbol{C}(\boldsymbol{G})$ be the set of all Colin-de-Verdière-matrices of the graph $\boldsymbol{G}$. Then*

$$\boldsymbol{c}(\boldsymbol{G}) := \max_{C \in \boldsymbol{C}(\boldsymbol{G})} \big(\dim\big(\ker(C)\big)\big)$$

*is called the* Colin-de-Verdière-number *of the graph $\boldsymbol{G}$.*

The Colin-de-Verdière-number thus indicates the maximal dimension of the kernel of all Colin-de-Verdière-matrices associated with a given graph. Since we will only consider Colin-de-Verdière-matrices with this maximal kernel dimension, we add the following definition from [Izm10, Def. 1.1, pp. 427–428]:

**Definition 4.3.** *A Colin-de-Verdière-matrix $C$ of the graph $\boldsymbol{G}$ is called* optimal *if*

$$\dim\big(\ker(C)\big) = \boldsymbol{c}(\boldsymbol{G}).$$

For the construction of optimal Colin-de-Verdière-matrices, we require a statement about the possible sizes of the Colin-de-Verdière-number of certain graphs. This is provided by the following theorem [HLS99, Num. 1.4, p. 35]:

**Theorem 4.4.** *The following statements hold for the Colin-de-Verdière-number $\boldsymbol{c}(\boldsymbol{G})$ of a graph $\boldsymbol{G}$:*

1. *$\boldsymbol{c}(\boldsymbol{G}) \leq 1$ if and only if $\boldsymbol{G}$ is a disjoint union of paths.*

2. *$\boldsymbol{c}(\boldsymbol{G}) \leq 2$ if and only if $\boldsymbol{G}$ is outerplanar.*

3. *$\boldsymbol{c}(\boldsymbol{G}) \leq 3$ if and only if $\boldsymbol{G}$ is planar.*

4. *$\boldsymbol{c}(\boldsymbol{G}) \leq 4$ if and only if $\boldsymbol{G}$ is loopless embeddable.*

The reference [HLS99] provides an overview of the history of proofs of these statements. The proofs of the first three statements can be found in [Col90] and [Col93]. The proof of the fourth statement is composed of results from several sources; an overview is given in [HLS99, p. 36]. However, in this work, statement four is only mentioned for completeness and will not be used later, so we omit a literature survey here.

The decisive advantage of Colin-de-Verdière-matrices is that with the Colin-de-Verdière-number for optimal matrices, we can make a statement about the spectrum of these matrices. For optimal Colin-de-Verdière-matrices, by property C2 there is exactly one negative eigenvalue and $\mathbf{c}(\mathbf{G})$ eigenvalues equal to zero. All other eigenvalues are positive.

In the next section, we construct optimal Colin-de-Verdière-matrices for graphs with one and two vertices using the statements collected in this section.

## 4.2 Colin-de-Verdière-matrices for Graphs with One or Two Vertices

The two cases considered in this section are the trivial cases of Colin-de-Verdière-matrices. Since we need these for the construction of generalized cubic B-spline subdivision algorithms, we describe their explicit construction. We begin with the graph consisting of one vertex and no edges, which can also be interpreted as a 0-polytope realization in $\mathbb{R}^d$:

**Lemma 4.5.** *Let $\boldsymbol{G} = (\boldsymbol{V}, \boldsymbol{E})$ be a graph with $\boldsymbol{V} = \{v\}$ and $\boldsymbol{E} = \emptyset$. Then the matrix*

$$C := [-1]$$

*is an optimal Colin-de-Verdière-matrix for **G**.*

*Proof.* The matrix $C$ is a Colin-de-Verdière-matrix because

- it is symmetric, since it consists of only one number,
- it has no off-diagonal entries, thus property C1 holds,
- the matrix has exactly one negative eigenvalue $-1$, so property C2 holds,
- since $0$ is the only number satisfying $0 \cdot (-1) = 0$, property C3 holds.

Furthermore, $C$ is optimal. Since $C$ consists of only one entry, it has only one eigenvalue. Thus,

$$\mathbf{c}(\mathbf{G}) = 0 = \dim(\ker(C)).$$

□

In the next step, we consider the slightly more complex case of a graph with two vertices and one edge, which can be interpreted as a realization in $\mathbb{R}^n$ as a 1-polytope. For this, we examine the following lemma:

**Lemma 4.6.** *Let $\boldsymbol{G} = (\boldsymbol{V}, \boldsymbol{E})$ be a graph with $\boldsymbol{V} = \{v_1, v_2\}$ and $\boldsymbol{E} = \{\{v_1, v_2\}\}$. Furthermore, let $p_1, p_2 \in \mathbb{R}^3$ with $p_1, p_2 \neq 0$ be the realization of the graph in $\mathbb{R}^3$ as a 1-polytope, and let $0$ be a point on the edge between $p_1$ and $p_2$. Then the matrix*

$$C = \begin{bmatrix} -\frac{|p_2|}{|p_1|} & -1 \\ -1 & -\frac{|p_1|}{|p_2|} \end{bmatrix}$$

*is an optimal Colin-de-Verdière-matrix of the graph **G**. Moreover,*

$$\begin{bmatrix} |p_1| \\ -|p_2| \end{bmatrix}$$

*is an eigenvector of $C$ corresponding to the eigenvalue $0$.*

Before starting the proof, it is worth noting that the above eigenvector can be interpreted as two points in $\mathbb{R}$. It thus contains the vertices of a 1-polytope in $\mathbb{R}$ that contains the origin.

*Proof.* We again verify the definition of Colin-de-Verdière-matrices. First, by definition, $C$ is symmetric. Moreover, all off-diagonal entries correspond to edges and are negative. Hence property C1 is satisfied.

To check the eigenvalues, we consider the characteristic polynomial of $C$:

$$\det(C - \lambda_i E_2) = \lambda_i \left( \lambda_i + \left( \frac{|p_2|}{|p_1|} + \frac{|p_1|}{|p_2|} \right) \right).$$

Thus, the eigenvalues of $C$ are exactly

$$\lambda_1 = 0 \quad \text{and} \quad \lambda_2 = -\left( \frac{|p_2|}{|p_1|} + \frac{|p_1|}{|p_2|} \right),$$

and $C$ has exactly one negative eigenvalue and one eigenvalue equal to zero, so property C2 is satisfied. Moreover,

$$C \cdot \begin{bmatrix} |p_1| \\ -|p_2| \end{bmatrix} = \begin{bmatrix} -\frac{|p_2|}{|p_1|}|p_1| + |p_2| \\ -|p_1| + \frac{|p_1|}{|p_2|}|p_2| \end{bmatrix} = \begin{bmatrix} 0 \\ 0 \end{bmatrix} = 0 \cdot \begin{bmatrix} |p_1| \\ -|p_2| \end{bmatrix}.$$

Hence, $[|p_1|, -|p_2|]^T$ is an eigenvector corresponding to the eigenvalue $0$.

Finally, any matrix $M \in \mathbb{R}^{2\times 2}$ that has zero entries on the diagonal and also zero entries where $C$ has nonzero off-diagonal entries must be the zero matrix. Thus, property C3 is also satisfied and $C$ is a Colin-de-Verdière-matrix.

Furthermore, $C$ can have at most one zero eigenvalue and is therefore optimal, as is clear from the first case of Theorem 4.4. Since **G** is a disjoint union of paths (here just a path with two vertices and one edge), the Colin-de-Verdière-number is at most 1. □

After these two trivial cases, we now consider Colin-de-Verdière-matrices for convex 2-polytopes in the next section.

## 4.3 Colin-de-Verdière-Matrices for 2-Polytopes

In this section, we aim to construct a Colin-de-Verdière-matrix for a 2-polytope. This is the most involved part of the chapter, since no direct blueprint for this case was found in the literature. However, we will use the following theorem, which appears in a more general form in [Izm10, Thm. 2.4, p. 434]:

**Theorem 4.7.** *Let **G** be an outerplanar graph consisting of a cycle with $n$ vertices and $n$ edges, and let **P** be a realization of **G** as a 2-polytope with $n$ vertices, that is,*

$$\boldsymbol{P} := \operatorname{conv}(P) \quad \textit{with} \quad P \in \mathbb{R}^{n\times 2}.$$

*Further, let the corresponding polar dual polytope*

$$\boldsymbol{P}^* = \{x \in \mathbb{R}^2 \mid P_{(i,:)}x \leq 1, \textit{ for } i = 1, \ldots, n\}$$

*be given by Theorem 3.7, associated with the dual graph $\boldsymbol{G}^*$. Define also the facet variation of the dual polytope*

$$\boldsymbol{P}^*(r) = \{x \in \mathbb{R}^2 \mid P_{(i,:)}x \leq r_i, \textit{ for } i = 1, \ldots, n\} \quad \textit{for} \quad r \in \mathbb{R}^n_{>0}.$$

*Then the matrix*

$$C_{(i,j)} := -\left.\frac{\partial^2 \operatorname{vol}(\boldsymbol{P}^*(r))}{\partial r_i \partial r_j}\right|_{r=\vec{1}}$$

*is a Colin-de-Verdière-matrix of the graph **G**. Moreover, the kernel dimension satisfies*

$$\dim\big(\ker(C)\big) = 2,$$

*and thus $C$ is optimal according to case 2 of Theorem 4.4.*

Our strategy is to construct, for the special case of a 2-polytope whose edges are tangent to the unit circle, a matrix whose entries coincide with those of $C$ from Theorem 4.7. This will prove that $C$ is a Colin-de-Verdière-matrix directly from the above theorem. Note that the 2-polytope does not have to be a regular $n$-gon. It suffices that its edges are tangent to the unit circle. This special case is sufficient for the continuation of this work.

**Theorem 4.8.** *Let **G** be an outerplanar graph consisting of a cycle with $n$ vertices and $n$ edges, and let **P** be a realization of **G**. We consider cyclic indices modulo $n$, i.e., $n+1 \equiv 1$ and $0 \equiv n$. For **P**, assume:*

- ***P** $:= \operatorname{conv}(P)$ with $P \in \mathbb{R}^{n\times 2}$.*
- ***P** is a 2-polytope with vertices $p_1 := P_{(1,:)}, \ldots, p_n := P_{(n,:)}$.*
- *For each $i \in \{1, \ldots, n\}$, there is an edge between $p_i$ and $p_{i+1}$, and these are all edges of the realization of the graph **G**.*
- ***P** contains the point $(0,0)$.*
- *The edges of **P** are tangent to the unit circle $\mathbb{S}^1$.*

*Then the matrix*

$$C_{(i,j)} := \begin{cases} -\frac{1}{|p_i \times p_j|} & \text{if } i \neq j \text{ and } p_i \text{ and } p_j \text{ share an edge,} \\ 0 & \text{if } i \neq j \text{ and } p_i \text{ and } p_j \text{ do not share an edge,} \\ \frac{\cos(\alpha)}{|p_i|^2 \sin(\alpha)} + \frac{\cos(\beta)}{|p_i|^2 \sin(\beta)} & \text{if } i = j, \end{cases}$$

*with angles $\alpha = \sphericalangle(p_{i-1}, p_i)$ and $\beta = \sphericalangle(p_i, p_{i+1})$, is an optimal Colin-de-Verdière-matrix and satisfies*

$$CP = C \begin{bmatrix} p_1 \\ \vdots \\ p_n \end{bmatrix} = 0.$$

*Proof.* To prove this theorem, we will show that the entries of $C$ correspond exactly to those from Theorem 4.7. First, we introduce the following notations and facts:

- Let $e_i$ denote the tangent points on the unit circle $\mathbb{S}^1$ lying on the edges between $p_i$ and $p_{i+1}$.
- For the dual polytope $\mathbf{P}^*$ of $\mathbf{P}$, the inequalities $\langle p_i, x \rangle \leq 1$ hold. The points $e_i$ satisfy

  $$\langle p_i, e_i \rangle = 1 \quad \text{and} \quad \langle p_{i+1}, e_i \rangle = 1.$$

  This follows since the triangle $(0, e_i, p_i)$ has a right angle at $e_i$, due to $(e_i, p_i)$ lying on the tangent to the unit circle at $e_i$. Hence,

  $$\langle p_i, e_i \rangle = |p_i| \cdot |e_i| \cos\big(\sphericalangle(p_i, e_i)\big) = |p_i| \cdot |e_i| \frac{|e_i|}{|p_i|} = 1.$$

  Analogously for $\langle p_{i+1}, e_i \rangle$. Thus, conv$(e_1, \ldots, e_n)$ is the polar dual polytope of $\mathbf{P}$, with vertices $e_i$.
- The projection of $p_i$ onto the line $\langle p_i, x \rangle = 1$ is given by

  $$\frac{1}{|p_i|^2} p_i, \quad \text{since} \quad \left\langle p_i, \frac{1}{|p_i|^2} p_i \right\rangle = 1.$$

  Denote this point by

  $$q_i := \frac{1}{|p_i|^2} p_i.$$

  Furthermore, the vectors $p_i$, $q_i$, and the normal vector to the line through $e_{i-1}$ and $e_i$ are all in the same direction, so the triangles

  $$(e_{i-1}, q_i, 0) \quad \text{and} \quad (e_i, q_i, 0)$$

  have right angles at $q_i$.

An illustrative figure of these notations and right angles is given in Figure 4.1.

To prove $C$ matches the matrix from Theorem 4.7, consider the area of the dual polytope. Decompose the polygon into $2n$ right triangles:

$$(0, e_i, q_i) \quad \text{and} \quad (0, e_i, q_{i+1}) \quad \text{for all} \quad i \in \{1, \ldots, n\},$$

where the right angle is at $q_i$. This sum equals the volume of the dual polytope (see Figure 4.2). If we vary the facets of the dual polytope as in Theorem 4.7, by changing the inequalities

$$\langle p_i, x \rangle \leq 1 \quad \text{to} \quad \langle p_i, x \rangle \leq r_i \quad \Leftrightarrow \quad \left\langle \frac{p_i}{r_i}, x \right\rangle \leq 1 \quad \text{with} \quad r_i \in \mathbb{R}_{>0} \quad \text{for all} \quad i \in \{1, \ldots, n\}$$

then we shift the edge $(e_{i-1}, e_i)$, since

$$\left\langle \frac{p_i}{r_i}, r_i q_i \right\rangle = \langle p_i, q_i \rangle = 1.$$

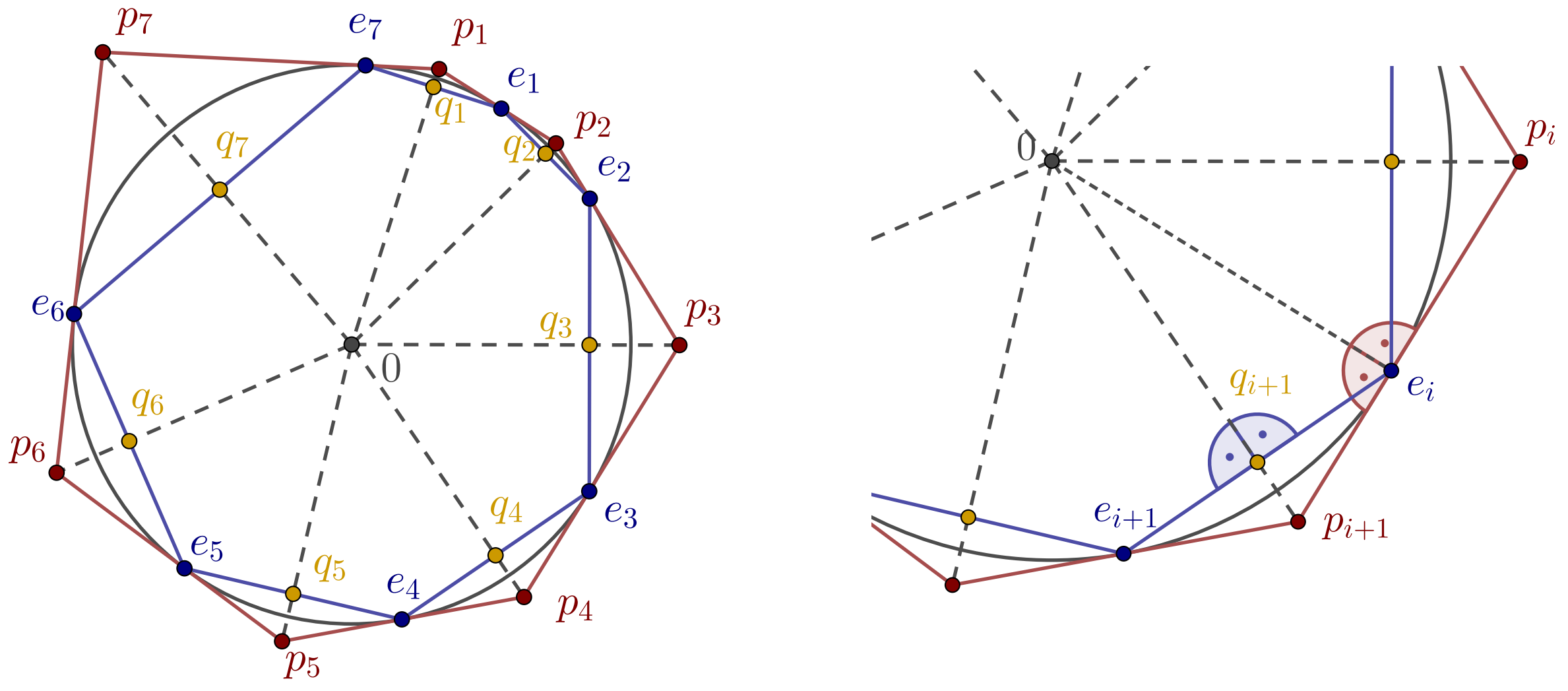

**Figure 4.1:** An example of a 2-polytope **P** (in red), as defined in Theorem 4.8, and its corresponding polar dual polytope $\mathbf{P}^*$ (in blue). The left image shows the labeled points, while the right image shows the right angles at the respective points.

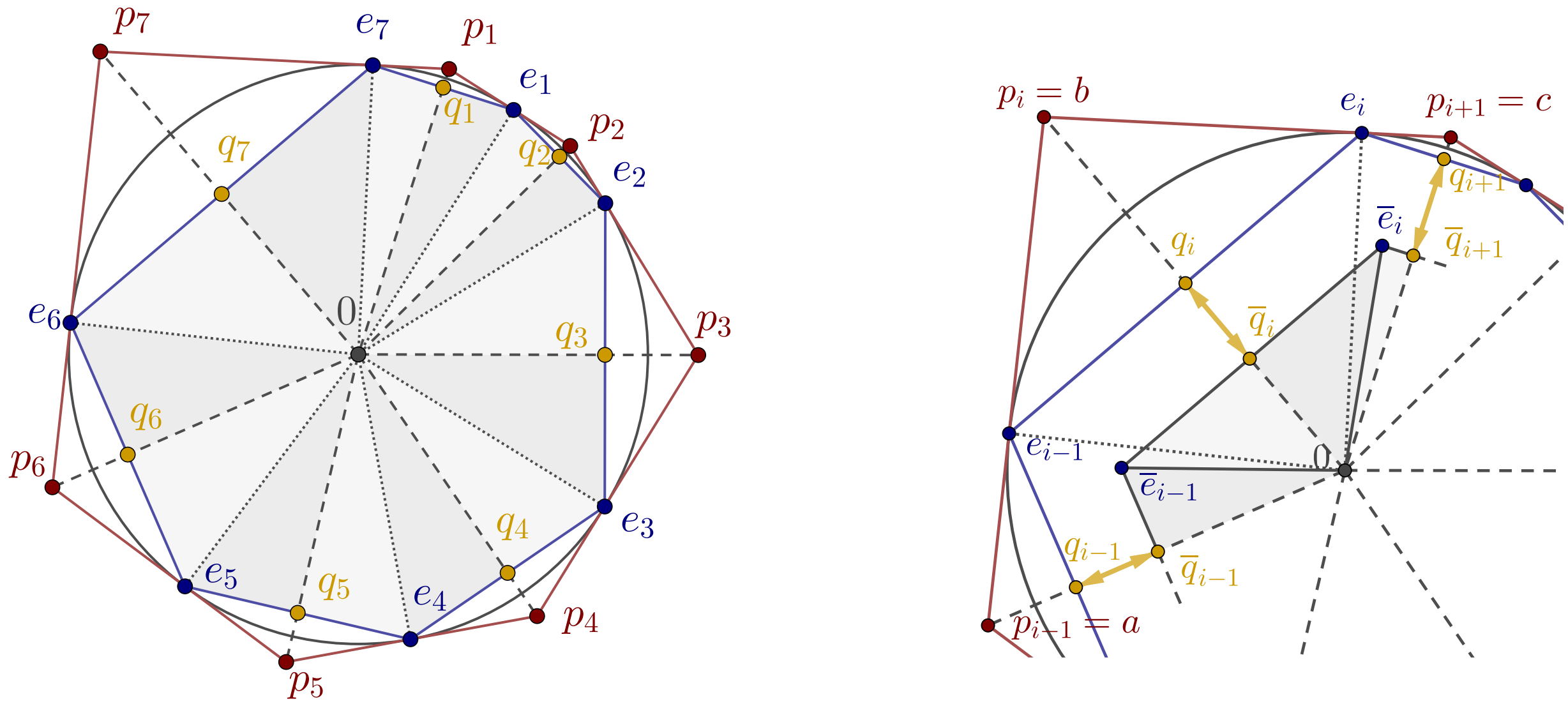

**Figure 4.2:** The $2n$ triangles of the dual polytope (left) and the variation of the facets (right), including the labeling of the individual points.

The question is how this affects the area of the polytope. For our application, only the variation around $r = \vec{1}$ is relevant, since the derivative of the volume with respect to the parameters $r_i$ in Theorem 4.7 only needs to be evaluated at $r = \vec{1}$. So it is enough to look at variations in a small neighborhood around $\vec{1}$.

If we fix an index $i$ and only vary the corresponding inequality slightly, this directly affects the point $q_i$ and the intersections of the corresponding half-space with its neighbors, i.e., the points $e_{i-1}$ and $e_i$. In terms of area, this influences the triangles

$$(0, q_{i-1}, e_{i-1}), \quad (0, q_i, e_{i-1}), \quad (0, q_i, e_i) \quad \text{and} \quad (0, q_{i+1}, e_i).$$

Conversely, if we fix the triangles

$$(0, q_{i-1}, e_{i-1}) \quad \text{and} \quad (0, q_i, e_{i-1}),$$

then they only depend on the variables $r_{i-1}$ and $r_i$, assuming the variation is close to $\vec{1}$. If we increase $i$ by one, we find that the area of the triangles

$$(0, q_i, e_i) \quad \text{and} \quad (0, q_{i+1}, e_i)$$

depends on $r_i$ and $r_{i+1}$.

Now, if we fix an index $i$, the four triangles affected by the variation are composed of the origin and five other points. We want to express them in terms of the variation of $r_{i-1}$, $r_i$, $r_{i+1}$ and the points $p_{i-1}$, $p_i$, $p_{i+1}$. To simplify notation, since we will work with the $x$- and $y$-coordinates of the points, we define

$$a := p_{i-1} = \begin{bmatrix} a_x \\ a_y \end{bmatrix}, \quad b := p_i = \begin{bmatrix} b_x \\ b_y \end{bmatrix}, \quad c := p_{i+1} = \begin{bmatrix} c_x \\ c_y \end{bmatrix}.$$

Using these definitions, we can write the five points relevant to the triangles affected by $r_i$ as

$$\overline{q}_{i-1} := r_{i-1} q_{i-1} = r_{i-1} \frac{a}{|a|^2}, \quad \overline{q}_i := r_i q_i = r_i \frac{b}{|b|^2}, \quad \overline{q}_{i+1} := r_{i+1} q_{i+1} = r_{i+1} \frac{c}{|c|^2}$$

and

$$\overline{e}_{i-1}, \text{ which depends on } r_{i-1} \text{ and } r_i, \qquad \overline{e}_i, \text{ which depends on } r_i \text{ and } r_{i+1}.$$

The points $\overline{e}_{i-1}$ and $\overline{e}_i$ can be computed using the linear systems

$$\begin{bmatrix} a^T \\ b^T \end{bmatrix} \overline{e}_{i-1} = \begin{bmatrix} r_{i-1} \\ r_i \end{bmatrix}, \qquad \begin{bmatrix} b^T \\ c^T \end{bmatrix} \overline{e}_i = \begin{bmatrix} r_i \\ r_{i+1} \end{bmatrix}.$$

This is because a point $e_i$ lies on two bounding lines of the dual polytope, and the varied boundary lines can be described by $\langle p_i, x \rangle = r_i$. From this, we find

$$\overline{e}_{i-1} = \left[ \frac{b_y r_{i-1} - a_y r_i}{a_x b_y - a_y b_x}, \frac{a_x r_i - b_x r_{i-1}}{a_x b_y - a_y b_x} \right]^T,$$
$$\overline{e}_i = \left[ \frac{c_y r_i - b_y r_{i+1}}{b_x c_y - b_y c_x}, \frac{b_x r_{i+1} - c_x r_i}{b_x c_y - b_y c_x} \right]^T.$$

An illustration of these five points is shown in Figure 4.2. With these points, we obtain the two varied triangles

$$(0, \overline{q}_{i-1}, \overline{e}_{i-1}) \quad \text{and} \quad (0, \overline{q}_i, \overline{e}_{i-1}),$$

whose area, aside from the points $a$, $b$, and $c$, only depends on $r_{i-1}$ and $r_i$. Their area can be calculated using the embedding of the points in $\mathbb{R}^3$ and the cross product as

$$\text{vol}_2 (r_{i-1}, r_i) = \frac{1}{2} \left| \overline{q}_{i-1} \times \overline{e}_{i-1} \right| + \frac{1}{2} \left| \overline{q}_i \times \overline{e}_{i-1} \right|$$

and further

$$\begin{aligned}\mathrm{vol}_2(r_{i-1}, r_i) &= \frac{1}{2}\left| r_{i-1}\frac{a_x}{|a|^2}\cdot\frac{a_x r_i - b_x r_{i-1}}{a_x b_y - a_y b_x} - r_{i-1}\frac{a_y}{|a|^2}\cdot\frac{b_y r_{i-1} - a_y r_i}{a_x b_y - a_y b_x}\right| \\ &+ \frac{1}{2}\left| r_i\frac{b_x}{|b|^2}\cdot\frac{a_x r_i - b_x r_{i-1}}{a_x b_y - a_y b_x} - r_i\frac{b_y}{|b|^2}\cdot\frac{b_y r_{i-1} - a_y r_i}{a_x b_y - a_y b_x}\right| \\ &= \frac{|r_{i-1}a_x(a_x r_i - b_x r_{i-1}) - r_{i-1}a_y(b_y r_{i-1} - a_y r_i)|}{2|a|^2\,|a_x b_y - a_y b_x|} \\ &+ \frac{|r_i b_x(a_x r_i - b_x r_{i-1}) - r_i b_y(b_y r_{i-1} - a_y r_i)|}{2|b|^2\,|a_x b_y - a_y b_x|} \\ &= \frac{\left|r_{i-1}r_i\langle a,a\rangle - r_{i-1}^2\langle a,b\rangle\right|}{2\langle a,a\rangle\,|a\times b|} + \frac{\left|r_i^2\langle a,b\rangle - r_{i-1}r_i\langle b,b\rangle\right|}{2\langle b,b\rangle\,|a\times b|}\end{aligned} \tag{4.1}$$

Similarly, the area of the other two triangles

$$(0, \overline{q}_i, \overline{e}_i) \quad \text{and} \quad (0, \overline{q}_{i+1}, \overline{e}_i)$$

is given by:

$$\begin{aligned}\mathrm{vol}_2(r_i, r_{i+1}) &= \frac{1}{2}\left|\overline{q}_i \times \overline{e}_i\right| + \frac{1}{2}\left|\overline{q}_{i+1}\times\overline{e}_i\right| \\ &= \frac{1}{2}\left| r_i\frac{b_x}{|b|^2}\cdot\frac{b_x r_{i+1} - c_x r_i}{b_x c_y - b_y c_x} - r_i\frac{b_y}{|b|^2}\cdot\frac{c_y r_i - b_y r_{i+1}}{b_x c_y - b_y c_x}\right| \\ &+ \frac{1}{2}\left| r_{i+1}\frac{c_x}{|c|^2}\cdot\frac{b_x r_{i+1} - c_x r_i}{b_x c_y - b_y c_x} - r_{i+1}\frac{c_y}{|c|^2}\cdot\frac{c_y r_i - b_y r_{i+1}}{b_x c_y - b_y c_x}\right| \\ &= \frac{|r_i b_x(b_x r_{i+1} - c_x r_i) - r_i b_y(c_y r_i - b_y r_{i+1})|}{2|b|^2\,|b_x c_y - b_y c_x|} \\ &+ \frac{|r_{i+1}c_x(b_x r_{i+1} - c_x r_i) - r_{i+1}c_y(c_y r_i - b_y r_{i+1})|}{2|c|^2\,|b_x c_y - b_y c_x|} \\ &= \frac{\left|r_i r_{i+1}\langle b,b\rangle - r_i^2\langle b,c\rangle\right|}{2\langle b,b\rangle\,|b\times c|} + \frac{\left|r_{i+1}^2\langle b,c\rangle - r_i r_{i+1}\langle c,c\rangle\right|}{2\langle c,c\rangle\,|b\times c|}.\end{aligned} \tag{4.2}$$

The total area of the varied dual 2-polytope $\mathbf{P}^*(r)$ is the sum over all such triangle areas.

Now we want to show that the second derivatives

$$\left.\frac{\partial^2\,\mathrm{vol}(\mathbf{P}^*(r))}{\partial r_i \partial r_j}\right|_{r=\vec{1}}$$

match the matrix entries in Theorem 4.7. To compute these derivatives, we use the fact that for an absolute value function:

$$\frac{\partial}{\partial r_{i-1}\partial r_i}\left|\boldsymbol{f}(r_{i-1}, r_i)\right| = \mathrm{sgn}\left(\boldsymbol{f}(r_{i-1}, r_i)\right)\frac{\partial}{\partial r_{i-1}\partial r_i}\boldsymbol{f}(r_{i-1}, r_i) \quad \text{and} \quad \frac{\partial}{\partial^2 r_i}\left|\boldsymbol{f}(r_i)\right| = \mathrm{sgn}\left(\boldsymbol{f}(r_i)\right)\frac{\partial}{\partial^2 r_i}\boldsymbol{f}(r_{i-1}, r_i).$$

We first consider the case $i \neq j$. If $j \notin \{i-1, i+1\}$, then there is no triangle that depends on both variables $r_i$ and $r_j$, so the derivative is zero.

Now assume $j = i - 1$ (without loss of generality). Then only the two triangles

$$(0, \overline{q}_{i-1}, \overline{e}_{i-1}) \quad \text{and} \quad (0, \overline{q}_i, \overline{e}_{i-1})$$

depend on both $r_{i-1}$ and $r_i$. Differentiating Equation (4.1) with respect to $r_{i-1}$ and $r_i$, we get:

$$2\,|a\times b|\left(\left.\frac{\partial^2\,\mathrm{vol}\left(\mathbf{P}^*(r)\right)}{\partial r_{i-1}\partial r_i}\right|_{r=\vec{1}}\right) = \left.\frac{\langle a,a\rangle}{\langle a,a\rangle}\,\mathrm{sgn}\left(r_{i-1}r_i\langle a,a\rangle - r_{i-1}^2\langle a,b\rangle\right) - \frac{\langle b,b\rangle}{\langle b,b\rangle}\,\mathrm{sgn}\left(r_i^2\langle a,b\rangle - r_{i-1}r_i\langle b,b\rangle\right)\right|_{r=\vec{1}}$$

And further:

$$\begin{aligned}
2\,|a\times b|\left(\left.\frac{\partial^2 \operatorname{vol}(\mathbf{P}^*(r))}{\partial r_{i-1}\partial r_i}\right|_{r=\vec{1}}\right) &= \operatorname{sgn}(\langle a,a\rangle-\langle a,b\rangle)-\operatorname{sgn}(\langle a,b\rangle-\langle b,b\rangle)\\
&= \operatorname{sgn}(\langle a,a\rangle-\langle a,b\rangle)+\operatorname{sgn}(\langle b,b\rangle-\langle a,b\rangle)\\
&= \operatorname{sgn}\left(|a|^2-|a||b|\cos(\alpha)\right)+\operatorname{sgn}\left(|b|^2-|a||b|\cos(\alpha)\right)\\
&= \operatorname{sgn}(|a|-|b|\cos(\alpha))+\operatorname{sgn}(|b|-|a|\cos(\alpha))\,,
\end{aligned}$$

with $\alpha = \sphericalangle(p_{i-1}, p_i)$ be the angle between $a$ and $b$. We now argue that both signs are positive. We consider several cases:

1. If $\cos(\alpha) \le 0$, both terms are positive.
2. $\cos(\alpha) = 1$ cannot happen, since this would mean that $a$ and $b$ are colinear, implying a degenerate polytope.
3. If $0 < \cos(\alpha) < 1$:
   a) If $|a| = |b|$, both terms are still positive because $\cos(\alpha) < 1$.
   b) $|a| \neq |b|$. Without loss of generality, we assume that $|b| < |a|$. Then the sign

   $$\operatorname{sgn}(|a|-|b|\cos(\alpha))$$

   is clearly $1$, and we now need to examine the sign of

   $$\operatorname{sgn}(|b|-|a|\cos(\alpha))\,.$$

   First, note that the lengths of $a$ and $b$ are both greater than 1, since they are vertices of the primal polytope, whose edges are tangent to the unit circle. This means both points lie outside the unit circle.

   Moreover, the point

   $$\frac{b}{|b|}|a|\cos(\alpha)$$

   is the orthogonal projection of $a$ onto the origin line in the direction of $b$. This projection has length $|a|\cos(\alpha)$. Since $a$ lies on the tangent through the point $e_{i-1}$, the orthogonal projection of $e_{i-1}$ must be longer than $|a|\cos(\alpha)$. As the projection of $e_{i-1}$ is precisely $q_i$, we conclude that

   $$|a|\cos(\alpha) < |q_i| = \frac{1}{|b|^2} < 1 < |b|,$$

   which implies that the second sign is also positive. An illustration of this projection is given in Figure 4.3.

We then conclude:

$$-\left.\frac{\partial^2 \operatorname{vol}(\mathbf{P}^*(r))}{\partial r_{i-1}\partial r_i}\right|_{r=\vec{1}} = -\frac{2}{2|a\times b|} = -\frac{1}{|p_{i-1}\times p_i|}.$$

Thus, the values for $i \neq j$ in Theorems 4.7 and 4.8 agree.

In the next step, we consider

$$\left.\frac{\partial^2 \operatorname{vol}(\mathbf{P}^*(r))}{\partial^2 r_i}\right|_{r=\vec{1}},$$

and analogously to Equations (4.1) and (4.2), we obtain:

$$\begin{aligned}
\left.\frac{\partial^2 \operatorname{vol}(\mathbf{P}^*(r))}{\partial^2 r_i}\right|_{r=\vec{1}} &= \left.\frac{2\langle a,b\rangle \operatorname{sgn}\left(r_i^2\langle a,b\rangle - r_{i-1}r_i\langle b,b\rangle\right)}{2\langle b,b\rangle|a\times b|} - \frac{2\langle b,c\rangle \operatorname{sgn}\left(r_{i+1}r_i\langle b,b\rangle - r_i^2\langle b,c\rangle\right)}{2\langle b,b\rangle|c\times b|}\right|_{r=\vec{1}}\\
&= \frac{\langle a,b\rangle \operatorname{sgn}(\langle a,b\rangle-\langle b,b\rangle)}{\langle b,b\rangle|a\times b|} - \frac{\langle b,c\rangle \operatorname{sgn}(\langle b,b\rangle-\langle b,c\rangle)}{\langle b,b\rangle|c\times b|}.
\end{aligned}$$

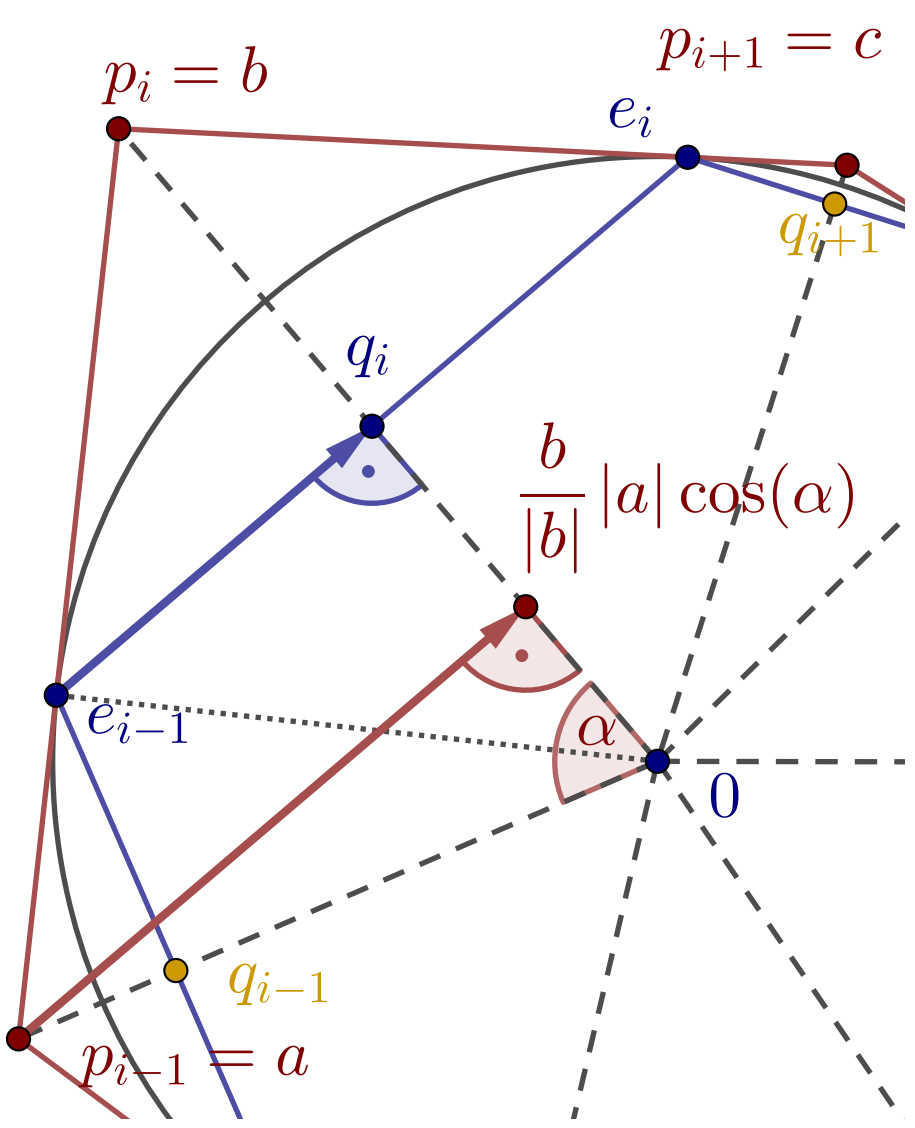

**Figure 4.3:** Orthogonal projection of $e_{i-1}$ and $a$ onto the line through $0$ and $b$. One can observe that for $\alpha \in (0, \pi/2) \cup (3\pi/2, 2\pi)$ the projection of $a$ is shorter than that of $e_{i-1}$.

Using the same sign argument as before, we find:

$$\left.\frac{\partial^2 \operatorname{vol}(\mathbf{P}^*(r))}{\partial^2 r_i}\right|_{r=\vec{1}} = -\frac{\langle a, b\rangle}{\langle b, b\rangle |a \times b|} - \frac{\langle b, c\rangle}{\langle b, b\rangle |c \times b|}.$$

Using the definitions of the scalar and cross product, we get:

$$\begin{aligned} -\left.\frac{\partial^2 \operatorname{vol}(\mathbf{P}^*(r))}{\partial^2 r_i}\right|_{r=\vec{1}} &= \frac{|a||b|\cos(\alpha)}{|b|^2|a||b|\sin(\alpha)} + \frac{|b||c|\cos(\beta)}{|b|^2|b||c|\sin(\beta)} \\ &= \frac{\cos(\alpha)}{|b|^2\sin(\alpha)} + \frac{\cos(\beta)}{|b|^2\sin(\beta)} \\ &= \frac{\cos(\alpha)}{|p_i|^2\sin(\alpha)} + \frac{\cos(\beta)}{|p_i|^2\sin(\beta)}. \end{aligned}$$

Thus, the matrix $C$ from Theorem 4.7 agrees with the matrix from Theorem 4.8, confirming that $C$ is an optimal Colin-de-Verdière-matrix.

Finally, we consider

$$C \cdot \begin{bmatrix} p_1^T \\ \vdots \\ p_n^T \end{bmatrix}.$$

For the $i$-th row of this product, we use the entries of the matrix $C$ and compute:

$$C_{(i,:)} \cdot \begin{bmatrix} p_1^T \\ \vdots \\ p_n^T \end{bmatrix} = -\frac{1}{|p_i \times p_{i-1}|} \cdot p_{i-1} + \left(\frac{\cos(\alpha)}{|p_i|^2\sin(\alpha)} + \frac{\cos(\beta)}{|p_i|^2\sin(\beta)}\right) \cdot p_i - \frac{1}{|p_i \times p_{i+1}|} \cdot p_{i+1}.$$

Since we want to argue using the individual coordinates of the four summands, we rotate the unit circle—without loss of generality—so that the point $p_i$ lies on the $x$-axis. This is possible because all values in the equation above involve distances and angles, which remain unchanged under rotation. With this rotation, the coordinate expression

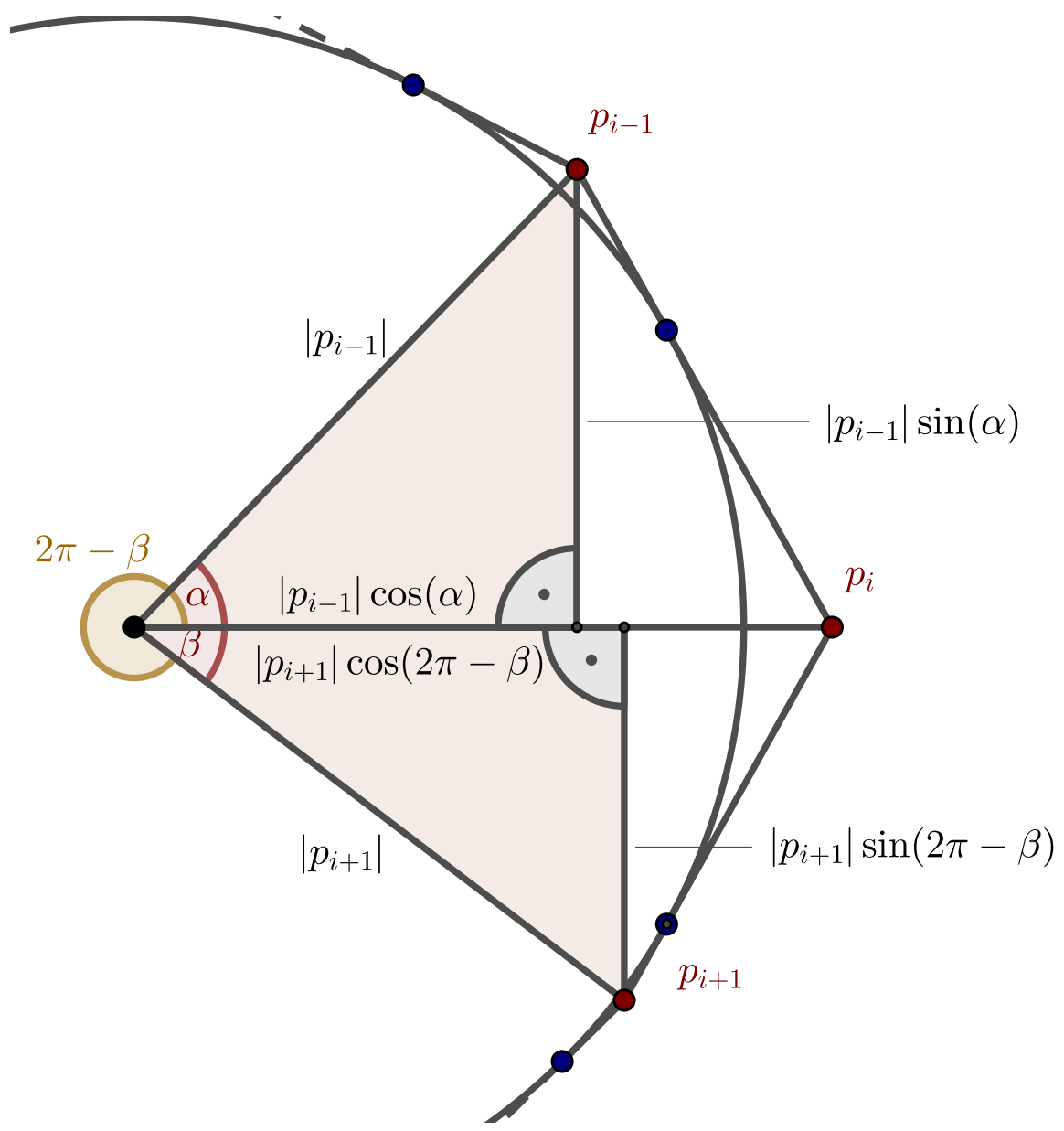

**Figure 4.4:** The 2-polytope, rotated so that $p_i$ lies on the $x$-axis. The respective lengths from which the $x$- and $y$-coordinates of the corresponding points can be derived are indicated.

becomes:

$$C_{(i,:)} \cdot \begin{bmatrix} p_1^T \\ \vdots \\ p_n^T \end{bmatrix} = -\begin{bmatrix} \frac{|p_{i-1}|\cos(\alpha)}{|p_i \times p_{i-1}|} \\ \frac{|p_{i-1}|\sin(\alpha)}{|p_i \times p_{i-1}|} \end{bmatrix}^T + \begin{bmatrix} \frac{\cos(\alpha)}{|p_i|^2 \sin(\alpha)} \cdot |p_i| \\ 0 \end{bmatrix}^T + \begin{bmatrix} \frac{\cos(\beta)}{|p_i|^2 \sin(\beta)} \cdot |p_i| \\ 0 \end{bmatrix}^T - \begin{bmatrix} \frac{|p_{i+1}|\cos(2\pi-\beta)}{|p_i \times p_{i+1}|} \\ \frac{|p_{i+1}|\sin(2\pi-\beta)}{|p_i \times p_{i+1}|} \end{bmatrix}^T .$$

In the last vector, the angle $2\pi - \beta$ appears because the points $p_{i-1}$ and $p_{i+1}$ lie on opposite sides of the $x$-axis. If this were not the case, the origin $\vec{0}$ would not lie inside the polytope. Since we need the oriented angle for the coordinates, we use the exterior angle $2\pi - \beta$ for $p_{i+1}$ instead of the interior angle $\beta$. An illustration of these coordinates is shown in Figure 4.4. Using the identities

$$\cos(2\pi-\beta) = \cos(\beta), \quad \sin(2\pi-\beta) = -\sin(\beta)$$

and the definition of the cross product (which requires the interior angle), we compute:

$$\begin{aligned} C_{(i,:)} \begin{bmatrix} p_1^T \\ \vdots \\ p_n^T \end{bmatrix} &= -\begin{bmatrix} \frac{|p_{i-1}|\cos(\alpha)}{|p_i||p_{i-1}|\sin(\alpha)} \\ \frac{|p_{i-1}|\sin(\alpha)}{|p_i||p_{i-1}|\sin(\alpha)} \end{bmatrix}^T + \begin{bmatrix} \frac{\cos(\alpha)}{|p_i|^2\sin(\alpha)}|p_i| \\ 0 \end{bmatrix}^T + \begin{bmatrix} \frac{\cos(\beta)}{|p_i|^2\sin(\beta)}|p_i| \\ 0 \end{bmatrix}^T - \begin{bmatrix} \frac{|p_{i+1}|\cos(2\pi-\beta)}{|p_i||p_{i+1}|\sin(\beta)} \\ \frac{|p_{i+1}|\sin(2\pi-\beta)}{|p_i||p_{i+1}|\sin(\beta)} \end{bmatrix}^T \\ &= -\begin{bmatrix} \frac{\cos(\alpha)}{|p_i|\sin(\alpha)} \\ \frac{\sin(\alpha)}{|p_i|\sin(\alpha)} \end{bmatrix}^T + \begin{bmatrix} \frac{\cos(\alpha)}{|p_i|\sin(\alpha)} \\ 0 \end{bmatrix}^T + \begin{bmatrix} \frac{\cos(\beta)}{|p_i|\sin(\beta)} \\ 0 \end{bmatrix}^T - \begin{bmatrix} \frac{\cos(\beta)}{|p_i|\sin(\beta)} \\ -\frac{\sin(\beta)}{|p_i|\sin(\beta)} \end{bmatrix}^T \\ &= -\begin{bmatrix} \frac{\cos(\alpha)}{|p_i|\sin(\alpha)} \\ \frac{1}{|p_i|} \end{bmatrix}^T + \begin{bmatrix} \frac{\cos(\alpha)}{|p_i|\sin(\alpha)} \\ 0 \end{bmatrix}^T + \begin{bmatrix} \frac{\cos(\beta)}{|p_i|\sin(\beta)} \\ 0 \end{bmatrix}^T - \begin{bmatrix} \frac{\cos(\beta)}{|p_i|\sin(\beta)} \\ -\frac{1}{|p_i|} \end{bmatrix}^T = \begin{bmatrix} 0 \\ 0 \end{bmatrix}^T . \end{aligned}$$

Hence, the columns of $P$ span the kernel of the matrix $C$. □

The proof above was quite technical, but the theorem gives precisely the special case we need in Chapters 5 and 6. We now add the following result, which we will need for normalizing the Colin-de-Verdière-matrices in the next two chapters:

**Proposition 4.9.** *Let $C$ be a Colin-de-Verdière-matrix as in Theorem 4.8. Then every row sum of $C$ is negative.*

*Proof.* Each row of $C$ has exactly three nonzero entries. Hence, for the $i$-th row, the row sum is

$$\begin{aligned}\sum_j C_{(i,j)} &= -\frac{1}{|p_i \times p_{i-1}|} + \frac{\cos(\alpha)}{|p_i|^2 \sin(\alpha)} + \frac{\cos(\beta)}{|p_i|^2 \sin(\beta)} - \frac{1}{|p_i \times p_{i+1}|} \\ &= -\frac{1}{|p_i \times p_{i-1}|} + \frac{\langle p_{i-1}, p_i \rangle}{\langle p_i, p_i \rangle |p_i \times p_{i-1}|} + \frac{\langle p_i, p_{i+1} \rangle}{\langle p_i, p_i \rangle |p_i \times p_{i+1}|} - \frac{1}{|p_i \times p_{i+1}|} \\ &= \frac{\langle p_{i-1}, p_i \rangle - \langle p_i, p_i \rangle}{\langle p_i, p_i \rangle |p_i \times p_{i-1}|} + \frac{\langle p_i, p_{i+1} \rangle - \langle p_i, p_i \rangle}{\langle p_i, p_i \rangle |p_i \times p_{i+1}|}.\end{aligned}$$

By the same argument as in the proof of Theorem 4.8, we have

$$\langle p_{i-1}, p_i \rangle < \langle p_i, p_i \rangle \quad \text{and} \quad \langle p_i, p_{i+1} \rangle < \langle p_i, p_i \rangle,$$

thus

$$\sum_j C_{(i,j)} < 0.$$

□

We now describe the construction of the Colin-de-Verdière-matrix of the polytope dual to **P**:

**Theorem 4.10.** *Let the notions be as in Theorem 4.8, and let* ***P***** *with vertices $e_1, \dots, e_n$ be the dual polytope to* ***P****. Then the matrix*

$$C_{(i,j)} := \begin{cases} -\frac{1}{|e_i \times e_j|} & \text{if } i \neq j \text{ and } e_i, e_j \text{ share an edge}, \\ 0 & \text{if } i \neq j \text{ and } e_i, e_j \text{ do not share an edge}, \\ \frac{\cos(\alpha)}{|e_i|^2 \sin(\alpha)} + \frac{\cos(\beta)}{|e_i|^2 \sin(\beta)} & \text{if } i = j, \end{cases}$$

*where $\alpha = \sphericalangle(e_{i-1}, e_i)$ and $\beta = \sphericalangle(e_i, e_{i+1})$, is an optimal Colin-de-Verdière-matrix for* ***P*****, and it holds that*

$$C \begin{bmatrix} e_1 \\ \vdots \\ e_n \end{bmatrix} = 0.$$

*Proof.* The proof proceeds analogously to that of Theorem 4.8, so we only highlight the main steps and abbreviate here. The primal polytope, which is the dual of the dual polytope, can be decomposed into the triangles

$$(0, e_i, p_i) \quad \text{and} \quad (0, e_i, p_{i+1}).$$

Varying the edges of the primal polytope by factors $r_i$, the varied edge points are

$$\overline{e}_i = r_i e_i.$$

Setting

$$a := e_{i-1}, \quad b := e_i, \quad c := e_{i+1},$$

the varied vertices are given by

$$\overline{p}_i = \begin{bmatrix} \frac{b_y r_{i-1} - a_y r_i}{a_x b_y - a_y b_x} \\ \frac{a_x r_i - b_x r_{i-1}}{a_x b_y - a_y b_x} \end{bmatrix}, \quad \overline{p}_{i+1} = \begin{bmatrix} \frac{c_y r_i - b_y r_{i+1}}{b_x c_y - b_y c_x} \\ \frac{b_x r_{i+1} - c_x r_i}{b_x c_y - b_y c_x} \end{bmatrix}.$$

The area of the varied primal polytope depending on $r$ is thus

$$\text{vol}_2(r_{i-1}, r_i) = \frac{1}{2} |\overline{e}_{i-1} \times \overline{p}_i| + \frac{1}{2} |\overline{e}_i \times \overline{p}_i|$$

and further

$$\begin{aligned}\mathrm{vol}_2(r_{i-1}, r_i) &= \frac{|r_{i-1}r_i\langle a, a\rangle - r_{i-1}^2\langle a, b\rangle|}{2\langle a, a\rangle|a \times b|} + \frac{|r_i^2\langle a, b\rangle - r_{i-1}r_i\langle b, b\rangle|}{2\langle b, b\rangle|a \times b|} \\ &= \frac{|r_{i-1}r_i - r_{i-1}^2\langle a, b\rangle|}{2|a \times b|} + \frac{|r_i^2\langle a, b\rangle - r_{i-1}r_i|}{2|a \times b|},\end{aligned}$$

and similarly

$$\begin{aligned}\mathrm{vol}_2(r_i, r_{i+1}) &= \frac{1}{2}|\overline{e}_i \times \overline{p}_{i+1}| + \frac{1}{2}|\overline{e}_{i+1} \times \overline{p}_{i+1}| \\ &= \frac{|r_i r_{i+1}\langle b, b\rangle - r_i^2\langle b, c\rangle|}{2\langle b, b\rangle|b \times c|} + \frac{|r_{i+1}^2\langle b, c\rangle - r_i r_{i+1}\langle c, c\rangle|}{2\langle c, c\rangle|b \times c|} \\ &= \frac{|r_i r_{i+1} - r_i^2\langle b, c\rangle|}{2|b \times c|} + \frac{|r_{i+1}^2\langle b, c\rangle - r_i r_{i+1}|}{2|b \times c|}.\end{aligned}$$

Here the simplification is valid since $|a| = |b| = |c| = 1$. Calculating the derivatives yields

$$2|a \times b| \left.\frac{\partial^2 \mathrm{vol}(\mathbf{P}^*(r))}{\partial r_{i-1}\partial r_i}\right|_{r=\mathbf{1}} = \mathrm{sgn}(1 - \cos\alpha) + \mathrm{sgn}(1 - \cos\alpha) = 1,$$

and

$$\left.\frac{\partial^2 \mathrm{vol}(\mathbf{P}^*(r))}{\partial r_i^2}\right|_{r=\mathbf{1}} = \frac{\langle a, b\rangle \mathrm{sgn}(\langle a, b\rangle - 1)}{|a \times b|} - \frac{\langle b, c\rangle \mathrm{sgn}(1 - \langle b, c\rangle)}{|c \times b|} = -\frac{\langle a, b\rangle}{|a \times b|} - \frac{\langle b, c\rangle}{|c \times b|}.$$

Thus the matrix entries of $C$ coincide with those described in the theorem, analogously to the proof of Theorem 4.8. Finally, consider

$$\begin{aligned}C_{(i,:)}\begin{bmatrix} e_1 \\ \vdots \\ e_n \end{bmatrix} &= -\frac{1}{|e_{i-1} \times e_i|}e_{i-1} - \frac{1}{|e_i \times e_{i+1}|}e_{i+1} + \frac{\cos(\alpha)}{|e_i|^2 \sin(\alpha)}e_i + \frac{\cos(\beta)}{|e_i|^2 \sin(\beta)}e_i \\ &= -\frac{e_{i-1}}{\sin(\alpha)} - \frac{e_{i+1}}{\sin(\beta)} + \frac{\cos(\alpha)}{\sin(\alpha)}e_i + \frac{\cos(\beta)}{\sin(\beta)}e_i \\ &= \frac{\cos(\alpha)e_i - e_{i-1}}{\sin(\alpha)} + \frac{\cos(\beta)e_i - e_{i+1}}{\sin(\beta)}.\end{aligned}$$

The vectors in the two summands point in opposite directions and have the same length (cf. Figure 4.4), hence

$$C_{(i,:)}\begin{bmatrix} e_1 \\ \vdots \\ e_n \end{bmatrix} = \vec{0}.$$

□

We also add the following theorem here:

**Proposition 4.11.** *The row sums of the matrix $C$ constructed as in Theorem 4.10 are all negative.*

*Proof.* We compute

$$-\frac{1}{|e_{i-1} \times e_i|} - \frac{1}{|e_i \times e_{i+1}|} + \frac{\cos(\alpha)}{|e_i|^2 \sin(\alpha)} + \frac{\cos(\beta)}{|e_i|^2 \sin(\beta)} = \frac{\cos(\alpha) - 1}{\sin(\alpha)} + \frac{\cos(\beta) - 1}{\sin(\beta)}.$$

Since the angles $\alpha$ and $\beta$ lie in $(0, \pi)$, the above expression is negative, proving the claim. □

In the next section, we finally consider the construction of Colin-de-Verdière-matrices for 3-polytopes.

## 4.4 Colin-de-Verdière-Matrices for 3-Polytopes

For the construction of Colin-de-Verdière-matrices for 3-polytopes, we follow the approach by Lovász [Lov01]. Theorem 4.7 by Izmestiev is formulated for general $d$-polytopes in [Izm10] and could also be used. Both constructions yield the same matrices (a proof can be found in [Izm10]). We choose Lovász's construction because it is more geometric and thus easier to understand. We start with the following definition [Lov01, p. 226]:

**Definition 4.12.** *A Colin-de-Verdière-matrix $C$ of a 3-connected planar graph **G** is called* appropriately scaled *if the vectors spanning the kernel of $C$ are the vertices of a realization of **G** as a convex polytope **P**.*

We summarize [Lov01, Section 5, pp. 228–231] and [Lov01, Thm. 6, p. 229] in the following theorem:

**Theorem 4.13.** *Let **G** be a 3-connected planar graph and let $\mathbf{P} \subset \mathbb{R}^3$ be a realization of the graph as a convex 3-polytope containing the origin. Let $F$ be the list of $m$ facets of **P**, and let $\mathbf{P}^*$ be the polar dual polytope with vertices $f_1, \ldots, f_m$. Furthermore, let $P$ be the matrix of vertices of **P** with rows $p_1 := P_{(1,:)}, \ldots, p_n := P_{(n,:)}$. Then the matrix*

$$C := \begin{cases} -\frac{|f_k - f_l|}{|p_i \times p_j|} & \text{for } i \neq j \text{ and } p_i \text{ and } p_j \text{ share an edge} \\ & \text{with } f_k, f_l \text{ as the dual points of the primal facets laying on the edge } \{p_i, p_j\} \\ 0 & \text{for } i \neq j \text{ and } p_i \text{ and } p_j \text{ do not share an edge} \\ -\frac{\left\langle p_i, \sum_{j, j \neq i} C_{(i,j)} p_j \right\rangle}{\langle p_i, p_i \rangle} & \text{for } i = j \end{cases}$$

*is an optimal Colin-de-Verdière-matrix of the graph **G**, and it holds that $CP = 0$. Hence, $C$ is appropriately scaled.*

*Proof.* The proof that $C$ is a Colin-de-Verdière-matrix can be found in [Lov01, pp. 229–230] and will not be repeated here. However, we describe the motivation behind the construction and, in particular, the reasoning for $CP = 0$. All steps given here are taken from [Lov01].

First, note that **P** does not necessarily have edges tangent to the unit sphere.

Suppose that for $i \neq j$ there is an edge between $p_i$ and $p_j$ which is adjacent to the facets corresponding to $f_k$ and $f_l$. Then, by the definition of the polar dual polytope, we have

$$\langle p_i, f_k \rangle = \langle p_i, f_l \rangle = \langle p_j, f_k \rangle = \langle p_j, f_l \rangle = 1,$$

since the dual points $f_k$ and $f_l$ lie on the dual facets generated by $p_i$ and $p_j$. Thus,

$$0 = \langle p_i, f_k \rangle - \langle p_i, f_l \rangle = \langle p_i, f_k - f_l \rangle, \quad \text{and} \quad 0 = \langle p_j, f_k \rangle - \langle p_j, f_l \rangle = \langle p_j, f_k - f_l \rangle.$$

The vector $f_k - f_l$ is therefore orthogonal to both $p_i$ and $p_j$, and thus (up to sign) parallel to $p_i \times p_j$. See Figure 4.5 for an illustration. Therefore, following [Lov01], we set

$$f_k - f_l = C_{(i,j)}(p_i \times p_j).$$

The value $C_{(i,j)}$ can be determined by comparing the lengths of $f_k - f_l$ and $p_i \times p_j$. The sign depends on the order of $p_i$ and $p_j$; we choose the order so that $C_{(i,j)}$ is negative. Hence,

$$C_{(i,j)} = -\frac{|f_k - f_l|}{|p_i \times p_j|}.$$

For $i \neq j$ with no edge between $p_i$ and $p_j$, set

$$C_{(i,j)} = 0.$$

Next, consider the diagonal entries $i = j$. Look at the cross product

$$\left( \sum_{j, j \neq i} C_{(i,j)} p_j \right) \times p_i = \sum_{j, j \neq i} C_{(i,j)} (p_j \times p_i) = \sum_{(k,l)} f_k - f_l,$$

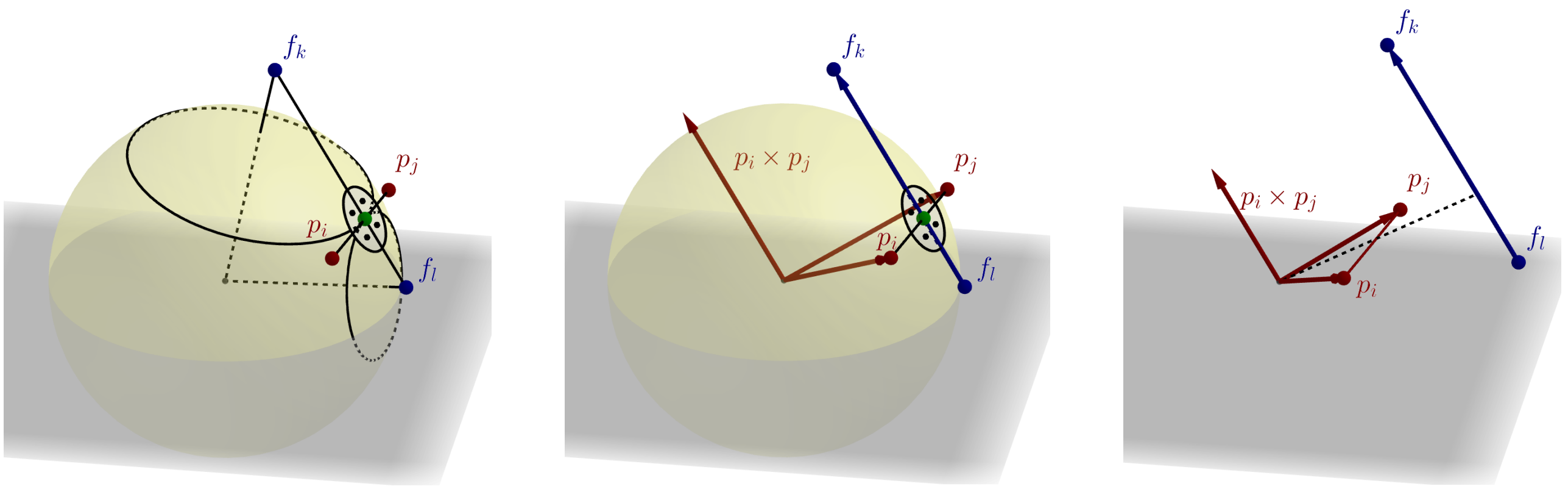

**Figure 4.5:** Orthogonality of primal and dual edges (left), parallel directions of vectors $f_k - f_l$ and $p_i \times p_j$ (center), and analogous picture for a polytope whose edges are not tangent to the unit sphere (right).

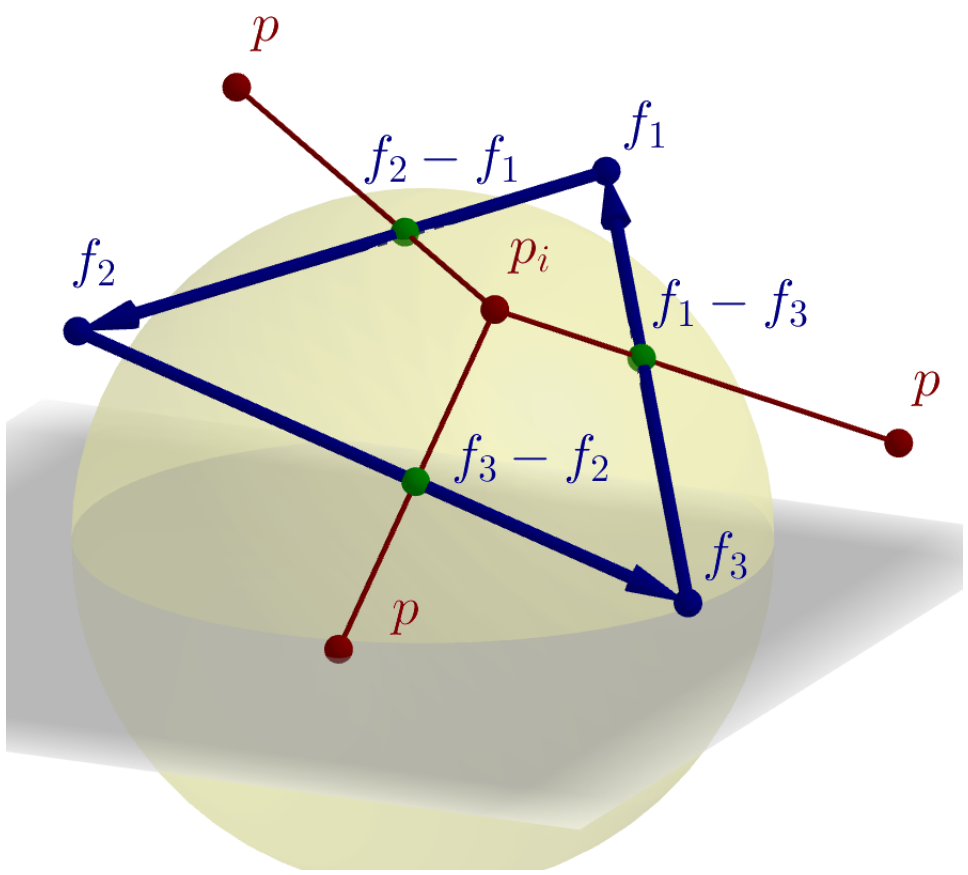

**Figure 4.6:** A primal vertex $p_i$ with three adjacent primal vertices. Each primal edge corresponds to a dual edge, so each dual vertex on these edges is counted twice with opposite signs.

where the last sum is over all edges of the facet dual to $p_i$. Fixing $p_i$ in the cross product induces a cyclic order on the vectors $f_k - f_l$, so each dual vertex is counted twice with opposite signs. Hence,

$$\left( \sum_{j, j \neq i} C_{(i,j)} p_j \right) \times p_i = \sum_{k,l} f_k - f_l = 0. \tag{4.3}$$

See Figure 4.6 for illustration. This shows that the vector $\sum_{j,\, j \neq i} C_{(i,j)} p_j$ points in the same direction as $p_i$, meaning the two vectors are parallel. According to [Lov01], we then obtain the following equation:

$$\sum_{j,\, j \neq i} C_{(i,j)} p_j = -C_{(i,i)} p_i, \tag{4.4}$$

which we can solve for $C_{(i,i)}$ as follows:

$$\begin{aligned} \sum_{j,\, j \neq i} C_{(i,j)} p_j = -C_{(i,i)} p_i \quad &\Leftrightarrow \quad p_i^T \sum_{j,\, j \neq i} C_{(i,j)} p_j = -p_i^T C_{(i,i)} p_i \\ &\Leftrightarrow \quad C_{(i,i)} = -\frac{\left\langle p_i, \sum_{j,\, j \neq i} C_{(i,j)} p_j \right\rangle}{\langle p_i, p_i \rangle}. \end{aligned} \tag{4.5}$$

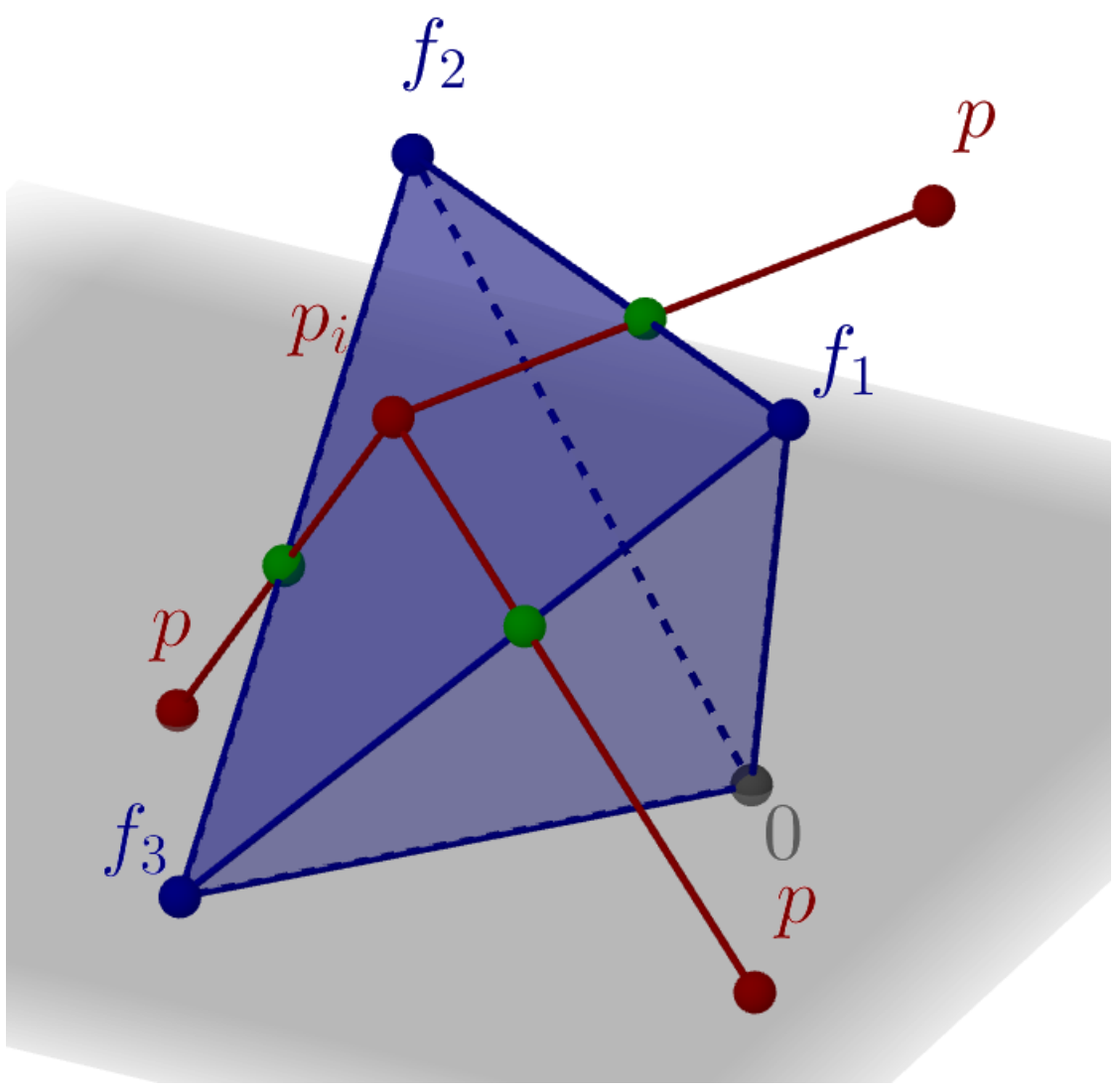

**Figure 4.7:** A primal vertex $p_i$ and the cone spanned by the corresponding dual facet and the origin.

This allows us to explicitly determine all entries of the matrix $C$. From Equation (4.4), we also directly obtain:

$$\sum_{j,\, j \neq i} C_{(i,j)} p_j = -C_{(i,i)} p_i \quad \Leftrightarrow \quad \sum_j C_{(i,j)} p_j = 0.$$

Therefore, we have $CP = 0$, and the columns of the matrix $P$, whose rows represent the vertices of the polytope, span the kernel of $C$. Since Theorem 4.4 tells us that the dimension of the kernel can be at most 3, the matrix $C$ is optimal. Moreover, because the columns of $P$ are linearly independent, the matrix $C$ is also properly scaled. □

Thus, we have described the construction of Colin-de-Verdière-matrices for all dimensions relevant to this work. We conclude this chapter with the following statement about Colin-de-Verdière-matrices of 3-polytopes, which we will need for the proofs in Chapters 5 and 6 [Lov01, Eq. 11, p. 234]:

**Theorem 4.14.** *Let $\boldsymbol{G} = (\boldsymbol{V}, \boldsymbol{E})$ be a 3-connected planar graph with $|\boldsymbol{V}| = n$. Let $\boldsymbol{P}$ be a realization of $\boldsymbol{G}$ containing the origin, and let $\boldsymbol{P}^*$ be its polar dual polytope. Furthermore, let $C$ be the Colin-de-Verdière-matrix constructed from $\boldsymbol{P}$ and $\boldsymbol{P}^*$ according to Theorem 4.13. Let $a_i$ denote the volume of the cone formed by the convex hull of the dual facet corresponding to $p_i$ and the origin. Then, for all $i \in \{1, \ldots, n\}$, it holds that*

$$\sum_j C_{(i,j)} = -6a_i.$$

We do not provide a proof here, but an illustration of the cone can be found in Figure 4.7.

In this chapter, we have gathered the theory of Colin-de-Verdière-matrices. We have worked out the construction for the two trivial cases and detailed the construction of Colin-de-Verdière-matrices for two-dimensional polytopes in the needed special cases. For 3-polytopes, we described the known construction from the literature. We will use these results for constructing subdivision matrices in the next chapter.

# 5 Generalized Quadratic B-Spline Subdivision

In this chapter, we consider the construction and properties of generalized quadratic B-spline subdivision algorithms, thus the case $g = 2$. Together, Chapters 5 and 6 form the main part of this work.

In Section 5.1, we explain the origin of generalized quadratic B-spline subdivision algorithms and briefly discuss the case $t = 1$. For the case $t = 2$, we describe the subdivision matrix for a regular initial element and discuss the Doo-Sabin algorithm [DS78], which is the established subdivision algorithm in the literature for this case. In particular, it provides refinement rules for irregular initial elements that we examine more closely. For $t = 3$, no construction of subdivision algorithms could be found in the literature, so in Section 5.1 we only describe the refinement rules for the regular case.

Section 5.2 then deals with the actual construction of our subdivision matrices for the cases $t = 2$ and $t = 3$. For this, we use the results from Chapters 3 and 4. We specifically present two different variants designed for different applications. The case $t = 3$ is particularly interesting here, since, as mentioned, no proposal for possible refinement rules seems to exist yet.

In Section 5.3, we address the quality criteria from Chapter 2. In particular, we show for which of our variants which criteria hold. It will not be possible to prove or disprove all criteria for all variants; therefore, we support the analysis with an empirical study. Specifically, we numerically test the 305,095 examples from Section 3.1.1 on the criteria for which this is possible.

The goal of this chapter is therefore, besides constructing the subdivision algorithms, also to make explicit statements about them in order to give a recommendation for their use.

## 5.1 Literature Overview and Refinement Rules for Regular Initial Elements

In this section, we briefly review the history and the currently known subdivision algorithms and refinements found in the literature.

We start with the one-dimensional case. The first subdivision algorithm for quadratic B-splines was developed by Chaikin in [Cha74]. The proposed refinement rules were initially independent of the idea of refining quadratic B-splines. The connection between the refinement rules and the refinement of quadratic B-splines was proven by Riesenfeld in [Rie75]. This can be considered the starting point of subdivision, as the idea of refining the basis of space curves with simple refinement rules was extended from these two works to other cases $t \in \{2, 3\}$ and $g = 3$.

The original refinement rules require three control points. However, they can be transformed so that only two control points are needed. Thus, according to [Cha74], for the case $t = 1$ and $g = 2$, we obtain the subdivision matrix

$$S^{(1)} := \frac{1}{4} \begin{bmatrix} 3 & 1 \\ 1 & 3 \end{bmatrix}.$$

An illustration of the refinement can be found in Figure 5.1.

Doo and Sabin generalized these rules in [DS78] to the bivariate case ($t = 2$) in order to construct quadratic B-spline surfaces. For the regular case, which is the tensor product of the $[3/4, 1/4]$-rule with itself, the corresponding

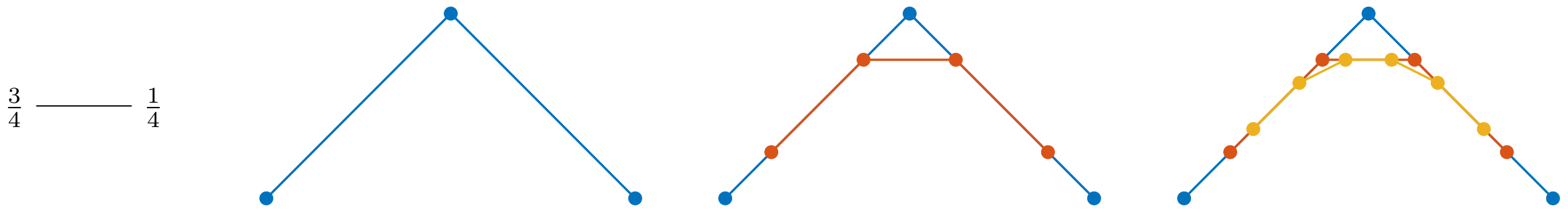

**Figure 5.1:** One of the two refinement rules for $t = 1$ (left) and an example configuration with two initial elements (second from left). Refining both initial elements produces four control points and three initial elements (second from right). Refining once more produces six control points and five initial elements (right).

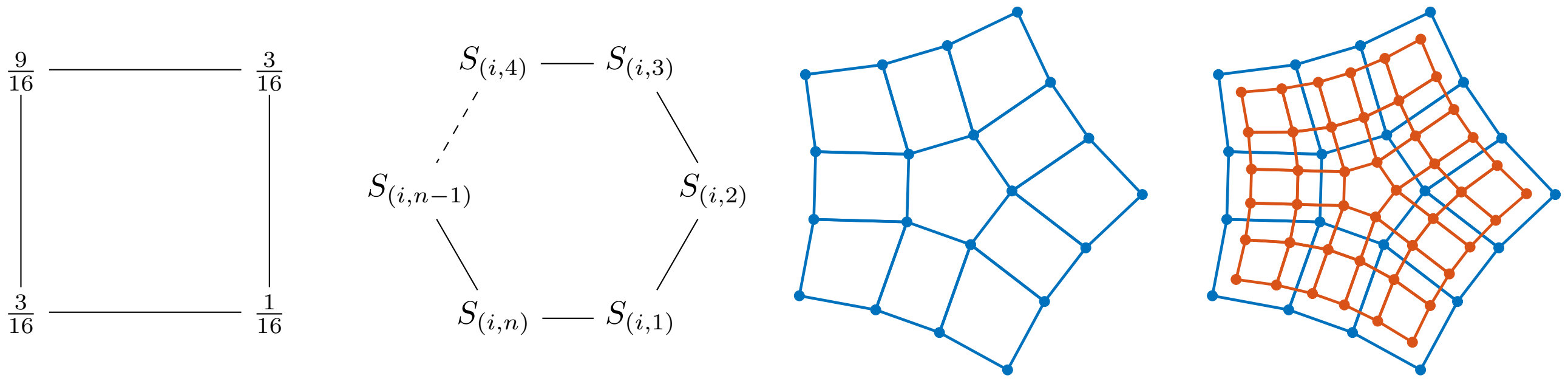

**Figure 5.2:** One of the four regular refinement rules (left) and one of the $n$ irregular refinement rules (second from left) for $t = 2$ and an example configuration with eleven initial elements (second from right). Refining these eleven initial elements produces 31 new initial elements (right).

matrix is

$$S^{(2)} := \operatorname{kron}\left(\frac{1}{4}\begin{bmatrix}3 & 1\\ 1 & 3\end{bmatrix}, \frac{1}{4}\begin{bmatrix}3 & 1\\ 1 & 3\end{bmatrix}\right) = \frac{1}{16}\begin{bmatrix}9 & 3 & 3 & 1\\ 3 & 9 & 1 & 3\\ 3 & 1 & 9 & 3\\ 1 & 3 & 3 & 9\end{bmatrix}.$$

Doo and Sabin also gave rules for the irregular case, that is, for general $n$-gons. For this, all vertices are first numbered counterclockwise. The subdivision matrices are then given by [DS78, p. 360] as

$$S_{(i,j)} := \frac{\delta_{i,j}}{4} + \frac{3 + 2\cos\left(\frac{2\pi(i-j)}{n}\right)}{4n} \quad \text{with} \quad \delta_{i,j} = \begin{cases}1, & \text{if } i = j\\ 0, & \text{otherwise}\end{cases}, \tag{5.1}$$

where $\delta_{i,j}$ is the *Kronecker delta*. An illustration of the regular and irregular refinement rules as well as an example refinement is shown in Figure 5.2.

The Doo-Sabin algorithm satisfies all quality criteria stated in this work. Many of the proofs can be found in [PR08]. In particular, the subdivision matrices from [DS78] fulfill quality criterion Q13 and thus construct surfaces that are $C^1$-smooth at the central points (see [PR08, Thm. 6.2, p. 119]). Moreover, the matrices are chosen such that their eigenvalues for the case of initial elements are

$$1, \quad \frac{1}{2}, \quad \frac{1}{2}, \quad \frac{1}{4}, \quad \dots, \quad \frac{1}{4}.$$

For $t = 3$ we obtain for regular initial elements the subdivision matrix

$$S^{(3)} := \operatorname{kron}\left(\frac{1}{4}\begin{bmatrix}3 & 1\\ 1 & 3\end{bmatrix}, \operatorname{kron}\left(\frac{1}{4}\begin{bmatrix}3 & 1\\ 1 & 3\end{bmatrix}, \frac{1}{4}\begin{bmatrix}3 & 1\\ 1 & 3\end{bmatrix}\right)\right)$$

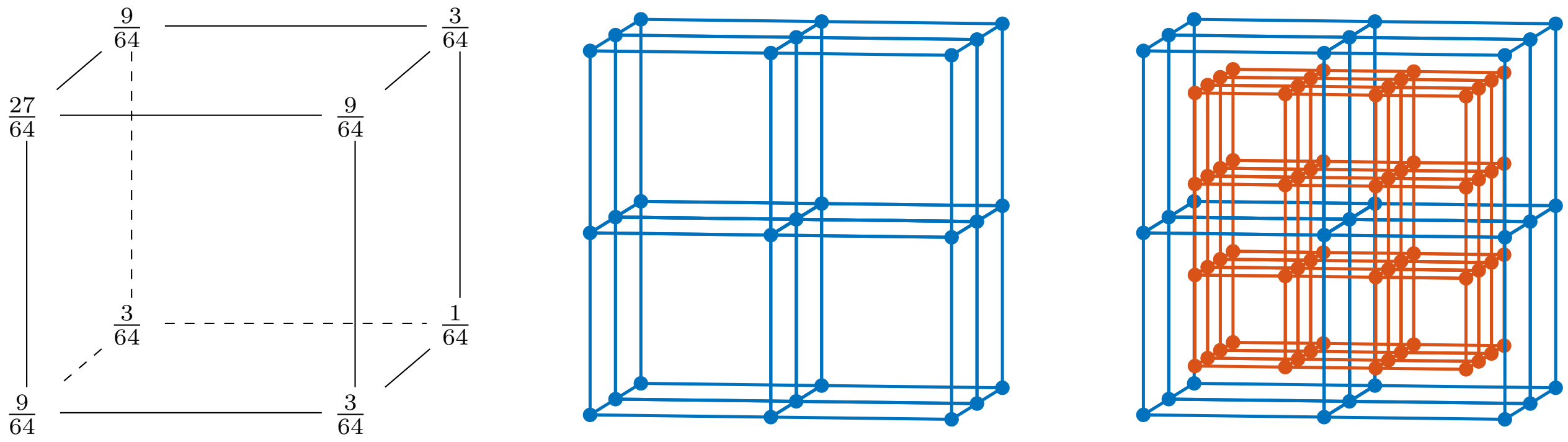

**Figure 5.3:** One of the eight regular refinement rules (left) and an example configuration with eight initial elements and 27 control points (middle). Refining the eight initial elements produces 27 initial elements and 64 control points (right).

and further

$$S^{(3)} = \text{kron}\left(\frac{1}{4}\begin{bmatrix}3 & 1\\ 1 & 3\end{bmatrix}, \frac{1}{16}\begin{bmatrix}9 & 3 & 3 & 1\\ 3 & 9 & 1 & 3\\ 3 & 1 & 9 & 3\\ 1 & 3 & 3 & 9\end{bmatrix}\right) = \frac{1}{64}\begin{bmatrix}27 & 9 & 9 & 3 & 9 & 3 & 3 & 1\\ 9 & 27 & 3 & 9 & 3 & 9 & 1 & 3\\ 9 & 3 & 27 & 9 & 3 & 1 & 9 & 3\\ 3 & 9 & 9 & 27 & 1 & 3 & 3 & 9\\ 9 & 3 & 3 & 1 & 27 & 9 & 9 & 3\\ 3 & 9 & 1 & 3 & 9 & 27 & 3 & 9\\ 3 & 1 & 9 & 3 & 9 & 3 & 27 & 9\\ 1 & 3 & 3 & 9 & 3 & 9 & 9 & 27\end{bmatrix}.$$

An illustration of a refinement rule and an example refinement can be found in Figure 5.3. With this, we have described all regular matrices for the case $g = 2$. These are the matrices of Definition 2.15 that enable the regular refinement and thus also the regular evaluation of quadratic B-spline elements.

For the irregular combinatorial configurations of type $t = 3$, to our knowledge, no refinement rules are known. Of course, for the semi-irregular case, it is possible to extend the two-dimensional Doo-Sabin rules to three dimensions by taking the tensor product with the one-dimensional case. However, the challenge lies in developing refinement rules for the other combinatorial configurations.

The goal of the next section is to generate a subdivision matrix for an arbitrary admissible three-dimensional combinatorial configuration. We will construct rules both for the two-dimensional and the three-dimensional cases.

## 5.2 Construction of the Subdivision Matrices

In this section, we describe the construction of subdivision matrices for initial elements. Since there are no overlapping initial elements for $g = 2$, the subdivision matrices of larger structures are composed of the subdivision matrices constructed here for the initial elements.

We construct two variants in this section. The first variant leads to a subdivision matrix for which we can determine the eigenvalues exactly. However, since this variant sacrifices some quality criteria, we construct a second variant that relinquishes control over the subsubdominant eigenvalue but satisfies the other criteria. Depending on the application, one or the other variant may be preferable. Before we begin the construction, the following remark is to be noted:

**Remark 5.1.** *The strategies described in this section regarding the construction of subdivision matrices are quite unusual. In common approaches in the literature, the strategy is mostly to give explicit rules for the matrix entries. This has the* *decisive advantage of simplicity. From a user's perspective, one would focus on the explicit formulas as these are easy to implement.*

*In our methods, we do not provide explicit formulas but rather describe an indirect construction of the subdivision matrices. First, we select the convex polytopes from Chapter 3 as bases of the subdominant eigenspace of the subdivision matrix and then construct a matrix with this eigenspace.*

*In Chapter 3 we have seen that constructing convex polytopes involves some (programming) effort. To compensate for this drawback, we provide in [Die25] a library containing a Matlab implementation of all algorithms described in this work. This can be used either as a blackbox or transferred to other programming languages.*

### 5.2.1 Variant 1: Exact Spectrum

The construction of the subdivision matrix of the first variant for an initial element can be found in Algorithm 10. All matrices and variables occurring in Section 5.2.1 refer to those of Algorithm 10. We describe the individual steps as follows:

**Algorithm 10:** GeneralizedQuadraticB-SplineSubdivisionMatrixVariant1

**Data:** valid combinatorial graph **G** of an initial element in the input form of Definition 3.10, type $t \in \{2,3\}$, subsubdominant eigenvalue $\mu$ with $0 \leq \mu < \frac{1}{2}$

**Result:** subdivision matrix $S_1^{(t)}$

1 **begin**
2   **if** $t == 2$ *or* **G** *is not a combinatorial graph of a prism* **then**
3     $P =$ (coordinates of the $t$-polytope corresponding to **G** constructed according to Chapter 3) with $P \in \mathbb{R}^{n \times t}$;
4     **if** $t == 3$ **then**
5       $P' = P - \frac{1}{n}\sum_{i=1}^{n} P_{(i,:)}$;
6       $\tilde{P} = \mathtt{orth}(P')$;
7     **else**
8       $\tilde{P} = P$;
9     $S_1^{(t)} := \left(\frac{\vec{1}\cdot\vec{1}^T}{|\vec{1}|^2} + \frac{1-2\mu}{2-2\mu}\left(\frac{\tilde{P}_{(:,1)}\cdot\tilde{P}_{(:,1)}^T}{|\tilde{P}_{(:,1)}|^2} + \cdots + \frac{\tilde{P}_{(:,t)}\cdot\tilde{P}_{(:,t)}^T}{|\tilde{P}_{(:,t)}|^2}\right)\right)(1-\mu) + \mu E_n$;
10   **else**
11     $n =$ number of corners of the base of the prism;
12     $S^{(2)} :=$ Doo-Sabin matrix for $n$ corners according to formula (5.1);
13     $S^{(1)} := \frac{1}{4}\begin{bmatrix} 3 & 1 \\ 1 & 3 \end{bmatrix}$;
14     $S_1^{(t)} := \mathrm{kron}(S^{(1)}, S^{(2)})$;

**Input and Convex Polytope** We start with a concrete initial element. By Definition 2.51, the initial element is valid if its combinatorial graph is a cycle for $t = 2$ and a planar 3-connected graph for $t = 3$. We assume this validity and take as input the combinatorial graph of the initial element.

For $t = 2$ this yields the structure of a cycle (in the graph-theoretical sense) with $n$ vertices and $n$ edges, and we construct, as described in Section 3.4, a convex 2-polytope as a regular $n$-gon with control points $P \in \mathbb{R}^{n \times 2}$.

For $t = 3$ we obtain a planar 3-connected graph and select the same input options as in Definition 3.10. Using the procedure described in Section 3.3, we then construct a 3-polytope with the properties given in Theorem 3.25.

**Correction of $P$** For this procedure, we require a polytope whose vertex centroid is at the origin. For $t = 2$, this is already the case. Summing the entries from Equation (3.16) yields that each entry

$$\frac{\sin\left(\frac{2\pi j}{n}\right)}{\sin\left(\frac{2\pi}{n}\right)} \quad \text{respectively} \quad \frac{\cos\left(\frac{2\pi j}{n}\right)}{\sin\left(\frac{2\pi}{n}\right)}$$

appears once positively and once negatively (with $j$ taken modulo $n$). Hence, all coordinate entries sum to zero. Thus, the centering correction step is only necessary for $t = 3$. Here, the 3-polytope constructed in Chapter 3 is chosen such that the edge centroid lies at the origin. Therefore, the polytope might have to be shifted to place the vertex centroid at the origin. We accomplish this by

$$P' := P - \frac{\sum_{i=1}^n P_{(i,:)}}{n} := P - \frac{1}{n}\begin{bmatrix}\sum_{i=1}^n P_{(i,:)} \\ \vdots \\ \sum_{i=1}^n P_{(i,:)}\end{bmatrix}. \tag{5.2}$$

The columns of $P'$ shall be contained in the eigenspace of the subdivision matrix corresponding to the eigenvalue $\frac{1}{2}$. Since the coordinate vectors are linearly independent, they span a $t$-dimensional subspace of $\mathbb{R}^n$. For this subspace, we require an orthogonal basis in a second step. Orthogonality is an essential part of the proof of Theorem 5.15. Without it, $S_1^{(t)}$ neither has the desired spectrum nor the desired eigenvectors.

For $t = 2$, the two coordinate vectors are already orthogonal. This follows geometrically from their construction. Since $(1, 0)$ is a tangent point of the 2-polytope, we obtain a tangent perpendicular to the $x$-axis, on which the two control points $P_{(1,:)}$ and $P_{(n,:)}$ lie. By symmetry, both have the same $x$-coordinate and opposite $y$-coordinates. Hence,

$$P_{(1,1)}P_{(1,2)} + P_{(n,1)}P_{(n,2)} = xy + x(-y) = 0.$$

By symmetry, this continues for the other pairs $P_{(2,:)}$ and $P_{(n-1,:)}$, $P_{(3,:)}$ and $P_{(n-2,:)}$, etc. If the number of control points is odd, one control point lies on the $x$-axis. Since the corresponding $y$-coordinate is zero, this part of the scalar product is also zero, and we obtain

$$\langle P_{(:,1)}, P_{(:,2)}\rangle = 0.$$

Geometrically, this relation can be seen exemplarily in Figure 3.16. Therefore, we define

$$\tilde{P} := P' := P \in \mathbb{R}^{n\times 2}.$$

For the convex 3-polytope, however, orthogonality does not hold. The coordinate vectors need not be orthogonal, and concrete examples can be constructed where they are not. Since they span a $t$-dimensional subspace of $\mathbb{R}^n$, we can construct an orthogonal basis for this subspace. We use the Matlab function `orth` (see [Mate]) and define

$$\tilde{P} := \texttt{orth}(P') \in \mathbb{R}^{n\times 3}.$$

Two remarks are in order. The `orth` command even returns an orthonormal basis. To clarify the process and keep consistency with the case $t = 2$, the normalization of the columns of $P$ is explicitly included in the formula for $S_1^{(t)}$ in Algorithm 10. Furthermore, we treat the `orth` command as a black box. A possible procedure to generate an orthonormal basis would be the Gram–Schmidt orthonormalization process from Theorem B.13.

Since both the columns of $P'$ and $\tilde{P}$ are linearly independent and span the same space, the `orth` command corresponds to a basis change given by

$$\tilde{P} = P'B \quad \text{with} \quad B \in \mathbb{R}^{t\times t}, \quad B \text{ invertible}, \tag{5.3}$$

and the rows of $\tilde{P}$ sum, just like those of $P'$, to zero, because for the shifted $P'$ it holds that

$$\sum_{i=1}^n \tilde{P}_{(i,j)} = \sum_{i=1}^n B_{(1,j)}P'_{(i,1)} + B_{(2,j)}P'_{(i,2)} + B_{(3,j)}P'_{(i,3)} = B_{(1,j)}\sum_{i=1}^n P'_{(i,1)} + B_{(2,j)}\sum_{i=1}^n P'_{(i,2)} + B_{(3,j)}\sum_{i=1}^n P'_{(i,3)} = 0,$$

for $j \in \{1, 2, 3\}$.

**Construction of the subdivision matrix** In the final step, we construct a matrix from the orthogonal coordinate vectors. This is given concretely by

$$S_1^{(t)} := \left( \frac{\vec{1} \cdot \vec{1}^T}{|\vec{1}|^2} + \frac{1-2\mu}{2-2\mu} \left( \frac{\tilde{P}_{(:,1)} \cdot \tilde{P}_{(:,1)}^T}{|\tilde{P}_{(:,1)}|^2} + \cdots + \frac{\tilde{P}_{(:,t)} \cdot \tilde{P}_{(:,t)}^T}{|\tilde{P}_{(:,t)}|^2} \right) \right) (1-\mu) + \mu E_n. \tag{5.4}$$

We have already proposed this variant in a modified form in [Die+23, p. 148]. In Theorem 5.15, we will show that the matrix $S_1^{(t)}$ from Equation (5.4) has the eigenvalues

$$1, \quad \underbrace{\frac{1}{2}, \quad \cdots, \quad \frac{1}{2}}_{t \text{ times}}, \quad \mu, \quad \cdots, \quad \mu.$$

Equation (5.4) also has a direct connection to the Doo-Sabin algorithm, which we describe in the next lemma and corollary:

**Lemma 5.2.** *For $t = 2$, $n \geq 3$, and $0 < \mu < 1/2$, the entries of the matrix $S_1^{(2)} \in \mathbb{R}^{n \times n}$ are given by*

$$\left(S_1^{(2)}\right)_{(i,j)} = (1-\mu) \frac{2 - 2\mu + (2-4\mu)\cos\left(\frac{2\pi(j-i)}{n}\right)}{n(2-2\mu)} + \mu\delta_{i,j}.$$

*Proof.* We consider the elements of Formula (5.4) for $\left(S_1^{(2)}\right)_{(i,j)}$ individually. First, we obtain

$$\frac{\vec{1} \cdot \vec{1}^T}{|\vec{1}|^2} = \frac{1}{n} \begin{bmatrix} 1 & \cdots & 1 \\ \vdots & & \vdots \\ 1 & \cdots & 1 \end{bmatrix}.$$

For the squared lengths of the coordinate vectors, using a computer algebra system, we find

$$\begin{aligned} \left|\tilde{P}_{(:,1)}\right|^2 &= \frac{1}{\sin^2\left(\frac{2\pi}{n}\right)} \sum_{i=1}^{n} \left( \sin\left(\frac{2\pi i}{n}\right) - \sin\left(\frac{2\pi(i+1)}{n}\right) \right)^2 \\ = \left|\tilde{P}_{(:,2)}\right|^2 &= \frac{1}{\sin^2\left(\frac{2\pi}{n}\right)} \sum_{i=1}^{n} \left( \cos\left(\frac{2\pi i}{n}\right) - \cos\left(\frac{2\pi(i+1)}{n}\right) \right)^2 = \frac{2n\sin^2\left(\frac{\pi}{n}\right)}{\sin^2\left(\frac{2\pi}{n}\right)}. \end{aligned} \tag{5.5}$$

Furthermore, for the product of the vectors, again with a computer algebra system, we have

$$\begin{aligned} \tilde{P}_{(i,1)} \cdot \tilde{P}_{(j,1)}^T + \tilde{P}_{(i,2)} \cdot \tilde{P}_{(j,2)}^T &= \frac{1}{\sin^2\left(\frac{2\pi}{n}\right)} \left( \sin\left(\frac{2\pi i}{n}\right) - \sin\left(\frac{2\pi(i+1)}{n}\right) \right) \cdot \left( \sin\left(\frac{2\pi j}{n}\right) - \sin\left(\frac{2\pi(j+1)}{n}\right) \right) \\ &+ \frac{1}{\sin^2\left(\frac{2\pi}{n}\right)} \left( \cos\left(\frac{2\pi i}{n}\right) - \cos\left(\frac{2\pi(i+1)}{n}\right) \right) \cdot \left( \cos\left(\frac{2\pi j}{n}\right) - \cos\left(\frac{2\pi(j+1)}{n}\right) \right) \\ &= \frac{1}{\sin^2\left(\frac{2\pi}{n}\right)} 4\sin^2\left(\frac{\pi}{n}\right) \cos\left(\frac{2\pi(j-i)}{n}\right) \end{aligned}$$

With normalization, this yields

$$\frac{\tilde{P}_{(i,1)} \cdot \tilde{P}_{(j,1)}^T}{|\tilde{P}_{(:,1)}|^2} + \frac{\tilde{P}_{(i,2)} \cdot \tilde{P}_{(j,2)}^T}{|\tilde{P}_{(:,2)}|^2} = \frac{2}{n} \cos\left(\frac{2\pi(j-i)}{n}\right).$$

The last element is trivially

$$(E_n)_{(i,j)} = \delta_{i,j}.$$

Inserting the components into Formula (5.4) yields

$$\left(S_1^{(2)}\right)_{(i,j)} = (1-\mu)\cdot\frac{1}{n} + (1-\mu)\cdot\frac{1-2\mu}{2-2\mu}\cdot\frac{2}{n}\cos\left(\frac{2\pi(j-i)}{n}\right) + \mu\delta_{i,j}$$
$$= (1-\mu)\frac{2-2\mu+(2-4\mu)\cos\left(\frac{2\pi(j-i)}{n}\right)}{n(2-2\mu)} + \mu\delta_{i,j}.$$

□

Thus, we immediately obtain the following corollary:

**Corollary 5.3.** *Let $t = 2$, $n \geq 3$, and $\mu = \frac{1}{4}$. Then Algorithm 10 with the input adjacency matrix of a graph-theoretic circle with $n$ vertices generates the rules of the Doo-Sabin algorithm given by Formula* (5.1).

*Proof.* Setting $\mu = \frac{1}{4}$ in the above lemma yields

$$\left(S_1^{(2)}\right)_{(i,j)} = \frac{3}{4}\cdot\frac{1}{n} + \frac{3}{4}\cdot\frac{1-\frac{2}{4}}{2-\frac{2}{4}}\cdot\frac{2\cos\left(\frac{2\pi(j-i)}{n}\right)}{n} + \frac{1}{4}\delta_{i,j} = \frac{3+2\cos\left(\frac{2\pi(j-i)}{n}\right)}{4n} + \frac{\delta_{i,j}}{4},$$

which corresponds exactly to Formula (5.1). □

Concerning the eigenvalues, we add the following remark:

**Remark 5.4.** *As explained in the section on Quality Criterion Q5, a subdominant eigenvalue of $\frac{1}{2}$ is meaningful for several reasons. For certain applications, it can be beneficial to have the subdominant eigenvalue as small as possible. A smaller subdominant eigenvalue causes the irregular region to shrink faster, since the surface or volume around the irregularity contracts more quickly (see Figure 2.24, left column).*

*If one wishes to choose the subdominant eigenvalue $\lambda$ freely in this construction, the above formula* (5.4) *must be modified to*

$$S_1^{(t)} := \left(\frac{\vec{1}\cdot\vec{1}^T}{|\vec{1}|^2} + \frac{\lambda-\mu}{1-\mu}\left(\frac{\tilde{P}_{(:,1)}\cdot\tilde{P}_{(:,1)}^T}{|\tilde{P}_{(:,1)}|^2} + \cdots + \frac{\tilde{P}_{(:,t)}\cdot\tilde{P}_{(:,t)}^T}{|\tilde{P}_{(:,t)}|^2}\right)\right)(1-\mu) + \mu E_n.$$

*For the prism case described in the next paragraph, this implies either giving up the tensor product structure and applying the above formula even in the prism case, or accepting that different subdivision matrices around irregular regions have different subdominant eigenvalues.*

It is also noteworthy that parallels can be found to the semi-regular case from [Pet20, Eq. (9), p. 9] and [YP22, Eq. (A5), p. 13]. Both works study generalized B-splines in the framework of singular parametrization. Their constructions likewise contain a stationary component (similar to the first summand in Equation (5.4)) and a component adapted to the valence of the prism (similar to the second summand in Equation (5.4)). Moreover, polytopes also appear in their constructions.

**The Prism Case** For $t = 3$, we define an alternative rule for the prism case. The reason is that we want to satisfy the quality criteria Q7 and Q8. However, this is not possible with the formula (5.4). To see this, consider the eigenvalues of $S^{(1)}$ and $S^{(2)}$:

$$\text{eig}\left(S^{(1)}\right) = \begin{bmatrix}1 & \frac{1}{2}\end{bmatrix} \quad \text{and} \quad \text{eig}\left(S^{(2)}\right) = \begin{bmatrix}1 & \frac{1}{2} & \frac{1}{2} & \frac{1}{4}\end{bmatrix}.$$

By Theorem B.32, the eigenvalues of the Kronecker product $\text{kron}(S^{(2)}, S^{(1)})$ are exactly

$$\text{eig}\left(S^{(3)}\right) = \text{eig}\left(\text{kron}(S^{(2)}, S^{(1)})\right) = \begin{bmatrix}1 & \frac{1}{2} & \frac{1}{2} & \frac{1}{2} & \frac{1}{4} & \frac{1}{4} & \frac{1}{4} & \frac{1}{8}\end{bmatrix}.$$

Since the matrix $S_1^{(t)}$ from (5.4) has eigenvalues

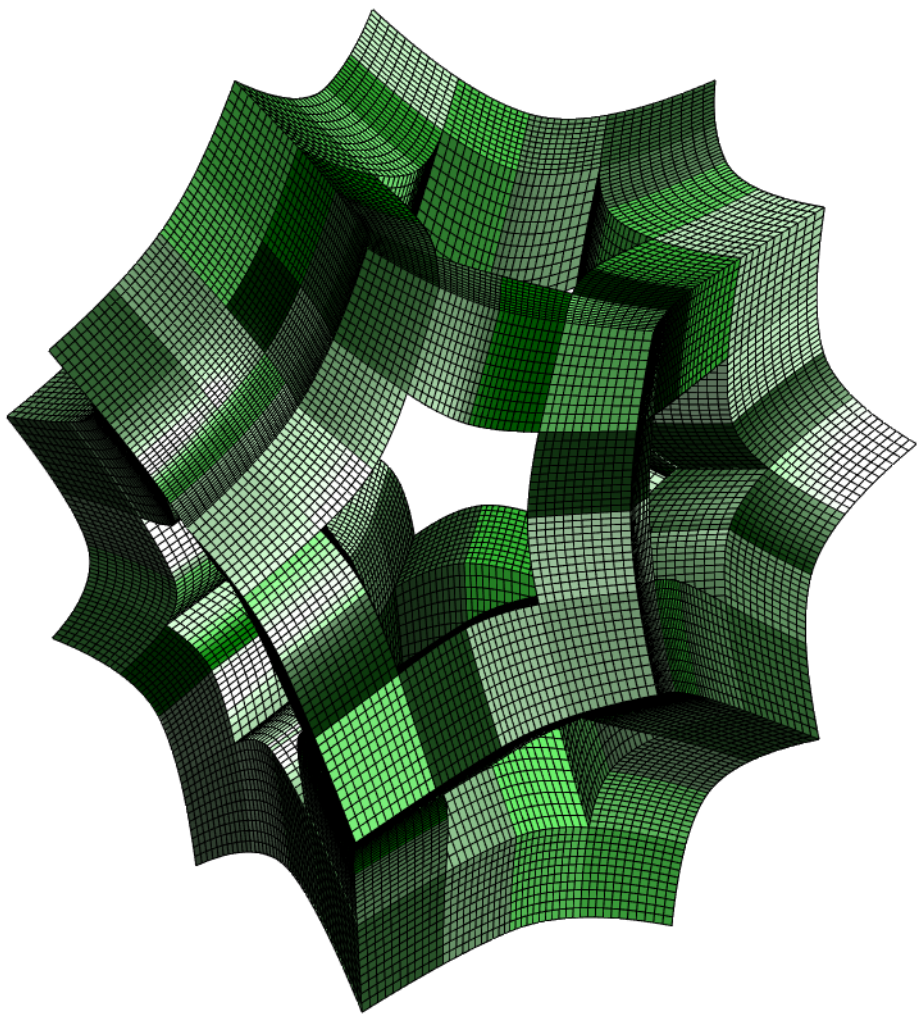

**Figure 5.4:** An evaluated element from the eigenspace of the subdominant eigenvalue of the subdivision matrix from Example 5.5 for the invariant shell case.

$$1, \quad \underbrace{\frac{1}{2}, \quad \cdots, \quad \frac{1}{2}}_{t \text{ times}}, \quad \mu, \quad \cdots, \quad \mu,$$

it is impossible to simultaneously achieve eigenvalues $1/4$ and $1/8$. The same applies analogously for all prisms with a base polygon having $n > 3$ edges. Therefore, for these cases, we choose the tensor product of the two-dimensional Doo-Sabin variant, which coincides with our construction for $\mu = \frac{1}{4}$, and the one-dimensional regular rule.

It is quite interesting that the subdivision matrix in the non-prism case has only three distinct eigenvalues. With the above construction, we fix the eigenspaces for the four eigenvalues $1, \frac{1}{2}, \frac{1}{2}$, and $\frac{1}{2}$. The remaining space is then assigned the eigenvalue $\mu$. To construct a natural tensor product extension, one must carefully consider the remaining eigenspace corresponding to $\mu$ and sensibly split it into (at least) two eigenspaces with different eigenvalues. However, such a construction is lacking because it is unclear how to choose the other eigenvectors meaningfully.

The variant in the next section focuses, among other things, exactly on this question. However, this comes at the price that the subsubdominant eigenvalue cannot be freely chosen. Before considering the second variant, we demonstrate the construction of the first variant by way of example, to give an impression of the generated objects:

**Example 5.5.** *We choose the facet matrix and the corresponding adjacency matrix as*

$$F := \begin{bmatrix} 1 & 2 & 3 & 4 & 5 & 6 & 7 \\ 1 & 8 & 9 & 2 & 0 & 0 & 0 \\ 1 & 7 & 14 & 18 & 8 & 0 & 0 \\ 6 & 7 & 14 & 13 & 0 & 0 & 0 \\ 5 & 6 & 13 & 17 & 12 & 0 & 0 \\ 4 & 5 & 12 & 16 & 11 & 0 & 0 \\ 4 & 3 & 10 & 15 & 11 & 0 & 0 \\ 2 & 3 & 10 & 9 & 0 & 0 & 0 \\ 8 & 9 & 10 & 15 & 19 & 22 & 18 \\ 14 & 13 & 17 & 21 & 22 & 18 & 0 \\ 20 & 19 & 22 & 21 & 0 & 0 & 0 \\ 12 & 16 & 20 & 21 & 17 & 0 & 0 \\ 11 & 15 & 19 & 20 & 16 & 0 & 0 \end{bmatrix}$$

*and*

$$A := \begin{bmatrix}
&1&&&&&1&1&&&&&&&&&&&&&& \\
1&&1&&&&&&1&&&&&&&&&&&&& \\
&1&&1&&&&&&1&&&&&&&&&&&& \\
&&1&&1&&&&&&1&&&&&&&&&&& \\
&&&1&&1&&&&&&1&&&&&&&&&& \\
&&&&1&&1&&&&&&1&&&&&&&&& \\
1&&&&&1&&&&&&&&1&&&&&&&& \\
1&&&&&&&&1&&&&&&&&&1&&&& \\
&1&&&&&&1&&1&&&&&&&&&&&& \\
&&1&&&&&&1&&&&&&1&&&&&&& \\
&&&1&&&&&&&&&&&1&1&&&&&& \\
&&&&1&&&&&&&&&&&1&1&&&&& \\
&&&&&1&&&&&&&&1&&&1&&&&& \\
&&&&&&1&&&&&&1&&&&&1&&&& \\
&&&&&&&&&1&1&&&&&&&&1&&& \\
&&&&&&&&&&1&1&&&&&&&&1&& \\
&&&&&&&&&&&1&1&&&&&&&&1& \\
&&&&&&&1&&&&&&1&&&&&&&&1 \\
&&&&&&&&&&&&&&1&&&&&1&&1 \\
&&&&&&&&&&&&&&&1&&&1&&1& \\
&&&&&&&&&&&&&&&&1&&&1&&1 \\
&&&&&&&&&&&&&&&&&1&1&&1&
\end{bmatrix}.$$

*This is the adjacency matrix for the example from Figure 3.13. For $\mu = 1/4$ we obtain*

$$S_1^{(3)} = \frac{1}{1000}\begin{bmatrix}
312 & 57 & 48 & 33 & 32 & 45 & 58 & 56 & 53 & 42 & 16 & 15 & 37 & 48 & 22 & 8 & 19 & 45 & 14 & 7 & 13 & 22 \\
57 & 311 & 59 & 47 & 33 & 35 & 46 & 52 & 58 & 54 & 32 & 15 & 21 & 32 & 37 & 16 & 9 & 33 & 20 & 11 & 7 & 16 \\
48 & 59 & 316 & 63 & 44 & 32 & 35 & 39 & 53 & 58 & 49 & 25 & 13 & 16 & 46 & 30 & 7 & 14 & 23 & 17 & 5 & 7 \\
33 & 47 & 63 & 328 & 67 & 44 & 30 & 16 & 33 & 45 & 63 & 49 & 20 & 8 & 41 & 48 & 22 & -8 & 18 & 26 & 10 & \mathbf{-4} \\
32 & 33 & 44 & 67 & 330 & 66 & 45 & 8 & 14 & 18 & 47 & 64 & 48 & 29 & 15 & 48 & 46 & -2 & 5 & 25 & 22 & -3 \\
45 & 35 & 32 & 44 & 66 & 322 & 62 & 25 & 19 & 11 & 21 & 51 & 62 & 52 & 0 & 28 & 49 & 24 & 1 & 16 & 25 & 10 \\
58 & 46 & 35 & 30 & 45 & 62 & 316 & 46 & 37 & 22 & 10 & 29 & 55 & 59 & 4 & 12 & 36 & 43 & 5 & 9 & 21 & 20 \\
56 & 52 & 39 & 16 & 8 & 25 & 46 & 315 & 59 & 48 & 14 & 3 & 28 & 48 & 35 & 7 & 16 & 62 & 34 & 19 & 24 & 44 \\
53 & 58 & 53 & 33 & 14 & 19 & 37 & 59 & 312 & 60 & 30 & 6 & 15 & 31 & 47 & 15 & 7 & 46 & 36 & 20 & 16 & 34 \\
42 & 54 & 58 & 45 & 18 & 11 & 22 & 48 & 60 & 316 & 47 & 13 & 4 & 14 & 61 & 29 & 5 & 32 & 44 & 29 & 18 & 31 \\
16 & 32 & 49 & 63 & 47 & 21 & 10 & 14 & 30 & 47 & 318 & 48 & 12 & 2 & 58 & 59 & 27 & 2 & 44 & 47 & 31 & 22 \\
15 & 15 & 25 & 49 & 64 & 51 & 29 & 3 & 6 & 13 & 48 & 317 & 48 & 28 & 24 & 58 & 57 & 9 & 28 & 46 & 44 & 22 \\
37 & 21 & 13 & 20 & 48 & 62 & 55 & 28 & 15 & 4 & 12 & 48 & 317 & 61 & 2 & 29 & 60 & 43 & 17 & 30 & 45 & 35 \\
48 & 32 & 16 & 8 & 29 & 52 & 59 & 48 & 31 & 14 & 2 & 28 & 61 & 317 & 6 & 13 & 47 & 61 & 21 & 23 & 40 & 44 \\
22 & 37 & 46 & 41 & 15 & 0 & 4 & 35 & 47 & 61 & 58 & 24 & 2 & 6 & 320 & 47 & 16 & 28 & 61 & 50 & 35 & 44 \\
8 & 16 & 30 & 48 & 48 & 28 & 12 & 7 & 15 & 29 & 59 & 58 & 29 & 13 & 47 & 313 & 46 & 10 & 47 & 56 & 47 & 33 \\
19 & 9 & 7 & 22 & 46 & 49 & 36 & 16 & 7 & 5 & 27 & 57 & 60 & 47 & 16 & 46 & 315 & 36 & 34 & 48 & 57 & 44 \\
45 & 33 & 14 & -8 & -2 & 24 & 43 & 62 & 46 & 32 & 2 & 9 & 43 & 61 & 28 & 10 & 36 & 329 & 46 & 33 & 48 & 67 \\
14 & 20 & 23 & 18 & 5 & 1 & 5 & 34 & 36 & 44 & 44 & 28 & 17 & 21 & 61 & 47 & 34 & 46 & 320 & 61 & 56 & 65 \\
7 & 11 & 17 & 26 & 25 & 16 & 9 & 19 & 20 & 29 & 47 & 46 & 30 & 23 & 50 & 56 & 48 & 33 & 61 & 313 & 59 & 56 \\
13 & 7 & 5 & 10 & 22 & 25 & 21 & 24 & 16 & 18 & 31 & 44 & 45 & 40 & 35 & 47 & 57 & 48 & 56 & 59 & 315 & 63 \\
22 & 16 & 7 & -4 & -3 & 10 & 20 & 44 & 34 & 31 & 22 & 22 & 35 & 44 & 44 & 33 & 44 & 67 & 65 & 56 & 63 & 327
\end{bmatrix}$$

*and for $\mu = 0$ we obtain*

$$S_1^{(3)} = \frac{1}{1000}\begin{bmatrix}
101 & 91 & 72 & 44 & 41 & 68 & 93 & 90 & 83 & 60 & 10 & 7 & 50 & 74 & 20 & -7 & 14 & 68 & 5 & -8 & 2 & 21 \\
91 & 100 & 95 & 71 & 43 & 47 & 69 & 82 & 92 & 85 & 41 & 7 & 20 & 41 & 51 & 9 & -5 & 43 & 17 & -1 & -8 & 10 \\
72 & 95 & 110 & 103 & 64 & 41 & 47 & 56 & 83 & 94 & 76 & 27 & 2 & 9 & 70 & 37 & -8 & 6 & 22 & 12 & -12 & -8 \\
44 & 71 & 103 & 133 & 111 & 64 & 38 & 9 & 44 & 68 & 102 & 76 & 18 & -6 & 60 & 74 & 21 & -38 & 13 & 28 & -2 & \mathbf{-30} \\
41 & 43 & 64 & 111 & 137 & 110 & 68 & -7 & 6 & 14 & 71 & 105 & 73 & 35 & 8 & 72 & 68 & -28 & -12 & 27 & 20 & -28 \\
68 & 47 & 41 & 64 & 110 & 121 & 101 & 28 & 16 & -1 & 20 & 79 & 101 & 82 & -22 & 34 & 75 & 25 & -21 & 9 & 27 & -3 \\
93 & 69 & 47 & 38 & 68 & 101 & 109 & 69 & 51 & 21 & -3 & 36 & 88 & 96 & -14 & 0 & 50 & 64 & -13 & -5 & 18 & 17 \\
90 & 82 & 56 & 9 & -7 & 28 & 69 & 108 & 96 & 74 & 4 & -16 & 34 & 73 & 47 & -9 & 9 & 102 & 45 & 15 & 26 & 66 \\
83 & 92 & 83 & 44 & 6 & 16 & 51 & 96 & 102 & 96 & 37 & -11 & 7 & 39 & 72 & 7 & -9 & 69 & 48 & 17 & 10 & 45 \\
60 & 85 & 94 & 68 & 14 & -1 & 21 & 74 & 96 & 109 & 72 & 3 & -16 & 6 & 99 & 35 & -14 & 40 & 66 & 36 & 12 & 39 \\
10 & 41 & 76 & 102 & 71 & 20 & -3 & 4 & 37 & 72 & 113 & 73 & 2 & -20 & 93 & 96 & 32 & -18 & 66 & 72 & 39 & 21 \\
7 & 7 & 27 & 76 & 105 & 79 & 36 & -16 & -11 & 3 & 73 & 111 & 73 & 33 & 26 & 93 & 91 & -5 & 33 & 69 & 66 & 22 \\
50 & 20 & 2 & 18 & 73 & 101 & 88 & 34 & 7 & -16 & 2 & 73 & 111 & 99 & -19 & 35 & 96 & 63 & 12 & 36 & 66 & 47 \\
74 & 41 & 9 & -6 & 35 & 82 & 96 & 73 & 39 & 6 & -20 & 33 & 99 & 112 & -11 & 4 & 72 & 99 & 19 & 22 & 57 & 65 \\
20 & 51 & 70 & 60 & 8 & -22 & -14 & 47 & 72 & 99 & 93 & 26 & -19 & -11 & 118 & 70 & 9 & 33 & 100 & 77 & 48 & 66 \\
-7 & 9 & 37 & 74 & 72 & 34 & 0 & -9 & 7 & 35 & 96 & 93 & 35 & 4 & 70 & 104 & 70 & -2 & 72 & 90 & 72 & 44 \\
14 & -5 & -8 & 21 & 68 & 75 & 50 & 9 & -9 & -14 & 32 & 91 & 96 & 72 & 9 & 70 & 107 & 48 & 45 & 73 & 91 & 64 \\
68 & 43 & 6 & -38 & -28 & 25 & 64 & 102 & 69 & 40 & -18 & -5 & 63 & 99 & 33 & -2 & 48 & 135 & 69 & 44 & 72 & 112 \\
5 & 17 & 22 & 13 & -12 & -21 & -13 & 45 & 48 & 66 & 66 & 33 & 12 & 19 & 100 & 72 & 45 & 69 & 118 & 99 & 89 & 108 \\
-8 & -1 & 12 & 28 & 27 & 9 & -5 & 15 & 17 & 36 & 72 & 69 & 36 & 22 & 77 & 90 & 73 & 44 & 99 & 103 & 95 & 89 \\
2 & -8 & -12 & -2 & 20 & 27 & 18 & 26 & 10 & 12 & 39 & 66 & 66 & 57 & 48 & 72 & 91 & 72 & 89 & 95 & 107 & 103 \\
21 & 10 & -8 & -30 & -28 & -3 & 17 & 66 & 45 & 39 & 21 & 22 & 47 & 65 & 66 & 44 & 64 & 112 & 108 & 89 & 103 & 131
\end{bmatrix}$$

*with values rounded to three decimal places. The evaluation of an element from the eigenspace of the subdominant eigenvalue corresponding to the shell matrix is shown in Figure 5.4.*

### 5.2.2 Variant 2: Colin-de-Verdière-Matrix and Matrix Exponential

Here we want to present, compared to the previous one, another variant for constructing generalized quadratic B-spline subdivision matrices. For this, we use the results from Chapters 3 and 4 and combine them in Algorithm 11. All variables and matrices in Section 5.2.2 refer to those of Algorithm 11.

**Algorithm 11:** GeneralizedQuadraticBSplineSubdivisionMatrixVariant2

**Data:** Valid combinatorial graph **G** of an initial element in the input form of Definition 3.10, type $t \in \{2, 3\}$
**Result:** Subdivision matrix $S_2^{(t)} \in \mathbb{R}^{n \times n}$

**begin**
- **if** $t == 2$ **then**
  - $(\mathbf{P}, \mathbf{P}^*)$ = Construct a primal and dual 2-polytope for **G** according to Section 3.4;
  - $C$ = Construct the Colin-de-Verdière-matrix realizing **P** for graph **G** and realizing **P**$^*$ for the dual graph **G**$^*$ according to Section 4.3;
- **else if** $t == 3$ **then**
  - $(\mathbf{P}, \mathbf{P}^*)$ = Construct a primal and dual 3-polytope for **G** according to Section 3.3;
  - $C$ = Construct the Colin-de-Verdière-matrix realizing **P** for graph **G** and realizing **P**$^*$ for the dual graph **G**$^*$ according to Section 4.4;
- $N := \texttt{diag}\left(1./\left(\sum_j C_{(:,j)}\right)\right)$;
- $S_2^{(t)} := \exp_2\left(NC - E_n\right)$;

The first two steps of the algorithm have been explained in detail in Chapters 3 and 4. Therefore, only the formula

$$S_2^{(t)} := \exp_2\left(NC - E_n\right) \quad \text{with} \quad C \in \mathbb{R}^{n \times n}$$

remains to be described. We will explain this in several steps.

**Normalization** $NC$ First, the definition of $N$ contains several commands not described before. The command `1./` comes from Matlab notation and means that the number 1 is divided by each entry of a vector or matrix. This is called *element-wise division*. The Matlab command `diag` (see [Matb]) either converts a vector into a diagonal matrix with that vector on the diagonal or returns the diagonal of a square matrix. Thus, the matrix $N$ is given by

$$N := \texttt{diag}\left(1./\left(\sum_j C_{(:,j)}\right)\right) := \begin{bmatrix} \frac{1}{\sum_j C_{(1,j)}} & 0 & \cdots & 0 \\ 0 & \ddots & \ddots & \vdots \\ \vdots & \ddots & \ddots & 0 \\ 0 & \cdots & 0 & \frac{1}{\sum_j C_{(n,j)}} \end{bmatrix}.$$

The matrix $N$ thus normalizes the rows of $C$ so that the rows of $NC$ sum to 1. Furthermore, the matrix $N$ is well-defined because, by Lemma 4.9 for $t = 2$ and Theorem 4.14 for $t = 3$, the row sums of $C$ are negative and therefore in particular nonzero.

We will discuss the spectrum of $S_2^{(t)}$ in detail in Section 5.3. Here, it should already be noted that the kernel of

$$\tilde{C} := NC \tag{5.6}$$

coincides with the kernel of $C$. Let $a$ be a vector in the kernel of $C$. Then

$$\tilde{C}a = NCa = N\vec{0} = \vec{0}. \tag{5.7}$$

Since $N$ is an invertible matrix, there is no other nonzero vector $b \neq 0$ with $Nb = 0$. Hence the kernels of $\tilde{C}$ and $C$

coincide. Moreover, by Lemma 2.36, $\tilde{C}$ has an eigenvalue $1$ with eigenvector $\vec{1}$ because all rows of $\tilde{C}$ sum to $1$.

**Shifting the spectrum of** $NC$ In this paragraph, we anticipate the results of Theorems 5.17 and 5.18 to motivate the shifting of the spectrum. Detailed justifications can be found in the proofs of these theorems. To better describe the spectrum shifting, we first argue that $\tilde{C}$ is diagonalizable.

For this, consider the matrix $C$. Since the Colin-de-Verdière-matrix $C$ is symmetric by Definition 4.1, Lemma B.43 implies that it is also Hermitian. Moreover, the matrix $N$ is a diagonal matrix with strictly negative diagonal entries. Thus, the matrix $-N$ is a diagonal matrix with strictly positive diagonal entries. Since $-N$ is diagonal, it is symmetric and therefore Hermitian as well. Its eigenvalues are exactly the diagonal entries of $-N$, hence all real and positive. By characterization B.47, the matrix $-N$ is positive definite.

By Theorem B.48, it follows that the matrix $-\tilde{C} = -NC$ is diagonalizable, and thus $\tilde{C}$ is also diagonalizable.

Now consider the expression

$$S_2^{(t)} = \exp_2\big(\tilde{C} - E_n\big).$$

Since $\tilde{C}$ is diagonalizable, it can be written as

$$\tilde{C} = V\tilde{D}V^{-1} \quad \text{with} \quad \tilde{D} \text{ diagonal.}$$

Hence,

$$\tilde{C} - E_n = V\tilde{D}V^{-1} - VE_nV^{-1} = V(\tilde{D} - E_n)V^{-1}.$$

Therefore,

$$S_2^{(t)} = \exp_2(\tilde{C} - E_n) = \exp\big(\ln(2)\cdot V(\tilde{D} - E_n)V^{-1}\big) = \exp\big(V\ln(2)(\tilde{D} - E_n)V^{-1}\big).$$

Applying Proposition B.53, we get

$$S_2^{(t)} = \exp\Big(V\ln(2)\Big(\tilde{D} - E_n\Big)V^{-1}\Big) = V\exp\Big(\ln(2)\Big(\tilde{D} - E_n\Big)\Big)V^{-1} = V\exp_2\Big(\tilde{D} - E_n\Big)V^{-1}.$$

Since $\tilde{D} - E_n$ is diagonal, Proposition B.54 implies

$$S_2^{(t)} = V\begin{bmatrix} \exp_2(\tilde{D}_{(1,1)} - 1) & 0 & \cdots & 0 \\ 0 & \ddots & \ddots & \vdots \\ \vdots & \ddots & \ddots & 0 \\ 0 & \cdots & 0 & \exp_2(\tilde{D}_{(n,n)} - 1) \end{bmatrix} V^{-1}. \tag{5.8}$$

Thus, $S_2^{(t)}$ has the same eigenvectors as $\tilde{C}$, and its eigenvalues are

$$\exp_2(\lambda_i - 1) = 2^{\lambda_i - 1}, \tag{5.9}$$

where $\lambda_i$ are the eigenvalues of $\tilde{C}$. Hence, for the eigenvalues $0$ and $1$ of $\tilde{C}$ (see Theorem 5.18), we get

$$2^{1-1} = 2^0 = 1$$

as the eigenvalue of $S_2^{(t)}$ corresponding to eigenvector $\vec{1}$, and

$$2^{0-1} = 2^{-1} = \frac{1}{2}$$

as an $t$-fold eigenvalue of $S_2^{(t)}$ corresponding to the eigenspace spanned by the convex polytope **P**.

This also clarifies the construction idea: first, we create the eigenspace structure we want using the polytope **P**, then compute the associated Colin-de-Verdière-matrix, constructing a matrix that has the polytope as a basis of an eigenspace, and finally shift the eigenvalues so that the correct eigenspaces correspond to the desired eigenvalues.

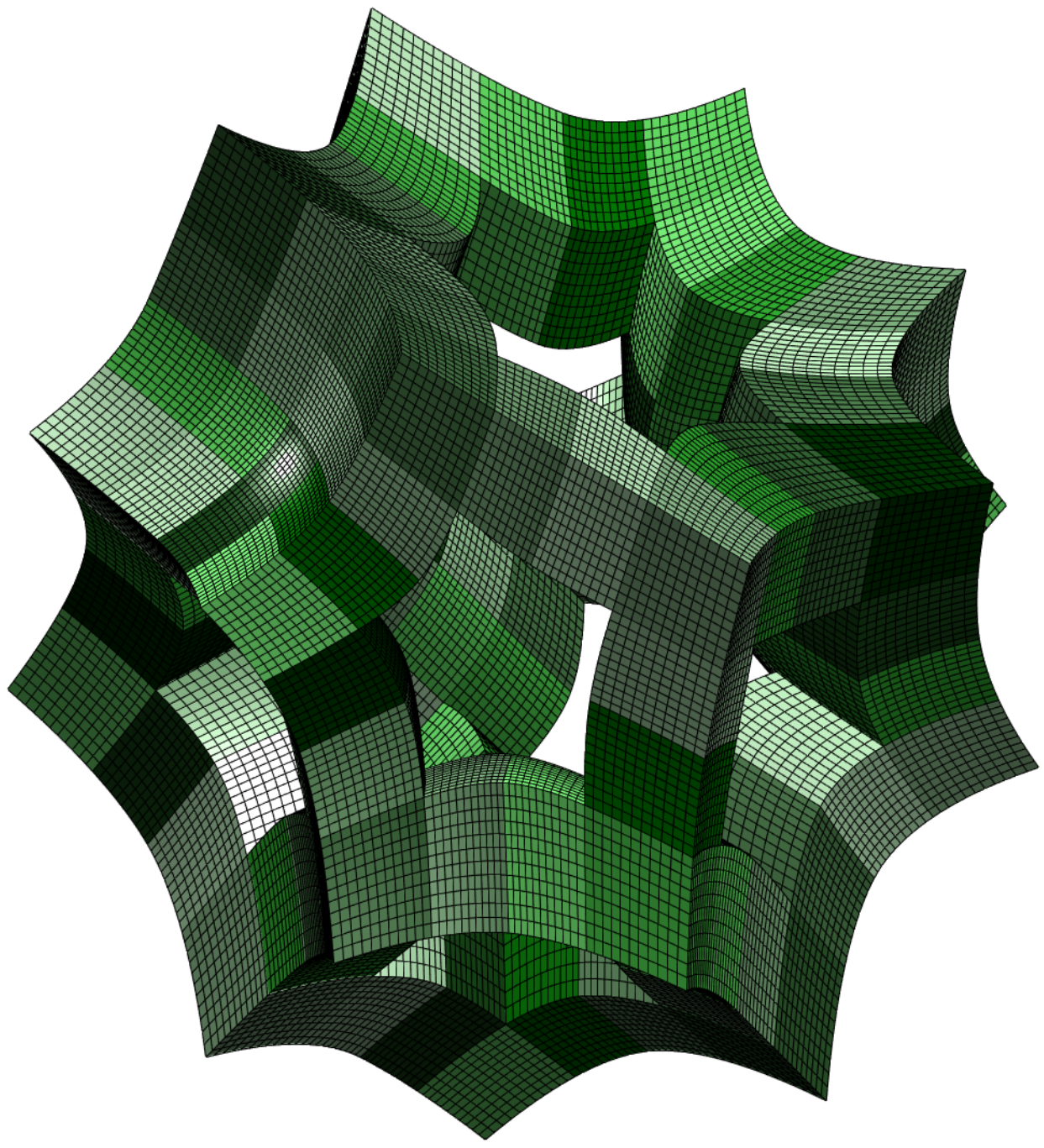

**Figure 5.5:** An evaluated element from the eigenspace of the subdominant eigenvalue of the subdivision matrix from Example 5.7 for the invariant shell case.

Analogous to Remark 5.4 we have:

**Remark 5.6.** *If we replace* $\ln(2)$ *by* $\ln(1/\lambda)$ *with* $0 \leq \lambda < 1$*, we can choose the subdominant eigenvalue* $\lambda$ *freely. However, if* $\lambda \neq 1/2$*, we lose the tensor product property in the (semi-)regular case, for the same reasons given in Remark 5.4.*

To give an impression of the matrices, we consider an example for the second variant as well:

**Example 5.7.** *Let* $F$ *and* $A$ *be as in Example 5.5. Then the associated Colin-de-Verdière-matrix for Variant 2 is*

$$C = \frac{1}{100}\begin{bmatrix}
520 & -244 & & & & & -246 & -135 & & & & & & & & & & & & & & \\
-244 & 532 & -239 & & & & & & -120 & & & & & & & & & & & & & \\
 & -239 & 476 & -212 & & & & & & -137 & & & & & & & & & & & & \\
 & & -212 & 376 & -184 & & & & & & -151 & & & & & & & & & & & \\
 & & & -184 & 374 & -204 & & & & & & -152 & & & & & & & & & & \\
 & & & & -204 & 423 & -202 & & & & & & -139 & & & & & & & & & \\
-246 & & & & & -202 & 467 & & & & & & & -137 & & & & & & & & \\
-135 & & & & & & & 483 & -254 & & & & & & & & & -205 & & & & \\
 & -120 & & & & & & -254 & 547 & -248 & & & & & & & & & & & & \\
 & & -137 & & & & & & -248 & 494 & & & & & -226 & & & & & & & \\
 & & & -151 & & & & & & & 352 & & & & -143 & -168 & & & & & & \\
 & & & & -152 & & & & & & & 365 & & & & -166 & -150 & & & & & \\
 & & & & & -139 & & & & & & & 418 & -181 & & & -200 & & & & & \\
 & & & & & & -137 & & & & & & -181 & 386 & & & & -166 & & & & \\
 & & & & & & & & & -226 & -143 & & & & 421 & & & & -215 & & & \\
 & & & & & & & & & & -168 & -166 & & & & 409 & & & & -172 & & \\
 & & & & & & & & & & & -150 & -200 & & & & 421 & & & & -194 & \\
 & & & & & & & -205 & & & & & & -166 & & & & 388 & & & & -215 \\
 & & & & & & & & & & & & & & 183 & & & & -377 & 124 & & 170 \\
 & & & & & & & & & & & & & & & 264 & & & 222 & -611 & 225 & \\
 & & & & & & & & & & & & & & & & 219 & & & 166 & -473 & 188 \\
 & & & & & & & & & & & & & & & & & 140 & 130 & & 108 & -279
\end{bmatrix}$$

*and the correspondingly normalized matrix*

$$\tilde{C} = \frac{1}{100}\begin{bmatrix}
-496 & 233 & & & & & 235 & 128 \\
342 & -745 & 335 & & & & & & 168 \\
& 213 & -425 & 189 & & & & & & 123 \\
& & 125 & -221 & 108 & & & & & & 89 \\
& & & 110 & -224 & 122 & & & & & & 91 \\
& & & & 167 & -346 & 166 & & & & & & 114 \\
208 & & & & & 171 & -394 & & & & & & & 116 \\
122 & & & & & & & -438 & 230 & & & & & & & & & 186 \\
& 159 & & & & & & 336 & -725 & 329 \\
& & 117 & & & & & & 211 & -419 & & & & & 192 \\
& & & 137 & & & & & & & -320 & & & & 130 & 152 \\
& & & & 149 & & & & & & & -357 & & & & 162 & 146 \\
& & & & & 136 & & & & & & & -408 & 177 & & & 195 \\
& & & & & & 138 & & & & & & 183 & -389 & & & & 168 \\
& & & & & & & & & 139 & 88 & & & & -259 & & & & 132 \\
& & & & & & & & & & 174 & 172 & & & & -425 & & & & 179 \\
& & & & & & & & & & & 122 & 163 & & & & -343 & & & & 158 \\
& & & & & & & 103 & & & & & & 84 & & & & -196 & & & & 109 \\
& & & & & & & & & & & & & & 183 & & & & -377 & 124 & & 170 \\
& & & & & & & & & & & & & & & 264 & & & 222 & -611 & 225 \\
& & & & & & & & & & & & & & & & 219 & & & 166 & -473 & 188 \\
& & & & & & & & & & & & & & & & & 140 & 130 & & 108 & -279
\end{bmatrix}$$

*rounded to two decimal places. From this, the subdivision matrix*

$$S_2^{(t)} = \frac{1}{1000}\begin{bmatrix}
136 & 76 & 74 & 44 & 35 & 66 & 125 & 91 & 57 & 54 & 9 & 7 & 30 & 51 & 23 & 3 & 11 & 77 & 7 & 2 & 5 & 18 \\
111 & 85 & 118 & 90 & 35 & 35 & 76 & 83 & 67 & 87 & 24 & 7 & 13 & 26 & 47 & 7 & 5 & 52 & 12 & 3 & 3 & 13 \\
69 & 75 & 145 & 161 & 61 & 24 & 37 & 51 & 56 & 97 & 49 & 14 & 6 & 10 & 67 & 16 & 5 & 24 & 17 & 6 & 3 & 9 \\
27 & 38 & 106 & 244 & 138 & 43 & 21 & 15 & 21 & 46 & 90 & 40 & 12 & 6 & 52 & 40 & 15 & 7 & 15 & 12 & 7 & 6 \\
22 & 15 & 41 & 141 & 239 & 114 & 51 & 7 & 6 & 13 & 43 & 86 & 44 & 23 & 16 & 41 & 48 & 10 & 7 & 12 & 15 & 7 \\
56 & 20 & 22 & 60 & 156 & 163 & 118 & 20 & 9 & 8 & 14 & 47 & 88 & 66 & 5 & 16 & 56 & 36 & 4 & 6 & 16 & 11 \\
110 & 46 & 35 & 30 & 71 & 122 & 160 & 53 & 27 & 20 & 6 & 18 & 68 & 85 & 8 & 5 & 31 & 71 & 5 & 3 & 9 & 17 \\
87 & 54 & 52 & 23 & 10 & 22 & 57 & 138 & 80 & 78 & 9 & 3 & 18 & 45 & 43 & 4 & 9 & 172 & 22 & 5 & 12 & 57 \\
79 & 64 & 83 & 47 & 14 & 15 & 42 & 116 & 90 & 124 & 24 & 3 & 9 & 23 & 88 & 7 & 4 & 96 & 28 & 6 & 6 & 31 \\
48 & 53 & 92 & 67 & 18 & 8 & 20 & 73 & 79 & 152 & 48 & 7 & 3 & 9 & 156 & 17 & 4 & 44 & 54 & 13 & 8 & 28 \\
9 & 16 & 50 & 140 & 64 & 15 & 6 & 9 & 16 & 51 & 154 & 54 & 8 & 3 & 129 & 98 & 23 & 9 & 56 & 40 & 22 & 26 \\
7 & 5 & 15 & 67 & 141 & 56 & 21 & 3 & 3 & 8 & 58 & 143 & 53 & 20 & 26 & 99 & 113 & 14 & 27 & 42 & 52 & 29 \\
31 & 9 & 7 & 19 & 71 & 105 & 79 & 20 & 6 & 4 & 8 & 53 & 140 & 102 & 6 & 21 & 128 & 75 & 13 & 16 & 49 & 38 \\
54 & 19 & 11 & 11 & 39 & 82 & 101 & 50 & 18 & 11 & 3 & 21 & 105 & 129 & 7 & 7 & 65 & 161 & 15 & 8 & 27 & 56 \\
15 & 21 & 46 & 54 & 16 & 4 & 6 & 29 & 41 & 113 & 87 & 16 & 3 & 5 & 228 & 43 & 12 & 30 & 112 & 34 & 23 & 62 \\
4 & 5 & 19 & 71 & 71 & 21 & 6 & 4 & 6 & 21 & 112 & 105 & 22 & 8 & 73 & 139 & 67 & 15 & 63 & 69 & 54 & 46 \\
10 & 3 & 4 & 20 & 65 & 56 & 30 & 8 & 3 & 4 & 21 & 94 & 107 & 52 & 15 & 53 & 179 & 44 & 33 & 40 & 93 & 66 \\
40 & 19 & 14 & 6 & 8 & 22 & 43 & 96 & 37 & 26 & 5 & 7 & 39 & 80 & 25 & 7 & 27 & 277 & 44 & 15 & 34 & 130 \\
6 & 7 & 16 & 22 & 10 & 4 & 5 & 20 & 18 & 54 & 53 & 23 & 11 & 13 & 155 & 51 & 35 & 73 & 153 & 61 & 63 & 145 \\
3 & 3 & 10 & 30 & 32 & 12 & 5 & 9 & 7 & 23 & 67 & 66 & 25 & 12 & 86 & 101 & 76 & 44 & 109 & 81 & 88 & 109 \\
5 & 2 & 4 & 12 & 28 & 22 & 12 & 14 & 5 & 10 & 28 & 60 & 56 & 30 & 43 & 59 & 128 & 76 & 83 & 64 & 116 & 141 \\
12 & 6 & 6 & 7 & 8 & 9 & 13 & 41 & 15 & 22 & 19 & 19 & 25 & 36 & 65 & 29 & 53 & 169 & 111 & 47 & 81 & 207
\end{bmatrix}, \tag{5.10}$$

*rounded to three decimal places. The evaluation of an element from the eigenspace of the subdominant eigenvalue of the corresponding shell matrix can be found in Figure 5.5.*

Thus, we have described the two variants for constructing generalized quadratic B-spline subdivision matrices and proceed in the next section to examine which quality criteria they satisfy.

## 5.3 Proofs and Quality Criteria of the Subdivision Matrices

This section forms the core of the chapter and is one of the two main parts of this work. Here, we analyze which quality criteria are met by the subdivision matrices constructed in the previous section and also present some related results. Furthermore, we consider eight different categories of subdivision matrices that differ in three aspects:

| | | $S_1^{(2)}$ | $\overline{S}_1^{(2)}$ | $S_2^{(2)}$ | $\overline{S}_2^{(2)}$ | $S_1^{(3)}$ | $\overline{S}_1^{(3)}$ | $S_2^{(3)}$ | $\overline{S}_2^{(3)}$ |
|---|---|---|---|---|---|---|---|---|---|
| Q1 | Affine Invariance | 🟩 | 🟩 | 🟩 | 🟩 | 🟩 | 🟩 | 🟩 | 🟩 |
| Q2 | Convex Hull | 🟩 | 🟩 | 🟩 | 🟩 | 🟥 | 🟥 | 🟩 | 🟩 |
| Q3 | Dominant Eigenvalue 1 | 🟩 | 🟩 | 🟩 | 🟩 | 🟩 | 🟨 | 🟩 | 🟨 |
| Q4 | $t$ Subdominant Eigenvalues | 🟩 | 🟩 | 🟩 | 🟩 | 🟩 | 🟨 | 🟩 | 🟨 |
| Q5 | $t$ Subdominant Eigenvalues 0.5 | 🟩 | 🟩 | 🟩 | 🟩 | 🟩 | 🟨 | 🟩 | 🟨 |
| Q6 | Subsubdominant Eigenvalue 0.25 | 🟩 | 🟩 | 🟥 | 🟨 | 🟩 | 🟨 | 🟥 | 🟥 |
| Q7 | Regular Case | 🟢 | 🟢 | 🟢 | 🟢 | 🟢 | 🟢 | 🟢 | 🟢 |
| Q8 | Semi-Regular Case | ▨ | ▨ | ▨ | ▨ | 🟢 | 🟢 | 🟢 | 🟢 |
| Q9 | Subdivision Matrix for every Input | 🟢 | 🟢 | 🟢 | 🟢 | 🟢 | 🟢 | 🟢 | 🟢 |
| Q10 | Symmetries | 🟢 | 🟢 | 🟢 | 🟢 | 🟨 | 🟨 | 🟩 | 🟩 |
| Q11 | Proper Support | 🟢 | 🟢 | 🟢 | 🟢 | 🟢 | 🟢 | 🟢 | 🟢 |
| Q12 | Self-Penetration-Free Structure | 🟩 | 🟧 | 🟩 | 🟧 | 🟩 | 🟧 | 🟩 | 🟧 |
| Q13 | Injective and Regular Characteristic Mapping | ▨ | 🟢 | ▨ | 🟧 | ▨ | 🟧 | ▨ | 🟧 |
| | Stochastic | 🟩 | 🟩 | 🟩 | 🟩 | 🟥 | 🟥 | 🟩 | 🟩 |
| | Symmetric | 🟢 | 🟥 | 🟢 | 🟥 | 🟢 | 🟥 | 🟥 | 🟥 |
| | Diagonalizable | 🟢 | 🟧 | 🟢 | 🟧 | 🟢 | 🟧 | 🟢 | 🟧 |
| | Real Eigenvalues | 🟢 | 🟢 | 🟢 | 🟢 | 🟢 | 🟧 | 🟢 | 🟧 |

**Table 5.1:** Overview of the quality criteria and other properties of the subdivision matrices from Definition 5.8.

**Definition 5.8.** *We name the eight categories of subdivision matrices using three different notations:*

- *Matrices $S$ represent subdivision matrices for initial elements, whereas matrices $\overline{S}$ represent matrices for rings and shells.*
- *Matrices $S_1$ are subdivision matrices from Variant 1, i.e., constructed by Algorithm 10, whereas matrices $S_2$ are subdivision matrices from Variant 2, constructed by Algorithm 11.*
- *Matrices $S^{(2)}$ are subdivision matrices of type 2, whereas matrices $S^{(3)}$ are subdivision matrices of type 3.*

*Overall, this yields eight different categories of subdivision matrices, which are also displayed in Table 5.1.*

We will not be able to find a proof for every combination of quality criterion and category of subdivision matrices. Therefore, for $t = 3$ we use the 305,095 random examples described in Section 3.1.1 to either verify that the theoretical results agree with the empirical calculations or, in the absence of a proof, to at least make a qualitative statement. For $t = 2$, we empirically examine the subdivision matrices for 2-polytopes with 3 to 100 vertices, i.e., for (ir)regularities with valence less or equal to 100. This results in six possible combinations of quality criterion and category, represented by the following color scheme:

| | | | |
|---|---|---|---|
| 🟩 dark green | proven and empirically verified | 🟢 green | proven |
| 🟨 yellow-green | not proven, but empirically verified | 🟧 orange | unclear |
| 🟥 red | disproved or counterexample constructed | ▨ gray | not applicable |

The results of this section are summarized clearly in Table 5.1. In the following, we discuss the quality criteria in detail and describe the corresponding theorems and results. Since the subdivision matrices are constructed by the algorithms from Section 5.2, we use the notation from that construction without redefining them here.

## Q1 Affine Invariance

This quality criterion is satisfied by all subdivision matrices, as shown by the following two theorems:

**Theorem 5.9.** *The subdivision matrices of the four categories for $S_1 \in \mathbb{R}^{n\times n}$ satisfy quality criterion Q1, i.e., all rows of all subdivision matrices sum to* 1.

*Proof.* We begin with the matrices of the initial elements and first consider the case where $S_1$ is defined either for $t = 2$ or for a combinatorial graph that does not represent a prism. In this case, we have

$$S_1^{(t)} = \left( \frac{\vec{1} \cdot \vec{1}^T}{|\vec{1}|^2} + \frac{1-2\mu}{2-2\mu} \left( \frac{\tilde{P}_{(:,1)} \cdot \tilde{P}^T_{(:,1)}}{|\tilde{P}_{(:,1)}|^2} + \cdots + \frac{\tilde{P}_{(:,t)} \cdot \tilde{P}^T_{(:,t)}}{|\tilde{P}_{(:,t)}|^2} \right) \right) (1-\mu) + \mu E_n.$$

Multiplying $S_1^{(t)}$ by the vector $\vec{1}$ gives

$$\begin{aligned} S_1^{(t)} \cdot \vec{1} &= \left( \left( \frac{\vec{1} \cdot \vec{1}^T}{|\vec{1}|^2} + \frac{1-2\mu}{2-2\mu} \left( \frac{\tilde{P}_{(:,1)} \cdot \tilde{P}^T_{(:,1)}}{|\tilde{P}_{(:,1)}|^2} + \cdots + \frac{\tilde{P}_{(:,t)} \cdot \tilde{P}^T_{(:,t)}}{|\tilde{P}_{(:,t)}|^2} \right) \right) (1-\mu) + \mu E_n \right) \vec{1} \\ &= \left( \frac{\vec{1} \cdot \vec{1}^T}{|\vec{1}|^2} \vec{1} + \frac{1-2\mu}{2-2\mu} \left( \frac{\tilde{P}_{(:,1)} \cdot \tilde{P}^T_{(:,1)}}{|\tilde{P}_{(:,1)}|^2} + \cdots + \frac{\tilde{P}_{(:,t)} \cdot \tilde{P}^T_{(:,t)}}{|\tilde{P}_{(:,t)}|^2} \right) \vec{1} \right) (1-\mu) + \mu E_n \vec{1}. \end{aligned}$$

Examining the summands individually, the first summand yields

$$\frac{\vec{1} \cdot \vec{1}^T}{|\vec{1}|^2} \vec{1} = \frac{1}{n} \begin{bmatrix} 1 & \cdots & 1 \\ \vdots & & \vdots \\ 1 & \cdots & 1 \end{bmatrix} \vec{1} = \vec{1}.$$

Since the vectors $\tilde{P}_{(:,i)}$ for $i \in \{1, \ldots, t\}$ are constructed such that they sum to zero, it follows that

$$\frac{\tilde{P}_{(:,i)} \cdot \tilde{P}^T_{(:,i)}}{|\tilde{P}_{(:,i)}|^2} \vec{1} = \frac{\tilde{P}_{(:,i)} \cdot \tilde{P}^T_{(:,i)} \vec{1}}{|\tilde{P}_{(:,i)}|^2} = \frac{\tilde{P}_{(:,i)} \cdot 0}{|\tilde{P}_{(:,i)}|^2} = \vec{0}. \tag{5.11}$$

Finally, for the last summand we have

$$\mu E_n \vec{1} = \mu \vec{1}.$$

Combining these results gives

$$\begin{aligned} S_1^{(t)} \cdot \vec{1} &= \left( \frac{\vec{1} \cdot \vec{1}^T}{|\vec{1}|^2} \vec{1} + \frac{1-2\mu}{2-2\mu} \left( \frac{\tilde{P}_{(:,1)} \cdot \tilde{P}^T_{(:,1)}}{|\tilde{P}_{(:,1)}|^2} + \cdots + \frac{\tilde{P}_{(:,t)} \cdot \tilde{P}^T_{(:,t)}}{|\tilde{P}_{(:,t)}|^2} \right) \vec{1} \right) (1-\mu) + \mu E_n \vec{1} \\ &= (\vec{1} + \vec{0})(1-\mu) + \mu \vec{1} = \vec{1}. \end{aligned}$$

This shows why centering the polytope was necessary: it ensures that the summand in Equation (5.11) becomes zero. Without centering, the overall expression would not yield the all-ones vector.

For the prism case with $t = 3$, we use Theorem B.32, which states that the eigenvalues and eigenvectors of the Kronecker product of two matrices are the Kronecker products of the eigenvalues and eigenvectors of the two matrices. Since both $S^{(1)}$ and $S^{(2)}$ have an eigenvalue 1 corresponding to the eigenvector $\vec{1}$, it follows that $S_1^{(3)}$ in the prism case also has an eigenvalue 1 with eigenvector $\vec{1}$. By Lemma 2.36, which explains the relation between having eigenvalue 1 with eigenvector $\vec{1}$ and the row sums of the matrix, every row of $S_1^{(3)}$ sums to 1. Thus, the statement holds for both categories of initial elements.

For the ring or shell cases, the same result holds. The ring or shell matrices are composed of the subdivision matrices of the initial elements. Each row of such a ring or shell matrix consists of entries from the subdivision matrices of initial elements, padded with zeros. Therefore, affine invariance transfers from $S_1$ to $\overline{S}_1$. □

For the matrices of the four categories corresponding to $S_2$ we obtain the same result with the following theorem:

**Theorem 5.10.** *The subdivision matrices of the four categories for $S_2 \in \mathbb{R}^{n\times n}$ satisfy Quality Criterion Q1, i.e., for all four categories the rows of the subdivision matrices sum up to* 1.

*Proof.* We also start here with the subdivision matrix of an initial element. The rows of the matrix $\tilde{C}$ from Equation (5.7) are normalized, so each row of $\tilde{C}$ sums to 1. By Lemma 2.36, $\tilde{C}$ therefore has an eigenvalue 1 with eigenvector $\vec{1}$.

Using Equation (5.8), where Theorem B.53 is used to factor out eigenvectors from the matrix exponential, both $\tilde{C}$ and $S_2^{(t)}$ share the same eigenvectors because combinations of a matrix and its inverse can be factored out of the exponential. Hence, both $S_2^{(t)}$ and $\tilde{C}$ are diagonalizable by the matrix $V$ and consequently share the same eigenvectors.

The corresponding eigenvalues of $S_2^{(t)}$ are, according to Equation (5.9), precisely $2^{\lambda_i - 1}$. Therefore, $S_2^{(t)}$ has an eigenvector $\vec{1}$ with eigenvalue $2^{1-1} = 1$, and consequently all rows of $S_2^{(t)}$ sum to 1 by Lemma 2.36.

The proof for the ring and shell matrices proceeds analogously to the proof of Theorem 5.9. Since the rows of the ring and shell matrices are composed of entries of the subdivision matrices of the initial elements, padded with zeros, Quality Criterion Q1 transfers from $S_2^{(t)}$ to $\overline{S}_2^{(t)}$. □

Besides the proofs, it is also interesting to empirically test how well this quality criterion is numerically satisfied for the examples generated in Section 3.1.1. The results for the two subdivision matrices of the initial elements for $t = 3$ can be found in Figure 5.6.

Here we observe that the error in the row sums for Variant 1 lies around $10^{-15}$, i.e., within machine precision, whereas the error for Variant 2 is approximately on the order of $10^{-13}$. Since the matrix $\tilde{C}$ has already been normalized, this is the magnitude of the error introduced by the operation

$$S_2^{(t)} = \exp_2\left(\tilde{C} - E_n\right).$$

Thus, we lose about two orders of magnitude in accuracy due to the exponentiation. However, this is entirely acceptable because the error refers not to individual matrix entries but to the sums of the rows, each consisting of between 4 and 100 entries.
For the eight categories, we obtain the following maximum and minimum values of the row sums over all examples:

| | $S_1^{(2)}$ | $\overline{S}_1^{(2)}$ | $S_2^{(2)}$ | $\overline{S}_2^{(2)}$ | $S_1^{(3)}$ | $\overline{S}_1^{(3)}$ | $S_2^{(3)}$ | $\overline{S}_2^{(3)}$ |
|---|---|---|---|---|---|---|---|---|
| $\max \sum -1$ | 8,88e−16 | 8,88e−16 | 7,70e−14 | 5,24e−14 | 1,33e−15 | 1,33e−15 | 3,03e−13 | 3,03e−13 |
| $\min \sum -1$ | −8,88e−16 | −8,88e−16 | −1,00e−13 | −6,71e−14 | −1,33e−15 | −9,99e−16 | −5,32e−13 | −5,32e−13 |

It is noticeable that the values of the ring and shell matrices are sometimes better than those of the matrices of the initial elements, even though the latter are part of the larger matrices. However, this phenomenon can be explained by general numerical inaccuracies. Overall, it can be stated that the quality criterion Q1 is satisfied both formally and empirically by all categories of subdivision matrices.

## Q2 Convex Hull

Here we obtain a more differentiated picture for the various categories. We begin with the following theorem:

**Theorem 5.11.** *For $t = 2$, the subdivision matrices $S_1^{(2)}$ and $\overline{S}_1^{(2)}$ have no negative entries and together with quality criterion Q1 thus satisfy quality criterion Q2 of the convex hull.*

*Proof.* With Theorem 5.9, $S_1^{(2)}$ and $\overline{S}_1^{(2)}$ satisfy quality criterion Q1. With Lemma 5.2, the entries of the matrix $S_1^{(2)}$ are

$$\left(S_1^{(2)}\right)_{(i,j)} = (1-\mu)\frac{2 - 2\mu + (2-4\mu)\cos\left(\frac{2\pi(j-i)}{n}\right)}{n(2-2\mu)} + \mu\delta_{i,j}.$$

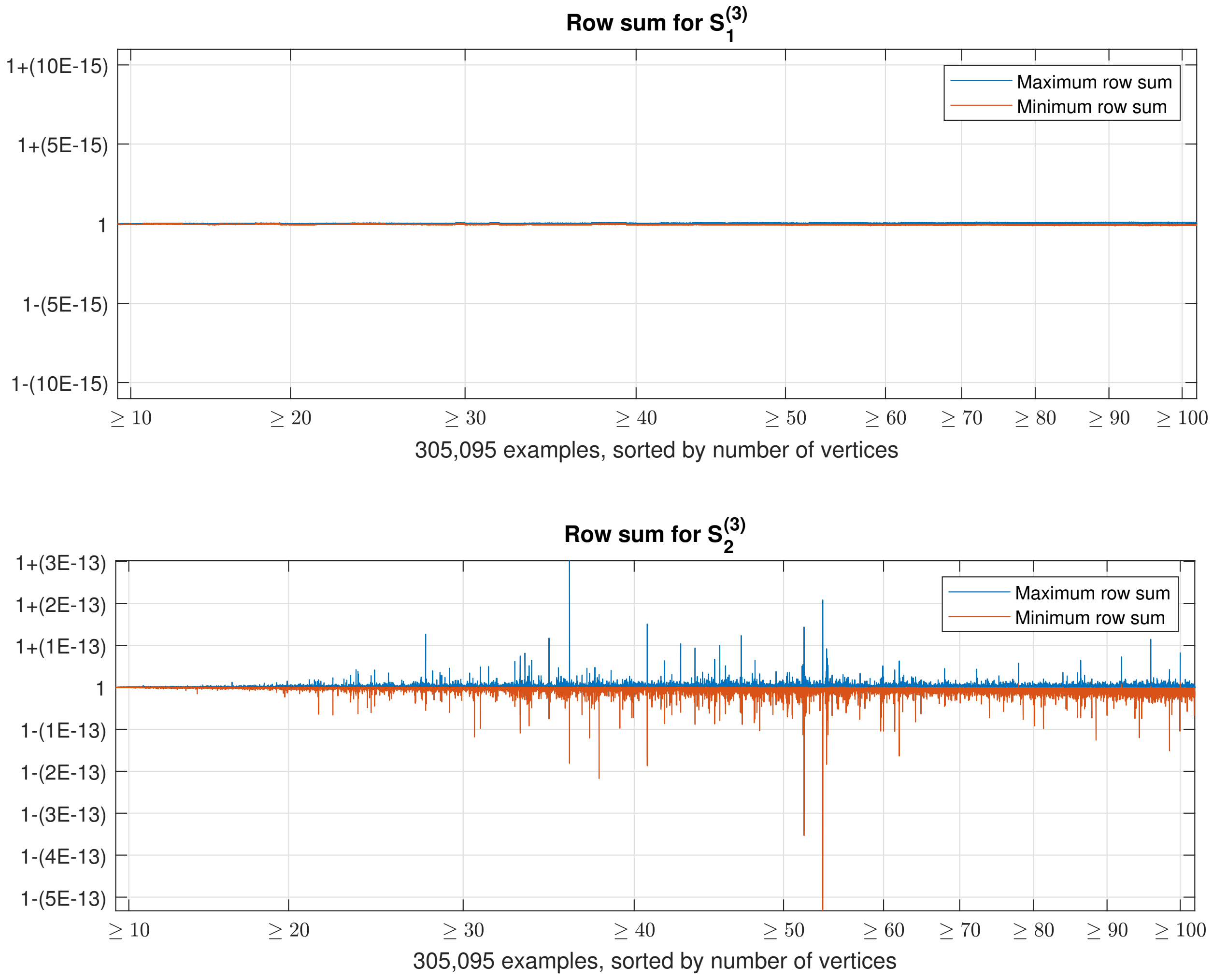

**Figure 5.6:** Maximum and minimum row sums of the subdivision matrices of the volumetric initial elements for Variant 1 (top) and Variant 2 (bottom).

Since for $0 \leq \mu < 1/2$

$$2 - 2\mu > 2 - 4\mu > 0 \quad \text{and} \quad -1 \leq \cos\left(\frac{2\pi(j-i)}{n}\right) \leq 1,$$

we have

$$2 - 2\mu + (2 - 4\mu)\cos\left(\frac{2\pi(j-i)}{n}\right) > 0.$$

Furthermore,

$$\mu\delta_{i,j} \geq 0 \quad \text{and} \quad 1 - \mu > 0,$$

and consequently,

$$\left(S_1^{(2)}\right)_{(i,j)} > 0.$$

Since the subdivision matrix $\overline{S}_1^{(2)}$ consists only of entries from various subdivision matrices $S_1^{(2)}$ and zero entries, all entries of $\overline{S}_1^{(2)}$ are also greater than or equal to zero. $\square$

For the case $t = 3$, however, quality criterion Q2 does not hold, as can be seen in Example 5.5. Both subdivision matrices for $\mu = 0$ and $\mu = 1/4$ contain negative entries. We can even make the following statement:

**Proposition 5.12.** *For $0 \leq \mu \leq 1/4$, the subdivision matrix corresponding to the adjacency matrix from Example 5.5 contains negative entries.*

*Proof.* The matrix entry $\left(S_1^{(3)}\right)_{(4,22)}$ is negative for both $\mu = 0$ and $\mu = 1/4$. It is computed by

$$\left(S_1^{(3)}\right)_{(4,22)} := \left(\frac{1}{22} + \frac{1-2\mu}{2-2\mu}\left(\frac{\tilde{P}_{(4,1)} \cdot \tilde{P}^T_{(22,1)}}{\left|\tilde{P}_{(:,1)}\right|^2} + \frac{\tilde{P}_{(4,2)} \cdot \tilde{P}^T_{(22,2)}}{\left|\tilde{P}_{(:,2)}\right|^2} + \frac{\tilde{P}_{(4,3)} \cdot \tilde{P}^T_{(22,3)}}{\left|\tilde{P}_{(:,3)}\right|^2}\right)\right)(1-\mu).$$

Define

$$a := \frac{\tilde{P}_{(4,1)} \cdot \tilde{P}^T_{(22,1)}}{\left|\tilde{P}_{(:,1)}\right|^2} + \frac{\tilde{P}_{(4,2)} \cdot \tilde{P}^T_{(22,2)}}{\left|\tilde{P}_{(:,2)}\right|^2} + \frac{\tilde{P}_{(4,3)} \cdot \tilde{P}^T_{(22,3)}}{\left|\tilde{P}_{(:,3)}\right|^2} < 0.$$

The function

$$\boldsymbol{f}(\mu) := \frac{1-2\mu}{2-2\mu} = \frac{1}{2(\mu-1)} + 1$$

is positive, continuous, and strictly decreasing on the interval $[0, 1/4]$. Thus, the function

$$\boldsymbol{f}_2(\mu) = \frac{1}{22} + \boldsymbol{f}(\mu)a = \frac{1}{22} + \frac{1-2\mu}{2-2\mu}a$$

is continuous and strictly increasing. Since $\left(S_1^{(3)}\right)_{(4,22)} = \boldsymbol{f}_2(\mu)(1-\mu)$ is negative at both $\mu = 0$ and $\mu = 1/4$, it follows that $\boldsymbol{f}_2(\mu)$ is negative at these values. Due to the monotonicity of $\boldsymbol{f}_2(\mu)$, it is therefore negative for all $\mu$ in the interval $[0, 1/4]$. Hence, $\left(S_1^{(3)}\right)_{(4,22)}$ is negative for all $0 \leq \mu \leq 1/4$. □

We could at this point investigate whether there exists a value $\mu$ for which all subdivision matrices have positive entries. However, the main advantage of the first variant is that the distance between $\lambda$ and $\mu$ is at least $\lambda/2$. If we were to enforce the convex hull criterion, we would lose this crucial advantage of this variant.

The situation is different for the second variant, for which we obtain the following theorem:

**Theorem 5.13.** *All entries of the subdivision matrices of the four categories for $S_2$ are non-negative. Thus, together with Quality Criterion Q1, they satisfy Quality Criterion Q2 of the convex hull.*

*Proof.* By Theorem 5.10, all four categories of $S_2$ satisfy Quality Criterion Q1. We again start with the subdivision matrices of the initial elements. The off-diagonal entries of the Colin-de-Verdière-Matrix $C$ for $S_2$ are all non-positive by Definition 4.1. All diagonal entries of the diagonal matrix $N$ are negative by Proposition 4.9 and Theorem 4.14. Therefore, the off-diagonal entries of $\tilde{C}$ and consequently also of $\ln(2)(\tilde{C} - E_n)$ are all non-negative.

By Lemma B.63, which considers the exponential of matrices with non-negative off-diagonal entries, the matrix

$$S_2^{(t)} = \exp_2(\tilde{C} - E_n)$$

is correspondingly non-negative.

Since the subdivision matrix $\overline{S}_2^{(2)}$ only consists of entries from different subdivision matrices $S_2^{(2)}$ and zeros, all entries of $\overline{S}_2^{(2)}$ are also greater than or equal to zero. □

Combining the two quality criteria Q1 and Q2, we obtain the following corollary:

**Corollary 5.14.** *The matrices $S_1^{(2)}$, $\overline{S}_1^{(2)}$, and all categories of $S_2$ are stochastic.*

*Proof.* All matrices of these six categories have row sums equal to 1 and only non-negative entries. This is precisely the definition B.39 of stochastic matrices. □

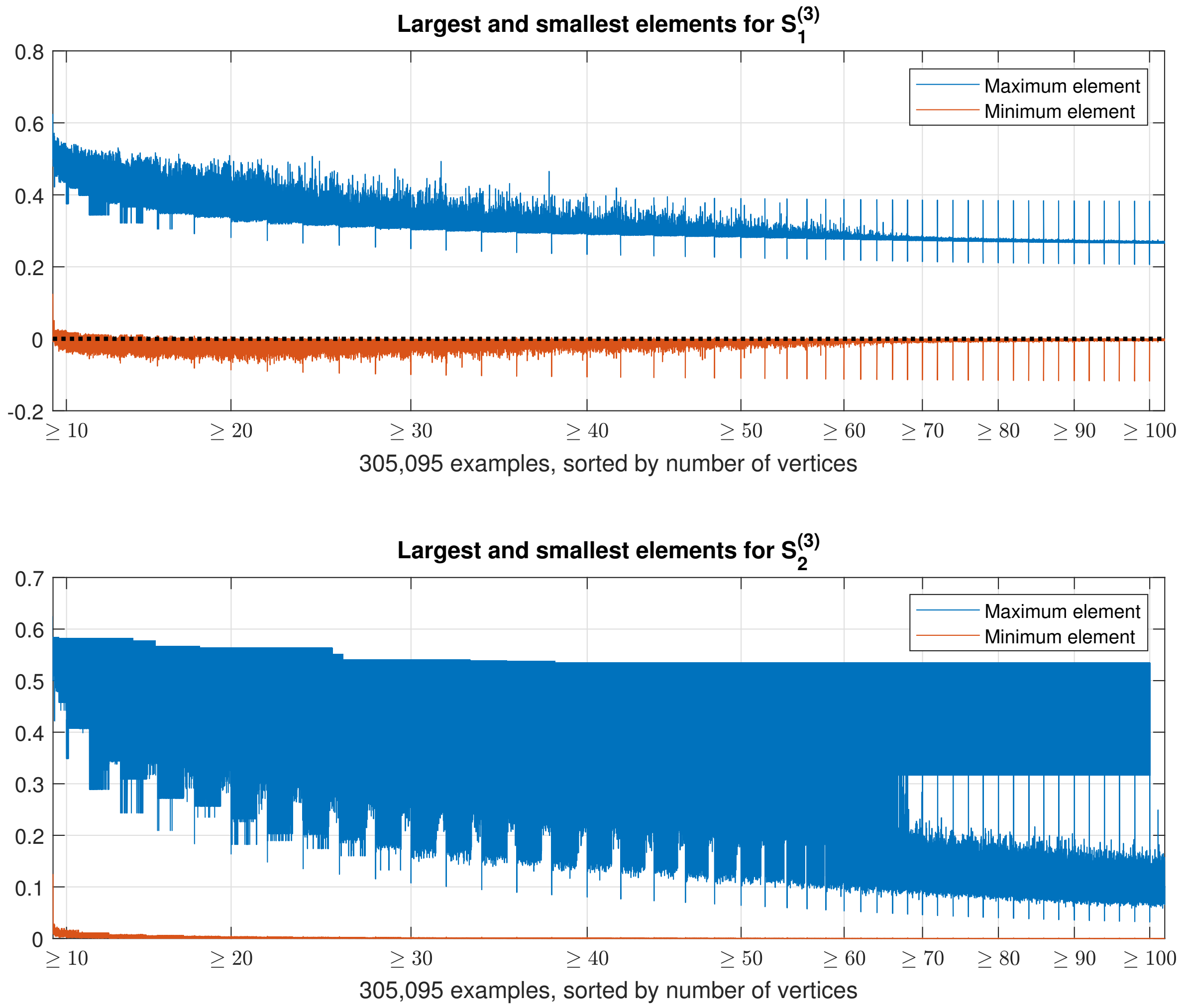

**Figure 5.7:** Largest and smallest entries of the subdivision matrices of the volumetric initial elements for variant 1 (above) and variant 2 (below).

Empirically, the entries of the three-dimensional initial elements are again of interest. The corresponding evaluation can be found in Figure 5.7.

For this evaluation, it is particularly noteworthy that negative entries are the rule for variant 1. The percentage of matrices with negative entries for the examples from Section 3.1.1 is given in the following table:

| | $S_1^{(2)}$ | $\overline{S}_1^{(2)}$ | $S_2^{(2)}$ | $\overline{S}_2^{(2)}$ | $S_1^{(3)}$ | $\overline{S}_1^{(3)}$ | $S_2^{(3)}$ | $\overline{S}_2^{(3)}$ |
|---|---|---|---|---|---|---|---|---|
| Examples with negative entries in % | 0 | 0 | 0 | 0 | 98.35 | 99.48 | 0 | 0 |

The fact that there are more matrices with negative entries in the category $\overline{S}_1^{(3)}$ than in $S_1^{(3)}$ can be explained by the presence of adjacency matrices of planar 3-connected graphs in the examples that do not represent a hexahedral structure. Such cases can contain initial elements for the shell case that are not prisms. The entries of the matrices of the initial elements for these non-prism structures may then possibly contain negative entries. This case is described in Chapter 7.

## Q3, Q4, Q5, Q6 and Q12 Spectrum of Eigenvalues

In this section, we summarize the statements about the eigenvalue spectrum of the subdivision matrices. We first consider the statements for the matrices of the initial elements and then address the ring and shell matrices. We start with variant 1 in the following theorem:

**Theorem 5.15.** *For the case $t = 2$ and for the case $t = 3$ where the associated graph **G** is not a combinatorial graph of a prism, the subdivision matrices $S_1^{(2)} \in \mathbb{R}^{n\times n}$ and $S_1^{(3)} \in \mathbb{R}^{n\times n}$ have an eigenvalue spectrum of*

$$1,\quad \underbrace{\frac{1}{2},\quad \cdots,\quad \frac{1}{2}}_{t\ \textit{times}},\quad \underbrace{\mu,\quad \cdots,\quad \mu}_{(n-t-1)\ \textit{times}}.$$

*For the case where $t = 3$ and **G** is the combinatorial graph of a prism, they have an eigenvalue spectrum of*

$$1,\quad \frac{1}{2},\quad \frac{1}{2},\quad \frac{1}{2},\quad \underbrace{\frac{1}{4},\quad \cdots,\quad \frac{1}{4}}_{(\frac{n}{2}-1)\ \textit{times}},\quad \underbrace{\frac{1}{8},\quad \cdots,\quad \frac{1}{8}}_{(\frac{n}{2}-3)\ \textit{times}}.$$

*For $\mu = 1/4$ they thus satisfy the quality criteria Q3, Q4, Q5 and Q6. Moreover, both subdivision matrices satisfy quality criterion Q12.*

*Proof.* We begin with the non-prism case. The matrix $S_1^{(t)}$ is computed according to Formula (5.4) from the $t+1$ vectors $\vec{1}$ and $\tilde{P}_{(:,1)}, \ldots, \tilde{P}_{(:,t)}$. For $t = 2$, the vectors $\tilde{P}_{(:,1)}$ and $\tilde{P}_{(:,2)}$ are orthogonal by construction, and for $t = 3$, the three vectors $\tilde{P}_{(:,1)}$, $\tilde{P}_{(:,2)}$, and $\tilde{P}_{(:,3)}$ are all mutually orthogonal due to Matlab's `orth` command. Orthogonality to the vector $\vec{1}$ follows from Equation (5.11). Normalizing the columns of $\tilde{P}$ and normalizing the vector $\vec{1}$ to $\vec{1}/\sqrt{n}$, these $t+1$ vectors form an orthonormal set.

These four vectors can be extended to a basis of $\mathbb{R}^n$ by the basis extension theorem B.12. Let the columns of the matrix

$$M = \begin{bmatrix} \frac{\vec{1}}{\sqrt{n}} & \frac{\tilde{P}_{(:,1)}}{|\tilde{P}_{(:,1)}|} & \cdots & \frac{\tilde{P}_{(:,t)}}{|\tilde{P}_{(:,t)}|} & \cdots \end{bmatrix}$$

be this basis. By applying Gram–Schmidt orthonormalization from theorem B.13, we can convert it into an orthonormal basis containing the vectors

$$\frac{\vec{1}}{\sqrt{n}},\quad \frac{\tilde{P}_{(:,1)}}{|\tilde{P}_{(:,1)}|}, \ldots, \frac{\tilde{P}_{(:,t)}}{|\tilde{P}_{(:,t)}|}$$

Let the columns of the matrix

$$\tilde{M} = \begin{bmatrix} \frac{\vec{1}}{\sqrt{n}} & \frac{\tilde{P}_{(:,1)}}{|\tilde{P}_{(:,1)}|} & \cdots & \frac{\tilde{P}_{(:,t)}}{|\tilde{P}_{(:,t)}|} & \cdots \end{bmatrix}$$

be this orthonormal basis. Note that for the proof, it is not necessary to explicitly compute the other entries of $\tilde{M}$; their existence suffices. Since all columns of $\tilde{M}$ are orthogonal and have length 1, we have $\tilde{M}\tilde{M}^T = E_n$ and thus $\tilde{M}$ is an orthogonal matrix.

Using this matrix, Formula (5.4) can be rewritten as

$$S_1^{(t)} = \left( \tilde{M} \begin{bmatrix} 1 & 0 & \cdots & 0 \\ 0 & 0 & \ddots & \vdots \\ \vdots & \ddots & \ddots & 0 \\ 0 & \cdots & 0 & 0 \end{bmatrix} \tilde{M}^T + \frac{1-2\mu}{2-2\mu} \tilde{M} \begin{bmatrix} 0 & \overbrace{0 \quad \cdots \quad \cdots}^{t \text{ columns}} & \cdots & \cdots & 0 \\ 0 & 1 \quad \ddots & & & \vdots \\ \vdots & \ddots \quad \ddots & \ddots & & \vdots \\ \vdots & \ddots & 1 & \ddots & \vdots \\ \vdots & & \ddots & 0 \quad \ddots & \vdots \\ \vdots & & & \ddots \quad \ddots & 0 \\ 0 & \cdots \quad \cdots & \cdots & \cdots \quad 0 & 0 \end{bmatrix} \tilde{M}^T \right) (1-\mu) + \mu \tilde{M} E_n \tilde{M}^T.$$

Distributing the factors yields

$$S_1^{(t)} = \tilde{M} \begin{bmatrix} 1-\mu & 0 & \cdots & 0 \\ 0 & 0 & \ddots & \vdots \\ \vdots & \ddots & \ddots & 0 \\ 0 & \cdots & 0 & 0 \end{bmatrix} \tilde{M}^T + \tilde{M} \begin{bmatrix} 0 & \overbrace{0 \quad \cdots \quad \cdots}^{t \text{ columns}} & & \cdots & \cdots & 0 \\ 0 & \frac{1-2\mu}{2-2\mu}(1-\mu) \quad \ddots & & & & \vdots \\ \vdots & \ddots \quad \ddots & \ddots & & & \vdots \\ \vdots & \ddots & \frac{1-2\mu}{2-2\mu}(1-\mu) & \ddots & & \vdots \\ \vdots & & \ddots & 0 & \ddots & \vdots \\ \vdots & & & \ddots & \ddots & 0 \\ 0 & \cdots \quad \cdots & \cdots & \cdots & 0 & 0 \end{bmatrix} \tilde{M}^T + \tilde{M} \mu E_n \tilde{M}^T.$$

Summing the matrices using associativity yields

$$S_1^{(t)} = \tilde{M} \begin{bmatrix} 1-\mu+\mu & \overbrace{0 \quad \cdots \quad \cdots}^{t \text{ columns}} & & \cdots & \cdots & 0 \\ 0 & \frac{1-2\mu}{2}+\mu \quad \ddots & & & & \vdots \\ \vdots & \ddots \quad \ddots & \ddots & & & \vdots \\ \vdots & \ddots & \frac{1-2\mu}{2}+\mu & \ddots & & \vdots \\ \vdots & & \ddots & \mu & \ddots & \vdots \\ \vdots & & & \ddots & \ddots & 0 \\ 0 & \cdots \quad \cdots & \cdots & \cdots & 0 & \mu \end{bmatrix} \tilde{M}^T = \tilde{M} \begin{bmatrix} 1 & \overbrace{0 \quad \cdots \quad \cdots}^{t \text{ columns}} & & \cdots & 0 & \\ 0 & \frac{1}{2} \quad \ddots & & & \vdots & \\ \vdots & \ddots \quad \ddots & \ddots & & \vdots & \\ \vdots & \ddots & \frac{1}{2} & \ddots & \vdots & \\ \vdots & & \ddots & \mu & \ddots & \vdots \\ \vdots & & & \ddots & \ddots & 0 \\ 0 & \cdots \quad \cdots & \cdots & \cdots & 0 & \mu \end{bmatrix} \tilde{M}^T. \tag{5.12}$$

Thus, 1 is the dominant eigenvalue with eigenvector $\vec{1}$, so quality criterion Q3 is satisfied. Moreover, $S_1^{(t)}$ has an eigenvalue $1/2$ of multiplicity $t$ corresponding to the eigenvectors $\tilde{P}$, so quality criteria Q4 and Q5 are fulfilled. Setting $\mu = 1/4$ also satisfies quality criterion Q6. The geometric multiplicity of the subdominant eigenvalue being equal to $t$ follows from the diagonalizability of $S_1^t$, which is directly evident from Equation (5.12).

This part of the proof reveals the necessity of the orthogonality of $\tilde{P}$. Without it, the matrix $\tilde{M}^T \neq \tilde{M}^{-1}$ in the above proof. The submatrix $\tilde{M}^{-1}_{(2:t+1,:)}$ is generally not equal to $\tilde{P}^T$ in this case. Hence, the above representation for $S_1^{(t)}$ would not be possible. Experiments also show that in this case $S_1^{(t)}$ neither exhibits the desired spectrum nor has the columns of $\tilde{P}$ as eigenvectors.

For quality criterion Q12, consider the expression

$$S_1^{(t)}\left(P - \frac{1}{n}\sum_{i=1}^{n} P_{(i,:)}\right).$$

The subtraction here is to be understood pointwise as in Equation (5.2). Using Equation (5.3), we have

$$S_1^{(t)}\left(P - \frac{1}{n}\sum_{i=1}^{n} P_{(i,:)}\right) = S_1^{(t)}\tilde{P}B^{-1} = \frac{1}{2}\tilde{P}B^{-1} = \frac{1}{2}\left(P - \frac{1}{n}\sum_{i=1}^{n} P_{(i,:)}\right).$$

Since $P$ is the matrix of vertices of a convex polytope, $P - \frac{1}{n}\sum_{i=1}^{n} P_{(i,:)}$ is also a convex polytope. Therefore, the subdominant eigenstructure of $S_1^{(t)}$ can be represented as a convex polytope, and hence quality criterion Q12 is satisfied.

Now consider the prism case for $S_1^{(3)} \in \mathbb{R}^{n\times n}$. We have already shown in Corollary 5.3 that for $t = 2$ the Doo-Sabin rules coincide with ours for $\mu = 1/4$. Hence, as shown in the first part of the proof, the eigenvalues of $S^{(2)}$ from Algorithm 10 are

$$1,\quad \frac{1}{2},\quad \frac{1}{2},\quad \underbrace{\frac{1}{4},\quad \dots,\quad \frac{1}{4}}_{(\frac{n}{2}-3)\text{-times}}.$$

The eigenvalues of $S^{(1)}$ are $1$ and $1/2$. Using Theorem B.32, which describes eigenvalues and eigenvectors of matrices obtained as Kronecker products, the eigenvalues of the prism matrix are the tensor product of the eigenvalues of $S^{(1)}$ and $S^{(2)}$, thus

$$1,\quad \frac{1}{2},\quad \frac{1}{2},\quad \frac{1}{2},\quad \underbrace{\frac{1}{4},\quad \dots,\quad \frac{1}{4}}_{(\frac{n}{2}-1)\text{-times}},\quad \underbrace{\frac{1}{8},\quad \dots,\quad \frac{1}{8}}_{(\frac{n}{2}-3)\text{-times}}.$$

The eigenvectors of $S^{(1)}$ are

$$\begin{bmatrix}1\\1\end{bmatrix}\quad \text{for eigenvalue } 1, \quad \text{and} \quad \begin{bmatrix}1\\-1\end{bmatrix}\quad \text{for eigenvalue } \frac{1}{2}.$$

The eigenvector of $S^{(2)}$ for eigenvalue $1$ is the vector $\vec{1}$. By Theorem B.32, $\vec{1} = \text{kron}(\vec{1},\vec{1})$ is an eigenvector to eigenvalue $1$ of $S_1^{(3)}$. For the prism case, quality criteria Q3, Q4, Q5 and Q6 are thus also satisfied.

Consider the eigenvectors corresponding to eigenvalue $1/2$ of the matrix $S_1^{(3)}$. The eigenvectors of the subdivision matrix $S^{(2)}$ to the eigenvalue $1/2$ generate a convex 2-polytope. Let this polytope be defined by the vertices $P^{(2)} \in \mathbb{R}^{\frac{n}{2}\times 2}$. Since the eigenvectors of $S_1^{(3)}$ are the Kronecker product of the eigenvectors of $S^{(1)}$ and $S^{(2)}$ by Theorem B.32, the eigenvectors of $S_1^{(3)}$ corresponding to eigenvalue $1/2$ are

$$\text{kron}\left(\begin{bmatrix}1\\1\end{bmatrix}, P^{(2)}_{(:,1)}\right) = \begin{bmatrix}P^{(2)}_{(:,1)}\\P^{(2)}_{(:,1)}\end{bmatrix},\quad \text{kron}\left(\begin{bmatrix}1\\1\end{bmatrix}, P^{(2)}_{(:,2)}\right) = \begin{bmatrix}P^{(2)}_{(:,2)}\\P^{(2)}_{(:,2)}\end{bmatrix},\quad \text{and}\quad \text{kron}\left(\begin{bmatrix}1\\-1\end{bmatrix}, \vec{1}\right) = \begin{bmatrix}\vec{1}\\-\vec{1}\end{bmatrix}.$$

These three vectors form a convex prism, and thus quality criterion Q12 is also satisfied for the prism case. Since they span a prism, they are also linearly independent. Hence the geometric multiplicity of the subdominant eigenvalue in the prism case is exactly $3 = t$. □

For the subdivision matrices $S_1^{(2)}$ and $S_1^{(3)}$ we also obtain the following corollary:

**Corollary 5.16.** *The matrices $S_1^{(2)} \in \mathbb{R}^{n\times n}$ and $S_1^{(3)} \in \mathbb{R}^{n\times n}$ are symmetric and therefore diagonalizable.*

*Proof.* For the non-prism case, $S_1^{(t)}$ according to Equation (5.4) is the sum of three matrices. The first matrix

$$\frac{\vec{1} \cdot \vec{1}^T}{|\vec{1}|^2}$$

is symmetric, since all entries are equal. The second matrix

$$\frac{\tilde{P}_{(:,1)} \cdot \tilde{P}_{(:,1)}^T}{|\tilde{P}_{(:,1)}|^2} + \cdots + \frac{\tilde{P}_{(:,t)} \cdot \tilde{P}_{(:,t)}^T}{|\tilde{P}_{(:,t)}|^2}$$

is symmetric because the product of two numbers commutes, and the third matrix is symmetric since it is a scalar multiple of the identity matrix. Hence $S_1^{(t)}$, being a scaled sum of symmetric matrices, is symmetric as well.

For the prism case, we obtain via the Kronecker product a matrix of the form

$$S_1^{(3)} = \begin{bmatrix} \frac{3}{4}S^{(2)} & \frac{1}{4}S^{(2)} \\ \frac{1}{4}S^{(2)} & \frac{3}{4}S^{(2)} \end{bmatrix}.$$

Since by Corollary 5.3 the matrices $S^{(2)}$ coincide with the matrices $S_1^{(2)}$ for $\mu = 1/4$ and the latter are symmetric by the above argument, it follows that

$$\left(S_1^{(3)}\right)^T = \begin{bmatrix} \frac{3}{4}S^{(2)} & \frac{1}{4}S^{(2)} \\ \frac{1}{4}S^{(2)} & \frac{3}{4}S^{(2)} \end{bmatrix}^T = \begin{bmatrix} \frac{3}{4}(S^{(2)})^T & \left(\frac{1}{4}S^{(2)}\right)^T \\ \frac{1}{4}(S^{(2)})^T & \left(\frac{3}{4}S^{(2)}\right)^T \end{bmatrix} = \begin{bmatrix} \frac{3}{4}S^{(2)} & \frac{1}{4}S^{(2)} \\ \frac{1}{4}S^{(2)} & \frac{3}{4}S^{(2)} \end{bmatrix} = S_1^{(3)},$$

thus $S_1^{(3)}$ is symmetric.

Since all entries of $S_1^{(t)}$ are real, Theorem B.45, which states the diagonalizability of real symmetric matrices, implies that $S_1^{(t)}$ is diagonalizable. □

For the matrices $S_2$ we first obtain diagonalizability analogous to the first variant with the following theorem:

**Theorem 5.17.** *The subdivision matrices $S_2^{(2)} \in \mathbb{R}^{n \times n}$ and $S_2^{(3)} \in \mathbb{R}^{n \times n}$ are diagonalizable.*

*Proof.* Since $C$ is a real symmetric matrix, it is diagonalizable by Theorem B.45. By Theorem B.48, which considers the diagonalizability of the product of a Hermitian and a positive definite Hermitian matrix, the matrix

$$\tilde{C} = -(-NC)$$

is diagonalizable. Let $\tilde{D}$ be the diagonal matrix similar to $\tilde{C}$. Then by Equation (5.8) we have

$$S_2 = V \begin{bmatrix} \exp_2\left(\tilde{D}_{(1,1)} - 1\right) & 0 & \cdots & 0 \\ 0 & \ddots & \ddots & \vdots \\ \vdots & \ddots & \ddots & 0 \\ 0 & \cdots & 0 & \exp_2\left(\tilde{D}_{(n,n)} - 1\right) \end{bmatrix} V^{-1}$$

and thus $S_2$ is diagonalizable as well. □

Furthermore, for the subdivision matrices of the initial elements $S_2$ we obtain almost the same result regarding eigenvalues as for the first variant:

**Theorem 5.18.** *The subdivision matrices $S_2^{(2)} \in \mathbb{R}^{n\times n}$ and $S_2^{(3)} \in \mathbb{R}^{n\times n}$ have an eigenvalue spectrum*

$$1, \quad \underbrace{\frac{1}{2}, \quad \cdots, \quad \frac{1}{2},}_{t \text{ times}} \quad > \quad \lambda_{t+1}, \quad \geq \quad \cdots, \quad \geq \quad \lambda_{n-1} > 0,$$

*with $\lambda_{t+1}, \ldots, \lambda_{n-1} \in \mathbb{R}$. They thus satisfy the quality criteria Q3, Q4 and Q5. Moreover, both matrices satisfy quality criterion Q12.*

*Proof.* We start with the Colin-de-Verdière-matrix $C$. By definition, it is symmetric and real, hence Hermitian. By Proposition B.44, the eigenvalues of Hermitian matrices are real, so the eigenvalues of $C$ are real. Moreover, by definition, the matrix $C$ has exactly one negative eigenvalue and, by Theorems 4.8 and 4.13, exactly $t$ eigenvalues equal to zero. All other eigenvalues of $C$ are positive.

The eigenvalues of the diagonal normalization matrix $N$ are its diagonal entries, which, by Theorem 4.9 and Theorem 4.14, are real and negative. Thus, $N$ is negative definite, and $-N$ is positive definite.

By Theorem B.48, the product $-NC$ of a Hermitian positive definite matrix $-N$ and a Hermitian matrix $C$ is diagonalizable with real eigenvalues. Moreover, the signs of the eigenvalues of $-NC$ coincide with those of $C$. For the matrix

$$\tilde{C} = NC = -(-NC),$$

the signs of the real eigenvalues are reversed accordingly. Hence, $\tilde{C}$ has one positive eigenvalue, exactly $t$ eigenvalues equal to zero, and $n-(t+1)$ negative eigenvalues. Since $\tilde{C}$ is normalized, the positive eigenvalue is exactly $1$ with eigenvector $\vec{1}$.

By Equation (5.9), shifting by the identity matrix $E_n$ and then exponentiating maps the eigenvalue $1$ to $1$, the eigenvalues $0$ to $1/2$, and all other eigenvalues to values in the interval $(0, 1/2)$. Since the eigenvectors remain unchanged in this process (see Equation (5.8)), these shifted eigenvalues are the eigenvalues of $S_2$. Therefore, the eigenvalues are

$$1, \quad \underbrace{\frac{1}{2}, \quad \cdots, \quad \frac{1}{2},}_{t \text{ times}} \quad > \quad \lambda_{t+1}, \quad \geq \quad \cdots, \quad \geq \quad \lambda_{n-1} > 0,$$

with $\lambda_{t+1}, \ldots, \lambda_{n-1} \in \mathbb{R}$, and thus the subdivision matrices $S_2^{(t)}$ satisfy the quality criteria Q3, Q4 and Q5.

Next, we consider quality criterion Q12. Let $P$ be the vertices of the convex polytope from which the matrix $C$ was constructed. By Theorems 4.8 and 4.13, we have $CP = \vec{0}$.

Multiplying by $N$ does not change the kernel of $C$, since

$$NCP = N \cdot \vec{0} = \vec{0}.$$

Thus, the columns of $P$ are eigenvectors of $\tilde{C}$ corresponding to the eigenvalue $0$, and therefore eigenvectors of $S_2$ corresponding to the eigenvalue $1/2$. Hence, the eigenspace of the subdominant eigenvalue of $S_2$ is spanned by a convex polytope, which means quality criterion Q12 is fulfilled. Since the columns of $P$ span a $t$-polytope, they are linearly independent, so the geometric multiplicity of the subdominant eigenvalue is exactly $t$. □

Furthermore, we obtain symmetry for $S_2^{(2)}$ with the following theorem:

**Proposition 5.19.** *The subdivision matrix $S_2^{(2)}$ is symmetric.*

*Proof.* The Colin-de-Verdière-matrix $C$ corresponding to $S_2^{(2)}$ is symmetric. Since the row sums of $C$ are all equal due to the symmetric arrangement of the 2-polytope, the matrix $NC$ is symmetric, and hence also $\ln(2)(NC - E_n)$ is symmetric. The matrix exponential is a power series in the matrix, and powers of symmetric matrices satisfy

$$(M^a)^T = (\underbrace{M \cdots M}_{a \text{ times}})^T = \underbrace{M^T \cdots M^T}_{a \text{ times}} = \underbrace{M \cdots M}_{a \text{ times}} = M^a \quad \text{for} \quad a \in \mathbb{N},$$

so these powers remain symmetric. Therefore, $S_2^{(2)}$ is symmetric. □

However, quality criterion Q6 does not hold for $S_2$. A counterexample for $t = 2$ is given by a combinatorial arrangement with the adjacency matrix $A$ below. The corresponding subdivision matrix has eigenvalues

$$1, \quad \tfrac{1}{2}, \quad \tfrac{1}{2}, \quad 0.1629, \quad \text{and} \quad 0.1629,$$

with

$$A = \begin{bmatrix} & 1 & & & 1 \\ 1 & & 1 & & \\ & 1 & & 1 & \\ & & 1 & & 1 \\ 1 & & & 1 & \end{bmatrix}.$$

A counterexample for $t = 3$ is the subdivision matrix from Example 5.7. Its eigenvalues (rounded to four decimal places) are

$$1, \quad \tfrac{1}{2}, \quad \tfrac{1}{2}, \quad \tfrac{1}{2}, \quad 0.2172, \quad \cdots .$$

Thus, counterexamples can be constructed for both cases.

Moreover, the subdivision matrix $S_2^{(3)}$ is not symmetric. A counterexample also appears in Example 5.7. The matrix from Equation (5.10) is not symmetric since the Colin-de-Verdière-matrix $C$ from Theorem 4.13 has unequal row sums here. Therefore, $NC$ is not symmetric, and consequently, $S_2^{(3)}$ is not symmetric either.

We now consider the empirical results for the subdivision matrices of the initial elements. The eigenvalues of the sample collection are shown in Figure 5.8 and are discussed below.

Initially, it is noticeable for all variants that the eigenvalues $\mu = \lambda_{t+1}$ and $\lambda_{t+2}$ in Figure 5.8 appear to be zero for examples with a low number of vertices. The two smallest possible examples for $t = 2$ are the triangle and the quadrilateral, and for $t = 3$ they are the tetrahedron and the four-sided pyramid. These each have only 3, 4, or 5 eigenvalues, respectively, so in the evaluation the values for $\mu = \lambda_{t+1}$ and $\lambda_{t+2}$ are accordingly set to zero.

For the accuracy, we consider the following table, which summarizes the relevant data:

| | $S_1^{(2)}$ | $S_2^{(2)}$ | $S_1^{(3)}$ | $S_2^{(3)}$ |
|---|---|---|---|---|
| $\max \lvert\lambda_0 - 1\rvert$ | 1,1102e−15 | 7,9270e−14 | 2,5535e−15 | 1,36e−13 |
| $\max \lvert\lambda_1 - 0.5\rvert$ | 5,5511e−16 | 4,2411e−14 | 1,3323e−15 | 1,2013e−13 |
| $\max \lvert\lambda_2 - 0.5\rvert$ | 6,1062e−16 | 5,4734e−14 | 8,3267e−16 | 8,4543e−14 |
| $\max \lvert\lambda_3 - 0.5\rvert$ | | | 1,3878e−15 | 1,8152e−13 |
| $\max \lvert\mu - 0.25\rvert$ | 9,9920e−16 | | 1,4988e−15 | |
| $\max \lvert\mu\rvert$ | | 0,25 | | 0,353 55 |
| $\max \lvert SP - 0.5P\rvert$ | 1,1258e−15 | 8,6098e−14 | 1,026e−13 | 1,3305e−12 |

Here, the maximum, as in the other tables, runs over all considered examples. We see that all errors lie within an acceptable range and thus the quality criteria are also empirically verified.

Particularly interesting are the maximal values for $\mu$ of the matrices of variant 2. For $t = 2$ these seem to decrease with increasing number of vertices, whereas for $t = 3$ this does not appear to hold. In Figure 5.8 it can be seen that on average the sub-subdominant eigenvalue decreases with increasing vertex count, but there are always examples where a high value is reached. Interestingly, the peaks in Figure 5.8 correspond to the trapezohedra discussed in Chapter 7. The measured maximum value for $S_2^{(3)}$ of approximately $0.35$ is indeed greater than $1/4$, but the distance to $1/2$ is still large enough so that the local asymptotic behavior of the subdominant eigenvalues dominates sufficiently quickly.

Considering the eigenvalues of the ring and shell matrices, significantly fewer statements can be made. However, we at least obtain the following transfer of the eigenvalues of the corresponding subdivision matrices of the initial elements:

**Proposition 5.20.** *Let $\overline{S}$ be a subdivision matrix of the form $\overline{S}^{(2)}$ or $\overline{S}^{(3)}$, i.e., a subdivision matrix of a ring or a shell, with a central initial element whose subdivision matrix is $S^{(2)}$ or $S^{(3)}$. Then the eigenvalues of $S^{(t)}$ form a subset of the eigenvalues of $\overline{S}^{(t)}$.*

*Proof.* The refinement rules of the initial elements depend, according to Definition 2.11 and Algorithms 10 and 11, only on the control points of the initial element. Since in the ring and shell case the central initial element maps to itself, all entries in the corresponding refinement rules, and thus in the corresponding rows of the matrix $\overline{S}$, that do not belong to $S$ are zero. With a suitable permutation $H$, the matrix $\overline{S}$ can be brought into the form

$$H\overline{S}H^T = \begin{bmatrix} M_1 & M_2 \\ 0 & S \end{bmatrix}.$$

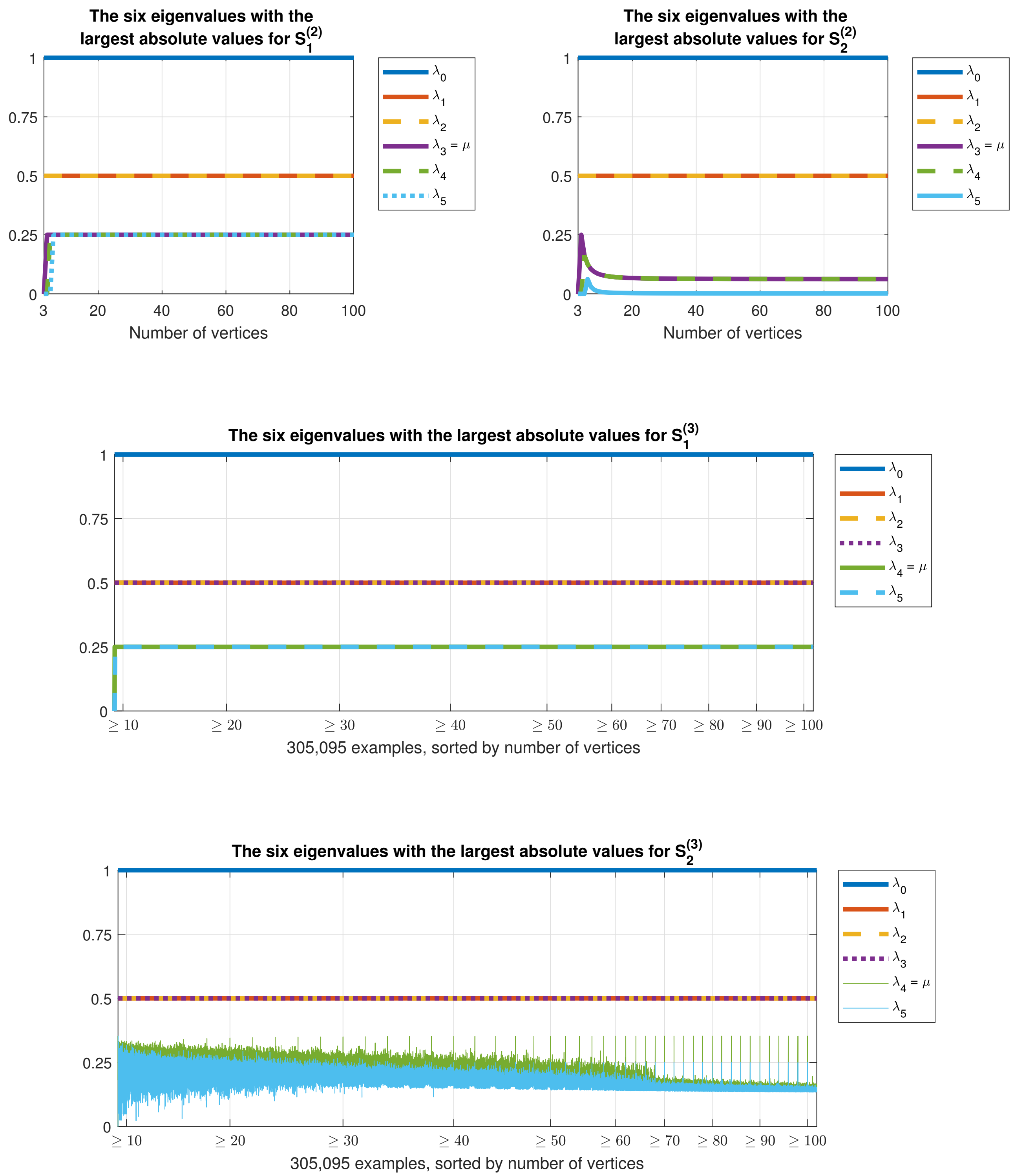

**Figure 5.8:** The eigenvalues of the matrices of the initial elements for variants 1 and 2.

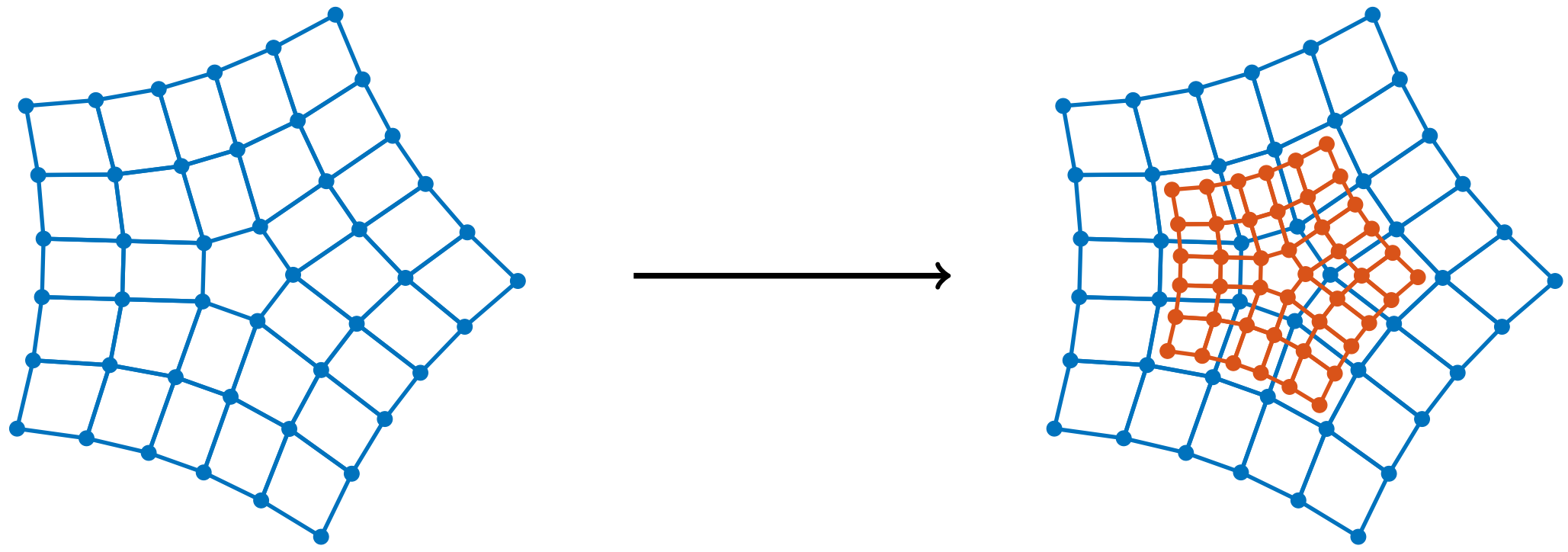

**Figure 5.9:** The lattice structure of control points refined by a ring matrix (left) and the invariant refinement (right). It can be seen that the outer ring of control points has no influence on the refinement. The corresponding columns in the subdivision matrix are therefore zero.

Since $H\overline{S}H^T$ is similar to $\overline{S}$, both matrices have the same eigenvalues by Corollary B.31. By Proposition B.34, the eigenvalues of $H\overline{S}H^T$ are the eigenvalues of $M_1$ and of $S$. Hence, the eigenvalues of $S$ form a subset of the eigenvalues of $\overline{S}$. □

Additionally, we can state the following about the ring matrices:

**Theorem 5.21.** *The matrices $\overline{S}_1^{(2)}$ and $\overline{S}_2^{(2)}$ satisfy the quality criteria Q3, Q4, and Q5. The matrix $\overline{S}_1^{(2)}$ furthermore satisfies quality criterion Q6. Furthermore, both matrices have only real eigenvalues.*

*Proof.* For the case $t = 2$, the subdivision matrices of the rings consist only of regular initial elements and the central initial element. This case is discussed in [PR08, Section 6.2, pp. 116–120]. Peters and Reif show that from the outer region only eigenvalues $1/4$, $1/8$, $1/16$, and $0$ appear (cf. [PR08, p. 117]). Since this outer region in [PR08] coincides with ours, the quality criteria Q3, Q4, and Q5 transfer from the matrices of the initial elements to the ring matrices.

Because for the subsubdominant eigenvalue $\mu$ of the subdivision matrix $S_1^{(2)}$ it holds that $\mu \leq 1/4$, quality criterion Q6 is also fulfilled for $\overline{S}_1^{(2)}$. Furthermore, by this reasoning, the eigenvalues of both variants are real. □

However, we have no access to the eigenvalues of $\overline{S}_1^{(3)}$ and $\overline{S}_2^{(3)}$. The central element for the analysis of the eigenstructure in the case $t = 2$ is the discrete Fourier transform. It is exploited that the subdivision matrices are cyclic, and the discrete Fourier transform is used to make statements about the structure. This cyclicity is missing in the case $t = 3$ (except for the prism case), as well as the general theoretical background.

What we can only state here is that the eigenvalues of $\overline{S}_2^{(3)}$ all have absolute value less than or equal to $1$. This follows from the fact that $\overline{S}_2^{(3)}$ is stochastic. Likewise, we cannot make statements about diagonalizability or, for $t = 3$, about the reality of the eigenvalues.

We can, however, exclude that the ring and shell matrices are symmetric. Due to the invariant structure, there exist columns in the subdivision matrices for both $t = 2$ and $t = 3$ that consist solely of zero entries. Since each row sums to $1$, the matrices cannot be symmetric. An illustration of this can be found in Figure 5.9.

It is all the more crucial here to rely on empirical evaluation. Analogous to the case of the initial elements, we obtain the eigenvalues summarized in Figure 5.10. It can be seen that the quality criteria cannot be formally proven here but are satisfied for the examples shown.

Also interesting is the behavior of the subsubdominant eigenvalue. For $t = 2$, the outer part of the ring structure produces eigenvalues of value $1/4$, so empirically the matrix $S_2^{(2)}$ also satisfies quality criterion Q6.

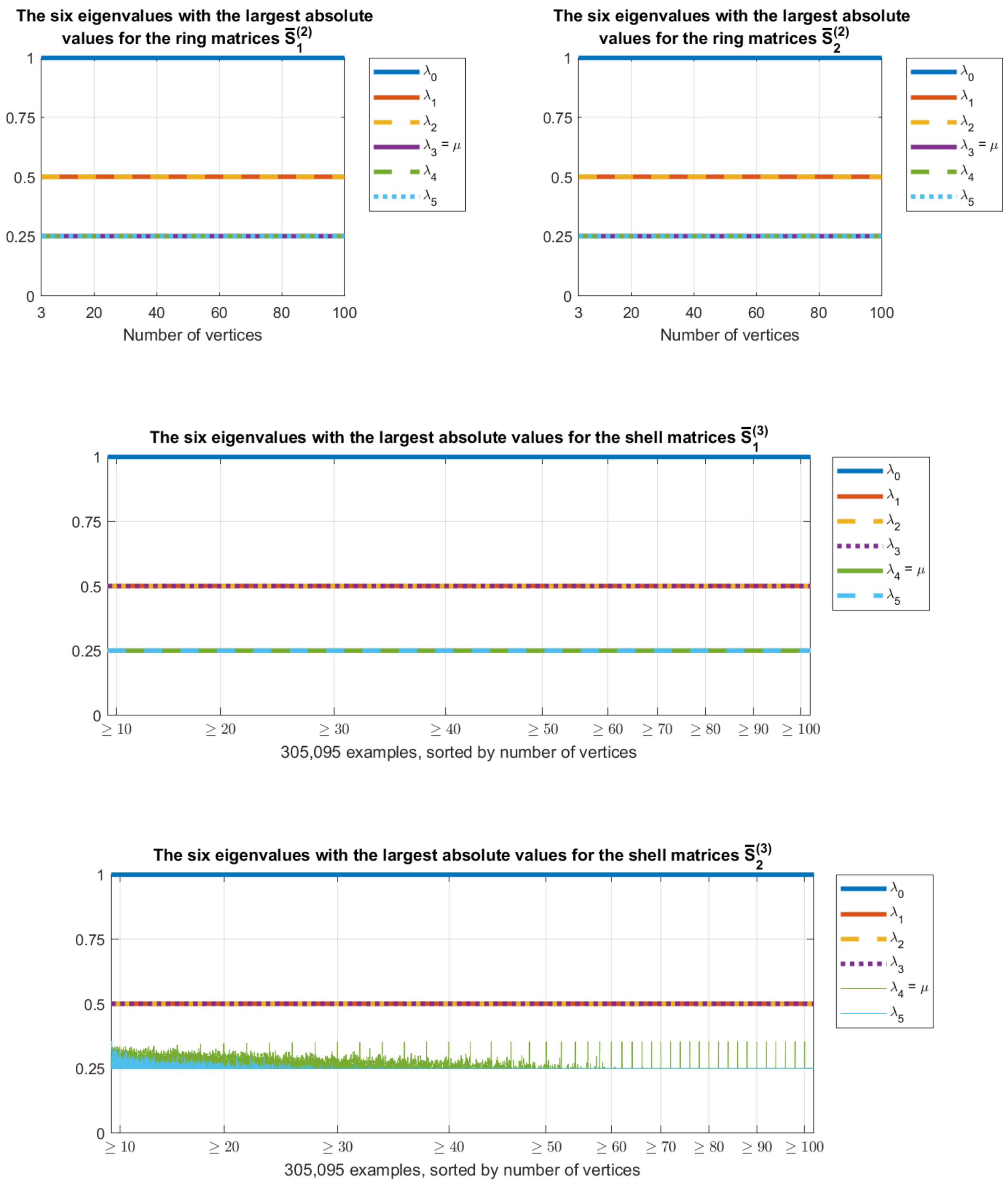

**Figure 5.10:** Eigenvalues of the ring and shell matrices for variants 1 and 2.

For $t = 3$, the criterion does not hold; however, it appears that eigenvalues of value $1/4$ are added from the outer part, so the subsubdominant eigenvalues in the figure are bounded below by $1/4$.

For the errors in the empirical evaluations, we obtain the following table analogous to the initial elements:

| | $\overline{S}_1^{(2)}$ | $\overline{S}_2^{(2)}$ | $\overline{S}_1^{(3)}$ | $\overline{S}_2^{(3)}$ |
|---|---|---|---|---|
| $\max\|\lambda_0 - 1\|$ | 3,1086e−15 | 6,7502e−14 | 8,8818e−15 | 1,3634e−13 |
| $\max\|\lambda_1 - 0.5\|$ | 8,8818e−16 | 3,2974e−14 | 6,4393e−15 | 1,1957e−13 |
| $\max\|\lambda_2 - 0.5\|$ | 1,6653e−15 | 3,2974e−14 | 4,3299e−15 | 8,3211e−14 |
| $\max\|\lambda_3 - 0.5\|$ | | | 6,2172e−15 | 1,808e−13 |
| $\max\|\mu - 0.25\|$ | 9,3708e−9 | 1,9429e−15 | 1,1702e−8 | |
| $\max\|\mu\|$ | | 0,25 | | 0,353 55 |

Thus, the quality criteria are at least empirically satisfied.

Regarding quality criterion Q12, no statement can be made for the ring and shell matrices. The structure of the central initial element must be identical to the subdivision matrix of the initial element. However, the outer part of the control points cannot be described. Empirically, there is also no way to quantify this criterion, as only a visual inspection could provide a qualitative assessment.

## Q7 and Q8 Tensor Product Structure

For the case $t = 2$, the only tensor product structure is the reproduction of the regular case as the Kronecker product of two one-dimensional cases. Variant 1 for $\mu = 1/4$ and Variant 2 fulfill this reproduction. This can be verified by straightforward substitution of the graph structure into the algorithms. Since for the regular case the ring matrices consist only of regular initial elements, the tensor product structure also arises for $\overline{S}^{(2)}$. Thus, for all variants of $S^{(2)}$ and $\overline{S}^{(2)}$, quality criterion Q7 is satisfied.

The interesting case in this context is the case $t = 3$. For initial elements, the graphs represent prisms in both the semi-regular and the regular case. Let us first consider Variant 1. As already described in Section 5.2.1, formula (5.4) cannot reproduce the semi-regular and the regular case because the subdivision matrix would need to have at least four different eigenvalues. We therefore enforce the tensor product structure by treating the prism case separately in Algorithm 10 and computing the subdivision matrix by

$$S_1^{(3)} := \text{kron}\left(S^{(1)}, S^{(2)}\right).$$

This directly yields quality criterion Q8 for $S_1^{(3)}$. The reproduction of the regular case and thus quality criterion Q7 also follows by direct substitution and is likewise fulfilled with this construction for $S_1^{(3)}$.

The two-part construction strategy, as described in Section 2.4 for quality criterion Q8, is indeed possible, but at this point it would be desirable to have a unified construction for all cases.

Variant 2 satisfies both quality criteria, however, with a unified construction of the subdivision matrices, which we explain in the following:

**Theorem 5.22.** *Let $A^{(2)}$ be the adjacency matrix of a two-dimensional initial element with $n$ vertices, and let*

$$A := \begin{bmatrix} A^{(2)} & E_n \\ E_n & A^{(2)} \end{bmatrix}$$

*be the adjacency matrix of a prism in the case $t = 3$. Then the subdivision matrix $S_2^{(3)}$ can be represented as a Kronecker product of two matrices, one of which is*

$$S^{(1)} = \frac{1}{4}\begin{bmatrix} 3 & 1 \\ 1 & 3 \end{bmatrix}$$

*and the other is described in the course of the following proof.*

The following proof is based on ideas and discussions with Kai Hormann, who developed the sketch for it.

*Proof.* To simplify the notation, we define for this proof

$$a := \cos\left(\frac{\pi}{n}\right) \quad \text{and} \quad b := \sin\left(\frac{\pi}{n}\right).$$

To access the concrete coordinates of the 3-polytope corresponding to the adjacency matrix $A$, we first construct the coordinates of a prism with an $n$-sided base whose edges are tangent to the unit sphere. We start with the edges of the lateral surface. The tangent points of the edges are placed equally spaced around the equator of the unit sphere. Specifically, we define the points as

$$e_i := \left(\cos\left(\frac{2\pi i}{n}\right), \sin\left(\frac{2\pi i}{n}\right), 0\right) \quad \text{for} \quad i = 1, \dots, n.$$

We set the direction of the corresponding edge of the 3-polytope as $[0, 0, 1]$. Thus, the corners of both base faces lie on these edges, and the $x$- and $y$-coordinates of the vertices of the 3-polytope are fixed. The two base faces are therefore regular $n$-gons with a circumradius of $1$. An illustration is shown in Figure 5.11.

In the next step, we need to determine the height of the two base faces so that their edges are also tangent to the unit sphere. For this, we must shift the base facet shown in Figure 5.11 upwards and downwards until its edges just touch the unit sphere.

By symmetry, the midpoint of the edges of both base faces lies tangent to the unit sphere. Since the circumradius of the base facet is exactly $1$, the distance from the tangent point to the center of the $n$-gon is

$$\cos\left(\frac{\frac{2\pi}{n}}{2}\right) = a.$$

An illustration can be found in Figure 5.11. Because the tangent points lie on the unit circle, the length of their position vectors is exactly $1$, so the height of the base facet is

$$\sqrt{1 - a^2} = b \quad \Rightarrow \quad z = \pm b.$$

An illustration of this can also be found in Figure 5.11. Thus, we obtain the following coordinates for the primal polytope:

$$p_i = \left(\cos\left(\frac{2\pi i}{n}\right), \sin\left(\frac{2\pi i}{n}\right), b\right) \qquad \text{für } i = 1, \dots, n \quad \text{und}$$
$$p_i = \left(\cos\left(\frac{2\pi i}{n}\right), \sin\left(\frac{2\pi i}{n}\right), -b\right) \qquad \text{für } i = n+1, \dots, 2n$$

We will refer to these points in the following. To keep the notation as simple as possible, we index the points using the variable $i$. For the top base facet, $i$ runs from $1$ to $n$. In this case, we identify $0 \equiv n$ and $n + 1 \equiv 1$. For the bottom base facet, $i$ runs from $n + 1$ to $2n$. In this case, we identify $n \equiv 2n$ and $2n + 1 \equiv n + 1$.

The points of the dual polytope can also be determined geometrically. The direction of the normals of the faces of the lateral surfaces is orthogonal to the equator of the unit sphere. Since the tangent points, as shown in Figure 5.11, have a circumradius of $1$, the projection of the dual points lies at

$$a\left(\cos\left(\frac{(2i+1)\pi}{n}\right), \sin\left(\frac{(2i+1)\pi}{n}\right), 0\right) \quad \text{for } i = 1, \dots, n,$$

and using Equation (3.12), we obtain for the dual points

$$f_i = \frac{1}{a}\left(\cos\left(\frac{(2i+1)\pi}{n}\right), \sin\left(\frac{(2i+1)\pi}{n}\right), 0\right) \quad \text{for } i = 1, \dots, n.$$

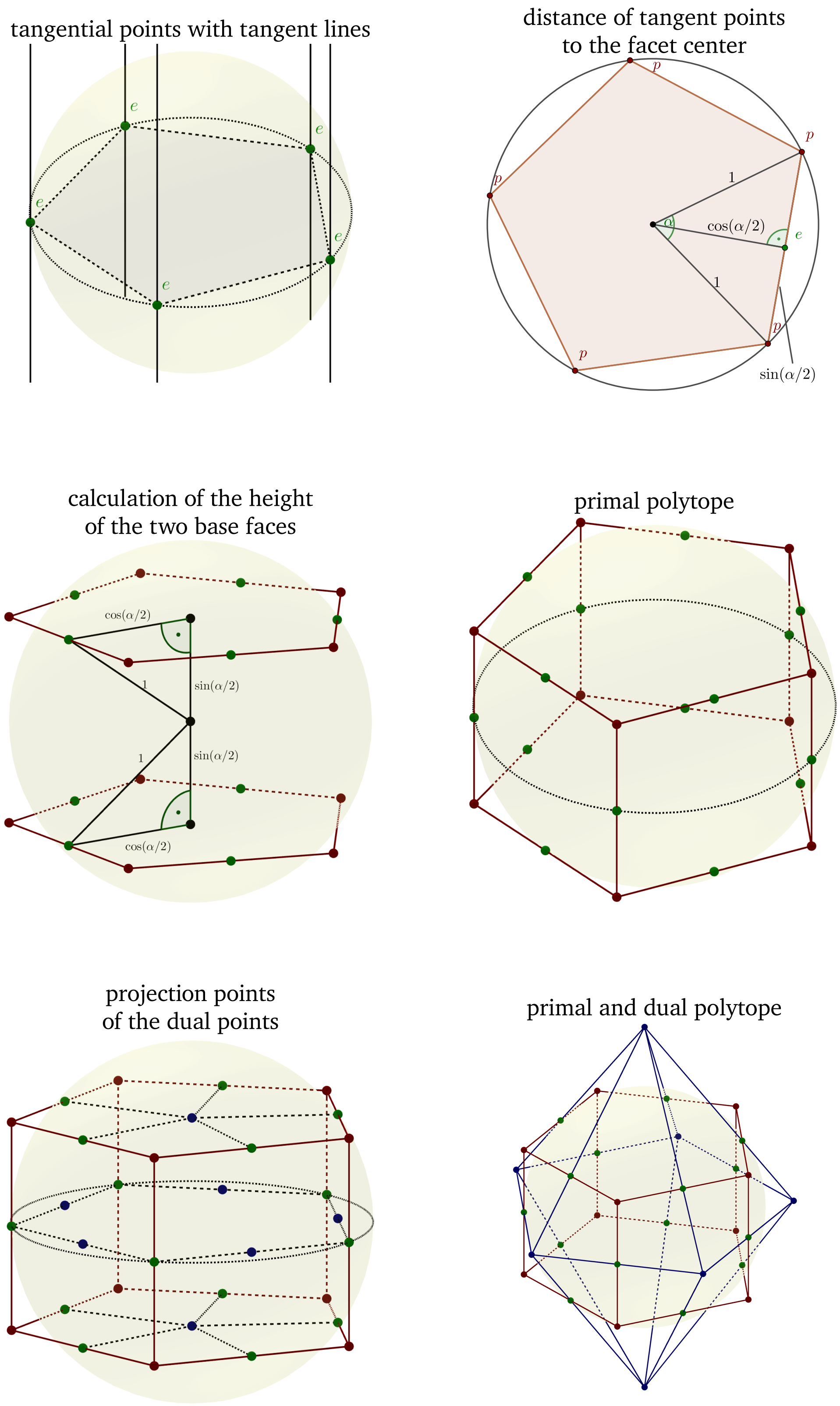

**Figure 5.11:** Different construction steps of the primal and dual points.

Similarly, for the dual points of the two base faces we have

$$f_0 = \left(0, 0, \frac{1}{b}\right) \quad \text{and} \quad f_{n+1} = \left(0, 0, -\frac{1}{b}\right).$$

An illustration is given in Figure 5.11.

With these coordinates, we can compute the Colin-de-Verdière-matrix of the polytope. To avoid lengthy transformations of sine and cosine terms, we use a computer algebra system for their calculation. For the off-diagonal entries, we get two different values. We start with the vertical edges and obtain

$$|p_i \times p_{i+n}| = \left\| \begin{bmatrix} -2\sin\left(\frac{2\pi i}{n}\right) b \\ 2\cos\left(\frac{2\pi i}{n}\right) b \\ 0 \end{bmatrix} \right\| = 2b,$$

and

$$|f_{i-1} - f_i| = \left\| \frac{1}{a} \begin{bmatrix} \cos\left(\frac{(2i-1)\pi}{n}\right) - \cos\left(\frac{(2i+1)\pi}{n}\right) \\ \sin\left(\frac{(2i-1)\pi}{n}\right) - \sin\left(\frac{(2i+1)\pi}{n}\right) \\ 0 \end{bmatrix} \right\| = 2\frac{b}{a}$$

for $i = 1, \ldots, n$. Thus, the non-zero entries are

$$C_{(i,i+n)} = C_{(i+n,i)} = -\frac{2\frac{b}{a}}{2b} = -\frac{1}{a}.$$

Next, we consider the horizontal edges. For these, we obtain

$$|p_i \times p_{i+1}| = \left\| \begin{bmatrix} \pm\sin\left(\frac{2\pi i}{n}\right) b \mp \sin\left(\frac{2\pi(i+1)}{n}\right) b \\ \pm\cos\left(\frac{2\pi i}{n}\right) b \mp \cos\left(\frac{2\pi(i+1)}{n}\right) b \\ \cos\left(\frac{2\pi i}{n}\right)\sin\left(\frac{2\pi(i+1)}{n}\right) - \sin\left(\frac{2\pi i}{n}\right)\cos\left(\frac{2\pi(i+1)}{n}\right) \end{bmatrix} \right\| = 2b,$$

for $i = 1, \ldots, n$ and $i = n+1, \ldots, 2n$. Furthermore, we have

$$|f_i - f_0| = |f_i - f_{n+1}| = \left\| \begin{bmatrix} \frac{1}{a}\cos\left(\frac{(2i+1)\pi}{n}\right) \\ \frac{1}{a}\sin\left(\frac{(2i+1)\pi}{n}\right) \\ \pm\frac{1}{b} \end{bmatrix} \right\| = \frac{1}{ba},$$

for $i = 1, \ldots, n$. Therefore, the non-zero entries are

$$C_{(i,i+1)} = C_{(i+1,i)} = -\frac{\frac{1}{ba}}{2b} = -\frac{1}{2b^2 a},$$

for $i = 1, \ldots, n$ and $i = n+1, \ldots, 2n$. It remains to compute the diagonal entries of $C$. Due to the symmetry of the prism, all diagonal entries are equal and given by formula (4.5) as

$$C_{(i,i)} = -\frac{\langle p_i, \sum_{j,j\neq i} C_{(i,j)} p_j \rangle}{\langle p_i, p_i \rangle}.$$

The sum in the numerator consists of three terms corresponding to the two adjacent nodes in the base facet and the adjacent node in the opposite base facet. Hence,

$$\sum_{j,j\neq i} C_{(i,j)}p_j = -\frac{1}{2b^2a}\left(\begin{bmatrix}\cos\left(\frac{2\pi(i-1)}{n}\right)\\ \sin\left(\frac{2\pi(i-1)}{n}\right)\\ \pm b\end{bmatrix} + \begin{bmatrix}\cos\left(\frac{2\pi(i+1)}{n}\right)\\ \sin\left(\frac{2\pi(i+1)}{n}\right)\\ \pm b\end{bmatrix}\right) - \frac{1}{a}\begin{bmatrix}\cos\left(\frac{2\pi i}{n}\right)\\ \sin\left(\frac{2\pi i}{n}\right)\\ \mp b\end{bmatrix}.$$

Using addition formulas, we find the length of the above expression to be

$$\left|\sum_{j,j\neq i} C_{(i,j)}p_j\right| = \frac{a}{b^2}\sqrt{1+b^2}.$$

The length of $p_i$ is

$$|p_i| = \sqrt{1+b^2}.$$

Because $p_i$ and $\sum_{j,j\neq i} C_{(i,j)}p_j$ point in the same direction up to sign (cf. Equation (4.3)), we get

$$|C_{(i,i)}| = \frac{\frac{a}{b^2}\sqrt{1+b^2}^2}{\sqrt{1+b^2}^2} = \frac{a}{b^2}.$$

Checking the sign gives

$$C_{(i,i)} = \frac{a}{b^2}.$$

With the above computations, all entries of the Colin-de-Verdière-matrix are explicitly determined and we get

$$C = \frac{1}{a}\begin{bmatrix} M & -E_n \\ -E_n & M\end{bmatrix} \quad \text{with} \quad M = \frac{1}{2b^2}\begin{bmatrix} 2a^2 & -1 & 0 & \cdots & 0 & -1 \\ -1 & 2a^2 & -1 & 0 & \cdots & 0 \\ 0 & \ddots & \ddots & \ddots & \ddots & \vdots \\ \vdots & \ddots & \ddots & \ddots & \ddots & 0 \\ 0 & \cdots & 0 & -1 & 2a^2 & -1 \\ -1 & 0 & \cdots & 0 & -1 & 2a^2 \end{bmatrix}. \tag{5.13}$$

Moreover, the sum of each row is

$$\sum_j C_{(i,j)} = -\frac{1}{a} - 2\cdot\frac{1}{2b^2a} + \frac{2a^2}{2b^2a} = \frac{2a^2-2b^2-2}{2b^2a} = \frac{2-2b^2-2b^2-2}{2b^2a} = \frac{-2b^2}{b^2a} = -\frac{2}{a}.$$

Now consider the formula for the subdivision matrix. Using Theorem B.53, it can be rewritten as

$$S_2^{(3)} = \exp_2(NC - E_{2n}) = \exp_2(NC)\exp_2(-E_{2n}) = \frac{1}{2}\exp_2(NC).$$

Since $N$ is a diagonal matrix with identical diagonal entries $-\frac{a}{2}$, we get

$$S_2^{(3)} = \frac{1}{2}\exp_2(NC) = \frac{1}{2}\exp_2\left(-\frac{a}{2}C\right).$$

We split $C$ into the sum of two matrices, defining

$$M_1 := \frac{1}{a}\begin{bmatrix} M & 0 \\ 0 & M\end{bmatrix} \quad \text{and} \quad M_2 := \frac{1}{a}\begin{bmatrix} 0 & -E_n \\ -E_n & 0\end{bmatrix}.$$

These matrices commute since

$$M_1 M_2 = \frac{1}{a}\begin{bmatrix} M & 0 \\ 0 & M \end{bmatrix}\frac{1}{a}\begin{bmatrix} 0 & -E_n \\ -E_n & 0 \end{bmatrix} = \frac{1}{a^2}\begin{bmatrix} 0 & -M \\ -M & 0 \end{bmatrix} = \frac{1}{a}\begin{bmatrix} 0 & -E_n \\ -E_n & 0 \end{bmatrix}\frac{1}{a}\begin{bmatrix} M & 0 \\ 0 & M \end{bmatrix} = M_2 M_1.$$

By Theorem B.53 we get

$$S_2^{(3)} = \frac{1}{2}\exp_2\left(-\frac{a}{2}(M_1 + M_2)\right) = \frac{1}{2}\exp_2\left(-\frac{a}{2}M_1\right)\exp_2\left(-\frac{a}{2}M_2\right).$$

Considering the exponentials separately, with Corollary B.65 we have

$$\exp_2\left(-\frac{a}{2}M_1\right) = \begin{bmatrix} \exp\left(-\frac{\ln(2)}{2}M\right) & 0 \\ 0 & \exp\left(-\frac{\ln(2)}{2}M\right) \end{bmatrix}.$$

With Corollary B.70 we get

$$\exp_2\left(-\frac{a}{2}M_2\right) = \begin{bmatrix} \cosh\left(\frac{\ln(2)}{2}\right)E_n & \sinh\left(\frac{\ln(2)}{2}\right)E_n \\ \sinh\left(\frac{\ln(2)}{2}\right)E_n & \cosh\left(\frac{\ln(2)}{2}\right)E_n \end{bmatrix}.$$

Evaluating the hyperbolic functions yields

$$\exp_2\left(-\frac{a}{2}M_2\right) = \begin{bmatrix} \frac{3}{2\sqrt{2}}E_n & \frac{1}{2\sqrt{2}}E_n \\ \frac{1}{2\sqrt{2}}E_n & \frac{3}{2\sqrt{2}}E_n \end{bmatrix}.$$

Multiplying these matrices gives

$$\begin{aligned} S_2^{(3)} &= \frac{1}{2}\begin{bmatrix} \exp\left(-\frac{\ln(2)}{2}M\right) & 0 \\ 0 & \exp\left(-\frac{\ln(2)}{2}M\right) \end{bmatrix}\begin{bmatrix} \frac{3}{2\sqrt{2}}E_n & \frac{1}{2\sqrt{2}}E_n \\ \frac{1}{2\sqrt{2}}E_n & \frac{3}{2\sqrt{2}}E_n \end{bmatrix} \\ &= \begin{bmatrix} \frac{3}{4\sqrt{2}}\exp\left(-\frac{\ln(2)}{2}M\right) & \frac{1}{4\sqrt{2}}\exp\left(-\frac{\ln(2)}{2}M\right) \\ \frac{1}{4\sqrt{2}}\exp\left(-\frac{\ln(2)}{2}M\right) & \frac{3}{4\sqrt{2}}\exp\left(-\frac{\ln(2)}{2}M\right) \end{bmatrix}. \end{aligned}$$

Thus, we obtain

$$S_2^{(3)} = \operatorname{kron}\left(\begin{bmatrix} \frac{3}{4} & \frac{1}{4} \\ \frac{1}{4} & \frac{3}{4} \end{bmatrix}, \frac{1}{\sqrt{2}}\exp\left(-\frac{\ln(2)}{2}M\right)\right) = \operatorname{kron}\left(\begin{bmatrix} \frac{3}{4} & \frac{1}{4} \\ \frac{1}{4} & \frac{3}{4} \end{bmatrix}, \frac{1}{\sqrt{2}}\exp_2\left(-\frac{1}{2}M\right)\right).$$

Hence, the matrix $S_2^{(3)}$ can be expressed as a Kronecker product of two matrices, one of which is the matrix $S^{(1)}$. □

Moreover, from the proof of the above theorem, we obtain the following corollary:

**Corollary 5.23.** *The matrix*

$$\frac{1}{\sqrt{2}}\exp_2\left(-\frac{1}{2}M\right) \in \mathbb{R}^{n\times n}, \tag{5.14}$$

*with $M$ as in equation* (5.13) *from the above proof, coincides with $S_2^{(2)}$ for the 2-polytope given as a regular $n$-gon.*

*Proof.* We define analogously to the above proof

$$a := \cos\left(\frac{\pi}{n}\right) \quad \text{and} \quad b := \sin\left(\frac{\pi}{n}\right).$$

We first construct the Colin-de-Verdière-matrix for $S_2^{(2)}$. By Theorem 4.8, we obtain for the off-diagonal entries

$$C_{(i,j)} = -\frac{1}{|p_i \times p_j|} = -\frac{1}{|p_i||p_j|\sin(\alpha)} = -\frac{1}{\frac{1}{a^2}\sin\left(\frac{2\pi}{n}\right)} = -\frac{a^2}{\sin\left(\frac{2\pi}{n}\right)},$$

where $p_i$ and $p_j$ share an edge and $\alpha$ is the angle between $p_i$ and $p_j$. An illustration of the concrete values is shown in Figure 5.12. For the diagonal entries, we obtain accordingly

$$C_{(i,i)} = \frac{2\cos\left(\frac{2\pi}{n}\right)}{\frac{1}{a^2}\sin\left(\frac{2\pi}{n}\right)} = \frac{2\cos\left(\frac{2\pi}{n}\right)a^2}{\sin\left(\frac{2\pi}{n}\right)}.$$

Thus, the Colin-de-Verdière-matrix is

$$C = \frac{a^2}{\sin\left(\frac{2\pi}{n}\right)} \begin{bmatrix} 2\cos\left(\frac{2\pi}{n}\right) & -1 & 0 & \cdots & 0 & -1 \\ -1 & 2\cos\left(\frac{2\pi}{n}\right) & -1 & 0 & \cdots & 0 \\ 0 & \ddots & \ddots & \ddots & \ddots & \vdots \\ \vdots & \ddots & \ddots & \ddots & \ddots & 0 \\ 0 & \cdots & 0 & -1 & 2\cos\left(\frac{2\pi}{n}\right) & -1 \\ -1 & 0 & \cdots & 0 & -1 & 2\cos\left(\frac{2\pi}{n}\right) \end{bmatrix}.$$

For the row sums of $C$, we get

$$\sum_j C_{(i,j)} = \frac{a^2}{\sin\left(\frac{2\pi}{n}\right)}\left(2\cos\left(\frac{2\pi}{n}\right) - 2\right).$$

Hence, the normalization of the Colin-de-Verdière-matrix yields

$$N \cdot C = \frac{1}{2\cos\left(\frac{2\pi}{n}\right) - 2} \begin{bmatrix} 2\cos\left(\frac{2\pi}{n}\right) & -1 & 0 & \cdots & 0 & -1 \\ -1 & 2\cos\left(\frac{2\pi}{n}\right) & -1 & 0 & \cdots & 0 \\ 0 & \ddots & \ddots & \ddots & \ddots & \vdots \\ \vdots & \ddots & \ddots & \ddots & \ddots & 0 \\ 0 & \cdots & 0 & -1 & 2\cos\left(\frac{2\pi}{n}\right) & -1 \\ -1 & 0 & \cdots & 0 & -1 & 2\cos\left(\frac{2\pi}{n}\right) \end{bmatrix}.$$

Using various addition formulas, we find

$$2\cos\left(\frac{2\pi}{n}\right) - 2 = 2\big(2a^2 - 1\big) - 2 = 4a^2 - 4 = -4b^2,$$

and

$$2\cos\left(\frac{2\pi}{n}\right) = 2\big(2a^2 - 1\big) = 4a^2 - 2.$$

Therefore, the normalized matrix becomes

$$N \cdot C = -\frac{1}{4b^2} \begin{bmatrix} 4a^2 - 2 & -1 & 0 & \cdots & 0 & -1 \\ -1 & 4a^2 - 2 & -1 & 0 & \cdots & 0 \\ 0 & \ddots & \ddots & \ddots & \ddots & \vdots \\ \vdots & \ddots & \ddots & \ddots & \ddots & 0 \\ 0 & \cdots & 0 & -1 & 4a^2 - 2 & -1 \\ -1 & 0 & \cdots & 0 & -1 & 4a^2 - 2 \end{bmatrix}.$$

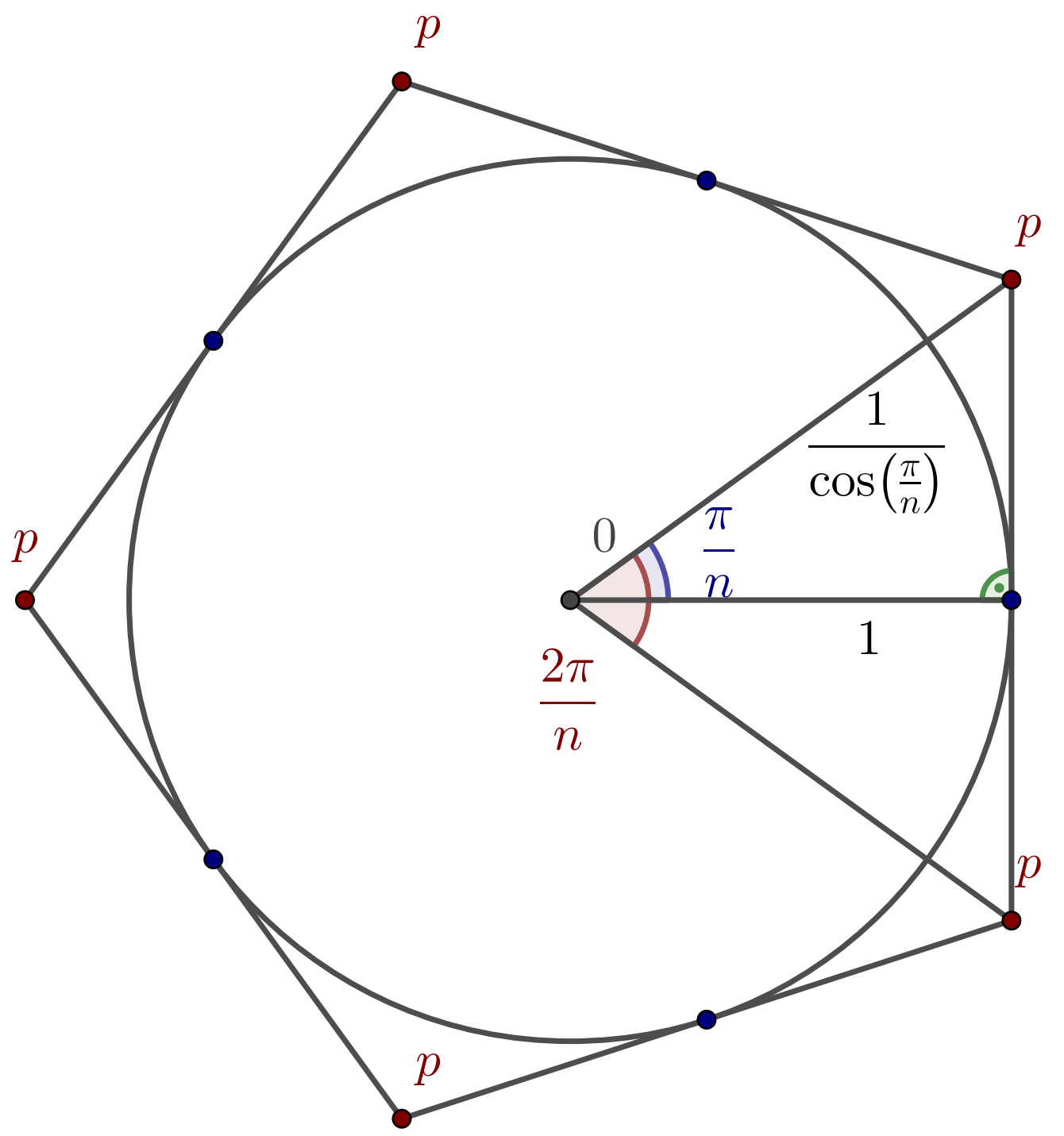

**Figure 5.12:** Illustration of the regular $n$-gon according to Section 3.4 with the required values.

For $N \cdot C - E_n$, we have

$$N \cdot C - E_n = -\frac{1}{4b^2} \begin{bmatrix} 4a^2-2+4b^2 & -1 & 0 & \cdots & 0 & -1 \\ -1 & 4a^2-2+4b^2 & -1 & 0 & \cdots & 0 \\ 0 & \ddots & \ddots & \ddots & \ddots & \vdots \\ \vdots & \ddots & \ddots & \ddots & \ddots & 0 \\ 0 & \cdots & 0 & -1 & 4a^2-2+4b^2 & -1 \\ -1 & 0 & \cdots & 0 & -1 & 4a^2-2+4b^2 \end{bmatrix}$$

$$= -\frac{1}{4b^2} \begin{bmatrix} 2 & -1 & 0 & \cdots & 0 & -1 \\ -1 & 2 & -1 & 0 & \cdots & 0 \\ 0 & \ddots & \ddots & \ddots & \ddots & \vdots \\ \vdots & \ddots & \ddots & \ddots & \ddots & 0 \\ 0 & \cdots & 0 & -1 & 2 & -1 \\ -1 & 0 & \cdots & 0 & -1 & 2 \end{bmatrix}.$$

Now consider the expression (5.14). We compute

$$\begin{aligned} \frac{1}{\sqrt{2}} \exp_2\left(-\frac{1}{2}M\right) &= \frac{1}{\sqrt{2}} \exp\left(-\frac{\ln(2)}{2}M\right) = \exp\left(\ln\left(\frac{1}{\sqrt{2}}\right)E_n\right) \exp\left(-\frac{\ln(2)}{2}M\right) \\ &= \exp\left(-\frac{\ln(2)}{2}E_n - \frac{\ln(2)}{2}M\right) \\ &= \exp\left(\ln(2)\left(-\frac{1}{2}(E_n + M)\right)\right) = \exp_2\left(-\frac{1}{2}(E_n + M)\right). \end{aligned}$$

We thus need to compare $N \cdot C - E_n$ with the expression $-\frac{1}{2}(E_n + M)$. For the latter, we have

$$-\frac{1}{2}(E_n+M) = -\frac{1}{4b^2}\begin{bmatrix} 2a^2+2b^2 & -1 & 0 & \cdots & 0 & -1 \\ -1 & 2a^2+2b^2 & -1 & 0 & \cdots & 0 \\ 0 & \ddots & \ddots & \ddots & \ddots & \vdots \\ \vdots & \ddots & \ddots & \ddots & \ddots & 0 \\ 0 & \cdots & 0 & -1 & 2a^2+2b^2 & -1 \\ -1 & 0 & \cdots & 0 & -1 & 2a^2+2b^2 \end{bmatrix}$$

$$= -\frac{1}{4b^2}\begin{bmatrix} 2 & -1 & 0 & \cdots & 0 & -1 \\ -1 & 2 & -1 & 0 & \cdots & 0 \\ 0 & \ddots & \ddots & \ddots & \ddots & \vdots \\ \vdots & \ddots & \ddots & \ddots & \ddots & 0 \\ 0 & \cdots & 0 & -1 & 2 & -1 \\ -1 & 0 & \cdots & 0 & -1 & 2 \end{bmatrix} = N \cdot C - E_n.$$

Thus, the two matrices before multiplying by ln(2) and taking the exponential coincide and are therefore identical. □

From these two statements, the following corollary can be derived:

**Corollary 5.24.** *The matrices $S_2^{(3)}$ satisfy the quality criteria Q7 and Q8.*

*Proof.* The regular case and thus quality criterion Q7 follow by simply substituting the corresponding adjacency matrix into the algorithm.

Since, by Theorem 5.22 and Corollary 5.23, the subdivision matrix $S_2^{(3)}$ is the Kronecker product of the one-dimensional and the associated two-dimensional rule, quality criterion Q8 also holds. □

Finally, we discuss the tensor product structure of the shell matrices with the following theorem:

**Theorem 5.25.** *The matrices $\overline{S}^{(3)}$ satisfy the quality criteria Q7 and Q8 for both variants.*

*Proofsketch.* Since in the regular case the subdivision matrices $\overline{S}^{(3)}$ consist exclusively of regular initial elements, quality criterion Q7 is satisfied.

In the semi-regular case, the subdivision matrices $\overline{S}^{(3)}$ consist of the subdivision matrix of the central semi-regular initial element, the subdivision matrices of the two opposite semi-regular initial elements, and otherwise only subdivision matrices of regular initial elements. Thus, the tensor product property transfers accordingly and quality criterion Q8 is also fulfilled. □

## Q10 Symmetry

For this criterion, we must show that for all permutations $H$ of the adjacency matrix $A$ with

$$HAH^T = A,$$

the corresponding subdivision matrices $S$ satisfy

$$HSH^T = S.$$

For $t = 2$ and Variant 1, we obtain the following result:

**Theorem 5.26.** *The subdivision matrix $S_1^{(2)}$ satisfies $HSH^T = S$ for any automorphism $H$ of the associated graph and thus fulfills quality criterion Q10.*

*Proof.* Substituting the subdivision matrix into the above condition, we obtain

$$\begin{aligned} HS_1^{(2)}H^T &= H\left(\left(\frac{\vec{1}\cdot\vec{1}^T}{|\vec{1}|^2} + \frac{1-2\mu}{2-2\mu}\left(\frac{\tilde{P}_{(:,1)}\cdot\tilde{P}_{(:,1)}^T}{|\tilde{P}_{(:,1)}|^2} + \frac{\tilde{P}_{(:,2)}\cdot\tilde{P}_{(:,2)}^T}{|\tilde{P}_{(:,2)}|^2}\right)\right)(1-\mu) + \mu E_n\right)H^T \\ &= \left(\frac{H\vec{1}\cdot\vec{1}^T H^T}{|\vec{1}|^2} + \frac{1-2\mu}{2-2\mu}\left(\frac{H\tilde{P}_{(:,1)}\cdot\tilde{P}_{(:,1)}^T H^T}{|\tilde{P}_{(:,1)}|^2} + \frac{H\tilde{P}_{(:,2)}\cdot\tilde{P}_{(:,2)}^T H^T}{|\tilde{P}_{(:,2)}|^2}\right)\right)(1-\mu) + \mu H E_n H^T. \end{aligned}$$

We examine the summands individually. The matrix $\vec{1}\cdot\vec{1}^T$ consists only of entries equal to 1, hence for any permutation it holds that

$$H\vec{1}\cdot\vec{1}^T H^T = \vec{1}\cdot\vec{1}^T.$$

Since by Theorem B.16 we have $H^T = H^{-1}$, it follows that

$$HE_nH^T = E_n.$$

Using equation (5.5), the lengths $|\tilde{P}_{(:,1)}|^2$ and $|\tilde{P}_{(:,2)}|^2$ are equal, so we get

$$\frac{H\tilde{P}_{(:,1)}\cdot\tilde{P}_{(:,1)}^T H^T}{|\tilde{P}_{(:,1)}|^2} + \frac{H\tilde{P}_{(:,2)}\cdot\tilde{P}_{(:,2)}^T H^T}{|\tilde{P}_{(:,2)}|^2} = \frac{H\tilde{P}_{(:,1)}\cdot\tilde{P}_{(:,1)}^T H^T + H\tilde{P}_{(:,2)}\cdot\tilde{P}_{(:,2)}^T H^T}{|\tilde{P}_{(:,1)}|^2} = \frac{H\tilde{P}\cdot\tilde{P}^T H^T}{|\tilde{P}_{(:,1)}|^2}.$$

Since $P = \tilde{P}$ as a regular $n$-gon satisfies all symmetries of the underlying graph structure and the scalar product depends only on relative quantities (lengths and angles), it holds that

$$H\tilde{P}\cdot\tilde{P}^T H^T = \tilde{P}\cdot\tilde{P}^T.$$

For the entire expression we then have

$$\begin{aligned} HS_1^{(2)}H^T &= \left(\frac{H\vec{1}\cdot\vec{1}^T H^T}{|\vec{1}|^2} + \frac{1-2\mu}{2-2\mu}\left(\frac{H\tilde{P}_{(:,1)}\cdot\tilde{P}_{(:,1)}^T H^T}{|\tilde{P}_{(:,1)}|^2} + \frac{H\tilde{P}_{(:,2)}\cdot\tilde{P}_{(:,2)}^T H^T}{|\tilde{P}_{(:,2)}|^2}\right)\right)(1-\mu) + \mu H E_n H^T \\ &= \left(\frac{\vec{1}\cdot\vec{1}^T}{|\vec{1}|^2} + \frac{1-2\mu}{2-2\mu}\left(\frac{\tilde{P}_{(:,1)}\cdot\tilde{P}_{(:,1)}^T}{|\tilde{P}_{(:,1)}|^2} + \frac{\tilde{P}_{(:,2)}\cdot\tilde{P}_{(:,2)}^T}{|\tilde{P}_{(:,2)}|^2}\right)\right)(1-\mu) + \mu E_n \\ &= S_1^{(2)}, \end{aligned}$$

and consequently $S_1^{(2)}$ satisfies all symmetries of the underlying structure. □

For $t = 3$, no statement about the symmetry of Variant 1 can be made here. It would need to be shown that

$$H\tilde{P}\cdot\tilde{P}^T H^T = \tilde{P}\cdot\tilde{P}^T$$

holds. However, we have no direct access to the values of $\tilde{P}$ since these were generated as an orthogonal basis of the space spanned by $P'$ using the `orth` command. Tests on individual examples suggest that the symmetry for Variant 1 is satisfied. A proof, however, remains outstanding at this point.

For Variant 2, we obtain the following result:

**Theorem 5.27.** *The subdivision matrices $S_2^{(t)}$ for $t \in \{2,3\}$ satisfy*

$$HS_2^{(t)}H^{-1} = S_2^{(t)}$$

*for any automorphism $H$ of the corresponding graph. Thus, $S_2^{(t)}$ fulfills Quality Criterion Q10.*

*Proof.* First, the constructed 2-polytope as a regular $n$-gon and the 3-polytope satisfy all symmetries of the underlying graph by Theorem 3.25. Since the entries of the Colin-de-Verdière-matrices from Theorems 4.8 and 4.13 only use relative information of the control points (cross products, lengths, and angles), the Colin-de-Verdière-matrices also respect all symmetries, i.e.,

$$HCH^{-1} = C.$$

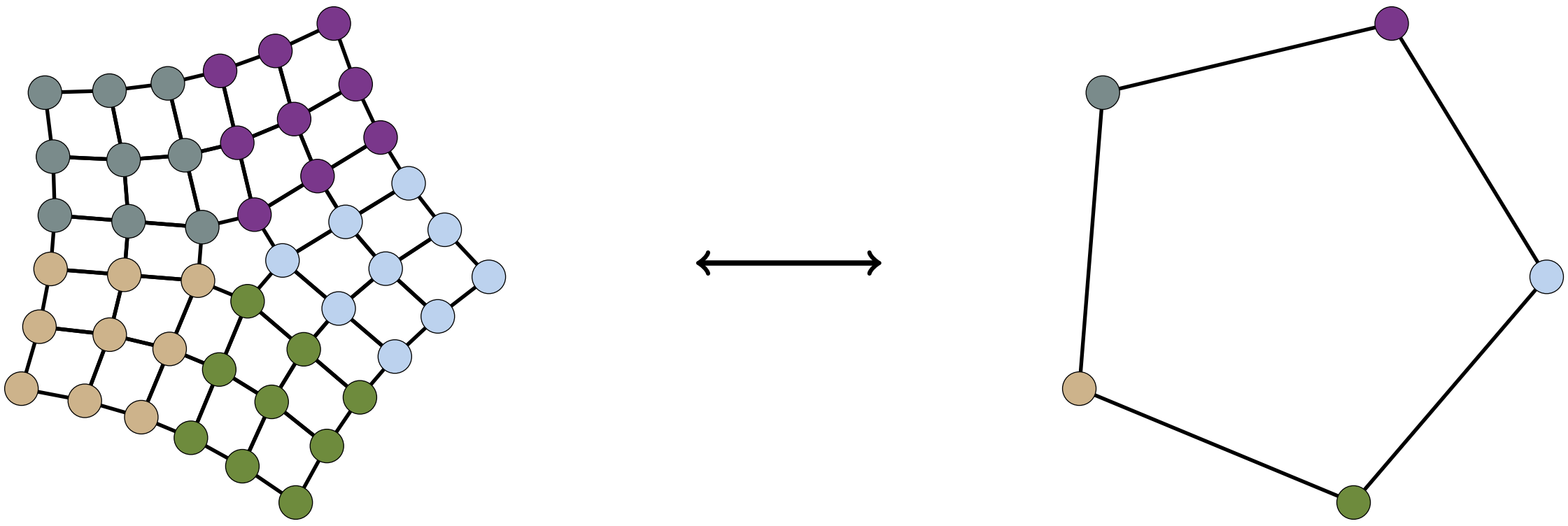

**Figure 5.13:** Central initial element consisting of five points (right) and control point structure of the corresponding ring matrix (left). For quality criterion Q10, the rows and columns of the left-colored $3 \times 3$ blocks in $\overline{S}^{(2)}$ must be mapped onto each other by the automorphisms induced by the right structure.

The same applies to the row sums of $C$, thus

$$HNH^{-1} = N.$$

For the entire expression and using Proposition B.53, which allows pulling invertible matrices into the matrix exponential, we obtain

$$\begin{aligned} HS_2H^{-1} &= H \exp_2(NC - E_n)H^{-1} \\ &= \exp_2\left(H(NC - E_n)H^{-1}\right) \\ &= \exp_2(HNCH^{-1} - HE_nH^{-1}) \\ &= \exp_2(HNH^{-1}HCH^{-1} - HE_nH^{-1}) \\ &= \exp_2(NC - E_n) \\ &= S_2, \end{aligned}$$

so that $S_2$ respects all symmetries of the underlying structure. □

The symmetry transfers directly to the subdivision matrices of rings and shells, as explained in the following theorem:

**Theorem 5.28.** *The subdivision matrices* $\overline{S}_2^{(t)}$ *for* $t \in \{2, 3\}$ *and the subdivision matrix* $\overline{S}_1^{(2)}$ *satisfy*

$$H\overline{S}H^{-1} = \overline{S}$$

*for an automorphism* $H$ *of the corresponding graph induced by the combinatorial graph of the central initial element. For these subdivision matrices, quality criterion Q10 is thus fulfilled.*

*Proofsketch.* The subdivision matrix of the ring and shell cases depends, due to the invariant structure and the descriptions in Section 2.3, only on the central initial element. Each vertex of the polytope thus corresponds to a $3 \times 3$ or $3 \times 3 \times 3$ structure whose matrix entries are computed equally for symmetric points.

By ”induced by the central initial element” we mean that the automorphism maps these triple blocks onto each other. It is not excluded here that there might be further automorphisms of the graph for the ring or shell. Relevant for the symmetry criterion at this point are only these induced automorphisms. An illustration can be found in Figure 5.13.

Since the matrix entries of the triple blocks are computed equally for symmetric points, the automorphism property of the initial elements transfers to the ring and shell matrices. □

Quantifying the symmetry property empirically is difficult, since the problem of determining the automorphism group of a graph lies in NP (cf. [LS16, pp. 13–15]) and computing this for the example set in the case $t = 3$ would be too costly. However, within the example collection there are adjacency matrices that are isomorphic to each other. These examples thus represent the same structure, but the numbering of the points differs. For these, we can check whether the isomorphism of the adjacency matrix implies an isomorphism of the subdivision matrix. The maximal deviations in the corresponding entries over all 36,336 isomorphic examples are found in the following table:

| | $S_1^{(3)}$ | $\overline{S}_1^{(3)}$ | $S_2^{(3)}$ | $\overline{S}_2^{(3)}$ |
|---|---|---|---|---|
| $\max \lvert S - HSH^T \rvert$ | 3,0698e−14 | 3,0698e−14 | 1,3703e−13 | 1,3703e−13 |

Thus, at least a qualitative statement about symmetry can be made.

## Q9, Q11 und Q13 Input, Support and Injectivity

The first two criteria of this section can be described quickly. The input of Algorithms 10 and 11 are graphs of 2- and 3-polytopes. This input corresponds to Definition 2.51, and thus for every valid input a subdivision matrix for an initial element can be generated. Since the ring and shell matrices consist only of initial elements, a subdivision matrix can be generated for every configuration. Consequently, quality criterion Q9 is fulfilled for all subdivision matrices presented in this chapter.

The same applies to the supports of the refinement rules. Every new control point in both variants depends only on the control points of the associated initial element, and thus quality criterion Q11 is also fulfilled for all subdivision matrices in this chapter.

Injectivity, however, as described in Chapter 2, is a long-term goal. For $\overline{S}_1^{(2)}$ we can make the following statement:

**Theorem 5.29.** *Let **D** be a spline definition domain of a ring, **g** the associated system of generating functions, and $\overline{S}_1^{(2)}$ with $\mu = 1/4$ the associated subdivision matrix. The subdivision algorithm $(\overline{S}_1^{(2)}, \boldsymbol{g})$ has an injective regular characteristic map; thus quality criterion Q13 is fulfilled.*

*Proofsketch.* The proof is based on the results from [PR08, Sec. 6.2, pp. 116–120] and refers to the theory in [PR08], which we do not elaborate on in this work.

Peters and Reif in [PR08] consider exactly the subdivision matrices $\overline{S}_1^{(2)}$ with $\mu = 1/4$, i.e., the refinement rules of the Doo-Sabin algorithm for invariant rings. Hence, their characteristic map is injective and regular by [PR08, Thm. 6.2, p. 119] and [PR08, Thm. 5.8, p. 88]. □

Quality criterion Q13 could likely also be verified for variant $\overline{S}_2^{(2)}$ using the methods from [PR08]. Since we do not cover the theory described there in this thesis, we do not pursue this further. However, this may serve as an avenue for future research.

No statement can be made about the injectivity of the shell matrices, since no approaches exist for their analysis. It also cannot be verified empirically, as for every example at least a visual inspection would be necessary. However, we could not construct any counterexamples for our variants. The counterexamples constructed in Section 2.5 for the algorithms [JM99] and [Baj+02] yield injective eigenshells for our variants, as illustrated in Figure 5.14.

After this analysis, looking again at Table 5.1, we see that both algorithms are suitable from both theoretical and practical viewpoints. The advantage of variant 1 is the exact determination of the subsubdominant eigenvalue at the cost of the non-negativity of entries and provability of symmetries. Variant 2 does not allow the exact determination of the subsubdominant eigenvalue, but empirical results show that the gap to the subdominant eigenvalue is sufficiently large.

Placing the analysis in context, it is the first time a three-dimensional variant for the generalized quadratic B-spline subdivision has been developed. Moreover, the described variants of the initial elements possess a $t$-fold subdominant eigenvalue of $1/2$, which previous subdivision algorithms in the literature do not exhibit. Furthermore, no counterexample to injectivity has been found for this construction, distinguishing it from previous subdivision algorithms. Overall, this constitutes a genuine improvement over the previously established subdivision algorithms.

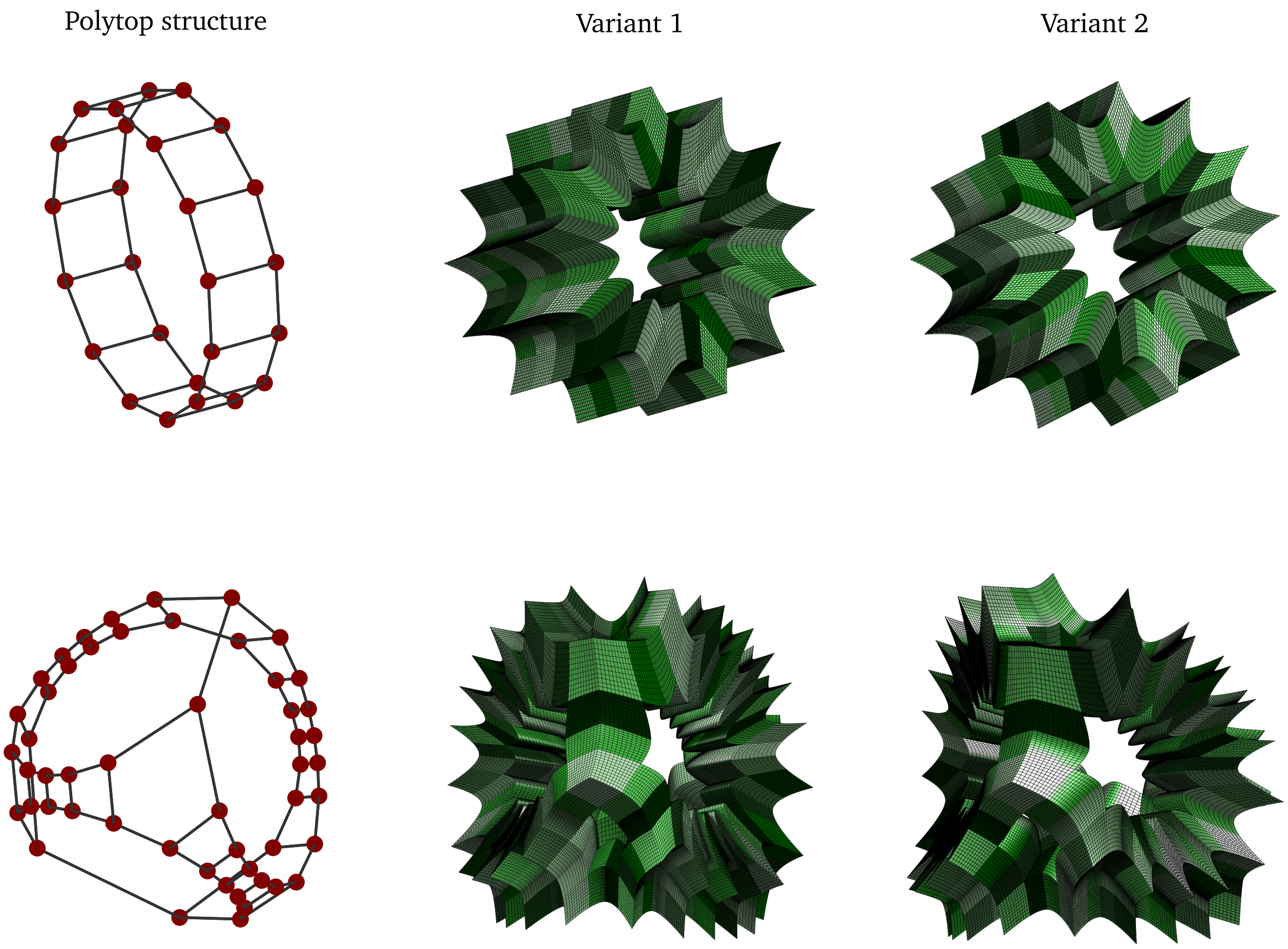

**Figure 5.14:** The examples from Figure 2.36 (top) and Figure 2.37 (bottom). On the left are the polygon structures, in the middle the eigenshells of variant one, and on the right those of variant two. It can be seen that all shells are injective.

We conclude this section with Figure 5.15, which shows a gallery consisting of images of the characteristic maps of various subdivision matrices generated with variant 2 of Algorithm 11.

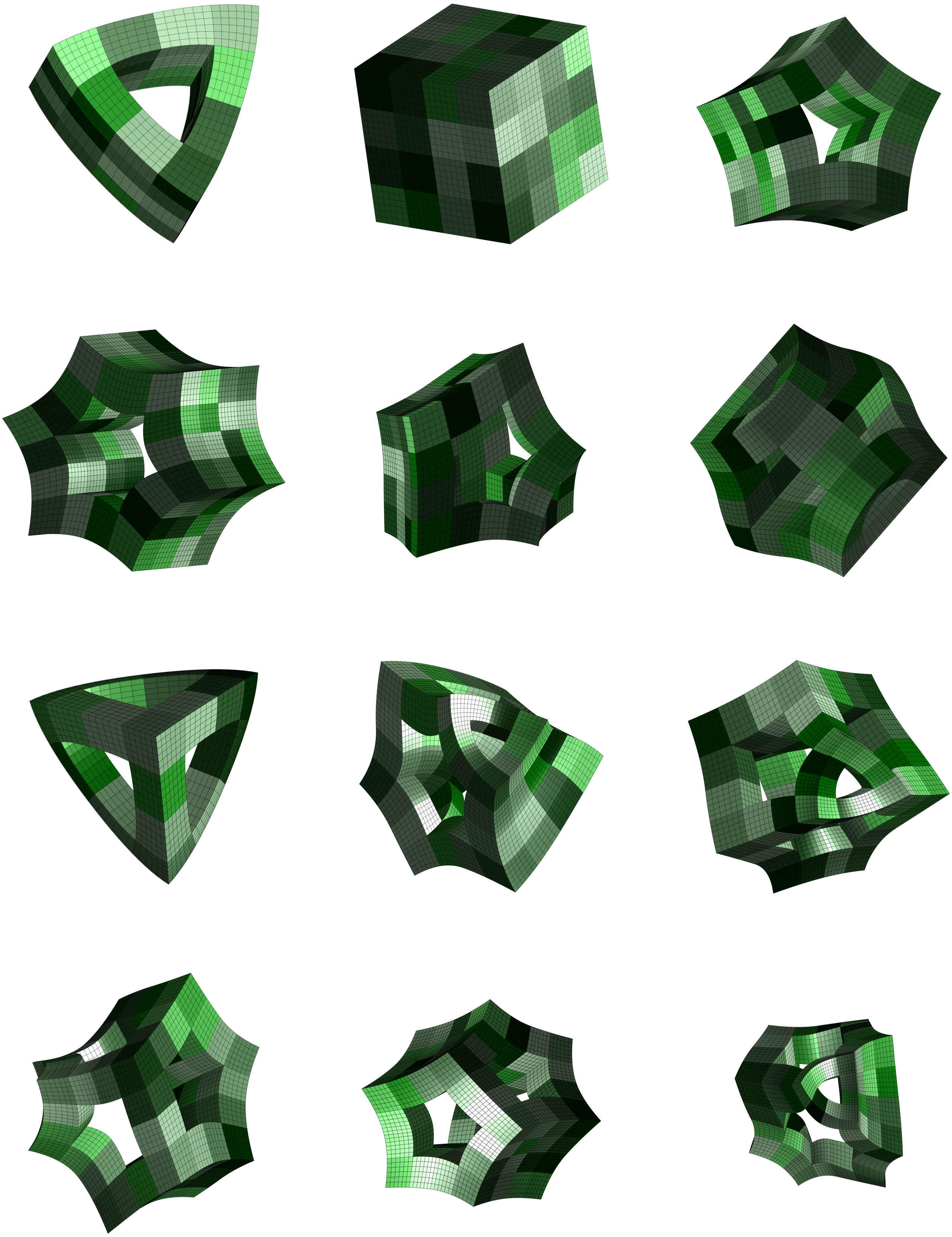

**Figure 5.15:** Gallery of example images of the characteristic map of subdivision matrices corresponding to the 3-polytopes from Figure 3.1, generated using variant 2 of Algorithm 11.

# 6 Generalized Cubic B-Spline Subdivision: An Experimental Approach

In contrast to the generalized quadratic B-spline subdivision, the literature on generalized cubic B-spline subdivision is more extensive. This may be partly due to the fact that the Catmull-Clark algorithm [CC78] generates $C_1^2$ surfaces. This notation means that the surfaces are $C^1$ at irregular points and $C^2$ everywhere else. In contrast, the Doo-Sabin algorithm for the surface case produces only $C_1^1$ surfaces, i.e., surfaces that are $C^1$ everywhere, including at irregular points. The theory of subdivision surfaces related to these algorithms is treated extensively in [PR08]. The smoothness theorems for both subdivision algorithms can be found especially in [PR08, Thm. 6.1, pp. 114–115, and Thm. 6.2, p. 119].

For the surface case, there are various variants or improvements of the Catmull-Clark algorithm. This area is known in the literature as „tuning“ . This suggests that the standard Catmull-Clark algorithm should be improved for surfaces. The motivation for tuning usually concerns the curvature of the subdivision surfaces. Since these surfaces in generalized cubic B-spline subdivision are almost everywhere $C^2$, the curvature of the surface—which typically lies in $\mathbb{R}^3$ for these applications—can also be defined almost everywhere. The behavior of the curvature at irregular points depends on the ratio $\lambda^2/\mu$. If this ratio is less than 1, both principal curvatures converge to zero at the irregular points, leading to *flat points*. If the ratio equals 1, both principal curvatures remain bounded, and typically one of them is nonzero. If the ratio is greater than 1, at least one of the principal curvatures diverges. An overview of the characterization of subdivision surface shapes, including statements on curvature behavior, is found in [PR04], [KPR04], and [PR08, Ch. 7].

Regarding tuning of the subdivision matrices with respect to the ratio $\lambda^2/\mu$, the works [ADS06] and [MM18] are notable. Both use alternative rules for generalized cubic B-spline subdivision, whose support we will discuss later.

Additionally, in the context of guided subdivision, there are several approaches to improving surfaces, especially to enhance reflection lines in irregular areas. Notable here are the variants [KP18] and [KP19]. The variants [KP17] and [KP24] also achieve a scaling behavior of $1/2$ in irregular regions.

For the volumetric case, compared to the generalized quadratic B-spline subdivision from the previous chapter, refinement rules exist. As already mentioned and discussed in Section 2.5, the works [JM99] and [Baj+02] are notable. It is thus quite surprising that proposals exist for refinement rules for the generalized cubic B-spline subdivision, whereas none exist for the generalized quadratic B-spline subdivision. This could be because initial variants of refinement rules can be derived as tensor products of the two-dimensional rules. Essentially, for a first proposal, only a refinement rule for irregular points must be constructed.

In this chapter, we present a construction of the generalized cubic B-spline subdivision, following Variant 2 from Chapter 5. Unlike the approach introduced in Chapter 5, this variant explicitly refers to a structure composed of quadrilaterals and hexahedra. Therefore, the construction presented here cannot, in particular, be generalized to the approaches discussed in Chapter 7.

Moreover, Quality Criterion Q11 poses a real challenge for the generalized cubic B-spline subdivision. Thus, the goal of this chapter is to present a construction that satisfies both Quality Criterion Q5 and Quality Criterion Q11.

As the title suggests, the approach presented here is of an experimental nature. This is because the $t$-fold subdominant eigenvalue $1/2$ could only be verified empirically and not proven. Additionally, two of the proofs in Section 6.3 are given only in sketch form.

To develop the approach, we first describe in Section 6.1 the regular case, the evaluable domain, and the challenges of the generalized cubic B-spline subdivision. Subsequently, in Section 6.2, we construct the subdivision matrices.

$\frac{1}{2}$ ——— $\frac{1}{2}$ $\qquad\qquad$ $\frac{1}{8}$ ——— $\frac{6}{8}$ ——— $\frac{1}{8}$

**Figure 6.1:** The refinement rules for $t = 1$ and $g = 3$. The left refinement rule creates a control point in the middle of two points, thus generating a control point from an edge, whereas the right refinement rule generates a control point from an existing control point.

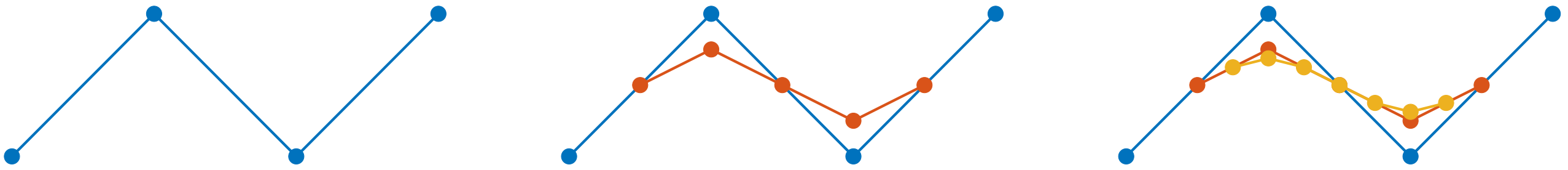

**Figure 6.2:** An example configuration with two initial elements (left). Refining both initial elements produces five control points and three initial elements (middle). Refining once more results in seven control points and five initial elements (right). It can also be seen that a new control point is inserted at each edge midpoint. This point arises from the $[1/2,\ 1/2]$ rule. The other control points are generated from existing control points by the $[1/8,\ 6/8,\ 1/8]$ rule.

This construction is divided into three parts. In the first part, Section 6.2.1, we establish the eigenstructure of the subdominant eigenvalues, which is derived by refining the eigenstructure from Chapter 5. Next, from this eigenstructure, we generate a Colin-de-Verdière-like matrix in Section 6.2.2. This matrix will be formed as the product of two matrices, each composed of various Colin-de-Verdière-matrices. Finally, in Section 6.2.3, we use the matrix exponential to produce the subdivision matrix. For this, we first decompose the Colin-de-Verdière-like matrix into the sum of two matrices and then multiply their matrix exponentials. This method can be applied equally to both the surface and volumetric cases. In Section 6.3, we conclude with an overview of which quality criteria are (not) satisfied by the constructed subdivision matrices. To support these statements, we additionally consider empirical results from the example set generated in Section 3.1.1.

## 6.1 Refinement rules for regular initial elements, evaluable domain, and challenges

The regular refinement rules for the one- and two-dimensional cases are described in [CC78]. Analogous to Definition 2.15 and [CC78, p. 351], the one-dimensional subdivision matrix for an initial element is given by

$$S^{(1)} := \frac{1}{8}\begin{bmatrix} 4 & 4 & 0 \\ 1 & 6 & 1 \\ 0 & 4 & 4 \end{bmatrix}.$$

An illustration of the refinement rules can be found in Figure 6.1. Here the first peculiarity becomes apparent: the support of the refinement rules differs depending on the type of control points. There are new points inserted on edges of the associated grid, which depend only on the two endpoints of the edges, and are therefore dependent on just two control points. New control points that arise from old control points, however, depend on three control points: the predecessor control point and its neighbors. An illustration can be found in Figure 6.2.

This difference to the generalized quadratic B-spline subdivision can be explained by the increased smoothness. Since the evaluation is piecewise polynomial, more generating functions per element are needed for evaluating the cells, which leads to larger initial elements compared to the generalized quadratic B-spline subdivision. Moreover, the generating functions have a larger support, which makes it necessary that the refinement rules involve more points.

Thus, already in the one-dimensional case we obtain different types of refined control points, namely control points that arise from vertices or control points through the $[1/8,\ 6/8,\ 1/8]$ rule, and points that arise from edges through the $[1/2,\ 1/2]$ rule. Both types have different supports with respect to the control points influencing the refinement.

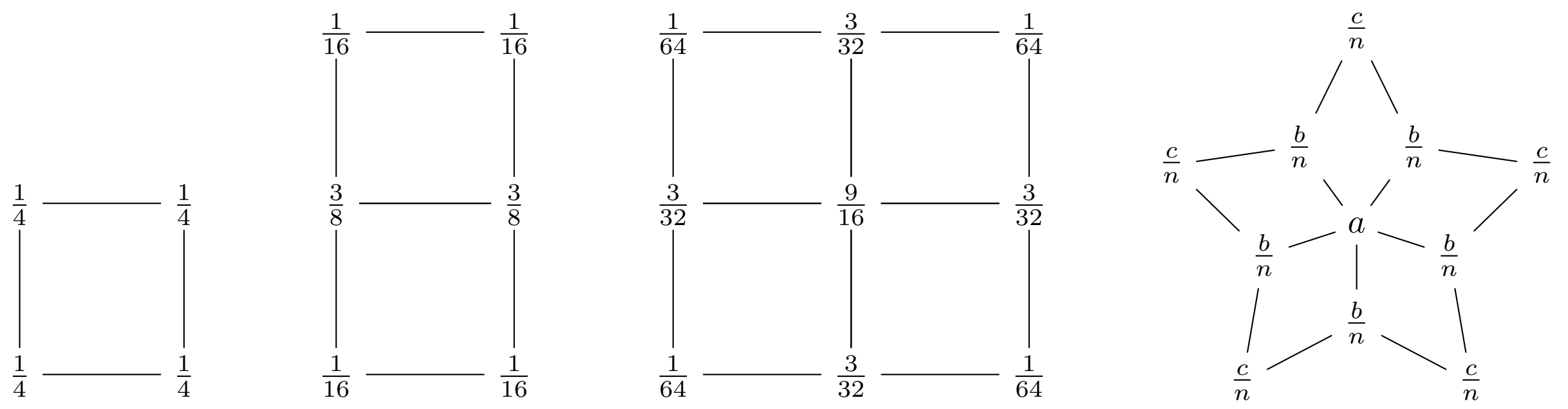

**Figure 6.3:** The refinement rules for $t = 2$ and $g = 3$ of the *Catmull-Clark algorithm*. The refinement rule for a facet point (left), for an edge point (second from left), for a regular vertex point (second from right), and for an extraordinary vertex point (right).

This behavior continues in the two-dimensional case. Here, for the regular case, according to [CC78] and using a different notation, the refinement matrix is given by

$$S^{(2)} := \frac{1}{64}\left[\begin{array}{ccc|ccc|ccc} 16 & 16 & & 16 & 16 & & & & \\ 4 & 24 & 4 & 4 & 24 & 4 & & & \\ & 16 & 16 & & 16 & 16 & & & \\ \hline 4 & 4 & & 24 & 24 & & 4 & 4 & \\ 1 & 6 & 1 & 6 & 36 & 6 & 1 & 6 & 1 \\ & 4 & 4 & & 24 & 24 & & 4 & 4 \\ \hline & & & 16 & 16 & & 16 & 16 & \\ & & & 4 & 24 & 4 & 4 & 24 & 4 \\ & & & & 16 & 16 & & 16 & 16 \end{array}\right] = \frac{1}{8}\begin{bmatrix} 4S^{(1)} & 4S^{(1)} & 0S^{(1)} \\ 1S^{(1)} & 6S^{(1)} & 1S^{(1)} \\ 0S^{(1)} & 4S^{(1)} & 4S^{(1)} \end{bmatrix}$$

of an initial element as the Kronecker product of $S^{(1)}$ with itself. An illustration of the refinement rules can be found in [PR08, Fig. 6.2, p. 110] in Figure 6.3.

Here we even see three different refinement rules. The first is the combination of two $[1/2, 1/2]$-rules and generates a new control point in the center of a facet. These facet points thus arise from faces. The combination of a $[1/2, 1/2]$- and a $[1/8, 6/8, 1/8]$-rule produces a new edge point, and the combination of two $[1/8, 6/8, 1/8]$-rules produces a new vertex point.

Since in the surface case more or fewer than four cells can meet at a vertex, initial elements with other combinatorial arrangements also arise. Catmull and Clark use the refinement rules of the regular case for facet and vertex points and for refining the vertex point they use the weights [CC78] cited according to [PR08, Eq. 6.2 and Fig. 6.2, p. 110]

$$a = 1 - \frac{7}{4n}, \quad b = \frac{3}{2n} \quad \text{and} \quad c = \frac{1}{4n},$$

where $n$ is the number of quads at the central vertex. The weight $a$ is used for the vertex point, the weight $b/n$ for control points connected to the vertex, and the weight $c/n$ for the remaining control points of the initial element. An illustration can also be found in Figure 6.3.

Altogether, these give the rules for each initial element. An example refinement is shown in Figure 6.4.

Combining the three regular rules from the two-dimensional case with the two one-dimensional rules results in four combinations for the three-dimensional case. The combination of three $[1/2, 1/2]$ rules constructs a control point from a volume, the combination of two $[1/2, 1/2]$ rules and one $[1/8, 6/8, 1/8]$ rule constructs a control point from a facet, the combination of one $[1/2, 1/2]$ rule and two $[1/8, 6/8, 1/8]$ rules constructs a control point from an edge, and the combination of three $[1/8, 6/8, 1/8]$ rules constructs a control point from a vertex or control point. An illustration of these rules can be found in Figure 6.6.

For a regular initial element, we thus obtain 27 control points, which are mapped to 27 control points by the above

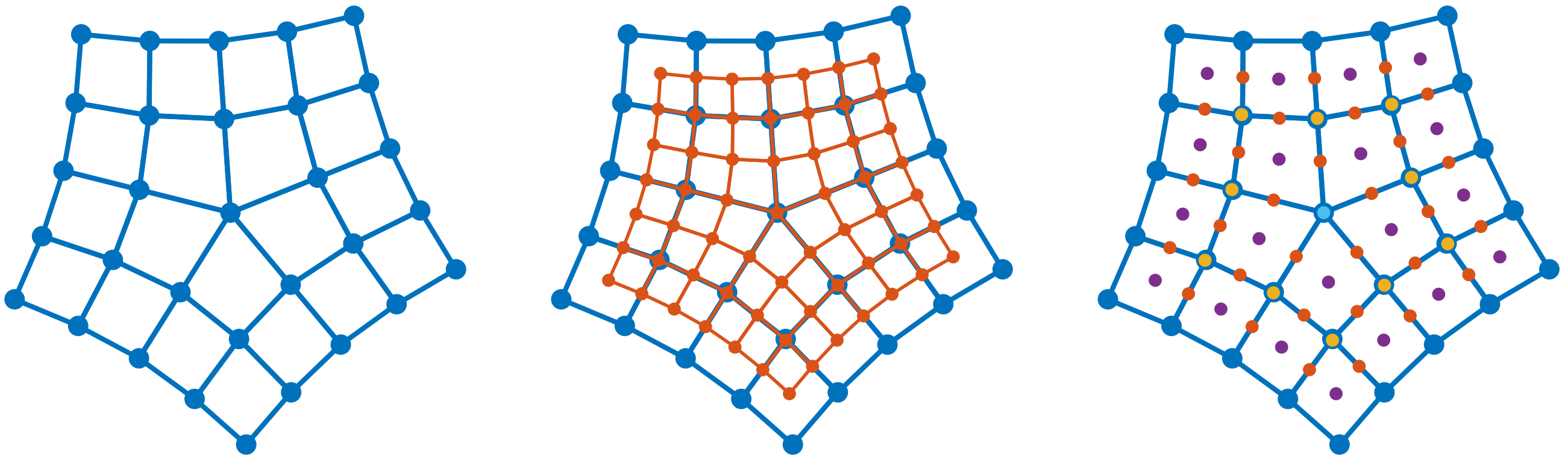

**Figure 6.4:** An example configuration with eleven initial elements and 31 control points (left). Refining all initial elements produces 61 control points and 31 initial elements (center). In the right image, the new control points are colored according to the respective rules from which they originated. The colors correspond to the rules shown in Figure 6.3 from left to right as purple, red, yellow, and light blue.

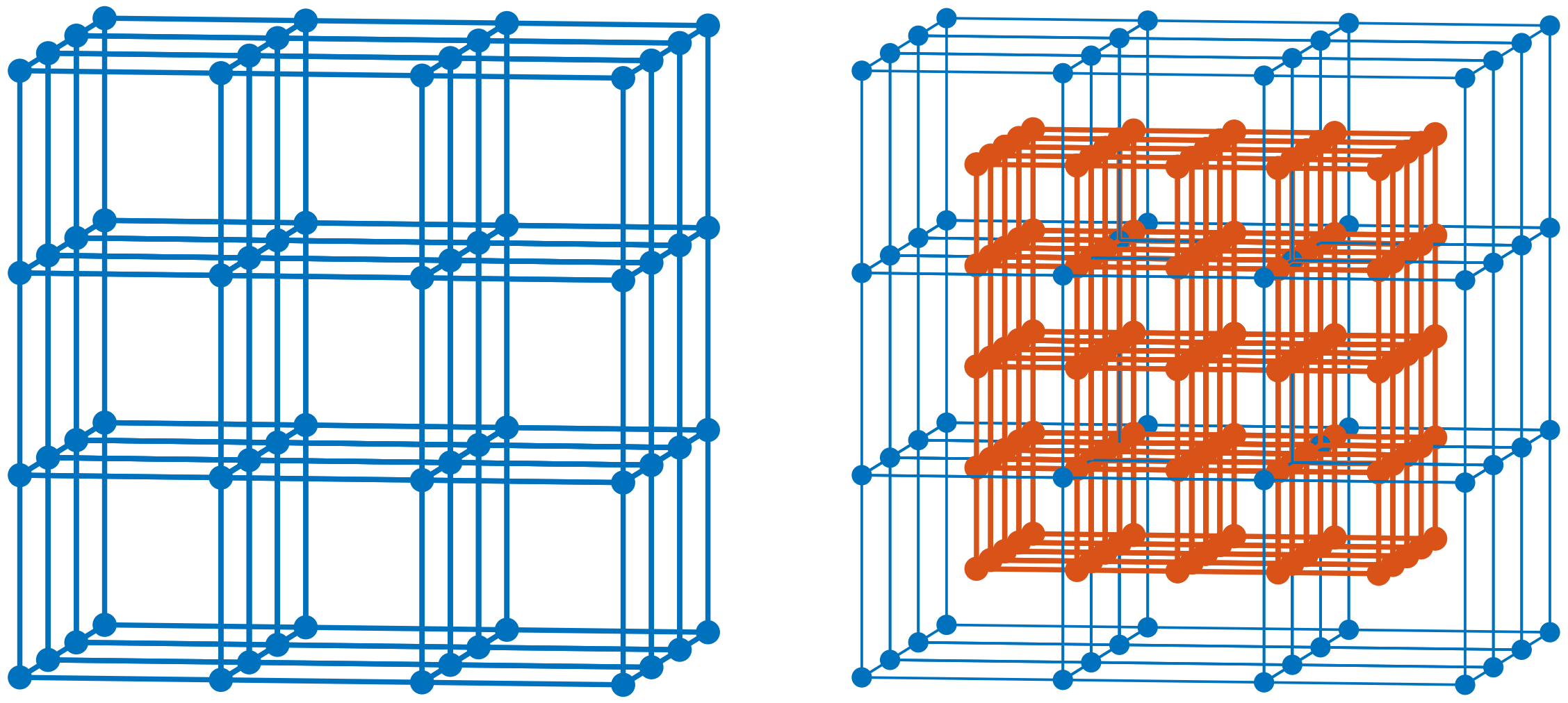

**Figure 6.5:** An example configuration with eight initial elements and 64 control points (left). Refining all initial elements produces 125 control points and 27 initial elements (right). The initial elements were refined using the subdivision matrix from Equation (6.1), whose refinement rules are illustrated in Figure 6.6.

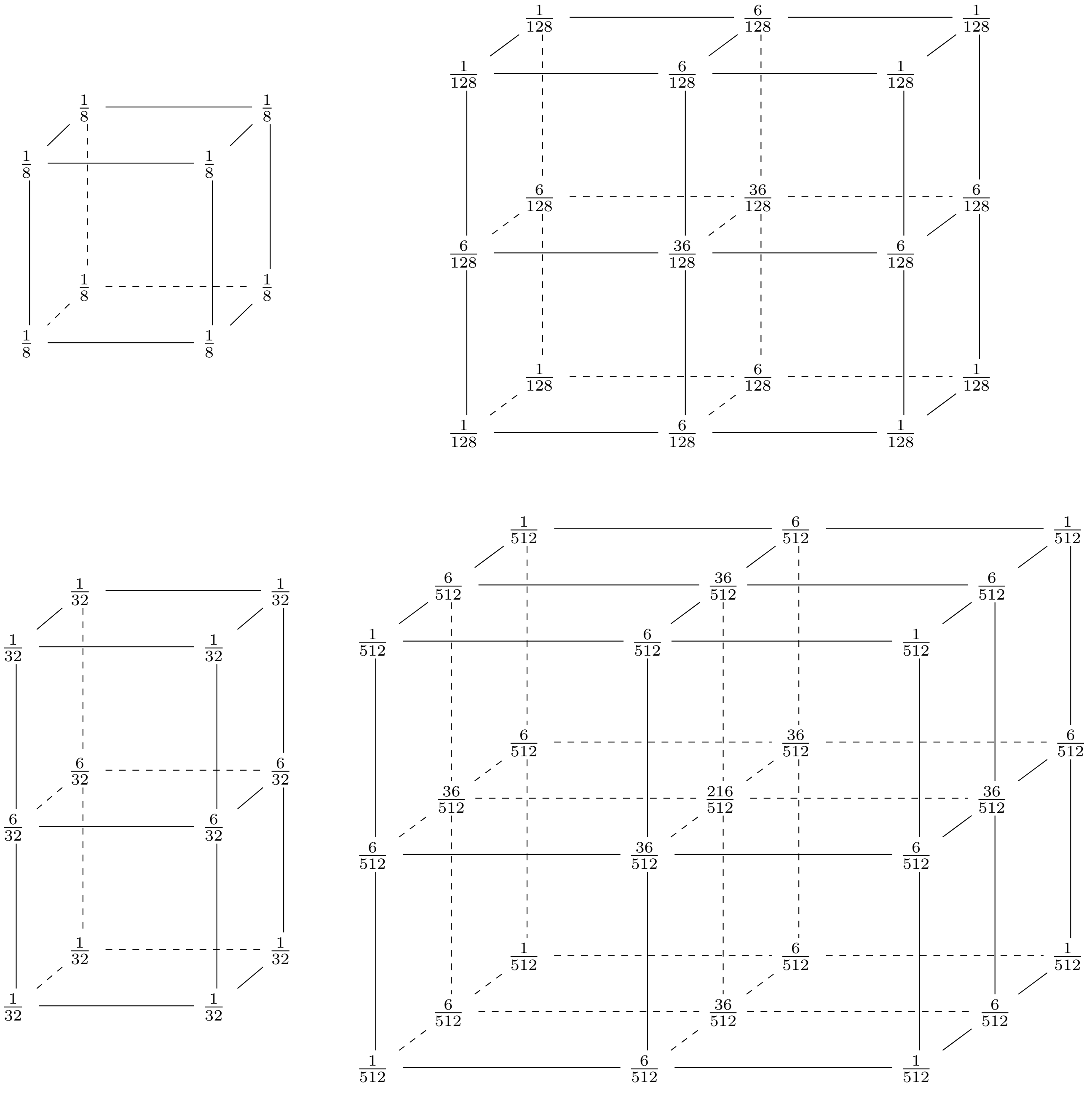

**Figure 6.6:** The refinement rules for $t = 3$ and $g = 3$ in the regular case. The masks generate new control points from volumes (top left), faces (bottom left), edges (top right), and vertices or control points (bottom right).

rules. This is realized by the matrix

$$S^{(3)} := \frac{1}{512}\left[\begin{array}{ccccccccc|ccccccccc|ccccccccc}
64 & 64 & & 64 & 64 & & & & & 64 & 64 & & 64 & 64 & & & & & & & & & & & & & \\
16 & 96 & 16 & 16 & 96 & 16 & & & & 16 & 96 & 16 & 16 & 96 & 16 & & & & & & & & & & & & \\
 & 64 & 64 & & 64 & 64 & & & & & 64 & 64 & & 64 & 64 & & & & & & & & & & & & \\
16 & 16 & & 96 & 96 & & 16 & 16 & & 16 & 16 & & 96 & 96 & & 16 & 16 & & & & & & & & & & \\
4 & 24 & 4 & 24 & 144 & 24 & 4 & 24 & 4 & 4 & 24 & 4 & 24 & 144 & 24 & 4 & 24 & 4 & & & & & & & & & \\
 & 16 & 16 & & 96 & 96 & & 16 & 16 & & 16 & 16 & & 96 & 96 & & 16 & 16 & & & & & & & & & \\
 & & & 64 & 64 & & 64 & 64 & & & & & 64 & 64 & & 64 & 64 & & & & & & & & & & \\
 & & & 16 & 96 & 16 & 16 & 96 & 16 & & & & 16 & 96 & 16 & 16 & 96 & 16 & & & & & & & & & \\
 & & & & 64 & 64 & & 64 & 64 & & & & & 64 & 64 & & 64 & 64 & & & & & & & & & \\
\hline
16 & 16 & & 16 & 16 & & & & & 96 & 96 & & 96 & 96 & & & & & 16 & 16 & & 16 & 16 & & & & \\
4 & 24 & 4 & 4 & 24 & 4 & & & & 24 & 144 & 24 & 24 & 144 & 24 & & & & 4 & 24 & 4 & 4 & 24 & 4 & & & \\
 & 16 & 16 & & 16 & 16 & & & & & 96 & 96 & & 96 & 96 & & & & & 16 & 16 & & 16 & 16 & & & \\
4 & 4 & & 24 & 24 & & 4 & 4 & & 24 & 24 & & 144 & 144 & & 24 & 24 & & 4 & 4 & & 24 & 24 & & 4 & 4 & \\
1 & 6 & 1 & 6 & 36 & 6 & 1 & 6 & 1 & 6 & 36 & 6 & 36 & 216 & 36 & 6 & 36 & 6 & 1 & 6 & 1 & 6 & 36 & 6 & 1 & 6 & 1 \\
 & 4 & 4 & & 24 & 24 & & 4 & 4 & & 24 & 24 & & 144 & 144 & & 24 & 24 & & 4 & 4 & & 24 & 24 & & 4 & 4 \\
 & & & 16 & 16 & & 16 & 16 & & & & & 96 & 96 & & 96 & 96 & & & & & 16 & 16 & & 16 & 16 & \\
 & & & 4 & 24 & 4 & 4 & 24 & 4 & & & & 24 & 144 & 24 & 24 & 144 & 24 & & & & 4 & 24 & 4 & 4 & 24 & 4 \\
 & & & & 16 & 16 & & 16 & 16 & & & & & 96 & 96 & & 96 & 96 & & & & & 16 & 16 & & 16 & 16 \\
\hline
 & & & & & & & & & 64 & 64 & & 64 & 64 & & & & & 64 & 64 & & 64 & 64 & & & & \\
 & & & & & & & & & 16 & 96 & 16 & 16 & 96 & 16 & & & & 16 & 96 & 16 & 16 & 96 & 16 & & & \\
 & & & & & & & & & & 64 & 64 & & 64 & 64 & & & & & 64 & 64 & & 64 & 64 & & & \\
 & & & & & & & & & 16 & 16 & & 96 & 96 & & 16 & 16 & & 16 & 16 & & 96 & 96 & & 16 & 16 & \\
 & & & & & & & & & 4 & 24 & 4 & 24 & 144 & 24 & 4 & 24 & 4 & 4 & 24 & 4 & 24 & 144 & 24 & 4 & 24 & 4 \\
 & & & & & & & & & & 16 & 16 & & 96 & 96 & & 16 & 16 & & 16 & 16 & & 96 & 96 & & 16 & 16 \\
 & & & & & & & & & & & & 64 & 64 & & 64 & 64 & & & & & 64 & 64 & & 64 & 64 & \\
 & & & & & & & & & & & & 16 & 96 & 16 & 16 & 96 & 16 & & & & 16 & 96 & 16 & 16 & 96 & 16 \\
 & & & & & & & & & & & & & 64 & 64 & & 64 & 64 & & & & & 64 & 64 & & 64 & 64
\end{array}\right] = \frac{1}{8}\begin{bmatrix} 4S^{(2)} & 4S^{(2)} & 0S^{(2)} \\ 1S^{(2)} & 6S^{(2)} & 1S^{(2)} \\ 0S^{(2)} & 4S^{(2)} & 4S^{(2)} \end{bmatrix}. \tag{6.1}$$

An example refinement can be found in Figure 6.5. With these three regular refinement matrices $S^{(1)}$, $S^{(2)}$, and $S^{(3)}$, we have explained the refinement of the regular case, that is, the concrete matrices for Definition 2.15.

At this point, we could also discuss the semi-regular and irregular configurations of the three-dimensional case. For the semi-regular case, one could, for example, combine the two-dimensional Catmull-Clark rules with the $[1/2, 1/2]$-rule and the $[1/8, 6/8, 1/8]$-rule, and for the irregular case, one could consider the subdivision matrices from [Baj+02] and [JM99]. However, since these, as explained in Section 2.5, do not provide a suitable spectrum, we omit this here and instead discuss the prerequisites for our variants. We consider different scenarios that deviate from this work to show what does not work in them.

The first central aspect under discussion is the evaluable domain and thus the definition of which combinatorial initial element is considered a regular initial element. For this discussion, in a first scenario, we initially consider the definition from [CC78], [Baj+02], and [JM99], which differs from the definition in this work.

As an example, consider a spline definition domain for $t = 2$ and $g = 3$ with an irregular corner, illustrated in Figure 6.7. For this spline definition domain, the initial elements according to Definition 2.12 are associated with the corners of the cells. With a suitable embedding, we obtain a structure of control points as illustrated in Figure 6.7. Using the rules of Catmull and Clark, all initial elements consisting of a $3 \times 3$ structure—that is, those adjacent to four cells—are regular according to Definition 2.15, since they have $S^{(2)}$ as subdivision matrix, even though according to Definition 2.17 they do not necessarily have to be regular. In the example of Figure 6.7, these correspondingly include all initial elements except the one at the central point. The regular and irregular initial elements corresponding to this idea are shown in Figure 6.8. Thus, Catmull and Clark have a stricter definition of which initial elements must necessarily be regular. Concretely, according to their understanding, in Definition 2.17 all initial elements with a regular structure are also necessarily regular initial elements.

Extending this idea further, according to Definition 2.20, all cells whose corners are associated with regular initial elements—that is, all cells whose corners are adjacent to exactly four cells—are evaluable. These are all cells that do not have an irregular corner. An illustration of the evaluable cells according to this idea and their evaluation is also found in Figure 6.8.

The refinement rules of Catmull and Clark are chosen so that the evaluable region is as large as possible, and this choice is indeed consistent as a tensor product extension of the one-dimensional case. The price for the (compared to the definition in this work) larger evaluable region is, however, that the subdominant eigenvalues of the subdivision matrices cannot be freely chosen. The argumentation for this can be found in [PR08, Section 6.1, pp. 109–120]. If one chooses the refinement rules for irregular initial elements as in [CC78], then the rules not belonging to the central point define most of the spectrum. In particular, the subdominant eigenvalues corresponding to the desired eigenvectors are determined by this extension of the regular rules. The part of the spectrum influenced by the rule of the central point, i.e., by the choice of $a$, $b$, and $c$, must explicitly not determine the subdominant eigenvalue; otherwise, the local asymptotic behavior no longer aligns with the desired eigenvectors. A graph of the eigenvalues for the Catmull-Clark algorithm is found in [PR08, Fig. 6.5, p. 115], along with a detailed explanation and analysis of this fact.

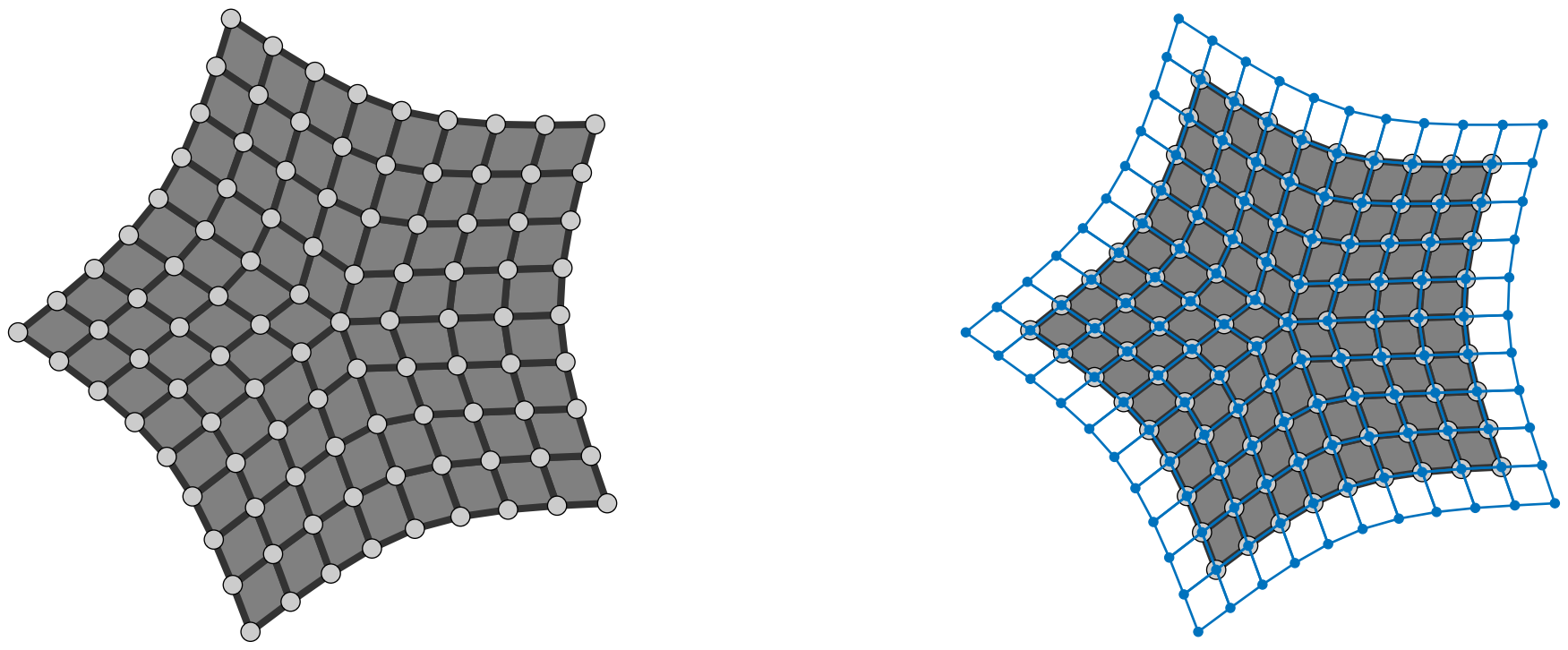

**Figure 6.7:** An exemplary illustration of the combinatorial arrangement of a spline definition domain for $t = 2$ and $g = 3$ with an irregular corner where five cells meet (left). Embedding the structure accordingly results in a structure of control points (in blue), shown in the right image.

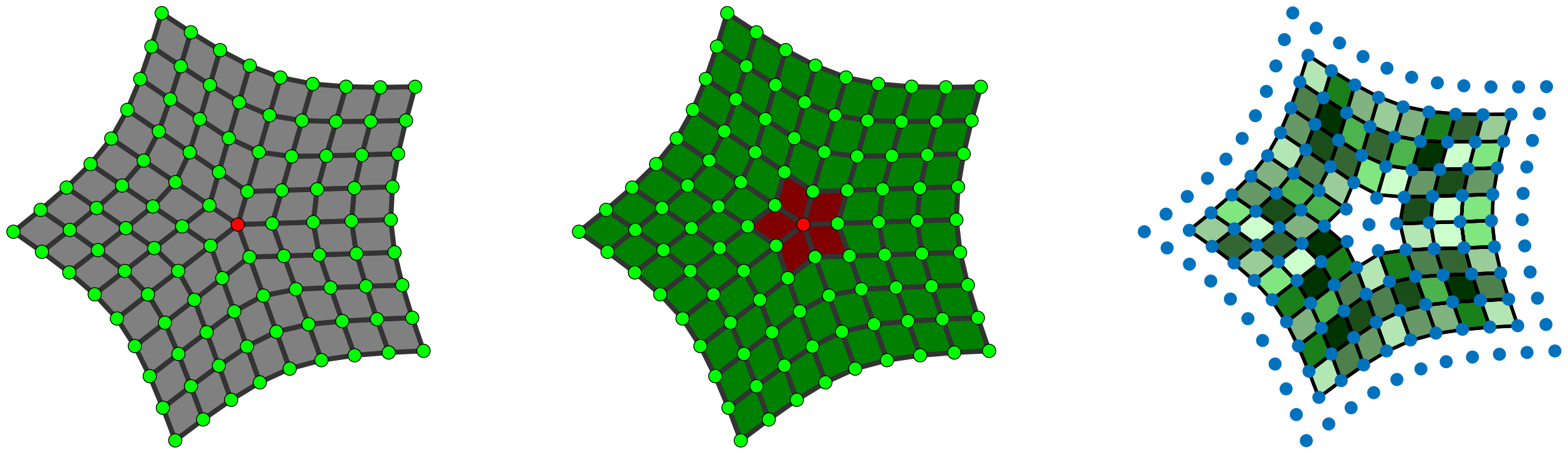

**Figure 6.8:** The spline definition domain from Figure 6.7. The initial elements are labeled with the subdivision matrices from [CC78]. The resulting regular initial elements are colored green, whereas the irregular ones are colored red (left). In the middle image, the evaluable cells are shown in green and the non-evaluable ones in red, which leads to the evaluation depicted in the right image. It is important to note that the regular initial elements and the evaluable regions arise from the rules of the Catmull-Clark algorithm and differ from the regions of the subdivision algorithms constructed in this work.

For the three-dimensional case, we obtain analogous results. The two algorithms [JM99] and [Baj+02] use the tensor products of the one- and two-dimensional cases as refinement rules for volumes and facets, i.e., the rules from Figure 6.6. For regular edges and regular points, both use the refinement rules from Figure 6.6 as well; only for irregular edges and points were custom rules constructed. This leads, as described in Section 2.5, to an unsuitable spectrum, both regarding eigenvalues and eigenvectors. Although the theory of the three-dimensional case is not yet fully developed, we have already explained in Section 2.5 that this behavior cannot be fixed by adjusting the refinement rule for irregular edges and points, since even the semi-irregular case as a tensor product of the irregular two-dimensional and the one-dimensional case does not produce three identical subdominant eigenvalues. Therefore, the regular refinement rules for irregular initial elements must also be abandoned here.

Conversely, this means that for the initial elements considered irregular in this work, the refinement rules for volumes, facets, and edges must be adapted to influence the eigenvalues accordingly. This leads to a series of changes and challenges that we must discuss individually.

By giving up the regular rules for irregular initial elements, we gain great freedom of choice, which in turn entails consequences for the behavior of the refinement.

Let us first consider a scenario in which no restrictions are imposed on the entries of the subdivision matrix of an irregular initial element. In this scenario, we can generate subdivision matrices with great flexibility and in particular control the spectrum very well. However, if a fully populated matrix arises—i.e., the matrix entries that are zero in the cases of [CC78], [JM99], and [Baj+02] are explicitly nonzero—then we violate quality criterion Q11. The consequences have already been described in Section 2.4. On the one hand, the structure of initial elements and their influence domains, as defined in this work, is dissolved; on the other hand, the supports of the generating functions become larger and develop inward corners. To what extent these consequences can be accepted may be subject to further research. For this work, we aim to construct subdivision algorithms that satisfy quality criterion Q11. Thus, we fix the sparse structure described in Section 2.4 as a restriction on the choice of the subdivision matrix.

With this restriction, for a new volume and facet point we obtain the same support of the refinement rule as shown in Figures 6.3 and 6.6. For edge and vertex points in two dimensions, we likewise obtain the support from Figure 6.3. In the three-dimensional case, the support of edge and vertex points comprises all control points of all hexahedra incident to the respective edge or vertex. The zero entries thus coincide with those of [CC78], [JM99], and [Baj+02]. The weights of the respective points, i.e., the nonzero entries, may be freely chosen. This restriction (but not the refinement rules) also agrees with the work [MM18] for the two-dimensional case. There, the double-ring structure appears, which we motivate below. It is worth noting that multi-ring structures have already been considered in the literature, for example in [MM18], as well as in [KP17], [KP18], and [KP19].

As already described in Definition 2.20, regular cells are those whose vertices are associated exclusively with regular initial elements. These are then evaluated as tensor product B-spline elements. The refinement matrices of the regular initial elements are chosen precisely so that the evaluation is reproduced after refinement. Hence, subdivision changes the parametrization of the surface or volume, but not its image.

If we refine the structure from Figure 6.8 with the rules of the Catmull-Clark algorithm, the evaluation of the already evaluable region is reproduced. Furthermore, the evaluable region increases by one ring, as shown in Figure 6.9.

Let us now consider a scenario in which the regular rules for the irregular initial element are modified. For the evaluation of cells, one could—differing from the approach in this work—take the position that for $g = 3$ every cell whose union of control points of its initial elements forms a $4 \times 4$ structure can be evaluated as a B-spline element. In this scenario, the concept of a *regular initial element* in Definition 2.20 would weaken to an *initial element with a regular neighborhood*, and the evaluation around an irregular vertex would initially be identical to the Catmull-Clark scenario from Figure 6.8.

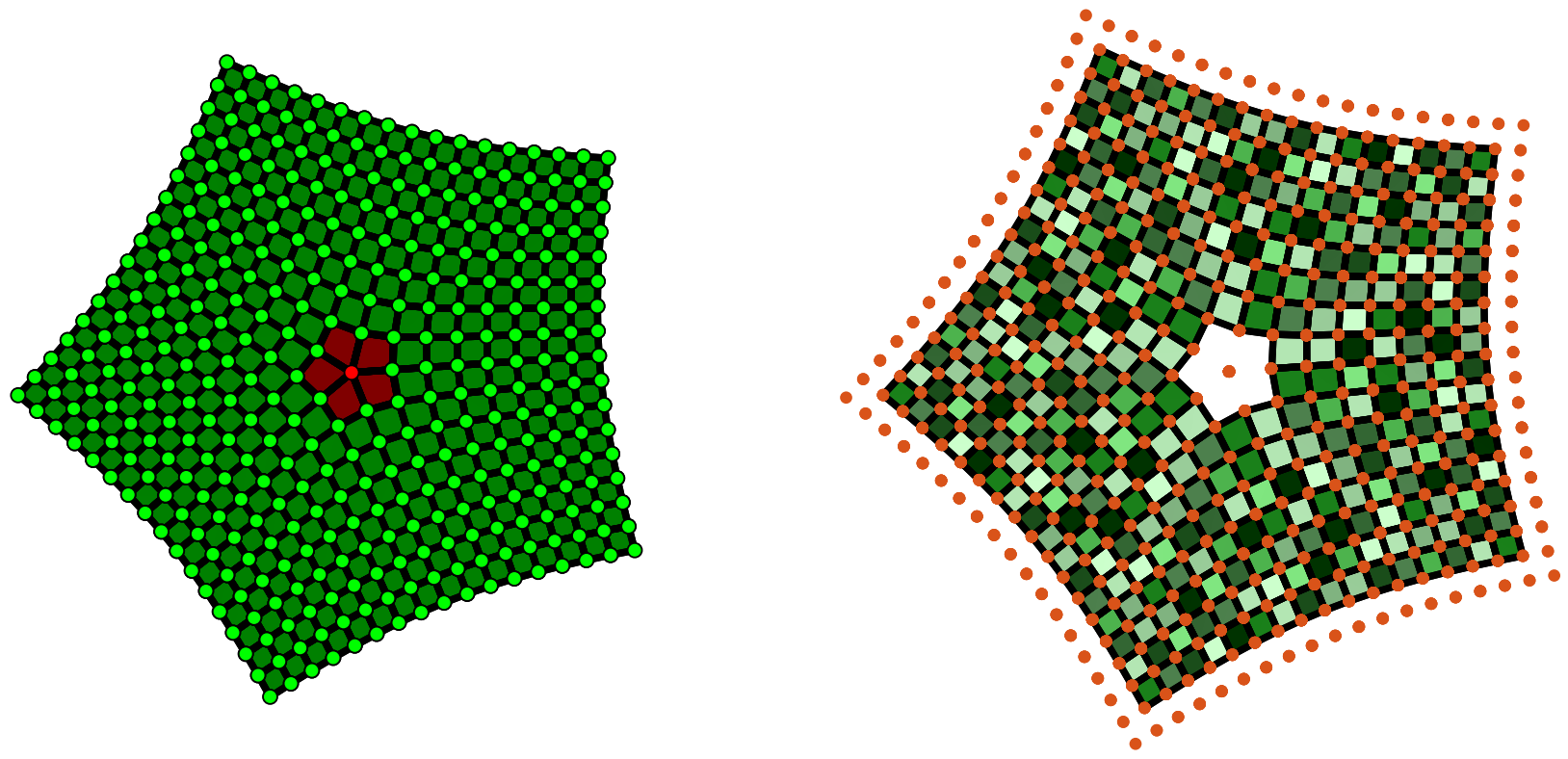

**Figure 6.9:** The spline definition domain from Figure 6.7, refined using the Catmull-Clark subdivision algorithm. The left image shows the refined combinatorial arrangement of the spline definition domain with the corresponding regular and irregular initial elements, as well as the evaluable and non-evaluable cells. On the right, the evaluation including the refined control points is shown.

Now, if we perform a global refinement, the control points of the refined irregular initial element deviate from those of the previous Catmull-Clark scenario. Consequently, all evaluated B-spline surfaces computed with these control points also differ from those in the previous scenario. Thus, on the one hand, the newly generated ring, depicted in yellow in Figure 6.10, changes. This is unproblematic, since the purpose of the subdivision algorithm is precisely to construct these new rings. On the other hand, in this scenario the B-spline elements outside the new ring, shown in blue in Figure 6.10, also change. This change means in particular that parts of the B-spline surface after refinement differ from before; that is, the evaluation is not reproduced by the refinement.

Hence, we recognize through this scenario the price to pay for the freedom in choosing the refinement rules for irregular initial elements. The non-evaluable region thus increases by one ring, and the definitions of mandatory regular initial elements in Definition 2.17 and the evaluation of cells in Definition 2.20 are sensibly chosen. Although these definitions differ from those in [CC78], [JM99] and [Baj+02], the refinement rules for irregular initial elements also differ among these approaches, which justifies the method of this work. The same behavior can analogously be transferred to the case $t = 3$.

After these deviating and non-productive scenarios, we now consider the evaluation structure intended in this work in combination with the freedom of the refinement rules. Subdivision matrices of irregular initial elements may thus be freely chosen, except for the necessary zero entries.

We thereby obtain the condition for the overlapping regions of the initial elements that the refinement rules of individual control points, which are part of the refinement of various initial elements, must be identical. These refinement rules are described in the next section. However, there we use only information from the initial elements as input to the algorithm, not the information of the surrounding environment including overlaps. Therefore, we define:

**Definition 6.1.** *If a control point is refined by different initial elements, the refinement rule is to be chosen according to the following priority list: The initial element is associated with a*

1. *irregular vertex,*
2. *semi-irregular vertex,*
3. *regular vertex.*

If two equally ranked initial elements propose different refinement rules for a control point, then the refinement must be selected manually. However, since the regular region grows larger during subdivision and the irregular regions become increasingly isolated, this conflict no longer occurs after sufficiently many subdivision steps and can be neglected.

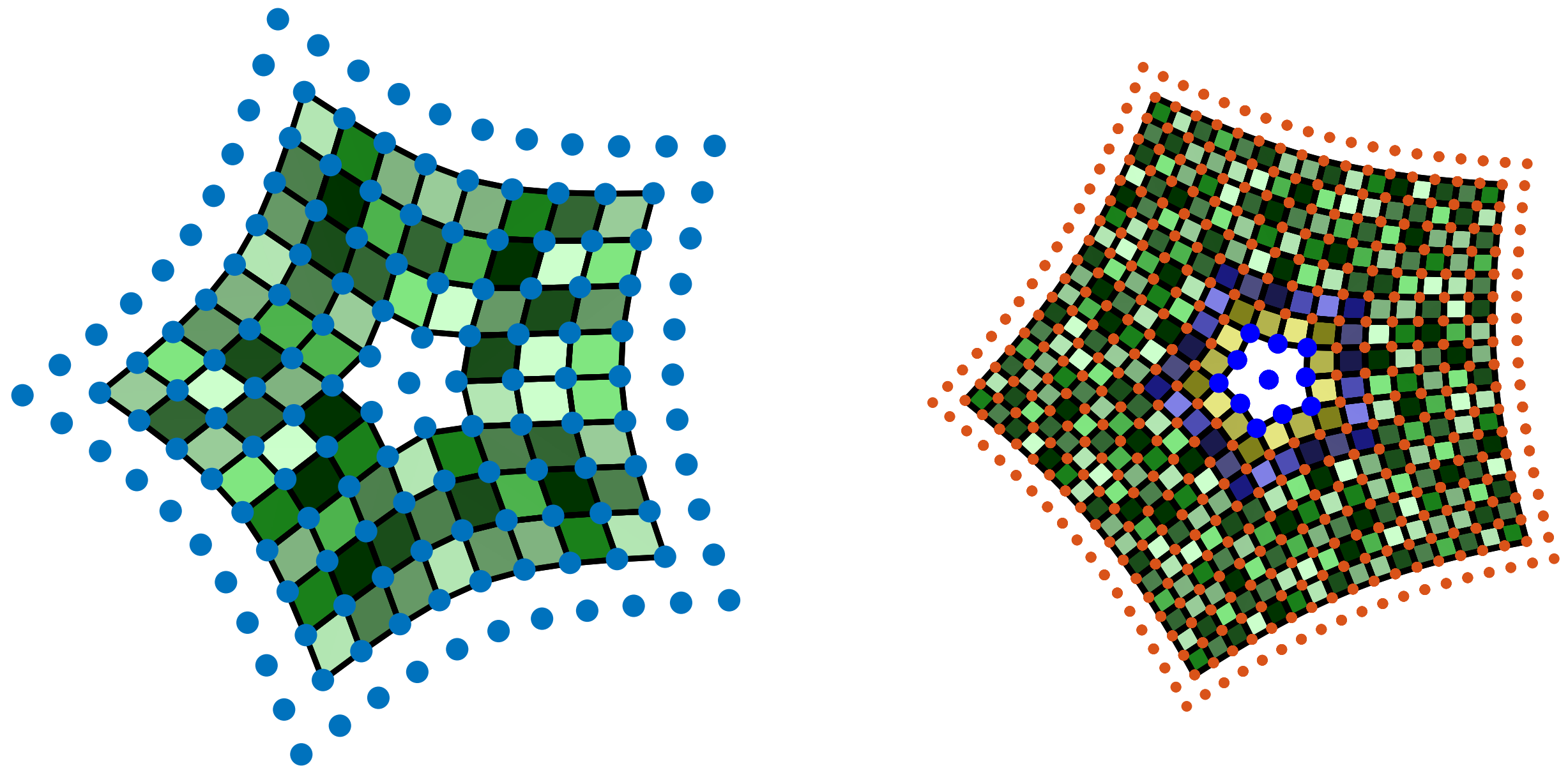

**Figure 6.10:** An experimental scenario intended to illustrate the problem of the evaluable region. The basis is the spline definition domain from Figure 6.7. The irregular initial element at the center receives alternative refinement rules for each point, causing the initial elements surrounding the center to become irregular as well. This is ignored during the evaluation shown in the left image. Every cell whose initial elements all have regular neighborhoods is evaluated, resulting in three evaluated rings. If the spline definition domain is refined globally and evaluated according to the same principle, the control points and evaluation in the right image are obtained. The control points generated by the irregular rules are colored blue. All evaluated elements that include these blue control points in their evaluation are colored yellow and blue. The yellow elements are new, while the blue elements are refinements of already existing elements from the left image, specifically part of the inner ring. Due to the deviation in the refinement rule, the blue evaluated elements do not reproduce the previous coarser structure. Therefore, the differing evaluation in this scenario is not meaningful.

This also explains why, in Definition 2.12, the subdivision matrix of an initial element depends on its neighbors. In case of conflict, the refinement rule can also depend, by priority, on another initial element and thus on its combinatorial structure. Examples of overlaps for $t = 2$ can be found in Figure 2.30.

With this definition, the regularity of the initial elements for the case $g = 3$ can be described. Since the irregular and semi-irregular refinement rules have precedence, their neighbors, which overlap with these initial elements' refinement rules, are automatically irregular. Therefore, for the necessarily regular region, an initial element must not only have a regular neighborhood, but the surrounding initial elements must also have regular neighborhoods, as described in Definition 2.17. As already explained, this enlarges the non-evaluable region by one element each time. For the example chosen in this section, the evaluable region corresponds to that shown in Figure 6.11.

To generate an invariant evaluable structure, we need a double ring or double shell structure. When we refine the spline definition domain globally, the evaluable region around the irregular corner grows by the size of one element of the old spline definition domain, which corresponds to the size of two elements in the new spline definition domain. Therefore, for the invariant structure, we must always map two rings or two shells onto each other. This is consistent with the description in [MM18] for the case $t = 2$, as already discussed. An illustration can be found in Figure 6.12. The three-dimensional case behaves analogously.

Thus, we have identified the challenges of this section as well as the objectives of the next section. The rows of a subdivision matrix for $g = 3$ are allowed to have nonzero entries only for those control points that appear in all overlapping initial elements. If different initial elements propose different refinement rules, then initial elements associated with irregular and semi-irregular corners take precedence. Furthermore, to map evaluable structures onto each other, we require subdivision matrices that map double rings and double shells onto double rings and double shells. In the next section, we will explain the construction of subdivision matrices for initial elements, from which the larger structures will be composed.

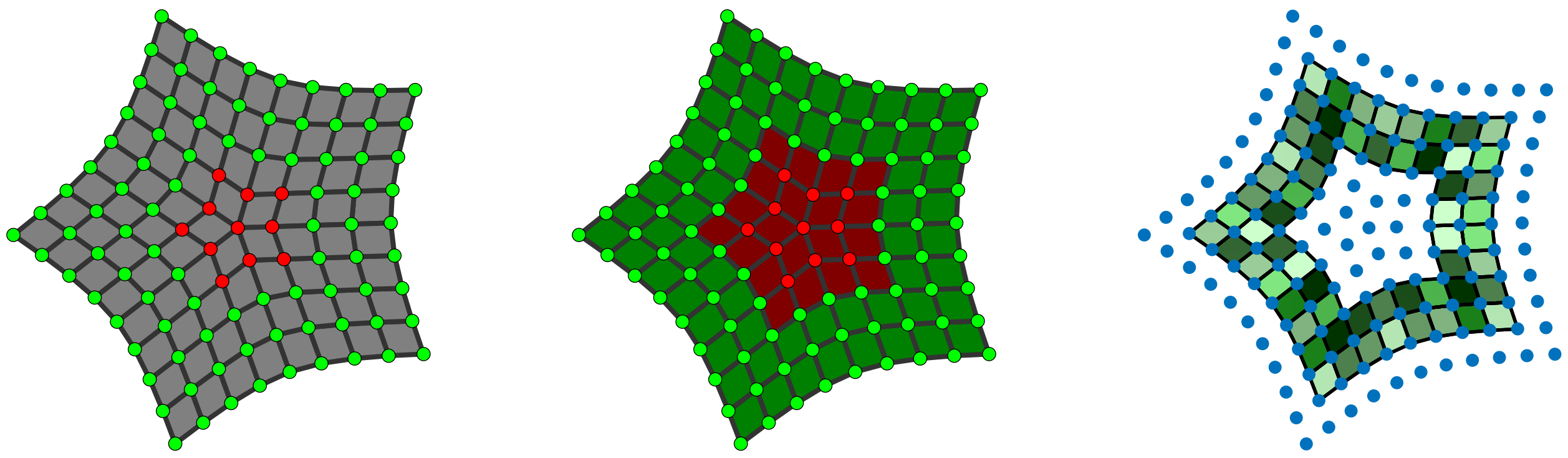

**Figure 6.11:** The spline definition domain from Figure 6.7. The initial elements are labeled with the subdivision matrices from this work. The resulting regular initial elements are colored green, whereas the irregular ones are colored red (left). In the middle image, the evaluable cells are shown in green and the non-evaluable cells in red, which leads to the evaluation shown on the right. Here, the non-evaluable region is enlarged by one ring compared to Figure 6.8.

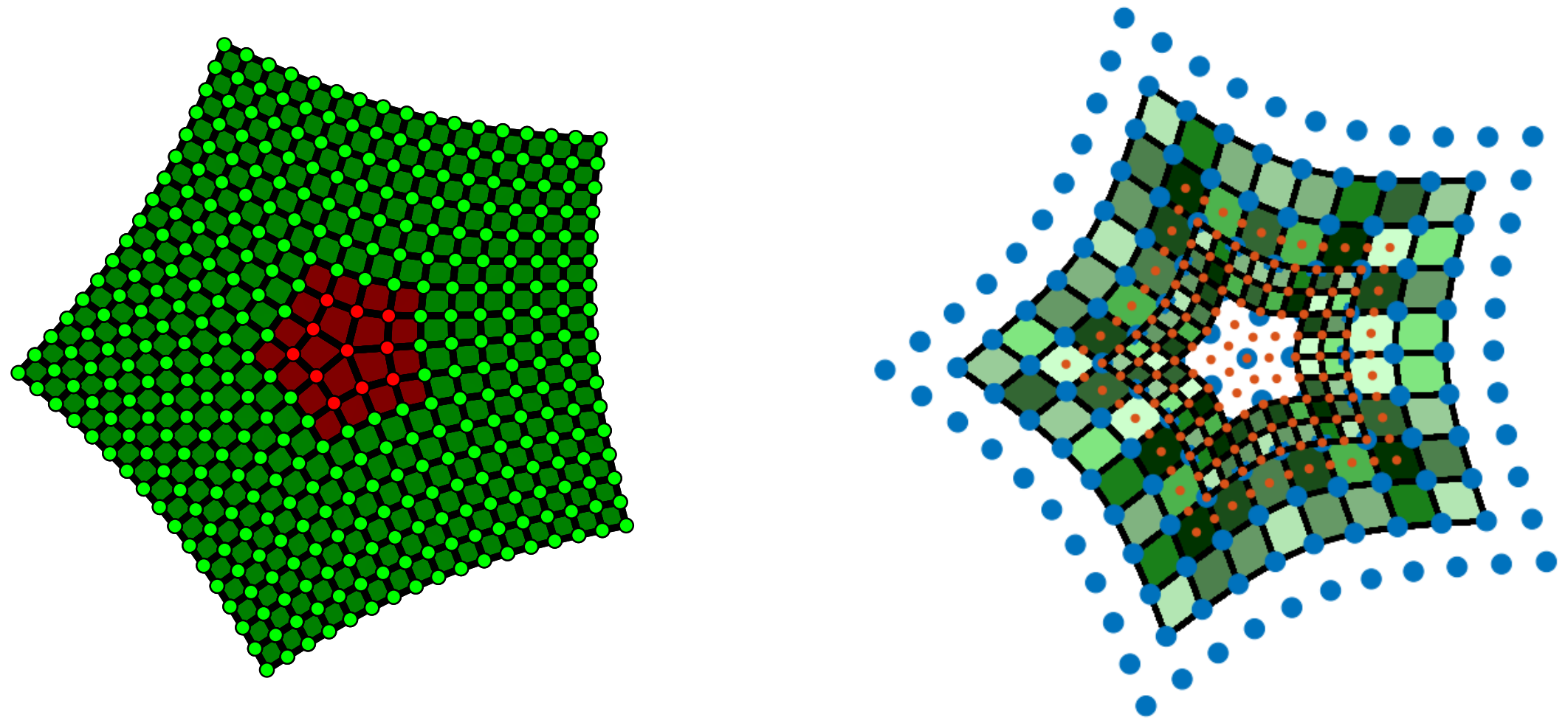

**Figure 6.12:** The spline definition domain from Figure 6.7, refined using the subdivision algorithm of this work. The left image shows a representation of the refined spline definition domain with the corresponding regular and irregular initial elements as well as the evaluable and non-evaluable cells. On the right, the evaluation of the original structure and the evaluation of the invariantly refined structure including control points are shown in one image. The new double ring structure fits precisely into the gap of the old one.

## 6.2 Construction of the Subdivision Matrices

The construction of subdivision matrices for the generalized cubic B-spline subdivision is more complex than that for the generalized quadratic B-spline subdivision. This is partly because the geometric objects we work with are no longer just polytopes, but refinements of polytopes, so we must invest considerable effort to apply the theory from Chapters 3 and 4.

Moreover, as described in the previous section, there are constraints on the matrix entries, expressed as necessary zero entries. This restriction is not compatible with Variant 1 from Section 5.2.1, which is why we do not use it in this chapter. Instead, we focus on adapting Variant 2 from Section 5.2.2 and only provide a brief note at the end about a variant that defines the spectrum exactly.

Because the construction is more involved, Algorithm 12 only gives a brief overview of the individual steps, which we explain in detail in the following subsections. Furthermore, we prove some lemmas in these subsections that are needed both for the construction and for the quality criteria in Section 6.3.

**Algorithm 12:** GeneralizedCubicB-SplineSubdivisionMatrixOverview

**Data:** Type $t \in \{2, 3\}$, valid combinatorial graph $\mathbf{G}_K = (\mathbf{V}_K, \mathbf{E}_K)$ of an initial element in the input form of Definition 3.10, where each node in $\mathbf{V}_K$ has exactly $t$ neighbors
**Result:** Subdivision matrix $S^{(t)}$

```
begin
    Construct, according to Section 6.2.1, from the combinatorial graph G_K a t-dimensional realization
      P̃ ∈ ℝ^{n×t} of the graph G of the initial element, which serves as a basis of the subdominant eigenspace;
    From the t-dimensional structure P̃, construct a Colin-de-Verdière-like matrix C̃ according to
      Section 6.2.2;
    Split the matrix C̃ according to Section 6.2.3 into the sum of two matrices. Compute the matrix
      exponentials of these two summands and multiply the two exponentials to obtain the subdivision matrix
      S^(t);
```

To follow the further course of this section, the following is essential: Each construction step in the subsections 6.2.1, 6.2.2, and 6.2.3 is described by an algorithm. In these subsections, these algorithms are explained, and some statements about the variables and matrices appearing in the algorithms are additionally proven. The notation, matrices, and variables in these subsections always refer to those of the respective algorithms.

### 6.2.1 Eigenstructure of the Subdominant Eigenvalues

To construct the eigenstructure of the subdominant eigenvalues, we must first consider the combinatorial structure of the initial elements. According to Definition 2.12, the graph **G** of an initial element consists of the structure of the cells adjacent to the corner associated with that initial element. Thus, the graph **G** represents a collection of combinatorial quadrilaterals for $t = 2$ or combinatorial hexahedra for $t = 3$. It is important to note here that throughout the construction we often refer to these quadrilaterals or hexahedra. By this we always and exclusively mean the quadrilaterals and hexahedra that can be associated with the corresponding cells.

Using Definition 2.51, we further restrict the structure of the initial elements. The combinatorial graph $\mathbf{G}_K$ of the initial element from Definition 2.50 can, for $g = 3$, be interpreted as the volumetric dual graph of the structure in **G**. This graph must be a cycle (circle) in the graph-theoretic sense for $t = 2$, and for $t = 3$ it must be a planar 3-connected graph. It is important to emphasize here that, for this chapter, we explicitly assume that the initial element arises from a quadrilateral or hexahedral structure. Thus, each node in $\mathbf{G}_K$ must have exactly $t$ neighbors. The form of the eigenstructure is initially completely open. We can only orient ourselves by the regular initial elements, as their subdivision matrices are the only ones already fixed. These eigenstructures are shown in Figure 6.13.

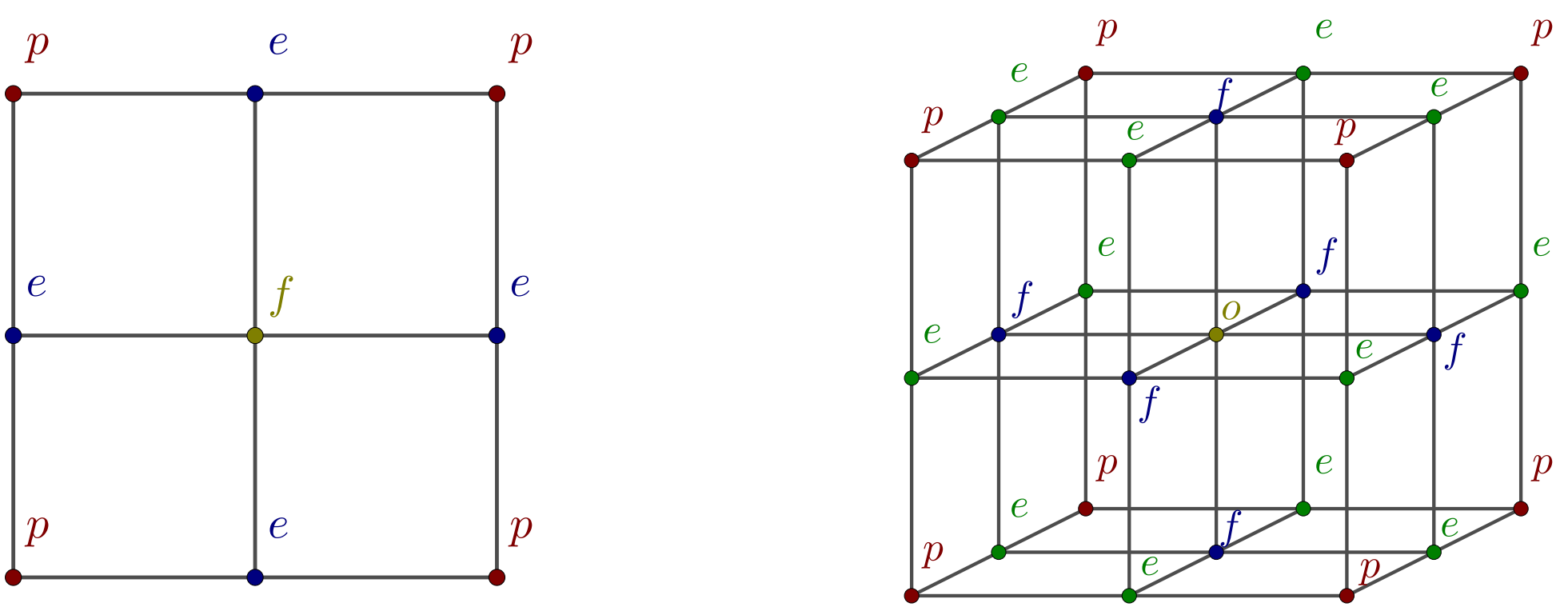

**Figure 6.13:** Eigenstructure of the regular initial elements for $t = 2$ (left) and $t = 3$ (right) with the corresponding categorization of the nodes.

At this point, it would be possible to use the eigenspaces of the Catmull-Clark algorithm for $t = 2$. However, the further construction will require certain properties of the eigenstructure that the eigenvectors of the Catmull-Clark algorithm do not possess. For $t = 3$, we have no template or orientation since the subdivision algorithms of [Baj+02] and [JM99] do not provide suitable eigenspaces, as we have seen in Section 2.5.

Therefore, we construct our own proposal for an eigenstructure. Since this differs significantly between $t = 2$ and $t = 3$, we split this section into the two cases $t = 2$ and $t = 3$. We begin with the first case.

### The case $t$=2

The construction of the subdominant eigenstructure for $t = 2$ is given in the following Algorithm 13, whose steps we explain in detail. As already mentioned, the variables and matrices in this subsection always refer to those in Algorithm 13.

**Algorithm 13:** EigenstructureGeneralizedCubicBSplineSubdivisionMatrix2D

**Data:** Valid combinatorial graph $\mathbf{G}_K = (\mathbf{V}_K, \mathbf{E}_K)$ of an initial element in the input form of Definition 3.10, where each node in $\mathbf{V}_K$ has exactly 2 neighbors

**Result:** Control points $\left[P^T, E^T, f^T\right]^T \in \mathbb{R}^{n\times 2}$, whose columns represent the subdominant eigenvectors of the subdivision matrix

**begin**
- Construct the graph $\mathbf{G}$ from the combinatorial graph $\mathbf{G}_K$ with the three node types $p$, $e$, and $f$;
- Create the primal and dual polytopes as realizations of $\mathbf{G}_K$ according to Section 3.4;
- Set the coordinates of the vertex points $P$ as the vertices of the primal polytope;
- Set the coordinates of the edge points $E$ as the vertices of the dual polytope;
- Set the coordinates of the facet point $f$ as $(0, 0)$;
- **return** $\left[P^T, E^T, f^T\right]^T$;

The combinatorial graph $\mathbf{G}_K$ is the input to Algorithm 12, from which we want to construct the eigenstructure for the graph $\mathbf{G}$. It is important to emphasize that the combinatorial graph $\mathbf{G}_K$ is not the graph $\mathbf{G}$ of the initial element.

**Construction of the graph G** We first describe the structure of the graph **G**. Linguistically, we use the terms point and node synonymously here, since we want to generate a realization in $\mathbb{R}^2$ from the abstract nodes of the graph **G**. For this, we divide the points of the realization of **G** into the following categories:

**Definition 6.2.** *Let **G** be the graph of an initial element for $t = 2$ and $g = 3$, which, by Definition 2.12, consists of an assembly of quadrilaterals all sharing a common vertex. The realization of this common vertex is called the* facet point *and denoted by $f \in \mathbb{R}^2$.*

*The realization of all nodes in **G** that share an edge with $f$ are called* edge points. *An edge point is denoted by $e \in \mathbb{R}^2$, and the matrix of all edge points of a graph is denoted by $E \in \mathbb{R}^{n\times 2}$.*

*All other points are called* vertex points. *A vertex point is denoted by $p \in \mathbb{R}^2$, and the matrix of all vertex points of a graph is denoted by $P \in \mathbb{R}^{m\times 2}$. Additionally, we define*

$$\tilde{P} = \begin{bmatrix} P \\ E \\ f \end{bmatrix}.$$

With this definition, the graph **G** contains only edges of the form $\{p, e\}$ and $\{e, f\}$. The motivation for these terms and the construction of the graph **G** from the combinatorial graph $\mathbf{G}_K$ will be explained below.

Since each quadrilateral has exactly one vertex, the structure of the combinatorial graph $\mathbf{G}_K$ can be used to generate the graph **G**. In a first step, we identify the nodes of the combinatorial graph $\mathbf{G}_K$ with the vertices $P$. If two nodes in the combinatorial graph $\mathbf{G}_K$ are adjacent, then the corresponding quadrilaterals of the graph **G** share an (internal) edge. On this (internal) edge lies an edge point $e$. The point $p$ does not lie on this (internal) edge, since the identified edges always contain the point $f$. Thus, we can extend the combinatorial graph by the points $E$ by splitting each edge into two edges and inserting the points $E$ at their midpoint. The graph **G** is then obtained in a final step by inserting the facet point $f$ and connecting it by an edge to each edge point $e$. Hence, the combinatorial graph $\mathbf{G}_K$ suffices to generate the graph **G**, and the points or nodes of the graph **G** can be partitioned into three disjoint sets.

**Construction of the coordinates** Since the points $P$ can be identified with the nodes of the graph $\mathbf{G}_K$, we can use this to generate the coordinates of the points $P$. For this purpose, we construct the primal polytope of $\mathbf{G}_K$ with the procedure from Section 3.4 and set the coordinates of $P$ as the vertices of this realization.

Realizing the graph $\mathbf{G}_K$ by these constructed points $P$, the edges of the realization of $\mathbf{G}_K$ lie tangent to the unit circle. The tangent points of these edges to the unit circle are the vertices of the polytope dual to $\mathbf{G}_K$, and we define the points $E$ as these dual tangent points. The two edges $\{p_1, e\}$ and $\{e, p_2\}$ thus replace the edge $\{p_1, p_2\}$ from $\mathbf{G}_K$.

We place the point $f$ at the center of the unit circle, i.e., at $(0, 0)$, and connect it by edges to the edge points $E$. This procedure is illustrated in Figure 6.14.

Of course, this procedure described for $t = 2$ is trivial. An $n$-gon is simply decomposed into $n$ quadrilaterals. We nevertheless describe this procedure in detail to both fully describe the construction and to convey the idea of the construction, which will later be transferred to the much more complex case $t = 3$. Therefore, the next two lemmas for the points $\left[P^T, E^T, f^T\right]^T$ will also be proven in detail:

**Lemma 6.3.** *Let $t = 2$ and $g = 3$, and let the realization of **G**, consisting of the points $\left[P^T, E^T, f^T\right]^T$, be constructed according to Algorithm 13. Then the line segments corresponding to the edges $\{p, e\}$ and $\{e, f\}$, which share the common point $e$, are orthogonal to each other.*

*Proof.* This lemma follows from the construction, but it can also be verified by calculation. We describe the segments by the vectors $p - e$ and $f - e$. The scalar product of these two vectors is

$$\langle p - e, f - e\rangle = \langle p - e, -e\rangle = \langle p, -e\rangle + \langle -e, -e\rangle = -\langle p, e\rangle + 1.$$

The last equality holds since the points $e$ lie on the unit circle. Because the points $p$ are constructed via the method in Section 3.4 using Equation (3.15), it holds that

$$\langle p, e\rangle = 1,$$

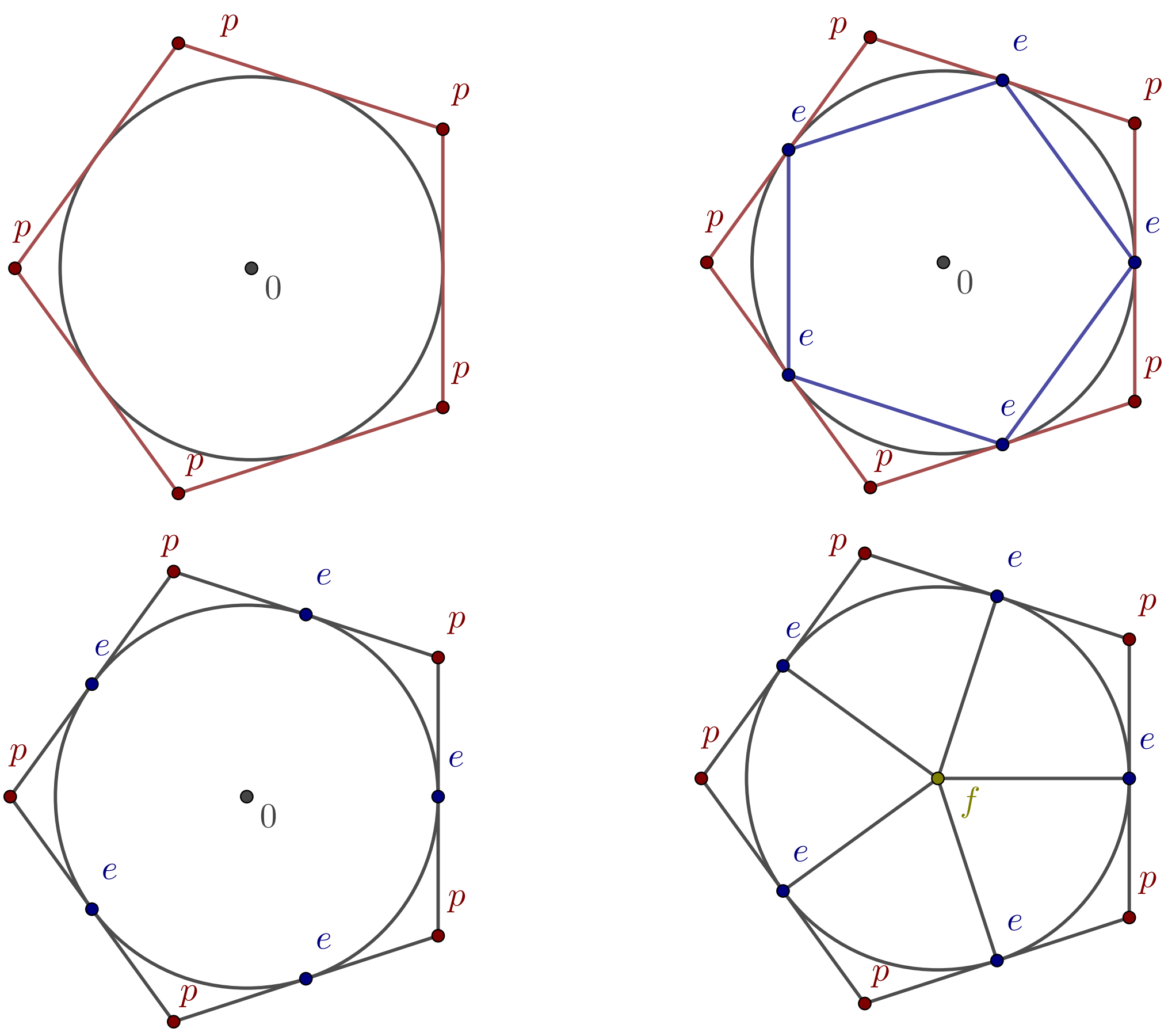

**Figure 6.14:** Construction steps of the subdominant eigenstructure for $t = 2$. First, the primal polytope of $\mathbf{G}_K$ is generated, and its vertices are associated with the points $P$ (top left). Then the dual polytope is generated, and its vertices are associated with the points $E$ (top right). Subsequently, the edges of the primal polytope are replaced by two edges (bottom left), and finally, the facet point $f$ is inserted at the origin and connected by edges to the points of $E$ (bottom right).

which implies

$$\langle p - e, f - e \rangle = -1 + 1 = 0.$$

Thus, the two segments are orthogonal to each other. □

**Lemma 6.4.** *Let $t = 2$ and $g = 3$, and let the realization of **G**, consisting of the points $\left[P^T, E^T, f^T\right]^T$, be constructed according to Algorithm 13. Then each quadrilateral of the graph **G**, formed by the points $p$, $e_1$, $e_2$, and $f$, is convex.*

*Proof.* Initially, the realization of the combinatorial graph $\mathbf{G}_K$, consisting of the points $P$, is constructed to be tangent to the unit circle and contains it. Therefore, by Lemma 3.45, which describes the convexity of exactly this object, it is convex. Moreover, it contains the quadrilateral described above.

The triangle $(f, e_1, e_2)$ is part of the dual polytope and is also convex and non-empty, since the three points do not lie on a line. The subgraph of **G**, consisting of the vertices $P$ and the tangent points $E$, forms a cycle in the graph-theoretical sense. On this cycle, there is always a vertex $p$ between two edge points $e_1$ and $e_2$. The orthogonal projection of $p$ onto the line $(e_1, e_2)$ in the direction of the origin lies on the segment $(e_1, e_2)$.

The segment $(e_1, e_2)$ is part of the dual polytope corresponding to the realization of $\mathbf{G}_K$ and is defined by the equation

$$\langle p, x \rangle = 1.$$

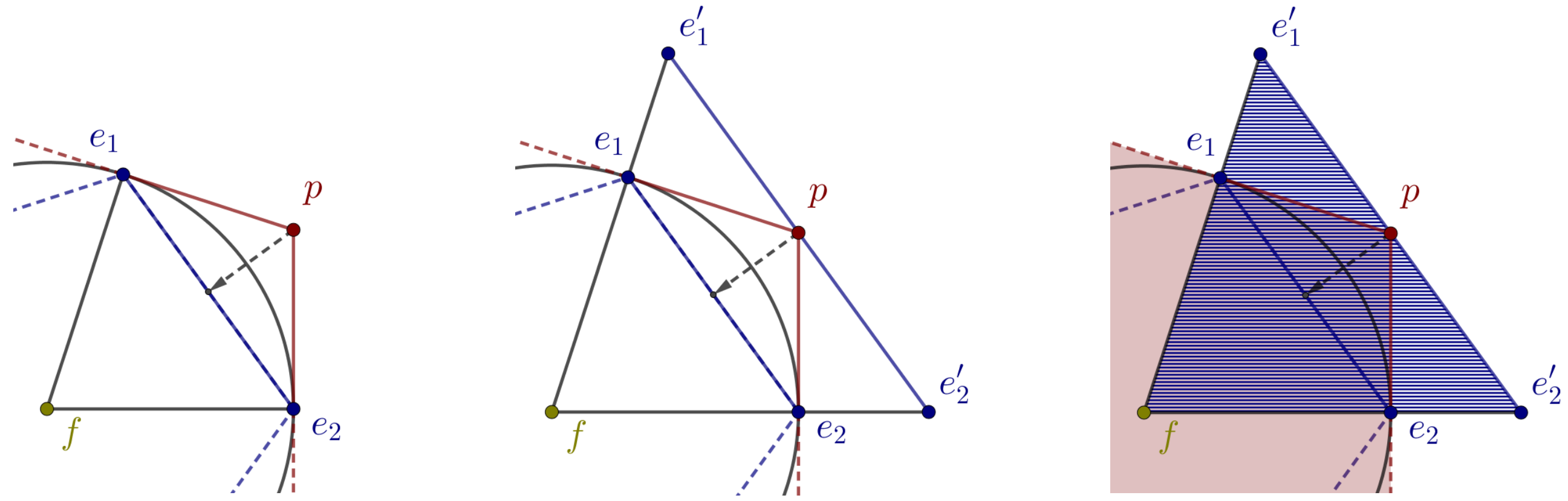

**Figure 6.15:** Illustration of the proof of Lemma 6.4. On the left, the four points of the quadrilateral and the projection of point $p$ onto the segment $(e_1, e_2)$ are shown. In the middle, the line spanned by $e_1$ and $e_2$ is shifted so that point $p$ lies on the new line. From this, the points $e_1'$ and $e_2'$ can be constructed. On the right, the two convex sets are shown whose intersection is the quadrilateral $(f, e_1, p, e_2)$.

Furthermore, $p$ is not contained in the dual polytope, since the dual polytope is a subset of the unit circle and the vertices $P$ lie outside the unit circle. We now construct a new line defined by

$$\langle p, x \rangle = |p|^2.$$

This line is parallel to the line $(e_1, e_2)$ and contains the point $p$, since $\langle p, p \rangle = |p|^2$. Intersecting this line with the two lines spanned by $(0, e_1)$ and $(0, e_2)$ creates a triangle $(0, e_1', e_2')$, which contains the quadrilateral $(p, e_1, f, e_2)$. Since this is a triangle, it is also convex.

The intersection of $(0, e_1', e_2')$ with the realization of $\mathbf{G}_K$ is thus the quadrilateral $(p, e_1, f, e_2)$. Since both sets in the intersection are convex, by Lemma 3.3, which considers the intersection of two convex sets, the quadrilateral $(p, e_1, f, e_2)$ is convex. An illustration of the proof is given in Figure 6.15.

□

This completes the construction of the eigenstructure for $t = 2$. In the next subsection, we consider the construction for $t = 3$.

### The case $t$=3

For the case $t = 3$, we proceed similarly to the previous case, but the construction of the eigenstructure is somewhat more complex. We describe it in Algorithm 14 and explain the individual steps of the algorithm below. All matrices and variables in this subsection refer to those in Algorithm 14.

**Construction of the graph G** We first describe the structure of the graph **G**. Here, the terms point and node are used synonymously. In the first step, we divide the points of the realization of **G** into the following categories:

**Definition 6.5.** *Let **G** be the graph of an initial element for $t = 3$ and $g = 3$, which, according to Definition 2.12, consists of a collection of hexahedra all sharing a common vertex. The realization of this common vertex is called the* volume point *and is denoted by $o \in \mathbb{R}^3$.*

*The realization of all nodes in **G** that share an edge with $o$ are called* facet points. *A facet point is denoted by $f \in \mathbb{R}^3$, and the matrix of all facet points of a graph is denoted by $F \in \mathbb{R}^{n \times 3}$.*

*The realization of all nodes in **G** that share an edge with a facet point, but are not the point $o$, are called* edge points. *An edge point is denoted by $e \in \mathbb{R}^3$, and the matrix of all edge points of a graph is denoted by $E \in \mathbb{R}^{m \times 3}$.*

**Algorithm 14:** EigenstructureGeneralizedCubicB-SplineSubdivisionMatrix3D

**Data:** Valid combinatorial graph $\mathbf{G}_K = (\mathbf{V}_K, \mathbf{E}_K)$ of an initial element in the input form of Definition 3.10, where each node in $\mathbf{V}_K$ has exactly 3 neighbors

**Result:** Control points $\left[P^T, E^T, F^T, o^T\right]^T \in \mathbb{R}^{n\times 3}$, whose columns represent the subdominant eigenvectors of the subdivision matrix

```
begin
    Construct the graph G from the combinatorial graph G_K with the four types of nodes p, e, f, and o;
    Create the primal and dual polytope as realizations of G_K according to Section 3.3;
    Set the coordinates of points P as the vertices of the primal polytope;
    Set the coordinates of points E as the tangent points of the primal polytope to the unit sphere, or
      equivalently, as the intersection points of the primal and dual polytopes;
    Project the vertices of the dual polytope onto the facets of the primal polytope and set the coordinates of
      points F as these projections;
    Set the coordinates of point o as (0, 0, 0)^T;
    return P := [P^T  E^T  F^T  o^T]^T;
```

*All remaining points are called* corner points. *A corner point is denoted by $p \in \mathbb{R}^3$, and the matrix of all corner points of a graph is denoted by $P \in \mathbb{R}^{k\times 3}$. Additionally, we define*

$$\tilde{P} = \begin{bmatrix} P \\ E \\ F \\ o \end{bmatrix} .$$

Thus, the graph **G** contains only edges of the form $\{p, e\}$, $\{e, f\}$, and $\{f, o\}$.

Since each hexahedron has exactly one corner point, we can use the structure of the combinatorial graph $\mathbf{G}_K$ to generate the graph **G**. In the first step, we identify the nodes of the combinatorial graph $\mathbf{G}_K$ with the corner points $P$. If two nodes in the combinatorial graph $\mathbf{G}_K$ are adjacent, then the corresponding hexahedra in the graph **G** share an (interior) facet. On this (interior) facet lie the central point $o$, two facet points $f_1$ and $f_2$, and one edge point $e$, which is connected to the respective corner points $p$.

We can extend the combinatorial graph by the points $E$ by splitting each edge into two edges and inserting the points $e$ in their middle. Since each edge point $e$ is connected to two facet points, we obtain the facet points by inserting them in the centers of the facets of $\mathbf{G}_K$ and connecting them to the respective edge points.

Finally, the graph **G** is obtained by adding the volume point $o$ and connecting it by an edge to each facet point $f$. Thus, the combinatorial graph $\mathbf{G}_K$ is also sufficient for $t = 3$ to generate the graph **G**, and the points or nodes of the graph **G** can be divided into four disjoint sets. An illustration of the regular eigenstructure including the naming is shown in Figure 6.13.

**Construction of the coordinates** Since the points $P$ can be identified with the graph $\mathbf{G}_K$, we can use it to generate the coordinates of the points $P$. For this, we use the method from Section 3.3 and set the coordinates $P$ as the vertices of the primal polytope. The tangent points are also generated during the construction of the primal polytope, so we obtain the coordinate points of the edge points $E$ also from the primal polytope.

The construction of the facet points is somewhat more involved. Here, an alternative approach would be to use the coordinates of the dual polytope directly. However, this would contradict the eigenstructure of the regular case, which we know from the matrix in Equation (6.1). But from the eigenstructure of the regular case, we see that the facet points lie in the facets of the primal polytope, which is spanned by the corner points $P$. Therefore, we project the points of the dual polytope using Equation (3.12) onto the facets of the primal polytope and obtain

$$f = \frac{1}{|a|^2} a,$$

where $a$ is a vertex of the dual polytope. Finally, we place the volume point $o$ at the origin $(0, 0, 0)$. This procedure is illustrated in Figure 6.16.

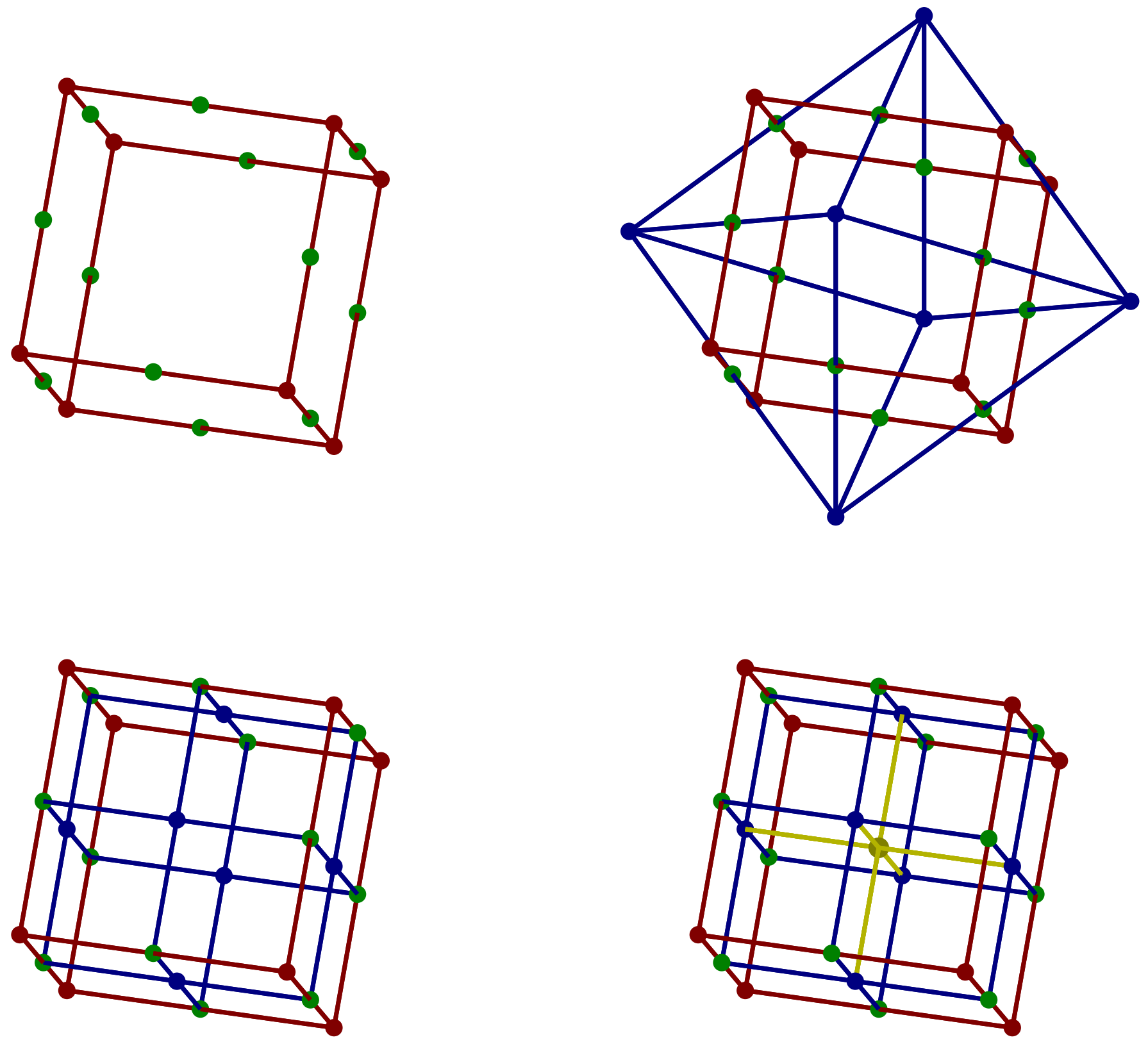

**Figure 6.16:** Construction steps of the subdominant eigenstructure for $t = 3$. The points $P$ are shown in red, the points $E$ in green, and the points $F$ in blue. Additionally, the point $o$ is shown in yellow. First, the primal polytope of $\mathbf{G}_K$ is generated. Its vertices are associated with the points $P$ and its tangent points with the points $E$ (top left). Next, the dual polytope is generated (top right). The vertices of the dual polytope are projected onto the facets of the primal polytope and associated with the points $F$ (bottom left). Finally, the volume point $o$ is inserted at $(0, 0, 0)$ (bottom right).

To establish a further connection, we note that the structure of the points $P$, $E$, and $F$ and their edges is precisely that of the restricted quad graph from Definition 3.27.

Analogous to the two-dimensional case, we discuss the following lemmas:

**Lemma 6.6.** *Let $t = 3$ and $g = 3$. Consider the realization of* **G***, consisting of the points $\left[P^T, E^T, F^T, o^T\right]^T$, constructed according to Algorithm 14. Then the segments corresponding to the edges $\{p, e\}$ and $\{e, f\}$, sharing the common point $e$, are orthogonal to each other. Moreover, the segments corresponding to the edges $\{e, f\}$ and $\{f, o\}$, sharing the common point $f$, are orthogonal to each other.*

*Proof.* We begin with the edge pair $\{p, e\}$ and $\{e, f\}$. The primal polytope was constructed via a circle packing. Thus, the intersection of each facet of the primal polytope with the unit sphere is a circle on which all tangent points of the facet lie. An illustration can be found in Figure 3.3. Furthermore, by Corollary 3.40, the projection of the dual point, which we denote by $f$, is the center of this circle.

If we translate the circle so that $f$ lies at the origin, the rest of this part of the proof follows analogously to the proof of Lemma 6.3. We get

$$\langle p - e, f - e \rangle = \langle p - e, -e \rangle = \langle p, -e \rangle + \langle -e, -e \rangle = -\langle p, e \rangle + |e|^2.$$

Since the point $p$ lies on the tangent line of the circle at the tangent point $e$, it follows, after the translation, that

$$\langle p, e\rangle = |e|^2,$$

because this is the equation of the tangent line with support point $e$. Thus,

$$\langle p - e, f - e\rangle = -|e|^2 + |e|^2 = 0.$$

Since the translation of the circle does not affect orthogonality, this also holds for the original points.

For the edge pair $\{e, f\}$ and $\{f, o\}$, we again use that the intersection of the unit sphere and the facet corresponding to $f$ is a circle whose center is $f$. The tangent point $e$ lies on this circle, so the segment corresponding to the edge $\{e, f\}$ lies in the plane spanned by the facet.

The segment corresponding to the edge $\{f, o\}$ can also be interpreted as the position vector of $f$, i.e., the center of the circle. By Lemma 3.39, this position vector points in the same direction as the normal vector of the plane containing the facet at $f$. Since the segment corresponding to the edge $\{e, f\}$ lies in this plane, the two segments corresponding to the edges $\{e, f\}$ and $\{f, o\}$ are orthogonal and form a right angle at the common point $f$. An illustration with different notation can be found in Figure 3.15. □

**Lemma 6.7.** *Let* **G** *be the graph realized by the points* $\left[P^T, E^T, F^T, o^T\right]^T$ *constructed according to Algorithm 14 for* $t = 3$ *and* $g = 3$*. Then each hexahedron of the graph* **G***, consisting of the points* $p$*,* $e_1$*,* $e_2$*,* $e_3$*,* $f_1$*,* $f_2$*,* $f_3$*, and* $o$*, is convex.*

*Proof.* We apply here a strategy analogous to the proof of Lemma 6.4. First, the realization of $\mathbf{G}_K$, consisting of the points $P$, is convex according to the final version of the Koebe-Andreev-Thurston theorem 3.25 and contains the hexahedron.

Let $f_1'$, $f_2'$, and $f_3'$ be the vertices of the polytope dual to the realization of $\mathbf{G}_K$, which correspond to the projection points $f_1$, $f_2$, and $f_3$, so that

$$f_i = \frac{1}{|f_i'|^2} f_i', \quad \text{for } i \in \{1, 2, 3\}.$$

The tetrahedron formed by the convex hull of the points $f_1'$, $f_2'$, $f_3'$, and $o$ is convex by definition. Since the points $f_1'$, $f_2'$, and $f_3'$ lie outside the unit sphere, the points $f_1$, $f_2$, and $f_3$ lie inside the unit sphere. Hence, the points $f_i$ for $i \in \{1, 2, 3\}$ lie on the line segments between $f_i'$ and $o$, and therefore inside the aforementioned tetrahedron.

The points $f_1'$, $f_2'$, and $f_3'$ form a facet of the polytope dual to the realization of $\mathbf{G}_K$. The intersection of this facet with the unit sphere is, due to the circle packing, a circle containing the points $e_1$, $e_2$, and $e_3$. Since the points $e_1$, $e_2$, and $e_3$ lie on the three edges of the described facet, the other three facets of the tetrahedron each contain a facet of the hexahedron. These are concretely $(o, f_1, e_2, f_2)$, $(o, f_2, e_3, f_3)$, and $(o, f_3, e_1, f_1)$. Thus, the intersection of the tetrahedron with the convex polytope consists exclusively of points of the hexahedron.

Furthermore, by Corollary 3.40, the projection of $p$ onto the facet spanned by $f_1'$, $f_2'$, and $f_3'$ is the center of the circle formed by the intersection of the facet with the unit sphere. This center can be expressed as

$$p' = \frac{1}{|p|^2} p,$$

and the plane containing the facet can be described, by Lemma 3.39, as

$$\langle p', x\rangle = |p'|^2.$$

If we shift this plane containing the points $p'$, $f_1$, $f_2$, and $f_3$ so that

$$\langle p', x\rangle = 1,$$

then $p$ lies on this shifted plane, since

$$\langle p', p\rangle = |p| \cdot |p'| = |p| \frac{1}{|p|^2} |p| = 1.$$

Using this shifted plane, we can construct a second tetrahedron. We start from the convex hull of $f_1'$, $f_2'$, $f_3'$, and $o$,

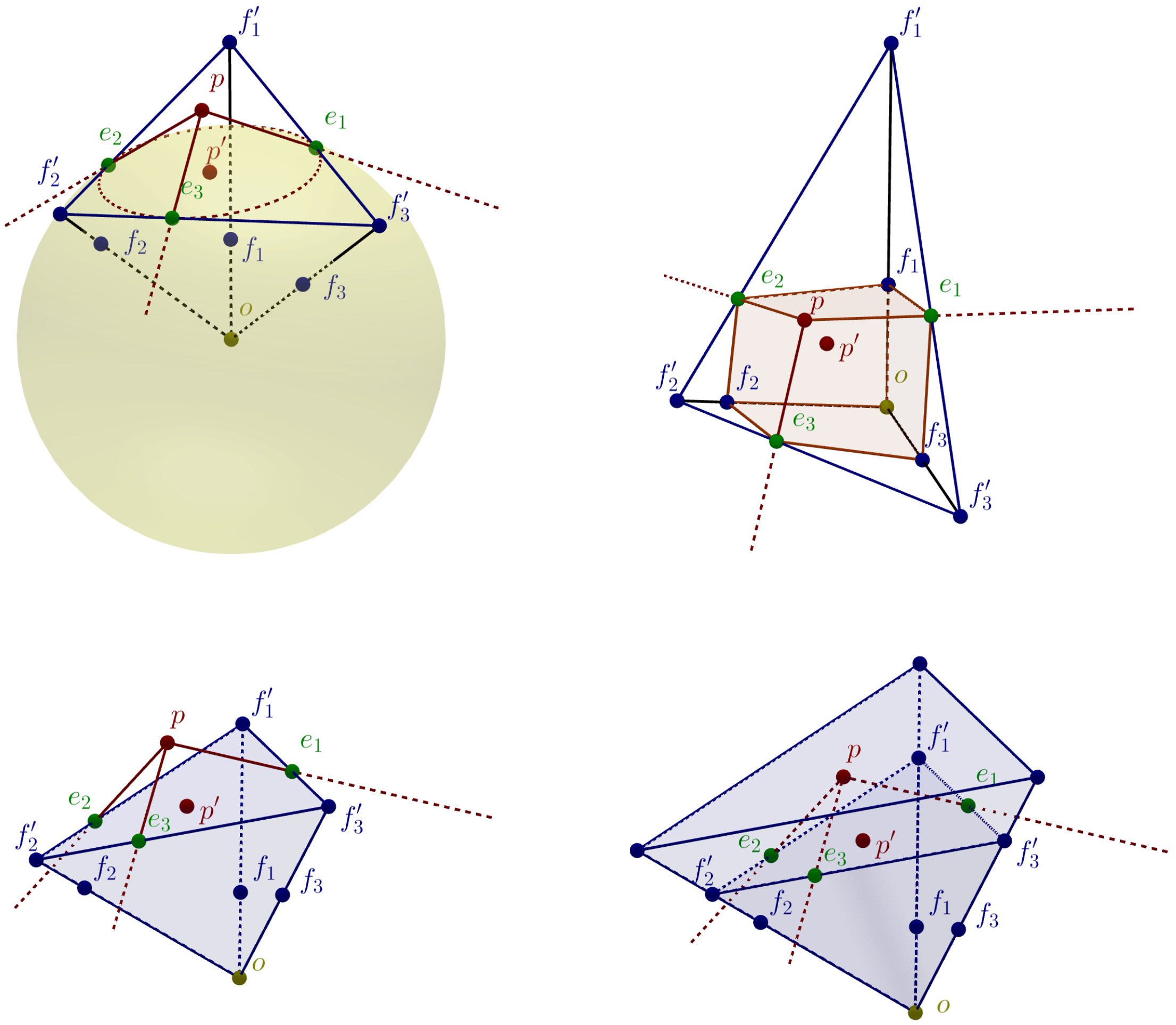

**Figure 6.17:** Illustration of the proof of Lemma 6.7. The top left shows, as an example, all the points described in the proof. The top right shows the hexahedron, and the bottom left shows the tetrahedron. The bottom right shows the tetrahedron with the shifted facet.

and then shift the plane spanned by $f_1'$, $f_2'$, and $f_3'$ so that it contains $p$. This new tetrahedron is convex and contains all eight vertices of the hexahedron and thus the entire hexahedron.

The intersection of the realization of $\mathbf{G}_K$, consisting of points $P$, with the constructed tetrahedron is the hexahedron composed of the eight mentioned points, because both the initial tetrahedron and the tetrahedron with the shifted facet contain only points of the realization of $\mathbf{G}_K$ that belong to the hexahedron. Since, by Lemma 3.3, the intersection of two convex sets is again convex, the hexahedron is convex. An illustration of the proof is given in Figure 6.17. □

With this construction, we have created a candidate for the eigenstructure of the subdivision matrices. In the next section, we describe the construction of a matrix that contains this eigenstructure.

### 6.2.2 Construction of a Colin-de-Verdière-like Matrix

In this section, we want to construct a Colin-de-Verdière-like matrix $\tilde{C}$ from the eigenstructure just created. To do so, we first need to clarify what Colin-de-Verdière-like means and how such a matrix should look.

We have already described in Chapter 4 what a Colin-de-Verdière-matrix is and how it can be generated from $d$-polytopes. However, the graph $\mathbf{G}$ is not the graph of a $d$-polytope, since for $t = 2$ it consists of several quadrilaterals and for $t = 3$ of several hexahedra.

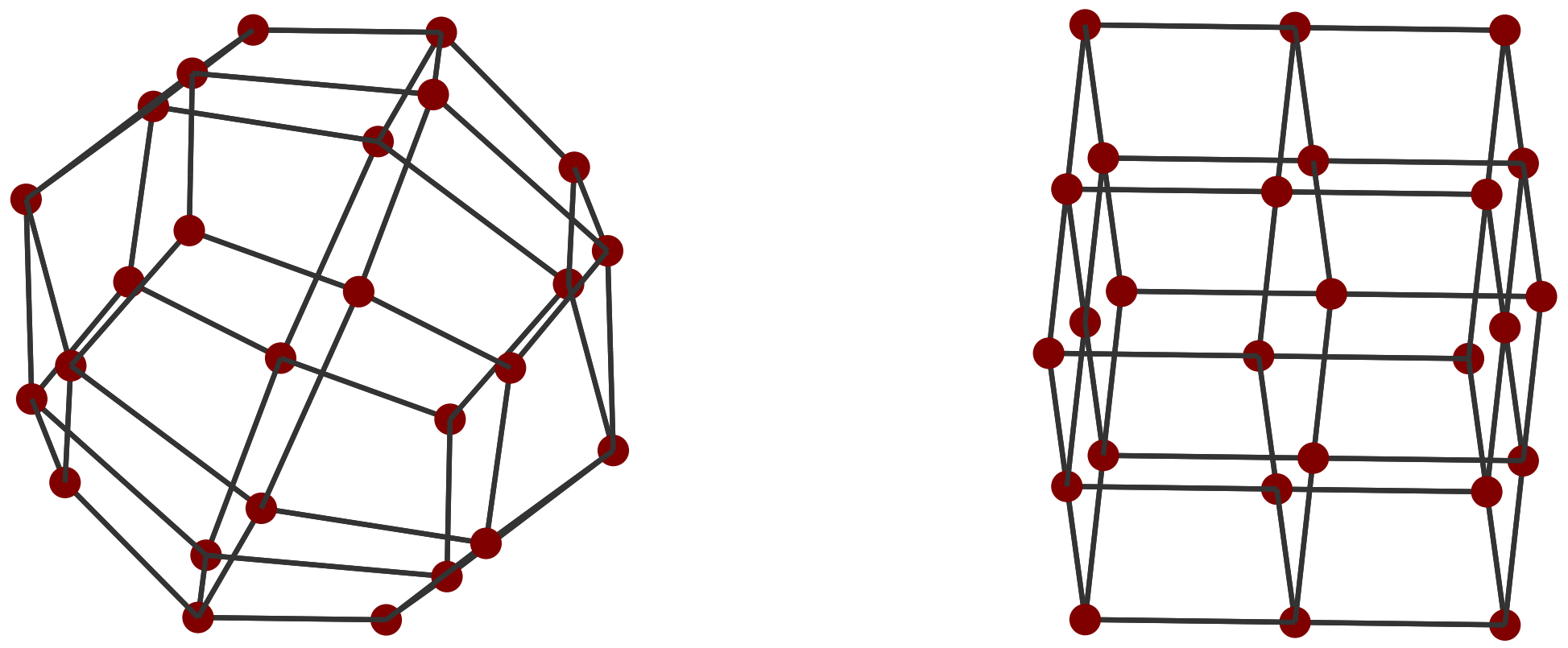

**Figure 6.18:** Convex polytope generated from the combinatorial arrangement of the union of the primal and dual polytope of the hexahedron (left), and eigenstructure of the regular case without the central point (right). Both realizations differ.

We will briefly discuss an alternative approach based on an obvious idea that, however, does not lead to a solution. If one removes the central point $f$ or $o$ from the graph **G** along with all edges incident to $f$ or $o$, one obtains a graph structure that for $t = 2$ is a cycle and for $t = 3$ is planar and 3-connected. For $t = 2$ this is immediately clear; for $t = 3$ it would require proof and discussion, but conducted tests support this statement.

For this graph, one could construct an eigenstructure by means of Chapter 3 and then generate a genuine Colin-de-Verdière-matrix using Chapter 4. However, two problems arise with this approach. The first problem is that it is unclear how to treat the central point. One might place it at the origin in the eigenstructure, but it is unclear how the matrix entries in the corresponding row and column of the Colin-de-Verdière-matrix should look.

This, however, is not a severe problem, as a solution could be found for it. Much more decisive is the second problem: the realization of the convex polytopes does not match those of the regular case. The differences for $t = 3$ are shown in Figure 6.18.

Disregarding that and taking the eigenstructure generated in the previous section, we could remove the central point from it and create a Colin-de-Verdière-matrix for this eigenstructure. However, the resulting input is not valid for constructing Colin-de-Verdière-matrices. The reason is that in the modified eigenstructure, multiple edges or facets lie in the same plane, as shown in the right illustration of Figure 6.18. Consequently, the facet points are not vertices of the polytope, and the facets lying in the same plane share identical dual points. Hence, the entire construction collapses. For $t = 3$, matrix entries corresponding to neighboring points on the off-diagonal are defined as

$$-\frac{|f_k - f_l|}{|p_i \times p_j|} = 0,$$

since in this case $f_k = f_l$. This contradicts Definition 4.1 of Colin-de-Verdière-matrices, and for this reason, this approach is not promising.

Therefore, we pursue a different strategy. We aim to construct a Colin-de-Verdière-like matrix $\tilde{C}$, which is modeled after the matrix $\tilde{C}$ in Equation (5.6) and should have the following properties:

**Definition 6.8.** *Let the control points $\tilde{P}$ and the graph $\boldsymbol{G} = (\boldsymbol{V}, \boldsymbol{E})$ be given as in Definition 6.2 for $t = 2$, and as in Definition 6.5 for $t = 3$. Let $n = |\boldsymbol{V}|$. A matrix $\tilde{C}$ is called a* Colin de Verdière-like matrix for the graph **G** *if the following four properties hold:*

*E1* $\tilde{C} \in \mathbb{R}^{n \times n}$, *where $n$ is the number of vertices of **G**. Each row and each column of $\tilde{C}$ therefore corresponds to one vertex of **G**.*

*E2* *Each row of $\tilde{C}$ sums to 1.*

*E3 The rows and columns of $\tilde{C}$ are arranged such that*

$$\tilde{C}\begin{bmatrix}P\\E\\f\end{bmatrix} = \frac{1}{2}\begin{bmatrix}P\\E\\f\end{bmatrix} \quad \textit{for } t = 2 \quad \textit{and} \quad \tilde{C}\begin{bmatrix}P\\E\\F\\o\end{bmatrix} = \frac{1}{2}\begin{bmatrix}P\\E\\F\\o\end{bmatrix} \quad \textit{for } t = 3$$

*holds. This defines both the order of the rows and columns of $\tilde{C}$, and it means that the columns of the eigenstructure from Section 6.2.1 are eigenvectors of $\tilde{C}$ with eigenvalue $\frac{1}{2}$.*

*E4 Let $A$ be the adjacency matrix of **G**. Then*

$$\tilde{C}_{(i,j)} = \begin{cases} > 0 & \textit{for } i \neq j \textit{ and } A_{(i,j)} = 1 \\ = 0 & \textit{for } i \neq j \textit{ and } A_{(i,j)} = 0 \end{cases}$$

*must hold. Except for the diagonal, which may be arbitrary, the matrix $\tilde{C}$ thus has the same non-zero structure as the adjacency matrix $A$.*

First, we need to explain some aspects. The matrix $\tilde{C}$ is not yet the subdivision matrix $S$, even though the above properties might suggest this. However, it is directly a row-stochastic matrix. Compared to Colin de Verdière matrices, it is not symmetric, just like the normalized matrix in equation (5.6) from Chapter 5 was no longer symmetric.

Another difference to the previous chapter is that the constructed eigenstructure spans the eigenspace corresponding to the eigenvalue $1/2$ rather than $0$. The reason is that we have to handle the spectrum shifting differently in the next section and therefore need a different spectrum for the matrix $\tilde{C}$.

Moreover, we cannot control the spectrum of the matrix $\tilde{C}$ here. Therefore, the properties C2 and C3 do not have to hold for the matrix $\tilde{C}$ in this chapter. Generally, many of the eigenvalue-related quality criteria described in Section 2.4 cannot be proven in Section 6.3, which is why the approach described here is experimental. However, with respect to the Colin de Verdière matrices, the (non-)zero pattern property C1 with inverted sign remains valid and is implemented in property E4.

The construction of the matrix $\tilde{C}$ is described in Algorithm 15. Although it may seem complex and not well motivated at first, we will need the properties of this construction in the following sections. We explain the individual steps below. Furthermore, all variables and matrices in this section refer to those appearing in Algorithm 15.

**General Structure and Initialization** We provide here a rough description of the algorithm's structure to give a first impression of its functioning. The individual steps will be detailed in the following paragraphs.

The algorithm is essentially divided into three steps: initialization, construction of the matrix $\overline{M}$, and construction of the matrix $\overline{B}$. In the initialization step, a distinction is first made between the cases $t = 2$ and $t = 3$. This distinction mainly concerns the points $\tilde{P}$ and is purely technical in nature, to avoid describing the algorithm twice. The points $\tilde{P}$ represent the realization of the graph **G** from the previous section and differ depending on the type. It is important to note that the vertices $P$ are explicitly used in some parts of the algorithm. The variable $P$ therefore always denotes the vertices and thus a part of $\tilde{P}$. Furthermore, we will consider the single elements of the structure **G**, namely the quadrilaterals or hexahedra. In the following, we will also call these *objects* to avoid constant double naming.

In the second part of the algorithm, the matrix $\overline{M}$ is constructed. For this, a Colin de Verdière matrix is generated for each object, i.e., for each quadrilateral when $t = 2$ and for each hexahedron when $t = 3$. These Colin de Verdière matrices are then combined into the matrix $\overline{M}$. This matrix has $ka > n$ rows. The quadrilaterals or hexahedra share points with other quadrilaterals or hexahedra. This duplication increases the number of rows in $\overline{M}$. Thus, when multiplying the matrix $\overline{M}$ by the control points $\tilde{P}$, some control points are mapped to multiple points. The vertices are mapped to one point only, the edge points to two points, the facet points to (number of vertices of the facet) points, and the volume point to (number of hexahedra) points.

This multiple mapping is then compensated in the last step by the matrix $\overline{B}$. The copies of the points are weighted such that together they map to half of the original point. We will explain this process and its justification at the appropriate place.

Since the constructions of the matrices $\overline{M}$ and $\overline{B}$ in Algorithm 15 are extensive, we describe them in the next two subsections. There, we will also prove several statements regarding the objects occurring in these subsections.

**Algorithm 15:** ConstructionOf$\tilde{C}$

**Data:** Graph **G** of an initial element, combinatorial graph $\mathbf{G}_K$ of the initial element, type $t$ of the initial element, and control points $\tilde{P} \in \mathbb{R}^{n \times t}$

**Result:** Matrix $\tilde{C} \in \mathbb{R}^{n \times n}$

1 **begin**
2   **if** $t == 2$ **then**
3     $a := 4$;
4     *Object* := quadrilateral;
5     $\tilde{P} = \left[P^T, E^T, f^T\right]^T$;
6   **else if** $t == 3$ **then**
7     $a := 8$;
8     *Object* := hexahedron;
9     $\tilde{P} = \left[P^T, E^T, F^T, o^T\right]^T$;
10   $k :=$ number of *Objects* in **G**;
11   $\overline{M} :=$ zero matrix of size $k \cdot a \times n$;
12   $\overline{I} :=$ zero matrix of size $k \cdot a \times 2$;
13   **for** $i = 1 : k$ **do**
14     $I :=$ construct $I$ as follows: determine the sorted row indices of the vertices of the $i$-th *Object* from $\tilde{P}$ with $I \in \{1, \dots, n\}^a$. Thus, $I_{(1)} = i$ is the index of the vertex, $I_{(2:t+1)}$ are the indices of the edge points, for $t = 3$: $I_{(5:7)}$ are the indices of the facet points, and $I_{(a)} = n$ is the index of the central facet or volume point;
15     $P' := \tilde{P}_{(I,:)} - \frac{1}{2}P_{(i,:)}$;
16     **if** $t == 2$ **then**
17       Scale $P'$ so that the circumcircle of the quadrilateral is the unit circle;
18     $M :=$ construct a Colin de Verdière matrix for the *Object* with vertices $P'$ according to Chapter 4;
19     $N := \mathtt{diag}\left(1./\left(\sum_j M_{(:,j)}\right)\right)$;
20     $M' := NM$;
21     $\overline{M}_{(a(i-1)+1:ai,I)} = M'$;
22     $\overline{I}_{(a(i-1)+1:ai,1)} = I$;
23     $\overline{I}_{(a(i-1)+1:ai,2)} = i$;
24   $\overline{B} :=$ zero matrix of size $n \times k \cdot a$;
25   **for** $j = 1 : n$ **do**
26     $K :=$ find indices of rows where $\overline{I}_{(:,1)} = j$;
27     $K' := \overline{I}_{(K,2)}$;
28     $\hat{P} :=$ translate the control points $P_{(K',:)}$ as described in the section and set $\hat{P}$ as these shifted and scaled points;
29     $B :=$ construct the Colin de Verdière matrix of the points $\hat{P}$ with the combinatorial structure of $\mathbf{G}_K$ according to Chapter 4. This can be a Colin de Verdière matrix of the following dimensions: dimension 0 for $\hat{P}$ representing a point in $\mathbf{G}_K$, dimension 1 representing a line in $\mathbf{G}_K$, dimension 2 representing a facet in $\mathbf{G}_K$, and for $t = 3$ dimension 3 representing the entire graph $\mathbf{G}_K$;
30     $B' := \sum_l B_{(l,:)}$;
31     $\overline{B}_{(j,K)} := \frac{1}{\sum B'} B'$;
32   $\tilde{C} := \overline{B} \cdot \overline{M}$;
33   **return** $\tilde{C}$;

### Construction of the Matrix $\overline{M}$

In this subsection, we describe the construction of the matrix $M$ from Algorithm 15, specifically lines 13–23 of Algorithm 15. All notation in this subsection refers to that of Algorithm 15.

We start with the index $i$ from line 13. This index runs through all vertices, i.e., all objects (quadrilaterals or hexahedra) of the graph **G**. For each $i \in \{1, \ldots, k\}$, the matrix $I$ is created in line 14. This matrix essentially contains the indices of the points of the quadrilateral or hexahedron. For a quadrilateral, these are four points consisting of one vertex, two edge points, and one facet point. For a hexahedron, there are eight points consisting of one vertex, three edge points, three facet points, and the volume point. These are denoted as

$$p, e_1, e_2 \text{ and } f \text{ for } t = 2 \quad \text{and} \quad p, e_1, e_2, e_3, f_1, f_2, f_3 \text{ and } o \text{ for } t = 3.$$

The control points of these objects are shifted in line 15 by the value $P_{(i,:)}/2 =: p/2$ (and for $t = 2$ additionally scaled in line 17). We denote these shifted (and possibly scaled) points by

$$P' := r\left(\tilde{P}_{(I,:)} - \frac{1}{2}P_{(i,:)}\right) \quad \text{with } r \in \mathbb{R}_{>0} \quad \text{for } t = 2, \quad \text{and} \quad P' := \tilde{P}_{(I,:)} - \frac{1}{2}P_{(i,:)} \quad \text{for } t = 3.$$

The above notation is to be understood row-wise. Each row of $\tilde{P}_{(I,:)}$ is added to the vector $-\frac{1}{2}P_{(i,:)}$.

With this, the origin lies inside the respective object, namely for $t = 2$ on the segment $\left(p - \frac{p}{2}, f - \frac{p}{2}\right)$, and for $t = 3$ on the segment $\left(p - \frac{p}{2}, o - \frac{p}{2}\right)$. We denote the shifted points as

$$p', e_1', e_2' \text{ and } f' \quad \text{for } t = 2, \quad \text{and} \quad p', e_1', e_2', e_3', f_1', f_2', f_3' \text{ and } o' \quad \text{for } t = 3.$$

An illustration for $t = 2$ can be found in Figure 6.19 and for $t = 3$ in Figure 6.20.

We then create in line 18 a Colin de Verdière matrix $M$ for this shifted object, that is, for the control points $P'$.

First, we consider the case $t = 2$. The shifted quadrilateral from line 15 is positioned so that the midpoint of the segment between $p'$ and $f'$ lies at the origin. In line 17, we scale the quadrilateral so that the distance from $f'$ and $p'$ to the origin is exactly 1. By Lemma 6.3, the original quadrilateral has a right angle at the two points $e_1$ and $e_2$. By Thales' theorem, the two shifted and scaled edge points also lie on the unit circle. Moreover, the shifted and scaled quadrilateral is convex, since by Lemma 6.4 the original quadrilateral is convex and translation and scaling preserve this property. An illustration can be found in Figure 6.22.

With this shifting, we can interpret the quadrilateral as the dual polytope from Theorem 4.10 and generate its Colin de Verdière matrix $M$. This is also a Colin de Verdière matrix for the unscaled but shifted quadrilateral consisting of the points $P'/r$ (with $r \in \mathbb{R}_{>0}$), since

$$M\frac{1}{r}P' = \frac{1}{r}MP' = 0 \cdot \frac{1}{r} = 0 \quad \text{for} \quad r \in \mathbb{R}_{>0}.$$

For $t = 3$, we can use the construction from Theorem 4.13. The shifted hexahedron is convex by Lemma 6.7 and contains the origin after shifting, so the conditions of Theorem 4.13 are satisfied. The corners of the dual polytope needed for the construction can be computed as follows.

Let the points $p'$, $e_1'$, $e_2'$, and $f'$ or the points $e'$, $f_1'$, $f_2'$, and $o'$ be the vertices of a facet of the hexahedron with coordinates taken from the rows of $P'$. Then the coordinates of the dual point of the facet are obtained from the linear system

$$\begin{bmatrix} p'^T \\ e_1'^T \\ e_2'^T \\ f'^T \end{bmatrix} \cdot \begin{bmatrix} x \\ y \\ z \end{bmatrix} = \begin{bmatrix} 1 \\ 1 \\ 1 \\ 1 \end{bmatrix} \quad \text{respectively} \quad \begin{bmatrix} e'^T \\ f_1'^T \\ f_2'^T \\ o'^T \end{bmatrix} \cdot \begin{bmatrix} x \\ y \\ z \end{bmatrix} = \begin{bmatrix} 1 \\ 1 \\ 1 \\ 1 \end{bmatrix}.$$

With this, we have all the points needed to generate the Colin de Verdière matrix $M$ for $t = 3$.

The Colin de Verdière matrix $M$ is then normalized in line 20 analogously to equation (5.6). The normalized matrix is denoted by $M'$. This normalization is well-defined because, according to Theorem 4.11 and Theorem 4.14, the row sums of $M$ are negative.

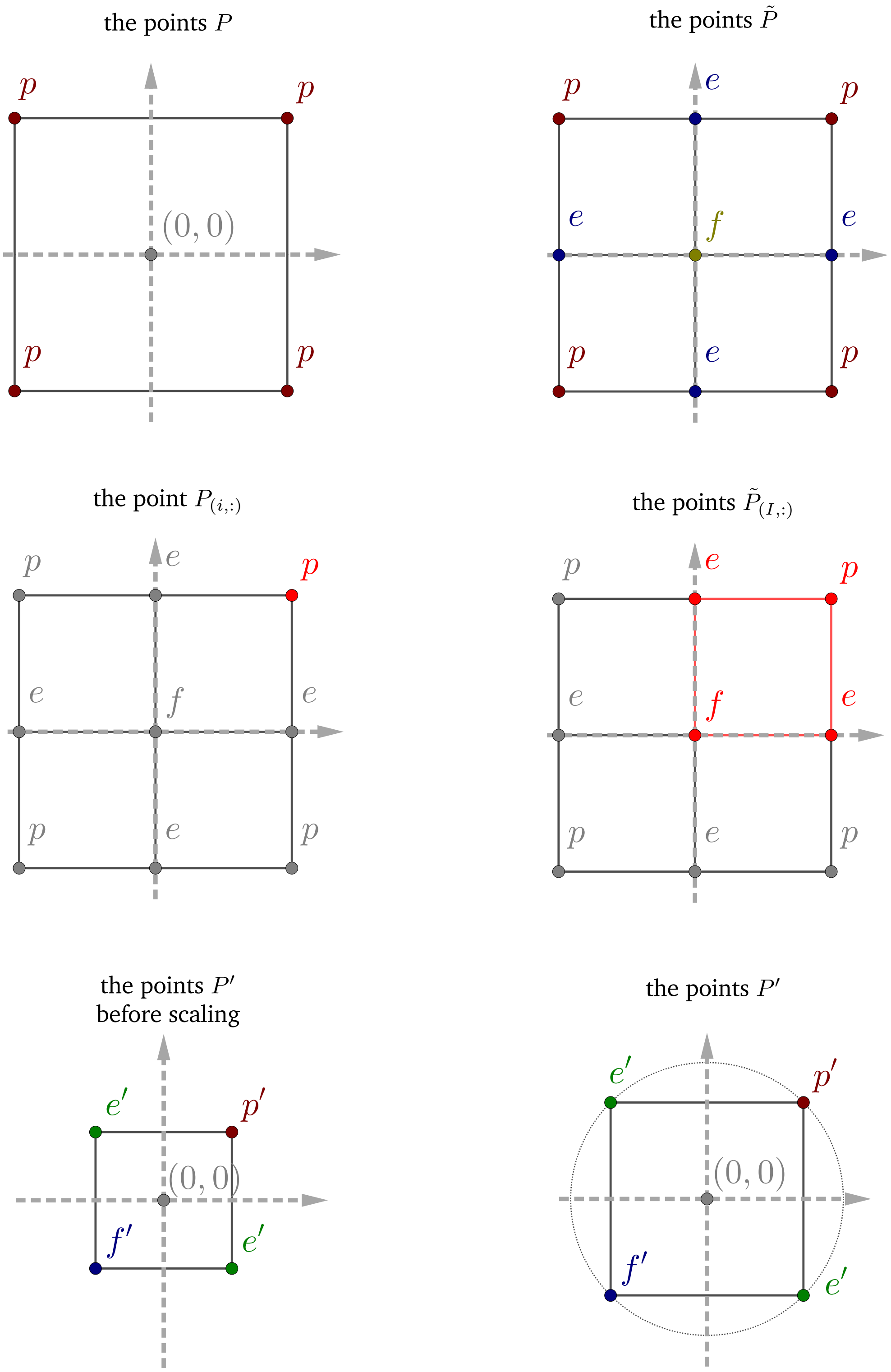

**Figure 6.19:** Example illustration of the different control points from Algorithm 15 for constructing the matrix $M$ for $t = 2$. The points $P$ are the vertices of the 2-polytope. The points $\tilde{P}$ are generated by Algorithm 13 and serve as the input for Algorithm 15. Using the for-loop from line 13, each point $P_{(I,:)}$ is accessed, and with the matrix $I$, the points $\tilde{P}_{(I,:)}$ of the quadrilateral corresponding to $P_{(I,:)}$ are determined. These points are then shifted and scaled, denoted as $P'$. Finally, the cdV-matrix $M$ is generated for these points $P'$.

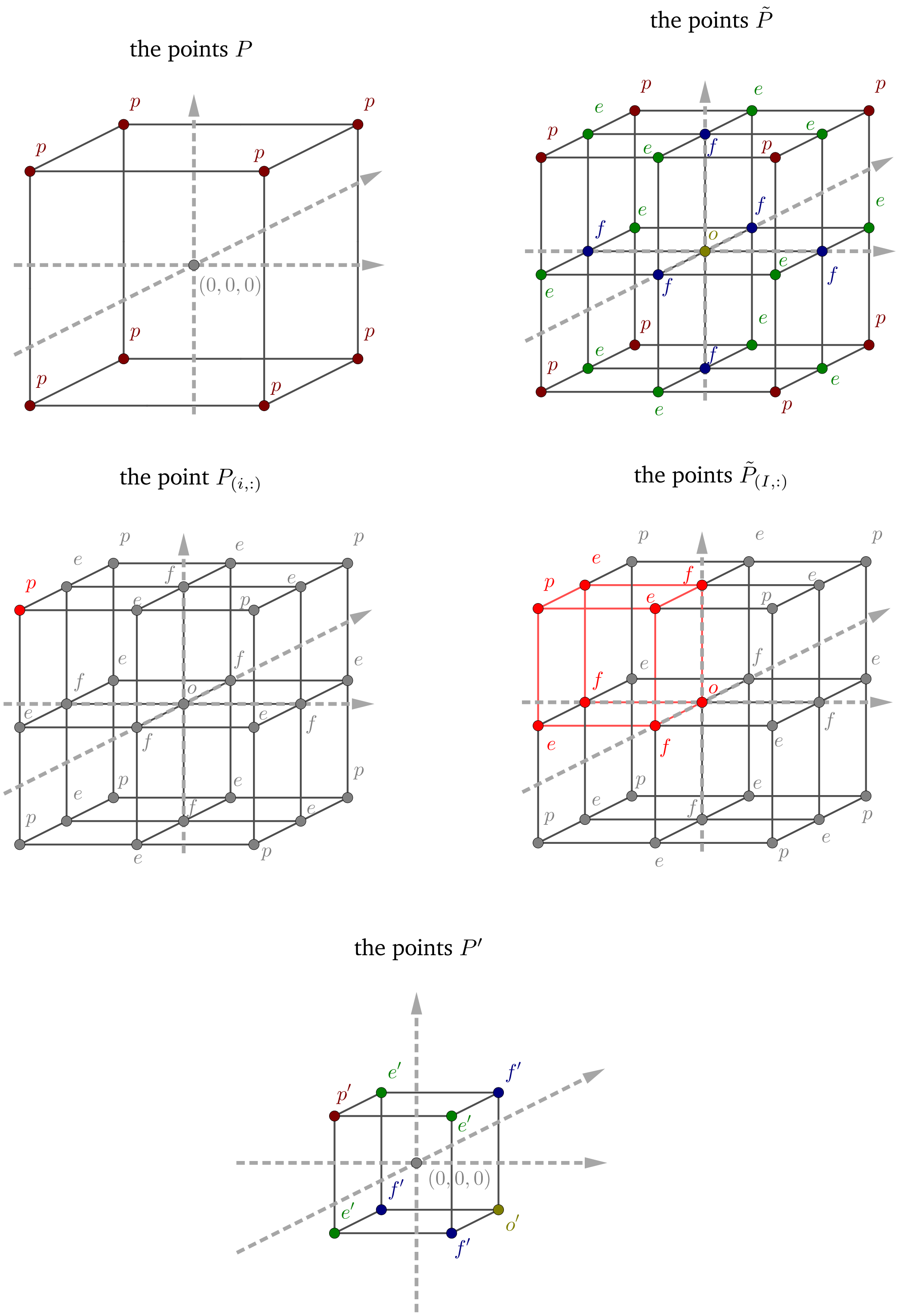

**Figure 6.20:** Analogous example illustration of the different control points for $t = 3$ corresponding to Figure 6.19. Here, however, the points $P'$ do not need to be scaled in the last step.

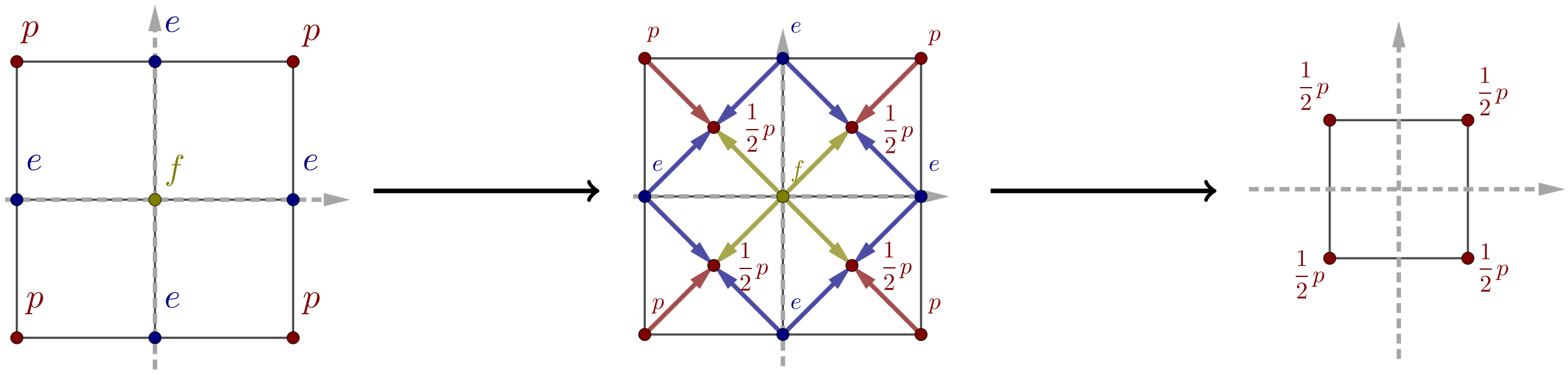

**Figure 6.21:** Exemplary illustration of the effect of the matrix $\overline{M}$. On the left, the points $\tilde{P}$ are shown as an example. These are mapped by $\overline{M}\tilde{P}$ onto the points in the right image. Here, multiple points lie on top of each other. The correspondence between the points $\tilde{P}$ and the points $\overline{M}\tilde{P}$, which is encoded in the index matrix $\overline{I}$, can be seen in the middle image.

This normalized matrix maps the unshifted object, i.e., the points $\tilde{P}_{(I,:)}$, to the respective point $P_{(i,:)}/2 = p^T/2$, since

$$M'\tilde{P}_{(I,:)} = M'\left(\tilde{P}_{(I,:)} - \frac{1}{2}\begin{bmatrix} p^T \\ \vdots \\ p^T \end{bmatrix} + \frac{1}{2}\begin{bmatrix} p^T \\ \vdots \\ p^T \end{bmatrix}\right) = M'\left(\tilde{P}_{(I,:)} - \frac{1}{2}\begin{bmatrix} p^T \\ \vdots \\ p^T \end{bmatrix}\right) + \frac{1}{2}M'\begin{bmatrix} p^T \\ \vdots \\ p^T \end{bmatrix} = M'P' + \frac{1}{2}M'\begin{bmatrix} p^T \\ \vdots \\ p^T \end{bmatrix}.$$

The first summand is $\vec{0}$ because the columns of $P'$ span the kernel of $M'$ (cf. Theorem 4.10 for $t = 2$ and Theorem 4.13 for $t = 3$), and the second summand equals $\frac{1}{2}[p \dots p]^T$, since the rows of $M'$ sum to 1. Hence, we obtain

$$M'\tilde{P}_{(I,:)} = \frac{1}{2}\begin{bmatrix} p^T \\ \vdots \\ p^T \end{bmatrix}.$$

From the normalized Colin de Verdière matrices $M'$ of all objects, we construct in line 21 the matrix $\overline{M}$. The rows of $\overline{M}$ correspond to the rows of the matrices $M'$, which are stacked vertically. For the columns, the matrices $M'$ are inserted into $\overline{M}$ so that for the product $\overline{M}\tilde{P}$, the corresponding entries of $\overline{M}$ are multiplied with the associated control points from $\tilde{P}$. Overall, we obtain

$$\overline{M}\tilde{P} = \frac{1}{2}\begin{bmatrix} P^T_{(1,:)} & \dots & P^T_{(1,:)} & \dots & P^T_{(k,:)} & \dots & P^T_{(k,:)} \end{bmatrix}^T.$$

Thus, the matrix $\overline{M}$ maps each object to its respective point $P_{(i,:)}$. Since the corner points and the edge points for $t = 2$ and the corner points, edge points, and volume point for $t = 3$ lie in multiple objects, these points are "duplicated" by the matrix $\overline{M}$. Hence, $\overline{M}$ has more rows than columns, and $\overline{M}\tilde{P}$ has more rows than $\tilde{P}$. Applying the matrix $\overline{M}$ to the control points $\tilde{P}$ thus produces a structure that corresponds to the combinatorial graph $\mathbf{G}_K$, i.e., the control points $P$ (up to scaling). In this image, several control points also lie on top of each other. An illustration for $t = 2$ can be found in Figure 6.21.

The matrix $\overline{I}$, whose values are set in lines 22 and 23, stores the information of the assignment. For each row in $\overline{I}$, the first entry notes to which control point from $\tilde{P}$ the corresponding row from $\overline{M}$ belongs, and the second entry notes from which object the row in $\overline{M}$ originated.

Thus, we have described the construction of the matrix $\overline{M}$. Before we consider the construction of the matrix $\overline{B}$ in the next subsection, we prove some statements about the normalized Colin de Verdière matrix $M'$, which we will need later.

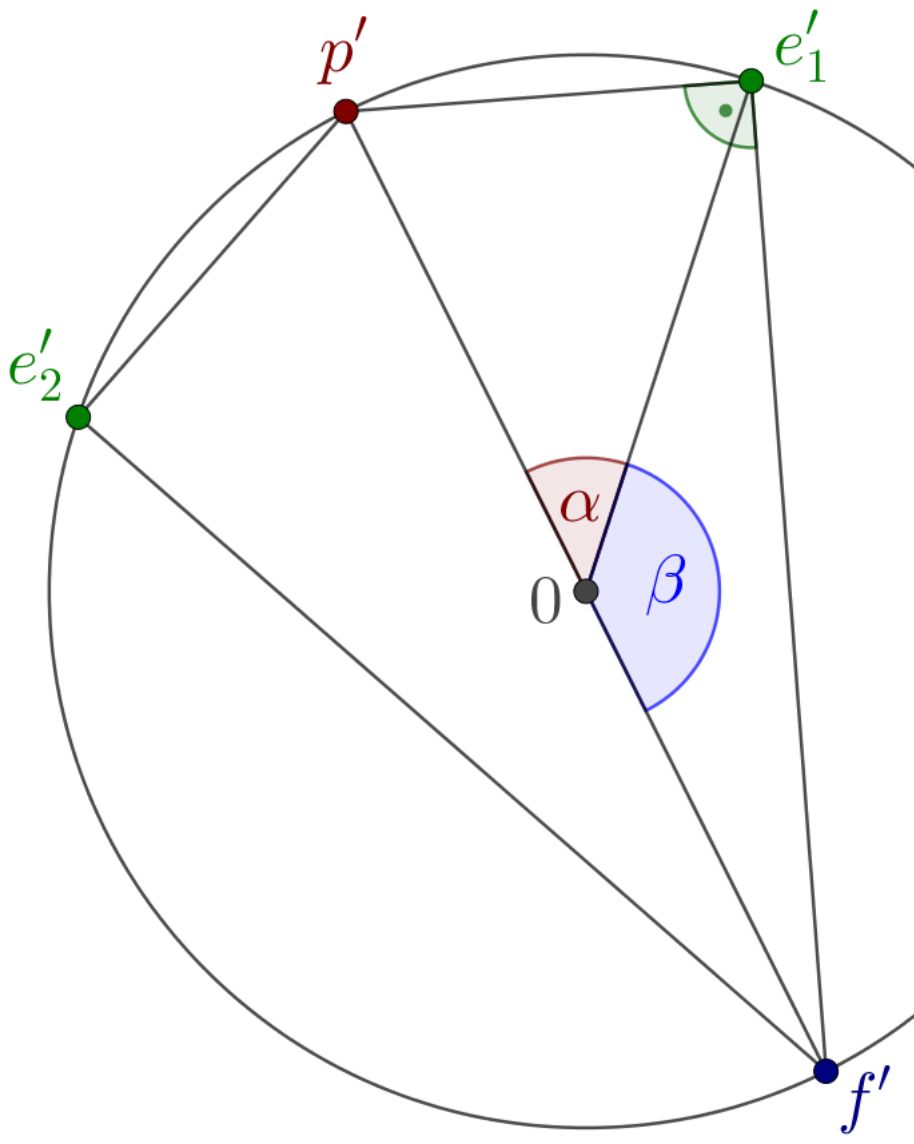

**Figure 6.22:** The elements used in the proof of Lemma 6.9, including the shifted quadrilateral.

Sorting the rows and columns of the matrix $M'$ for $t = 2$ in the order $p$, $e_1$, $e_2$, $f$, it has the form

$$M' = \left[\begin{array}{c|cc|c} ? & ? & ? & 0 \\ \hline a & ? & 0 & b \\ a & 0 & ? & b \\ \hline 0 & ? & ? & ? \end{array}\right],$$

for which the following statement holds:

**Lemma 6.9.** *Let $M'$ be a normalized Colin de Verdière matrix of the above form for $t = 2$, constructed according to Algorithm 15. Then it holds that $a = b = \frac{1}{2}$.*

*Proof.* A quadrilateral, for which a Colin de Verdière matrix is created, has the four vertices $p$, $e_1$, $e_2$, and $f$. The corners of the shifted and scaled quadrilateral, shifted by $-\frac{1}{2}p$ and scaled by a factor $r \in \mathbb{R}_{>0}$, are denoted by $p'$, $e'_1$, $e'_2$, and $f'$.

We first consider the value $a$ for the second row of the matrix $M'$. The value $a$ for the third row is computed analogously. First, we have

$$f' = \left(0 - \frac{1}{2}p\right) r = -p'.$$

Thus, the midpoint of the segment between $p'$ and $f'$ is the origin, and the position vectors of both points have the same length. The triangle $(p', e'_1, f')$ has a right angle at $e'_1$ by construction of the eigenstructure from the previous section. By Thales' theorem, it follows that the length of the position vector $e'_1$ satisfies

$$|e'_1| = |p'|.$$

Hence, the position vectors of the three points $p'$, $e'_1$, and $f'$ all have the same length.

Let $\alpha = \sphericalangle(p', e'_1)$ and $\beta = \sphericalangle(e'_1, f')$. Since the segment $(p', f')$ contains the origin, we have $\alpha + \beta = \pi$.

The row sum $c$ of the second row of $M$ is by construction in Theorem 4.10

$$c := -\frac{1}{|p' \times e'_1|} - \frac{1}{|e'_1 \times f'|} + \frac{\cos(\alpha)}{|e'_1|^2 \sin(\alpha)} + \frac{\cos(\beta)}{|e'_1|^2 \sin(\beta)}.$$

Using the identical lengths of the position vectors $p'$, $e_1'$, and $f'$, we obtain

$$c = \frac{\sin(\beta)(\cos(\alpha)-1)+\sin(\alpha)(\cos(\beta)-1)}{|e_1'|^2\sin(\alpha)\sin(\beta)}.$$

Hence, the entry $a$ in the second row of $M'$ is

$$\begin{aligned}
a &= \frac{-\frac{1}{|e_1'\times p'|}}{c}\\
&= -\frac{|e_1'|^2\sin(\alpha)\sin(\beta)}{|p'||e_1'|\sin(\alpha)\left(\sin(\beta)(\cos(\alpha)-1)+\sin(\alpha)(\cos(\beta)-1)\right)}\\
&= -\frac{\sin(\beta)}{\sin(\beta)(\cos(\alpha)-1)+\sin(\alpha)(\cos(\beta)-1)}\\
&= -\frac{\sin(\pi-\alpha)}{\sin(\pi-\alpha)(\cos(\alpha)-1)+\sin(\alpha)(\cos(\pi-\alpha)-1)}.
\end{aligned}$$

Using a computer algebra system, we find

$$a = \frac{1}{2}.$$

The value $b$ is obtained analogously by replacing $|e_1' \times p'|$ with $|e_1' \times f'|$. An illustration of the various points and angles can be found in Figure 6.22. □

Sorting the rows and columns of the matrix $M'$ for $t = 3$ in the order $p$, $e_1$, $e_2$, $e_3$, $f_1$, $f_2$, $f_3$, $o$, it has the form

$$M' = \left[\begin{array}{c|ccc|ccc|c}
? & ? & ? & ? & 0 & 0 & 0 & 0\\
\hline
a & ? & 0 & 0 & ? & ? & ? & 0\\
a & 0 & ? & 0 & ? & ? & ? & 0\\
a & 0 & 0 & ? & ? & ? & ? & 0\\
\hline
0 & ? & ? & ? & ? & 0 & 0 & b\\
0 & ? & ? & ? & 0 & ? & 0 & b\\
0 & ? & ? & ? & 0 & 0 & ? & b\\
\hline
0 & 0 & 0 & 0 & ? & ? & ? & ?
\end{array}\right], \tag{6.2}$$

for which the following statement holds:

**Lemma 6.10.** *Let $t = 3$ and $M'$ be a normalized Colin de Verdière matrix in the above form, constructed according to Algorithm 15. Then $a = b = \frac{1}{2}$.*

*Proof.* For the proof of this lemma, we need some notations which we explain below. The original hexahedron consists of eight points: one vertex point, three edge points, three facet points, and one volume point. We denote them by

$$p, e_1, e_2, e_3, f_1, f_2, f_3, \text{ and } o.$$

These points are shifted by $-p/2$, and for these shifted points the Colin de Verdière matrix $M$ is constructed. The shifted points are denoted by

$$p', e_1', e_2', e_3', f_1', f_2', f_3', \text{ and } o'.$$

The shifted hexahedron has six faces

$$(p', e_1', f_1', e_2'),\quad (p', e_2', f_2', e_3'),\quad (p', e_3', f_3', e_1'),\quad (o', f_1', e_2', f_2'),\quad (o', f_2', e_3', f_3'),\quad (o', f_3', e_1', f_1').$$

For the construction of the Colin de Verdière matrix, we need the dual polytope to the shifted hexahedron, which has six vertices corresponding to the six faces above. We denote these vertices by $c_1, \ldots, c_6$, in the given order. Furthermore, we consider the tetrahedron given by the convex hull

$$\mathbf{T} := \text{conv}(0, c_1, c_3, c_6).$$

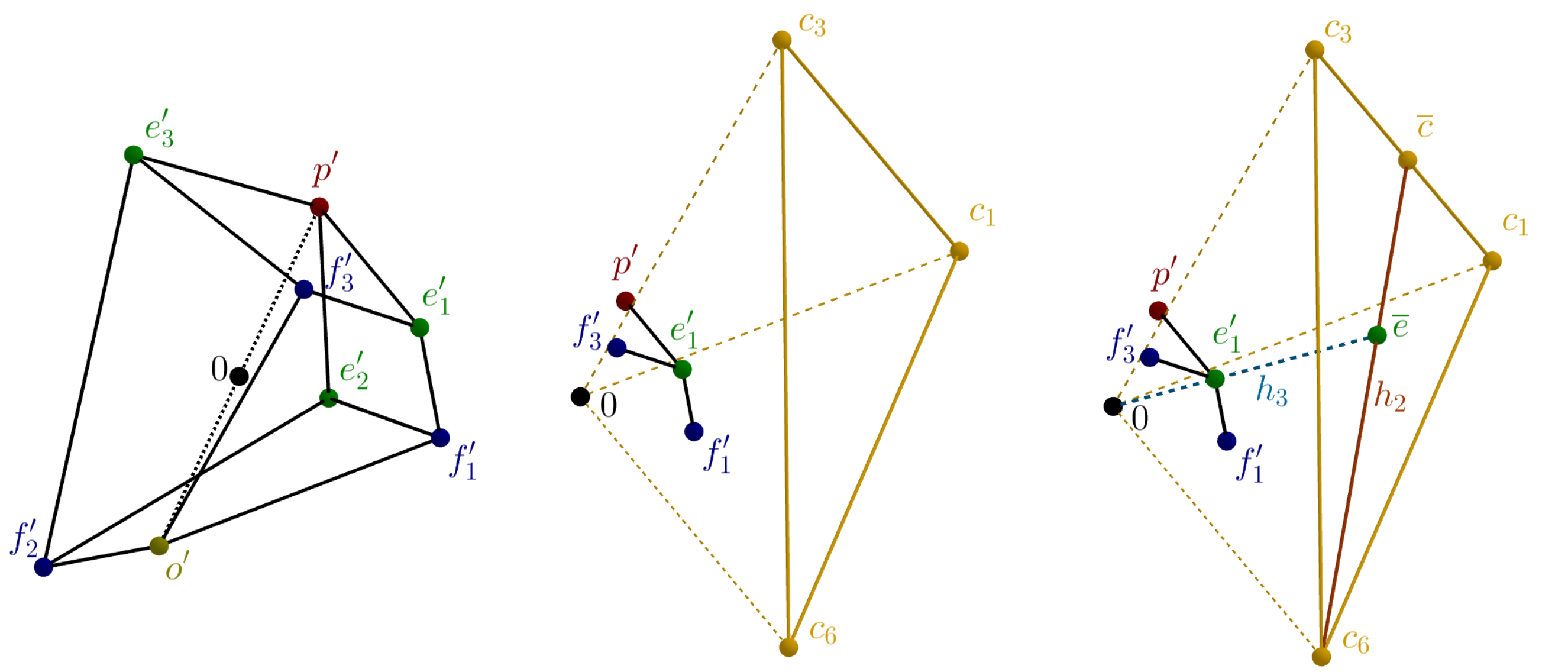

**Figure 6.23:** Shifted hexahedron with the eight shifted points (left), a part of the shifted hexahedron and the tetrahedron **T** at the edge point $e_1'$ (center), the heights $h_2$ and $h_3$ as well as the points $\bar{e}$ and $\bar{c}$ (right).

We start the calculation of $a$ by choosing the second row of $M'$ corresponding to the edge point $e_1$. The arguments for $e_2$ and $e_3$ are analogous. By Theorem 4.13, the value $a$ of the matrix $M$ is

$$-\frac{|c_1 - c_3|}{|p' \times e_1'|}.$$

This value is normalized by the matrix $N$. According to Theorem 4.14, the row sum of $M_{(2,:)}$ equals $-6\,\mathrm{vol}(\mathbf{T})$, since $c_1, c_3, c_6$ are dual points to the faces adjacent to $e_1'$. Hence,

$$a = \frac{|c_1 - c_3|}{6\,\mathrm{vol}(\mathbf{T})\,|p' \times e_1'|}. \tag{6.3}$$

An illustration of the points and tetrahedron **T** relevant to $a$ can be found in Figure 6.23. Next, define the triangle

$$\mathbf{M} := \mathrm{conv}(c_1, c_3, c_6),$$

and let $h_3$ be the height of the perpendicular from the origin to the triangle **M** inside the tetrahedron **T**. Applying the volume formula for a tetrahedron gives

$$\mathrm{vol}(\mathbf{T}) = \frac{1}{3} h_3\,\mathrm{vol}(\mathbf{M}).$$

Also define $h_2$ as the height of the perpendicular from $c_6$ to the segment $(c_1, c_3)$. Then the area of **M** is

$$\mathrm{vol}(\mathbf{M}) = \frac{1}{2}|c_1 - c_3| h_2.$$

Substituting into (6.3) yields

$$a = \frac{|c_1 - c_3|}{6\,\mathrm{vol}(\mathbf{T})\,|p' \times e_1'|} = \frac{|c_1 - c_3|}{6\frac{1}{3}h_3\,\mathrm{vol}(\mathbf{M})\,|p' \times e_1'|} = \frac{|c_1 - c_3|}{6\frac{1}{3}h_3\frac{1}{2}\,|c_1 - c_3|\,h_2\,|p' \times e_1'|} = \frac{1}{h_3 h_2\,|p' \times e_1'|}.$$

An illustration of the heights $h_2$ and $h_3$ is shown in Figure 6.23. The plane **M** containing $c_1, c_3, c_6$ is given by the dual polytope definition (Definition 3.6) as

$$\langle e_1', x \rangle = 1.$$

By the Hessian normal form, the distance from the origin to **M** (and thus $h_3$) is

$$h_3 = \frac{1}{|e_1'|}.$$

Define the angle $\alpha := \sphericalangle(p', e_1')$ between the vectors $p'$ and $e_1'$. Then

$$a = \frac{|c_1 - c_3|}{6\,\mathrm{vol}(\mathbf{T})\,|p' \times e_1'|} = \frac{1}{h_3 h_2\,|p' \times e_1'|} = \frac{1}{h_3 h_2 |p'||e_1'|\sin(\alpha)} = \frac{1}{\frac{1}{|e_1'|} h_2 |p'||e_1'|\sin(\alpha)} = \frac{1}{h_2 |p'|\sin(\alpha)}. \tag{6.4}$$

It remains to determine $h_2$. Consider the plane

$$\mathbf{M}' := \{x \in \mathbb{R}^3 : \langle p' \times e_1', x\rangle = 0\}.$$

All points relevant for $h_2$ lie in $\mathbf{M}'$, as shown next:

$$\langle p' \times e_1', p'\rangle = 0, \quad \langle p' \times e_1', e_1'\rangle = 0, \quad \langle p' \times e_1', o'\rangle = \langle p' \times e_1', -p'\rangle = 0, \quad \langle p' \times e_1', 0\rangle = 0.$$

Hence, the points $0, e_1', o', p'$ all lie in $\mathbf{M}'$. The vector $p' - e_1'$ is orthogonal to another plane spanned by $e_1', f_1', f_3'$, so the dual point $c_6$ of the facet $(o', f_3', e_1', f_1')$ can be written as

$$c_6 = \varphi(p' - e_1'), \quad \varphi \in \mathbb{R}.$$

Plugging into the plane equation,

$$\langle p' \times e_1', c_6\rangle = \varphi\langle p' \times e_1', p' - e_1'\rangle = \varphi(0 - 0) = 0,$$

so $c_6$ lies in $\mathbf{M}'$.

The vectors $c_1 - c_3$ and $p' \times e_1'$ point in the same direction (see proof of Theorem 4.13). Since $c_6 \in \mathbf{M}'$, the perpendicular from $c_6$ onto the segment $(c_1, c_3)$, which defines $h_2$, also lies in $\mathbf{M}'$. Denote the foot of this perpendicular by $\bar{c}$.

Additionally,

$$\bar{e} := \frac{1}{|e_1'|^2} e_1'$$

lies in $\mathbf{M}'$ and in the plane spanned by $c_1, c_3, c_6$. Thus, $\bar{e}$ lies on both segments $(0, \bar{e})$ and $(c_6, \bar{c})$. Since the vector $e_1'$ points in the direction of the normal to the plane **M**, these segments are orthogonal. With this, we have all the information needed to compute $h_2$. An illustration of the points and all terms used in the following can be found in Figure 6.24.

We now want to determine the length of $h_2$ using the points just described in the plane $\mathbf{M}'$. To do this, we first discuss the orthogonality of various line segments. First, by the construction of the dual polytope, the segment $(c_1, c_3)$ (which does not lie in the plane $\mathbf{M}'$) is orthogonal to the segment $(p', e_1')$. This is clearly shown in Figure 4.5, although with different labels.

In addition, the segment $(p', e_1')$ is orthogonal to the segment $(0, c_1)$ and to the segment $(0, c_3)$, because $(p', e_1')$ is part of the facets corresponding to $c_1$ and $c_3$. This means that the segment $(p', e_1')$ is orthogonal to the plane spanned by the three points $0$, $c_1$, and $c_3$. As a result, the segment $(0, \bar{c})$, which lies in this plane, is also orthogonal to the segment $(p', e_1')$.

The segment $(p', e_1')$ is also orthogonal to the segment $(o', e_1')$. This follows from the construction of the eigenspace in the previous section, where the vector $e_1 - 0 = e_1$ is orthogonal to the segment $(p', e_1')$. The reason is that $(p', e_1')$ lies on the tangent plane from $e_1$ to the unit circle in the construction of the primal and dual polytopes. If we shift all points by $-p/2$, this orthogonality does not change.

Since the two points $e_1'$ and $o'$ lie in the facet $(o', f_3', e_1', f_1')$, the segment $(e_1', o')$ is also orthogonal to the segment $(0, c_6)$. Altogether, we have

$$(0, \bar{c}) \perp (p', e_1') \perp (e_1', o') \perp (0, c_6)$$

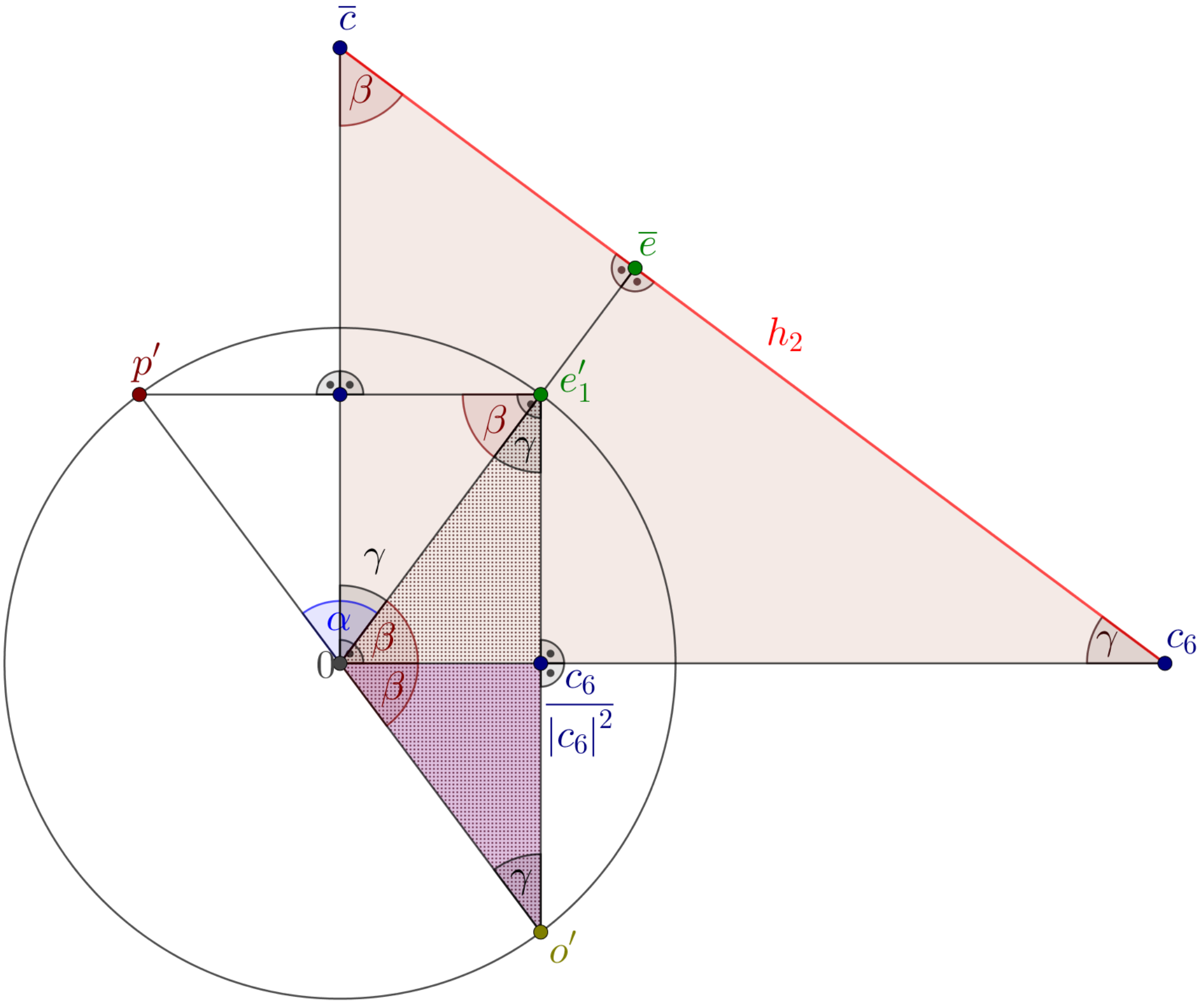

**Figure 6.24:** The plane $\mathbf{M}'$ including all points, segments, and angles that are needed to compute $h_2$.

and since all points lie in one plane, it follows directly that

$$(0, \bar{c}) \perp (0, c_6).$$

Therefore, the triangle $(0, \bar{c}, c_6)$ has a right angle at the origin. Its area can be expressed in two ways:

$$\frac{1}{2}|\bar{c}||c_6| = \text{vol}(0, \bar{c}, c_6) = \frac{1}{2}|\bar{e}|h_2,$$

which implies

$$h_2 = \frac{|\bar{c}||c_6|}{|\bar{e}|}.$$

Now consider these three lengths: 0 is the midpoint of $(p', o')$. Since the triangle $(p', e'_1, o')$ has a right angle at $e'_1$, by Thales' theorem,

$$|e'_1| = |p'| = |o'|.$$

For $|c_6|$, consider the projection of $c_6$ onto the facet $(o', f'_3, e'_1, f'_1)$, which is $\frac{c_6}{|c_6|^2}$ with length $\frac{1}{|c_6|}$. This lies in the triangle $(0, e'_1, o')$. By the law of sines,

$$\frac{|(e'_1, o')|}{\sin(\pi - \alpha)} = \frac{|(e'_1, o')||e'_1||o'|}{2\,\text{vol}(0, e'_1, o')} = \frac{|(e'_1, o')||e'_1||o'|}{2 \cdot \frac{1}{2}|(e'_1, o')|\frac{1}{|c_6|}}.$$

Rearranging for $|c_6|$ gives

$$|c_6| = \frac{|(e'_1, o')|}{\sin(\alpha)|e'_1||o'|}.$$

For $|\overline{c}|$, apply the intercept theorem (cf. Euclid [Euc86, Book 6, Prop. 2, p. 111] and [AF15, Prop. 1, p. 1]). The triangles $(0, \overline{c}, c_6)$ and $\left(\frac{c_6}{|c_6|^2}, 0, o'\right)$ have the same angles, so

$$\frac{|\overline{c}|}{\frac{|c_6|}{|c_6|^2}} = \frac{|c_6|}{|(o', \frac{c_6}{|c_6|^2})|} \quad \Leftrightarrow \quad |\overline{c}| = \frac{1}{|(o', \frac{c_6}{|c_6|^2})|} = \frac{2}{|(o', e_1')|}.$$

Altogether, for $h_2$ we have

$$h_2 = \frac{\frac{2}{|(o', e_1')|} \cdot \frac{|(e_1', o')|}{\sin(\alpha)|e_1'||o'|}}{\frac{1}{|e_1'|}} = \frac{2}{\sin(\alpha)|o'|}.$$

Substituting into (6.4), we obtain

$$a = \frac{1}{\frac{2}{\sin(\alpha)|o'|}|p'|\sin(\alpha)} = \frac{1}{2}.$$

The analogous case for $b$ is obtained by swapping

$$p' \leftrightarrow o', \quad f' \leftrightarrow e',$$

and the proof proceeds in exactly the same manner. □

### Construction of the matrix $\overline{B}$

In this subsection, we consider the construction of the matrix $\overline{B}$ from Algorithm 15, specifically lines 25–31. The notation in this subsection explicitly refers to that of Algorithm 15. We begin again with the index $j$, which runs in the for-loop of line 25 over all points of $\tilde{P}$. For each control point, we want to construct one row of $\overline{B}$.

The matrix $\overline{M}$ initially maps the control points $\tilde{P}$ onto the respective points $\frac{1}{2}p$ that lie in the corresponding objects. From the structure of $\overline{M}$, it can be seen that the product $\overline{M}\tilde{P}$ has more rows, and thus more control points, than $\tilde{P}$. As already mentioned, this is because the matrix $\overline{M}$ partially creates multiple copies of a control point from $\tilde{P}$. The row indices of the copies of the point with index $j$ are encoded in the matrix $K$ of line 26 for this part of the algorithm. This is done by filtering which rows in $\overline{I}$ correspond to the point with index $j$. Subsequently, in the matrix $K'$ from line 27, the objects that generated these copies are encoded.

The points $\hat{P}$ are more complicated to describe. Essentially, these are initially the points onto which the point with index $j$ is mapped by the matrix $\overline{M}$. Since $\overline{M}$ maps the points $\tilde{P}$ (including duplicates) onto the points $P/2$, i.e., scaled versions of the vertex points, we choose for $\hat{P}$ the unscaled points. These can be identified via the indices in $K'$, which correspond to the indices of the objects containing the respective vertex point. An exemplary illustration for the case $t = 2$ can be found in Figure 6.25, and for the case $t = 3$ in Figure 6.26.

The points $\hat{P}$ are then shifted, scaled, and rotated so that a Colin-de-Verdière-matrix $B$ of appropriate dimension can be constructed from them in line 29. Subsequently, in line 30 of Algorithm 15, the rows of $B$ are summed to yield the row vector $B'$. This vector is normalized in line 31 of Algorithm 15, which results in the corresponding row of the matrix $\overline{B}$.

Thus, the matrix $\overline{B}$ averages the copies created by the matrix $\overline{M}$. The matrix

$$\tilde{C} = \overline{B} \cdot \overline{M},$$

described in line 32, arises from combining the matrix $\overline{M}$, which creates the copies, and the matrix $\overline{B}$, which averages them. This already establishes the first property E1 of $\tilde{C}$. The matrix $\tilde{C}$ is a $n \times n$ matrix, and each row corresponds to a control point from $\tilde{P}$.

After this overview, we now consider the construction of the rows of $\overline{B}$ and $\tilde{C}$ for the different types of control points $p$, $e$, $f$, and $o$ separately and in detail, directly referring to the properties E2–E4 of the corresponding rows of $\tilde{C}$.

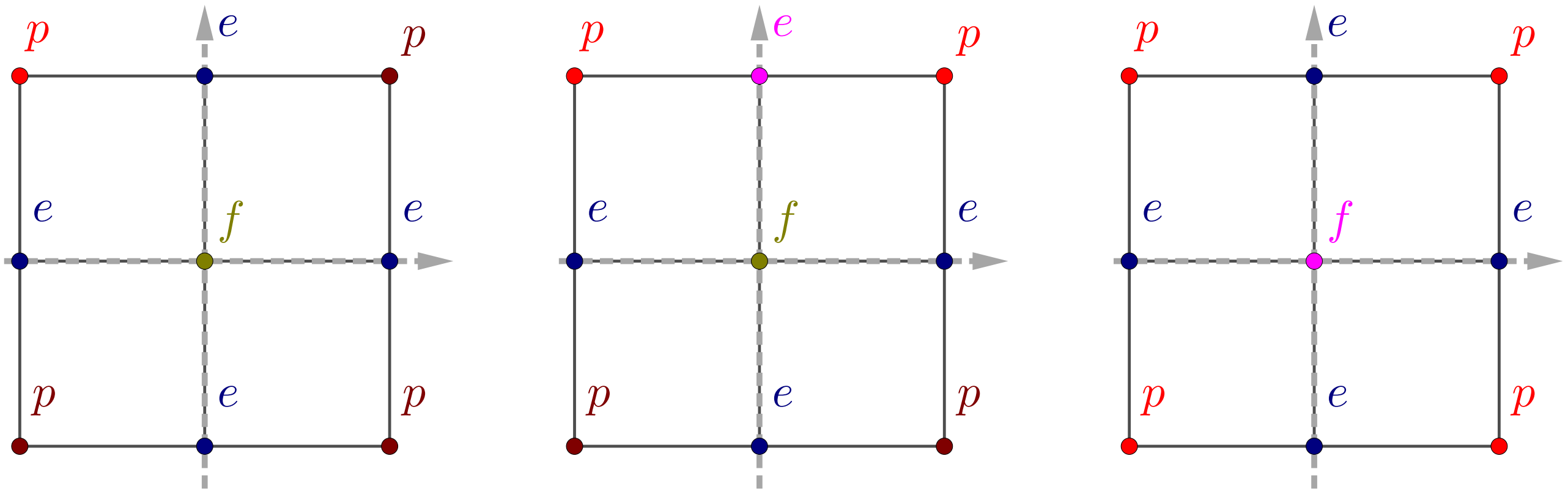

**Figure 6.25:** Control point (pink) and associated control points in $\hat{P}$ (red) for a vertex point (left), an edge point (middle), and the central point (right) for $t = 2$. In the left image, the pink and red points coincide, which is why it is shown only in red.

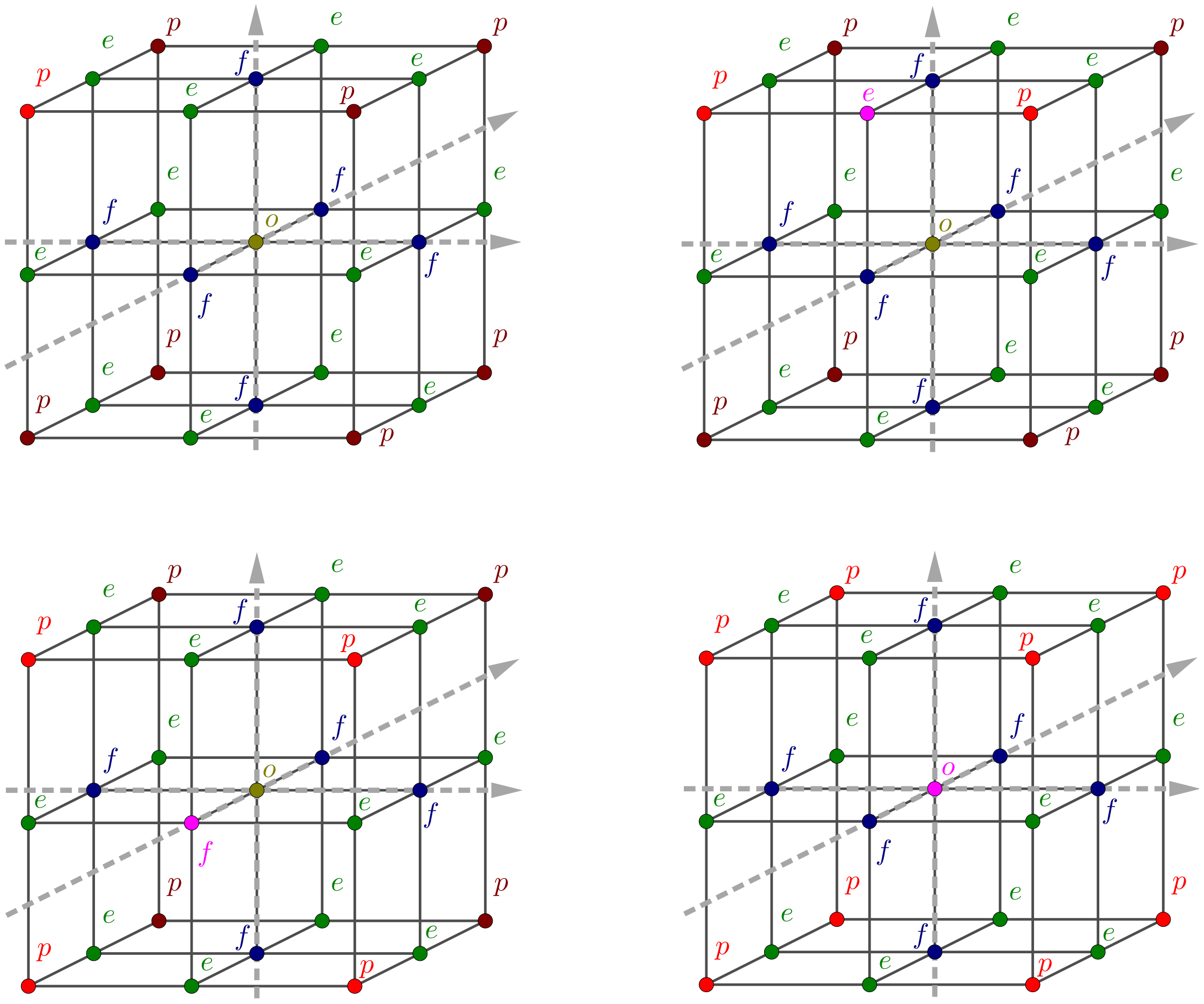

**Figure 6.26:** Control point (pink) and associated control points in $\hat{P}$ (red) for a vertex point (top left), an edge point (top right), a facet point (bottom left), and the central point (bottom right) for $t = 3$. In the top left image, the pink and red points coincide, which is why it is shown only in red.

**Vertex Points $p$** The simplest case is given by the control points from $P$. Each of these control points appears in only one object. Therefore, for the construction of the Colin-de-Verdière-matrix $B$ for this structure, which we interpret as a graph with a single node and no edges, we use Lemma 4.5 and obtain

$$B = \begin{bmatrix}-1\end{bmatrix} = B'.$$

Thus, the "matrix" in this case consists of only one entry. By normalization, we get

$$\frac{1}{\sum B'} B' = \frac{1}{-1} B' = \begin{bmatrix}1\end{bmatrix}.$$

Hence, the corresponding row in $\overline{B}$ contains only a single entry 1, with all other entries zero. With this description and the product of $\overline{B}$ and $\overline{M}$, we obtain the corresponding row $\tilde{C}_{(j,:)}$.

Let $j$ be the row number of the row constructed in $\overline{B}$. We now verify the properties of the corresponding row $\tilde{C}_{(j,:)}$. First, we consider which entries of $\tilde{C}_{(j,:)}$ are nonzero. Since

$$\tilde{C}_{(j,:)} = \overline{B}_{(j,:)}\overline{M} = 1 \cdot \overline{M}_{(K,:)},$$

the $j$-th row of $\tilde{C}$ is identical to the $K$-th row of $\overline{M}$. Because $\overline{M}_{(K,:)}$ is the normalized row of a Colin-de-Verdière-matrix, the nonzero off-diagonal entries of $\tilde{C}_{(j,:)}$ correspond to the neighbors of the control point $\tilde{P}_{(j,:)}$.

The off-diagonal entries of Colin-de-Verdière-matrices $M$ are negative by Definition 4.1. Since the row sums of $M$ are negative by Satz 4.9 and Theorem 4.14, these off-diagonal entries become positive after normalization. Because these entries are multiplied by 1 in $\tilde{C}$, this property transfers to $\tilde{C}_{(j,:)}$. Thus, the row $\tilde{C}_{(j,:)}$ satisfies property E4.

Multiplying the row $\tilde{C}_{(j,:)}$ by the vector of all ones $\vec{1}$ yields

$$\tilde{C}_{(j,:)}\vec{1} = \overline{B}_{(j,:)}\overline{M}\vec{1} = 1 \cdot \overline{M}_{(K,:)}\vec{1}.$$

Since the rows of $\overline{M}$ are normalized, this implies

$$\tilde{C}_{(j,:)}\vec{1} = 1,$$

so the row sums to 1, fulfilling property E2.

For the product of $\tilde{C}$ and $\tilde{P}$, we have

$$\tilde{C}_{(j,:)}\tilde{P} = \overline{B}_{(j,:)}\overline{M}\tilde{P} = 1 \cdot \overline{M}_{(K,:)}\tilde{P} = \frac{1}{2}\tilde{P}_{(j,:)},$$

thus the row $\tilde{C}_{(j,:)}$ also satisfies property E3. The corresponding vertex point $p$ is mapped to $p/2$. A schematic illustration of the mapping of $p$ to $p/2$ by the matrices $\overline{M}$ and $\overline{B}$ can be found in Figure 6.27.

**Edge points $e$** The construction of the matrix entries for the edge points from $E$ is somewhat more complex. We denote by $e$ the corresponding edge point for the row $\tilde{C}_{(j,:)}$. This edge point lies on exactly two objects. We label the vertex points of these two objects as $p_1$ and $p_2$. Accordingly, there are two entries in $\overline{I}_{(:,1)}$ equal to $j$, and the vector $K$ consists of two entries, which we denote as $K_1$ and $K_2$.

Thus, the matrix $\overline{M}$ creates two copies of the point $e$. One is mapped by $\overline{M}$ to $p_1/2$, and the other to $p_2/2$. The segment between $p_1/2$ and $p_2/2$ contains by construction the point $\frac{1}{2}e$. We therefore define the points

$$p_1' = \frac{1}{2}p_1 - \frac{1}{2}e \quad \text{and} \quad p_2' = \frac{1}{2}p_2 - \frac{1}{2}e.$$

The segment between these two points contains the origin. In the structural graph $\mathbf{G}_K$, the points $p_1$ and $p_2$ share an edge. Therefore, by Lemma 4.6 and using the points $p_1'$ and $p_2'$, we construct a Colin-de-Verdière-matrix $B \in \mathbb{R}^{2\times 2}$

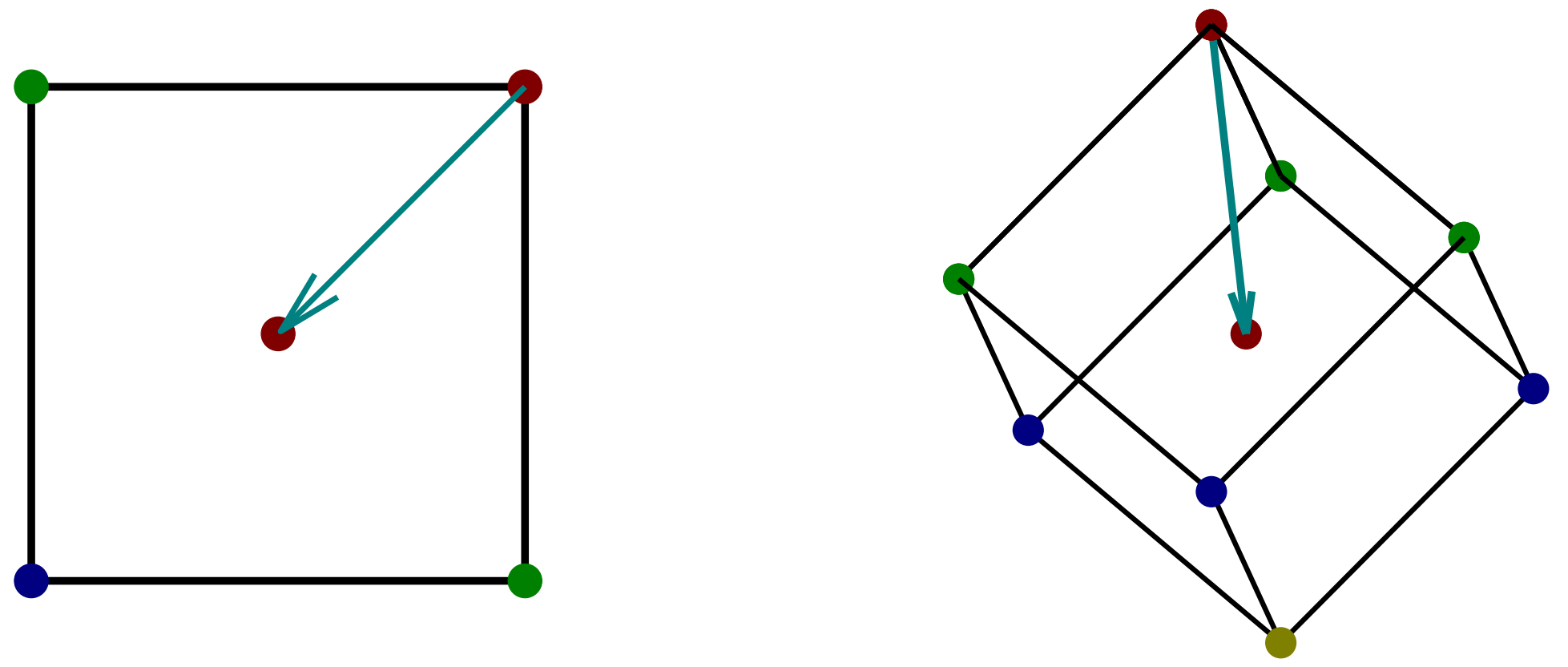

**Figure 6.27:** Schematic illustration of the effect of the matrices $\overline{M}$ (turquoise) and $\overline{B}$ (magenta) on a vertex point $p$ for $t = 2$ (left) and $t = 3$ (right). The matrix $\overline{M}$ creates a copy of $p$ and maps it to $p/2$. The matrix $\overline{B}$ maps this point onto itself, so overall $p$ is mapped to $p/2$. Therefore, the effect of $\overline{B}$ is not visible in this illustration, and accordingly the magenta depiction is omitted.

for a graph consisting of two nodes and one edge, for which initially

$$B \begin{bmatrix} p_1'^T \\ p_2'^T \end{bmatrix} = \begin{bmatrix} -\frac{|p_2'|}{|p_1'|} p_1'^T - p_2'^T \\ -p_1'^T - \frac{|p_1'|}{|p_2'|} p_2'^T \end{bmatrix}$$

holds. Since both points $p_1'$ and $p_2'$ lie on a line and the segment between them contains the origin, it follows that

$$p_1' = -\frac{|p_1'|}{|p_2'|} p_2' \quad \Leftrightarrow \quad p_2' = -\frac{|p_2'|}{|p_1'|} p_1'.$$

Hence,

$$B \begin{bmatrix} p_1'^T \\ p_2'^T \end{bmatrix} = \begin{bmatrix} p_2'^T - p_2'^T \\ -p_1'^T + p_1'^T \end{bmatrix} = \begin{bmatrix} \vec{0}^T \\ \vec{0}^T \end{bmatrix}.$$

Moreover, by construction and as seen in Lemma 4.6, the row and column sums of $B$ are negative. Thus, the matrix $B'$ consists of two negative entries, and the matrix $\frac{1}{\sum B'} B'$ consists of two positive entries summing to 1. With these descriptions, for the row $\tilde{C}_{(j,:)}$ it holds that

$$\tilde{C}_{(j,:)} = \overline{B}_{(j,:)} \overline{M} = \overline{B}_{(j,K_1)} \overline{M}_{(K_1,:)} + \overline{B}_{(j,K_2)} \overline{M}_{(K_2,:)}.$$

Considering the entries of $\tilde{C}_{(j,:)}$, these consist of a linear combination of the rows $\overline{M}_{(K_1,:)}$ and $\overline{M}_{(K_2,:)}$. These two rows contain only entries that can be nonzero exclusively for $e$ and the neighbors of $e$. Each row covers only part of the neighbors with nonzero entries, but together they cover all neighbors of $e$. Moreover, the entries corresponding to the neighbors of $e$ are non-negative.

Since the entries of $\frac{1}{\sum B'} B'$ are positive, the two entries $\overline{B}_{(j,K_1)}$ and $\overline{B}_{(j,K_2)}$ are positive. Together with the structure of $\overline{M}_{(K_1,:)}$ and $\overline{M}_{(K_2,:)}$, it follows that the entries of the row $\tilde{C}_{(j,:)}$ corresponding to the neighbors of $e$ are positive, and except for the diagonal entry, all other entries are zero. Hence, property E4 holds.

Multiplying the row $\tilde{C}_{(j,:)}$ by the all-ones vector $\vec{1}$, using the normalization of rows in $\overline{M}$ and $\overline{B}$, we obtain

$$\tilde{C}_{(j,:)} \vec{1} = \overline{B}_{(j,:)} \overline{M} \vec{1} = \overline{B}_{(j,K_1)} \overline{M}_{(K_1,:)} \vec{1} + \overline{B}_{(j,K_2)} \overline{M}_{(K_2,:)} \vec{1} = \overline{B}_{(j,K_1)} + \overline{B}_{(j,K_2)} = 1,$$

so the row $\tilde{C}_{(j,:)}$ sums to 1 and thus property E2 holds.

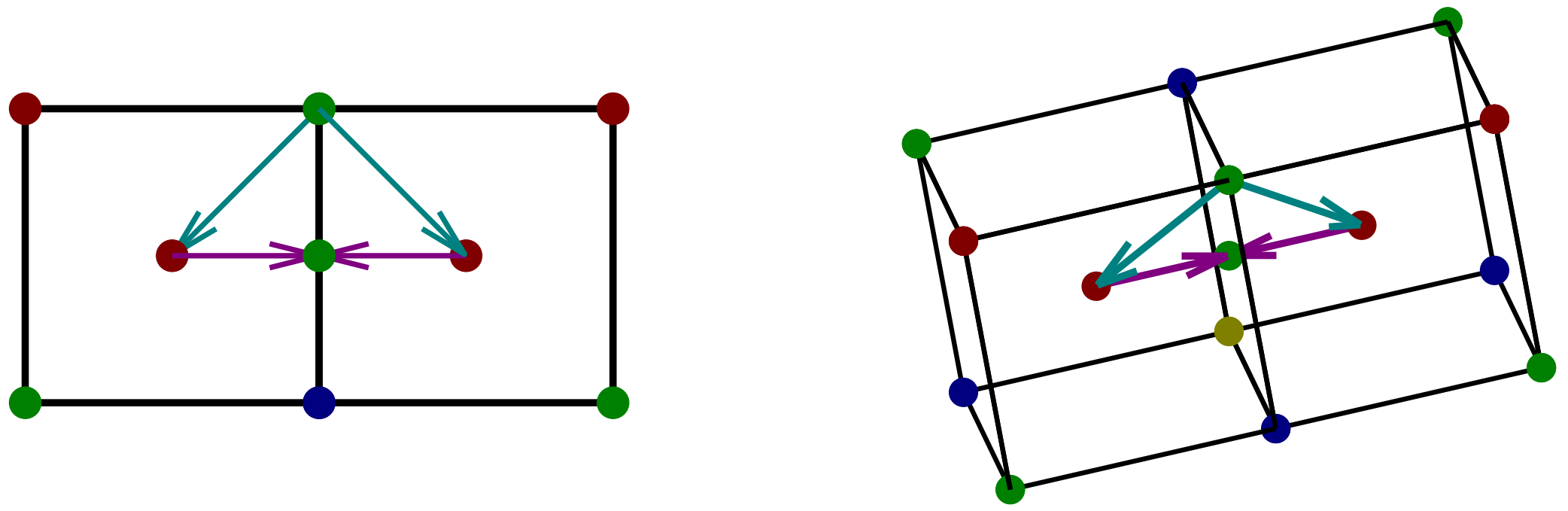

**Figure 6.28:** Schematic illustration of the effect of the matrices $\overline{M}$ (turquoise) and $\overline{B}$ (magenta) on an edge point $e$ for $t = 2$ (left) and $t = 3$ (right). The matrix $\overline{M}$ creates two copies of $e$ and maps them to $p_1/2$ and $p_2/2$. The matrix $\overline{B}$ maps these two points to $e/2$, so that overall $e$ is mapped to $e/2$.

For the multiplication of the row $\tilde{C}_{(j,:)}$ with the control points, we have

$$\tilde{C}_{(j,:)}\tilde{P} = \overline{B}_{(j,:)}\overline{M}\tilde{P} = \overline{B}_{(j,K_1)}\overline{M}_{(K_1,:)}\tilde{P} + \overline{B}_{(j,K_2)}\overline{M}_{(K_2,:)}\tilde{P} = \overline{B}_{(j,K_1)}\frac{1}{2}p_1 + \overline{B}_{(j,K_2)}\frac{1}{2}p_2.$$

The entries $\overline{B}_{(j,K_1)}$ and $\overline{B}_{(j,K_2)}$ are given by

$$\begin{aligned}\overline{B}_{(j,K_1)} &= \frac{B_{(1,1)} + B_{(2,1)}}{B_{(1,1)} + B_{(1,2)} + B_{(2,1)} + B_{(2,2)}} =: \frac{B_{(1,1)} + B_{(2,1)}}{b},\\ \overline{B}_{(j,K_2)} &= \frac{B_{(1,2)} + B_{(2,2)}}{B_{(1,1)} + B_{(1,2)} + B_{(2,1)} + B_{(2,2)}} =: \frac{B_{(1,2)} + B_{(2,2)}}{b},\end{aligned}$$

and thus we obtain

$$\begin{aligned}\tilde{C}_{(j,:)}\tilde{P} &= \frac{B_{(1,1)} + B_{(2,1)}}{b}\frac{1}{2}p_1 + \frac{B_{(1,2)} + B_{(2,2)}}{b}\frac{1}{2}p_2 \\ &= \left(\frac{B_{(1,1)}}{b}\frac{1}{2}p_1 + \frac{B_{(1,2)}}{b}\frac{1}{2}p_2\right) + \left(\frac{B_{(2,1)}}{b}\frac{1}{2}p_1 + \frac{B_{(2,2)}}{b}\frac{1}{2}p_2\right) \\ &= \left(\frac{B_{(1,1)}}{b}\left(p_1' + \frac{1}{2}e\right) + \frac{B_{(1,2)}}{b}\left(p_2' + \frac{1}{2}e\right)\right) + \left(\frac{B_{(2,1)}}{b}\left(p_1' + \frac{1}{2}e\right) + \frac{B_{(2,2)}}{b}\left(p_2' + \frac{1}{2}e\right)\right) \\ &= \left(\frac{B_{(1,1)}}{b}\frac{1}{2}e + \frac{B_{(1,2)}}{b}\frac{1}{2}e\right) + \left(\frac{B_{(2,1)}}{b}\frac{1}{2}e + \frac{B_{(2,2)}}{b}\frac{1}{2}e\right) \\ &= \frac{1}{2}e.\end{aligned} \tag{6.5}$$

Hence, the row $\tilde{C}_{(j,:)}$ fulfills property E3. A schematic illustration of the mapping of $e$ to $e/2$ by the matrices $\overline{M}$ and $\overline{B}$ can be found in Figure 6.28. For the row $\tilde{C}_{(j,:)}$, which corresponds to an edge point $e$, we now prove the following lemmas:

**Lemma 6.11.** *Let $\tilde{C}_{(j,:)}$ be the row of $\tilde{C}$ corresponding to the edge point $e$. Then it holds that*

$$\tilde{C}_{(j,:)}\begin{bmatrix} & P & \\ 0 & & 0 \\ \vdots & & \vdots \\ 0 & & 0\end{bmatrix} = \frac{1}{2}e \quad \textit{for } t = 2, \quad \textit{and} \quad \tilde{C}_{(j,:)}\begin{bmatrix} & P & \\ 0 & 0 & 0 \\ & \vdots & \\ 0 & 0 & 0\end{bmatrix} = \frac{1}{2}e \quad \textit{for } t = 3.$$

*Proof.* The edge point $e$ is adjacent in the eigenstructure to two vertex points, which we denote by $p_1$ and $p_2$. We further define the two points

$$p_1' = \frac{1}{2}p_1 - \frac{1}{2}e \quad \text{and} \quad p_2' = \frac{1}{2}p_2 - \frac{1}{2}e.$$

The row $\overline{B}_{(j,:)}$ has two nonzero entries, which we denote by $b_1$ and $b_2$. By construction of the matrix $\overline{B}$, it holds that

$$b_1 p_1' + b_2 p_2' = 0 \quad \text{and} \quad b_1 + b_2 = 1.$$

We also know that the entries of the matrix $\overline{M}$ corresponding to the two points $p_1$ and $p_2$ with respect to the row $\tilde{C}_{(j,:)}$ are exactly $\frac{1}{2}$ by Lemma 6.9 for a quadrilateral and by Lemma 6.10 for a hexahedron. Hence, for $t = 3$ we obtain

$$\tilde{C}_{(j,:)} \begin{bmatrix} & P & \\ 0 & 0 & 0 \\ & \vdots & \\ 0 & 0 & 0 \end{bmatrix} = \overline{B}_{(j,:)}\overline{M} \begin{bmatrix} & P & \\ 0 & 0 & 0 \\ & \vdots & \\ 0 & 0 & 0 \end{bmatrix} = b_1\frac{1}{2}p_1 + b_2\frac{1}{2}p_2 = \frac{1}{2}\left(b_1(2p_1' + e) + b_2(2p_2' + e)\right) = \frac{1}{2}\left(b_1 e + b_2 e\right) = \frac{1}{2}e.$$

The above line can be formulated analogously for the case $t = 2$. □

This directly leads to the following lemma:

**Lemma 6.12.** *Let $\tilde{C}_{(j,:)}$ be the row of $\tilde{C}$ corresponding to the edge point $e$. Then it holds that*

$$\tilde{C}_{(j,:)} \begin{bmatrix} 0 & 0 \\ \vdots & \vdots \\ 0 & 0 \\ & E \\ & f \end{bmatrix} = [0,0] = 0 \cdot e \quad \textit{for } t = 2 \qquad \textit{and} \qquad \tilde{C}_{(j,:)} \begin{bmatrix} 0 & 0 & 0 \\ & \vdots & \\ 0 & 0 & 0 \\ & E & \\ & F & \\ & o & \end{bmatrix} = [0,0,0] = 0 \cdot e \quad \textit{for } t = 3.$$

*Proof.* Using equation (6.5) for $t = 2$, we have

$$\tilde{C}_{(j,:)}\tilde{P} = \tilde{C}_{(j,:)} \begin{bmatrix} & P \\ 0 & 0 \\ \vdots & \vdots \\ 0 & 0 \end{bmatrix} + \tilde{C}_{(j,:)} \begin{bmatrix} 0 & 0 \\ \vdots & \vdots \\ 0 & 0 \\ & E \\ & f \end{bmatrix} = \frac{1}{2}e.$$

From Lemma 6.11 it follows directly that

$$\tilde{C}_{(j,:)} \begin{bmatrix} & P \\ 0 & 0 \\ \vdots & \vdots \\ 0 & 0 \end{bmatrix} + \tilde{C}_{(j,:)} \begin{bmatrix} 0 & 0 \\ \vdots & \vdots \\ 0 & 0 \\ & E \\ & f \end{bmatrix} = \frac{1}{2}e \quad \Leftrightarrow \quad \frac{1}{2}e + \tilde{C}_{(j,:)} \begin{bmatrix} 0 & 0 \\ \vdots & \vdots \\ 0 & 0 \\ & E \\ & f \end{bmatrix} = \frac{1}{2}e \quad \Leftrightarrow \quad \tilde{C}_{(j,:)} \begin{bmatrix} 0 & 0 \\ \vdots & \vdots \\ 0 & 0 \\ & E \\ & f \end{bmatrix} = [0,0] = 0 \cdot e.$$

The case $t = 3$ follows analogously. □

The statement $[0,0] = 0 \cdot e$ is indeed trivial but serves as preparation for Lemmas 6.14 and 6.15.

**Facet points *f*** Here we consider the case of the facet points $f$ or $F$. We denote by $f$ the corresponding facet point associated with the row $\tilde{C}_{(j,:)}$. This point lies on $m$ objects, where $m$ is the number of vertices of the corresponding facet in the combinatorial graph $\mathbf{G}_K$. The vertices of these objects are denoted by $p_1, \ldots, p_m$. The entries of the matrix $K$, which contain the information about the rows of $\overline{M}$ corresponding to the copies of $f$, are denoted by $K_1, \ldots, K_m$.

The copies of $f$ are mapped by the matrix $\overline{M}$ onto the points $p_1/2, \ldots, p_m/2$. Since the points $p_1, \ldots, p_m$ all lie on one facet in the combinatorial graph, the points $p_1/2, \ldots, p_m/2$ lie in a plane which also contains the point $\frac{1}{2}f$. This facet has, because it is constructed via a circle packing, an incircle with center $f/2$. With this information, we define the points

$$p'_l = \frac{1}{2}p_l - \frac{1}{2}f \quad \text{for} \quad l \in \{1, \ldots, m\}.$$

These points also lie in a plane which contains the origin due to the shift. The origin is furthermore the center of the incircle for the shifted points.

For constructing the Colin de Verdière matrix $B$, we want to use the shifted points $p'_l$. To do this, we first scale these points by a factor $a$ so that the radius of the incircle is 1. For the case $t = 3$, in order to apply the theorems from Section 4.3, we also rotate the points $ap'_l$ so that they lie in $\mathbb{R}^2$. We use an appropriate rotation matrix $G \in \mathbb{R}^{3\times 3}$ for this purpose. In the case $t = 2$, the points already lie in $\mathbb{R}^2$, so we set $G = E_2$. From the rotated and scaled points

$$a \begin{bmatrix} p'_1 \\ \vdots \\ p'_m \end{bmatrix} G$$

we construct a Colin de Verdière matrix $B$ according to Theorem 4.8. For this matrix, it holds that

$$B \begin{bmatrix} p'^T_1 \\ \vdots \\ p'^T_m \end{bmatrix} = \frac{1}{a} Ba \begin{bmatrix} p'^T_1 \\ \vdots \\ p'^T_m \end{bmatrix} GG^{-1} = \frac{1}{a}\vec{0}G^{-1} = \vec{0},$$

i.e., the vectors from the points $p'$ lie in the kernel of the matrix $B$. Moreover, as with all Colin de Verdière matrices constructed in this work, the row and column sums of $B$ are negative by Satz 4.9. Hence, the matrix $B'$ consists analogously to previous cases of $m$ negative entries and the matrix $\frac{1}{\sum B'}B'$ consists of $m$ positive entries summing to 1. With these descriptions, the row $\tilde{C}_{(j,:)}$ can be expressed as

$$\tilde{C}_{(j,:)} = \sum_{l=1}^{m} \overline{B}_{(j,K_l)} \overline{M}_{(K_l,:)}.$$

Next, we examine the properties of the row $\tilde{C}_{(j,:)}$. Using the same argument as in the case of edge points, the row $\tilde{C}_{(j,:)}$ contains only the diagonal entry and positive non-zero entries corresponding to the neighbors of $f$. Hence, property E4 also holds for the facet case.

Furthermore, we have

$$\tilde{C}_{(j,:)}\vec{1} = \sum_{l=1}^{m} \overline{B}_{(j,K_l)} \overline{M}_{(K_l,:)}\vec{1} = \sum_{l=1}^{m} \overline{B}_{(j,K_l)} \cdot 1 = 1,$$

so the entries of $\tilde{C}_{(j,:)}$ sum to 1 and property E2 is satisfied. For the multiplication with $\tilde{P}$ it follows that

$$\tilde{C}_{(j,:)}\tilde{P} = \sum_{l=1}^{m} \overline{B}_{(j,K_l)} \overline{M}_{(K_l,:)}\tilde{P} = \sum_{l=1}^{m} \overline{B}_{(j,K_l)} \frac{1}{2}p_l.$$

The entries $\overline{B}_{(j,K_l)}$ can be expressed as

$$\overline{B}_{(j,K_l)} = \frac{\sum_{c=1}^{m} B_{(c,l)}}{\sum_{l=1}^{m}\sum_{c=1}^{m} B_{(c,l)}} =: \frac{\sum_{c=1}^{m} B_{(c,l)}}{b}.$$

Thus,

$$\tilde{C}_{(j,:)}\tilde{P} = \sum_{l=1}^{m} \frac{\sum_{c=1}^{m} B_{(c,l)}}{b} \frac{1}{2}p_l = \frac{1}{b}\sum_{c=1}^{m}\sum_{l=1}^{m} B_{(c,l)} \left(p'_l + \frac{1}{2}f\right) = \frac{\sum_{c=1}^{m}\sum_{l=1}^{m} B_{(c,l)}}{b}\frac{1}{2}f = \frac{1}{2}f,$$

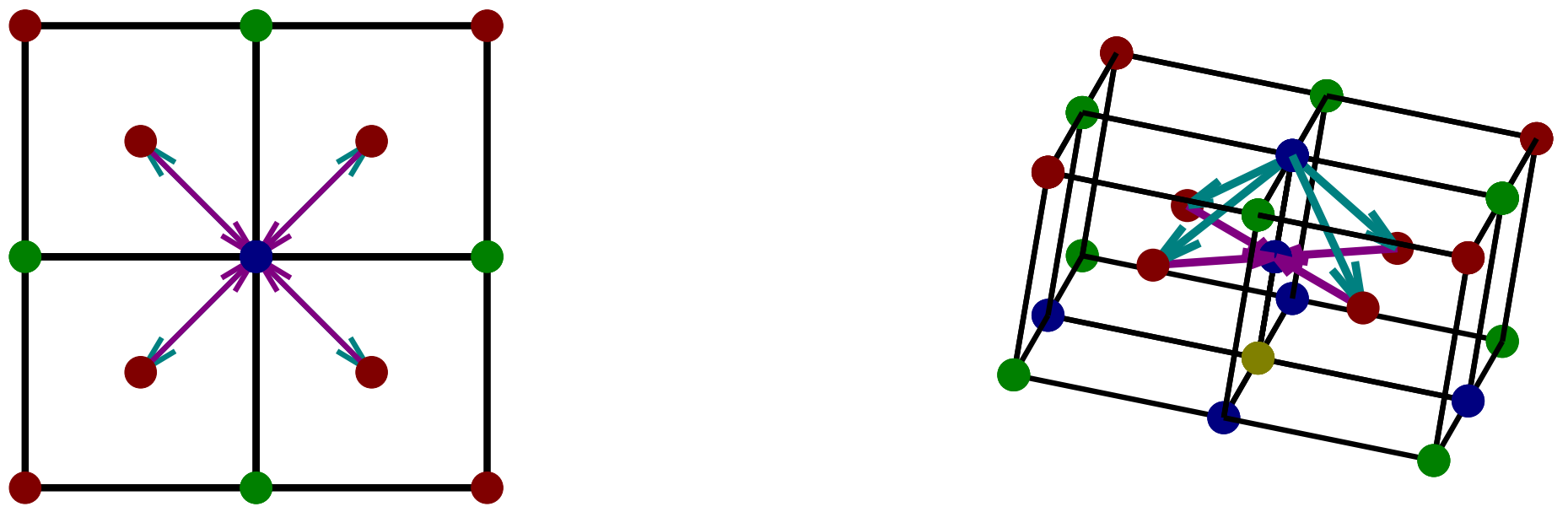

**Figure 6.29:** Schematic illustration of the effect of the matrices $\overline{M}$ (turquoise) and $\overline{B}$ (magenta) on a facet point $f$ for $t = 2$ (left) and $t = 3$ (right). The matrix $\overline{M}$ creates $m$ copies of $f$ and maps these onto $p_1/2, \dots, p_m/2$. The matrix $\overline{B}$ maps these $m$ points onto $f/2$, so that overall $f$ is mapped to $f/2$.

and so the facet point $f$ is mapped to $f/2$. Hence, property E3 is also satisfied. A schematic illustration of the mapping of $f$ to $f/2$ by the matrices $\overline{M}$ and $\overline{B}$ can be found in Figure 6.29.

**Volume Point *o*** This case occurs only for $t = 3$. The volume point $o$ is part of every hexahedron. Therefore, it is copied exactly $k$ times by the matrix $\overline{M}$ and mapped to all corner points $p_1/2, \dots, p_k/2$. The corresponding indices of $\overline{M}$ are stored in the matrix $K$. We denote these entries by $K_1, \dots, K_k$. The matrix $B$ is thus the Colin-de-Verdière-matrix of the 3-polytope scaled by the factor $1/2$ with corners $P$, constructed according to Theorem 4.13, and it holds that

$$BP = 2B\frac{1}{2}P = 2\vec{0} = \vec{0}.$$

Since the row and column sums of $B$ are negative by Theorem 4.14, the entries of $B'$ are negative. Thus, the entries of $\frac{1}{\sum B'}B'$ are positive and sum to 1. With these descriptions, the row $\tilde{C}_{(j,:)}$ can be written as

$$\tilde{C}_{(j,:)} = \sum_{l=1}^{k} \overline{B}_{(j,K_l)} \overline{M}_{(K_l,:)}.$$

We again check the properties of the row $\tilde{C}_{(j,:)}$. Property E4 behaves analogously to the edge point case and is fulfilled. Multiplying the row $\tilde{C}_{(j,:)}$ by the vector $\vec{1}$ yields again

$$\tilde{C}_{(j,:)}\vec{1} = \sum_{l=1}^{k} \overline{B}_{(j,K_l)} \overline{M}_{(K_l,:)}\vec{1} = \sum_{l=1}^{k} \overline{B}_{(j,K_l)} \cdot 1 = 1,$$

and thus property E2 is fulfilled. Multiplying the row $\tilde{C}_{(j,:)}$ with the control points $\tilde{P}$ gives

$$\tilde{C}_{(j,:)}\tilde{P} = \sum_{l=1}^{k} \overline{B}_{(j,K_l)} \overline{M}_{(K_l,:)}\tilde{P} = \sum_{l=1}^{k} \overline{B}_{(j,K_l)} \frac{1}{2} p_l.$$

With

$$\overline{B}_{(j,K_l)} = \frac{\sum_{c=1}^{k} B_{(c,l)}}{\sum_{l=1}^{k}\sum_{c=1}^{k} B_{(c,l)}} =: \frac{\sum_{c=1}^{k} B_{(c,l)}}{b},$$

we obtain, analogously to the facet case,

$$\tilde{C}_{(j,:)}\tilde{P} = \sum_{l=1}^{k} \frac{\sum_{c=1}^{k} B_{(c,l)}}{b} \frac{1}{2} p_l = \frac{1}{b} \sum_{c=1}^{k}\sum_{l=1}^{k} B_{(c,l)} \frac{1}{2} p_l = [0,0,0]^T = \frac{1}{2} o.$$

Thus, this row also satisfies property E3. A schematic illustration of the mapping of $o$ to $o/2$ by the matrices $\overline{M}$ and $\overline{B}$ can be found in Figure 6.30.

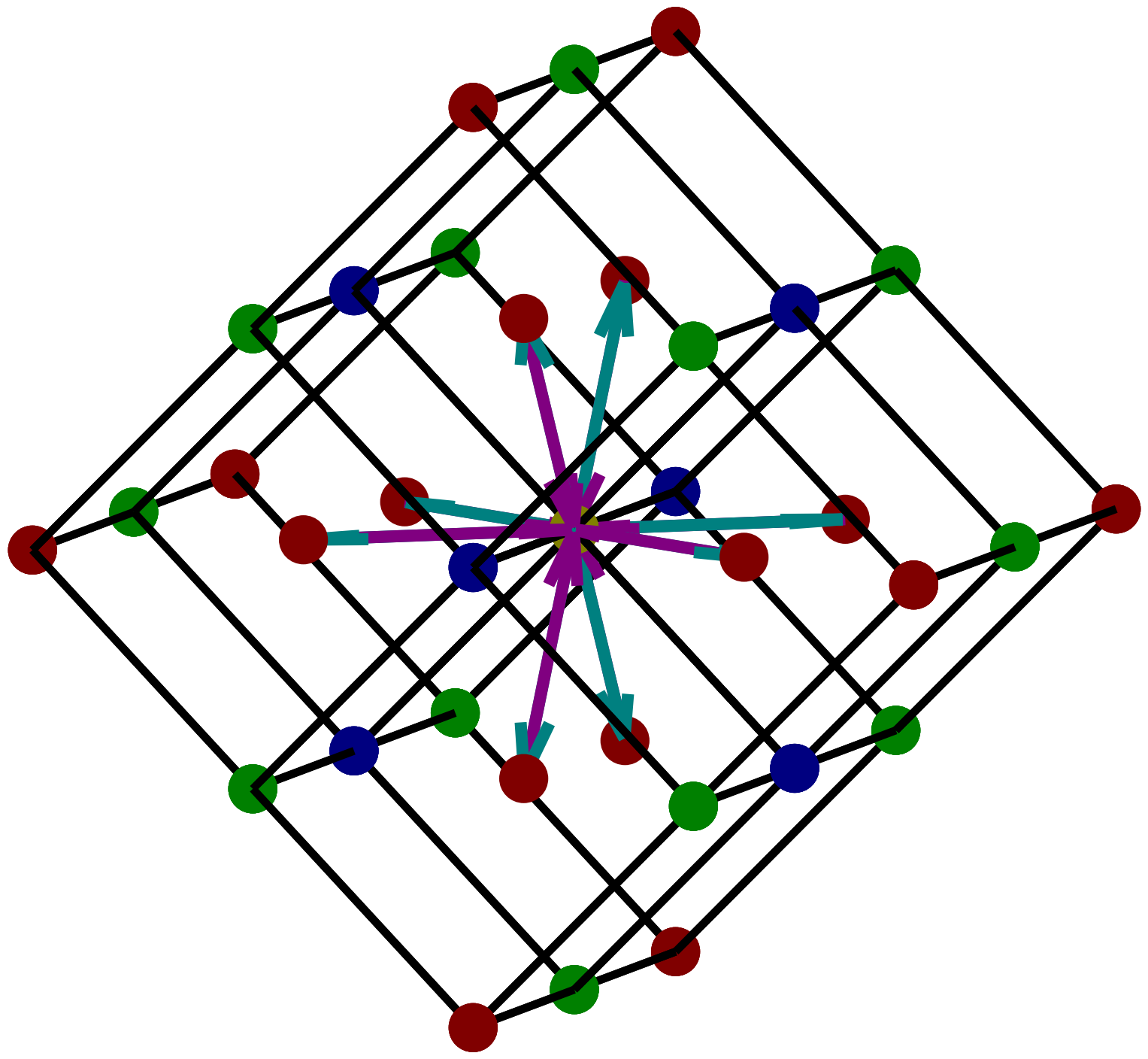

**Figure 6.30:** Schematic illustration of the effect of the matrices $\overline{M}$ (turquoise) and $\overline{B}$ (magenta) on the volume point $o$ for $t = 3$. The matrix $\overline{M}$ creates $k$ copies of $o$ and maps them onto $p_1/2, \ldots, p_k/2$. The matrix $\overline{B}$ maps these $k$ points back onto $o/2 = o$, so that overall $o$ is mapped to $o/2$.

As a summary of this subsection, we can state the following theorem:

**Theorem 6.13.** *The matrix $\tilde{C}$, constructed by Algorithm 15, has the properties E1–E4. Therefore, it is a Colin de Verdière-like matrix for the graph **G**.*

In the next subsection, we use the matrix $\tilde{C}$ to generate the final subdivision matrix $S$.

### 6.2.3 Decomposition of $\tilde{C}$ and the Matrix Exponential

In this step, we use the matrix exponential, based on Variant 2 from Section 5.2.2, to transform the matrix $\tilde{C}$ into a subdivision matrix. Before describing the exact transformation, we take a look at the backward construction in a scenario for $t = 3$ of the only known case, the regular case. This is the only combinatorial arrangement for which the refinement rules are known; all others must first be constructed and therefore depend on the construction.

Although this scenario will lead to a dead end, it clearly shows why we need an alternative construction for this chapter compared to Chapter 5, and it can possibly serve as a starting point for further research.

We first permute the matrix $S^{(3)}$ from Equation (6.1) such that

$$HS^{(3)}H^{-1}\begin{bmatrix} P \\ E \\ F \\ o \end{bmatrix} = \frac{1}{2}\begin{bmatrix} P \\ E \\ F \\ o \end{bmatrix}.$$

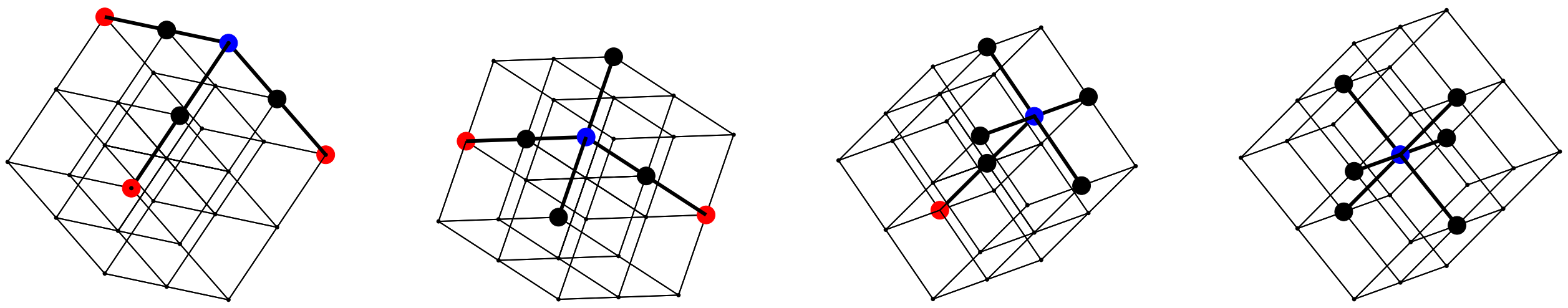

**Figure 6.31:** Illustration of the nonzero entries of the matrix $\tilde{M}'$, divided by point types: vertex (left), edge (middle left), facet (middle right), and volume (right). The colors correspond to the colored entries in Equation (6.6).

For this we get

$$\tilde{S} := HS^{(3)}H^{-1} := \frac{1}{512}\left[\begin{array}{cccccccc|cccccccccccc|cccccc|c}
64 & & & & & & & & 64 & 64 & 64 & & & & & & & & & & 64 & 64 & 64 & & & & 64 \\
 & 64 & & & & & & & 64 & & & 64 & 64 & & & & & & & & 64 & 64 & & 64 & & & 64 \\
 & & 64 & & & & & & & & & 64 & & 64 & 64 & & & & & & 64 & & & 64 & 64 & & 64 \\
 & & & 64 & & & & & & 64 & & & & 64 & & 64 & & & & & 64 & & 64 & & 64 & & 64 \\
 & & & & 64 & & & & & & 64 & & & & & & 64 & 64 & & & & 64 & 64 & & & 64 & 64 \\
 & & & & & 64 & & & & & & & 64 & & & & 64 & & 64 & & & 64 & & 64 & & 64 & 64 \\
 & & & & & & 64 & & & & & & & & 64 & & & & 64 & 64 & & & & 64 & 64 & 64 & 64 \\
 & & & & & & & 64 & & & & & & & & 64 & & 64 & & 64 & & & 64 & & 64 & 64 & 64 \\
\hline
16 & 16 & & & & & & & 96 & 16 & 16 & 16 & 16 & & & & & & & & 96 & 96 & 16 & 16 & & & 96 \\
16 & & & 16 & & & & & 16 & 96 & 16 & & & 16 & & 16 & & & & & 96 & 16 & 96 & & 16 & & 96 \\
16 & & & & 16 & & & & 16 & 16 & 96 & & & & & & 16 & 16 & & & 16 & 96 & 96 & & & 16 & 96 \\
 & 16 & 16 & & & & & & 16 & & & 96 & 16 & 16 & 16 & & & & & & 96 & 16 & & 96 & 16 & & 96 \\
 & 16 & & & & 16 & & & 16 & & & 16 & 96 & & & & 16 & & 16 & & 16 & 96 & & 96 & & 16 & 96 \\
 & & 16 & 16 & & & & & & 16 & & 16 & & 96 & 16 & 16 & & & & & 96 & & 16 & 16 & 96 & & 96 \\
 & & 16 & & & & 16 & & & & & 16 & & 16 & 96 & & & & 16 & 16 & 16 & & & 96 & 96 & 16 & 96 \\
 & & & 16 & & & & 16 & & 16 & & & & 16 & & 96 & & 16 & & 16 & 16 & & 96 & & 96 & 16 & 96 \\
 & & & & 16 & 16 & & & & & 16 & & 16 & & & & 96 & 16 & 16 & & & 96 & 16 & 16 & & 96 & 96 \\
 & & & & 16 & & & 16 & & & 16 & & & & & 16 & 16 & 96 & & 16 & & 16 & 96 & & 16 & 96 & 96 \\
 & & & & & 16 & 16 & & & & & & 16 & & 16 & & 16 & & 96 & 16 & & 16 & & 96 & 16 & 96 & 96 \\
 & & & & & & 16 & 16 & & & & & & & 16 & 16 & & 16 & 16 & 96 & & & 16 & 16 & 96 & 96 & 96 \\
\hline
4 & 4 & 4 & 4 & & & & & 24 & 24 & 4 & 24 & 4 & 24 & 4 & 4 & & & & & 144 & 24 & 24 & 24 & 24 & & 144 \\
4 & 4 & & & 4 & 4 & & & 24 & 4 & 24 & 4 & 24 & & & & 24 & 4 & 4 & & 24 & 144 & 24 & 24 & & 24 & 144 \\
4 & & & 4 & 4 & & & 4 & 4 & 24 & 24 & & & 4 & & 24 & 4 & 24 & & 4 & 24 & 24 & 144 & & 24 & 24 & 144 \\
 & 4 & 4 & & & 4 & 4 & & 4 & & & 24 & 24 & 4 & 24 & & 4 & & 24 & 4 & 24 & 24 & & 144 & 24 & 24 & 144 \\
 & & 4 & 4 & & & 4 & 4 & & 4 & & 4 & & 24 & 24 & 24 & & 4 & 4 & 24 & 24 & & 24 & 24 & 144 & 24 & 144 \\
 & & & & 4 & 4 & 4 & 4 & & & 4 & & 4 & & 4 & 4 & 24 & 24 & 24 & 24 & & 24 & 24 & 24 & 24 & 144 & 144 \\
\hline
1 & 1 & 1 & 1 & 1 & 1 & 1 & 1 & 6 & 6 & 6 & 6 & 6 & 6 & 6 & 6 & 6 & 6 & 6 & 6 & 36 & 36 & 36 & 36 & 36 & 36 & 216
\end{array}\right].$$

We can apply the matrix logarithm to this matrix, as described in Definition B.55, because all eigenvalues of $\tilde{S}$ are positive. Therefore, by Theorem B.56, a unique matrix logarithm exists whose eigenvalues have imaginary parts between $-\pi$ and $\pi$. Using this matrix logarithm, we hope to find an idea of the corresponding matrix that the matrix $\tilde{C}$ from the previous section should take. We define

$$\tilde{M} := \frac{\ln\left(\tilde{S}\right)}{\ln(2)} + E,$$

and obtain

$$\tilde{M} := \frac{1}{6}\left[\begin{array}{cccccccc|cccccccccccc|cccccc|c}
-15 & -1 & & -1 & -1 & & & & 8 & 8 & 8 & & & & & & & & & & & & & & & & \\
-1 & -15 & -1 & & & -1 & & & 8 & & & 8 & 8 & & & & & & & & & & & & & & \\
 & -1 & -15 & -1 & & & -1 & & & & & 8 & & 8 & 8 & & & & & & & & & & & & \\
-1 & & -1 & -15 & & & & -1 & & 8 & & & & 8 & & 8 & & & & & & & & & & & \\
-1 & & & & -15 & -1 & & -1 & & & 8 & & & & & & 8 & 8 & & & & & & & & & \\
 & -1 & & & -1 & -15 & -1 & & & & & & 8 & & & & 8 & & 8 & & & & & & & & \\
 & & -1 & & & -1 & -15 & -1 & & & & & & & 8 & & & & 8 & 8 & & & & & & & \\
 & & & -1 & -1 & & -1 & -15 & & & & & & & & 8 & & 8 & & 8 & & & & & & & \\
\hline
2 & 2 & & & & & & & -12 & & & & & -1 & & & -1 & & & & 8 & 8 & & & & & \\
2 & & & 2 & & & & & & -12 & & -1 & & & & & & -1 & & & 8 & & 8 & & & & \\
2 & & & & 2 & & & & & & -12 & & -1 & & & -1 & & & & & & 8 & 8 & & & & \\
 & 2 & 2 & & & & & & & -1 & & -12 & & & & & & & -1 & & 8 & & & 8 & & & \\
 & 2 & & & & 2 & & & & & -1 & & -12 & & -1 & & & & & & & 8 & & 8 & & & \\
 & & 2 & 2 & & & & & -1 & & & & & -12 & & & & & & -1 & 8 & & & & 8 & & \\
 & & 2 & & & & 2 & & & & & & -1 & & -12 & -1 & & & & & & & & 8 & 8 & & \\
 & & & 2 & & & & 2 & & & -1 & & & & -1 & -12 & & & & & & & 8 & & 8 & & \\
 & & & & 2 & 2 & & & -1 & & & & & & & & -12 & & & -1 & & 8 & & & & 8 & \\
 & & & & 2 & & & 2 & & -1 & & & & & & & & -12 & -1 & & & & 8 & & & 8 & \\
 & & & & & 2 & 2 & & & & & -1 & & & & & & -1 & -12 & & & & & 8 & & 8 & \\
 & & & & & & 2 & 2 & & & & & & -1 & & & -1 & & & -12 & & & & & 8 & 8 & \\
\hline
 & & & & & & & & 2 & 2 & & 2 & & 2 & & & & & & & -9 & & & & & -1 & 8 \\
 & & & & & & & & 2 & & 2 & & 2 & & & & 2 & & & & & -9 & & & -1 & & 8 \\
 & & & & & & & & & 2 & 2 & & & & & 2 & & 2 & & & & & -9 & -1 & & & 8 \\
 & & & & & & & & & & & 2 & 2 & & 2 & & & & 2 & & & & -1 & -9 & & & 8 \\
 & & & & & & & & & & & & & 2 & 2 & 2 & & & & 2 & & -1 & & & -9 & & 8 \\
 & & & & & & & & & & & & & & & & 2 & 2 & 2 & 2 & -1 & & & & & -9 & 8 \\
\hline
 & & & & & & & & & & & & & & & & & & & & 2 & 2 & 2 & 2 & 2 & 2 & -6
\end{array}\right].$$

This matrix is normalized, but from it a symmetric matrix can be constructed by multiplying the rows corresponding to the edge points by the factor 4, the rows corresponding to the facet points by the factor 16, and the row corresponding to the volume point by the factor 64. This leads to

$$\tilde{M}' := \frac{1}{6}\left[\begin{array}{cccccccc|cccccccccccc|cccccc|c}
-15 & -1 & & -1 & -1 & & & & 8 & 8 & 8 & & & & & & & & & & & & & & & & \\
-1 & -15 & -1 & & & -1 & & & 8 & & & 8 & 8 & & & & & & & & & & & & & & \\
 & -1 & -15 & -1 & & & -1 & & & & & 8 & & 8 & 8 & & & & & & & & & & & & \\
-1 & & -1 & -15 & & & & -1 & & 8 & & & & 8 & & 8 & & & & & & & & & & & \\
-1 & & & & -15 & -1 & & -1 & & & 8 & & & & & & 8 & 8 & & & & & & & & & \\
 & -1 & & & -1 & -15 & -1 & & & & & & 8 & & & & 8 & & 8 & & & & & & & & \\
 & & -1 & & & -1 & -15 & -1 & & & & & & & 8 & & & & 8 & 8 & & & & & & & \\
 & & & -1 & -1 & & -1 & -15 & & & & & & & & 8 & & 8 & & 8 & & & & & & & \\
\hline
8 & 8 & & & & & & & -48 & & & & & -4 & & & -4 & & & & 32 & 32 & & & & & \\
8 & & & 8 & & & & & & -48 & & -4 & & & & & & -4 & & & 32 & & 32 & & & & \\
8 & & & & 8 & & & & & & -48 & & -4 & & & -4 & & & & & & 32 & 32 & & & & \\
 & 8 & 8 & & & & & & & -4 & & -48 & & & & & & & -4 & & 32 & & & 32 & & & \\
 & 8 & & & & 8 & & & & & -4 & & -48 & & -4 & & & & & & & 32 & & 32 & & & \\
 & & 8 & 8 & & & & & -4 & & & & & -48 & & & & & & -4 & 32 & & & & 32 & & \\
 & & 8 & & & & 8 & & & & & & -4 & & -48 & -4 & & & & & & & & 32 & 32 & & \\
 & & & 8 & & & & 8 & & & -4 & & & & -4 & -48 & & & & & & & 32 & & 32 & & \\
 & & & & 8 & 8 & & & -4 & & & & & & & & -48 & & & -4 & & 32 & & & & 32 & \\
 & & & & 8 & & & 8 & & -4 & & & & & & & & -48 & -4 & & & & 32 & & & 32 & \\
 & & & & & 8 & 8 & & & & & -4 & & & & & & -4 & -48 & & & & & 32 & & 32 & \\
 & & & & & & 8 & 8 & & & & & & -4 & & & -4 & & & -48 & & & & & 32 & 32 & \\
\hline
 & & & & & & & & 32 & 32 & & 32 & & 32 & & & & & & & -144 & & & & & -16 & 128 \\
 & & & & & & & & 32 & & 32 & & 32 & & & & 32 & & & & & -144 & & & -16 & & 128 \\
 & & & & & & & & & 32 & 32 & & & & & 32 & & 32 & & & & & -144 & -16 & & & 128 \\
 & & & & & & & & & & & 32 & 32 & & 32 & & & & 32 & & & & -16 & -144 & & & 128 \\
 & & & & & & & & & & & & & 32 & 32 & 32 & & & & 32 & & -16 & & & -144 & & 128 \\
 & & & & & & & & & & & & & & & & 32 & 32 & 32 & 32 & -16 & & & & & -144 & 128 \\
\hline
 & & & & & & & & & & & & & & & & & & & & 128 & 128 & 128 & 128 & 128 & 128 & -384
\end{array}\right]. \tag{6.6}$$

This matrix has a very interesting structure. First, all entries of the matrix $\tilde{M}'$ for which the corresponding entries of the adjacency matrix are equal to $1$ are positive. This is as expected, because if you multiply the matrix $\tilde{M}'$ by $-1$, then all entries become negative, matching condition C1 of the Colin-de-Verdière-matrices from Definition 4.1.

However, there are additional off-diagonal entries that are nonzero. These are marked in red in the matrix $\tilde{M}'$. An overview of the geometric assignment can be found in Figure 6.31. It is completely unclear what the criterion for the red entries is. It could be something like "an entry is nonzero if it lies on the line between a neighboring point and the original point". It could also be something like "an entry $(i, j)$ is zero if a shortest path from node $i$ to node $j$ contains a right angle." The problem here is that we only have one example from which we would have to extrapolate the rules. Therefore, this scenario currently leads to a dead end, but further research on this approach would be interesting.

The consequence of this is that we cannot simply apply the matrix exponential to the matrix $\tilde{C}$ from Algorithm 15 in the same way as in Section 5.2.2. Therefore, we need a different strategy, which we describe in Algorithm 16. We will carry out the individual steps below. All terms used in the following section refer to those in Algorithm 16.

**Construction of $\tilde{L}$ and $\tilde{R}$** The rows of the matrix $\tilde{C}$ all sum up to $1$, as we saw in the previous section with Theorem 6.13. As a first step, we multiply the matrix $\tilde{C}$ by the factor $2$. This step appears only implicitly in the above Algorithm 16.

We then decompose the matrix $2\tilde{C}$ into a lower triangular matrix $\tilde{L}$ and an upper triangular matrix $\tilde{R}$. The off-diagonal entries are taken over exactly as they are. The diagonal entries of $2\tilde{C}$ are split according to lines 11 and 12 of Algorithm 16. Since the rows of $2\tilde{C}$ sum up to $2$, we obtain

$$\mathtt{diag}\left(\tilde{L}\right) + \mathtt{diag}\left(\tilde{R}\right) = \begin{bmatrix} 1 - \sum_{j=1,j\neq 1}^{n} \tilde{L}_{(1,j)} + 1 - \sum_{j=1,j\neq 1}^{n} \tilde{R}_{(1,j)} \\ \vdots \\ 1 - \sum_{j=1,j\neq n}^{n} \tilde{L}_{(n,j)} + 1 - \sum_{j=1,j\neq n}^{n} \tilde{R}_{(n,j)} \end{bmatrix} = \begin{bmatrix} 2 - \sum_{j=1,j\neq 1}^{n} \tilde{C}_{(1,j)} \\ \vdots \\ 2 - \sum_{j=1,j\neq n}^{n} \tilde{C}_{(n,j)} \end{bmatrix} = \mathtt{diag}\left(2\tilde{C}\right) \tag{6.7}$$

and

$$\tilde{L} + \tilde{R} = 2\tilde{C}.$$

Some statements about the entries of $\tilde{L}$ and $\tilde{R}$ can be made, which we formulate and prove in the following:

**Lemma 6.14.** *Let $t = 3$, and let the rows of $\tilde{C}$, constructed according to Algorithm 15, be arranged such that property E3 holds. Moreover, let $|P|$ be the number of vertex points, $|E|$ the number of edge points, $|F|$ the number of facet points,* *and* $|o| = 1$ *the number of volume points of the control points* $\tilde{P} = \left[P^T, E^T, F^T, o^T\right]^T$. *We define the index sets*

$$I_P = [1, \dots, |P|], \quad I_E = [|P|+1, \dots, |P|+|E|], \quad I_F = [|P|+|E|+1, \dots, |P|+|E|+|F|], \quad I_o = |P|+|E|+|F|+1$$

*as the indices of the corresponding rows and columns of the matrix* $\tilde{C}$ *associated to the points* $\tilde{P}$. *Let* $\tilde{L}$ *and* $\tilde{R}$ *be constructed from the input* $\tilde{C}$ *according to Algorithm 16, with rows and columns of* $\tilde{C}$, $\tilde{L}$, *and* $\tilde{R}$ *arranged consistently. Then the matrices have the block structures*

$$\tilde{L} = \left[\begin{array}{ccc|ccc|ccc|c}
1 & & 0 & 0 & \cdots & 0 & 0 & \cdots & 0 & 0 \\
 & \ddots & & \vdots & & \vdots & \vdots & & \vdots & \vdots \\
0 & & 1 & 0 & \cdots & 0 & 0 & \cdots & 0 & 0 \\
\hline
 & & & 0 & & 0 & 0 & \cdots & 0 & 0 \\
\multicolumn{3}{c|}{2\tilde{C}_{(I_E,I_P)}} & & \ddots & & \vdots & & \vdots & \vdots \\
 & & & 0 & & 0 & 0 & \cdots & 0 & 0 \\
\hline
0 & \cdots & 0 & & & & 2\tilde{C}_{\left((I_F)_1,(I_F)_1\right)} & & 0 & 0 \\
\vdots & & \vdots & \multicolumn{3}{c|}{2\tilde{C}_{(I_F,I_E)}} & & \ddots & & \vdots \\
0 & \cdots & 0 & & & & 0 & & 2\tilde{C}_{\left((I_F)_{|F|},(I_F)_{|F|}\right)} & 0 \\
\hline
0 & \cdots & 0 & 0 & \cdots & 0 & \multicolumn{3}{c|}{2\tilde{C}_{(I_o,I_F)}} & 2\tilde{C}_{(I_o,I_o)} - 1
\end{array}\right]$$

*and*

$$\tilde{R} = \left[\begin{array}{ccc|ccc|ccc|c}
2\tilde{C}_{(1,1)} - 1 & & 0 & & & & 0 & \cdots & 0 & 0 \\
 & \ddots & & \multicolumn{3}{c|}{2\tilde{C}_{(I_P,I_E)}} & \vdots & & \vdots & \vdots \\
0 & & 2\tilde{C}_{(|P|,|P|)} - 1 & & & & 0 & \cdots & 0 & 0 \\
\hline
0 & \cdots & 0 & 2\tilde{C}_{\left((I_E)_1,(I_E)_1\right)} & & 0 & & & & 0 \\
\vdots & & \vdots & & \ddots & & \multicolumn{3}{c|}{2\tilde{C}_{(I_E,I_F)}} & \vdots \\
0 & \cdots & 0 & 0 & & 2\tilde{C}_{\left((I_E)_{|E|},(I_E)_{|E|}\right)} & & & & 0 \\
\hline
0 & \cdots & 0 & 0 & \cdots & 0 & 0 & & 0 & 1 \\
\vdots & & \vdots & \vdots & & \vdots & & \ddots & & \vdots \\
0 & \cdots & 0 & 0 & \cdots & 0 & 0 & & 0 & 1 \\
\hline
0 & \cdots & 0 & 0 & \cdots & 0 & 0 & & 0 & 1
\end{array}\right].$$

*In particular,* $\tilde{L}$ *is a lower triangular matrix and* $\tilde{R}$ *is an upper triangular matrix. Moreover, the rows of both* $\tilde{L}$ *and* $\tilde{R}$ *each sum to* $1$. *Additionally, it holds that*

$$\tilde{L}\tilde{P} = \tilde{P} \quad \textit{and} \quad \tilde{R}\tilde{P} = 0 \cdot \tilde{P},$$

*so the three columns of* $\tilde{P}$ *are eigenvectors corresponding to the eigenvalue* $1$ *of* $\tilde{L}$ *and eigenvalue* $0$ *of* $\tilde{R}$.

*Proof.* The matrices $\tilde{L}$ and $\tilde{R}$ can be divided, as shown above, into 16 blocks, which we consider in the following:

- $I_P \times I_P$ and $I_o \times I_o$: The corresponding block in $\tilde{C}$ for $I_P \times I_P$ contains only diagonal entries, since there are no edges between vertex points in the graph **G**. Therefore, all entries in the rows $\tilde{L}_{(I_P,:)}$ except the diagonal are zero. Consequently, by Algorithm 16, the diagonal entries in this block satisfy

$$\vec{1} - \sum_{j=1}^{n} \tilde{L}_{(I_P,j)} = \vec{1}.$$

**Algorithm 16:** ConstructionOf$S$

**Data:** Colin-de-Verdière-like matrix $\tilde{C} \in \mathbb{R}^{n \times n}$ constructed according to Algorithm 15
**Result:** Subdivision matrix $S \in \mathbb{R}^{n \times n}$ for $g = 3$

**begin**
  $\tilde{L} :=$ zero matrix of dimension $n \times n$;
  $\tilde{R} :=$ zero matrix of dimension $n \times n$;
  **for** $i = 1 : n$ **do**
    **for** $j = 1 : n$ **do**
      **if** $i > j$ **then**
        $\tilde{L}_{(i,j)} := 2\tilde{C}_{(i,j)}$;
      **else if** $i < j$ **then**
        $\tilde{R}_{(i,j)} := 2\tilde{C}_{(i,j)}$;
  **for** $i = 1 : n$ **do**
    $\tilde{L}_{(i,i)} := 1 - \sum_{j=1}^{n} \tilde{L}_{(i,j)}$;
    $\tilde{R}_{(i,i)} := 1 - \sum_{j=1}^{n} \tilde{R}_{(i,j)}$;
  $L' := \ln(2) \cdot (\tilde{L} - E_n)$;
  $R' := \ln(2) \cdot (\tilde{R} - E_n)$;
  $L := \exp(L')$;
  $R := \exp(R')$;
  $S := L \cdot R$;
  **return** $S$;

By equation (6.7), the diagonal of $2\tilde{C}$ equals the sum of the diagonals of $\tilde{L}$ and $\tilde{R}$. Thus, the diagonal entries of $\tilde{R}$ satisfy

$$\tilde{R}_{(i,i)} = 2\tilde{C}_{(i,i)} - 1 \quad \text{for } i \in I_P.$$

The same applies analogously to $I_o \times I_o$, swapping $\tilde{L}$ and $\tilde{R}$ in the above argument.

- $I_P \times I_E$, $I_E \times I_P$, $I_E \times I_F$, $I_F \times I_E$, and $I_o \times I_F$: Due to the splitting into lower and upper triangular matrices, the corresponding block in one matrix is zero while in the other it equals the corresponding block of $2\tilde{C}$.

- $I_P \times I_F$, $I_F \times I_P$, $I_P \times I_o$, $I_o \times I_P$, $I_E \times I_o$, and $I_o \times I_E$: There are no connections between these categories of points in the graph **G**, so the corresponding blocks in $\tilde{C}$ are zero, and thus the corresponding blocks in $\tilde{L}$ and $\tilde{R}$ are zero as well.

- $I_E \times I_E$: First, the block $I_E \times I_E$ in the matrix $\tilde{C}$ can only be nonzero on the diagonal, since there are no edges between two edge points in the graph **G**.

  The block $I_P \times I_E$ of the matrix $\tilde{C}$ has a special structure. The entries of $\overline{M}$ that are multiplied by the matrix $\overline{B}$ to form the entries of this block are, by Lemma 6.10, equal to $1/2$. Since each edge point is connected to two vertex points, there are two rows in $\overline{M}$ per edge point that are combined into the corresponding rows of $\tilde{C}$. Each such row contains a value $1/2$ for the respective vertex point. The corresponding row of $\overline{B}$ therefore has only two nonzero entries, and since the rows of $\overline{B}$ sum to 1, we obtain

$$\sum_{j \in I_P} \tilde{C}_{(i,j)} = \frac{1}{2}\overline{B}_{(i,k)} + \frac{1}{2}\overline{B}_{(i,l)} = \frac{1}{2}\left(\overline{B}_{(i,k)} + \overline{B}_{(i,l)}\right) = \frac{1}{2} \quad \text{for} \quad i \in I_E. \tag{6.8}$$

  Hence,

$$\sum_{j \in I_P} 2\tilde{C}_{(i,j)} = \sum_{j \in I_P} \tilde{L}_{(i,j)} = 1 \quad \text{for} \quad i \in I_E,$$

and thus

$$\tilde{L}_{(i,i)} = 1 - \sum_{j=1}^{n} \tilde{L}_{(i,j)} = 1 - 1 = 0.$$

Therefore, the block $I_E \times I_E$ of $\tilde{L}$ is everywhere zero, and the block $I_E \times I_E$ of $\tilde{R}$ equals the corresponding block of $2\tilde{C}$.

- $I_F \times I_o$: The volume point $o$ is present in all hexahedra. By Lemma 6.10, the corresponding entries in $\overline{M}$ equal $1/2$, and by analogy to equation (6.8), these sum to $1/2$ by multiplication with $\overline{B}$. Thus,

$$\tilde{C}_{(I_F,I_o)} = \frac{\vec{1}}{2} \quad \text{and} \quad \tilde{R}_{(I_F,I_o)} = 2\tilde{C}_{(I_F,I_o)} = \vec{1}.$$

  Due to the splitting into upper and lower triangular matrices, $\tilde{L}_{(I_F,I_o)} = \vec{0}$.

- $I_F \times I_F$: First, the block $I_F \times I_F$ in $\tilde{C}$ can only be nonzero on the diagonal, since there are no edges between two facet points in the graph **G**. Since $\tilde{R}_{(I_F,I_o)} = \vec{1}$, it follows that

$$\tilde{R}_{(i,i)} = 1 - 1 = 0 \quad \text{and} \quad \tilde{L}_{(i,i)} = 2\tilde{C}_{(i,i)} \quad \text{for } i \in I_F.$$

The matrices $\tilde{L}$ and $\tilde{R}$ are, by definition, triangular matrices whose rows each sum to 1. In the next step, we consider the multiplication of these matrices with the points $\tilde{P}$ and obtain the following statements. First, from Theorem 6.13 it follows that

$$\tilde{C}\tilde{P} = \frac{1}{2}\tilde{P} \quad \Leftrightarrow \quad 2\tilde{C}\tilde{P} = \tilde{P}.$$

Considering the matrix $\tilde{L}$, we obtain for the vertex points

$$\tilde{L}_{(I_P,:)}\tilde{P} = E_{|P|}P = P,$$

where $E_{|P|}$ denotes the identity matrix of size $|P| \times |P|$. For the edge points, by Lemma 6.11, we have

$$\tilde{L}_{(I_E,:)}\tilde{P} = 2\tilde{C}_{(I_E,:)} \begin{bmatrix} & P & \\ 0 & 0 & 0 \\ & \vdots & \\ 0 & 0 & 0 \end{bmatrix} = E,$$

where $E$ denotes the coordinate matrix of the edge points. For the facet points we obtain

$$\tilde{L}_{(I_F,:)}\tilde{P} = 2\tilde{C}_{(I_F,:)} \begin{bmatrix} & P & \\ & E & \\ & F & \\ 0 & 0 & 0 \end{bmatrix} = 2\tilde{C}_{(I_F,:)} \begin{bmatrix} P \\ E \\ F \\ o \end{bmatrix} - 2\tilde{C}_{(I_F,:)} \begin{bmatrix} 0 & 0 & 0 \\ & \vdots & \\ 0 & 0 & 0 \\ & o & \end{bmatrix} = F - \begin{bmatrix} o \\ \vdots \\ o \end{bmatrix} = F,$$

and for the volume point it follows

$$\tilde{L}_{(I_o,:)}\tilde{P} = 2\tilde{C}_{(I_o,:)} \begin{bmatrix} P \\ E \\ F \\ o \end{bmatrix} - o = o - o = [0,0,0] = o.$$

Correspondingly, for the matrix $\tilde{R}$ we have for the vertex points

$$\tilde{R}_{(I_P,:)}\tilde{P} = 2\tilde{C}_{(I_P,:)}\tilde{P} - P = P - P = 0 \cdot P.$$

For the edge points, by Lemma 6.11 and Lemma 6.12, we get

$$\tilde{R}_{(I_E,:)}\tilde{P} = 2\tilde{C}_{(I_E,:)}\tilde{P} - 2\tilde{C}_{(I_E,:)} \begin{bmatrix} & P & \\ 0 & 0 & 0 \\ & \vdots & \\ 0 & 0 & 0 \end{bmatrix} = E - E = 0 \cdot E.$$

For the facet points and volume point, we obtain

$$\tilde{R}_{(I_F,:)}\tilde{P} = \begin{bmatrix} o \\ \vdots \\ o \end{bmatrix} = \begin{bmatrix} 0 & 0 & 0 \\ & \vdots & \\ 0 & 0 & 0 \end{bmatrix} = 0 \cdot F, \quad \text{and} \quad \tilde{R}_{(I_o,:)}\tilde{P} = o = [0,0,0] = 0 \cdot o.$$

Overall, we thus obtain the above statement

$$\tilde{L}\tilde{P} = \tilde{P} \quad \text{and} \quad \tilde{R}\tilde{P} = 0 \cdot \tilde{P}.$$

□

For $t = 2$ we obtain analogously the following lemma:

**Lemma 6.15.** *Let $t = 2$ and let the rows of $\tilde{C}$, constructed according to Algorithm 15, be arranged so that property E3 holds. Further, let $|P|$ be the number of vertex points, $|E|$ the number of edge points, and $|f| = 1$ the number of facet points of the control points $\tilde{P} = \left[P^T, E^T, f^T\right]^T$. Define*

$$I_P = [1, \ldots, |P|], \quad I_E = [|P|+1, \ldots, |P|+|E|], \quad \textit{and} \quad I_F = |P| + |E| + 1$$

*as the indices of the respective rows and columns of the matrix $\tilde{C}$ corresponding to the points $\tilde{P}$. Let $\tilde{L}$ and $\tilde{R}$ be constructed from the input $\tilde{C}$ according to Algorithm 16, with the row and column order of $\tilde{C}$, $\tilde{L}$ and $\tilde{R}$ coinciding. Then*

$$\tilde{L} = \left[\begin{array}{ccc|ccc|c} 1 & & 0 & 0 & \ldots & 0 & 0 \\ & \ddots & & \vdots & & \vdots & \vdots \\ 0 & & 1 & 0 & \ldots & 0 & 0 \\ \hline & & & 0 & & 0 & 0 \\ & 2\tilde{C}_{(I_E,I_P)} & & & \ddots & & \vdots \\ & & & 0 & & 0 & 0 \\ \hline 0 & \ldots & 0 & & 2\tilde{C}_{(I_f,I_E)} & & 2\tilde{C}_{(I_f,I_f)} - 1 \end{array}\right] \quad \textit{and} \quad \tilde{R} = \left[\begin{array}{ccc|ccc|c} 2\tilde{C}_{(1,1)} - 1 & & 0 & & & & 0 \\ & \ddots & & & 2\tilde{C}_{(I_P,I_E)} & & \vdots \\ 0 & & 2\tilde{C}_{(|P|,|P|)} - 1 & & & & 0 \\ \hline 0 & \ldots & 0 & 0 & & 0 & 1 \\ \vdots & & \vdots & & \ddots & & \vdots \\ 0 & \ldots & 0 & 0 & & 0 & 1 \\ \hline 0 & \ldots & 0 & 0 & & 0 & 1 \end{array}\right].$$

*In particular, $\tilde{L}$ is a lower and $\tilde{R}$ an upper triangular matrix. Moreover, the rows of $\tilde{L}$ and $\tilde{R}$ each sum up to 1. Additionally,*

$$\tilde{L}\tilde{P} = \tilde{P} \quad \textit{and} \quad \tilde{R}\tilde{P} = 0 \cdot \tilde{P},$$

*so the two columns of $\tilde{P}$ are eigenvectors both to the eigenvalue 1 of the matrix $\tilde{L}$ and to the eigenvalue 0 of the matrix $\tilde{R}$.*

*Proof.* The proof proceeds entirely analogously to the proof of Lemma 6.14. It only requires replacing Lemma 6.10 by Lemma 6.9. □

**Construction of the Matrix $S$** Finally, we describe the construction of the matrix $S$ from Algorithm 16. To this end, we shift the spectrum of the matrices $\tilde{L}$ and $\tilde{R}$ by first subtracting the identity matrix from them in lines 13–16 of Algorithm 16 and multiplying the result by the factor $\ln(2)$. Afterwards, we compute the matrix exponential of the shifted matrices $L'$ and $R'$, and finally, in line 17 of Algorithm 16, we construct the matrix $S$ as the product of these two matrix exponentials $L$ and $R$.

Thus, the construction of the subdivision matrix $S$ is complete. We would like to remark the following at this point:

**Remark 6.16.** *For the construction of the matrix S, we used two essential core points that enable the statements made so far. The first core point is the specially constructed structure of the points $\tilde{P}$ with the right angles between the edges $\{p, e\}$ and $\{e, f\}$ and, additionally for $t = 3$, between the edges $\{e, f\}$ and $\{f, o\}$. Due to this orthogonality, many terms simplified and we were able to explicitly determine some entries of the matrix $\overline{M}$.*

*The second core point is the splitting of the matrix $\tilde{C}$ into the two matrices $\tilde{L}$ and $\tilde{R}$. Here, the ordering of the points $\tilde{P} = \left[P^T, E^T, f^T\right]^T$ or $\tilde{P} = \left[P^T, E^T, F^T, o^T\right]^T$ is insignificant, but simplifies the notation considerably. When splitting, it is only relevant that the entries corresponding to the respective point types are distributed to the two matrices. If this is satisfied, the split matrices can be permuted arbitrarily (but in the same way).*

**Remark 6.17.** *We will see in the next section that the subdominant eigenvalue of the constructed subdivision matrix $S$ is not $1/4$. Moreover, for examples with a high number of nodes, this eigenvalue tends to be close to $1/2$, which partially contradicts the goal of this work*

> *Construct a trivariate generalization of Catmull-Clark [Doo-Sabin] with triple subdominant eigenvalue $\lambda = \frac{1}{2}$ and smallest possible $\mu$*

*It remains completely open and requires further research whether for $g = 3$ a subdivision matrix with a suitable support of the refinement rules according to quality criterion Q11 and a „better“ subdominant eigenvalue can be constructed.*

*Alternatively to the construction of $S$ presented here, one could also use the eigenstructure from Section 6.2.1 to create a construction of the subdivision matrix $S$ based on variant 1 from Section 5.2.1. It would also be possible to decompose the constructed matrix $S$ into its eigenvalues and eigenvectors, adjust the spectrum as desired, and thereby generate an alternative subdivision matrix. This might be useful for some applications. However, since this approach surely loses the quality criterion Q11, we do not pursue it further here.*

Before we address the quality criteria in the next section, we consider an example for the construction of a subdivision matrix $S$.

**Example 6.18.** *First, we note that all matrices in this example have been rounded to two decimal places. The matrices $L$, $R$, and $S$ have also been rounded up slightly so that small entries are not rounded down to zero.*

*As input for $t = 3$, we choose the adjacency matrix*

$$A := \begin{bmatrix} & 1 & & & 1 & 1 & & \\ 1 & & 1 & & & & 1 & \\ & 1 & & 1 & & & 1 & \\ & & 1 & & 1 & & & 1 \\ 1 & & & 1 & & 1 & & \\ 1 & & & & 1 & & & 1 \\ & 1 & 1 & & & & & 1 \\ & & & 1 & & 1 & 1 & \end{bmatrix}$$

*of the combinatorial graph $\boldsymbol{G}_K$. For this, we first construct the primal and the dual polytope and then create the control points $\tilde{P}$ by projecting the dual points onto the facets of the primal polytope and by inserting the volume point $o$ at $[0, 0, 0]$. Both steps are shown in Figure 6.32.*

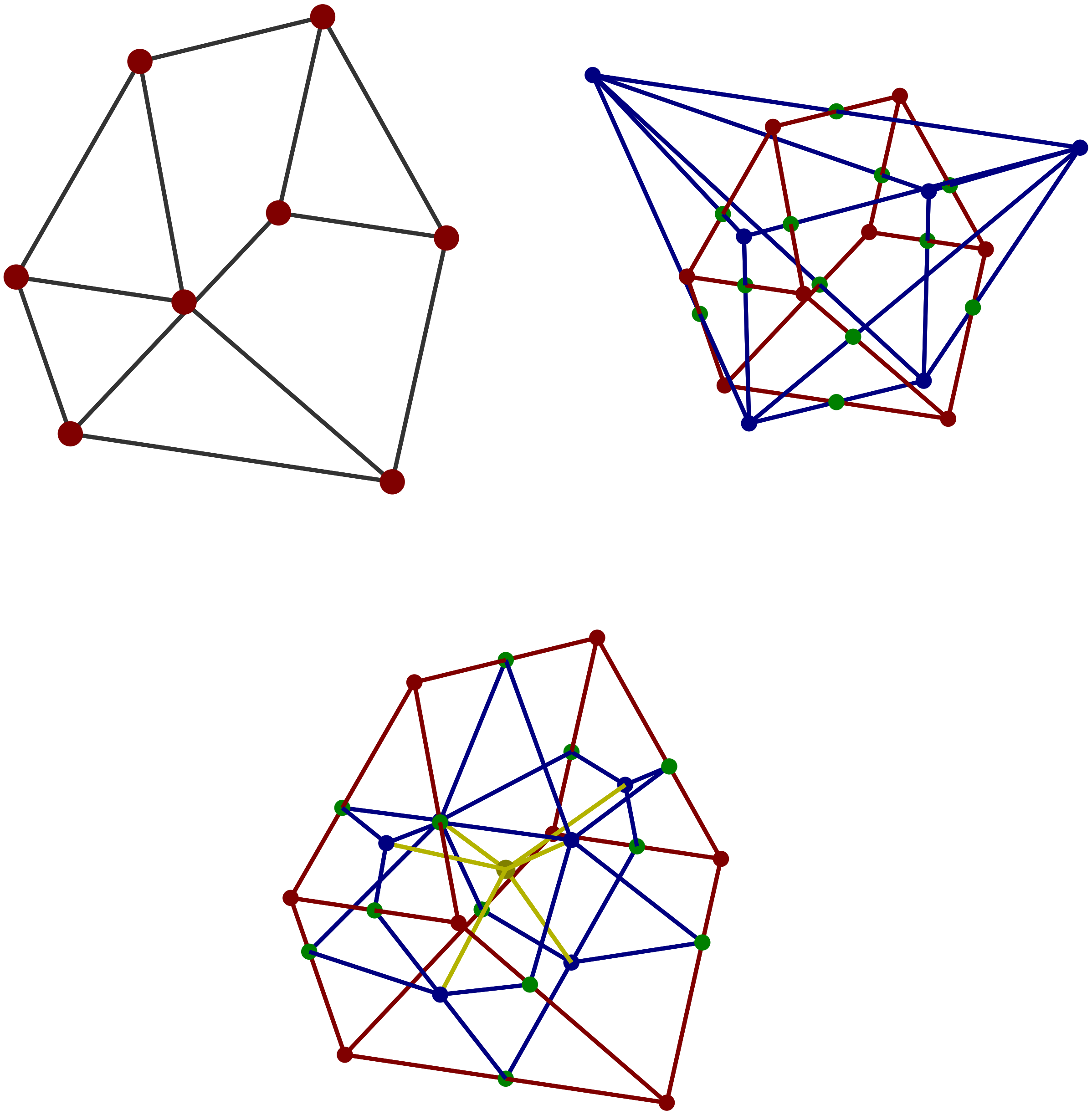

**Figure 6.32:** The primal polytope corresponding to the adjacency matrix $A$ from Example 6.18 (left), the primal and the associated dual polytope (middle), as well as the associated eigenstructure constructed according to Section 6.2.1 (right).

*With this eigenstructure, we obtain for the two matrices $\overline{M}$ and $\overline{B}$*

$$\overline{M} = \frac{1}{100} \begin{bmatrix}
-53 &&&&& &&& 71 & 41 & 41 &&&&& &&&&& &&&&& & \\
50 &&&&& &&& -11 &&&&& &&&&& &&&&& & 31 & 31 & \\
50 &&&&& &&&& -71 &&&&& &&&&& & 68 &&&& 53 && \\
50 &&&&& &&&&& -71 &&&&& &&&&& 68 &&&&& 53 & \\
&&&&& &&& 40 & 40 &&&&& &&&&& &&&&& -30 && 50 \\
&&&&& &&& 40 && 40 &&&&& &&&&& &&&&& -30 & 50 \\
&&&&& &&&& 143 & 143 &&&&& &&&&& -236 &&&&& & 50 \\
&&&&& &&&&& &&&&& &&&&& 38 &&&& 107 & 107 & -152 \\
& -53 &&&&& && 71 &&& 41 & 41 &&&&& &&&&& &&&& \\
& 50 &&&&& && -11 &&&&& &&&&& &&&&& & 31 & 31 & \\
& 50 &&&&& &&&&& -71 &&&&& &&&&& 68 &&& 53 && \\
& 50 &&&&& &&&&& & -71 &&&&& &&&& 68 &&&& 53 & \\
&&&&& &&& 40 &&& 40 &&&&& &&&&& &&& -30 && 50 \\
&&&&& &&& 40 &&&& 40 &&&&& &&&&& &&& -30 & 50 \\
&&&&& &&&&& & 143 & 143 &&&&& &&&& -236 &&&&& 50 \\
&&&&& &&&&& &&&&& &&&&& & 38 &&& 107 & 107 & -152 \\
&& -150 &&&&& &&&& 90 && 113 & 48 &&&&& &&&&& && \\
&& 50 &&&&& &&&& -71 &&&&& &&&&& 68 &&& 53 && \\
&& 50 &&&&& &&&&& && -155 &&&&& && 127 & 77 &&&& \\
&& 50 &&&&& &&&&& & -25 &&&&& &&&& 33 && 42 && \\
&&&&& &&&&& & 86 &&& 86 &&&&& && -122 &&&&& 50 \\
&&&&& &&&&& & 33 && 33 &&&&& &&&&& & -15 && 50 \\
&&&&& &&&&& &&& 38 & 38 &&&&& &&& -26 &&&& 50 \\
&&&&& &&&&& &&&&& &&&&& & 31 & 43 && 64 && -38 \\
&&& -5 &&&&& &&&&& 40 && 40 & 25 &&&&& &&&&& \\
&&& 50 &&&&& &&&&& -25 &&&&& &&&& 33 && 42 && \\
&&& 50 &&&&& &&&&& &&& -55 &&&&& & 52 & 52 &&& \\
&&& 50 &&&&& &&&&& && -25 &&&&& &&& 33 & 42 && \\
&&&&& &&&&& &&& 73 &&& 73 &&&&& & -96 &&&& 50 \\
&&&&& &&&&& &&& 53 && 53 &&&&& &&&& -56 && 50 \\
&&&&& &&&&& &&&&& 73 & 73 &&&&& && -96 &&& 50 \\
&&&&& &&&&& &&&&& &&&&& && 70 & 70 & 123 && -163 \\
&&&& -150 &&&&& 90 &&&&& & 113 && 48 &&&&& &&&& \\
&&&& 50 &&&&& -71 &&&&& &&&&& & 68 &&&& 53 && \\
&&&& 50 &&&&& &&&&& &&& -155 &&& 127 &&& 77 &&& \\
&&&& 50 &&&&& &&&&& & -25 &&&&& &&& 33 & 42 && \\
&&&&& &&&& 86 &&&&& &&& 86 &&& -122 &&&&& & 50 \\
&&&&& &&&& 33 &&&&& & 33 &&&&& &&&& -15 && 50 \\
&&&&& &&&&& &&&&& 38 && 38 &&&&& & -26 &&& 50 \\
&&&&& &&&&& &&&&& &&&&& 31 &&& 43 & 64 && -38 \\
&&&&& -150 &&&&& 90 &&&&& && 48 & 113 &&&&& &&& \\
&&&&& 50 &&&&& -71 &&&&& &&&&& 68 &&&&& 53 & \\
&&&&& 50 &&&&& &&&&& && -155 &&& 127 &&& 77 &&& \\
&&&&& 50 &&&&& &&&&& &&& -25 &&&&& 33 && 42 & \\
&&&&& &&&&& 86 &&&&& && 86 &&& -122 &&&&& & 50 \\
&&&&& &&&&& 33 &&&&& &&& 33 &&&&& && -15 & 50 \\
&&&&& &&&&& &&&&& && 38 & 38 &&&&& -26 &&& 50 \\
&&&&& &&&&& &&&&& &&&&& 31 &&& 43 && 64 & -38 \\
&&&&& & -150 &&&&& & 90 && 48 &&&&& 113 &&&&& && \\
&&&&& & 50 &&&&& & -71 &&&&& &&&& 68 &&&& 53 & \\
&&&&& & 50 &&&&& &&& -155 &&&&& && 127 & 77 &&&& \\
&&&&& & 50 &&&&& &&&&& &&& -25 &&& 33 &&& 42 & \\
&&&&& &&&&& && 86 && 86 &&&&& && -122 &&&&& 50 \\
&&&&& &&&&& && 33 &&&&& && 33 &&&&& & -15 & 50 \\
&&&&& &&&&& &&&& 38 &&&&& 38 &&& -26 &&&& 50 \\
&&&&& &&&&& &&&&& &&&&& & 31 & 43 &&& 64 & -38 \\
&&&&& && -5 &&&&& &&&& 25 && 40 & 40 &&&&& && \\
&&&&& && 50 &&&&& &&&& -55 &&&&& & 52 & 52 &&& \\
&&&&& && 50 &&&&& &&&&& && -25 &&& 33 &&& 42 & \\
&&&&& && 50 &&&&& &&&&& & -25 &&&&& 33 && 42 & \\
&&&&& &&&&& &&&&& & 73 &&& 73 &&& -96 &&&& 50 \\
&&&&& &&&&& &&&&& & 73 && 73 &&&&& -96 &&& 50 \\
&&&&& &&&&& &&&&& &&& 53 & 53 &&&&& & -56 & 50 \\
&&&&& &&&&& &&&&& &&&&& && 70 & 70 && 123 & -163
\end{bmatrix}$$

*and* $100 \cdot \overline{B} =$

$$\begin{bmatrix}
100 &&&&& &&&&& &&&&& &&&&& &&&&& &&&&& &&&&& &&&&& &&&&& &&&&& &&&&& &&&&& &&& \\
&&&&& &&& 100 &&&&& &&&&& &&&&& &&&&& &&&&& &&&&& &&&&& &&&&& &&&&& &&&&& &&&&& \\
&&&&& &&&&& &&&&& & 100 &&&&& &&&&& &&&&& &&&&& &&&&& &&&&& &&&&& &&&&& &&&&& && \\
&&&&& &&&&& &&&&& &&&&& &&&& 100 &&&&& &&&&& &&&&& &&&&& &&&&& &&&&& &&&&& &&&& \\
&&&&& &&&&& &&&&& &&&&& &&&&& &&&&& && 100 &&&&& &&&&& &&&&& &&&&& &&&&& &&&&& & \\
&&&&& &&&&& &&&&& &&&&& &&&&& &&&&& &&&&& &&&&& 100 &&&&& &&&&& &&&&& &&&&& &&& \\
&&&&& &&&&& &&&&& &&&&& &&&&& &&&&& &&&&& &&&&& &&&&& &&& 100 &&&&& &&&&& &&&&& \\
&&&&& &&&&& &&&&& &&&&& &&&&& &&&&& &&&&& &&&&& &&&&& &&&&& &&&&& & 100 &&&&& && \\
& 50 &&&&& &&& 50 &&&&& &&&&& &&&&& &&&&& &&&&& &&&&& &&&&& &&&&& &&&&& &&&&& &&&& \\
&& 42 &&&&& &&&&& &&&&& &&&&& &&&&& &&&&& & 58 &&&&& &&&&& &&&&& &&&&& &&&&& &&&&& \\
&&& 42 &&&&& &&&&& &&&&& &&&&& &&&&& &&&&& &&&&& &&& 58 &&&&& &&&&& &&&&& &&&&& && \\
&&&&& &&&&& 42 &&&&& && 58 &&&&& &&&&& &&&&& &&&&& &&&&& &&&&& &&&&& &&&&& &&&&& & \\
&&&&& &&&&& & 42 &&&&& &&&&& &&&&& &&&&& &&&&& &&&&& &&&&& &&& 58 &&&&& &&&&& &&&& \\
&&&&& &&&&& &&&&& &&&& 66 &&&&& & 34 &&&&& &&&&& &&&&& &&&&& &&&&& &&&&& &&&&& &&& \\
&&&&& &&&&& &&&&& &&& 50 &&&&& &&&&& &&&&& &&&&& &&&&& &&&&& && 50 &&&&& &&&&& &&& \\
&&&&& &&&&& &&&&& &&&&& &&&&& && 34 &&&&& &&& 66 &&&&& &&&&& &&&&& &&&&& &&&&& &&& \\
&&&&& &&&&& &&&&& &&&&& &&&&& & 50 &&&&& &&&&& &&&&& &&&&& &&&&& &&&&& & 50 &&&&& & \\
&&&&& &&&&& &&&&& &&&&& &&&&& &&&&& &&&& 50 &&&&& &&& 50 &&&&& &&&&& &&&&& &&&&& & \\
&&&&& &&&&& &&&&& &&&&& &&&&& &&&&& &&&&& &&&&& &&& 66 &&&&& &&&&& &&&&& & 34 &&&& \\
&&&&& &&&&& &&&&& &&&&& &&&&& &&&&& &&&&& &&&&& &&&&& &&&&& & 66 &&&&& && 34 &&&&& \\
&&&&& & 29 &&&&& &&&&& &&&&& &&&&& &&&&& &&&&& 35 &&&&& &&& 35 &&&&& &&&&& &&&&& &&&& \\
&&&&& &&&&& &&&& 29 &&&&& & 35 &&&&& &&&&& &&&&& &&&&& &&&&& &&&&& && 35 &&&&& &&&&& & \\
&&&&& &&&&& &&&&& &&&&& && 25 &&&&& & 25 &&&&& &&&&& &&&&& &&&&& &&&&& & 25 &&&&& & 25 &&& \\
&&&&& &&&&& &&&&& &&&&& &&&&& &&&&& 25 &&&&& &&& 25 &&&&& &&& 25 &&&&& &&&&& &&&&& 25 && \\
&&&& 21 &&&&& &&& 21 &&&&& &&&& 18 &&&&& &&& 22 &&&&& &&& 18 &&&&& &&&&& &&&&& &&&&& &&&&& & \\
&&&&& 21 &&&&& &&& 21 &&&&& &&&&& &&&&& &&&&& &&&&& && 18 &&&&& &&& 18 &&&&& &&&& 22 & \\
&&&&& && 16 &&&&& &&& 16 &&&&& &&& 10 &&&&& &&& 15 &&&&& &&& 10 &&&&& &&& 10 &&&&& &&& 10 &&&&& &&& 15
\end{bmatrix}.$$

*In the matrix $\overline{M}$, the entries of $1/2$ that we proved in Lemma 6.10 are clearly visible. Multiplying the two matrices, we*

*obtain*

$$
\tilde{C} = \frac{1}{100} \left[ \begin{array}{cccccccc|cccccccccccc|cccccc|c}
-53 &&&&&&&& 71 & 41 & 41 &&&&&&&&&&&&&&&& \\
& -53 &&&&&&& 71 &&& 41 & 41 &&&&&&&&&&&&&& \\
&& -150 &&&&&&&&& 90 && 113 & 48 &&&&&&&&&&&& \\
&&& -5 &&&&&&&&&& 40 && 40 & 25 &&&&&&&&&& \\
&&&& -150 &&&&& 90 &&&&&& 113 && 48 &&&&&&&&& \\
&&&&& -150 &&&&& 90 &&&&&&& 48 & 113 &&&&&&&& \\
&&&&&& -150 &&&&&& 90 && 48 &&&&& 113 &&&&&&& \\
&&&&&&& -5 &&&&&&&&& 25 && 40 & 40 &&&&&&& \\
\hline
25 & 25 &&&&&&& -11 &&&&&&&&&&&&&&&& 31 & 31 & \\
21 &&&& 29 &&&&& -71 &&&&&&&&&&& 68 &&&& 53 && \\
21 &&&&& 29 &&&&& -71 &&&&&&&&&& 68 &&&&& 53 & \\
& 21 & 29 &&&&&&&&& -71 &&&&&&&&&& 68 &&& 53 && \\
& 21 &&&&& 29 &&&&&& -71 &&&&&&&&& 68 &&&& 53 & \\
&& 33 & 17 &&&&&&&&&& -25 &&&&&&&&& 33 && 42 && \\
&& 25 &&&& 25 &&&&&&&& -155 &&&&&&& 127 & 77 &&&& \\
&&& 17 & 33 &&&&&&&&&&& -25 &&&&&&&& 33 & 42 && \\
&&& 25 &&&& 25 &&&&&&&&& -55 &&&&&& 52 & 52 &&& \\
&&&& 25 & 25 &&&&&&&&&&&& -155 &&& 127 &&& 77 &&& \\
&&&&& 33 && 17 &&&&&&&&&&& -25 &&&&& 33 && 42 & \\
&&&&&& 33 & 17 &&&&&&&&&&&& -25 &&& 33 &&& 42 & \\
\hline
&&&&&&&&& 72 & 72 &&&&&&& 61 &&& -156 &&&&&& 50 \\
&&&&&&&&&&& 72 & 72 && 61 &&&&&&& -156 &&&&& 50 \\
&&&&&&&&&&&&& 28 & 19 && 36 &&& 28 &&& -61 &&&& 50 \\
&&&&&&&&&&&&&&& 28 & 36 & 19 & 28 &&&&& -61 &&& 50 \\
&&&&&&&& 17 & 14 && 14 && 17 && 17 &&&&&&&&& -30 && 50 \\
&&&&&&&& 17 && 14 && 14 &&&&&& 17 & 17 &&&&&& -30 & 50 \\
\hline
&&&&&&&&&&&&&&&&&&&& 12 & 12 & 29 & 29 & 64 & 64 & -110
\end{array} \right] .
$$

*We multiply this matrix by 2 and then split it into the matrices $\tilde{L}$ and $\tilde{R}$. We obtain*

$$
\tilde{L} = \frac{1}{100} \left[ \begin{array}{cccccccc|cccccccccccc|cccccc|c}
100 &&&&&&&&&&&&&&&&&&&&&&&&&& \\
& 100 &&&&&&&&&&&&&&&&&&&&&&&&& \\
&& 100 &&&&&&&&&&&&&&&&&&&&&&&& \\
&&& 100 &&&&&&&&&&&&&&&&&&&&&&& \\
&&&& 100 &&&&&&&&&&&&&&&&&&&&&& \\
&&&&& 100 &&&&&&&&&&&&&&&&&&&&& \\
&&&&&& 100 &&&&&&&&&&&&&&&&&&&& \\
&&&&&&& 100 &&&&&&&&&&&&&&&&&&& \\
\hline
50 & 50 &&&&&&&&&&&&&&&&&&&&&&&&& \\
42 &&&& 58 &&&&&&&&&&&&&&&&&&&&&& \\
42 &&&&& 58 &&&&&&&&&&&&&&&&&&&&& \\
& 42 & 58 &&&&&&&&&&&&&&&&&&&&&&&& \\
& 42 &&&&& 58 &&&&&&&&&&&&&&&&&&&& \\
&& 66 & 34 &&&&&&&&&&&&&&&&&&&&&&& \\
&& 50 &&&& 50 &&&&&&&&&&&&&&&&&&&& \\
&&& 34 & 66 &&&&&&&&&&&&&&&&&&&&&& \\
&&& 50 &&&& 50 &&&&&&&&&&&&&&&&&&& \\
&&&& 50 & 50 &&&&&&&&&&&&&&&&&&&&& \\
&&&&& 66 && 34 &&&&&&&&&&&&&&&&&&& \\
&&&&&& 66 & 34 &&&&&&&&&&&&&&&&&&& \\
\hline
&&&&&&&&& 145 & 145 &&&&&&& 121 &&& -311 &&&&&& \\
&&&&&&&&&&& 145 & 145 && 121 &&&&&&& -311 &&&&& \\
&&&&&&&&&&&&& 55 & 38 && 73 &&& 55 &&& -122 &&&& \\
&&&&&&&&&&&&&&& 55 & 73 & 38 & 55 &&&&& -122 &&& \\
&&&&&&&& 33 & 29 && 29 && 35 && 35 &&&&&&&&& -60 && \\
&&&&&&&& 33 && 29 && 29 &&&&&& 35 & 35 &&&&&& -60 & \\
\hline
&&&&&&&&&&&&&&&&&&&& 24 & 24 & 58 & 58 & 128 & 128 & -320
\end{array} \right]
$$

*and*

$$
\tilde{R} = \frac{1}{100} \left[ \begin{array}{cccccccc|cccccccccccc|cccccc|c}
-206 &&&&&&&& 142 & 82 & 82 &&&&&&&&&&&&&&&& \\
& -206 &&&&&&& 142 &&& 82 & 82 &&&&&&&&&&&&&& \\
&& -401 &&&&&&&&& 180 && 225 & 96 &&&&&&&&&&&& \\
&&& -109 &&&&&&&&&& 80 && 80 & 50 &&&&&&&&&& \\
&&&& -401 &&&&& 180 &&&&&& 225 && 96 &&&&&&&&& \\
&&&&& -401 &&&&& 180 &&&&&&& 96 & 225 &&&&&&&& \\
&&&&&& -401 &&&&&& 180 && 96 &&&&& 225 &&&&&&& \\
&&&&&&& -109 &&&&&&&&& 50 && 80 & 80 &&&&&&& \\
\hline
&&&&&&&& -23 &&&&&&&&&&&&&&&& 61 & 61 & \\
&&&&&&&&& -142 &&&&&&&&&&& 136 &&&& 106 && \\
&&&&&&&&&& -142 &&&&&&&&&& 136 &&&&& 106 & \\
&&&&&&&&&&& -142 &&&&&&&&&& 136 &&& 106 && \\
&&&&&&&&&&&& -142 &&&&&&&&& 136 &&&& 106 & \\
&&&&&&&&&&&&& -50 &&&&&&&&& 66 && 84 && \\
&&&&&&&&&&&&&& -309 &&&&&&& 255 & 155 &&&& \\
&&&&&&&&&&&&&&& -50 &&&&&&&& 66 & 84 && \\
&&&&&&&&&&&&&&&& -109 &&&&&& 105 & 105 &&& \\
&&&&&&&&&&&&&&&&& -309 &&& 255 &&& 155 &&& \\
&&&&&&&&&&&&&&&&&& -50 &&&&& 66 && 84 & \\
&&&&&&&&&&&&&&&&&&& -50 &&& 66 &&& 84 & \\
\hline
&&&&&&&&&&&&&&&&&&&&&&&&&& 100 \\
&&&&&&&&&&&&&&&&&&&&&&&&&& 100 \\
&&&&&&&&&&&&&&&&&&&&&&&&&& 100 \\
&&&&&&&&&&&&&&&&&&&&&&&&&& 100 \\
&&&&&&&&&&&&&&&&&&&&&&&&&& 100 \\
&&&&&&&&&&&&&&&&&&&&&&&&&& 100 \\
\hline
&&&&&&&&&&&&&&&&&&&&&&&&&& 100
\end{array} \right] .
$$

*From the two matrices, we recognize the structure that we showed in Lemma 6.14. In the next step, we compute their*

*exponentials and obtain*

$$L = \frac{1}{100}\left[\begin{array}{cccccccc|cccccccccccc|cccccc|c}
100&&&&&&& & &&&&&&&&&&& & &&&&& & \\
&100&&&&&& & &&&&&&&&&&& & &&&&& & \\
&&100&&&&& & &&&&&&&&&&& & &&&&& & \\
&&&100&&&& & &&&&&&&&&&& & &&&&& & \\
&&&&100&&& & &&&&&&&&&&& & &&&&& & \\
&&&&&100&& & &&&&&&&&&&& & &&&&& & \\
&&&&&&100& & &&&&&&&&&&& & &&&&& & \\
&&&&&&&100 & &&&&&&&&&&& & &&&&& & \\
\hline
25&25&&&&&& & 50&&&&&&&&&&& & &&&&& & \\
21&&&&30&&& & &50&&&&&&&&&& & &&&&& & \\
21&&&&&30&& & &&50&&&&&&&&& & &&&&& & \\
&21&30&&&&& & &&&50&&&&&&&& & &&&&& & \\
&21&&&&&30& & &&&&50&&&&&&& & &&&&& & \\
&&33&18&&&& & &&&&&50&&&&&& & &&&&& & \\
&&25&&&&26& & &&&&&&50&&&&& & &&&&& & \\
&&&18&33&&& & &&&&&&&50&&&& & &&&&& & \\
&&&25&&&&25 & &&&&&&&&50&&& & &&&&& & \\
&&&&26&26&& & &&&&&&&&&50&& & &&&&& & \\
&&&&&33&&18 & &&&&&&&&&&50& & &&&&& & \\
&&&&&&33&18 & &&&&&&&&&&&50 & &&&&& & \\
\hline
11&&&&13&13&& & &21&21&&&&&&&18&& & 6&&&&& & \\
&11&13&&&&13& & &&&21&21&&18&&&&& & &6&&&& & \\
&&7&7&&&7&7 & &&&&&13&9&&18&&&13 & &&22&&& & \\
&&&7&7&7&&7 & &&&&&&&13&18&9&13& & &&&22&& & \\
4&4&6&4&6&&& & 10&9&&9&&10&&10&&&& & &&&&33& & \\
4&4&&&&6&6&4 & 10&&9&&9&&&&&&10&10 & &&&&&33 & \\
\hline
2&2&2&2&2&2&2&2 & 5&4&4&4&4&5&3&5&5&3&5&5 & 1&1&5&5&14&14 & 6
\end{array}\right]$$

*and*

$$R = \frac{1}{100}\left[\begin{array}{cccccccc|cccccccccccc|cccccc|c}
12&&&&&&& & 24&9&9&&&&&&&&& & 13&&&&12&12 & 12 \\
&12&&&&&& & 24&&&9&9&&&&&&& & &13&&&12&12 & 12 \\
&&4&&&&& & &&&11&&21&3&&&&& & &17&12&&18& & 18 \\
&&&24&&&& & &&&&&16&&16&9&&& & &&9&9&12& & 9 \\
&&&&4&&& & &11&&&&&&21&&3&& & 17&&&12&18& & 18 \\
&&&&&4&& & &&11&&&&&&&3&21& & 17&&&12&&18 & 18 \\
&&&&&&4& & &&&&11&&3&&&&&21 & &17&12&&&18 & 18 \\
&&&&&&&24 & &&&&&&&&9&&16&16 & &&9&9&&12 & 9 \\
\hline
&&&&&&& & 43&&&&&&&&&&& & &&&&20&20 & 18 \\
&&&&&&& & &19&&&&&&&&&& & 30&&&&24& & 28 \\
&&&&&&& & &&19&&&&&&&&& & 30&&&&&24 & 28 \\
&&&&&&& & &&&19&&&&&&&& & &30&&&24& & 28 \\
&&&&&&& & &&&&19&&&&&&& & &30&&&&24 & 28 \\
&&&&&&& & &&&&&36&&&&&& & &&20&&25& & 21 \\
&&&&&&& & &&&&&&6&&&&& & &37&23&&& & 36 \\
&&&&&&& & &&&&&&&36&&&& & &&&20&25& & 21 \\
&&&&&&& & &&&&&&&&24&&& & &&26&26&& & 26 \\
&&&&&&& & &&&&&&&&&6&& & 37&&&23&& & 36 \\
&&&&&&& & &&&&&&&&&&36& & &&&20&&25 & 21 \\
&&&&&&& & &&&&&&&&&&&36 & &&20&&&25 & 21 \\
\hline
&&&&&&& & &&&&&&&&&&& & 50&&&&& & 50 \\
&&&&&&& & &&&&&&&&&&& & &50&&&& & 50 \\
&&&&&&& & &&&&&&&&&&& & &&50&&& & 50 \\
&&&&&&& & &&&&&&&&&&& & &&&50&& & 50 \\
&&&&&&& & &&&&&&&&&&& & &&&&50& & 50 \\
&&&&&&& & &&&&&&&&&&& & &&&&&50 & 50 \\
\hline
&&&&&&& & &&&&&&&&&&& & &&&&& & 100
\end{array}\right].$$

*Multiplying the two matrices $L$ and $R$ together, we obtain the subdivision matrix*

$$S = \frac{1}{100}\left[\begin{array}{cccccccc|cccccccccccc|cccccc|c}
12&&&&&&& & 24&9&9&&&&&&&&& & 13&&&&12&12 & 12 \\
&12&&&&&& & 24&&&9&9&&&&&&& & &13&&&12&12 & 12 \\
&&4&&&&& & &&&11&&21&3&&&&& & &17&12&&18& & 18 \\
&&&24&&&& & &&&&&16&&16&9&&& & &&9&9&12& & 9 \\
&&&&4&&& & &11&&&&&&21&&3&& & 17&&&12&18& & 18 \\
&&&&&4&& & &&11&&&&&&&3&21& & 17&&&12&&18 & 18 \\
&&&&&&4& & &&&&11&&3&&&&&21 & &17&12&&&18 & 18 \\
&&&&&&&24 & &&&&&&&&9&&16&16 & &&9&9&&12 & 9 \\
\hline
3&3&&&&&& & 34&3&3&3&3&&&&&&& & 4&4&&&16&16 & 15 \\
3&&&&1&&& & 5&15&2&&&&&7&&1&& & 23&&&4&20&3 & 22 \\
3&&&&&1&& & 5&2&15&&&&&&&1&7& & 23&&&4&3&20 & 22 \\
&3&1&&&&& & 5&&&15&2&7&1&&&&& & &23&4&&20&3 & 22 \\
&3&&&&&1& & 5&&&2&15&&1&&&&&7 & &23&4&&3&20 & 22 \\
&&2&5&&&& & &&&4&&28&1&3&2&&& & &6&15&2&20& & 18 \\
&&1&&&&1& & &&&3&3&6&5&&&&&6 & &27&17&&5&5 & 27 \\
&&&5&2&&& & &4&&&&3&&28&2&1&& & 6&&2&15&20& & 18 \\
&&&6&&&&6 & &&&&&4&&4&16&&4&4 & &&17&17&3&3 & 18 \\
&&&&1&1&& & &3&3&&&&&6&&5&6& & 27&&&17&5&5 & 27 \\
&&&&&2&&5 & &&4&&&&&&2&1&28&3 & 6&&2&15&&20 & 18 \\
&&&&&&2&5 & &&&&4&&1&&2&&3&28 & &6&15&2&&20 & 18 \\
\hline
2&&&&1&1&& & 3&7&7&&&&&3&&2&3& & 28&&&7&9&9 & 27 \\
&2&1&&&&1& & 3&&&7&7&3&2&&&&&3 & &28&7&&9&9 & 27 \\
&&1&2&&&1&2 & &&&1&1&8&1&2&6&&2&8 & &6&25&6&6&6 & 28 \\
&&&2&1&1&&2 & &1&1&&&2&&8&6&1&8&2 & 6&&6&25&6&6 & 28 \\
1&1&1&1&1&&& & 6&3&1&3&1&6&1&6&1&1&& & 4&4&3&3&31&3 & 30 \\
1&1&&&&1&1&1 & 6&1&3&1&3&&1&&1&1&6&6 & 4&4&3&3&3&31 & 30 \\
\hline
1&1&1&1&1&1&1&1 & 3&1&1&1&1&3&1&3&2&1&3&3 & 4&4&7&7&13&13 & 37
\end{array}\right],$$

*where the sparse structure required for the quality criterion Q11 can be clearly seen. An illustration of the double eigenstructure of the associated matrix $\overline{S}$ can be found in Figure 6.33.*

We have thus covered all aspects of the construction and will discuss in the next section the quality criteria with respect to the matrix $S$ constructed in this way.

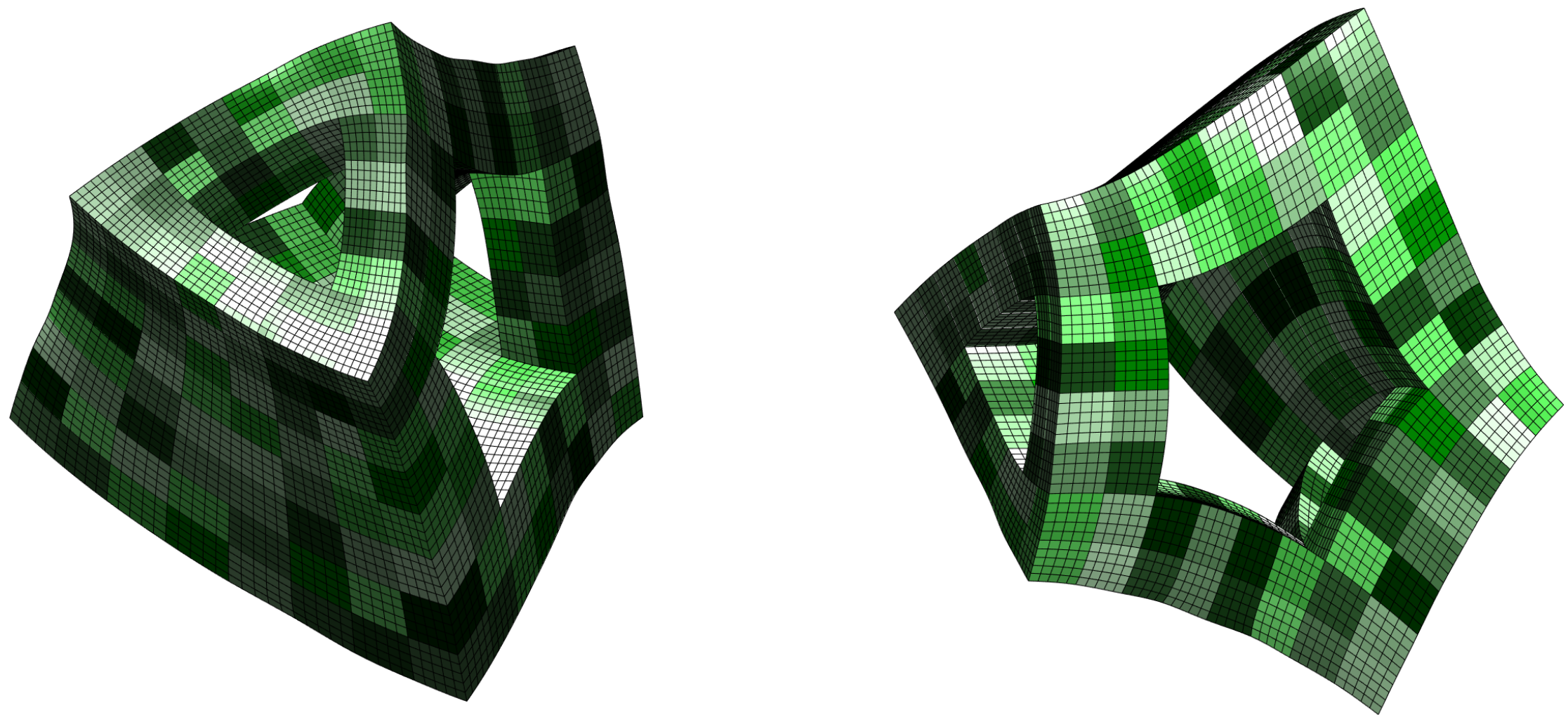

**Figure 6.33:** Evaluation of the double shell of the eigenstructure of the matrix $\overline{S}$, which was generated from the adjacency matrix of Example 6.18, shown here from two different perspectives.

## 6.3 Proofs and Quality Criteria of the Subdivision Matrices

As the title of this chapter already indicates, the generalized cubic B-spline subdivision matrices constructed here are of an experimental nature, since only limited statements can be made about these constructions based on some quality criteria. Some statements are within reach but would still need to be worked out in detail. Therefore, in the course of this section, we will describe which statements we have proven, which statements can basically be proven, and which statements are still completely open. We thus leave some room here for further research projects.

First, analogous to Chapter 5, we classify the variants of the subdivision matrices into the following four categories:

**Definition 6.19.** *We name the four categories of subdivision matrices for $g = 3$ using two different notations:*

- *Matrices $S$ represent subdivision matrices for initial elements, whereas matrices $\overline{S}$ represent matrices for double rings and double shells.*
- *Matrices $S^{(2)}$ are subdivision matrices of type 2, whereas matrices $S^{(3)}$ are subdivision matrices of type 3.*

*In total, this gives us four different categories of subdivision matrices, which are also shown in Table 6.1.*

Table 6.1 shows the current status of the quality criteria. We use the color coding defined in Section 5.3 again, resulting in the following color scheme:

| | | | |
|---|---|---|---|
| dark green | proven and empirically verified | green | proven |
| yellow | not proven but empirically verified | orange | unclear |
| red | disproven or counterexample constructed | gray | not applicable |

We also accompany the results with empirical tests. Since in the construction of Section 6.2 for $t = 3$ we explicitly assumed combinatorial graphs arising from a structure of hexahedra, we use for the empirical analysis only structures consisting of hexahedra. According to Section 3.1.1, there are exactly 186,506 such examples, of which 158,527 are unique examples and 27,979 are isomorphisms of already existing structures. For $t = 2$ we again empirically consider the subdivision matrices for (ir)regularities with valence less than or equal to 100.

For the proofs, we use the notation for variables and matrices from Section 6.2 without reintroducing them. These are clearly recognizable in the respective algorithms.

| | | $S^{(2)}$ | $\overline{S}^{(2)}$ | $S^{(3)}$ | $\overline{S}^{(3)}$ |
|---|---|---|---|---|---|
| Q1 | Affine invariance | | | | |
| Q2 | Convex hull property | | | | |
| Q3 | Dominant eigenvalue 1 | | | | |
| Q4 | $t$ subdominant eigenvalues | | | | |
| Q5 | $t$ subdominant eigenvalues $0.5$ | | | | |
| Q6 | Subsubdominant eigenvalue $0.25$ | | | | |
| Q7 | Regular case | | | | |
| Q8 | Semi-regular case | | | | |
| Q9 | Subdivision matrix for every input | | | | |
| Q10 | Symmetry | | | | |
| Q11 | Proper support | | | | |
| Q12 | Self-intersection-free structure | | | | |
| Q13 | Injective and regular characteristic map | | | | |
| | stochastic | | | | |
| | irreducible | | | | |

**Table 6.1:** Overview of quality criteria and other properties of the individual subdivision matrices from Definition 6.19.

## Q1 Affine Invariance

We start with the investigation of the partition of unity with the following theorem:

**Theorem 6.20.** *The rows of the subdivision matrices of all four categories sum to* $1$*, i.e., they satisfy quality criterion Q1.*

*Proof.* By Lemma 6.15 for $t = 2$ and Lemma 6.14 for $t = 3$, the rows of the two triangular matrices $\tilde{L}$ and $\tilde{R}$ sum to $1$. Hence, both matrices have an eigenvector $\vec{1}$ corresponding to the eigenvalue $1$, by Lemma 2.36.

The two matrices

$$\ln(2)\left(\tilde{L} - E_n\right) \quad \text{and} \quad \ln(2)\left(\tilde{R} - E_n\right)$$

are also triangular matrices. Let $J$ be such that $\tilde{L} = VJV^{-1}$ is the Jordan decomposition of $\tilde{L}$. Then

$$\tilde{L} - E_n = VJV^{-1} - VE_nV^{-1} = V(J - E_n)V^{-1}.$$

Thus, $J - E_n$ is the Jordan decomposition of $\tilde{L} - E_n$, which has eigenvalue $0$ with eigenvector $\vec{1}$. Moreover,

$$\ln(2)(\tilde{L} - E_n)\vec{1} = \ln(2) \cdot 0 \cdot \vec{1} = 0 \cdot \vec{1},$$

so the matrix $\ln(2)(\tilde{L} - E_n)$ also has eigenvalue $0$ with eigenvector $\vec{1}$. Using Lemma B.59 and Lemma B.60, which describe eigenvalues and eigenvectors of exponentials of triangular matrices, we obtain

$$L \cdot \vec{1} = \exp\left(\ln(2)(\tilde{L} - E_n)\right)\vec{1} = \exp(0) \cdot \vec{1} = 1 \cdot \vec{1}.$$

Therefore, $L$ has eigenvalue $1$ with eigenvector $\vec{1}$. The same holds analogously for $R$.

Consequently, for $S^{(2)}$ and $S^{(3)}$ we have

$$S^{(t)} \cdot \vec{1} = (LR) \cdot \vec{1} = L \cdot \vec{1} = \vec{1} \quad \text{for} \quad t \in \{2, 3\},$$

so the matrices $S^{(2)}$ and $S^{(3)}$ have eigenvalue $1$ with eigenvector $\vec{1}$. By Lemma 2.36, this implies the rows of $S^{(2)}$ and $S^{(3)}$ sum to $1$, fulfilling quality criterion Q1.

Since the rows of $\overline{S}$ consist only of entries from the subdivision matrices $S$ of the initial elements padded with zeros, the affine invariance shown for the subdivision matrices $S$ carries over to the subdivision matrices $\overline{S}$. Thus, they also satisfy quality criterion Q1. □

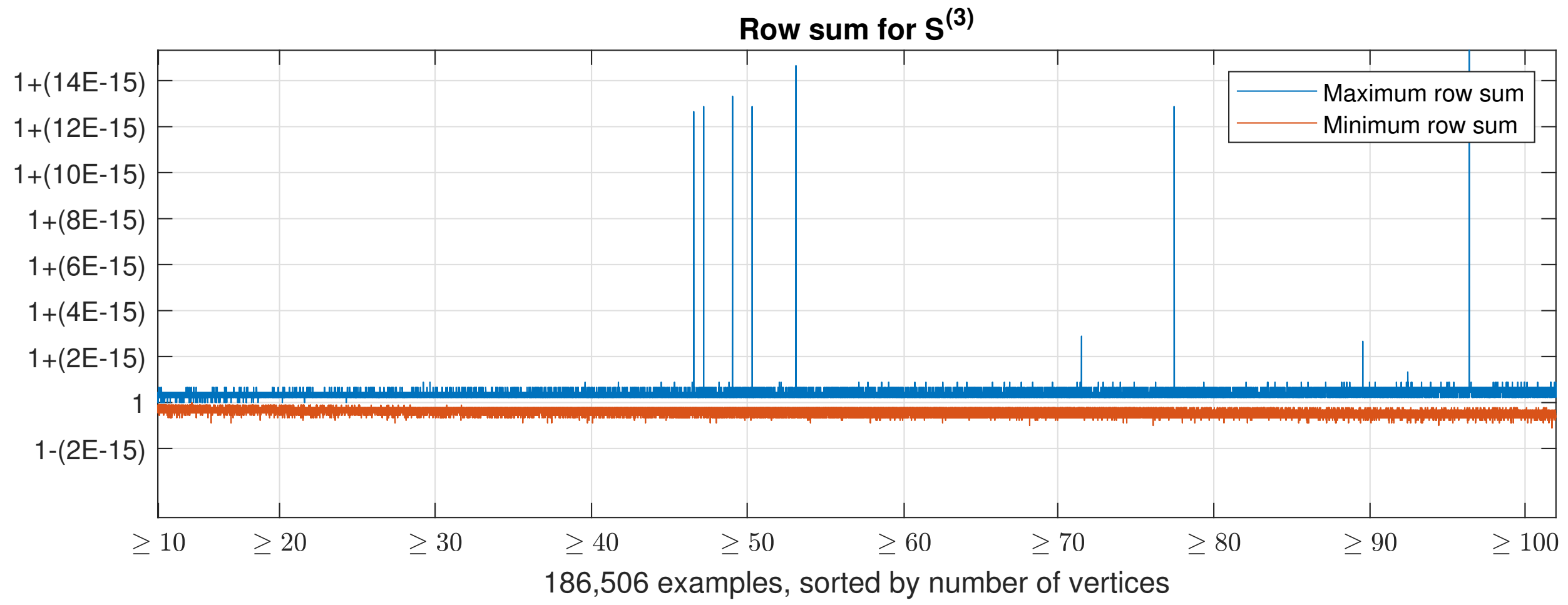

**Figure 6.34:** Maximum and minimum row sums of the subdivision matrices of the three-dimensional initial elements.

This result is also confirmed empirically. The maximum and minimum row sums of the 186,506 examples for the subdivision matrices of the initial elements are shown in Figure 6.34. The respective maximum values of all four categories are found in the following table:

| | $S^{(2)}$ | $\overline{S}^{(2)}$ | $S^{(3)}$ | $\overline{S}^{(3)}$ |
|---|---|---|---|---|
| $\max \sum -1$ | $1{,}7986\text{e}{-14}$ | $1{,}7986\text{e}{-14}$ | $1{,}5321\text{e}{-14}$ | $1{,}5321\text{e}{-14}$ |
| $\min \sum -1$ | $-8{,}8818\text{e}{-16}$ | $-8{,}8818\text{e}{-16}$ | $-1{,}1102\text{e}{-15}$ | $-1{,}1102\text{e}{-15}$ |

Overall, it can be observed that the numerical results yield values close to machine precision, thus confirming that Quality Criterion Q1 is also empirically satisfied.

## Q2 Convex Hull

For the convex hull property, we obtain a similar result with the following theorem:

**Theorem 6.21.** *All entries of the subdivision matrices $S$ of all four categories are non-negative. Thus, together with Quality Criterion Q1, they satisfy Quality Criterion Q2 of the convex hull property.*

*Proof.* By Theorem 6.20, all four categories of $S$ satisfy Quality Criterion Q1. By Theorem 6.13, the matrix $\tilde{C}$ fulfills property E4, meaning all off-diagonal entries of $\tilde{C}$ are non-negative. Since the off-diagonal entries of $\tilde{L}$ and $\tilde{R}$ are either entries from $\tilde{C}$ or zeros, all off-diagonal entries of $\tilde{L}$ and $\tilde{R}$ are also non-negative. The same holds for the two matrices

$$\ln(2)\left(\tilde{L} - E_n\right) \quad \text{and} \quad \ln(2)\left(\tilde{R} - E_n\right).$$

By Lemma B.63, which states the non-negativity of the exponential of matrices with non-negative off-diagonal entries, it follows that all entries of the matrices

$$L = \exp\left(\ln(2)\left(\tilde{L} - E_n\right)\right) \quad \text{and} \quad R = \exp\left(\ln(2)\left(\tilde{R} - E_n\right)\right)$$

are non-negative. Since

$$S^{(t)} = LR \quad \text{with} \quad t \in \{2, 3\},$$

all entries of $S^{(t)}$ are non-negative because the product of two non-negative matrices is again non-negative. Thus, the matrices $S^{(2)}$ and $S^{(3)}$ satisfy Quality Criterion Q2.

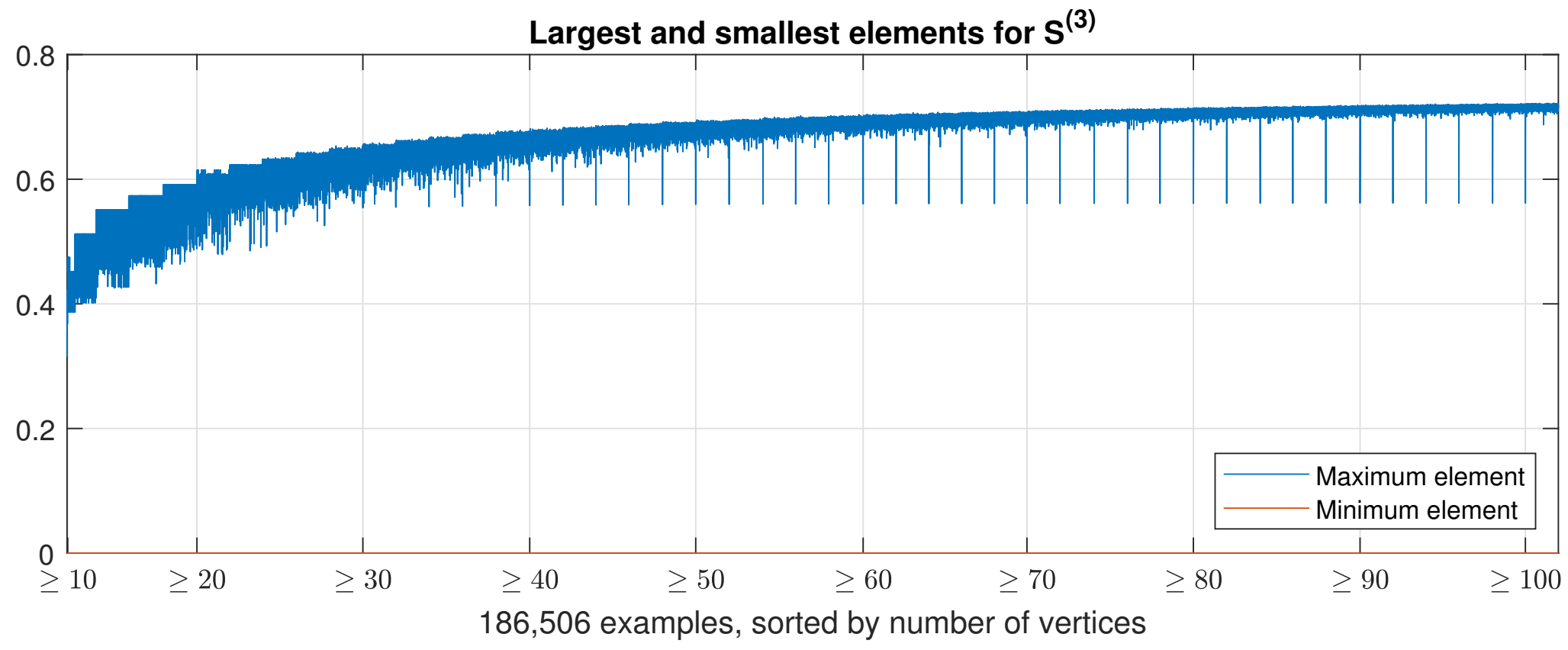

**Figure 6.35:** Maximum and minimum elements of the subdivision matrices of the three-dimensional initial elements.

Analogous to Theorem 6.20, this non-negativity property transfers from the matrices of the initial elements $S$ to the matrices $\overline{S}$, since the rows of $\overline{S}$ consist of rows of subdivision matrices of the initial elements padded with zeros. Therefore, Quality Criterion Q2 is also satisfied for the matrices $\overline{S}^{(2)}$ and $\overline{S}^{(3)}$. □

From the affine invariance of the previous section and the non-negativity shown here, the following corollary directly follows:

**Corollary 6.22.** *All four categories of the subdivision matrices are stochastic.*

The non-negativity can also be empirically verified. The maximum and minimum entries of the example matrices for the case $S^{(3)}$ are shown in Figure 6.35. The maximum entries for this category tend to increase with the number of vertices. However, this trend is interrupted by outliers belonging to the semi-regular cases. Thus, this trend appears to be related to the (random) selection of examples and cannot be generalized in principle.

The maximum and minimum elements of all tested examples across the four categories are summarized in the following table:

| | $S^{(2)}$ | $\overline{S}^{(2)}$ | $S^{(3)}$ | $\overline{S}^{(3)}$ |
|---|---|---|---|---|
| $\max(\max(S))$ | 0,749 75 | 0,749 75 | 0,722 31 | 0,722 31 |
| $\min(\min(S))$ | 0 | 0 | 0 | 0 |

The quality criterion Q2 can thus also be empirically verified.

## Q11 Appropriate Support of the Refinement Rules

In contrast to the generalized quadratic B-spline subdivision, this quality criterion is not satisfied by construction and must therefore be proven. We begin with the two-dimensional case, which we consider in the following theorem:

**Theorem 6.23.** *The subdivision matrix $S^{(2)} \in \mathbb{R}^{n\times n}$ satisfies quality criterion Q11, i.e., the refinement rules have an appropriate support.*

*Proof.* We first define $\tilde{P} := \left[P^T, E^T, f^T\right]^T$ as the control points from which $S^{(2)}$ is generated. The rows of $S^{(2)}$ are thus arranged such that $S^{(2)}\tilde{P}$ corresponds to the refinement of $\tilde{P}$. Moreover, the structure of the control points $\tilde{P}$ is encoded in the graph **G**, so $\tilde{P}$ is the realization of **G** and describes how the control points are adjacent. In this proof, we use the nodes of the graph **G** and the realization of the graph by the control points $\tilde{P}$ synonymously. Each node in **G** is therefore associated with a control point from $\tilde{P}$.

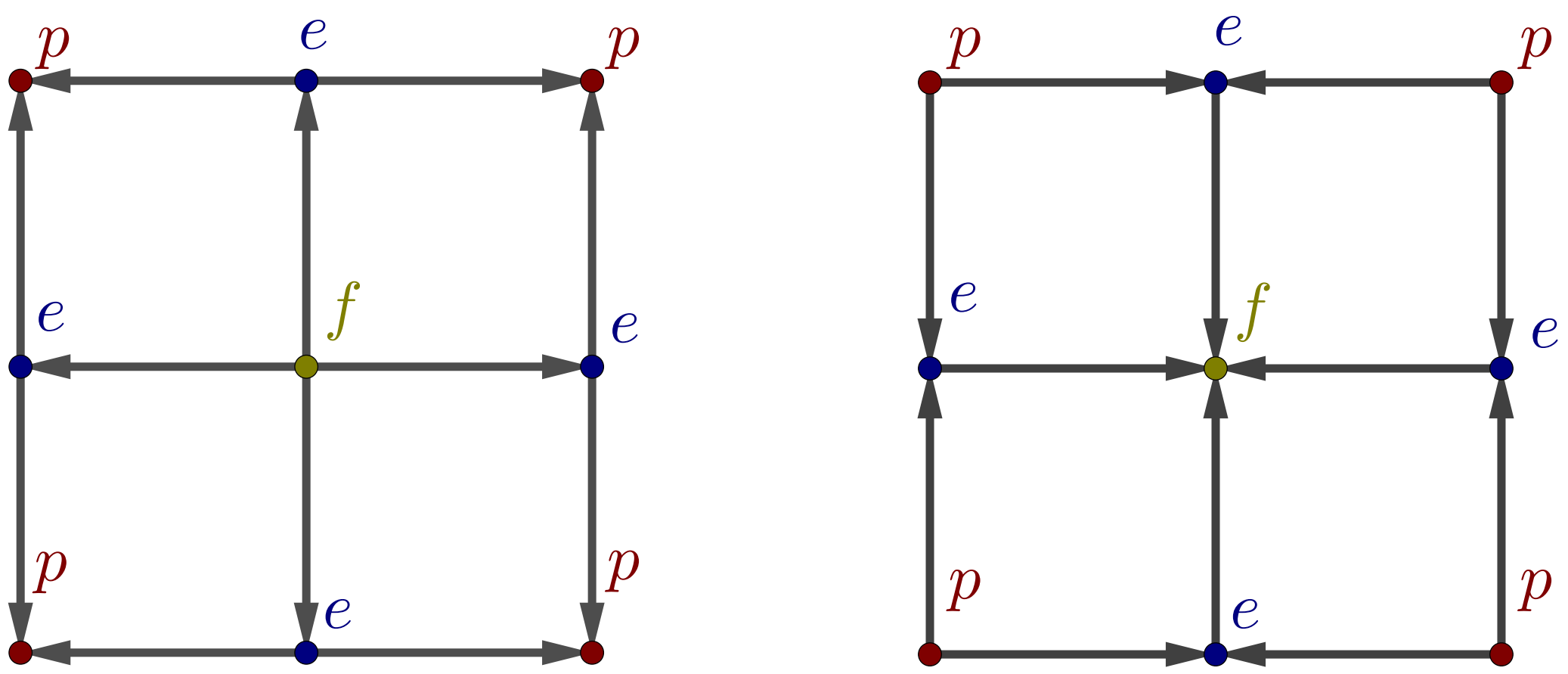

**Figure 6.36:** Example illustration of the realizations of the directed graphs $\mathbf{G}_{\tilde{L}}$ (left) and $\mathbf{G}_{\tilde{R}}$ (right) for $t = 2$.

The subdivision matrix satisfies the quality criterion exactly when only those matrix entries of $S^{(2)}$ are nonzero for which the corresponding control points lie in a common quadrilateral. Thus, for $S^{(2)}$ we have three categories of points:

- Refinement of vertex points: The refinement of vertex points overlaps with other initial elements in such a way that only those matrix entries of $S^{(2)}$ can be nonzero which lie on the quadrilateral of the corresponding vertex point. These are the vertex point itself, the two adjacent edge points, and the central facet point. An illustration of these points can be found in Figure 6.27.

- Refinement of edge points: The refinement of edge points overlaps with other initial elements so that only those matrix entries of $S^{(2)}$ can be nonzero which lie on the two quadrilaterals where the respective edge point lies. These are two vertex points, the edge point itself, two further edge points, and the central facet point. An illustration of these points can be found in Figure 6.28.

- Refinement of the facet point: Since the facet point lies on all quadrilaterals, all entries in the corresponding row of $S^{(2)}$ can be nonzero. A representative illustration of these points can be found in Figure 6.29.

For the proof of this statement, we first consider the matrices $\tilde{L}$ and $\tilde{R}$. Since the off-diagonal entries of the matrix $\tilde{C}$ are all non-negative by Theorem 6.13, the same holds for the matrices $\tilde{L}$ and $\tilde{R}$. This property also transfers to the matrices

$$L' := \ln(2)\left(\tilde{L} - E_n\right) \quad \text{and} \quad R' := \ln(2)\left(\tilde{R} - E_n\right).$$

Let us examine the entries of the matrices $L'$ and $R'$ more closely. The matrix $\tilde{C}$ is positive on the off-diagonal exactly when there is an edge between the two control points, according to property E4.

The splitting into the matrices $\tilde{L}$ and $\tilde{R}$ preserves this property, but only for the lower-left and upper-right parts of the matrix respectively. Thus, the graph **G** associated with $\tilde{L}$ and $\tilde{R}$ can be split into two directed graphs.

In the graph $\mathbf{G}_{\tilde{L}}$ corresponding to the matrix $\tilde{L}$, edges are directed from the facet point $f$ to points in $E$ and from points in $E$ to points in $P$.

In the graph $\mathbf{G}_{\tilde{R}}$ corresponding to the matrix $\tilde{R}$, edges are directed from points in $P$ to points in $E$ and from points in $E$ to the facet point $f$. A representative illustration can be found in Figure 6.36. This property transfers analogously to the matrices $L'$ and $R'$.

We want to use Theorem B.61 and Corollary B.62 in the next step, which describe the (non-)zero entries of powers and exponentials of such matrices. However, these statements require non-negative matrices. Therefore, we define

$$a := \min\left(\mathtt{diag}\left(L'\right)\right) \quad \text{and} \quad b := \min\left(\mathtt{diag}\left(R'\right)\right).$$

Thus, we can write $L'$ and $R'$ as

$$L' = L' + aE_n - aE_n \quad \text{and} \quad R' = R' + bE_n - bE_n.$$

Using the same approach as in Lemma B.63, which originates from [Ang14], we get

$$L = \exp(L') = \exp\left(L' + aE_n - aE_n\right) = \exp\left(L' + aE_n\right)\exp(-aE_n) = \exp\left(L' + aE_n\right)\exp(-a),$$

and similarly

$$R = \exp(R') = \exp\left(R' + bE_n - bE_n\right) = \exp\left(R' + bE_n\right)\exp(-bE_n) = \exp\left(R' + bE_n\right)\exp(-b).$$

Here, we have used that the matrices $L' + aE_n$ and $-aE_n$ commute, as do $R' + bE_n$ and $-bE_n$. Hence, Theorem B.53 applies, allowing the exponential to be split.

Since the matrices $L' + aE_n$ and $R' + bE_n$ are non-negative, Corollary B.62 applies to their exponentials. Moreover, since $L$ and $R$ differ only by positive factors from $\exp(L' + aE_n)$ and $\exp(R' + bE_n)$ respectively, Corollary B.62 implies that

$$L_{(i,j)} = \begin{cases} > 0 & \text{if there exists a path between } \tilde{P}_{(i,:)} \text{ and } \tilde{P}_{(j,:)} \text{ in } \mathbf{G}_{\tilde{L}}, \\ = 0 & \text{otherwise,} \end{cases}$$

and similarly,

$$R_{(i,j)} = \begin{cases} > 0 & \text{if there exists a path between } \tilde{P}_{(i,:)} \text{ and } \tilde{P}_{(j,:)} \text{ in } \mathbf{G}_{\tilde{R}}, \\ = 0 & \text{otherwise.} \end{cases}$$

For notational and verbal convenience, we define that there is a path of length zero between $\tilde{P}_{(i,:)}$ and itself in both graphs $\mathbf{G}_{\tilde{L}}$ and $\mathbf{G}_{\tilde{R}}$. Thus, the diagonal entries of both matrices are positive, which follows from the fact that the matrix exponentials contain $L'^0 = R'^0 = E_n$ as a summand. Hence, the diagonal entries are positive and $L$ and $R$ are non-negative matrices as described above.

For the entries of the matrix $S^{(2)}$, we have

$$S^{(2)}_{(i,j)} = \sum_{l=1}^{n} L_{(i,l)} R_{(l,j)} \quad \text{for} \quad i, j \in \{1, \ldots, n\}.$$

Since $L$ and $R$ are non-negative, it suffices for the positivity of $S^{(2)}_{(i,j)}$ that at least one summand is greater than zero.

Hence, a matrix entry $S^{(2)}_{(i,j)}$ is greater than zero if and only if there exists a control point $\tilde{P}_{(l,:)}$ such that there is a path in $\mathbf{G}_{\tilde{L}}$ from $\tilde{P}_{(i,:)}$ to $\tilde{P}_{(l,:)}$ and a path in $\mathbf{G}_{\tilde{R}}$ from $\tilde{P}_{(l,:)}$ to $\tilde{P}_{(j,:)}$.

Without loss of generality, it suffices to consider paths to and from corner points in $P$. Paths in $\mathbf{G}_{\tilde{L}}$ that end in a facet or edge point can be extended to a path to a corner point. From this corner point, a backward path can be constructed in $\mathbf{G}_{\tilde{R}}$ to the corresponding facet or edge point, and then extended arbitrarily.

Since in $\mathbf{G}_{\tilde{L}}$ there are only edges from the facet point $f$ to edge points $E$ and from edge points $E$ to corner points $P$, there exist paths in $\mathbf{G}_{\tilde{L}}$ from a point to a corner point in $P$ if and only if both points lie on a common quad. Conversely, in $\mathbf{G}_{\tilde{R}}$ there exist paths from a corner point in $P$ to a control point if and only if the target control point lies on the quad of the corner point in $P$.

Thus, we obtain

$$S^{(2)}_{(i,j)} > 0 \quad \text{with} \quad i, j \in \{1, \ldots, n\} \quad \Leftrightarrow \quad \tilde{P}_{(i,:)} \text{ and } \tilde{P}_{(j,:)} \text{ lie on a common quad}$$

and consequently, quality criterion Q11 is fulfilled for $S^{(2)}$.

□

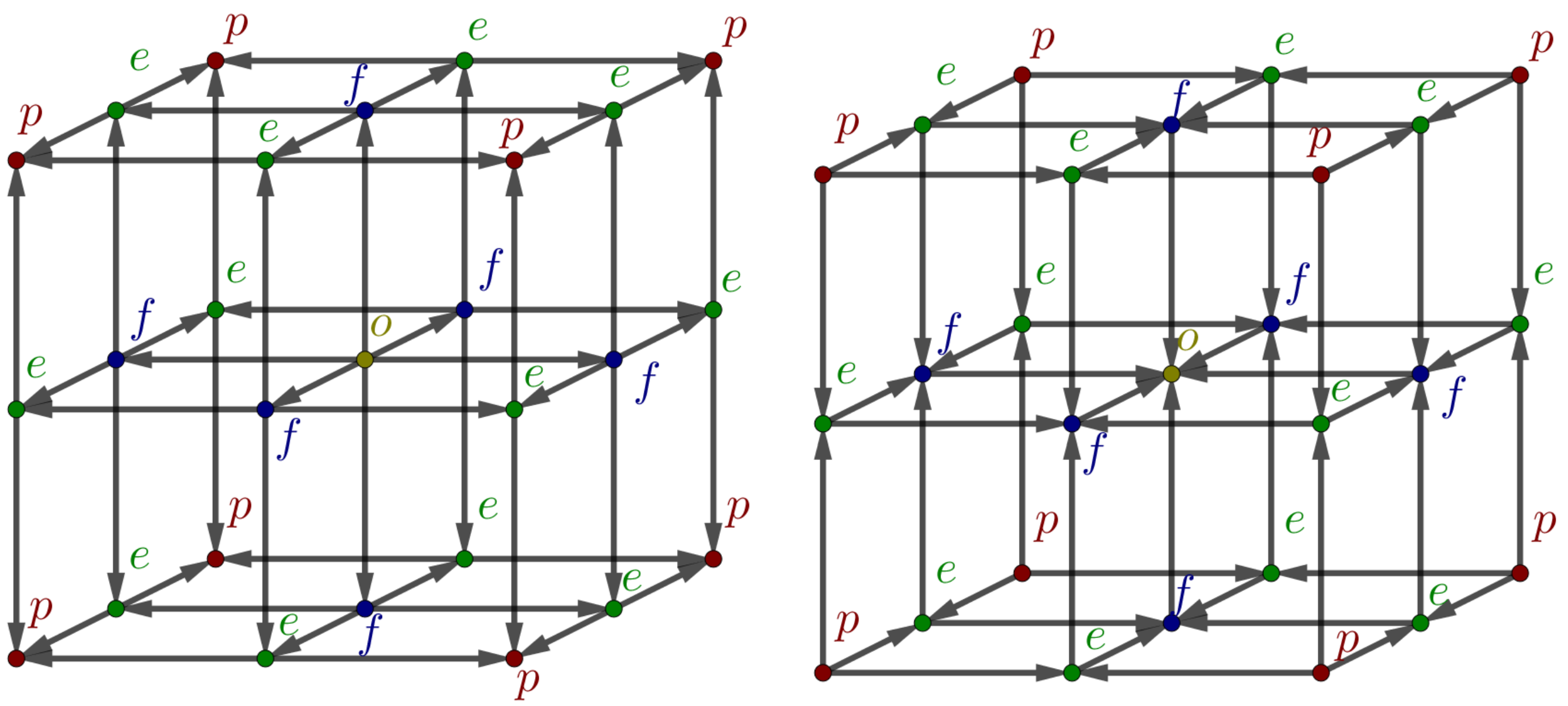

**Figure 6.37:** Example illustration of the realizations of the directed graphs $\mathbf{G}_{\tilde{L}}$ (left) and $\mathbf{G}_{\tilde{R}}$ (right) for $t = 3$.

We obtain the analogous result for the volumetric case with the following theorem:

**Theorem 6.24.** *The subdivision matrix $S^{(3)} \in \mathbb{R}^{n \times n}$ satisfies quality criterion Q11, meaning the refinement rules have a proper support.*

*Proof.* The proof proceeds analogously to the proof of Theorem 6.23, so here we only address the linguistic differences.

We first define $\tilde{P} := \left[P^T, E^T, F^T, o^T\right]^T$ as the control points from which $S^{(3)}$ is generated. The ordering of the rows and the association with the graph **G** is done analogously to Theorem 6.23.

The subdivision matrix satisfies quality criterion Q11 if and only if only those matrix entries of $S^{(3)}$ are nonzero for which the two associated control points lie on a common hexahedron. Thus, for $S^{(3)}$ we have four different categories of points:

- Refinement of vertex points: The refinement of vertex points overlaps with other initial elements such that only the matrix entries of the subdivision matrix $S^{(3)}$ that lie on the hexahedron of the respective vertex point may be nonzero. An illustration of these points can be found in Figure 6.27.
- Refinement of edge points: The refinement of edge points overlaps with other initial elements such that only the matrix entries of the subdivision matrix $S^{(3)}$ that lie on the two hexahedra to which the respective edge point belongs may be nonzero. An illustration of these points can be found in Figure 6.28.
- Refinement of facet points: The refinement of facet points overlaps with other initial elements such that only the matrix entries of the subdivision matrix $S^{(3)}$ that lie on the hexahedra to which the respective facet point belongs may be nonzero. An exemplary illustration of these points is shown in Figure 6.29.
- Refinement of the volume point: Since the volume point lies on all hexahedra, all entries in the corresponding row of $S^{(3)}$ may be nonzero. An exemplary illustration of these points can be found in Figure 6.30.

For the directed graph $\mathbf{G}_{\tilde{L}}$ we obtain only edges from the point $o$ to points in $F$, from points in $F$ to points in $E$, and from points in $E$ to points in $P$.

For the directed graph $\mathbf{G}_{\tilde{R}}$ we correspondingly obtain only edges from points in $P$ to points in $E$, from points in $E$ to points in $F$, and from points in $F$ to the point $o$. An exemplary illustration is found in Figure 6.37.

For the matrix entries of $S^{(3)}$, we have

$$S^{(3)}_{(i,j)} = \sum_{l=1}^{n} L_{(i,l)} R_{(l,j)} \quad \text{for} \quad i, j \in \{1, \dots, n\}.$$

Since $L$ and $R$ are non-negative, it suffices for the positivity of $S^{(3)}_{(i,j)}$ that at least one summand is greater than zero.

Hence, a matrix entry $S^{(3)}_{(i,j)}$ is greater than zero if and only if there exists a control point $\tilde{P}_{(l,:)}$ such that there is a path in $\mathbf{G}_{\tilde{L}}$ from $\tilde{P}_{(i,:)}$ to $\tilde{P}_{(l,:)}$ and a path in $\mathbf{G}_{\tilde{R}}$ from $\tilde{P}_{(l,:)}$ to $\tilde{P}_{(j,:)}$.

Without loss of generality, it suffices to consider paths to and from corner points in $P$. Paths in $\mathbf{G}_{\tilde{L}}$ that end in a volume, facet, or edge point can be continued to a path to a corner point. From this corner point, a backward path in $\mathbf{G}_{\tilde{R}}$ to the corresponding volume, facet, or edge point can be constructed and continued arbitrarily from there.

Since in $\mathbf{G}_{\tilde{L}}$ there are only edges from the volume point $o$ to facet points $F$, from facet points $F$ to edge points $E$, and from edge points $E$ to corner points $P$, there exist paths in $\mathbf{G}_{\tilde{L}}$ from points to corner points $P$ if and only if both lie in a common hexahedron. Conversely, in $\mathbf{G}_{\tilde{R}}$ there exist paths from a corner point in $P$ to a control point if and only if the target control point lies in the hexahedron of the corner point in $P$.

Thus, we obtain

$$S^{(3)}_{(i,j)} > 0 \quad \text{with} \quad i, j \in \{1, \dots, n\} \quad \Leftrightarrow \quad \tilde{P}_{(i,:)} \text{ and } \tilde{P}_{(j,:)} \text{ lie in a common hexahedron}$$

and consequently, Quality Criterion Q11 is satisfied for $S^{(3)}$.

□

From these two theorems we obtain the following corollary:

**Corollary 6.25.** *The matrices $S^{(2)} \in \mathbb{R}^{n \times n}$ and $S^{(3)} \in \mathbb{R}^{n \times n}$ are irreducible.*

*Proof.* For $t \in \{2, 3\}$, we define the directed graph $\mathbf{G}_S$ corresponding to the adjacency matrix

$$A_{(i,j)} := \begin{cases} 1 & \text{if } S^{(t)}_{(i,j)} \neq 0, \\ 0 & \text{otherwise.} \end{cases}$$

It is important to note that this graph $\mathbf{G}_S$ does not coincide with the graph $\mathbf{G}$. Since the central point $f$ for $t = 2$ and the central point $o$ for $t = 3$ lie in all quads or hexahedra respectively, it follows that

$$S^{(t)}_{(i,n)} \neq 0 \quad \text{and} \quad S^{(t)}_{(n,i)} \neq 0 \quad \text{for all} \quad i \in \{1, \dots, n\}.$$

Hence, in the directed graph $\mathbf{G}_S$ there exists an edge and thus a path from every control point to the point $f$ or $o$, and vice versa from $f$ or $o$ to every other control point. Consequently, a path from any point $\tilde{P}_{(i,:)}$ to any other point $\tilde{P}_{(j,:)}$ can be constructed via the two edges $(\tilde{P}_{(i,:)}, \tilde{P}_{(n,:)})$ and $(\tilde{P}_{(n,:)}, \tilde{P}_{(j,:)})$. Therefore, the graph $\mathbf{G}_S$ is strongly connected according to Definition A.16. By Characterization B.36, matrices whose zero structure induces strongly connected graphs are irreducible, hence $S^{(t)}$ is irreducible. □

The quality criterion Q11 transfers from the subdivision matrices of the initial elements to the matrices of the double rings and double shells, which we explain in the following corollary:

**Corollary 6.26.** *The matrices $\overline{S}^{(2)} \in \mathbb{R}^{n \times n}$ and $\overline{S}^{(3)} \in \mathbb{R}^{n \times n}$ satisfy quality criterion Q11.*

*Proof.* Analogously to Theorem 6.20, this property transfers from the subdivision matrices of the initial elements $S$ to the matrices $\overline{S}$, since the rows of $\overline{S}$ consist of rows of subdivision matrices of initial elements padded with zero entries. Because each initial element satisfies quality criterion Q11, it is also satisfied in the double ring and double shell matrices $\overline{S}$ for all overlap regions. □

The property of irreducibility, however, does not transfer, as discussed in the following theorem:

**Proposition 6.27.** *The matrices* $\overline{S}^{(2)} \in \mathbb{R}^{n\times n}$ *and* $\overline{S}^{(3)} \in \mathbb{R}^{n\times n}$ *are reducible.*

*Proof.* The central initial element is used unchanged in the double ring and double shell matrices, and the remaining entries of these rows are padded with zero entries. If we permute the rows and columns so that the rows and columns of the central initial element are arranged at the far right or bottom, then $\overline{S}$ can be arranged as

$$\overline{S} = \begin{bmatrix} M_1 & M_2 \\ 0 & S \end{bmatrix},$$

where the matrices $S$ and $M_1$ are square. Therefore, by Definition B.35, $\overline{S}$ is reducible. □

## Q3 Dominant Eigenvalue 1

To verify this criterion for the matrices of the initial elements, we first need the following lemma:

**Lemma 6.28.** *The diagonal entries of* $S^{(2)} \in \mathbb{R}^{n\times n}$ *and* $S^{(3)} \in \mathbb{R}^{n\times n}$ *are positive.*

*Proof.* As already described, the two matrices

$$\ln(2)(\tilde{L} - E_n) \quad \text{and} \quad \ln(2)(\tilde{R} - E_n)$$

are triangular matrices consisting of real entries. By Lemma B.58, which considers the exponential of triangular matrices, it holds for the diagonal entries of the matrix exponentials that

$$L_{(i,i)} = \exp\big(\ln(2)(\tilde{L}_{(i,i)} - 1)\big) \quad \text{and} \quad R_{(i,i)} = \exp\big(\ln(2)(\tilde{R}_{(i,i)} - 1)\big) \quad \text{for} \quad i \in \{1, \dots, n\}.$$

Since the exponential of a real number is always positive, the diagonal entries of $L$ and $R$ are positive.

The diagonal entries of $S$ result from multiplying the $i$-th row of $L$ with the $i$-th column of $R$. Due to the triangular structure, we have

$$S^{(t)}_{(i,i)} = L_{(i,i)} R_{(i,i)} > 0 \quad \text{for} \quad i \in \{1, \dots, n\}, \quad \text{and} \quad t \in \{2, 3\}.$$

Thus, the diagonal entries of $S^{(2)}$ and $S^{(3)}$ are positive. □

With this lemma, in the next step we prove the quality criterion Q3 for the subdivision matrices of the initial elements:

**Theorem 6.29.** *The subdivision matrices* $S^{(2)}$ *and* $S^{(3)}$ *have a simple eigenvalue* $1$ *corresponding to the eigenvector* $\vec{1}$*, and for all other eigenvalues* $\lambda_i$ *it holds that* $|\lambda_i| < 1$*. Thus, the eigenvalue* $1$ *is the dominant eigenvalue of both matrices, and the subdivision matrices* $S^{(2)}$ *and* $S^{(3)}$ *satisfy quality criterion Q3.*

*Proof.* The matrices $S^{(2)}$ and $S^{(3)}$ are stochastic by Corollary 6.22 and, according to Lemma 6.28, all diagonal entries are positive real numbers.

By Theorem B.40, which is a special case of the Gerschgorin circle theorem B.38, all eigenvalues of both matrices have magnitude less than or equal to 1, and furthermore $1$ is the only eigenvalue on the unit circle.

Since $S^{(2)}$ and $S^{(3)}$ are irreducible by Corollary 6.25, the Perron-Frobenius theorem B.37 implies that the algebraic multiplicity of the eigenvalue $1$ is exactly $1$. These steps can also be summarized using Corollary B.41. Hence, $1$ is the dominant eigenvalue of $S^{(2)}$ and $S^{(3)}$.

Since all rows of $S^{(2)}$ and $S^{(3)}$ sum to $1$ by Theorem 6.20, the eigenvector corresponding to the eigenvalue 1 is, by Lemma 2.36, exactly $\vec{1}$. □

We cannot transfer this proof to the matrices $\overline{S}$ because they are neither irreducible nor have all diagonal entries positive. However, since the matrices $\overline{S}$ are stochastic, for every eigenvalue $\lambda_i$ of $\overline{S}$ it holds that

$$|\lambda_i| \leq 1.$$

Since all rows of $\overline{S}$ sum to 1, by Lemma 2.36 the value 1 is an eigenvalue of $\overline{S}$ with eigenvector $\vec{1}$. We could not make any statements about other eigenvalues with $|\lambda_i| = 1$ or about the algebraic multiplicity of the eigenvalue 1.

Therefore, the empirical analysis of the matrices $\overline{S}$ is all the more crucial. This will be examined together with the other eigenvalues in the next section.

## Q4, Q5, Q6 und Q12 Spectrum of Eigenvalues

Regarding the eigenvalue spectrum, unfortunately only few results could be proven for the subdivision matrices of this chapter, since the subdivision matrices in the final construction step arise as the product of two matrices. There is only limited theory about the eigenvalues of a product of two matrices. Many statements provide bounds for eigenvalues, but these were too imprecise for our purposes.

For this reason, the construction described in this chapter is of experimental nature. Compared to other subdivision algorithms for cubic B-spline volumes, this is not a disadvantage, since no results about the spectrum exist for those algorithms either. Proving that the matrices constructed here have the desired spectrum can be the subject of further research.

However, we can make statements about the existence of a $t$-fold eigenvalue $1/2$. We start with the two-dimensional case of the initial elements:

**Theorem 6.30.** *Let $\tilde{P} := \left[P^T, E^T, f^T\right]^T \in \mathbb{R}^{n\times 2}$ be the eigenstructure from which the matrix $S^{(2)}$ is constructed. Then it holds that*

$$S^{(2)}\tilde{P} = \tfrac{1}{2}\tilde{P},$$

*that is, the matrix $S^{(2)}$ has a double eigenvalue $1/2$ with eigenvectors given by the columns of $\tilde{P}$.*

*Proof.* By Lemma 6.15, we have

$$\tilde{L}\tilde{P} = \tilde{P} \quad \text{and} \quad \tilde{R}\tilde{P} = 0 \cdot \tilde{P}.$$

Thus, for the matrices $L'$ and $R'$ it holds that

$$L'\tilde{P} = \ln(2)\left(\tilde{L} - E_n\right)\tilde{P} = \ln(2)\left(\tilde{L}\tilde{P} - E_n\tilde{P}\right) = \ln(2)(\tilde{P} - \tilde{P}) = 0 \cdot \tilde{P}$$

and

$$R'\tilde{P} = \ln(2)\left(\tilde{R} - E_n\right)\tilde{P} = \ln(2)\left(\tilde{R}\tilde{P} - E_n\tilde{P}\right) = \ln(2)(-\tilde{P}) = -\ln(2)\cdot\tilde{P}.$$

By Lemma B.60, which treats eigenvalues and eigenvectors of exponentials of triangular matrices, we get for the exponentials $L$ and $R$

$$L\tilde{P} = \exp(L')\tilde{P} = \exp(0)\tilde{P} = \tilde{P}$$

and

$$R\tilde{P} = \exp(R')\tilde{P} = \exp(-\ln(2))\tilde{P} = \frac{1}{2}\tilde{P}.$$

Consequently, for $S^{(2)}$ we have

$$S^{(2)}\tilde{P} = LR\tilde{P} = \frac{1}{2}\tilde{P},$$

so the columns of $\tilde{P}$ are eigenvectors of $S^{(2)}$ corresponding to the eigenvalue $1/2$. □

Analogously, we obtain the following theorem for the matrix $S^{(3)}$:

**Theorem 6.31.** *Let $\tilde{P} := \left[P^T, E^T, F^T, o^T\right]^T \in \mathbb{R}^{n\times 3}$ be the eigenstructure from which the matrix $S^{(3)}$ is constructed. Then it holds that*

$$S^{(3)}\tilde{P} = \frac{1}{2}\tilde{P},$$

*so the matrix $S^{(3)}$ has a triple eigenvalue $1/2$ with eigenvectors given by the columns of $\tilde{P}$.*

*Proof.* Replacing Lemma 6.15 by Lemma 6.14 in the proof of Theorem 6.30, the proof proceeds completely analogously. □

A proof that $1/2$ is the subdominant eigenvalue with algebraic multiplicity 2 or 3 must be left open. This is, as already mentioned, because little is known about the eigenvalues of the product of two matrices. Further research is necessary here. However, if $1/2$ is a $t$-fold subdominant eigenvalue, then the corresponding eigenstructure is self-intersection-free due to the construction of $\tilde{P}$ as a refinement of a 2- or 3-polytope whose quads or hexahedra are all convex. In this case, Quality Criterion Q12 would be fulfilled for $S^{(t)}$.

We consider the transfer of existence to the double ring and double shell matrices in the following theorem:

**Theorem 6.32.** *The matrices* $\overline{S}^{(t)} \in \mathbb{R}^{n\times n}$ *have a t-fold eigenvalue* $1/2$ *for* $t \in \{2,3\}$.

*Proof.* Since the rows of $S^{(t)}$ in $\overline{S}^{(t)}$ are only padded with zero entries, the matrix can be brought into the form

$$H\overline{S}^{(t)}H^T = \begin{bmatrix} M_1 & M_2 \\ 0 & S^{(t)} \end{bmatrix}$$

by a permutation matrix $H \in \{0,1\}^{n\times n}$. Because the permutation of rows and columns maps a matrix to a similar matrix, both forms have the same eigenvalues by Corollary B.31. By Theorem B.34, which considers eigenvalues of matrices in the above form, the eigenvalues of $S^{(t)}$ transfer to the matrix $\overline{S}^{(t)}$. □

The spectral issues of the matrices $S$ naturally transfer to the matrices $\overline{S}$. Additionally, for $\overline{S}$, independent of the spectral problem, no statement can be made about a self-intersection-free structure, since, analogous to Section 5, we have no access to eigenvectors of $\overline{S}$ and no theoretical approaches to verify the quality criterion.

With the above statements, we could show the existence of the desired part of the spectrum, but it could not be proven that no eigenvalues lie in the interval $(\frac{1}{2}, 1)$ in absolute value.

The eigenvalues of the matrices $\tilde{L}$ and $\tilde{R}$, and by Lemma B.59 also the eigenvalues of the matrices $L$ and $R$, can be determined and have a favorable structure; however, this structure could not be exploited for the product matrix $L \cdot R$.

Therefore, empirical analysis is all the more important for these quality criteria. The six eigenvalues with largest absolute value of the matrices in the four categories can be found in Figure 6.38, with the concrete maximal values listed in the following table:

| | $S^{(2)}$ | $\overline{S}^{(2)}$ | $S^{(3)}$ | $\overline{S}^{(3)}$ |
|---|---|---|---|---|
| $\max\lvert\lambda_0 - 1\rvert$ | 3,7748e−15 | 3,5527e−15 | 6,4393e−15 | 8,4377e−15 |
| $\max\lvert\lambda_1 - 0.5\rvert$ | 2,7756e−15 | 5,7732e−15 | 2,7318e−12 | 2,7194e−12 |
| $\max\lvert\lambda_2 - 0.5\rvert$ | 3,1086e−15 | 5,3291e−15 | 8,9928e−14 | 8,2268e−14 |
| $\max\lvert\lambda_3 - 0.5\rvert$ | | | 1,7625e−12 | 1,7608e−12 |
| $\max\lvert\mu\rvert$ | 0,499 26 | 0,499 26 | 0,497 05 | 0,497 05 |
| $\max\lvert S\tilde{P} - 0.5\tilde{P}\rvert$ | 1,3878e−14 | | 8,9589e−10 | |

Several notable observations arise here. Empirically, for all four categories, it is shown that 1 is the (simple) dominant eigenvalue of the subdivision matrices and that $1/2$ is the $t$-fold subdominant eigenvalue. Thus, the quality criteria Q3, Q4, and Q5 are empirically fulfilled. However, the subsubdominant eigenvalue is greater than $1/4$, so quality criterion Q6 is not satisfied for all four categories.

In contrast to the generalized quadratic B-spline subdivision from Section 5, the subsubdominant eigenvalue empirically approaches the value $1/2$ for an increasing number of vertices of the primal polytope in all four categories. This behavior is rather unfavorable but corresponds to that of the Catmull-Clark algorithm as described in [PR08, Sec. 6.1, pp. 109–116].

Our unproven conjecture is that this is related to the zero-structure of quality criterion Q11. The locality of the outer refinement rules could, at least for the case $t = 2$, lead to rings with higher frequency—that is, rings that wrap multiple times around the central point—not being attracted towards the central point during refinement. Whether there are approaches that improve the ratio $\lambda^2/\mu$ while simultaneously fulfilling quality criteria Q4, Q5, and Q11 can be a subject of further research.

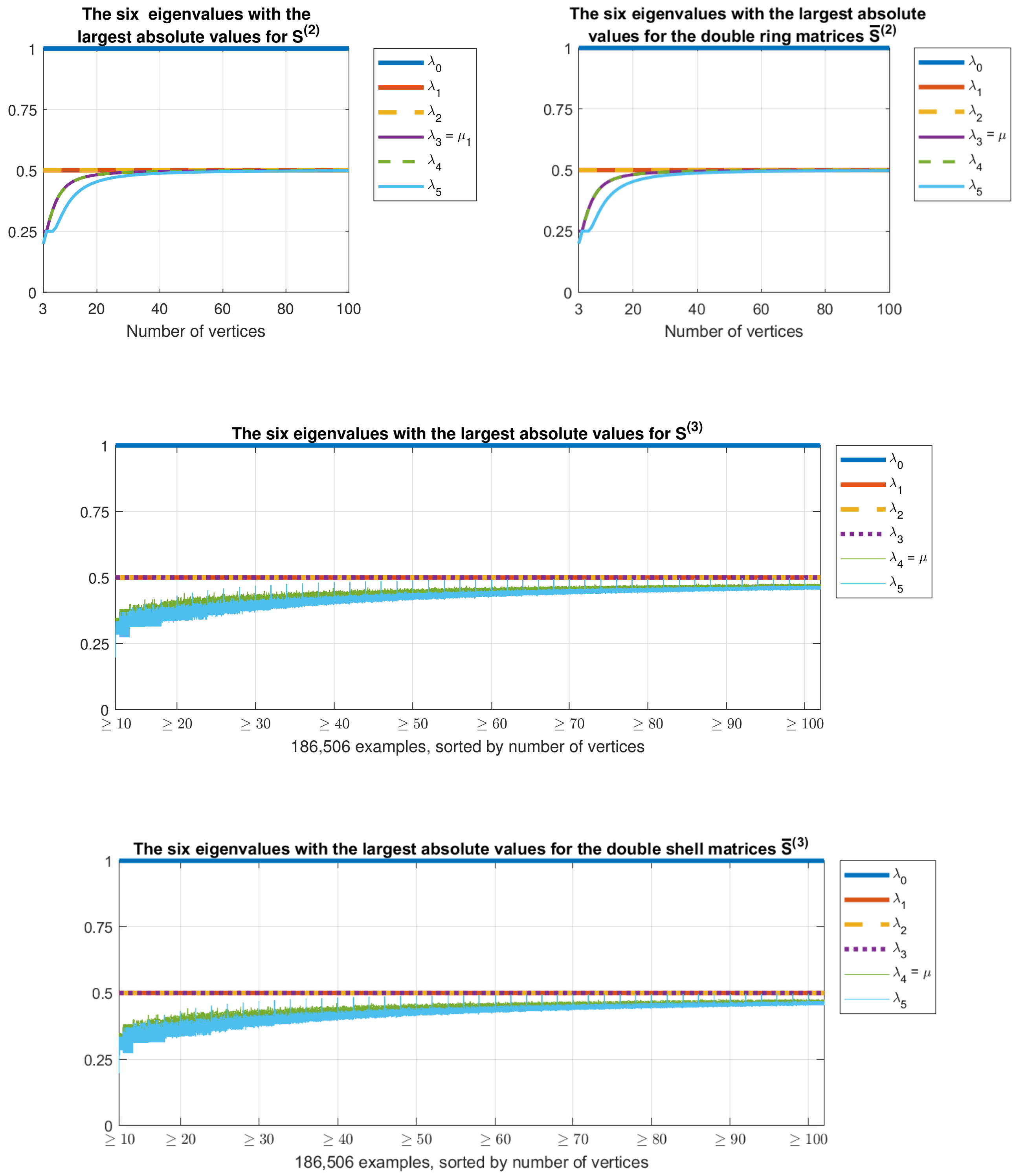

**Figure 6.38:** Eigenvalues of the matrices of the initial elements and of the double ring and double shell matrices for $t = 2$ (bottom) and for $t = 3$ (top). The examples are each sorted according to the number of vertices of the primal polytope.

Another notable point is that in the graphs of Figure 6.38, the cases of $S$ and $\overline{S}$ appear to be identical. This is also confirmed numerically. The deviations are shown in the following table:

| | $\max\|S^{(2)} - \overline{S}^{(2)}\|$ | $\max\|S^{(3)} - \overline{S}^{(3)}\|$ |
|---|---|---|
| $\lambda_0$ | 3,5527e−15 | 9,5479e−15 |
| $\lambda_1$ | 5,7732e−15 | 8,582e−14 |
| $\lambda_2$ | 6,2172e−15 | 3,7192e−14 |
| $\lambda_3$ | | 5,9397e−14 |
| $\mu$ | 3,6637e−15 | 1,0214e−14 |

Thus, empirically, the relevant part of the spectrum of the double-ring and double-shell matrices derives from the initial-element part of the matrix.

It is also noticeable that the empirical results of the subsubdominant eigenvalues for $t = 3$ exhibit outliers on the higher side. These outliers belong to the semi-irregular cases, which is directly evident from the graphics of the surface case. The eigenvalues of the tensor product case arise via Corollary 6.36 and Theorem B.32 from the eigenvalues of the two-dimensional case combined with the values $[1, 1/2, \ldots]$. Thus, the eigenvalues of $S^{(2)}$ transfer to the semi-irregular matrices of $S^{(3)}$ and are therefore particularly large.

Finally, it can be noted that the eigenstructure for the case of the initial elements is self-penetration free, provided that the matrices $S$ have a $t$-fold eigenvalue $1/2$. This follows from the structure of $\tilde{P}$, which for $t = 2$ is an assembly of convex quadrilaterals and for $t = 3$ of convex hexahedra, which by construction do not intersect themselves. The condition on the eigenvalues is at least empirically fulfilled, and the condition on the eigenvectors has been empirically verified. The deviation of $S\tilde{P}$ from $\tilde{P}/2$ is empirically on the order of $10^{-14}$ for $t = 2$ and on the order of $10^{-10}$ for $t = 3$, thus quality criterion Q12 is at least empirically fulfilled. However, for the structure of the double-ring and double-shell matrices, no empirical statement can be made.

Fundamentally, proofs regarding the spectrum of the surface case are within reach. With the help of the discrete Fourier transform, the spectrum could be explicitly determined analogously to [PR08, Sec. 6.1, pp. 109–116]. Since the related theory and associated terminology have not been introduced here, we leave the detailed elaboration of this sketch to further research.

## Q10 Symmetry

For the generalized cubic B-spline subdivision, we must restrict the automorphisms of the graph **G** to those induced by the graph $\mathbf{G}_K$. Hence, for the subdivision matrices of the initial elements $S$, we only consider automorphisms that map quadrilaterals to quadrilaterals or hexahedra to hexahedra.

These automorphisms are therefore designed so that the respective categories of points are mapped onto the corresponding categories. This leads to the following theorem for $S$:

**Theorem 6.33.** *Let **G** be the graph corresponding to the subdivision matrix $S^{(t)}$ with $t \in \{2, 3\}$, and let*

$$\tilde{P} := \begin{cases} \left[P^T, E^T, f^T\right]^T & \text{for } t = 2, \\ \left[P^T, E^T, F^T, o^T\right]^T & \text{for } t = 3, \end{cases}$$

*be the eigenstructure from which the subdivision matrix $S^{(t)}$ is constructed. Then it holds that*

$$HS^{(t)}H^{-1} = S^{(t)}$$

*for an automorphism $H$ of **G** that maps nodes corresponding to $P$ to nodes corresponding to $P$, nodes corresponding to $E$ to nodes corresponding to $E$, and, for $t = 3$, nodes corresponding to $F$ to nodes corresponding to $F$. Thus, the subdivision matrices $S^{(t)}$ satisfy quality criterion Q10.*

*Proof.* We first consider the symmetries of the realization of $\mathbf{G}_K$, i.e., the points $P$ as a convex $t$-polytope in $\mathbb{R}^t$. For $t = 2$, the realization of $\mathbf{G}_K$ is a regular $n$-gon, which satisfies all symmetries of $\mathbf{G}_K$. For $t = 3$, the primal and dual polytopes satisfy all symmetries of the graph $\mathbf{G}_K$ by Theorem 3.25. Since an automorphism of $P$ maps edges to edges and, for $t = 3$, facets to facets, the symmetry extends to all points of the realization $\tilde{P}$ of $\mathbf{G}$. In particular, symmetric quadrilaterals and hexahedra have the same shape up to rotation and reflection.

Identical shapes produce identical matrices $M$ and $M'$ in Algorithm 15. The matrices $B$ operate on the primal polytope $P$, which contains all symmetries of $\mathbf{G}_K$. Hence, the weights of the rows of $\overline{B}$ for symmetric points are identical. Since these weights weight identically generated rows in $\overline{M}$ identically, the matrix $\tilde{C}$ satisfies all symmetries of the underlying structure, i.e.,

$$H\tilde{C}H^{-1} = \tilde{C}.$$

The matrices $\tilde{L}$ and $\tilde{R}$ partition the rows of $\tilde{C}$ according to the respective categories. Because the automorphisms, as specified in this theorem, only permute the matrices within these categories, the symmetry property of $\tilde{C}$ carries over to $\tilde{L}$ and $\tilde{R}$:

$$H\tilde{L}H^{-1} = \tilde{L} \quad \text{and} \quad H\tilde{R}H^{-1} = \tilde{R}.$$

Using Theorem B.53, which allows taking invertible matrix pairs in and out of the matrix exponential, we get

$$\begin{aligned}
HS^{(t)}H^{-1} &= HLRH^{-1} \\
&= HLH^{-1}HRH^{-1} \\
&= H \exp\left(\ln(2)\left(\tilde{L} - E_n\right)\right) H^{-1} H \exp\left(\ln(2)\left(\tilde{R} - E_n\right)\right) H^{-1} \\
&= \exp\left(H \ln(2)\left(\tilde{L} - E_n\right) H^{-1}\right) \exp\left(H \ln(2)\left(\tilde{R} - E_n\right) H^{-1}\right) \\
&= \exp\left(\ln(2)\left(H\tilde{L}H^{-1} - E_n\right)\right) \exp\left(\ln(2)\left(H\tilde{R}H^{-1} - E_n\right)\right) \\
&= \exp\left(\ln(2)\left(\tilde{L} - E_n\right)\right) \exp\left(\ln(2)\left(\tilde{R} - E_n\right)\right) \\
&= LR \\
&= S^{(t)}.
\end{aligned}$$

Therefore, the matrices $S^{(t)}$ satisfy all automorphisms of the underlying graph $\mathbf{G}$ induced by $\mathbf{G}_K$, and hence fulfill quality criterion Q10. □

The property transfers, analogously to Chapter 5, in a meaningful way to the double ring and double shell matrices, as explained in the following theorem:

**Theorem 6.34.** *The matrices $\overline{S}$ satisfy quality criterion Q10.*

*Proofsketch.* First, it must be discussed which automorphisms of the matrices $\overline{S}$ are induced by the graph $\mathbf{G}_K$. Each node of the graph $\mathbf{G}_K$ is represented in $\overline{S}^{(t)}$ by a combination of $5 \times 5$ control points for $t = 2$ and by a combination of $5 \times 5 \times 5$ control points for $t = 3$. Thus, we only need to consider automorphisms that map these fivefold structures onto each other.

The initial elements in the outer region are either semi-regular or regular. Semi-regular initial elements occur only along the facets. Since these depend on the degree of the facets and automorphisms of $\mathbf{G}_K$ only map facets of equal degree onto each other, only identical refinement rules of semi-regular initial elements are mapped onto each other. The same applies to the remaining regular initial elements, all of which have identical rules.

Therefore, the symmetry property transfers from the matrices of the initial elements to the double ring and double shell matrices, and quality criterion Q10 is fulfilled for $\overline{S}$. □

For the empirical analysis, we proceed analogously to Chapter 5. For the 27,979 isomorphisms in the empirical evaluation of the initial elements, we obtain the following values:

| | $S^{(3)}$ |
|---|---|
| $\max \lvert S - HSH^T \rvert$ | $4{,}1134\text{e}{-12}$ |

Thus, for the case of initial elements, a qualitative statement about the symmetry can also be made. For the double shell matrices, however, this difference was not checked, since finding the isomorphism of two large adjacency matrices was computationally too expensive in a technical sense.

## Q7 and Q8 Tensor Product Structure

Analogous to Chapter 5, we first consider the surface case here. For this case, only quality criterion Q7 is relevant. For $S^{(2)}$ this is obtained simply by inserting the corresponding adjacency matrix. The result must then be permuted with an appropriate permutation matrix. Thus, quality criterion Q7 is fulfilled for $S^{(2)}$.

This also transfers to the double ring matrix $\overline{S}^{(2)}$, since in the regular case it consists exclusively of subdivision matrices of regular initial elements.

For the tensor product structure, analogous to Chapter 5, the case $t = 3$ is the interesting one. We consider this in the following and begin with the next theorem:

**Theorem 6.35.** *Let $A^{(2)} \in \{0,1\}^{m\times m}$ be the adjacency matrix of a two-dimensional initial element with $n$ vertices, and let $m := 2n + 1$. Then define*

$$A := \begin{bmatrix} A^{(2)} & E_m & 0 \\ E_m & A^{(2)} & E_m \\ 0 & E_m & A^{(2)} \end{bmatrix}$$

*as the permuted adjacency matrix of the graph **G**, arising from the combinatorial graph $\boldsymbol{G}_K$ of a prism. Then the subdivision matrix $S_2^{(3)}$ in the permuted form corresponding to the adjacency matrix $A$ can be represented as a Kronecker product of the matrix*

$$S^{(1)} = \frac{1}{8}\begin{bmatrix} 4 & 4 & 0 \\ 1 & 6 & 1 \\ 0 & 4 & 4 \end{bmatrix}$$

*and another matrix.*

*Proof.* Deviating from the construction of Algorithm 15, we reorder the rows and columns of the adjacency matrix $A$ and later also of the matrix $\tilde{C}$ differently, in order to be able to show the tensor product structure at the end of the proof. This has, apart from the permutation, no influence on the matrices $\tilde{L}$, $\tilde{R}$, and $S^{(3)}$, as long as we separate the values according to point categories.

We first divide the points $\tilde{P}$ into three sets. Set 1 contains the points of the upper facet, set 3 contains those of the lower facet, and set 2 contains all other points. Within the sets, we sort the points so that they match the sorting of the two-dimensional case. We thus obtain the sorted sets or vectors

$$M_1 = \begin{bmatrix} p \\ \vdots \\ p \\ e \\ \vdots \\ e \\ f \end{bmatrix}, \quad M_2 = \begin{bmatrix} e \\ \vdots \\ e \\ f \\ \vdots \\ f \\ o \end{bmatrix} \quad \text{and} \quad M_3 = \begin{bmatrix} p \\ \vdots \\ p \\ e \\ \vdots \\ e \\ f \end{bmatrix} \tag{6.9}$$

with the respective points that (do not) belong to the respective facets. We arrange the points $\tilde{P}$ and thus the rows

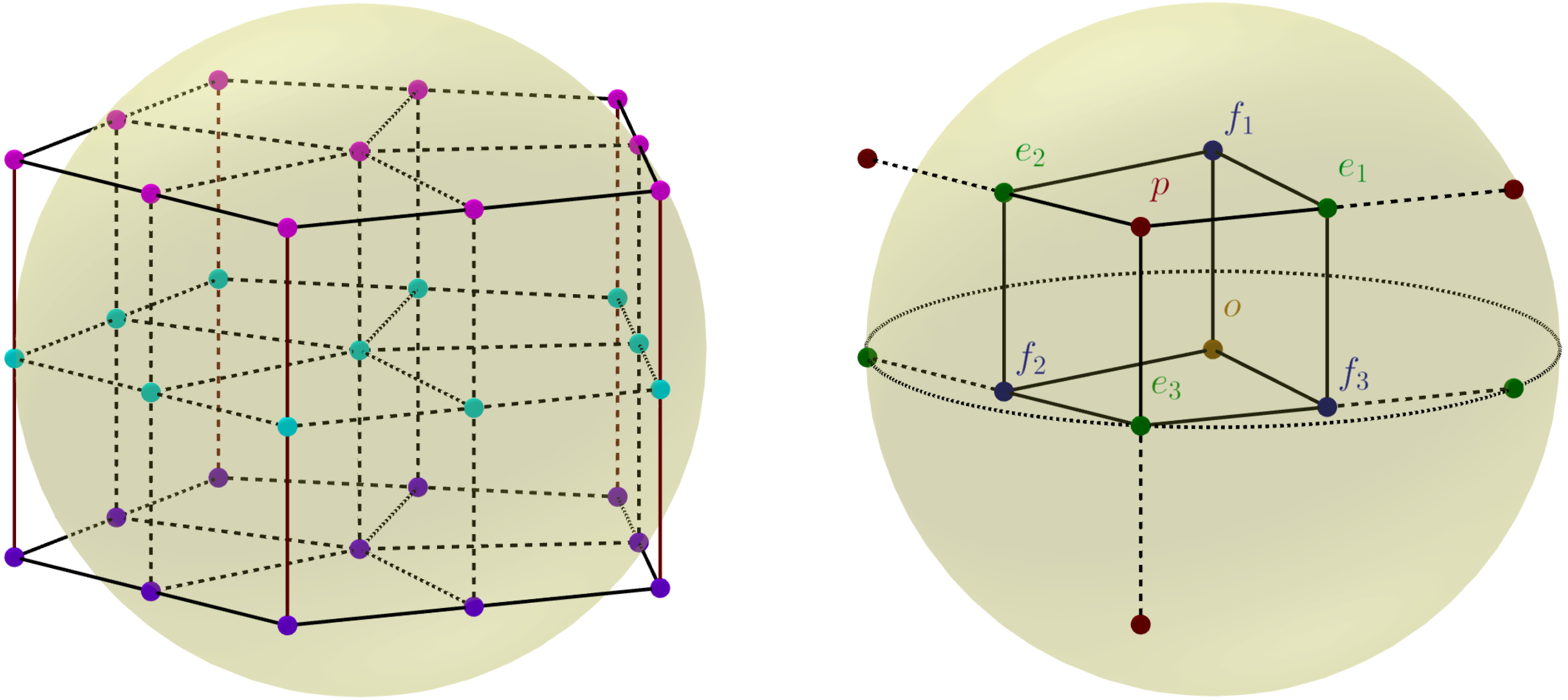

**Figure 6.39:** On the left, the partition of the points $\tilde{P}$ into the sets $M_1$ (pink), $M_2$ (cyan), and $M_3$ (purple) is shown. On the right, a hexahedron of the refinement of a prism is shown with the notation of points used in this proof.

and columns of the following matrices in the form

$$\tilde{P} := \begin{bmatrix} M_1 \\ M_2 \\ M_3 \end{bmatrix}.$$

An overview of the three sets can be found in Figure 6.39.

For the proof, we begin with the matrix $M'$ from Algorithm 15. Due to the symmetry we showed in Theorem 6.33, every hexahedron in the prism case has the same shape up to rotations and reflections. Since only relative information of the polytope is used for the construction of the Colin-de-Verdière-matrix in Theorem 4.13, the Colin-de-Verdière-matrices $M$ and thus also the normalized matrices $M'$ are all identical.

We explicitly construct the entries of the matrix $M'$, as described in equation (6.2), for the prism case depending on $n$.

To do this, we use the construction of the control points from Theorem 5.22 and, without loss of generality, choose a specific hexahedron on the upper hemisphere. We denote the points of the hexahedron by

$$p, e_1, e_2, e_3, f_1, f_2, f_3, \text{ and } o.$$

An exemplary illustration including the notation can be found in Figure 6.39.

We also denote

$$a := \cos\left(\frac{\pi}{n}\right) \quad \text{and} \quad b := \sin\left(\frac{\pi}{n}\right)$$

to simplify the notation. Using the coordinates already computed in the proof of Theorem 5.22, we obtain for a specific hexahedron

$$\begin{aligned} p &= (1, 0, b), & f_1 &= (0, 0, b), \\ e_1 &= \left(a^2, -ab, b\right), & f_2 &= \left(a^2, ab, 0\right), \\ e_2 &= \left(a^2, ab, b\right), & f_3 &= \left(a^2, -ab, 0\right), \\ e_3 &= (1, 0, 0), & o &= (0, 0, 0). \end{aligned}$$

Shifting the points by $-p/2$, we obtain, using Theorem 4.13 and with the help of a CAS, the normalized Colin-de-Verdière-matrix $M'$ as

$$M' = \left[\begin{array}{c|ccc|ccc|c}
\frac{1}{2}-\frac{1}{2b^2} & \frac{1}{4b^2} & \frac{1}{4b^2} & \frac{1}{2} & 0 & 0 & 0 & 0 \\ \hline
\frac{1}{2} & -\frac{1}{2} & 0 & 0 & \frac{1}{2} & 0 & \frac{1}{2} & 0 \\
\frac{1}{2} & 0 & -\frac{1}{2} & 0 & \frac{1}{2} & \frac{1}{2} & 0 & 0 \\
\frac{1}{2} & 0 & 0 & \frac{1}{2}-\frac{1}{2b^2} & 0 & \frac{1}{4b^2} & \frac{1}{4b^2} & 0 \\ \hline
0 & \frac{1}{4a^2} & \frac{1}{4a^2} & 0 & \frac{1}{2}-\frac{1}{2a^2} & 0 & 0 & \frac{1}{2} \\
0 & 0 & \frac{1}{2} & \frac{1}{2} & 0 & -\frac{1}{2} & 0 & \frac{1}{2} \\
0 & \frac{1}{2} & 0 & \frac{1}{2} & 0 & 0 & -\frac{1}{2} & \frac{1}{2} \\ \hline
0 & 0 & 0 & 0 & \frac{1}{2} & \frac{1}{4a^2} & \frac{1}{4a^2} & \frac{1}{2}-\frac{1}{2a^2}
\end{array}\right].$$

Next, we consider the entries of the matrix $\overline{B}$. For points in $P$, the corresponding row of $\overline{B}$ contains exactly one entry equal to 1 and zeros elsewhere.

For points in $E$, due to symmetry, the tangent point lies exactly in the middle between the two associated corner points. Thus, the corresponding row of the matrix $\overline{B}$ for points from $E$ contains two entries equal to $1/2$ and zeros elsewhere.

The upper and lower facets of the realization are, due to symmetry, regular $n$-gons. Hence, the corresponding row of the matrix $\overline{B}$ contains exactly $n$ entries of value $1/n$ and zeros elsewhere.

The same applies to the side facets, which consist of four points with right angles. The side lengths are each $2\sin\left(\frac{\pi}{n}\right)$, as shown in Figure 5.11. Thus, the corresponding rows of the matrix $\overline{B}$ contain four non-zero entries, each with value $1/4$.

The central point $o$ lies on $2n$ hexahedra. Due to symmetry, the corresponding row of $\overline{B}$ consists of $2n$ entries each equal to $1/(2n)$.

With this, in the next step, we can explicitly determine the matrix $\tilde{C}$ as the product of the matrices $\overline{M}$ and $\overline{B}$. Let

$$\tilde{C}_2 := \left[\begin{array}{ccccc|ccccc|c}
\frac{1}{2}-\frac{1}{2b^2} & & & & & \frac{1}{4b^2} & \frac{1}{4b^2} & 0 & \cdots & 0 & 0 \\
 & \ddots & & & & 0 & \ddots & \ddots & \ddots & \vdots & \vdots \\
 & & \ddots & & & \vdots & \ddots & \ddots & \ddots & 0 & \vdots \\
 & & & \ddots & & 0 & 0 & \ddots & \ddots & \frac{1}{4b^2} & \vdots \\
 & & & & \frac{1}{2}-\frac{1}{2b^2} & \frac{1}{4b^2} & 0 & \cdots & 0 & \frac{1}{4b^2} & 0 \\ \hline
\frac{1}{4} & 0 & \cdots & 0 & \frac{1}{4} & -\frac{1}{2} & & & & & \frac{1}{2} \\
\frac{1}{4} & \ddots & \ddots & 0 & 0 & & \ddots & & & & \vdots \\
0 & \ddots & \ddots & \ddots & \vdots & & & \ddots & & & \vdots \\
\vdots & \ddots & \ddots & \ddots & 0 & & & & \ddots & & \vdots \\
0 & \cdots & 0 & \frac{1}{4} & \frac{1}{4} & & & & & -\frac{1}{2} & \frac{1}{2} \\ \hline
0 & \cdots & \cdots & \cdots & 0 & \frac{1}{2na^2} & \cdots & \cdots & \cdots & \frac{1}{2na^2} & \frac{1}{2}-\frac{1}{2a^2}
\end{array}\right].$$

Combining the corresponding rows from $M'$ or $\overline{M}$ with the described values of the matrix $\overline{B}$, we get

$$\tilde{C} = \begin{bmatrix} \tilde{C}_2 & \frac{1}{2}E_m & 0 \\ \frac{1}{4}E_m & \tilde{C}_2 & \frac{1}{4}E_m \\ 0 & \frac{1}{2}E_m & \tilde{C}_2 \end{bmatrix}$$

and

$$2\tilde{C} = \begin{bmatrix} 2\tilde{C}_2 & E_m & 0 \\ \frac{1}{2}E_m & 2\tilde{C}_2 & \frac{1}{2}E_m \\ 0 & E_m & 2\tilde{C}_2 \end{bmatrix} = \begin{bmatrix} \left(2\tilde{C}_2 + E_m\right) - E_m & E_m & 0 \\ \frac{1}{2}E_m & \left(2\tilde{C}_2 + E_m\right) - E_m & \frac{1}{2}E_m \\ 0 & E_m & \left(2\tilde{C}_2 + E_m\right) - E_m \end{bmatrix}. \tag{6.10}$$

Next, we want to decompose $2\tilde{C}$ into the matrices $\tilde{L}$ and $\tilde{R}$ as in Algorithm 16. We first split the matrix $2\tilde{C}_2 + E_m$, whose rows sum to 2, using the approach of Algorithm 16. Here, the matrix $2\tilde{C}_2 + E_m$ can be interpreted as a matrix with the structure of a facet. The points are arranged as in equation (6.9) such that the matrix $2\tilde{C}$ can be decomposed into the two triangular matrices

$$\tilde{L}_2 = \left[\begin{array}{ccccc|ccccc|c}
1 & & & & & 0 & \dots & \dots & \dots & 0 & 0 \\
& \ddots & & & & \vdots & & & & \vdots & \vdots \\
& & \ddots & & & \vdots & & & & \vdots & \vdots \\
& & & \ddots & & \vdots & & & & \vdots & \vdots \\
& & & & 1 & 0 & \dots & \dots & \dots & 0 & 0 \\
\hline
\frac{1}{2} & 0 & \dots & 0 & \frac{1}{2} & 0 & & & & & 0 \\
\frac{1}{2} & \ddots & \ddots & 0 & 0 & & \ddots & & & & \vdots \\
0 & \ddots & \ddots & \ddots & \vdots & & & \ddots & & & \vdots \\
\vdots & \ddots & \ddots & \ddots & 0 & & & & \ddots & & \vdots \\
0 & \dots & 0 & \frac{1}{2} & \frac{1}{2} & & & & & 0 & 0 \\
\hline
0 & \dots & \dots & \dots & 0 & \frac{1}{na^2} & \dots & \dots & \dots & \frac{1}{na^2} & 1 - \frac{1}{a^2}
\end{array}\right]$$

and

$$\tilde{R}_2 = \left[\begin{array}{ccccc|ccccc|c}
1 - \frac{1}{b^2} & & & & & \frac{1}{2b^2} & \frac{1}{2b^2} & 0 & \dots & 0 & 0 \\
& \ddots & & & & 0 & \ddots & \ddots & \ddots & \vdots & \vdots \\
& & \ddots & & & \vdots & \ddots & \ddots & \ddots & 0 & \vdots \\
& & & \ddots & & 0 & 0 & \ddots & \ddots & \frac{1}{2b^2} & \vdots \\
& & & & 1 - \frac{1}{b^2} & \frac{1}{2b^2} & 0 & \dots & 0 & \frac{1}{2b^2} & 0 \\
\hline
0 & \dots & \dots & \dots & 0 & 0 & & & & & 1 \\
\vdots & & & & \vdots & & \ddots & & & & \vdots \\
\vdots & & & & \vdots & & & \ddots & & & \vdots \\
\vdots & & & & \vdots & & & & \ddots & & \vdots \\
0 & \dots & \dots & \dots & 0 & & & & & 0 & 1 \\
\hline
0 & \dots & \dots & \dots & 0 & 0 & \dots & \dots & \dots & 0 & 1
\end{array}\right]$$

which completes the decomposition.

Now we consider the decomposition of $2\tilde{C}$. Due to the alternative arrangement, the decomposition into $\tilde{L}$ and $\tilde{R}$ does not produce two triangular matrices and therefore must be described component-wise.

The blocks $2\tilde{C}_2 + E_m$ are decomposed into the matrices $\tilde{L}_2$ and $\tilde{R}_2$. An illustration of the corresponding directed edges can be found in Figure 6.37. The edges belonging to $\tilde{L}_2$ are the horizontal edges in the left image, and the edges belonging to $\tilde{R}_2$ are the horizontal edges in the right image.

The matrices $E_m$ in the blocks $(1,2)$ and $(3,2)$ of the matrix $2\tilde{C}$ in Equation (6.10) correspond to directed edges going from points of category $p$ to $e$, from $e$ to $f$, and from $f$ to $o$. An illustration is shown in the right image of Figure 6.37. There, these edges are the vertical edges pointing toward the middle layer. These two blocks thus belong to the matrix $\tilde{R}$ in the decomposition.

Conversely, the matrices $\frac{1}{2}E_m$ in blocks $(2,1)$ and $(2,3)$ correspond to edges going from points of category $o$ to $f$, from $f$ to $e$, and from $e$ to $p$. An illustration is shown in the left image of Figure 6.37. There, these edges are also vertical edges. These two blocks thus belong to the matrix $\tilde{L}$ in the decomposition.

With this decomposition and the knowledge that the diagonal entries of both matrices equal one minus the sum of the other entries, we obtain

$$\tilde{L} = \begin{bmatrix} \tilde{L}_2 & 0 & 0 \\ \frac{1}{2}E_m & \tilde{L}_2 - E_m & \frac{1}{2}E_m \\ 0 & 0 & \tilde{L}_2 \end{bmatrix} \quad \text{and} \quad \tilde{R} = \begin{bmatrix} \tilde{R}_2 - E_m & E_m & 0 \\ 0 & \tilde{R}_2 & 0 \\ 0 & E_m & \tilde{R}_2 - E_m \end{bmatrix}.$$

The matrices $L'$ and $R'$ are then given by

$$L' = \ln(2) \begin{bmatrix} \tilde{L}_2 - E_m & 0 & 0 \\ \frac{1}{2}E_m & \tilde{L}_2 - 2E_m & \frac{1}{2}E_m \\ 0 & 0 & \tilde{L}_2 - E_m \end{bmatrix} \quad \text{and} \quad R' = \ln(2) \begin{bmatrix} \tilde{R}_2 - 2E_m & E_m & 0 \\ 0 & \tilde{R}_2 - E_m & 0 \\ 0 & E_m & \tilde{R}_2 - 2E_m \end{bmatrix}.$$

We split these two matrices into two sums. This gives

$$L' = \ln(2) \begin{bmatrix} \tilde{L}_2 - E_m & 0 & 0 \\ 0 & \tilde{L}_2 - E_m & 0 \\ 0 & 0 & \tilde{L}_2 - E_m \end{bmatrix} + \ln(2) \begin{bmatrix} 0 & 0 & 0 \\ \frac{1}{2}E_m & -E_m & \frac{1}{2}E_m \\ 0 & 0 & 0 \end{bmatrix}$$

and

$$R' = \ln(2) \begin{bmatrix} \tilde{R}_2 - E_m & 0 & 0 \\ 0 & \tilde{R}_2 - E_m & 0 \\ 0 & 0 & \tilde{R}_2 - E_m \end{bmatrix} + \ln(2) \begin{bmatrix} -E_m & E_m & 0 \\ 0 & 0 & 0 \\ 0 & E_m & -E_m \end{bmatrix}.$$

These summands commute, because

$$\begin{aligned} &\ln(2) \begin{bmatrix} \tilde{L}_2 - E_m & 0 & 0 \\ 0 & \tilde{L}_2 - E_m & 0 \\ 0 & 0 & \tilde{L}_2 - E_m \end{bmatrix} \cdot \ln(2) \begin{bmatrix} 0 & 0 & 0 \\ \frac{1}{2}E_m & -E_m & \frac{1}{2}E_m \\ 0 & 0 & 0 \end{bmatrix} \\ &= \ln(2)^2 \begin{bmatrix} 0 & 0 & 0 \\ \frac{1}{2}(\tilde{L}_2 - E_m) & -(\tilde{L}_2 - E_m) & \frac{1}{2}(\tilde{L}_2 - E_m) \\ 0 & 0 & 0 \end{bmatrix} \\ &= \ln(2) \begin{bmatrix} 0 & 0 & 0 \\ \frac{1}{2}E_m & -E_m & \frac{1}{2}E_m \\ 0 & 0 & 0 \end{bmatrix} \cdot \ln(2) \begin{bmatrix} \tilde{L}_2 - E_m & 0 & 0 \\ 0 & \tilde{L}_2 - E_m & 0 \\ 0 & 0 & \tilde{L}_2 - E_m \end{bmatrix} \end{aligned}$$

and

$$\begin{aligned}
&\ln(2)\begin{bmatrix}\tilde{R}_2-E_m & 0 & 0\\ 0 & \tilde{R}_2-E_m & 0\\ 0 & 0 & \tilde{R}_2-E_m\end{bmatrix}\cdot\ln(2)\begin{bmatrix}-E_m & E_m & 0\\ 0 & 0 & 0\\ 0 & E_m & -E_m\end{bmatrix}\\
&=\ln(2)^2\begin{bmatrix}-(\tilde{R}_2-E_m) & (\tilde{R}_2-E_m) & 0\\ 0 & 0 & 0\\ 0 & (\tilde{R}_2-E_m) & -(\tilde{R}_2-E_m)\end{bmatrix}\\
&=\ln(2)\begin{bmatrix}-E_m & E_m & 0\\ 0 & 0 & 0\\ 0 & E_m & -E_m\end{bmatrix}\cdot\ln(2)\begin{bmatrix}\tilde{R}_2-E_m & 0 & 0\\ 0 & \tilde{R}_2-E_m & 0\\ 0 & 0 & \tilde{R}_2-E_m\end{bmatrix}.
\end{aligned}$$

Now, we have all prerequisites to express the subdivision matrix $S := S^{(3)}$. We get

$$\begin{aligned}
S &= \exp(L')\cdot\exp(R')\\
&=\exp\left(\ln(2)\begin{bmatrix}\tilde{L}_2-E_m & 0 & 0\\ 0 & \tilde{L}_2-E_m & 0\\ 0 & 0 & \tilde{L}_2-E_m\end{bmatrix}+\ln(2)\begin{bmatrix}0 & 0 & 0\\ \frac{1}{2}E_m & -E_m & \frac{1}{2}E_m\\ 0 & 0 & 0\end{bmatrix}\right)\\
&\quad\cdot\exp\left(\ln(2)\begin{bmatrix}\tilde{R}_2-E_m & 0 & 0\\ 0 & \tilde{R}_2-E_m & 0\\ 0 & 0 & \tilde{R}_2-E_m\end{bmatrix}+\ln(2)\begin{bmatrix}-E_m & E_m & 0\\ 0 & 0 & 0\\ 0 & E_m & -E_m\end{bmatrix}\right).
\end{aligned}$$

Since the summands commute, we can use Theorem B.53 to write the exponential of sums as products:

$$\begin{aligned}
S &= \exp\left(\ln(2)\begin{bmatrix}\tilde{L}_2-E_m & 0 & 0\\ 0 & \tilde{L}_2-E_m & 0\\ 0 & 0 & \tilde{L}_2-E_m\end{bmatrix}\right)\exp\left(\ln(2)\begin{bmatrix}0 & 0 & 0\\ \frac{1}{2}E_m & -E_m & \frac{1}{2}E_m\\ 0 & 0 & 0\end{bmatrix}\right)\\
&\quad\cdot\exp\left(\ln(2)\begin{bmatrix}-E_m & E_m & 0\\ 0 & 0 & 0\\ 0 & E_m & -E_m\end{bmatrix}\right)\exp\left(\ln(2)\begin{bmatrix}\tilde{R}_2-E_m & 0 & 0\\ 0 & \tilde{R}_2-E_m & 0\\ 0 & 0 & \tilde{R}_2-E_m\end{bmatrix}\right).
\end{aligned}$$

Using Lemma B.66 for the exponential of the block diagonal matrices, and Corollaries B.73 and B.76 for the other exponentials, we get

$$\begin{aligned}
S &= \begin{bmatrix}\exp\left(\ln(2)(\tilde{L}_2-E_m)\right) & 0 & 0\\ 0 & \exp\left(\ln(2)(\tilde{L}_2-E_m)\right) & 0\\ 0 & 0 & \exp\left(\ln(2)(\tilde{L}_2-E_m)\right)\end{bmatrix}\begin{bmatrix}E_m & 0 & 0\\ \frac{1}{4}E_m & \frac{1}{2}E_m & \frac{1}{4}E_m\\ 0 & 0 & E_m\end{bmatrix}\\
&\quad\cdot\begin{bmatrix}\frac{1}{2}E_m & \frac{1}{2}E_m & 0\\ 0 & E_m & 0\\ 0 & \frac{1}{2}E_m & \frac{1}{2}E_m\end{bmatrix}\begin{bmatrix}\exp\left(\ln(2)(\tilde{R}_2-E_m)\right) & 0 & 0\\ 0 & \exp\left(\ln(2)(\tilde{R}_2-E_m)\right) & 0\\ 0 & 0 & \exp\left(\ln(2)(\tilde{R}_2-E_m)\right)\end{bmatrix}.
\end{aligned}$$

Multiplying the two inner matrices yields

$$\begin{bmatrix}E_m & 0 & 0\\ \frac{1}{4}E_m & \frac{1}{2}E_m & \frac{1}{4}E_m\\ 0 & 0 & E_m\end{bmatrix}\cdot\begin{bmatrix}\frac{1}{2}E_m & \frac{1}{2}E_m & 0\\ 0 & E_m & 0\\ 0 & \frac{1}{2}E_m & \frac{1}{2}E_m\end{bmatrix}=\begin{bmatrix}\frac{1}{2}E_m & \frac{1}{2}E_m & 0\\ \frac{1}{8}E_m & \frac{3}{4}E_m & \frac{1}{8}E_m\\ 0 & \frac{1}{2}E_m & \frac{1}{2}E_m\end{bmatrix}.$$

Since this matrix commutes with the block diagonal matrix consisting of $\exp(\ln(2)(\tilde{L}_2 - E_m))$, we define

$$S_2 := \exp\left(\ln(2)(\tilde{L}_2 - E_m)\right) \exp\left(\ln(2)(\tilde{R}_2 - E_m)\right) \tag{6.11}$$

Therefore, the matrix $S$ can be written as

$$S = \begin{bmatrix} \frac{1}{2}E_m & \frac{1}{2}E_m & 0 \\ \frac{1}{8}E_m & \frac{3}{4}E_m & \frac{1}{8}E_m \\ 0 & \frac{1}{2}E_m & \frac{1}{2}E_m \end{bmatrix} \begin{bmatrix} S_2 & 0 & 0 \\ 0 & S_2 & 0 \\ 0 & 0 & S_2 \end{bmatrix} = \begin{bmatrix} \frac{1}{2}S_2 & \frac{1}{2}S_2 & 0 \\ \frac{1}{8}S_2 & \frac{3}{4}S_2 & \frac{1}{8}S_2 \\ 0 & \frac{1}{2}S_2 & \frac{1}{2}S_2 \end{bmatrix}.$$

This is exactly the Kronecker product of the matrix

$$\begin{bmatrix} \frac{1}{2} & \frac{1}{2} & 0 \\ \frac{1}{8} & \frac{3}{4} & \frac{1}{8} \\ 0 & \frac{1}{2} & \frac{1}{2} \end{bmatrix}$$

and the matrix $S_2$, i.e.,

$$S = \text{kron}\left(\begin{bmatrix} \frac{1}{2} & \frac{1}{2} & 0 \\ \frac{1}{8} & \frac{3}{4} & \frac{1}{8} \\ 0 & \frac{1}{2} & \frac{1}{2} \end{bmatrix}, S_2\right).$$

□

The relation between the matrix $S_2$ and the matrix $S^{(2)}$ is discussed in the following corollary:

**Corollary 6.36.** *The matrix $S_2 \in \mathbb{R}^{(2n+1)\times(2n+1)}$ from Equation* (6.11) *coincides with the subdivision matrix $S^{(2)}$ for $n$ quadrilaterals. Thus, the subdivision matrices of the volumetric semi-regular case arise from the Kronecker product of the two-dimensional case and the regular one-dimensional rule.*

*Proof.* We explicitly determine the matrix $\tilde{C}$ from Algorithm 15 for the two-dimensional subdivision matrix $S^{(2)}$. As in the above proof, all quadrilaterals have the same shape. Choosing a quadrilateral depending on $n$, we obtain from Equation (3.16) for $p$ with $j = 0$, and from Equation (3.14) for $e_1$ with $j = 0$ and for $e_2$ with $j = 1$ the points

$$p = \left[1, \frac{1-\cos\left(\frac{2\pi}{n}\right)}{\sin\left(\frac{2\pi}{n}\right)}\right], \qquad e_2 = \left[\cos\left(\frac{2\pi}{n}\right), \sin\left(\frac{2\pi}{n}\right)\right],$$

$$e_1 = [1, 0], \qquad f = [0, 0].$$

This quadrilateral is shifted by $-p/2$ and then the corresponding Colin-de-Verdière-matrix is constructed. We first define

$$a := \cos\left(\frac{\pi}{n}\right), \quad \text{and} \quad b := \sin\left(\frac{\pi}{n}\right).$$

Using a computer algebra system (CAS), we obtain for the matrix $M'$ from line 10 of Algorithm 15

$$M' = \begin{bmatrix} 1 - \frac{1}{2b^2} & \frac{1}{4b^2} & \frac{1}{4b^2} & 0 \\ \frac{1}{2} & 0 & 0 & \frac{1}{2} \\ \frac{1}{2} & 0 & 0 & \frac{1}{2} \\ 0 & \frac{1}{4a^2} & \frac{1}{4a^2} & 1 - \frac{1}{2a^2} \end{bmatrix}.$$

For the entries of the rows of $\overline{B}$, as in the above proof, for vertex points there is one entry of 1, for edge points there are two entries of $1/2$, and for the central facet point there are $n$ entries of $1/n$. Hence, we obtain

$$\tilde{C} = \left[\begin{array}{ccccc|ccccc|c}
1-\frac{1}{2b^2} & & & & & \frac{1}{4b^2} & \frac{1}{4b^2} & 0 & \dots & 0 & 0 \\
 & \ddots & & & & 0 & \ddots & \ddots & \ddots & \vdots & \vdots \\
 & & \ddots & & & \vdots & \ddots & \ddots & \ddots & 0 & \vdots \\
 & & & \ddots & & 0 & 0 & \ddots & \ddots & \frac{1}{4b^2} & \vdots \\
 & & & & 1-\frac{1}{2b^2} & \frac{1}{4b^2} & 0 & \dots & 0 & \frac{1}{4b^2} & 0 \\
\hline
\frac{1}{4} & 0 & \dots & 0 & \frac{1}{4} & 0 & & & & & \frac{1}{2} \\
\frac{1}{4} & \ddots & \ddots & 0 & 0 & & \ddots & & & & \vdots \\
0 & \ddots & \ddots & \ddots & \vdots & & & \ddots & & & \vdots \\
\vdots & \ddots & \ddots & \ddots & 0 & & & & \ddots & & \vdots \\
0 & \dots & 0 & \frac{1}{4} & \frac{1}{4} & & & & & 0 & \frac{1}{2} \\
\hline
0 & \dots & \dots & \dots & 0 & \frac{1}{2na^2} & \dots & \dots & \dots & \frac{1}{2na^2} & 1-\frac{1}{2a^2}
\end{array}\right].$$

If we split the matrix $2\tilde{C}$ into the matrices $\tilde{L}$ and $\tilde{R}$, we obtain

$$\tilde{L} = \left[\begin{array}{ccccc|ccccc|c}
1 & & & & & 0 & \dots & \dots & \dots & 0 & 0 \\
 & \ddots & & & & \vdots & & & & \vdots & \vdots \\
 & & \ddots & & & \vdots & & & & \vdots & \vdots \\
 & & & \ddots & & \vdots & & & & \vdots & \vdots \\
 & & & & 1 & 0 & \dots & \dots & \dots & 0 & 0 \\
\hline
\frac{1}{2} & 0 & \dots & 0 & \frac{1}{2} & 0 & & & & & 0 \\
\frac{1}{2} & \ddots & \ddots & 0 & 0 & & \ddots & & & & \vdots \\
0 & \ddots & \ddots & \ddots & \vdots & & & \ddots & & & \vdots \\
\vdots & \ddots & \ddots & \ddots & 0 & & & & \ddots & & \vdots \\
0 & \dots & 0 & \frac{1}{2} & \frac{1}{2} & & & & & 0 & 0 \\
\hline
0 & \dots & \dots & \dots & 0 & \frac{1}{na^2} & \dots & \dots & \dots & \frac{1}{na^2} & 1-\frac{1}{a^2}
\end{array}\right]$$

and

$$\tilde{R} = \left[\begin{array}{ccccc|ccccc|c}
1-\frac{1}{b^2} & & & & & \frac{1}{2b^2} & \frac{1}{2b^2} & 0 & \dots & 0 & 0 \\
 & \ddots & & & & 0 & \ddots & \ddots & \ddots & \vdots & \vdots \\
 & & \ddots & & & \vdots & \ddots & \ddots & \ddots & 0 & \vdots \\
 & & & \ddots & & 0 & 0 & \ddots & \ddots & \frac{1}{2b^2} & \vdots \\
 & & & & 1-\frac{1}{b^2} & \frac{1}{2b^2} & 0 & \dots & 0 & \frac{1}{2b^2} & 0 \\
\hline
0 & \dots & \dots & \dots & 0 & 0 & & & & & 1 \\
\vdots & & & & \vdots & & \ddots & & & & \vdots \\
\vdots & & & & \vdots & & & \ddots & & & \vdots \\
\vdots & & & & \vdots & & & & \ddots & & \vdots \\
0 & \dots & \dots & \dots & 0 & & & & & 0 & 1 \\
\hline
0 & \dots & \dots & \dots & 0 & 0 & \dots & \dots & \dots & 0 & 1
\end{array}\right].$$

The matrices $\tilde{L}$ and $\tilde{R}$ coincide with the two matrices $\tilde{L}_2$ and $\tilde{R}_2$. Thus, for $S_2$ from equation (6.11) we obtain

$$S_2 = \exp\left(\ln(2)(\tilde{L}_2 - E_n)\right)\exp\left(\ln(2)(\tilde{R}_2 - E_n)\right) = \exp\left(\ln(2)(\tilde{L} - E_n)\right)\exp\left(\ln(2)(\tilde{R} - E_n)\right) = S^{(2)},$$

and therefore $S_2$ and $S^{(2)}$ coincide. □

From these two statements, the following corollary can be derived analogously to Chapter 5:

**Corollary 6.37.** *The matrices $S^{(3)}$ satisfy the quality criteria Q7 and Q8.*

*Proof.* The regular case, and thus quality criterion Q7, follows by simply inserting the corresponding adjacency matrix into the algorithm. Afterwards, the result must be permuted with an appropriate permutation matrix to obtain the Kronecker product matrix.

Since by Theorem 6.35 and Corollary 6.36 the subdivision matrix $S^{(3)}$ is the Kronecker product of the one-dimensional and the corresponding two-dimensional rule, quality criterion Q8 also holds. □

Finally, we discuss the tensor product structure of the double ring and double shell matrices with the following theorem:

**Theorem 6.38.** *The matrices $\overline{S}^{(3)}$ satisfy the quality criteria Q7 and Q8.*

*Proofsketch.* Since in the regular case the subdivision matrix consists only of regular initial elements, quality criterion Q7 is satisfied.

For the semi-regular case, the subdivision matrices consist of the central semi-regular initial element, two times four opposite semi-regular initial elements, and otherwise only regular initial elements. Thus, the tensor product property transfers accordingly, and quality criterion Q8 is also satisfied. □

## Q9 and Q13 Input and Injectivity

Since the input in Algorithm 12 is a valid combinatorial graph $\mathbf{G}_K$, which for $t = 2$ must be a cycle and for $t = 3$ must be a 3-connected and planar graph, the input corresponds to the validity concept from Definition 2.51. The double ring and double shell matrices are composed accordingly from the information of the initial elements, and since these all come from valid combinatorial graphs, every double ring and double shell matrix can be generated with a central valid initial element. Therefore, quality criterion Q9 is fulfilled for all variants.

It is important to note here that we assumed the hexahedral structure of the cells for the case $t = 3$ as a fundamental prerequisite and used this in the construction. It is therefore not easily possible to extend the construction to planar 3-connected graphs with nodes where more than three edges meet. The reasons for this are discussed in Chapter 7.

The injectivity for the generalized cubic B-spline subdivision remains open. For $t = 2$ it is basically provable. For this, one would first need to determine the matrix entries explicitly. Since by Theorem 6.33 and Theorem 6.34 all symmetries hold, the discrete Fourier transform could be applied. Using this, it would first have to be proven that $1/2$ is the subdominant eigenvalue of the subdivision matrix. Then injectivity could be checked with the same strategy as in [PR08, Sec. 6.1, pp. 109–116]. However, since this requires many terms and theories not used in this work, we leave the elaboration of this sketch to future research.

For $t = 3$, as in Chapter 5, injectivity and regularity are completely open, and there is no idea yet on how to prove them. Empirically, the same situation as in Chapter 5 applies here. To make a qualitative statement about injectivity, at least a visual inspection of the characteristic shell would be necessary.

For the two counterexamples from Section 2.5, we obtain the characteristic shells shown in Figure 6.40. For these examples, the subdivision algorithm presented in this chapter provides satisfactory results.

Looking at the set of quality criteria as a whole, it has been possible to construct subdivision matrices that satisfy Quality Criterion Q11. Although the desired eigenstructure could not be proven for these subdivision matrices, empirical results suggest that it is indeed constructed. Moreover, the presented approach produces acceptable injective eigenstructures for the counterexamples to the algorithms from [Baj+02] and [JM99] discussed in Section 2.5. Thus, a genuine improvement compared to previously established methods in the literature has been achieved here as well.

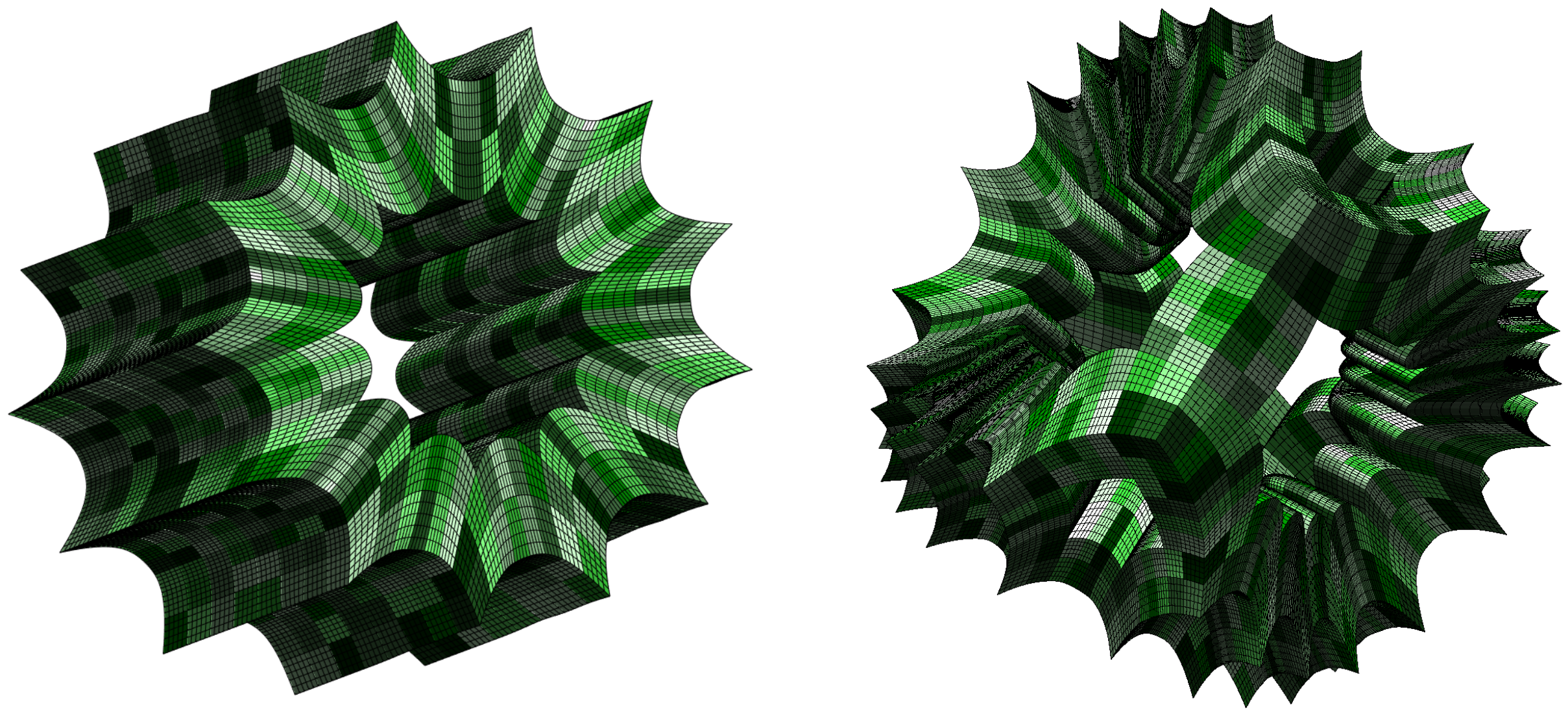

**Figure 6.40:** The examples from Figure 2.36 (left) and Figure 2.37 (right), created with the algorithm of this chapter. It can be seen that both shells are injective.

Criticism of the presented construction is that the desired eigenstructure has not yet been proven, and that the subsubdominant eigenvalue approaches the value $1/2$. Counterexamples or proof ideas for the first criticism, and alternative constructions for the second, can be topics for further research.

Overall, it can nevertheless be stated that the use of the subdivision matrices presented here is preferable, especially in comparison to established methods.

We conclude this section with Figure 6.41, which shows a gallery consisting of images of the characteristic map of various subdivision matrices generated with Algorithm 12.

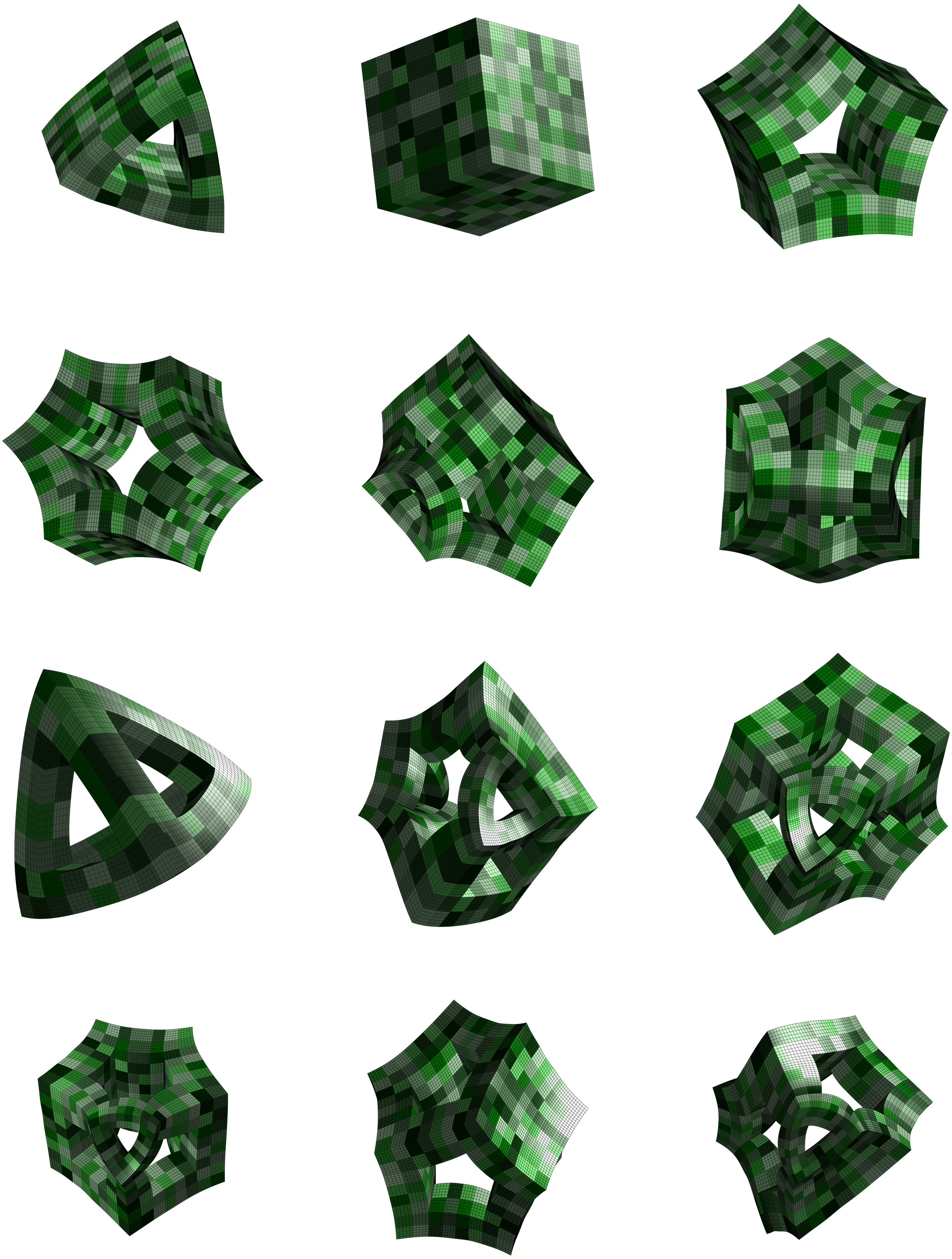

**Figure 6.41:** Gallery of exemplary images of the characteristic map of subdivision matrices for the 3-polytopes from Figure 3.1, generated with Algorithm 12.

# 7 Research Approaches for Arbitrary Volumetric Meshes

At first glance into the work of Catmull and Clark, the goal of their paper becomes clear right in the first sentence of the abstract [CC78, p. 350]:

> This paper describes a method for recursively generating surfaces that approximate points lying-on a mesh of arbitrary topology.

The key word in this sentence is „arbitrary". Catmull and Clark generate refinements of arbitrary two-dimensional meshes. In a first step, they convert any mesh into a quad mesh by subdividing each facet of the mesh into a set of quadrilaterals (cf. [CC78, pp. 351–352]). An exemplary subdivision is shown in Figure 7.1. This allows the fundamental concepts of the two-dimensional case, starting from the cells, to be defined on a quadrilateral base structure.

In this chapter, we want to explore to what extent this generality of meshes can be transferred to the volumetric case, and therefore focus exclusively on the case $t = 3$. Examples are given only for the case $g = 2$, since for $g = 3$ essential aspects of the approach described here are still missing. We point out the background information at the appropriate places.

This chapter should be understood only as a sketch of ideas and not as a fully developed theory. In particular, it is descriptive and exemplary. However, the presented outline may and should serve as a starting point for further research.

Before we start with the main content, we need some definitions. We begin with the structure of an $n$-gon:

**Definition 7.1.** *Let $n \in \mathbb{N}$ with $n \geq 3$. A graph $\boldsymbol{G} = (\boldsymbol{V}, \boldsymbol{E})$ with $|\boldsymbol{V}| = n$, whose nodes can be arranged so that the adjacency matrix takes the form*

$$T_n := A = \begin{bmatrix} 0 & 1 & 0 & \dots & \dots & 0 & 1 \\ 1 & 0 & 1 & 0 & \dots & \dots & 0 \\ 0 & 1 & 0 & 1 & 0 & \dots & 0 \\ \vdots & \ddots & \ddots & \ddots & \ddots & \ddots & \vdots \\ 0 & \dots & 0 & 1 & 0 & 1 & 0 \\ 0 & \dots & \dots & 0 & 1 & 0 & 1 \\ 1 & 0 & \dots & \dots & 0 & 1 & 0 \end{bmatrix} \in \{0,1\}^{n \times n}$$

*is called the* graph of an $n$-gon.

With this structure, we can describe a special family of 3-polytopes: the trapezohedra. A geometric description can be found, for example, in [CR61, p. 117]. We describe the structure using graphs as follows:

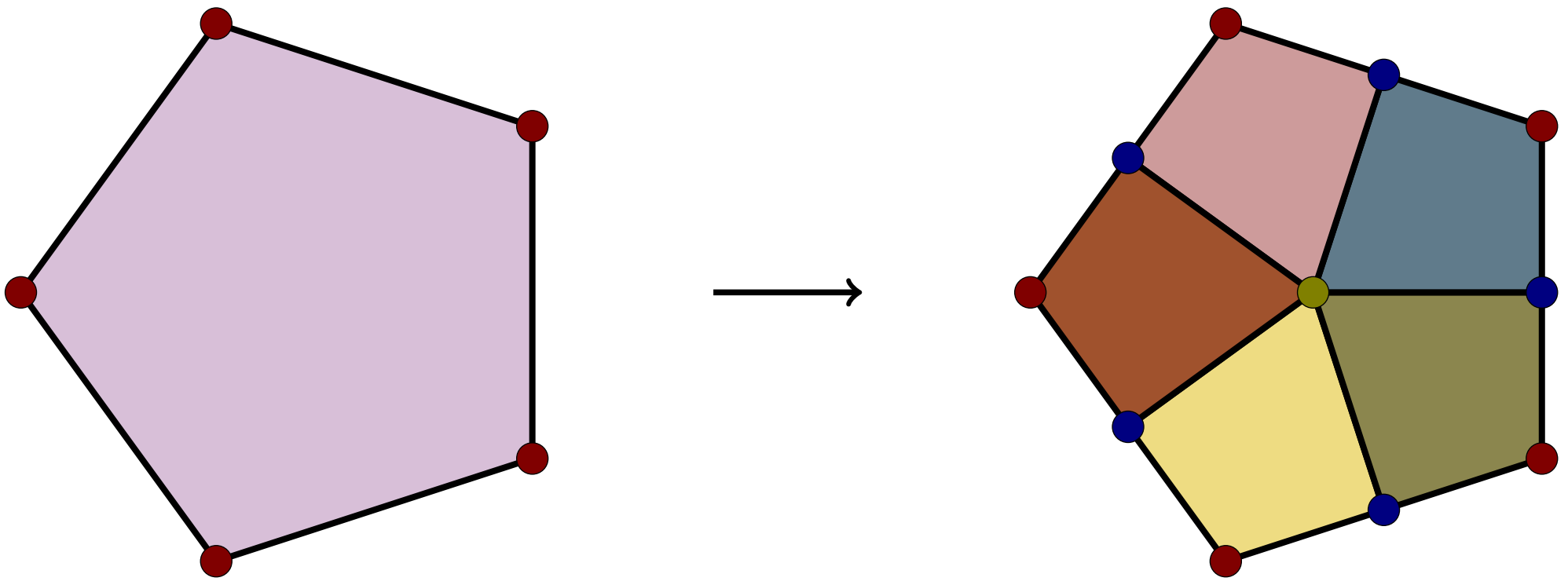

**Figure 7.1:** Example decomposition of a pentagon into five quadrilaterals.

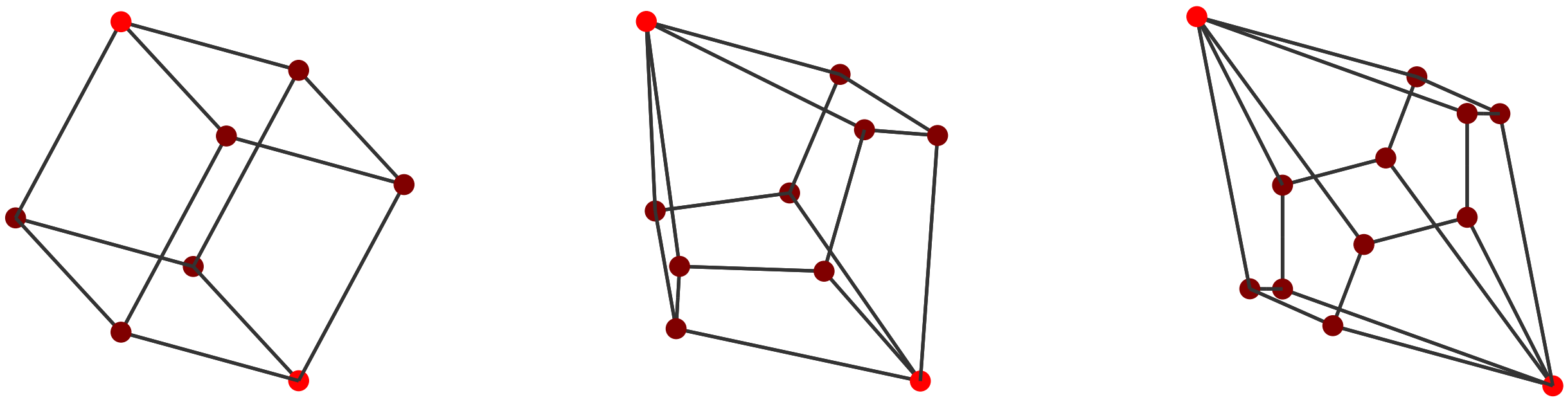

**Figure 7.2:** From left to right: illustrations of a 3-gonal, 4-gonal, and 5-gonal trapezohedron with highlighted $n$-fold nodes (light red).

**Definition 7.2.** *Let $n \in \mathbb{N}$ with $n \geq 3$. A graph $\boldsymbol{G} = (\boldsymbol{V}, \boldsymbol{E})$ with $|\boldsymbol{V}| = 2n + 2$ and $\boldsymbol{V} = \{v_1, \ldots, v_{2n+2}\}$, whose nodes can be arranged so that the adjacency matrix takes the form*

$$A = \left[\begin{array}{c|ccccccc|c} 0 & 1 & 0 & 1 & \ldots & 0 & 1 & 0 & 0 \\ \hline 1 & 0 & 1 & 0 & 0 & \ldots & 0 & 1 & 0 \\ 0 & 1 & 0 & 1 & 0 & 0 & \ldots & 0 & 1 \\ 1 & 0 & 1 & 0 & 1 & 0 & \ldots & 0 & 0 \\ \vdots & \vdots & \ddots & \ddots & \ddots & \ddots & \ddots & \vdots & \vdots \\ 0 & 0 & \ldots & 0 & 1 & 0 & 1 & 0 & 1 \\ 1 & 0 & 0 & \ldots & 0 & 1 & 0 & 1 & 0 \\ 0 & 1 & 0 & \ldots & 0 & 0 & 1 & 0 & 1 \\ \hline 0 & 0 & 1 & 0 & \ldots & 1 & 0 & 1 & 0 \end{array}\right] = \left[\begin{array}{c|ccccccc|c} 0 & 1 & 0 & 1 & \ldots & 0 & 1 & 0 & 0 \\ \hline 1 & & & & & & & & 0 \\ 0 & & & & & & & & 1 \\ 1 & & & & & & & & 0 \\ \vdots & & & & T_{2n} & & & & \vdots \\ 0 & & & & & & & & 1 \\ 1 & & & & & & & & 0 \\ 0 & & & & & & & & 1 \\ \hline 0 & 0 & 1 & 0 & \ldots & 1 & 0 & 1 & 0 \end{array}\right]$$

*has the name* graph of an $n$-gonal trapezohedron. *The nodes $v_1$ and $v_{2n+2}$ are called* $n$-fold nodes. *The realization of the graph as a convex 3-polytope is called a* trapezohedron.

A trapezohedron can thus be described by three components. First, the nodes $v_2, \ldots, v_{2n+1}$ form a cycle in the graph-theoretic sense. The nodes of this cycle are alternately connected to the two $n$-fold nodes. It is immediately clear that the 3-gonal trapezohedron is the hexahedron. Examples of various trapezohedra are shown in Figure 7.2. Trapezohedra are also 3-polytopes, which follows indirectly from [CR61, p. 117]. With the description of trapezohedra, the following claim can be formulated:

**Claim 7.3.** *Every 3-connected planar graph $\boldsymbol{G}' = (\boldsymbol{V}', \boldsymbol{E}')$ can be decomposed, by inserting nodes and edges, into a graph $\boldsymbol{G}$ which is the union of $|\boldsymbol{V}'|$ graphs of trapezohedra.*

*Proofsketch.* The construction of the graph **G** proceeds analogously to the construction of the graph **G** in Algorithm 14. For each node, edge, facet, and volume contained in **G**′, a node is created in **G**. These are called vertex nodes, edge nodes, facet nodes, and volume nodes.

These nodes are connected by edges according to the structure of **G**′. The vertex nodes in **G**, whose counterparts in **G**′ share an edge, are connected to the corresponding edge node in **G**. A facet node in **G** is connected to all edge nodes whose counterparts are edges of the corresponding facet in **G**′, and all facet nodes are connected to the volume node in **G**.

This results in a trapezohedron for each node of **G**′. It consists of the corresponding vertex node in **G**, the edge nodes connected to this vertex node, the facet nodes connected to these edge nodes, and the volume node. □

Considering this claim, we look at the notion of a cell from Definition 2.2. For a „mesh of arbitrary topology“, this concept must be generalized. Suppose a cell is an arbitrary convex 3-polytope in $\mathbb{R}^3$ with the properties from Theorem 3.25. It thus satisfies all symmetries, and the centroid of the tangent points lies at the origin. Then we obtain the next claim:

**Claim 7.4.** *Every 3-polytope with $n$ vertices and the properties from Theorem 3.25 can be decomposed into $n$ trapezohedra.*

*Proofsketch.* We proceed analogously to Lemma 6.7. We form the corresponding dual polytope and project its vertices onto the facets of the primal polytope. That the resulting trapezohedra are convex must be shown analogously to Lemma 6.7. □

Thus, a cell of the form of a 3-polytope decomposes in a first refinement step into trapezohedra. An exemplary illustration of the two claims can be found in Figure 7.3. If we generalize the initial notion of a cell from Chapter 2—from a hexahedron to a general 3-polytope—then by the decomposition just described, these 3-polytopes can be subdivided into trapezohedra in an initial refinement step. Hence, we achieve the required generality for a „mesh of arbitrary topology“ in the case $t = 3$ by defining the necessary objects—starting with the cells—on the basis of trapezohedra. It is worth noting that in [CC78] analogous considerations are made for the case $t = 2$, where every $n$-gon is decomposed into $n$ quadrilaterals, as already illustrated in Figure 7.1.

The trapezohedra obtained after this first step can be shaped differently depending on the form of the original polytope for the same combinatorial arrangement. At this point, a *standard trapezohedron* is therefore needed. For this, we choose the trapezohedron with the properties from Theorem 3.25. Additionally, it is rotated so that the two $n$-fold vertices lie on the $z$-axis. The trapezohedra arising from the 3-polytopes must still be mapped bijectively to the *standard trapezohedra* for the notion of a cell in this first step.

With these cells consisting of trapezohedra, the other concepts described in Chapter 2 can be analogously defined or extended. The concepts generalize as follows: Initially, nothing changes in the spline definition domain except for the modification of the cells. Control points can also be associated analogously to Chapter 2 with the cells for $g = 2$ and with the vertices of the cells for $g = 3$. Initial elements can likewise be defined analogously. However, the graphs belonging to the initial elements can now be more diverse.

The notion of regularity remains unchanged. However, cells that are $n$-gonal trapezohedra for $n > 3$ cannot be regular. This follows from the fact that the cells at the vertices corresponding to the $n$-fold nodes cannot represent a $2 \times 2 \times 2$ structure. Regular cells must, conversely, always be hexahedra; thus, only hexahedral cells can still be evaluated by tensor-product B-splines. This is consistent since an $n$-gonal trapezohedron for $n \neq 3$ does not have a tensor product structure.

Refinement can also proceed analogously. Using the decomposition described in Claim 7.4, an $n$-gonal trapezohedron is subdivided into $n$ hexahedra and two $n$-gonal trapezohedra. Consequently, we observe the behavior shown in Figure 7.4 for evaluability when $g = 2$. It should be noted for Figure 7.4 that the maximal evaluable region is plotted, assuming the trapezoidal cell is embedded in a way that maximizes the evaluable area. This differs from the similar illustration in Figure 2.9.

For the trapezoidal cell, we also see that the evaluable region grows with refinement. However, a new type of irregularity arises because the non-evaluable region extends through the increasingly refined trapezohedron (which is not a hexahedron). This forms a chain of trapezohedra whose length increases with refinement (in terms of the number of trapezohedra) but whose volume becomes thinner. Therefore, through this generalization, in addition to the irregular vertices and edges described in Definition 2.29, we obtain irregular volumes:

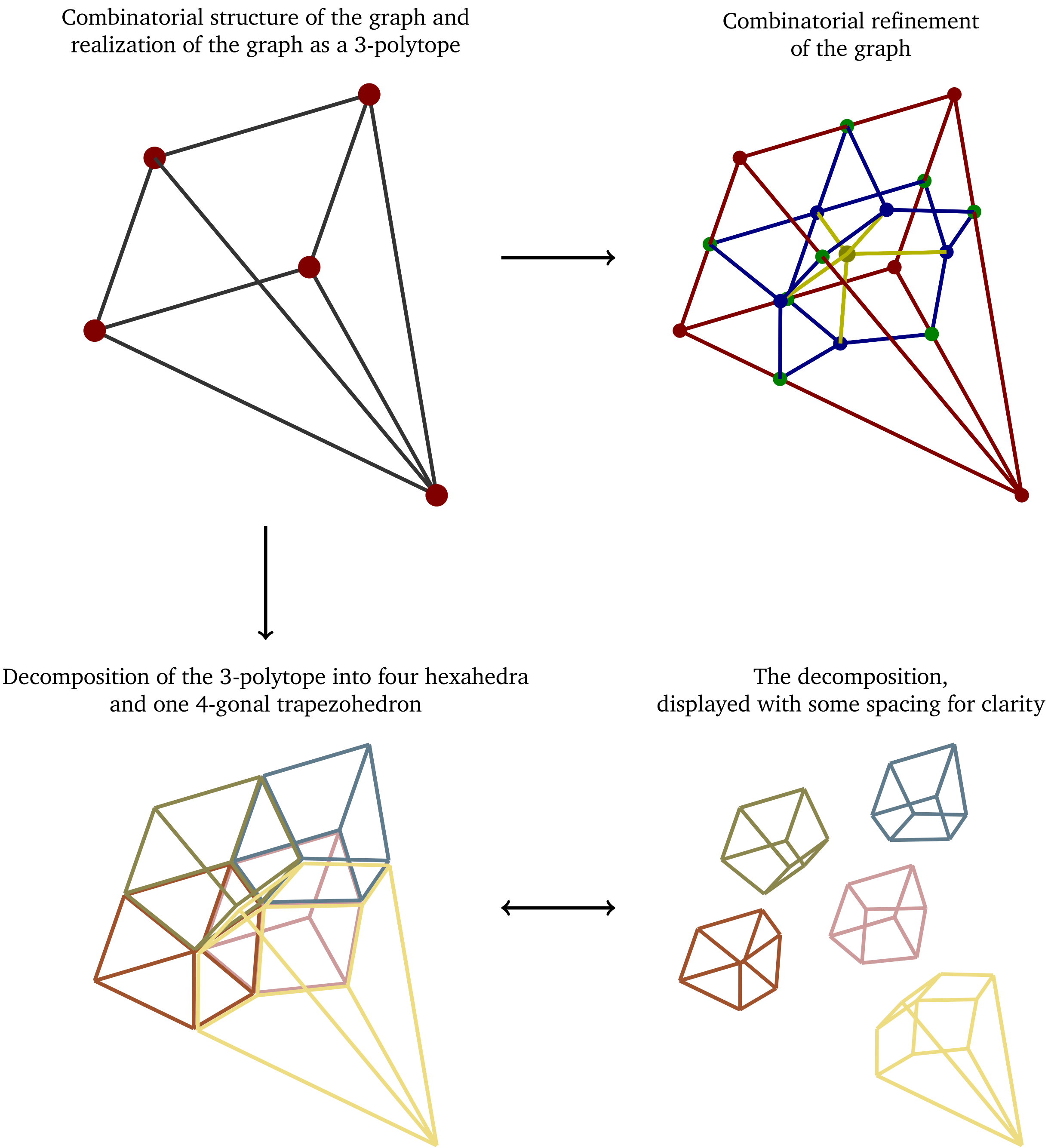

**Figure 7.3:** Example illustration of Claims 7.3 and 7.4 using the example of a four-sided pyramid. The pyramid as a combinatorial graph is shown in the top left. In the top right, it is refined combinatorially into a union of the graphs of five trapezohedra (four of which are hexahedra). The vertex nodes are colored red, the edge nodes green, the facet nodes blue, and the volume node is yellow. The bottom left image shows the pyramid as a 3-polytope decomposed into five trapezohedra (four hexahedra). These are shown with some spacing in the bottom right image for better visualization.

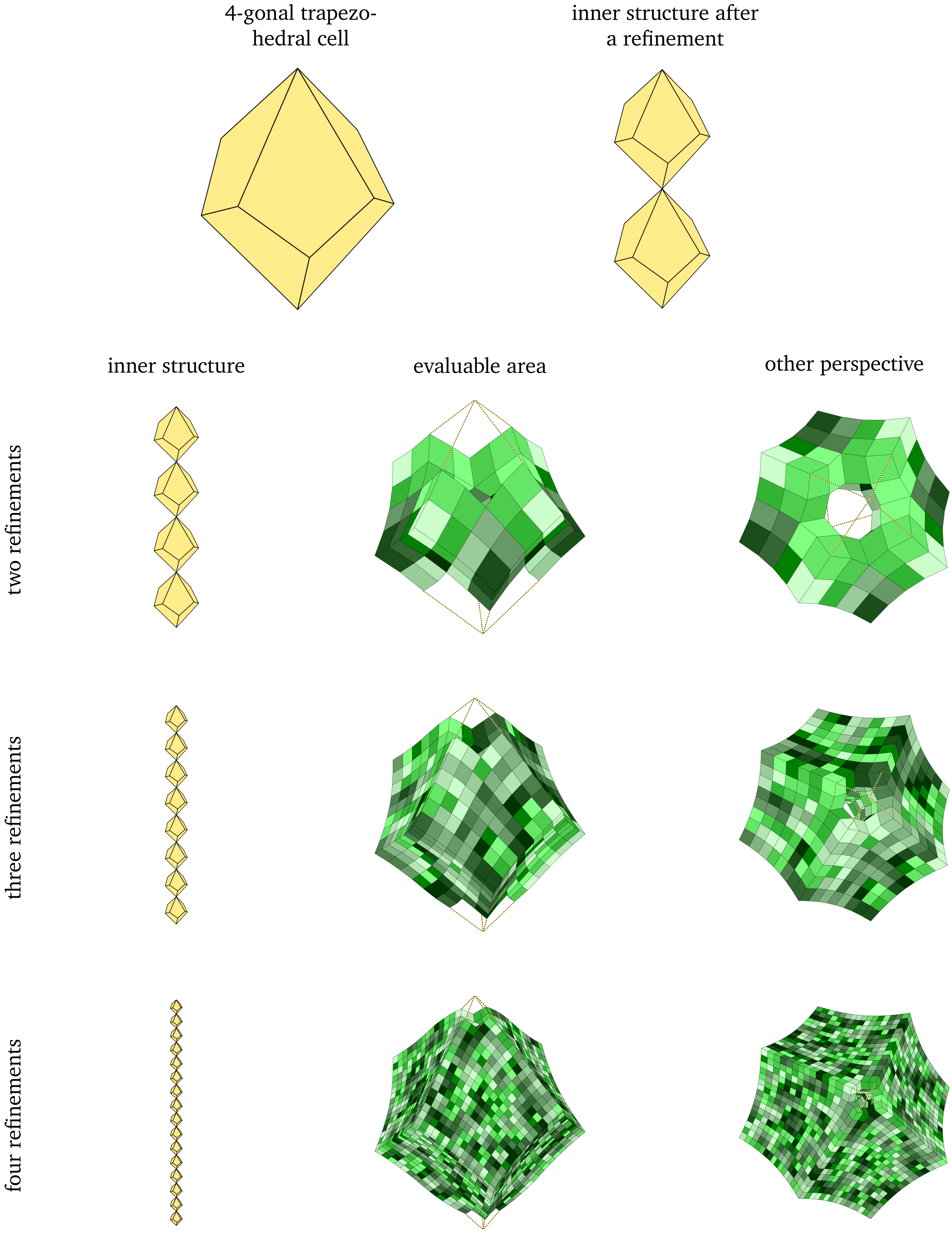

**Figure 7.4:** A 4-gonal trapezohedral cell, its inner structure, and its evaluable region (assuming an embedding where as many regions as possible are evaluable). On the top right and in the left column, the inner structure can be seen, which consists of a chain of trapezohedra. In the two right columns, the evaluable regions for each refinement step are shown. The evaluable regions are plotted as concrete B-spline evaluations and can thus be seen as a "proof of concept" for evaluability. In the top right image, the vertex shared by both trapezohedra is a semi-volume-irregular vertex according to Definition 7.6.

**Definition 7.5.** *Let $s$ be a cell that is not a hexahedron. Then we call $s$ an* irregular volume.

Such an irregular volume is the subject of further research in the context of the generalization described here. Studying such irregular volumes is already interesting on its own, since the irregularity raises many questions. For example, the limit of the refinement process and the analysis of the resulting generating functions are of interest.

Most notably, the resulting invariant structures arise. These first appear similarly to Section 2.3, producing again (double) shells. These shells also have holes in the visualization that go through the volumes themselves. This leads to the following categorization of vertices, which extends Definition 2.29:

**Definition 7.6.** *Let $t = 3$ and **D** be a spline definition domain with trapezoidal cells. A vertex of a cell is called*

1. regular *if it lies on $2 \times 2 \times 2$ hexahedra,*
2. semi-edge-irregular *if it lies on $2 \times n$, with $n \in \{3, 5, 6, \ldots\}$, hexahedra,*
3. edge-irregular *if it lies only on hexahedra but neither case 1 nor 2 applies,*
4. semi-volume-irregular *if it lies on two opposite $n$-gonal trapezohedra and $2n$ hexahedra, but case 1 does not apply,*
5. volume-irregular *if all edges connected to the vertex lie on exactly four cells, but neither case 1 nor 4 applies,*
6. super-irregular *if none of the above cases applies.*

Cases 1–3 were already treated extensively in this work. We call them (semi-)edge-irregular because only refinement allows evaluation in the region along irregular edges of the cells. In contrast, volume-irregular vertices appear.

Case 4 is analogous to the prism case (case 2) for edge-irregular points. This case describes the object that arises when an $n$-gonal trapezohedron refines into two $n$-gonal trapezohedra and $n$ hexahedra. This structure continues similarly to the prism case. This behavior can be seen in Figure 7.4. Case 5 describes the structure where all edges are regular but volumes can be irregular; this is analogous to case 3. Case 6 finally describes the structure that includes both types of irregularity. Examples for structures of cases 4, 5, and 6 can be found in Figure 7.5.

It is interesting how these invariant (double) shell structures can be described. To this end, we again consider the combinatorial graph from Definition 2.50, which can be defined analogously. The notion of validity from Definition 2.51 can also be transferred. However, here the extension outlined in Example 2.52 arises. While in the hexahedral case the combinatorial graph was a 3-connected planar graph whose vertices have exactly three neighbors, this concept generalizes to a 3-connected planar graph whose vertices have at least three neighbors. This is an important extension which we will consider separately for cases $g = 2$ and $g = 3$.

The subdivision algorithms described in Chapter 5 can easily handle this extension, since nowhere in their construction was it required that the nodes of the graph must have exactly three neighbors. The 3-polytopes can be constructed, the Colin-de-Verdière-matrix can be generated for variant 2, and the resulting matrices can be formed. Thus, the extension described in this chapter can be performed with the algorithms constructed in Chapter 5. The resulting B-spline volumes can also be generated and evaluated. Hence, this extension is practically possible and can be thoroughly analyzed and studied in future research. In particular, the quality criteria shown in Section 5.3 apply here as well. Furthermore, the empirical analysis already contained 118,589 examples of cases 4, 5, and 6. Using Quality Criterion Q3, we obtain especially the consistency for this extension; that is, the limit of the control points for $\lim_{m\to\infty} S^m P$ exists. Examples of evaluated shells for $t = 2$ for cases 4, 5, and 6 can also be found in Figure 7.5. Because irregular volumes lead to large regions that can only be evaluated after refinement, global refinements of the characteristic shells were plotted for better illustration, to show the characteristic shell more clearly.

However, the subdivision algorithm described in Chapter 6 cannot yet handle this extension. The crucial point is that the construction presented in Chapter 6 does not generate a subdivision matrix for all initial elements. This is due to the combinatorial graph $\mathbf{G}_K$, which describes the dual structure of the cells.

To explain this, we consider two different cases. Assume that the graph **G** of an initial element for $g = 3$ contains an $n$-gonal trapezohedron for $n > 3$. If one of the $n$-fold points of the trapezohedron is a vertex, then the other $n$-fold point is the central point. For this first case, a subdivision matrix can be created. This holds, for example, for the pyramid shown in Figure 7.3. This case occurs in the 305,095 examples from Section 3.1.1 for adjacency matrices where nodes of the associated graphs have more than three neighbors. There are exactly 118,589 such examples. The subdivision matrices of the initial elements of these examples can be generated without problems, and these have been numerically verified. The eigenvalue results can be found in Figure 7.6. There, one can see that the eigenvalues have the desired spectrum.

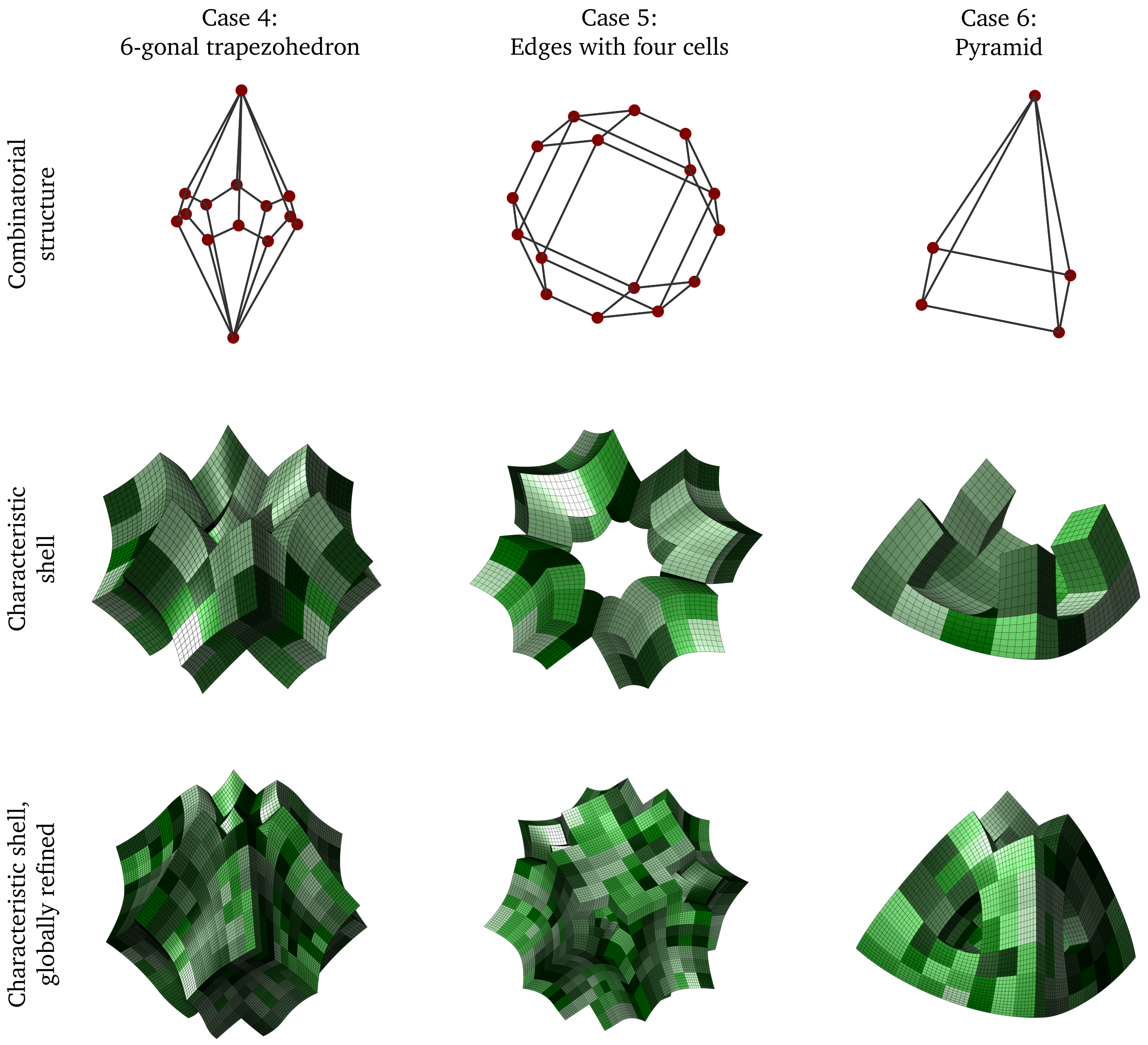

**Figure 7.5:** Illustration of cases 4, 5 and 6 from Definition 7.6. The first row shows the combinatorial structures. The second row plots the evaluable region of the characteristic shell for $t = 2$. The third row shows the evaluable region of the same characteristic shell after a global refinement.

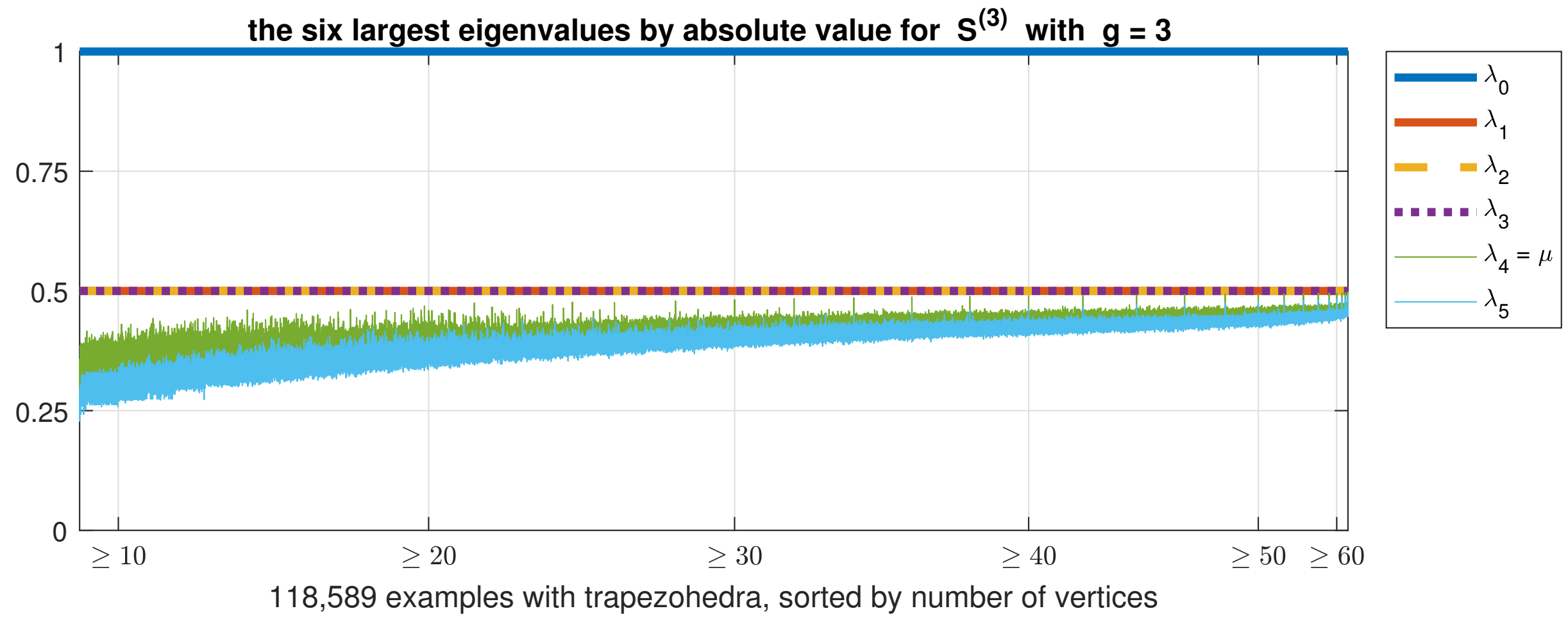

**Figure 7.6:** The six largest absolute eigenvalues of the 118,589 examples from Section 3.1.1 whose structure for $g = 3$ contains trapezohedra. It can be seen that both the construction of subdivision matrices was possible and that the constructed subdivision matrices have the desired spectrum.

For the second case, assume again that the graph **G** of an initial element with $g = 3$ for $n > 3$ contains an $n$-gonal trapezohedron. If one of the two $n$-fold points of the trapezohedron lies in an edge point, then the other $n$-fold point lies in a facet point. This case occurs for initial elements near those described in case 1. For this second case, the combinatorial graph $\mathbf{G}_K$ reflects the structure given by the cells, but when the graph **G** is generated from $\mathbf{G}_K$ by Algorithm 14, the generated graph **G** does not match the structure specified by the cells. This is exemplified in Figure 7.7. The structure shown there contains a 4-gonal trapezohedron. However, the combinatorial graph $\mathbf{G}_K$ indicated by dashed lines is that of the regular case. Algorithm 14 accordingly produces the graph **G** of the regular case, which has fewer nodes than the depicted structure. Therefore, the construction described in Chapter 6 does not work here, since the structure of the graph **G** cannot be generated from the combinatorial graph $\mathbf{G}_K$. This also explains why the hexahedral structure had to be assumed in Chapter 6. If this structure is present, the graph **G** can be generated from the combinatorial graph $\mathbf{G}_K$. If not, there exist combinatorial structures (as described by case 2) for which $\mathbf{G}_K$ is not sufficient to describe the graph.

Since initial elements of the second case lie close to those of the first, no subdivision matrix for double shells can be generated for the trapezohedral examples with $g = 3$. However, if subdivision matrices could be generated for the described second case, then the extension described in this chapter could also be performed for $g = 3$. This should be the subject of further research.

In summary, it can be stated that the generalization to arbitrary combinatorial structures is theoretically possible. However, a new category of irregularity arises that still needs to be examined in more detail. For the case $g = 2$, we were able to create a „proof of concept“ for this generalization. Since both the theoretical structure could be described and the practical application could be carried out, this shows that it may be worthwhile to study this generalization in more detail.

For the case $g = 3$, this was not possible. Here, a first step is to be able to create a subdivision matrix for each structure.

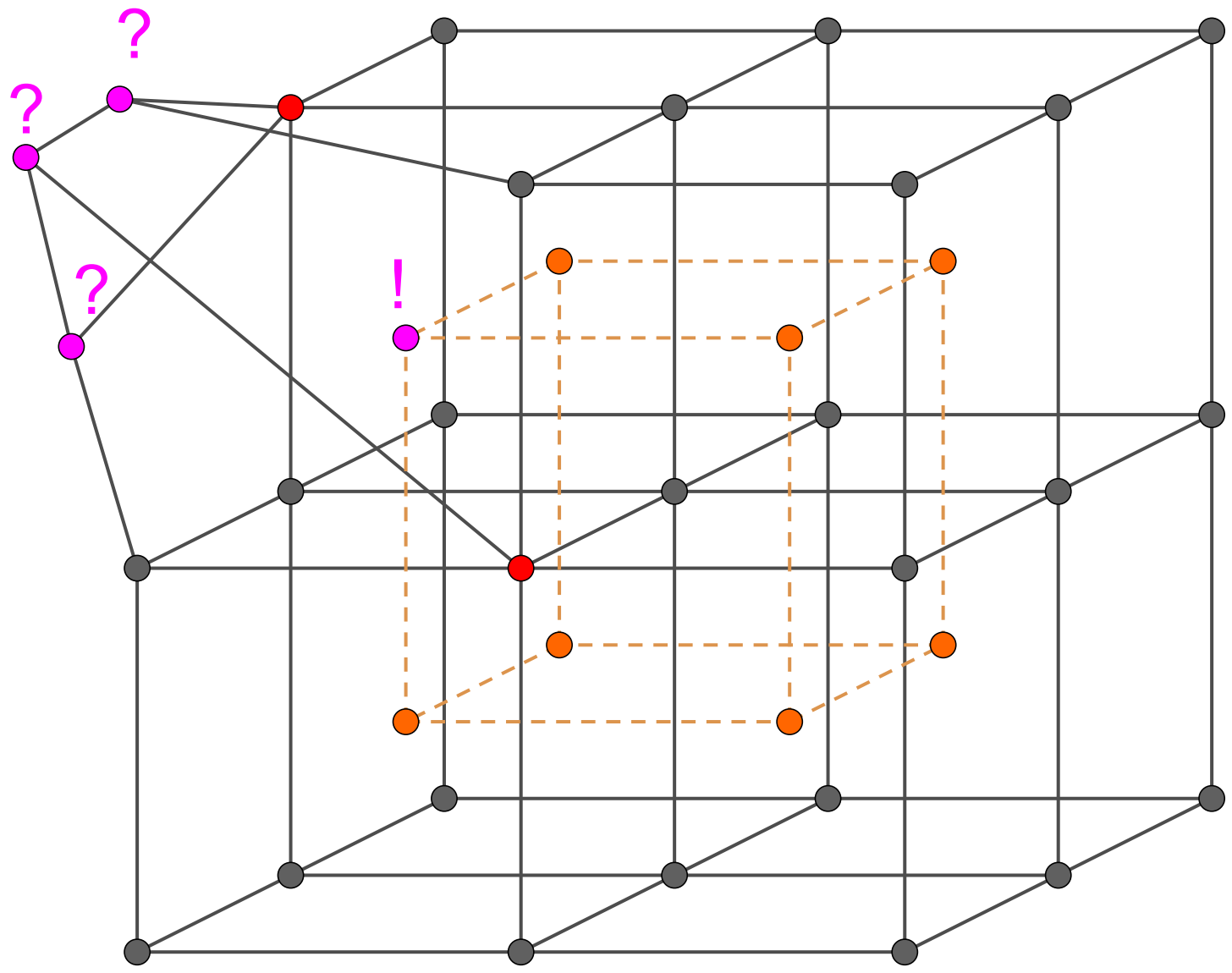

**Figure 7.7:** An example visualization of a graph **G** consisting of seven hexahedra and one 4-gonal trapezohedron. The 4-fold nodes (shown in red) lie on an edge point and a facet point. The associated combinatorial graph $\mathbf{G}_K$ is shown in orange. It can be seen that the pink node of the combinatorial graph cannot represent the structure of the pink nodes in **G**.

# 8 Conclusion and Outlook

The research goal of this thesis was to develop quality criteria for generalized quadratic ($g = 2$) and cubic ($g = 3$) B-spline subdivision, both for subdivision surfaces ($t = 2$) and subdivision volumes ($t = 3$). Subsequently, subdivision algorithms were to be developed for the four combinations of $t$ and $g$, which ideally meet these criteria. In particular, they should have a $t$-fold subdominant eigenvalue of $1/2$, i.e., a spectrum of the form

$$1, \quad \underbrace{\frac{1}{2}, \quad \cdots, \quad \frac{1}{2}}_{t \text{ times}}, \quad > \quad |\lambda_{t+1}| \quad \geq \quad \ldots \quad \geq \quad |\lambda_{n-1}|$$

for a $n \times n$ subdivision matrix. The initial question was answered comprehensively, and the results are summarized below.

A first central insight is the development and description of the quality criteria themselves, which address various aspects of the subdivision algorithms and matrices. In particular, the necessity of a $t$-fold subdominant eigenvalue $1/2$ was described.

Another insight is that the Catmull-Clark algorithm [CC78] is not suitable for a generalization from bivariate to trivariate subdivision algorithms, and that the already established subdivision algorithms from [JM99] and [Baj+02] generally do not provide an appropriate spectrum. This was previously unknown in the literature.

Based on these insights, a new approach to constructing subdivision algorithms was developed. For $g = 2$, this is based on the construction of 3-polytopes, the subsequent construction of a Colin de Verdière matrix, and finally shifting the spectrum. Using this new method, subdivision matrices were produced that, among other properties, have the required spectrum, contain no non-negative entries, and respect the tensor product structure (if present). Additionally, a variant was presented that guarantees a sub-subdominant eigenvalue of $1/4$ but loses the property of non-negative entries. The detailed results for $g = 2$ are clearly summarized in Table 5.1.

The approach presented for $g = 3$ also started with the construction of 3-polytopes. These were refined in a second step, and from this refined structure, a Colin de Verdière-like matrix was created. This matrix was then split into the sum of two matrices, whose exponentials were finally multiplied. Here, it was possible to empirically demonstrate the desired spectrum. Among other things, the subdivision matrices contain only non-negative entries, respect the tensor product structure (if present), and have the desired sparsity pattern. Thus, the subdivision matrices are notably sparse. The detailed results for $g = 3$ are summarized in Table 6.1.

For $t = 2$ and $g = 2$, no significant improvement over known subdivision algorithms from the literature was achieved. The first variant reproduces the refinement rules of the Doo-Sabin algorithm [DS78], while the second variant does not offer additional quality criteria. However, the first variant provides an alternative access to the rules from [DS78], and the second variant can be generalized to $t = 3$ in terms of tensor product structure. For both variants, the double subdominant eigenvalue $1/2$ was proven.

For $t = 3$ and $g = 2$, a subdivision algorithm was constructed for the first time that can generate generalized quadratic B-spline volumes. Both the construction and the detailed proofs of the quality criteria were presented. In particular, the triple subdominant eigenvalue $1/2$ was proven for subdivision matrices of initial elements and empirically verified for subdivision matrices of shells.

For $t = 2$ and $g = 3$, an alternative to the Catmull-Clark subdivision algorithm [CC78] was created. The double subdominant eigenvalue $1/2$ was also demonstrated empirically. Thus, the alternative constructed in this thesis exhibits uniform scaling around irregular vertices with respect to the asymptotic behavior.

For $t = 3$ and $g = 3$, an alternative to the already described subdivision algorithms was created. Although the triple subdominant eigenvalue $1/2$ was only demonstrated empirically, the analysis indicates that the construction is functional regarding the spectrum. If there is also a triple subdominant eigenvalue $1/2$, the corresponding eigenvectors form a suitable structure to describe the asymptotic behavior. Thus, concrete improvements were achieved in terms of eigenvalues and eigenvectors compared to the subdivision algorithms from [JM99] and [Baj+02].

This thesis not only presented the construction of subdivision matrices but also systematically examined them against the developed quality criteria. Where possible, these were proven or disproven. To make well-founded qualitative statements, the criteria were additionally tested on 305,095 randomly generated examples.

A disadvantage of the subdivision algorithms introduced here lies in their comparatively complex construction. While in [JM99] and [Baj+02] the entries of the subdivision matrices can be given directly by explicit formulas, the approach of this work proceeds indirectly. Algorithmic generation of the subdivision matrices requires a deeper understanding of the stepwise construction. To address this issue, the most demanding part—the construction of the 3-polytopes—was detailed both theoretically and practically. Moreover, all algorithmically relevant steps of the construction were described with pseudocode. The related Matlab implementation was also published in [Die25].

Finally, the presented approach for $t = 3$ and $g = 2$ was sketched for a generalization to arbitrary volumetric meshes. Hexahedral cells were generalized to trapezoidal ones. This generalization opens a new research field because it leads to a previously unknown form of irregularity: the so-called irregular volumes.

Throughout this thesis, various avenues for further research have been suggested. These are already reflected in the quality criteria themselves. In particular, for $t = 3$, the list of described quality criteria forms a meaningful and comprehensive elaboration of individual aspects of subdivision algorithms. However, since the theory for the case $t = 3$ has not yet been fully developed, it remains open which criteria—for example, regarding the smoothness of volumes at irregularities—are decisive. Both the theory of the volumetric case and the further development of quality criteria thus offer a wide field for future research.

The same applies to the quality criteria concerning the subdivision algorithms presented here. The results that have only been demonstrated empirically require theoretical consideration. This especially concerns Quality Criterion Q5 for $g = 3$. The challenge here is that the subdivision matrix in this case is composed of the product of two matrices. Since little is known in the literature about the eigenvalues of such matrix products, it would be natural to develop an alternative construction that avoids the final matrix product. Such a construction could build on the insights from Section 6.2.3.

For $g = 3$, it was also observed that, as the number of cells around an irregular vertex increases, the sub-subdominant eigenvalue $\mu$ of the subdivision matrix approaches the subdominant eigenvalue $\lambda$. This raises the question whether alternative subdivision matrices can be constructed that maintain the already established quality criteria but achieve a larger gap between $\lambda$ and $\mu$.

Another important research topic is the proof of injective and regular characteristic maps for $t = 3$, which represents the long-term goal of generalized trivariate B-spline subdivision algorithms.

Moreover, the sketched generalization to trapezoidal structures for $t = 3$ described in this work offers significant potential for future research. Here, the described sketches first need to be worked out in detail, clarified, and further developed, especially regarding the definition of various terms. Additionally, the presented subdivision algorithm for $t = 3$ and $g = 3$ must be generalized to these structures. A central challenge remains how the still missing refinement rules can be formed.

The described irregular volumes with the irregular regions they induce also open promising possibilities for further research. A thorough study could significantly extend the possibilities for describing volumetric structures via generalized B-spline subdivision and lead to a more versatile description of volumes and volumetric structures.

Furthermore, the initial application of the subdivision algorithms in the context of isogeometric analysis is of great interest. It must be investigated whether the subdivision algorithms presented here are suitable for simulation and how the approximations behave compared to other subdivision algorithms. While this comparison can be carried out with relatively little effort for $t = 2$, fundamental questions must first be clarified for $t = 3$, especially regarding the integration of the irregular regions.

The new approach to constructing subdivision algorithms presented in this thesis thus offers considerable potential for improving existing subdivision algorithms and for applications in various fields, such as simulation. Moreover, it can serve as a foundation for many further research directions.

# Appendix

# A Graphs

In this section, we briefly describe the fundamentals of graph theory relevant for this thesis. We mostly follow the presentation of Diestel [Die17] and begin with the definition of a graph according to [Die17, p. 1–2]:

**Definition A.1.** *Let $\boldsymbol{V}$ be a set and let $[\boldsymbol{V}]^2$ be the set of all two-element subsets of $\boldsymbol{V}$.*

1. *A* graph *is a tuple $\boldsymbol{G} = (\boldsymbol{V}, \boldsymbol{E})$ consisting of two sets such that $\boldsymbol{E} \subseteq [\boldsymbol{V}]^2$. Thus, the elements of $\boldsymbol{E}$ are two-element subsets of $\boldsymbol{V}$.*
2. *The elements of $\boldsymbol{V}$ are called* vertices *of the graph $\boldsymbol{G}$.*
3. *The elements of $\boldsymbol{E}$ are called* edges *of the graph $\boldsymbol{G}$.*

We remark the following regarding this definition:

**Remark/Definition A.2.** *In the context of graph theory, there are many terms and definitions to describe specific types or properties of graphs. The above definition implicitly excludes the following three concepts by its structure:*

- Loops: *A* loop *in a graph $\boldsymbol{G} = (\boldsymbol{V}, \boldsymbol{E})$ is an edge in $\boldsymbol{E}$ connecting a vertex to itself. Since in the above definition an edge is a two-element subset of $\boldsymbol{V}$, loops are not possible.*
- Multiple edges: *A graph has* multiple edges *if there exist two distinct edges in $\boldsymbol{E}$ connecting the same pair of vertices. Since $\boldsymbol{E}$ is defined as a set, multiple edges cannot occur.*
- Directed edges: *A directed edge has a start and an end vertex. For vertices $v_i, v_j \in \boldsymbol{V}$, edges $(v_i, v_j)$ and $(v_j, v_i)$ are distinct. Since edges in $\boldsymbol{E}$ are sets, this excludes directed edges.*

*According to Diestel's definition, graphs cannot have loops, multiple edges, or directed edges. Such graphs are called* simple graphs. *Unless explicitly stated otherwise (e.g., in Definition A.34), all graphs in this thesis are simple graphs. In some definitions, we refer explicitly to simplicity to emphasize this property.*

In the next definition, we describe some fundamental terms of graph theory following [Die17, pp. 2, 6–10]:

**Definition A.3.** *Let $\boldsymbol{G} = (\boldsymbol{V}, \boldsymbol{E})$ be a graph.*

1. *We denote the number of vertices of $\boldsymbol{G}$ by $|\boldsymbol{V}| \in \mathbb{N}_0$ and the number of edges by $|\boldsymbol{E}| \in \mathbb{N}_0$.*
2. *The two graphs with $|\boldsymbol{V}| < 2$, i.e., the empty graph and the graph with one vertex, are called* trivial graphs.
3. *A* path *is a non-empty graph $\boldsymbol{G} = (\boldsymbol{V}, \boldsymbol{E})$ of the form*
$$\boldsymbol{V} = \{v_0, v_1, \ldots, v_n\}, \quad \boldsymbol{E} = \{\{v_0, v_1\}, \{v_1, v_2\}, \ldots, \{v_{n-1}, v_n\}\} \quad \textit{with} \quad n \in \mathbb{N},$$
*where all $v_i$ for $i \in \{0, \ldots, n\}$ are distinct. If $\boldsymbol{G}$ is a subgraph of a graph $\boldsymbol{G}'$, then $\boldsymbol{G}$ is a path in $\boldsymbol{G}'$.*
4. *Let $\boldsymbol{G}$ be a path with $n \geq 2$. Then the graph $\boldsymbol{G}' = (\boldsymbol{V}', \boldsymbol{E}')$ with*
$$\boldsymbol{V}' = \boldsymbol{V} \quad \textit{and} \quad \boldsymbol{E}' = \boldsymbol{E} \cup \{\{v_n, v_0\}\}$$
*is called a* cycle. *If $\boldsymbol{G}'$ is a subgraph of a graph $\boldsymbol{G}''$, then $\boldsymbol{G}'$ is a cycle in $\boldsymbol{G}''$.*

5. *A* walk *of length $k \in \mathbb{N}$ in a non-empty graph $\boldsymbol{G} = (\boldsymbol{V}, \boldsymbol{E})$ is a non-empty alternating sequence of vertices and edges of the form*

$$v_0, e_0, v_1, \ldots, e_{k-1}, v_k \quad \textit{with} \quad v_i \in \boldsymbol{V}, \quad e_i \in \boldsymbol{E}, \quad e_i = \{v_i, v_{i+1}\}, \quad i \in \{0, \ldots, k-1\}.$$

*The vertices in a walk do not necessarily have to be distinct.*

Next, we define the adjacency matrix of a graph following [Die17, p. 27]:

**Definition A.4.** *Let $\boldsymbol{G} = (\boldsymbol{V}, \boldsymbol{E})$ be a graph with $|\boldsymbol{V}| = n$. The matrix $A \in \{0, 1\}^{n \times n}$ defined by*

$$A_{(i,j)} := \begin{cases} 1 & \textit{if } \{v_i, v_j\} \in \boldsymbol{E} \\ 0 & \textit{otherwise} \end{cases}$$

*is called the* adjacency matrix *of* $\boldsymbol{G}$.

We extend the concept of graphs by introducing directed graphs. These play only a minor role in this work but are necessary for certain concepts. The definition follows [Die17, pp. 27–28]:

**Definition A.5.** *A* directed graph *is a tuple $(\boldsymbol{V}, \boldsymbol{E})$ consisting of two sets of vertices $\boldsymbol{V}$ and edges $\boldsymbol{E}$, together with two mappings*

$$\text{start} : \boldsymbol{E} \to \boldsymbol{V} \quad \textit{and} \quad \text{target} : \boldsymbol{E} \to \boldsymbol{V},$$

*that assign to each edge $e \in \boldsymbol{E}$ a start vertex* start$(e) \in \boldsymbol{V}$ *and a target vertex* target$(e) \in \boldsymbol{V}$. *In particular, there may be multiple edges between two vertices, and loops are allowed and possible. For directed graphs, each element $e$ of the set $\boldsymbol{E}$ is represented as* (start$(e)$, target$(e)$) *and called a* directed edge. *For two vertices $v_i, v_j \in \boldsymbol{V}$, both $(v_i, v_j)$ and $(v_j, v_i)$ can be contained in $\boldsymbol{E}$. Accordingly, the adjacency matrix $A$ of a directed graph satisfies*

$$A_{(i,j)} := \begin{cases} 1 & \textit{if } (v_i, v_j) \in \boldsymbol{E} \\ 0 & \textit{otherwise} \end{cases}$$

With these definitions, we have described all fundamental terms and will consider various concepts of graph theory in the following sections.

## A.1 Automorphisms of Graphs

Automorphisms of graphs are of particular interest in this work. We begin with the definition according to [Die17, p. 3] and then explain how and for what purpose we use the concept.

**Definition A.6.** *Let $\boldsymbol{G} = (\boldsymbol{V}, \boldsymbol{E})$ and $\boldsymbol{G}' = (\boldsymbol{V}', \boldsymbol{E}')$ be two graphs. A mapping $\boldsymbol{h} : \boldsymbol{V} \to \boldsymbol{V}'$ is a* homomorphism from $\mathbf{G}$ to $\mathbf{G}'$ *if it preserves adjacency relations of vertices. Concretely, this means:*

$$\{v_i, v_j\} \in \boldsymbol{E} \implies \{\boldsymbol{h}(v_i), \boldsymbol{h}(v_j)\} \in \boldsymbol{E}' \quad \textit{for all } v_i, v_j \in \boldsymbol{V}.$$

*If $\boldsymbol{h}$ is bijective and its inverse $\boldsymbol{h}^{-1}$ is also a homomorphism, i.e., if*

$$\{\boldsymbol{h}(v_i), \boldsymbol{h}(v_j)\} \in \boldsymbol{E}' \iff \{v_i, v_j\} \in \boldsymbol{E} \quad \textit{for all } v_i, v_j \in \boldsymbol{V},$$

*then we call $\boldsymbol{h}$ an* isomorphism, *and the graphs $\boldsymbol{G}$ and $\boldsymbol{G}'$* isomorphic, *writing $\boldsymbol{G} \cong \boldsymbol{G}'$. An isomorphism from $\boldsymbol{G}$ to itself is called an* automorphism *of* $\boldsymbol{G}$.

Next, we link the concept of isomorphism with the adjacency matrices of the corresponding graphs to obtain a characterization. This characterization is taken from [GR01, Lem. 8.1.1, p. 164], where it is given for directed graphs. Since every undirected graph can be expressed as a directed graph, we can transfer this characterization accordingly.

**Characterization A.7.** *Let **G** and **G**$'$ be two undirected graphs with the same vertex set, and let $A$ and $A'$ be their adjacency matrices. The graphs **G** and **G**$'$ are isomorphic if and only if there exists a permutation matrix (see Definition B.15) $H$ such that*

$$H^T AH = A' \quad \Leftrightarrow \quad A = HA'H^T.$$

The characterization for automorphisms according to [LS16, Lem. 2.8, p. 30] follows as a special case for $\mathbf{G}' := \mathbf{G}$:

**Characterization A.8.** *Let $\boldsymbol{G} = (\boldsymbol{V}, \boldsymbol{E})$ be a graph with adjacency matrix $A$ and let $\boldsymbol{h}$ be a permutation on $\boldsymbol{V}$ (see Definition B.14) with orthogonal permutation matrix $H$ (see Theorem B.16). The permutation $\boldsymbol{h}$ is an automorphism of $\boldsymbol{G}$ if and only if*

$$H^T AH = A.$$

We illustrate the difference with the following example:

**Example A.9.** *Let **G** be a graph consisting of the four vertices $\boldsymbol{V} = \{v_1, v_2, v_3, v_4\}$ and the adjacency matrix*

$$A = \begin{bmatrix} 0 & 1 & 1 & 1 \\ 1 & 0 & 1 & 0 \\ 1 & 1 & 0 & 1 \\ 1 & 0 & 1 & 0 \end{bmatrix}.$$

*Furthermore, let the permutations*

$$\boldsymbol{h}_1 = \begin{pmatrix} v_1 & v_2 & v_3 & v_4 \\ v_2 & v_3 & v_4 & v_1 \end{pmatrix} \quad \text{and} \quad \boldsymbol{h}_2 = \begin{pmatrix} v_1 & v_2 & v_3 & v_4 \\ v_3 & v_4 & v_1 & v_2 \end{pmatrix}$$

*with permutation matrices*

$$H_1 = \begin{bmatrix} 0 & 1 & 0 & 0 \\ 0 & 0 & 1 & 0 \\ 0 & 0 & 0 & 1 \\ 1 & 0 & 0 & 0 \end{bmatrix} \quad \text{and} \quad H_2 = \begin{bmatrix} 0 & 0 & 1 & 0 \\ 0 & 0 & 0 & 1 \\ 1 & 0 & 0 & 0 \\ 0 & 1 & 0 & 0 \end{bmatrix}.$$

*Then we have*

$$H_1^T AH_1 = \begin{bmatrix} 0 & 1 & 0 & 1 \\ 1 & 0 & 1 & 1 \\ 0 & 1 & 0 & 1 \\ 1 & 1 & 1 & 0 \end{bmatrix} =: A' \quad \text{and} \quad H_2^T AH_2 = \begin{bmatrix} 0 & 1 & 1 & 1 \\ 1 & 0 & 1 & 0 \\ 1 & 1 & 0 & 1 \\ 1 & 0 & 1 & 0 \end{bmatrix} = A.$$

*If **G**$'$ is the graph corresponding to $A'$ on the vertex set **V**, then **G**$'$ and **G** are isomorphic and $\boldsymbol{h}_1$ is an isomorphism. In contrast, $\boldsymbol{h}_2$ is an automorphism of **G**. Both permutations are illustrated in Figure A.1.*

Isomorphisms on the same vertex set can thus be understood as reorderings of the vertices. Automorphisms, as illustrated in Figure A.1, map the graph onto itself. They therefore represent the symmetries of the graph (cf. also [BM08, p. 15] and [LS16, p. 18]). Moreover, the set of automorphisms has the following algebraic structure, see [GR01, p. 4]:

**Definition/Lemma A.10.** *Let **G** be a graph. We denote the set of all automorphisms of **G** by*

$$\mathrm{Aut}(\boldsymbol{G}).$$

*This set forms a group under the composition of automorphisms.*

After these fundamentals on the various types of mappings between graphs, we now proceed to the concept of connectivity in the next section.

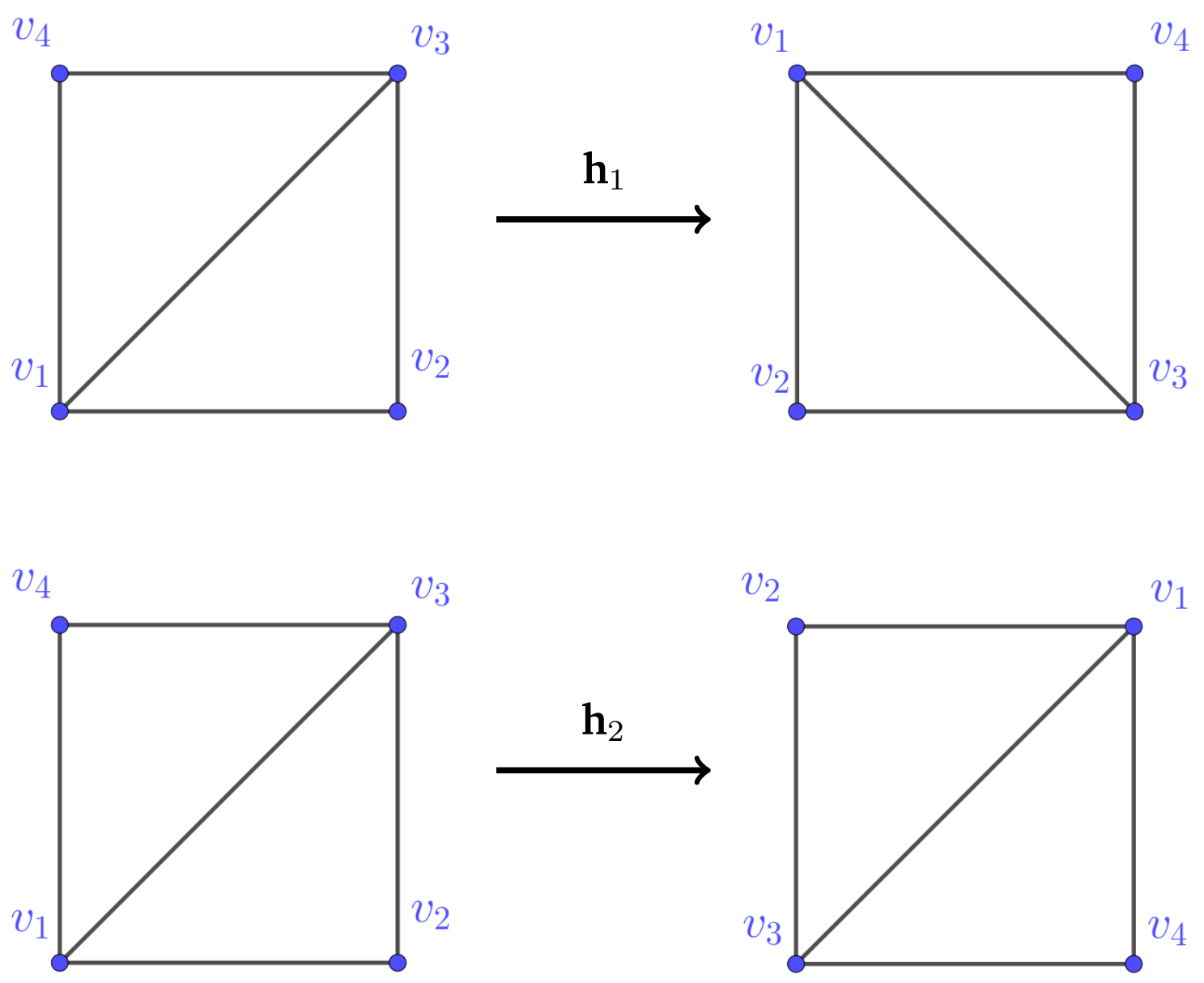

**Figure A.1:** Illustration of the graph $\mathbf{G}$ and the two permutations $\mathbf{h}_1$ and $\mathbf{h}_2$ from Example A.9.

## A.2 Connectivity

The concept of connectivity is central for this work, since many important theorems, such as Steinitz's theorem 3.8, require a certain connectivity as a premise. We begin with the definition of connectedness according to [Die17, p. 10]:

**Definition A.11.** *A graph $\boldsymbol{G} = (\boldsymbol{V}, \boldsymbol{E})$ is called* connected, *if $|\boldsymbol{V}| > 0$ and every pair of vertices is connected by a path in $\boldsymbol{G}$.*

We can use this definition to explain the concept of (vertex) connectivity following [Die17, p. 11–12]:

**Definition A.12.** *Let $k \in \mathbb{N}$. A graph $\boldsymbol{G} = (\boldsymbol{V}, \boldsymbol{E})$ is called $k$-*connected *(sometimes also called $k$-*vertex-connected*) if every subgraph $\boldsymbol{G}' = (\boldsymbol{V}', \boldsymbol{E}') \subset \boldsymbol{G}$ with*

$$\boldsymbol{V}' \subset \boldsymbol{V}, \quad |\boldsymbol{V}'| > |\boldsymbol{V}| - k, \quad \textit{and} \quad \boldsymbol{E}' = \{\{v_i, v_j\} \in \boldsymbol{E} \mid v_i, v_j \in \boldsymbol{V}'\}$$

*is connected. The largest natural number $k$ for which the graph $\boldsymbol{G}$ is $k$-connected is called the* connectivity *of $\boldsymbol{G}$.*

This means that one can remove up to $k-1$ vertices from the graph without losing connectivity. Next we define edge-connectivity according to [Die17, p. 12]:

**Definition A.13.** *Let $k \in \mathbb{N}$ and let $\boldsymbol{G} = (\boldsymbol{V}, \boldsymbol{E})$ be a graph. The graph $\boldsymbol{G}$ is called $k$-*edge-connected *if $|\boldsymbol{V}| > 1$ and every subgraph $\boldsymbol{G}' = (\boldsymbol{V}', \boldsymbol{E}') \subset \boldsymbol{G}$ with*

$$\boldsymbol{V}' = \boldsymbol{V}, \quad \boldsymbol{E}' \subset \boldsymbol{E}, \quad \textit{and} \quad |\boldsymbol{E}'| > |\boldsymbol{E}| - k$$

*is connected. This means analogously that up to $k-1$ edges can be removed without losing connectivity.*

*The largest natural number $k$ for which $\boldsymbol{G}$ is $k$-edge-connected is called the* edge-connectivity *of $\boldsymbol{G}$.*

For these two notions we have the following theorem, cf. [Die17, Prop. 1.4.2, p. 12]:

**Proposition A.14.** *Let $\boldsymbol{G}$ be a nontrivial graph. Then the vertex connectivity of $\boldsymbol{G}$ is less than or equal to the edge connectivity of $\boldsymbol{G}$.*

In other words, every $k$-connected graph is also $k$-edge-connected. A characterization of connectivity was developed by Menger in [Men27]. We state here the version from [Die17, Thm. 3.3.6, p. 72]:

**Theorem A.15** (Global version of Menger's theorem)**.** *Let $\boldsymbol{G} = (\boldsymbol{V}, \boldsymbol{E})$ be a graph and $k \in \mathbb{N}$.*

1. *The graph $\boldsymbol{G}$ is $k$-connected if and only if between every pair of vertices there exist $k$ vertex-disjoint paths.*
2. *The graph $\boldsymbol{G}$ is $k$-edge-connected if and only if between every pair of vertices there exist $k$ edge-disjoint paths.*

The next definition follows [Mey00, p. 671] and generalizes connectedness to directed graphs. Since every undirected graph can also be viewed as a directed graph, we define the following notion for both types. For undirected graphs, this definition coincides with Definition A.11.

**Definition A.16.** *A (directed) graph $\boldsymbol{G} = (\boldsymbol{V}, \boldsymbol{E})$ is called* strongly connected *if for every pair of vertices $v_i, v_j \in \boldsymbol{V}$ there exists a directed path from $v_i$ to $v_j$.*

## A.3 Planarity

The next concept is planarity. We first describe the notion of an arc following [Die17, p. 90]:

**Definition A.17.** *An* arc *is a subset $\boldsymbol{M} \subset \mathbb{R}^2$ which is a union of finitely many line segments and is homeomorphic to the interval $[0, 1]$.*

With this we can define planar-embedded graphs according to [Die17, p. 92, 102]:

**Definition A.18.** *A* planar-embedded graph *is a tuple $(\boldsymbol{V}, \boldsymbol{E})$ of vertices and edges such that the following hold:*

1. $\boldsymbol{V} \subseteq \mathbb{R}^2$.
2. *Each edge is an arc connecting two vertices.*
3. *Distinct edges have distinct sets of vertices.*
4. *The interior of any edge contains no vertex and no point of any other edge.*

*A graph $\boldsymbol{G} = (\boldsymbol{V}, \boldsymbol{E})$ is called* planar *if it can be drawn in $\mathbb{R}^2$ as a planar-embedded graph (i.e. it is isomorphic to a planar-embedded graph).*

We place this definition in context with the following remark:

**Remark A.19.** *Graphs as in Definition A.1 are initially described abstractly and can be interpreted abstractly. If the vertex set is considered as an abstract set of $|\boldsymbol{V}|$ elements, then a graph can be constructed from the adjacency matrix data alone. Thus, an abstractly interpreted graph encodes only the structural relationships between objects, here concretely between vertices. We also call such a graph an* abstract graph*.*

*Vertices can be given meaning by identifying them with concrete elements, or more precisely, by defining the vertex set $\boldsymbol{V}$ as these concrete elements. Thus, it is straightforward from the graph definition to realize the vertex set as a subset of $\mathbb{R}^2$ or $\mathbb{R}^3$. Such a graph is also called an* embedded graph *or a* realized graph*.*

*This work, especially in Chapter 3, deals with constructing an abstract graph from an adjacency matrix and then realizing it with vertices in $\mathbb{R}^2$ or $\mathbb{R}^3$. Since this realization is merely an isomorphism between abstract and concrete vertices, we use the abstract graph and its realization(s) synonymously. This considerably simplifies notation and makes the related concepts clearer.*

*The edges of a graph with vertices in $\mathbb{R}^2$ or $\mathbb{R}^3$ are initially, according to Definition A.1, just two-element sets and not lines or arcs. Therefore, Diestel defines planar-embedded graphs as tuples rather than graphs in order to include arcs as subsets of $\mathbb{R}^2$. In our realizations, edges are the shortest paths between control points with respect to the Euclidean norm.*

The concept of planarity leads to the notion of outerplanarity, defined after [Col93, Def. 5.5, p. 146] and [Die17, Exercise 23, p. 115] as follows:

**Definition A.20.** *A graph $\boldsymbol{G} = (\boldsymbol{V}, \boldsymbol{E})$ is called* outerplanar *if it is planar and there exists a planar-embedded isomorphic graph in $\mathbb{R}^2$ for which all vertices lie on the outer facet of the embedding.*

With the concepts of planarity and connectivity, we can proceed in the next section to consider the concept of facets.

## A.4 Facets

We can first approach the notion of a facet intuitively. In three-dimensional polytopes, facets are intended to represent the two-dimensional boundary faces of the polytopes. For example, a six-sided die has six facets, each displaying one of the numbers from one to six. However, on the graph-theoretic level, this is not so straightforward, since graphs are initially abstract structures. Therefore, we need an abstract definition of facets, which we will now introduce and motivate.

In graph theory, there are two concepts used to describe facets for 3-connected planar graphs. The first concept traces back to works by Tutte [Tut60; Tut63] and characterizes facets as cycles with at most one bridge. Since we will not pursue this concept here, we omit a discussion of bridges.

The second concept, found for example in [Die17], describes facets as induced non-separating cycles.

Both concepts are used equivalently in the literature, although explicit proofs of their equivalence are hard to find. A reasonably close reference regarding equivalence for 3-connected graphs can be found in [Bru04, Lemma 10, p. 242].

Before defining the notion of a facet, we consider the following definition of special cycles from [Die17, p. 8]:

**Definition A.21.** *Let $G = (V, E)$ be a graph and $G' \subset G$ be a cycle in $G$.*

1. *An edge of $G$ connecting two vertices of the cycle $G'$, but not part of $G'$, is called a* chord *of $G'$.*
2. *The cycle $G'$ is called an* induced cycle *if it has no chords in $G$.*
3. *The cycle $G'$ is called a* non-separating cycle *if the graph $G$ without $G'$ remains connected.*

We use this definition to describe facets. Before that, we define the notion of a facet according to [Die17, p. 90, 92] as follows:

**Definition A.22.** *A set $M \subset \mathbb{R}^2$ is called a* region *if any two points in $M$ can be connected by an arc lying entirely within $M$. Let $G = (V, E)$ be a planar-embedded graph with $V \subset \mathbb{R}^2$. The maximal regions of the complement $\mathbb{R}^2 \setminus G$ are called* facets *of $G$ and are open subsets of $\mathbb{R}^2$.*

Note that this definition always includes an outer region, which we call the *outer facet* (cf. [Die17, p. 92]).

Here we see that the notion of a facet is initially defined for planar-embedded graphs. However, we want to construct a more general definition that detaches from graphs embedded in $\mathbb{R}^2$ and defines facets in a combinatorial sense.

To create such a definition, we must first discuss whether an abstract definition of facets is well-defined, since an abstract graph may admit multiple planar embeddings. For this, we need the following definition from [Die17, p. 98–102]:

**Definition A.23.** *Let $G$ be a graph and let $G' = (V', E')$ and $G'' = (V'', E'')$ be two planar-embedded graphs realizing $G$, with facet sets $F'$ and $F''$. Let $h'$ and $h''$ be the isomorphisms between $G$ and $G'$, and between $G$ and $G''$, respectively.*

*The two graphs $G'$ and $G''$ are called* equivalent *if the composition $h'' \circ h'^{-1}$ is a bijection mapping $V' \cup E' \cup F'$ onto $V'' \cup E'' \cup F''$, i.e., a bijection of vertices, edges, and facets.*

With this definition, we can state Whitney's theorem from [Whi33], using the version in [Die17, Thm. 4.3.2, p. 102]:

**Theorem A.24.** *Any two planar embeddings of a 3-connected planar graph are equivalent.*

Well-definedness thus follows from Theorem A.24. Since every embedding of a planar graph in $\mathbb{R}^2$ is equivalent, we obtain the same set of facets up to deformation and rearrangement of the subsets of $\mathbb{R}^2$.

Thus, it is in principle possible to decouple the notion of a facet from the concrete embedding. The abstract description of facets for 3-connected planar graphs is given by the following theorem from [Die17, Prop. 4.2.7, p. 95]:

**Theorem A.25.** *The boundaries of the facets of a 3-connected planar embedded graph are its induced non-separating cycles.*

With this theorem, we obtain a characterization of the concept of a facet, which we can transfer to planar graphs by the following definition:

**Definition A.26.** *Let $\boldsymbol{G} = (\boldsymbol{V}, \boldsymbol{E})$ be a 3-connected planar graph. An induced non-separating cycle $\boldsymbol{G}' \subset \boldsymbol{G}$ of $\boldsymbol{G}$ is called an* abstract facet *of $\boldsymbol{G}$. The set of all abstract facets of $\boldsymbol{G}$ is denoted by $\boldsymbol{F}$.*

As already noted in Remark A.19, we use abstract and embedded graphs synonymously. Therefore, in the following, the terms abstract facets and facets of embeddings are used interchangeably.

With this definition, it becomes possible to compute the abstract facets of a graph. For this, we require some further results, which we now explain. We start with Euler's formula for graphs from [Die17, Thm. 4.2.9, p. 97]:

**Theorem A.27** (Euler's Formula for Graphs)**.** *Let $\boldsymbol{G} = (\boldsymbol{V}, \boldsymbol{E})$ be a connected planar embedded graph with $|\boldsymbol{F}|$ facets. Then*

$$|\boldsymbol{V}| - |\boldsymbol{E}| + |\boldsymbol{F}| = 2.$$

From this theorem, we obtain a concrete formula for the number of facets. An algorithm to construct the facets of a planar 3-connected graph, such as Algorithm 3, can terminate once $2 - |\mathbf{V}| + |\mathbf{E}|$ facets have been found.

A trivial corollary from Euler's formula and also from Theorem A.24 follows from [BM08, Corollary 10.20, p. 259]:

**Corollary A.28.** *All planar embeddings of a connected planar graph have the same number of facets.*

Euler's formula for graphs also coincides with Euler's formula for 3-polytopes, which is a more specialized formulation of Theorem A.27 for 3-connected planar graphs.

A well-developed overview including the historical background of Euler's formula for 3-polytopes can be found, for example, in [Ric08]. Euler himself stated the formula in [Eul58]. The first proof by Legendre appears in [Leg94] and in the form [LB86]. We use here the formulation from [Ric08, p. 66]:

**Theorem A.29** (Euler's Formula for 3-Polytopes)**.** *For a 3-polytope with $|\boldsymbol{V}|$ vertices, $|\boldsymbol{E}|$ edges, and $|\boldsymbol{F}|$ facets, it holds that*

$$|\boldsymbol{V}| - |\boldsymbol{E}| + |\boldsymbol{F}| = 2.$$

Next, we consider two results about facets in graphs. The first originates from Kelmans in [Kel81, Thm. 6.7, p. 354] and [Kel81, Thm. 10.6, p. 373]. We use here the formulation from [Die17, Thm. 4.5.2, p. 109]:

**Theorem A.30.** *A 3-connected graph is planar if and only if each edge lies in exactly two induced non-separating cycles.*

We can also state the following proposition:

**Proposition A.31.** *Two (abstract) facets of a 3-connected planar graph have at most one common edge.*

*Proof.* To prove this proposition, we need the theory of planar embedded two-dimensional duality of graphs from the next section.

Suppose that a graph has two facets sharing more than one common edge. Then the vertices corresponding to these two facets in the dual graph are connected by more than one edge. It follows that the dual graph is not simple. This contradicts Theorem A.43, and thus two facets share at most one common edge. □

Regarding planarity, we conclude with another statement. This is a corollary of Euler's formula A.27 and is taken from [Die17, Cor. 4.2.10, p. 97]:

**Corollary A.32.** *A planar graph with $n \geq 3$ vertices has at most $3n - 6$ edges.*

We use this corollary in Algorithm 2 to test planarity, enabling a quick exclusion of planarity in some cases.

The concept of facets allows us to introduce the concept of two-dimensional duality in the next section.

## A.5 Two-Dimensional Duality

The concept of duality is used in many different contexts in mathematics. In this work, we use two different forms of duality and therefore must distinguish between three-dimensional duality and two-dimensional duality. The latter is described in this section, where we first introduce the concept of a multigraph according to [Die17, p. 28]:

**Definition A.33.** *A* multigraph *is a tuple* $(\boldsymbol{V}, \boldsymbol{E})$ *consisting of two disjoint sets together with a mapping*

$$\boldsymbol{E} \to \boldsymbol{V} \cup [\boldsymbol{V}]^2,$$

*that assigns to each edge an element from* $\boldsymbol{V}$ *or* $[\boldsymbol{V}]^2$*, representing its endpoints.*

This definition simply states that a multigraph may contain both loops and multiple edges. Following this, we can define planar embedded multigraphs according to [Die17, p. 110]:

**Definition A.34.** *A* planar embedded multigraph *is a tuple* $\boldsymbol{G} = (\boldsymbol{V}, \boldsymbol{E})$ *consisting of a set of vertices and a set of edges satisfying the following conditions:*

1. $\boldsymbol{V} \subseteq \mathbb{R}^2$.
2. *Each edge is either an arc between two distinct vertices or an arc from a vertex to itself that contains only that vertex. Such an arc is a loop in the graph-theoretical sense.*
3. *Except for the endpoints, an edge contains no point of any other edge.*

For such a planar embedded multigraph, it is possible to define a planar embedded two-dimensional dual graph as follows (see [Die17, p. 110–111]):

**Definition A.35.** *Let* $\boldsymbol{G} = (\boldsymbol{V}, \boldsymbol{E})$ *and* $\boldsymbol{G}^* = (\boldsymbol{V}^*, \boldsymbol{E}^*)$ *be two planar embedded multigraphs with sets of facets* $\boldsymbol{F}$ *and* $\boldsymbol{F}^*$ *respectively. Let* $\mathring{e}$ *denote the interior of an edge* $e$*. We call* $\boldsymbol{G}^*$ *a* planar embedded two-dimensional dual graph *of* $\boldsymbol{G}$ *if there exists a bijection*

$$\begin{array}{ccc} \boldsymbol{F} \to \boldsymbol{V}^* & \boldsymbol{E} \to \boldsymbol{E}^* & \boldsymbol{V} \to \boldsymbol{F}^* \\ f \mapsto v^*(f) & e \mapsto e^* & v \mapsto f^*(v) \end{array}$$

*satisfying the following conditions:*

1. $v^*(f) \in f$ *for all* $f \in \boldsymbol{F}$.
2. $|e^* \cap \boldsymbol{G}| = |\mathring{e}^* \cap \mathring{e}| = |e \cap \boldsymbol{G}^*| = 1$ *for all* $e \in \boldsymbol{E}$*. Furthermore, this intersection point is part of a segment in both* $e$ *and* $e^*$.
3. $v \in f^*(v)$ *for all* $v \in \boldsymbol{V}$.

This somewhat technical definition can be intuitively understood as follows: The dual object to a planar embedded multigraph is again a planar embedded multigraph. Each vertex in **G** corresponds to a facet in $\mathbf{G}^*$, each edge in **G** corresponds to an edge in $\mathbf{G}^*$, and each facet in **G** corresponds to a vertex in $\mathbf{G}^*$. Moreover, the vertices associated with facets lie within those facets. Two corresponding edges intersect in exactly one point, which must be a proper intersection rather than merely a tangential contact.

Similar to facets, we also want to define an abstract version of the concept of duality. For this, we introduce the following notions according to [Die17, pp. 1, 24]:

**Definition A.36.** *Let* $\boldsymbol{G} = (\boldsymbol{V}, \boldsymbol{E})$ *be a graph.*

1. *Let* $\boldsymbol{V}_1, \boldsymbol{V}_2 \subset \boldsymbol{V}$*. The set* $\{\boldsymbol{V}_1, \boldsymbol{V}_2\}$ *is called a* partition *of* $\boldsymbol{V}$ *if* $\boldsymbol{V}_1 \cup \boldsymbol{V}_2 = \boldsymbol{V}$ *and* $\boldsymbol{V}_1 \cap \boldsymbol{V}_2 = \emptyset$.
2. *Given a partition* $\{\boldsymbol{V}_1, \boldsymbol{V}_2\}$ *of* $\boldsymbol{V}$*, denote by* $\boldsymbol{E}(\boldsymbol{V}_1, \boldsymbol{V}_2)$ *the set of all edges having one endpoint in* $\boldsymbol{V}_1$ *and the other in* $\boldsymbol{V}_2$.
3. *A subset* $\boldsymbol{E}' \subset \boldsymbol{E}$ *is called a* cut *of* $\boldsymbol{G}$ *if there exists a partition* $\{\boldsymbol{V}_1, \boldsymbol{V}_2\}$ *of* $\boldsymbol{V}$ *such that* $\boldsymbol{E}' = \boldsymbol{E}(\boldsymbol{V}_1, \boldsymbol{V}_2)$.
4. *A minimal nonempty cut of a graph* $\boldsymbol{G}$ *is called a* bond *of* $\boldsymbol{G}$.

With the notion of bonds, we can formulate the following theorem from [Die17, Prop. 4.6.1, p. 111]:

**Proposition A.37.** *Let* $\boldsymbol{G}$ *be a connected planar embedded multigraph and* $\boldsymbol{G}^*$ *its planar embedded two-dimensional dual multigraph. A subset* $\boldsymbol{E}' \subset \boldsymbol{E}$ *is the edge set of a cycle in* $\boldsymbol{G}$ *if and only if* $\boldsymbol{E}'^* := \{e^* \mid e \in \boldsymbol{E}'\}$ *is a bond in* $\boldsymbol{G}^*$.

With this proposition, it is possible to abstract the notion of planar embedded two-dimensional duality as follows (cf. [Die17, p. 112]):

**Definition A.38.** *We call a multigraph $G^* = (V^*, E^*)$ the* two-dimensional abstract dual graph *of a multigraph $G = (V, E)$ if $E = E^*$ and the bonds in $G^*$ are exactly the edge sets of the cycles of $G$. In the following, we refer to the two-dimensional abstract dual graph of $G$ simply as the* dual graph *of $G$.*

The notation $\mathbf{E} = \mathbf{E}^*$ here can be misleading if one thinks of edges as fixed entities. According to Diestel, the set of edges of a multigraph is initially just a set, to which a corresponding assignment of two endpoints is given. The definition means that while the set $\mathbf{E}$ equals $\mathbf{E}^*$ as sets, the endpoints assigned to these edges may differ.

Abstract duality thus means: For the edges of a multigraph $\mathbf{G}$, first find a new set of vertices $\mathbf{V}^*$. Assign to the edges $\mathbf{E}^* = \mathbf{E}$ pairs of vertices from $\mathbf{V}^* \cup [\mathbf{V}^*]^2$. If then the bonds of $\mathbf{G}^* = (\mathbf{V}^*, \mathbf{E}^*)$ coincide with the cycles of $\mathbf{G}$, with the same associated (or in Diestel's notation identical) edges, then $\mathbf{G}^*$ is the abstract dual graph of $\mathbf{G}$.

Similar to planarity and facets, it is thus also possible to abstract from the concrete embedding in $\mathbb{R}^2$ for the notion of duality.

It should be noted that (simple) graphs are special cases of multigraphs. Therefore, the concept of duality naturally extends to them.

After this definition, several questions arise regarding duality: First, it must be clarified for which graphs a dual graph exists. If such a dual graph exists, we must then consider when it is unique and finally when it is simple. We begin with existence by considering the following theorem from [Whi32, Thm. 29, p. 357], known as Whitney's criterion for planarity. We use the version from [Die17, Thm. 4.6.3, p. 113]:

**Theorem A.39** (Whitney's Planarity Criterion)**.** *A graph is planar if and only if it has an abstract dual graph.*

This settles the question of existence. For uniqueness, we add the property of 3-connectivity to planarity. Since by Theorem A.24 all planar embeddings of a planar 3-connected graph are equivalent, it follows directly and according to [BM08, Cor. 10.29, p. 267]:

**Corollary A.40.** *Every (simple) 3-connected planar graph has a unique dual (multi)graph.*

In the formulation of [BM08], graphs are mentioned. To be sure that the definition of graph is the same as in this work, we consider the following theorem from [Bal97, Thm. 8.64, p. 229] (reformulated in our notation):

**Proposition A.41.** *The planar embedded dual graph of a planar embedded 3-edge-connected graph is simple.*

Since all 3-connected graphs are also 3-edge-connected by Proposition A.14, it follows that:

**Corollary A.42.** *The planar embedded dual graph of a planar embedded 3-connected graph is simple.*

Combining these statements, we obtain:

**Theorem A.43.** *The dual graph of a (simple) 3-connected planar graph exists, is unique, and is simple again.*

Thus all necessary basics of graph theory have been described. In the next section, we present the matrix theory foundations needed for this work.

# B Matrices

In this section, we summarize the matrix fundamentals needed for this work as concisely as possible. Although we primarily deal with real matrices, the definitions and statements here follow the literature, and therefore some concepts are formulated for real matrices and others for complex matrices.

We start with some general definitions and statements. Subsequently, we cover basics related to eigenvalues, eigenvectors, and the matrix exponential. Finally, we compute some concrete exponentials relevant for us. We begin with the notion of square matrices:

**Definition B.1.** *A matrix $M \in \mathbb{C}^{n\times n}$ with $n \in \mathbb{N}$ is called a* square matrix. *The entries $M_{(i,i)}$ with $i \in \{1, \dots, n\}$ are called* diagonal entries, *and all other entries are called* off-diagonal entries.

In the following, we do not explicitly emphasize that a matrix is square since this is implied by the notation $M \in \mathbb{C}^{n\times n}$ or $M \in \mathbb{R}^{n\times n}$. Next, we define the identity matrix:

**Definition B.2.** *The matrix $E \in \mathbb{C}^{n\times n}$ with entries*

$$E_{(i,j)} := \begin{cases} 1 & \textit{if } i = j, \\ 0 & \textit{otherwise}, \end{cases} \qquad i, j \in \{1, \dots, n\}$$

*is called the* identity matrix. *To emphasize the size, we write $E_n$. We also define $0_n$ as the $n \times n$ matrix whose entries are all zero, called the* zero matrix.

We then introduce the notion of diagonal matrices:

**Definition B.3.** *A matrix $M \in \mathbb{C}^{n\times n}$ whose off-diagonal entries are all zero is called a* diagonal matrix.

Next, we define non-negative matrices:

**Definition B.4.** *A matrix $M \in \mathbb{R}^{n\times n}$ is called* non-negative *if all its entries are greater than or equal to zero.*

Matrices may have an inverse, which we describe as follows according to [LM15, p. 45]:

**Definition B.5.** *Let $M \in \mathbb{C}^{n\times n}$ be a matrix. If there exists a matrix $M^{-1}$ such that*

$$MM^{-1} = M^{-1}M = E,$$

*then $M$ is called* invertible *and $M^{-1}$ is the* inverse of $M$.

The set of invertible matrices forms a group, as stated in the following theorem from [LM15, Thm. 4.11, p. 46]:

**Theorem/Definition B.6.** *The set of all invertible matrices in $\mathbb{C}^{n\times n}$ forms a group under matrix multiplication. This group is called the* general linear group *and is denoted by $\boldsymbol{GL}_n(\mathbb{C})$.*

Next, we describe the transpose of a matrix according to [LM15, p. 42]:

**Definition B.7.** *Let $M \in \mathbb{C}^{n\times m}$ with $n, m \in \mathbb{N}$ be a matrix or a vector. Then*

$$M^T \in \mathbb{C}^{m\times n} \quad \textit{with} \quad M^T_{(i,j)} := M_{(j,i)} \quad \textit{for all} \quad i \in \{1, \dots, m\} \quad \textit{and} \quad j \in \{1, \dots, n\}$$

*is called the* transpose matrix *of $M$.*

In particular, $M'$ is not the transpose of $M$. Regarding two matrices, the Kronecker product is defined according to [HJ91, Def. 4.2.1, p. 243] as follows:

**Definition B.8.** *Let $M^{(1)} \in \mathbb{C}^{m\times n}$ and $M^{(2)} \in \mathbb{C}^{i\times j}$ be two matrices. Then the* Kronecker product *of $M^{(1)}$ and $M^{(2)}$ is defined as*

$$\text{kron}\left(M^{(1)}, M^{(2)}\right) := \begin{bmatrix} M^{(1)}_{(1,1)}M^{(2)} & \dots & M^{(1)}_{(1,n)}M^{(2)} \\ \vdots & & \vdots \\ M^{(1)}_{(m,1)}M^{(2)} & \dots & M^{(1)}_{(m,n)}M^{(2)} \end{bmatrix} \in \mathbb{C}^{(m\cdot i)\times(n\cdot j)}.$$

For the Kronecker product, the following holds according to [HJ91, Lem. 4.2.10, p. 244]:

**Lemma B.9.** *Let $M_1 \in \mathbb{C}^{m\times n}$, $M_2 \in \mathbb{C}^{i\times j}$, $M_3 \in \mathbb{C}^{n\times k}$, and $M_4 \in \mathbb{C}^{j\times l}$. Then*

$$\text{kron}(M_1, M_2) \cdot \text{kron}(M_3, M_4) = \text{kron}(M_1 \cdot M_3, M_2 \cdot M_4).$$

Moreover, according to [HJ91, Cor. 4.2.11, pp. 244–245]:

**Lemma B.10.** *If $M_1 \in \mathbb{C}^{n\times n}$ and $M_2 \in \mathbb{C}^{m\times m}$ are invertible matrices, then* kron$(M_1, M_2)$ *is invertible and*

$$(\text{kron}(M_1, M_2))^{-1} = \text{kron}(M_1^{-1}, M_2^{-1}).$$

In the following, we consider some statements regarding the concept of a basis of a vector space. We begin with the following definition according to [FS20, Def. in Sec. 2.5.1, p. 105 and Def. in Sec. 6.5.4, pp. 340–342]:

**Definition B.11.** *A family $\boldsymbol{M} = \{a_i \in \mathbb{R}^n,\ i = 1, \dots, m\}$ with $m, n \in \mathbb{N}$ and $m \geq n$ is called a* generating system *of $\mathbb{R}^n$ if*

$$\mathbb{R}^n = \text{span}(a_i)_{i\in\{1,\dots,m\}},$$

*that is, if every $a \in \mathbb{R}^n$ is a linear combination of finitely many $a_i$.*

*The family **M** is called a* basis *of $\mathbb{R}^n$ if it is a linearly independent generating system. A basis **M** is called an* orthonormal basis *if all vectors in **M** have length 1 and are mutually orthogonal.*

The following theorem applies to the construction of a basis according to [FS20, Sec. 2.5.5, p. 110]:

**Proposition B.12** (Basis Extension Theorem)**.** *Let $a_1, \dots, a_m$ be linearly independent vectors in $\mathbb{R}^n$ with $m, n \in \mathbb{N}$ and $m \leq n$. Then one can find vectors $b_{m+1}, \dots, b_n$ in $\mathbb{R}^n$ such that*

$$\boldsymbol{M} = \{a_1, \dots, a_m, b_{m+1}, \dots, b_n\}$$

*is a basis of $\mathbb{R}^n$.*

For the explicit construction of an orthonormal basis, the following theorem from [KB17, Construction pp. 120–122 and Thm. 1.112, p. 122] can be used. The procedure is known as the Gram–Schmidt orthonormalization:

**Proposition B.13** (Gram–Schmidt Orthonormalization)**.** *Let $\{a_1, \dots, a_n\}$ be a basis of $\mathbb{R}^n$. Define*

$$b_i' = a_i - \sum_{j=1}^{i-1} \langle b_j, a_i \rangle b_j$$

*and*

$$b_i = \frac{b_i'}{|b_i'|}.$$

*Then $\{b_1, \dots, b_n\}$ is an orthonormal basis of $\mathbb{R}^n$. If $a_1, \dots, a_m$ with $m \leq n$ are already orthonormal, then*

$$b_i = a_i \quad \textit{for} \quad i \in \{1, \dots, m\}.$$

*Proof.* We show here only the reproduction of the orthonormal vectors by induction:
*Base case*:

$$b_1 = \frac{a_1}{|a_1|} = a_1.$$

*Induction hypothesis*: Assume $b_i = a_i$ for all $i \leq k < m$.
*Induction step*: Inserting the formula for $b'_{k+1}$, we get

$$b'_{k+1} = a_{k+1} - \sum_{j=1}^{k} \langle b_j, a_{k+1} \rangle b_j.$$

Since $b_j = a_j$ in the above sum and due to the orthogonality of the $a_j$ we have $\langle a_j, a_{k+1} \rangle = 0$, it follows that

$$b'_{k+1} = a_{k+1} \quad \text{and hence} \quad b_{k+1} = \frac{a_{k+1}}{|a_{k+1}|} = a_{k+1}.$$

□

After these general statements, we consider in the next section concepts related to eigenvalues and eigenvectors.

## B.1 Eigenvalues and Eigenvectors

In this section, we develop the basics regarding eigenvalues and eigenvectors. We start by defining the necessary concepts and then consider statements about eigenvalues of special matrices, which we introduce throughout this section. We begin with the definition of a permutation according to [LM15, Def. 7.1, p. 81]:

**Definition B.14.** *Let $n \in \mathbb{N}$. A bijective mapping*

$$\boldsymbol{h} : \{1, \dots, n\} \to \{1, \dots, n\}, \quad i \mapsto \boldsymbol{h}(i)$$

*is called a* permutation *of the numbers $\{1, \dots, n\}$. The set of all permutations of the numbers $\{1, \dots, n\}$ is denoted by $\boldsymbol{H}_n$.*

For permutations one can generate matrices that represent these permutations, which we describe according to [LM15, Def. 4.15, p. 49] as follows:

**Definition B.15.** *A matrix $H \in \mathbb{C}^{n \times n}$ is called a* permutation matrix *if in every row and every column exactly one entry is $1$ and all other entries are $0$.*

For a permutation matrix, we obtain the following theorem from [LM15, Thm. 4.16, p. 50]:

**Theorem B.16.** *The set of all $n \times n$ permutation matrices forms a subgroup of $\boldsymbol{GL}_n(\mathbb{C})$. In particular, if $H \in \mathbb{C}^{n \times n}$ is a permutation matrix, then $H$ is invertible and its inverse is $H^{-1} = H^T$.*

We use permutations to introduce the concept of the determinant. For this, we first define the sign of a permutation according to [LM15, Def. 7.2, p. 82]:

**Definition B.17.** *Let $n \geq 2$ and $\boldsymbol{h} \in \boldsymbol{H}_n$ be a permutation. A pair $\big(\boldsymbol{h}(i), \boldsymbol{h}(j)\big)$ with $1 \leq i < j \leq n$ and $\boldsymbol{h}(i) > \boldsymbol{h}(j)$ is called an* inversion *of $\boldsymbol{h}$. Let $k$ be the number of all inversions of $\boldsymbol{h}$. Then we define*

$$\operatorname{sgn}(\boldsymbol{h}) := (-1)^k$$

*as the* sign *of the permutation $\boldsymbol{h}$.*

Using this sign, we define the determinant of a matrix according to [LM15, Def. 7.4, p. 82]:

**Definition B.18.** *Let $M \in \mathbb{C}^{n\times n}$ be a matrix. The mapping*

$$\det : \mathbb{C}^{n\times n} \to \mathbb{C}, \quad M \mapsto \det(M) := \sum_{\boldsymbol{h}\in \boldsymbol{H}_n} \left( \operatorname{sgn}(\boldsymbol{h}) \prod_{i=1}^{n} M_{(i,\boldsymbol{h}(i))} \right)$$

*is called the* determinant *of the matrix $M$.*

With the determinant, we can define the characteristic polynomial as follows according to [LM15, Def. 8.1, p. 102]:

**Definition B.19.** *Let $M \in \mathbb{C}^{n\times n}$ be a matrix and $\lambda_i \in \mathbb{C}$. Then*

$$\det(M - \lambda_i E)$$

*is called the* characteristic polynomial *of the matrix $M$.*

With these definitions, we have introduced all concepts necessary for the characterization of eigenvalues. We describe them according to [LM15, Def. 8.7, p. 106] in the following definition:

**Definition B.20.** *Let $M \in \mathbb{C}^{n\times n}$ be a matrix. If there exists $\lambda_i \in \mathbb{C}$ and a vector $a \in \mathbb{C}^n \setminus \{\mathbf{0}\}$ such that*

$$Ma = \lambda_i a,$$

*then we call $\lambda_i$ an* eigenvalue *and $a$ an* eigenvector *corresponding to the eigenvalue $\lambda_i$ of the matrix $M$.*

The eigenvalues of a matrix can be characterized as follows according to [LM15, Thm. 8.8, p. 107]:

**Characterization B.21.** *Let $M \in \mathbb{C}^{n\times n}$ be a matrix. The scalar $\lambda_i \in \mathbb{C}$ is an eigenvalue of $M$ if and only if $\lambda_i$ is a root of the characteristic polynomial, that is,*

$$\det(M - \lambda_i E) = 0.$$

The eigenvalues of a matrix can be assigned two multiplicities, the algebraic and the geometric multiplicity, which we describe next. The algebraic multiplicity is defined according to [Bär18, Def. 6.62, p. 298]:

**Definition B.22.** *Let $M \in \mathbb{C}^{n\times n}$ be a matrix and $\det(M - \lambda_i E)$ its characteristic polynomial. With roots $\lambda_0, \dots, \lambda_k \in \mathbb{C}$ and multiplicities $m_0, \dots, m_k \in \mathbb{N}$, the characteristic polynomial can be factored as*

$$\det(M - \lambda_i E) = \prod_{j=0}^{k} (\lambda_i - \lambda_j)^{m_j}.$$

*We call*

$$\operatorname{alg}(\lambda_j) = m_j$$

*the* algebraic multiplicity *of the eigenvalue $\lambda_j$.*

For the geometric multiplicity, we first need the notion of kernel, defined according to [Ser10, p. 23]:

**Definition B.23.** *Let $M \in \mathbb{C}^{n\times n}$ be a matrix. We define the* kernel *of $M$ as*

$$\ker(M) := \{a \in \mathbb{C}^n : Ma = \vec{0}\}.$$

Using this, we define the geometric multiplicity according to [Mey00, p. 510]:

**Definition B.24.** *Let $M \in \mathbb{C}^{n\times n}$ be a matrix and $\lambda_i$ an eigenvalue of $M$. We call*

$$\operatorname{geo}(\lambda_i) := \dim\big(\ker(M - \lambda_i E)\big)$$

*the* geometric multiplicity *of the eigenvalue $\lambda_i$.*

Building on the concept of eigenvalues, we introduce the following two notions. The first is defined according to [Ser10, p. 43]:

**Definition B.25.** *Let $M \in \mathbb{C}^{n\times n}$ be a matrix. The* spectrum *of $M$ is the set of its eigenvalues and is denoted by* $\mathrm{sp}(M)$*.*

The second notion is defined according to [Ser10, Def. 7.1, p. 127]:

**Definition B.26.** *Let $M \in \mathbb{C}^{n\times n}$ be a matrix. The* spectral radius *of $M$ is*

$$\boldsymbol{\rho}(M) := \max\{|\lambda_i| : \lambda_i \in \mathrm{sp}(M)\},$$

*that is, the largest absolute value among the eigenvalues.*

Using eigenvalues and eigenvectors, one can describe the Jordan decomposition of a matrix. We approach this concept by first defining similarity according to [LM15, Def. 8.11, p. 108]:

**Definition B.27.** *Let $M, M' \in \mathbb{C}^{n\times n}$ be two matrices. The matrices $M$ and $M'$ are called* similar *if there exists a matrix $V \in \boldsymbol{GL}_n(\mathbb{C})$ such that*

$$M = VM'V^{-1}.$$

For the Jordan decomposition, we first define Jordan blocks according to [Bos21, p. 300]:

**Definition B.28.** *Let $\lambda_i \in \mathbb{C}$. A matrix $J(\lambda_i, n) \in \mathbb{C}^{n\times n}$ with*

$$J(\lambda_i, n) = \begin{bmatrix} \lambda_i & 1 & 0 & \dots & 0 \\ 0 & \ddots & \ddots & \ddots & \vdots \\ \vdots & \ddots & \ddots & \ddots & 0 \\ \vdots & & \ddots & \ddots & 1 \\ 0 & \dots & \dots & 0 & \lambda_i \end{bmatrix}$$

*is called a* Jordan block*.*

From these Jordan blocks, one can define a Jordan matrix according to [Bos21, Thm. 15, p. 303] as follows:

**Theorem/Definition B.29.** *Let $M \in \mathbb{C}^{n\times n}$ be a matrix. Then there exists a matrix*

$$J := \begin{bmatrix} J(\lambda_0, n_0) & 0 & \dots & 0 \\ 0 & \ddots & \ddots & \vdots \\ \vdots & \ddots & \ddots & 0 \\ 0 & \dots & 0 & J(\lambda_k, n_k) \end{bmatrix} \in \mathbb{C}^{n\times n},$$

*similar to $M$, i.e.*

$$M = VJV^{-1} \quad \Leftrightarrow \quad J = V^{-1}MV$$

*with $V \in \boldsymbol{GL}_n(\mathbb{C})$. The matrix $J$ is called the* Jordan matrix *of $M$. The block matrices $J(\lambda_0, n_0), \dots, J(\lambda_k, n_k)$ are called* Jordan blocks*. The pairs $(\lambda_i, n_i)$ are uniquely determined up to ordering, where $\lambda_i$ (with possible multiplicities) are exactly the eigenvalues of $M$.*

Diagonal matrices are thus special Jordan matrices. For this special case, we define the following concept according to [Bos21, p. 249]:

**Definition B.30.** *A matrix $M \in \mathbb{C}^{n\times n}$ is called* diagonalizable *if it is similar to a diagonal matrix.*

Hence, a matrix is diagonalizable if and only if its Jordan form is a diagonal matrix.

Next, we consider statements about eigenvalues and eigenvectors of special matrices. We start with eigenvalues of two similar matrices according to [HJ12, Cor. 1.3.4, p. 58]:

**Corollary B.31.** *Two similar matrices $M_1, M_2 \in \mathbb{C}^{n\times n}$ have the same eigenvalues.*

Regarding eigenvalues and eigenvectors of the Kronecker product of two matrices, we obtain according to [HJ91, Thm. 4.2.12, p. 245]:

**Theorem B.32.** *Let $M^{(1)} \in \mathbb{C}^{n\times n}$ and $M^{(2)} \in \mathbb{C}^{m\times m}$. Let the eigenvalues of $M^{(1)}$ be $\lambda_1^{(1)}, \dots, \lambda_n^{(1)}$ and those of $M^{(2)}$ be $\lambda_1^{(2)}, \dots, \lambda_m^{(2)}$. Then the eigenvalues of*

$$\mathrm{kron}(M^{(1)}, M^{(2)})$$

*are exactly the products*

$$\{\lambda_j^{(1)} \cdot \lambda_k^{(2)} \mid j = 1, \dots, n,\ k = 1, \dots, m\}.$$

*If $V_{(:,j)}^{(1)}$ is an eigenvector corresponding to $\lambda_j^{(1)}$ and $V_{(:,k)}^{(2)}$ an eigenvector corresponding to $\lambda_k^{(2)}$, then*

$$\mathrm{kron}(V_{(:,j)}^{(1)}, V_{(:,k)}^{(2)})$$

*is an eigenvector of* $\mathrm{kron}(M^{(1)}, M^{(2)})$ *corresponding to the eigenvalue $\lambda_j^{(1)} \cdot \lambda_k^{(2)}$.*

Next, we want to state a result about eigenvalues of block matrices. For this, we first recall the following theorem on the determinant of block matrices from [HK71, Formula 5–19, p. 157]:

**Proposition B.33.** *Let $M_1 \in \mathbb{R}^{n\times n}$, $M_2 \in \mathbb{R}^{n\times m}$, and $M_3 \in \mathbb{R}^{m\times m}$. Then*

$$\det \begin{pmatrix} M_1 & M_2 \\ 0 & M_3 \end{pmatrix} = \det(M_1) \cdot \det(M_3).$$

From this, it follows directly, as noted in [Ser10, p. 59] and [HJ12, Exercise p. 52]:

**Proposition B.34.** *Let $M_1 \in \mathbb{R}^{n\times n}$, $M_2 \in \mathbb{R}^{n\times m}$, and $M_3 \in \mathbb{R}^{m\times m}$. Then the eigenvalues of*

$$\begin{pmatrix} M_1 & M_2 \\ 0 & M_3 \end{pmatrix}$$

*are exactly the eigenvalues of $M_1$ and $M_3$ counted with their algebraic multiplicities.*

The following theorem is known as the Perron-Frobenius theorem and traces back to the works of Perron and Frobenius. Perron considered the case of positive matrices in [Per07] (cf. [Per07, Prop., p. 261]), Frobenius extended it to positive matrices in [Fro08] and [Fro09] and to non-negative matrices in [Fro12]. Here, we use the formulation from [Mey00] and focus on the part relevant for this work. First, we define irreducible matrices after [Mey00, p. 671]:

**Definition B.35.** *Let $M \in \mathbb{R}^{n\times n}$ be a matrix. The matrix $M$ is called* reducible *if there exists a permutation matrix $H \in \mathbb{R}^{n\times n}$ such that*

$$H^T M H = \begin{pmatrix} M_1 & M_2 \\ 0 & M_3 \end{pmatrix},$$

*where $M_1$ and $M_3$ are square matrices. Otherwise, $M$ is called* irreducible.

For the notion of irreducible matrices, the following characterization can be derived from [Mey00, p. 209, 671, Solutions p. 36–37]:

**Characterization B.36.** *Let $M \in \mathbb{R}^{n\times n}$ be a matrix. The matrix $M$ is irreducible if and only if the directed graph corresponding to the adjacency matrix $A \in \mathbb{R}^{n\times n}$ defined by*

$$A_{(i,j)} := \begin{cases} 1 & \text{if } M_{(i,j)} \neq 0, \\ 0 & \text{otherwise}, \end{cases}$$

*is strongly connected.*

Under these conditions, we can formulate the relevant part of the Perron-Frobenius theorem for non-negative matrices as given in [Mey00, p. 673]:

**Theorem B.37** (Perron-Frobenius Theorem)**.** *Let $M \in \mathbb{R}^{n\times n}$ be a non-negative irreducible matrix. Then the spectral radius $\rho(M)$ is an eigenvalue of $M$ with algebraic multiplicity one.*

Another central theorem concerning eigenvalues of matrices is the Gerschgorin circle theorem, which we state following [Ger31, Thm. 2, p. 751]:

**Theorem B.38** (Gerschgorin Circle Theorem)**.** *Let $M \in \mathbb{R}^{n\times n}$ be a matrix. Define the sets*

$$\boldsymbol{M}_i := \left\{ a \in \mathbb{C} \ : \ |a - M_{(i,i)}| \leq \sum_{j=1, j\neq i}^{n} |M_{(i,j)}| \right\} \quad \textit{for} \quad i \in \{1, \ldots, n\}.$$

*Each $\boldsymbol{M}_i$ is a disk in the complex plane with center $M_{(i,i)}$ and radius $\sum_{j\neq i} |M_{(i,j)}|$. All eigenvalues of $M$ lie in the union*

$$\bigcup_{i=1}^{n} \boldsymbol{M}_i.$$

We use this theorem to describe eigenvalues of stochastic matrices, which we define following [Ser10, Def. 8.3, p. 156]:

**Definition B.39.** *Let $M \in \mathbb{R}^{n\times n}$ be a matrix. The matrix $M$ is called* stochastic *if it is non-negative and, additionally,*

$$\sum_{j=1}^{n} M_{(i,j)} = 1$$

*for all $i = 1, \ldots, n$.*

Using the Gerschgorin Circle Theorem B.38, it follows directly from [FHW79, Main Thm. 2.6, p. 15–16]:

**Theorem B.40.** *Let $M \in \mathbb{R}^{n\times n}$ be a stochastic matrix. Then the following hold:*

1. *If $\lambda_i$ is an eigenvalue of $M$, then $|\lambda_i| \leq 1$.*
2. *If $\min_i M_{(i,i)} > 0$, then $1$ is the only eigenvalue of $M$ with magnitude $1$.*

From Theorem B.40 and the Perron-Frobenius Theorem B.37 follows directly the next corollary:

**Corollary B.41.** *Let $M \in \mathbb{R}^{n\times n}$ be a stochastic irreducible matrix with eigenvalues*

$$\lambda_0, \ldots, \lambda_{n-1} \in \mathbb{C}, \quad \textit{with} \quad |\lambda_0| \geq \cdots \geq |\lambda_{n-1}|.$$

*If $\min_i M_{(i,i)} > 0$, then $1$ is an eigenvalue of $M$ with algebraic multiplicity $1$, and*

$$1 = \lambda_0 > |\lambda_i| \quad \textit{for all} \quad i \in \{1, \ldots, n-1\}.$$

Next, we consider eigenvalues of symmetric and Hermitian matrices, which we introduce in the following definition:

**Definition B.42.** *A matrix*

- *$M \in \mathbb{R}^{n\times n}$ is called* symmetric *if $M = M^T$ (cf. [Ser10, Def. 5.2, p. 84]).*
- *$M \in \mathbb{C}^{n\times n}$ is called* Hermitian *if $M$ equals its conjugate transpose (cf. [Ser10, Prop. 5.1, p. 84]).*
- *$M \in \mathbb{C}^{n\times n}$ is called* orthogonal *if $M^T M = M M^T = E$, equivalently $M^T = M^{-1}$ (cf. [Ser10, p. 23]).*

From this, the following lemma follows directly (cf. [Ser10, p. 84]):

**Lemma B.43.** *Every symmetric matrix is Hermitian.*

In this work, we focus almost exclusively on real matrices. However, many theorems in the literature are formulated for Hermitian matrices, so we adopt those formulations. Since every symmetric matrix is also Hermitian, the statements transfer directly to symmetric matrices.
In the following, we formulate some theorems for Hermitian matrices, beginning with the next theorem after [Ser10, Thm. 5.4, p. 87]:

**Theorem B.44.** *The eigenvalues of Hermitian and symmetric matrices are real.*

Moreover, we obtain the following statement from [Ser10, Cor. 5.3, p. 93]:

**Theorem B.45.** *Symmetric matrices are diagonalizable by orthogonal matrices over $\mathbb{R}$. More precisely, let $M \in \mathbb{R}^{n\times n}$ be a symmetric matrix, then there exists an orthogonal matrix $B \in \mathbb{R}^{n\times n}$ such that $BMB^{-1}$ is diagonal.*

We now add the concept of definiteness after [HJ12, p. 429]:

**Definition B.46.** *A Hermitian matrix $M \in \mathbb{C}^{n\times n}$ is called* positive definite *if*

$$\operatorname{kon}(a)^T Ma > 0 \quad \textit{for all} \quad a \in \mathbb{C}^n \setminus \{\vec{0}\}.$$

For positive definite matrices, we have the following characterization after [HJ12, Thm. 7.2.1, p. 438] and [HJ12, Thm. 4.1.8, p. 230]:

**Characterization B.47.** *A Hermitian matrix is positive definite if and only if all of its eigenvalues are positive.*

Finally, we formulate the important theorem for this work after [Ser10, Prop. 6.1, p. 109]:

**Theorem B.48.** *Let $B \in \mathbb{C}^{n\times n}$ be a Hermitian positive definite matrix and $M \in \mathbb{C}^{n\times n}$ be a Hermitian matrix. Then the products $BM$ and $MB$ are diagonalizable with real eigenvalues. Moreover, the number of positive and negative eigenvalues of $BM$ equals the number of positive and negative eigenvalues of $M$.*

We conclude by considering the eigenvalues of triangular matrices, introduced next following [HJ12, Sec. 0.9.3, p. 31]:

**Definition B.49.** *Let $M \in \mathbb{C}^{n\times n}$ be a matrix. If*

$$M_{(i,j)} = 0 \quad \textit{for all } i < j, \quad i, j \in \{1, \dots, n\},$$

*then $M$ is called a* lower triangular matrix. *If*

$$M_{(i,j)} = 0 \quad \textit{for all } i > j, \quad i, j \in \{1, \dots, n\},$$

*then $M$ is called an* upper triangular matrix.

For these matrices, we have the following lemma from [Lan86, Exercise VIII, §2, 2, p. 249 and Solution VIII, §2, 2, p. 284]:

**Lemma B.50.** *Let $M \in \mathbb{R}^{n\times n}$ be an upper or lower triangular matrix. Then the eigenvalues of $M$ are exactly its diagonal entries.*

Thus, we have described the concepts and statements needed in the context of eigenvalues and now proceed to consider the matrix exponential in the next section.

## B.2 Matrix Exponential

In this section, we consider the matrix exponential and related statements that are needed for this work. We start with the definition according to [Hal15, Section 2.1, p. 31]:

**Definition B.51.** *Let $M \in \mathbb{C}^{n\times n}$ be a matrix. The* matrix exponential $\exp(M) \in \mathbb{C}^{n\times n}$ *of $M$ is defined as*

$$\exp(M) = \sum_{i=0}^{\infty} \frac{M^i}{i!} = I + M + \frac{M^2}{2!} + \frac{M^3}{3!} + \cdots .$$

For the well-definedness of this concept, we use the following theorem from [Hal15, Prop. 2.1, p. 31]:

**Theorem B.52.** *The series*

$$\sum_{i=0}^{\infty} \frac{M^i}{i!}$$

*converges for all $M \in \mathbb{C}^{n\times n}$.*

For the matrix exponential, several properties hold. We start with the following theorem from [Hal15, Prop. 2.3, p. 32]:

**Theorem B.53.** *Let $M, M' \in \mathbb{C}^{n\times n}$ be two matrices, and let $0_n \in \mathbb{C}^{n\times n}$ be the zero matrix. Then the following hold:*

1. *The exponential of the zero matrix is*
$$\exp(0_n) = E_n.$$
2. *The matrix $\exp(M)$ is invertible and its inverse satisfies*
$$(\exp(M))^{-1} = \exp(-M).$$
3. *For scalars $a, b \in \mathbb{C}$, we have*
$$\exp\big((a+b)M\big) = \exp(aM)\exp(bM).$$
4. *If $M$ and $M'$ commute, i.e. $MM' = M'M$, then*
$$\exp(M + M') = \exp(M)\exp(M') = \exp(M')\exp(M).$$
5. *For $V \in \mathrm{GL}_n(\mathbb{C})$ (the general linear group), we have*
$$\exp(VMV^{-1}) = V\exp(M)V^{-1}.$$

The next statement considers the matrix exponential of diagonal matrices and is as follows according to [Hal15, Section 2.2, p. 34]:

**Theorem B.54.** *Let $D \in \mathbb{C}^{n\times n}$ be a diagonal matrix. Then*

$$\exp(D) = \begin{bmatrix} \exp\left(D_{(1,1)}\right) & 0 & \cdots & 0 \\ 0 & \ddots & \ddots & \vdots \\ \vdots & \ddots & \ddots & 0 \\ 0 & \cdots & 0 & \exp\left(D_{(n,n)}\right) \end{bmatrix}.$$

*Proof.* First, we have

$$\exp(D) = \sum_{i=0}^{\infty} \frac{D^i}{i!} = \sum_{i=0}^{\infty} \begin{bmatrix} \frac{D_{(1,1)}^i}{i!} & 0 & \cdots & 0 \\ 0 & \ddots & \ddots & \vdots \\ \vdots & \ddots & \ddots & 0 \\ 0 & \cdots & 0 & \frac{D_{(n,n)}^i}{i!} \end{bmatrix} = \begin{bmatrix} \sum_{i=0}^{\infty} \frac{D_{(1,1)}^i}{i!} & 0 & \cdots & 0 \\ 0 & \ddots & \ddots & \vdots \\ \vdots & \ddots & \ddots & 0 \\ 0 & \cdots & 0 & \sum_{i=0}^{\infty} \frac{D_{(n,n)}^i}{i!} \end{bmatrix}.$$

Since each diagonal entry is an exponential series, it follows that

$$\exp(D) = \begin{bmatrix} \exp\left(D_{(1,1)}\right) & 0 & \cdots & 0 \\ 0 & \ddots & \ddots & \vdots \\ \vdots & \ddots & \ddots & 0 \\ 0 & \cdots & 0 & \exp\left(D_{(n,n)}\right) \end{bmatrix}.$$

□

Next, we want to define the logarithm of a matrix. This concept plays only a minor role in this work and is therefore treated here only briefly. We start with the definition according to [Hal15, Def. 2.7, p. 37]:

**Definition B.55.** *Let $M \in \mathbb{C}^{n\times n}$ be a matrix. The* matrix logarithm *is defined as*

$$\ln(M) := \sum_{i=1}^{\infty} (-1)^{i+1} \frac{(M-E)^i}{i},$$

*provided that the series converges.*

A criterion for convergence is given in [Hig08, Thm. 1.31, p. 20]:

**Theorem B.56.** *Let $M \in \mathbb{C}^{n\times n}$ be a matrix that has no eigenvalues in the negative real axis $\mathbb{R}_{<0}$. Then there exists a unique logarithm $\ln(M)$ of $M$ whose eigenvalues lie in the region*

$$\{a \in \mathbb{C} \mid -\pi < \text{im}(a) < \pi\}.$$

*If $M$ is a real matrix, then the logarithm with eigenvalues in the stated region is also real.*

With these basics, we can now state some properties of the matrix exponential for special matrices. We start with triangular matrices according to [HJ12, p. 31]:

**Lemma B.57.** *Let $M \in \mathbb{R}^{n\times n}$ be an upper or lower triangular matrix. Then, for any $k \in \mathbb{N}$, the matrix $M^k$ is again an upper or lower triangular matrix, respectively. Moreover,*

$$(M^k)_{(i,i)} = (M_{(i,i)})^k$$

*holds for all $i \in \{1, \ldots, n\}$.*

From this we immediately obtain the following lemma:

**Lemma B.58.** *Let $M \in \mathbb{R}^{n\times n}$ be an upper or lower triangular matrix. Then*

$$\exp(M)$$

*is again an upper or lower triangular matrix, respectively. Furthermore, for the diagonal entries, we have*

$$(\exp(M))_{(i,i)} = \exp(M_{(i,i)})$$

*for $i \in \{1, \ldots, n\}$.*

*Proof.* The matrix exponential is a series of matrix powers. Each power of a triangular matrix is again a triangular matrix by Lemma B.57, and the sum of triangular matrices is again triangular. Therefore, $\exp(M)$ is again a triangular matrix.

For the diagonal entries, we have

$$(\exp(M))_{(i,i)} = \sum_{j=0}^{\infty} \frac{(M^j)_{(i,i)}}{j!} = \sum_{j=0}^{\infty} \frac{(M_{(i,i)})^j}{j!} = \exp(M_{(i,i)}).$$

□

We can use this lemma to formulate the following statement about the eigenvalues of the matrix exponential of triangular matrices:

**Lemma B.59.** *Let $M \in \mathbb{R}^{n\times n}$ be an upper or lower triangular matrix with eigenvalues $\lambda_1, \dots, \lambda_n$. Then the eigenvalues of* $\exp(M)$ *are exactly*

$$\exp(\lambda_1), \dots, \exp(\lambda_n).$$

*Proof.* By Lemma B.50, the eigenvalues of an upper or lower triangular matrix are its diagonal entries. By Lemma B.58, the exponential of a triangular matrix is again triangular, and its diagonal entries, which are also the eigenvalues, are given by

$$\exp(\lambda_1), \dots, \exp(\lambda_n).$$

□

For the eigenvectors of the exponential of triangular matrices, we have:

**Lemma B.60.** *Let $M \in \mathbb{R}^{n\times n}$ be an upper or lower triangular matrix and $\lambda_i$ an eigenvalue with eigenvector $V_{(:,i)}$. Then $V_{(:,i)}$ is an eigenvector of* $\exp(M)$ *with eigenvalue* $\exp(\lambda_i)$.

*Proof.* From the definition of the matrix exponential B.51, we get

$$\exp(M)V_{(:,i)} = \left(\sum_{j=0}^{\infty} \frac{M^j}{j!}\right) V_{(:,i)} = \sum_{j=0}^{\infty} \frac{M^j V_{(:,i)}}{j!} = \sum_{j=0}^{\infty} \frac{\lambda_i^j}{j!} V_{(:,i)} = \left(\sum_{j=0}^{\infty} \frac{\lambda_i^j}{j!}\right) V_{(:,i)} = \exp(\lambda_i)V_{(:,i)}.$$

□

The next theorem concerns matrices with a special structure related to zero entries. This theorem is based on the powers of the adjacency matrix, which can be found, among others, in [Die17, Exercise 49, p. 33]:

**Theorem B.61.** *Let $\boldsymbol{G} = (\boldsymbol{V}, \boldsymbol{E})$ be a directed graph possibly with loops, where $\boldsymbol{V} = \{v_1, \dots, v_n\}$ and $|\boldsymbol{V}| = n$ is the number of vertices. Let $M \in \mathbb{R}^{n\times n}_{\geq 0}$ be a non-negative matrix with*

$$M_{(i,j)} = \begin{cases} > 0 & \text{if } (v_i, v_j) \in \boldsymbol{E}, \\ = 0 & \text{if } (v_i, v_j) \notin \boldsymbol{E}. \end{cases}$$

*Then for any $k \in \mathbb{N}$, it holds that*

$$(M^k)_{(i,j)} \neq 0 \quad \Leftrightarrow \quad \text{There exists a walk of length } k \text{ in } \boldsymbol{G} \text{ from } v_i \text{ to } v_j.$$

*Proof.* First, note that a vertex can be adjacent to itself if there is a loop at that vertex. We prove the statement by induction:

*Base case*: For $k = 1$, an entry of $M$ is nonzero exactly when there is an edge between the corresponding vertices. Since an edge is a walk of length 1, the statement holds.

*Induction hypothesis*:

$$(M^k)_{(i,j)} \neq 0 \quad \Leftrightarrow \quad \text{There is a walk of length } k \text{ in } \mathbf{G} \text{ from } v_i \text{ to } v_j.$$

*Induction step*: We prove the equivalence by showing both implications:

"⇒": Assume $(M^{k+1})_{(i,j)} \neq 0$. Then

$$(M^{k+1})_{(i,j)} = (M^k)_{(i,:)}M_{(:,j)} = \sum_{\ell=1}^{n} (M^k)_{(i,\ell)}M_{(\ell,j)}.$$

Since all entries of $M$ and hence of $M^k$ are non-negative, the sum consists only of non-negative terms. If $(M^{k+1})_{(i,j)} \neq 0$, then there exists at least one summand $(M^k)_{(i,m)}M_{(m,j)}$ with both factors positive. By the induction hypothesis,

$$(M^k)_{(i,m)} \neq 0 \quad \Rightarrow \quad \text{There is a walk of length } k \text{ in } \mathbf{G} \text{ from } v_i \text{ to } v_m.$$

and

$$M_{(m,j)} \neq 0 \quad \Rightarrow \quad \text{There is a walk of length } 1 \text{ in } \mathbf{G} \text{ from } v_m \text{ to } v_j.$$

Therefore, there is a walk

$$\underbrace{v_i, \ldots, v_m}_{k \text{ vertices}}, v_j$$

of length $k+1$ from $v_i$ to $v_j$.

"$\Leftarrow$": Assume there is a walk of length $k+1$ from $v_i$ to $v_j$. Then there exists a walk of length $k$ from $v_i$ to a neighbor $v_m$ of $v_j$. By the induction hypothesis,

$$(M^k)_{(i,m)} \neq 0 \quad \Leftarrow \quad \text{There is a walk of length } k \text{ in } \mathbf{G} \text{ from } v_i \text{ to } v_m.$$

and

$$M_{(m,j)} \neq 0 \quad \Leftarrow \quad \text{There is a walk of length } 1 \text{ in } \mathbf{G} \text{ from } v_m \text{ to } v_j.$$

For the entry $(M^{k+1})_{(i,j)}$, we get

$$(M^{k+1})_{(i,j)} = (M^k)_{(i,:)}M_{(:,j)} = \sum_{\ell=1}^{n}(M^k)_{(i,\ell)}M_{(\ell,j)}.$$

Since $(M^k)_{(i,m)} \cdot M_{(m,j)} \neq 0$ is a summand of the above sum, and all terms are non-negative, the sum is positive, so

$$(M^{k+1})_{(i,j)} \neq 0.$$

□

From this follows the next corollary:

**Corollary B.62.** *Let $\boldsymbol{G} = (\boldsymbol{V}, \boldsymbol{E})$ be a directed graph possibly with loops, where $\boldsymbol{V} = \{v_1, \ldots, v_n\}$ and $|\boldsymbol{V}| = n$ is the number of vertices. Let $M \in \mathbb{R}_{\geq 0}^{n \times n}$ be a non-negative matrix with*

$$M_{(i,j)} \begin{cases} > 0 & \text{if } (v_i, v_j) \in \boldsymbol{E}, \\ = 0 & \text{if } (v_i, v_j) \notin \boldsymbol{E}, \end{cases}$$

*for $i, j \in \{1, \ldots, n\}$. Then*

$$\exp(M)_{(i,j)} \begin{cases} > 0 & \text{if } i = j, \\ > 0 & \text{if } i \neq j \text{ and there exists a walk between } v_i \text{ and } v_j, \\ = 0 & \text{if } i \neq j \text{ and there exists no walk between } v_i \text{ and } v_j. \end{cases}$$

*Proof.* By Definition B.51 we have

$$\exp(M)_{(i,j)} = \sum_{k=0}^{\infty} \frac{(M^k)_{(i,j)}}{k!}.$$

Since $M$ has only non-negative entries, all powers of $M$ also consist solely of non-negative entries.

Because $M^0 = E_n$, the term with $k=0$ for $i=j$ is positive, hence $\exp(M)_{(i,i)}$ is positive.

If for $i \neq j$ there is no walk between vertices $v_i$ and $v_j$, then in particular no walk of length $k$ exists between $v_i$ and $v_j$ for any $k$. By Theorem B.61 all summands in the above series are zero, so $\exp(M)_{(i,j)} = 0$.

Conversely, if for $i \neq j$ there exists a walk between $v_i$ and $v_j$, then by Theorem B.61 at least one summand in the series is positive. Since no power of $M$ has negative entries, $\exp(M)_{(i,j)}$ is positive. □

We conclude this section with a statement about the matrix exponential for matrices whose off-diagonal entries are all non-negative. The lemma is taken from [Ang14][1]:

**Lemma B.63.** *Let $M \in \mathbb{R}^{n\times n}$ be a matrix whose off-diagonal entries are all non-negative. Then every entry of* $\exp(M)$ *is non-negative.*

*Proof.* Let $a$ be the smallest diagonal entry of the matrix $M$, and define $M' := M + aE_n$. Then all entries of $M'$ satisfy $\geq 0$, and $M = M' - aE_n$. Since the matrices $M'$ and $aE_n$ commute, by Proposition B.53 we get

$$\exp(M) = \exp(M' - aE_n) = \exp(M')\exp(-aE_n) = \exp(M')\exp(-a).$$

By definition, the matrix exponential is a series of powers of the respective matrix. The matrix exponential of a non-negative matrix is again non-negative. Hence $\exp(M')$ is a non-negative matrix multiplied by the positive scalar $\exp(-a)$. Thus the product, and therefore $\exp(M)$, is non-negative. □

With this, we have described all necessary statements regarding the matrix exponential. In the next section, we consider concrete matrix exponentials needed for this work.

## B.3 Concrete Matrix Exponentials

For some statements in Chapters 5 and 6, we need the exponentials of specific matrices, which we compute in the following. We start with the following block matrix:

**Lemma B.64.** *Let $M \in \mathbb{R}^{n\times n}$, $i \in \mathbb{N}_0$, and define*

$$\overline{M} := \begin{bmatrix} M & 0 \\ 0 & M \end{bmatrix}.$$

*Then it holds that*

$$\overline{M}^i = \begin{bmatrix} M & 0 \\ 0 & M \end{bmatrix}^i = \begin{bmatrix} M^i & 0 \\ 0 & M^i \end{bmatrix}.$$

*Proof.* We prove this lemma by induction.
*Base case*:

$$\overline{M}^0 = \begin{bmatrix} M & 0 \\ 0 & M \end{bmatrix}^0 = \begin{bmatrix} E & 0 \\ 0 & E \end{bmatrix} = \begin{bmatrix} M^0 & 0 \\ 0 & M^0 \end{bmatrix} \quad \text{and} \quad \overline{M}^1 = \begin{bmatrix} M & 0 \\ 0 & M \end{bmatrix}^1 = \begin{bmatrix} M^1 & 0 \\ 0 & M^1 \end{bmatrix}.$$

*Induction hypothesis*:

$$\overline{M}^i = \begin{bmatrix} M & 0 \\ 0 & M \end{bmatrix}^i = \begin{bmatrix} M^i & 0 \\ 0 & M^i \end{bmatrix}.$$

*Induction step*:

$$\overline{M}^{i+1} = \overline{M}^i\overline{M} = \begin{bmatrix} M^i & 0 \\ 0 & M^i \end{bmatrix}\begin{bmatrix} M & 0 \\ 0 & M \end{bmatrix} = \begin{bmatrix} M^{i+1} & 0 \\ 0 & M^{i+1} \end{bmatrix}.$$

□

From this lemma, we directly obtain the following corollary:

[1] The lemma and its proof come from a forum post on the website „math.stackexchange.com“. Unfortunately, no corresponding reference could be found in the literature, and the author himself did not know of any. Presumably, the lemma is too specialized. The author agreed upon request that the source may be cited as indicated.

**Corollary B.65.** *Let* $M \in \mathbb{R}^{n\times n}$ *and*

$$\overline{M} := \begin{bmatrix} M & 0 \\ 0 & M \end{bmatrix}.$$

*Then it holds that*

$$\exp\left(\overline{M}\right) = \begin{bmatrix} \exp(M) & 0 \\ 0 & \exp(M) \end{bmatrix}.$$

*Proof.* By the previous lemma, we have

$$\exp\left(\overline{M}\right) = \sum_{i=0}^{\infty} \frac{\overline{M}^i}{i!} = \sum_{i=0}^{\infty} \frac{1}{i!} \begin{bmatrix} M & 0 \\ 0 & M \end{bmatrix}^i = \sum_{i=0}^{\infty} \frac{1}{i!} \begin{bmatrix} M^i & 0 \\ 0 & M^i \end{bmatrix} = \begin{bmatrix} \sum_{i=0}^{\infty} \frac{M^i}{i!} & 0 \\ 0 & \sum_{i=0}^{\infty} \frac{M^i}{i!} \end{bmatrix} = \begin{bmatrix} \exp(M) & 0 \\ 0 & \exp(M) \end{bmatrix}.$$

□

The same statement holds for matrices with three blocks:

**Lemma B.66.** *Let* $M \in \mathbb{R}^{n\times n}$, $i \in \mathbb{N}_0$, *and define*

$$\overline{M} := \begin{bmatrix} M & 0 & 0 \\ 0 & M & 0 \\ 0 & 0 & M \end{bmatrix}.$$

*Then it holds that*

$$\overline{M}^i = \begin{bmatrix} M & 0 & 0 \\ 0 & M & 0 \\ 0 & 0 & M \end{bmatrix}^i = \begin{bmatrix} M^i & 0 & 0 \\ 0 & M^i & 0 \\ 0 & 0 & M^i \end{bmatrix}.$$

*Proof.* The proof proceeds analogously to Lemma B.64. □

From this follows immediately the next corollary:

**Corollary B.67.** *Let* $M \in \mathbb{R}^{n\times n}$ *and*

$$\overline{M} := \begin{bmatrix} M & 0 & 0 \\ 0 & M & 0 \\ 0 & 0 & M \end{bmatrix}.$$

*Then*

$$\exp\left(\overline{M}\right) = \begin{bmatrix} \exp(M) & 0 & 0 \\ 0 & \exp(M) & 0 \\ 0 & 0 & \exp(M) \end{bmatrix}.$$

*Proof.* The proof proceeds analogously to Corollary B.65. □

For matrices with two blocks on the off-diagonal, the picture is more differentiated. We first obtain the following lemma:

**Lemma B.68.** *Let* $M \in \mathbb{R}^{n\times n}$, $i \in \mathbb{N}_0$, *and define*

$$\overline{M} := \begin{bmatrix} 0 & M \\ M & 0 \end{bmatrix}.$$

*Then it holds that*

$$\overline{M}^{2i} = \begin{bmatrix} 0 & M \\ M & 0 \end{bmatrix}^{2i} = \begin{bmatrix} M^{2i} & 0 \\ 0 & M^{2i} \end{bmatrix}.$$

*Proof.* We prove this lemma by induction.
*Base case*:

$$\overline{M}^0 = \begin{bmatrix} 0 & M \\ M & 0 \end{bmatrix}^0 = \begin{bmatrix} E & 0 \\ 0 & E \end{bmatrix} = \begin{bmatrix} M^0 & 0 \\ 0 & M^0 \end{bmatrix} \quad \text{and} \quad \overline{M}^2 = \begin{bmatrix} 0 & M \\ M & 0 \end{bmatrix}^2 = \begin{bmatrix} 0 & M \\ M & 0 \end{bmatrix}\begin{bmatrix} 0 & M \\ M & 0 \end{bmatrix} = \begin{bmatrix} M^2 & 0 \\ 0 & M^2 \end{bmatrix}.$$

*Induction hypothesis*:

$$\overline{M}^{2i} = \begin{bmatrix} 0 & M \\ M & 0 \end{bmatrix}^{2i} = \begin{bmatrix} M^{2i} & 0 \\ 0 & M^{2i} \end{bmatrix}.$$

*Induction step*:

$$\overline{M}^{2i+2} = \overline{M}^{2i}\overline{M}^{2} = \begin{bmatrix} M^{2i} & 0 \\ 0 & M^{2i} \end{bmatrix} \begin{bmatrix} M^{2} & 0 \\ 0 & M^{2} \end{bmatrix} = \begin{bmatrix} M^{2i+2} & 0 \\ 0 & M^{2i+2} \end{bmatrix}.$$

□

The other powers are given by the following lemma:

**Lemma B.69.** *Let $M \in \mathbb{R}^{n\times n}$, $i \in \mathbb{N}_0$, and define*

$$\overline{M} := \begin{bmatrix} 0 & M \\ M & 0 \end{bmatrix}.$$

*Then it holds that*

$$\overline{M}^{2i+1} = \begin{bmatrix} 0 & M \\ M & 0 \end{bmatrix}^{2i+1} = \begin{bmatrix} 0 & M^{2i+1} \\ M^{2i+1} & 0 \end{bmatrix}.$$

*Proof.* We prove this lemma by induction.
*Base case*:

$$\overline{M}^{2\cdot 0+1} = \begin{bmatrix} 0 & M \\ M & 0 \end{bmatrix}^{2\cdot 0+1} = \begin{bmatrix} 0 & M \\ M & 0 \end{bmatrix}.$$

*Induction hypothesis*: Assume that

$$\overline{M}^{2i+1} = \begin{bmatrix} 0 & M \\ M & 0 \end{bmatrix}^{2i+1} = \begin{bmatrix} 0 & M^{2i+1} \\ M^{2i+1} & 0 \end{bmatrix}.$$

*Induction step*:

$$\overline{M}^{2i+3} = \begin{bmatrix} 0 & M \\ M & 0 \end{bmatrix}^{2i+3} = \begin{bmatrix} 0 & M \\ M & 0 \end{bmatrix}^{2i+1} \begin{bmatrix} 0 & M \\ M & 0 \end{bmatrix}^{2} = \begin{bmatrix} 0 & M^{2i+1} \\ M^{2i+1} & 0 \end{bmatrix} \begin{bmatrix} M^{2} & 0 \\ 0 & M^{2} \end{bmatrix} = \begin{bmatrix} 0 & M^{2i+3} \\ M^{2i+3} & 0 \end{bmatrix}.$$

□

From Lemmas B.68 and B.69 we derive the following corollary:

**Corollary B.70.** *Let $a \in \mathbb{R}$ and $n \in \mathbb{N}$. Then*

$$\exp\left(\begin{bmatrix} 0 & a\cdot E_n \\ a\cdot E_n & 0 \end{bmatrix}\right) = \begin{bmatrix} \cosh(a)E_n & \sinh(a)E_n \\ \sinh(a)E_n & \cosh(a)E_n \end{bmatrix}.$$

*Proof.* Using the definition of the matrix exponential, we obtain

$$\exp\left(\begin{bmatrix} 0 & a\cdot E_n \\ a\cdot E_n & 0 \end{bmatrix}\right) = \sum_{i=0}^{\infty} \frac{1}{i!} \begin{bmatrix} 0 & a\cdot E_n \\ a\cdot E_n & 0 \end{bmatrix}^{i}.$$

By Lemmas B.68 and B.69, this equals

$$\exp\left(\begin{bmatrix} 0 & a\cdot E_n \\ a\cdot E_n & 0 \end{bmatrix}\right) = \sum_{i=0}^{\infty} \begin{bmatrix} \frac{(a\cdot E_n)^{2i}}{(2i)!} & \frac{(a\cdot E_n)^{2i+1}}{(2i+1)!} \\ \frac{(a\cdot E_n)^{2i+1}}{(2i+1)!} & \frac{(a\cdot E_n)^{2i}}{(2i)!} \end{bmatrix} = \begin{bmatrix} \sum_{i=0}^{\infty} \frac{a^{2i}}{(2i)!}E_n & \sum_{i=0}^{\infty} \frac{a^{2i+1}}{(2i+1)!}E_n \\ \sum_{i=0}^{\infty} \frac{a^{2i+1}}{(2i+1)!}E_n & \sum_{i=0}^{\infty} \frac{a^{2i}}{(2i)!}E_n \end{bmatrix}.$$

Using

$$\sinh(a) = \sum_{i=0}^{\infty} \frac{a^{2i+1}}{(2i+1)!} \quad \text{and} \quad \cosh(a) = \sum_{i=0}^{\infty} \frac{a^{2i}}{(2i)!},$$

we conclude

$$\exp\left(\begin{bmatrix} 0 & a \cdot E_n \\ a \cdot E_n & 0 \end{bmatrix}\right) = \begin{bmatrix} \cosh(a)E_n & \sinh(a)E_n \\ \sinh(a)E_n & \cosh(a)E_n \end{bmatrix}.$$

□

A similar statement is needed for special matrices with a block structure composed of $3 \times 3$ blocks. We begin with the following lemma:

**Lemma B.71.** *Let $E_n \in \mathbb{R}^{n \times n}$ be the identity matrix, $0_n$ the $n \times n$ zero matrix, and $i \in \mathbb{N}_0$. Then*

$$\begin{bmatrix} 0_n & 0_n & 0_n \\ \frac{1}{2}E_n & -E_n & \frac{1}{2}E_n \\ 0_n & 0_n & 0_n \end{bmatrix}^{2i+1} = \begin{bmatrix} 0_n & 0_n & 0_n \\ \frac{1}{2}E_n & -E_n & \frac{1}{2}E_n \\ 0_n & 0_n & 0_n \end{bmatrix}.$$

*Proof.* We prove the statement by induction:
*Base case*:

$$\begin{bmatrix} 0_n & 0_n & 0_n \\ \frac{1}{2}E_n & -E_n & \frac{1}{2}E_n \\ 0_n & 0_n & 0_n \end{bmatrix}^{2\cdot 0+1} = \begin{bmatrix} 0_n & 0_n & 0_n \\ \frac{1}{2}E_n & -E_n & \frac{1}{2}E_n \\ 0_n & 0_n & 0_n \end{bmatrix}^{1} = \begin{bmatrix} 0_n & 0_n & 0_n \\ \frac{1}{2}E_n & -E_n & \frac{1}{2}E_n \\ 0_n & 0_n & 0_n \end{bmatrix}$$

*Induction hypothesis*:

$$\begin{bmatrix} 0_n & 0_n & 0_n \\ \frac{1}{2}E_n & -E_n & \frac{1}{2}E_n \\ 0_n & 0_n & 0_n \end{bmatrix}^{2i+1} = \begin{bmatrix} 0_n & 0_n & 0_n \\ \frac{1}{2}E_n & -E_n & \frac{1}{2}E_n \\ 0_n & 0_n & 0_n \end{bmatrix}.$$

*Induction step*:

$$\begin{aligned}
\begin{bmatrix} 0_n & 0_n & 0_n \\ \frac{1}{2}E_n & -E_n & \frac{1}{2}E_n \\ 0_n & 0_n & 0_n \end{bmatrix}^{2(i+1)+1} &= \begin{bmatrix} 0_n & 0_n & 0_n \\ \frac{1}{2}E_n & -E_n & \frac{1}{2}E_n \\ 0_n & 0_n & 0_n \end{bmatrix}^{2i+1} \begin{bmatrix} 0_n & 0_n & 0_n \\ \frac{1}{2}E_n & -E_n & \frac{1}{2}E_n \\ 0_n & 0_n & 0_n \end{bmatrix}^{2} \\
&= \begin{bmatrix} 0_n & 0_n & 0_n \\ \frac{1}{2}E_n & -E_n & \frac{1}{2}E_n \\ 0_n & 0_n & 0_n \end{bmatrix} \begin{bmatrix} 0_n & 0_n & 0_n \\ -\frac{1}{2}E_n & E_n & -\frac{1}{2}E_n \\ 0_n & 0_n & 0_n \end{bmatrix} \\
&= \begin{bmatrix} 0_n & 0_n & 0_n \\ \frac{1}{2}E_n & -E_n & \frac{1}{2}E_n \\ 0_n & 0_n & 0_n \end{bmatrix}.
\end{aligned}$$

□

For the other powers, we have the following lemma:

**Lemma B.72.** *Let $E_n \in \mathbb{R}^{n \times n}$ be the identity matrix, $0_n$ the $n \times n$ zero matrix, and $i \in \mathbb{N}$. Then*

$$\begin{bmatrix} 0_n & 0_n & 0_n \\ \frac{1}{2}E_n & -E_n & \frac{1}{2}E_n \\ 0_n & 0_n & 0_n \end{bmatrix}^{2i} = \begin{bmatrix} 0_n & 0_n & 0_n \\ -\frac{1}{2}E_n & E_n & -\frac{1}{2}E_n \\ 0_n & 0_n & 0_n \end{bmatrix}.$$

*Proof.* We prove the statement by induction:
*Base case*:

$$\begin{bmatrix} 0_n & 0_n & 0_n \\ \frac{1}{2}E_n & -E_n & \frac{1}{2}E_n \\ 0_n & 0_n & 0_n \end{bmatrix}^{2\cdot 1} = \begin{bmatrix} 0_n & 0_n & 0_n \\ \frac{1}{2}E_n & -E_n & \frac{1}{2}E_n \\ 0_n & 0_n & 0_n \end{bmatrix} \cdot \begin{bmatrix} 0_n & 0_n & 0_n \\ \frac{1}{2}E_n & -E_n & \frac{1}{2}E_n \\ 0_n & 0_n & 0_n \end{bmatrix} = \begin{bmatrix} 0_n & 0_n & 0_n \\ -\frac{1}{2}E_n & E_n & -\frac{1}{2}E_n \\ 0_n & 0_n & 0_n \end{bmatrix}$$

*Induction hypothesis*: Assume that

$$\begin{bmatrix} 0_n & 0_n & 0_n \\ \frac{1}{2}E_n & -E_n & \frac{1}{2}E_n \\ 0_n & 0_n & 0_n \end{bmatrix}^{2i} = \begin{bmatrix} 0_n & 0_n & 0_n \\ -\frac{1}{2}E_n & E_n & -\frac{1}{2}E_n \\ 0_n & 0_n & 0_n \end{bmatrix}.$$

*Induction step*:

$$\begin{aligned} \begin{bmatrix} 0_n & 0_n & 0_n \\ \frac{1}{2}E_n & -E_n & \frac{1}{2}E_n \\ 0_n & 0_n & 0_n \end{bmatrix}^{2i+2} &= \begin{bmatrix} 0_n & 0_n & 0_n \\ \frac{1}{2}E_n & -E_n & \frac{1}{2}E_n \\ 0_n & 0_n & 0_n \end{bmatrix}^{2i} \cdot \begin{bmatrix} 0_n & 0_n & 0_n \\ \frac{1}{2}E_n & -E_n & \frac{1}{2}E_n \\ 0_n & 0_n & 0_n \end{bmatrix}^{2} \\ &= \begin{bmatrix} 0_n & 0_n & 0_n \\ -\frac{1}{2}E_n & E_n & -\frac{1}{2}E_n \\ 0_n & 0_n & 0_n \end{bmatrix} \cdot \begin{bmatrix} 0_n & 0_n & 0_n \\ -\frac{1}{2}E_n & E_n & -\frac{1}{2}E_n \\ 0_n & 0_n & 0_n \end{bmatrix} \\ &= \begin{bmatrix} 0_n & 0_n & 0_n \\ -\frac{1}{2}E_n & E_n & -\frac{1}{2}E_n \\ 0_n & 0_n & 0_n \end{bmatrix}. \end{aligned}$$

□

Combining these two lemmas, we obtain the following corollary:

**Corollary B.73.** *Let $E_n \in \mathbb{R}^{n\times n}$ be the identity matrix and $0_n$ the $n \times n$ zero matrix. Then*

$$\exp\left(\ln(2) \cdot \begin{bmatrix} 0_n & 0_n & 0_n \\ \frac{1}{2}E_n & -E_n & \frac{1}{2}E_n \\ 0_n & 0_n & 0_n \end{bmatrix}\right) = \begin{bmatrix} E_n & 0_n & 0_n \\ \frac{1}{4}E_n & \frac{1}{2}E_n & \frac{1}{4}E_n \\ 0_n & 0_n & E_n \end{bmatrix}.$$

*Proof.* We start with the definition of the matrix exponential B.51 and get

$$\begin{aligned} \exp\left(\ln(2) \cdot \begin{bmatrix} 0_n & 0_n & 0_n \\ \frac{1}{2}E_n & -E_n & \frac{1}{2}E_n \\ 0_n & 0_n & 0_n \end{bmatrix}\right) &= \begin{bmatrix} E_n & 0_n & 0_n \\ 0_n & E_n & 0_n \\ 0_n & 0_n & E_n \end{bmatrix} + \sum_{i=1}^{\infty} \frac{\left(\ln(2) \cdot \begin{bmatrix} 0_n & 0_n & 0_n \\ \frac{1}{2}E_n & -E_n & \frac{1}{2}E_n \\ 0_n & 0_n & 0_n \end{bmatrix}\right)^i}{i!} \\ &=: \begin{bmatrix} E_n & 0_n & 0_n \\ 0_n & E_n & 0_n \\ 0_n & 0_n & E_n \end{bmatrix} + M. \end{aligned}$$

Looking at the sum $M$, we write

$$M = \sum_{i=0}^{\infty} \frac{\ln(2)^{2i+1} \cdot \begin{bmatrix} 0_n & 0_n & 0_n \\ \frac{1}{2}E_n & -E_n & \frac{1}{2}E_n \\ 0_n & 0_n & 0_n \end{bmatrix}^{2i+1}}{(2i+1)!} + \frac{\ln(2)^{2i+2} \cdot \begin{bmatrix} 0_n & 0_n & 0_n \\ \frac{1}{2}E_n & -E_n & \frac{1}{2}E_n \\ 0_n & 0_n & 0_n \end{bmatrix}^{2i+2}}{(2i+2)!}.$$

Using Lemmas B.71 and B.72 we get

$$M = \sum_{i=0}^{\infty} \frac{\ln(2)^{2i+1} \cdot \begin{bmatrix} 0_n & 0_n & 0_n \\ \frac{1}{2}E_n & -E_n & \frac{1}{2}E_n \\ 0_n & 0_n & 0_n \end{bmatrix}}{(2i+1)!} + \frac{\ln(2)^{2i+2} \cdot \begin{bmatrix} 0_n & 0_n & 0_n \\ -\frac{1}{2}E_n & E_n & -\frac{1}{2}E_n \\ 0_n & 0_n & 0_n \end{bmatrix}}{(2i+2)!}.$$

Since the matrices no longer depend on the summation index, we write

$$M = \begin{bmatrix} 0_n & 0_n & 0_n \\ \frac{1}{2}E_n & -E_n & \frac{1}{2}E_n \\ 0_n & 0_n & 0_n \end{bmatrix} \sum_{i=0}^{\infty} \frac{\ln(2)^{2i+1}}{(2i+1)!} - \frac{\ln(2)^{2i+2}\cdot}{(2i+2)!}.$$

Using a computer algebra system (CAS), we evaluate the series and get

$$M = \begin{bmatrix} 0_n & 0_n & 0_n \\ \frac{1}{2}E_n & -E_n & \frac{1}{2}E_n \\ 0_n & 0_n & 0_n \end{bmatrix} \cdot \frac{1}{2} = \begin{bmatrix} 0_n & 0_n & 0_n \\ \frac{1}{4}E_n & -\frac{1}{2}E_n & \frac{1}{4}E_n \\ 0_n & 0_n & 0_n \end{bmatrix}.$$

Adding this to the identity matrix yields

$$\exp\left(\ln(2) \cdot \begin{bmatrix} 0_n & 0_n & 0_n \\ \frac{1}{2}E_n & -E_n & \frac{1}{2}E_n \\ 0_n & 0_n & 0_n \end{bmatrix}\right) = \begin{bmatrix} E_n & 0_n & 0_n \\ \frac{1}{4}E_n & \frac{1}{2}E_n & \frac{1}{4}E_n \\ 0_n & 0_n & E_n \end{bmatrix}.$$

□

We now consider the same statements for another special matrix and begin with the following lemma:

**Lemma B.74.** *Let $E_n \in \mathbb{R}^{n\times n}$ be the identity matrix, $0_n$ the $n \times n$ matrix consisting only of zeros, and let $i \in \mathbb{N}_0$. Then*

$$\begin{bmatrix} -E_n & E_n & 0_n \\ 0_n & 0_n & 0_n \\ 0_n & E_n & -E_n \end{bmatrix}^{2i+1} = \begin{bmatrix} -E_n & E_n & 0_n \\ 0_n & 0_n & 0_n \\ 0_n & E_n & -E_n \end{bmatrix}.$$

*Proof.* We prove the statement by induction:
*Base case*:

$$\begin{bmatrix} -E_n & E_n & 0_n \\ 0_n & 0_n & 0_n \\ 0_n & E_n & -E_n \end{bmatrix}^{2\cdot 0+1} = \begin{bmatrix} -E_n & E_n & 0_n \\ 0_n & 0_n & 0_n \\ 0_n & E_n & -E_n \end{bmatrix}$$

*Inductive hypothesis*: Assume

$$\begin{bmatrix} -E_n & E_n & 0_n \\ 0_n & 0_n & 0_n \\ 0_n & E_n & -E_n \end{bmatrix}^{2i+1} = \begin{bmatrix} -E_n & E_n & 0_n \\ 0_n & 0_n & 0_n \\ 0_n & E_n & -E_n \end{bmatrix}.$$

*Inductive step*:

$$\begin{aligned} \begin{bmatrix} -E_n & E_n & 0_n \\ 0_n & 0_n & 0_n \\ 0_n & E_n & -E_n \end{bmatrix}^{2(i+1)+1} &= \begin{bmatrix} -E_n & E_n & 0_n \\ 0_n & 0_n & 0_n \\ 0_n & E_n & -E_n \end{bmatrix}^{2i+1} \begin{bmatrix} -E_n & E_n & 0_n \\ 0_n & 0_n & 0_n \\ 0_n & E_n & -E_n \end{bmatrix}^{2} \\ &= \begin{bmatrix} -E_n & E_n & 0_n \\ 0_n & 0_n & 0_n \\ 0_n & E_n & -E_n \end{bmatrix} \begin{bmatrix} E_n & -E_n & 0_n \\ 0_n & 0_n & 0_n \\ 0_n & -E_n & E_n \end{bmatrix} \\ &= \begin{bmatrix} -E_n & E_n & 0_n \\ 0_n & 0_n & 0_n \\ 0_n & E_n & -E_n \end{bmatrix} \end{aligned}$$

□

The other powers follow from the next lemma:

**Lemma B.75.** *Let $E_n \in \mathbb{R}^{n\times n}$ be the identity matrix, $0_n$ the $n \times n$ zero matrix, and $i \in \mathbb{N}$. Then*

$$\begin{bmatrix} -E_n & E_n & 0_n \\ 0_n & 0_n & 0_n \\ 0_n & E_n & -E_n \end{bmatrix}^{2i} = \begin{bmatrix} E_n & -E_n & 0_n \\ 0_n & 0_n & 0_n \\ 0_n & -E_n & E_n \end{bmatrix}.$$

*Proof.* We prove the statement by induction:
*Base case*:

$$\begin{bmatrix} -E_n & E_n & 0_n \\ 0_n & 0_n & 0_n \\ 0_n & E_n & -E_n \end{bmatrix}^{2} = \begin{bmatrix} E_n & -E_n & 0_n \\ 0_n & 0_n & 0_n \\ 0_n & -E_n & E_n \end{bmatrix}$$

*Inductive hypothesis*: Assume

$$\begin{bmatrix} -E_n & E_n & 0_n \\ 0_n & 0_n & 0_n \\ 0_n & E_n & -E_n \end{bmatrix}^{2i} = \begin{bmatrix} E_n & -E_n & 0_n \\ 0_n & 0_n & 0_n \\ 0_n & -E_n & E_n \end{bmatrix}.$$

*Inductive step*:

$$\begin{aligned}
\begin{bmatrix} -E_n & E_n & 0_n \\ 0_n & 0_n & 0_n \\ 0_n & E_n & -E_n \end{bmatrix}^{2i+2} &= \begin{bmatrix} -E_n & E_n & 0_n \\ 0_n & 0_n & 0_n \\ 0_n & E_n & -E_n \end{bmatrix}^{2i} \cdot \begin{bmatrix} -E_n & E_n & 0_n \\ 0_n & 0_n & 0_n \\ 0_n & E_n & -E_n \end{bmatrix}^{2} \\
&= \begin{bmatrix} E_n & -E_n & 0_n \\ 0_n & 0_n & 0_n \\ 0_n & -E_n & E_n \end{bmatrix} \cdot \begin{bmatrix} E_n & -E_n & 0_n \\ 0_n & 0_n & 0_n \\ 0_n & -E_n & E_n \end{bmatrix} \\
&= \begin{bmatrix} E_n & -E_n & 0_n \\ 0_n & 0_n & 0_n \\ 0_n & -E_n & E_n \end{bmatrix}
\end{aligned}$$

□

We use these two lemmas for the following corollary:

**Corollary B.76.** *Let $E_n \in \mathbb{R}^{n\times n}$ be the identity matrix and $0_n$ the $n \times n$ matrix consisting only of zeros. Then*

$$\exp\left(\ln(2) \cdot \begin{bmatrix} -E_n & E_n & 0_n \\ 0_n & 0_n & 0_n \\ 0_n & E_n & -E_n \end{bmatrix}\right) = \begin{bmatrix} \frac{1}{2}E_n & \frac{1}{2}E_n & 0_n \\ 0_n & E_n & 0_n \\ 0_n & \frac{1}{2}E_n & \frac{1}{2}E_n \end{bmatrix}.$$

*Proof.* We begin by using the definition of the matrix exponential B.51, which gives us

$$\begin{aligned}
\exp\left(\ln(2) \cdot \begin{bmatrix} -E_n & E_n & 0_n \\ 0_n & 0_n & 0_n \\ 0_n & E_n & -E_n \end{bmatrix}\right) &= \begin{bmatrix} E_n & 0_n & 0_n \\ 0_n & E_n & 0_n \\ 0_n & 0_n & E_n \end{bmatrix} + \sum_{i=1}^{\infty} \frac{\left(\ln(2) \cdot \begin{bmatrix} -E_n & E_n & 0_n \\ 0_n & 0_n & 0_n \\ 0_n & E_n & -E_n \end{bmatrix}\right)^i}{i!} \\
&=: \begin{bmatrix} E_n & 0_n & 0_n \\ 0_n & E_n & 0_n \\ 0_n & 0_n & E_n \end{bmatrix} + M.
\end{aligned}$$

Now we consider the second term $M$, for which we have

$$M = \sum_{i=0}^{\infty} \frac{\ln(2)^{2i+1} \cdot \begin{bmatrix} -E_n & E_n & 0_n \\ 0_n & 0_n & 0_n \\ 0_n & E_n & -E_n \end{bmatrix}^{2i+1}}{(2i+1)!} + \frac{\ln(2)^{2i+2} \cdot \begin{bmatrix} -E_n & E_n & 0_n \\ 0_n & 0_n & 0_n \\ 0_n & E_n & -E_n \end{bmatrix}^{2i+2}}{(2i+2)!}.$$

Using Lemmas B.74 and B.75, we obtain

$$M=\sum_{i=0}^{\infty}\frac{\ln(2)^{2i+1}\cdot\begin{bmatrix}-E_n & E_n & 0_n\\ 0_n & 0_n & 0_n\\ 0_n & E_n & -E_n\end{bmatrix}}{(2i+1)!}+\frac{\ln(2)^{2i+2}\cdot\begin{bmatrix}E_n & -E_n & 0_n\\ 0_n & 0_n & 0_n\\ 0_n & -E_n & E_n\end{bmatrix}}{(2i+2)!}.$$

Since the matrices are no longer dependent on the summation index, we get

$$M=\begin{bmatrix}-E_n & E_n & 0_n\\ 0_n & 0_n & 0_n\\ 0_n & E_n & -E_n\end{bmatrix}\sum_{i=0}^{\infty}\frac{\ln(2)^{2i+1}}{(2i+1)!}-\frac{\ln(2)^{2i+2}\cdot}{(2i+2)!}.$$

Using a computer algebra system, we find the sums evaluate to

$$M=\begin{bmatrix}-E_n & E_n & 0_n\\ 0_n & 0_n & 0_n\\ 0_n & E_n & -E_n\end{bmatrix}\cdot\frac{1}{2}=\begin{bmatrix}-\frac{1}{2}E_n & \frac{1}{2}E_n & 0_n\\ 0_n & 0_n & 0_n\\ 0_n & \frac{1}{2}E_n & -\frac{1}{2}E_n\end{bmatrix}.$$

Finally, adding this matrix to the identity matrix results in

$$\exp\left(\ln(2)\cdot\begin{bmatrix}-E_n & E_n & 0_n\\ 0_n & 0_n & 0_n\\ 0_n & E_n & -E_n\end{bmatrix}\right)=\begin{bmatrix}\frac{1}{2}E_n & \frac{1}{2}E_n & 0_n\\ 0_n & E_n & 0_n\\ 0_n & \frac{1}{2}E_n & \frac{1}{2}E_n\end{bmatrix}.$$

□

# Index

# Nomenclature

**Number Spaces**

$\mathbb{C}$ Set of complex numbers

$\mathbb{N}$ Set of natural numbers (excluding zero)

$\mathbb{N}_0$ Set of natural numbers (including zero)

$\mathbb{R}$ Set of real numbers

$\mathbb{R}^d$ $d$-dimensional vector space over the real numbers

$\mathbb{R}_{\geq 0}$ Set of non-negative real numbers

$\mathbb{S}^1$ Unit circle

$\mathbb{S}^2$ Unit sphere

**Variables**

$a$ Universal variable; used differently depending on context

$b$ Universal variable; used differently depending on context

$c$ Universal variable; used differently depending on context

$d$ Dimension of a space or mapping; typically 1, 2, or 3 in this work

$e$ Edge of a graph $\mathbf{G} = (\mathbf{V}, \mathbf{E})$ with $e \in \mathbf{E}$; also used to denote tangent nodes in a quadrilateral graph or edge points in subdivisions of polytopes

$f$ Facet of a planar 3-connected graph $\mathbf{G}$; also used to denote dual nodes in a quadrilateral graph, primal points of dual polytopes, and facet points

$g$ Degree of a B-spline or subdivision element; in this work 2 for quadratic and 3 for cubic elements

$h$ Height

$i$ General counting variable from the natural numbers $\mathbb{N}$ or $\mathbb{N}_0$

$j$ General counting variable from the natural numbers $\mathbb{N}$ or $\mathbb{N}_0$

$k$ General counting variable from the natural numbers $\mathbb{N}$ or $\mathbb{N}_0$

$l$ General counting variable from the natural numbers $\mathbb{N}$ or $\mathbb{N}_0$

$m$ General counting variable from the natural numbers $\mathbb{N}$ or $\mathbb{N}_0$

$n$ General counting variable from the natural numbers $\mathbb{N}$ or $\mathbb{N}_0$

$o$ Volume point or node of a graph of an initial element for $t = 3$ and $g = 3$

$p$ Vertex of a primal polytope in $\mathbb{R}^d$; also used to denote (primal) nodes in graphs

| | |
|---|---|
| $q$ | Coordinate point in $\mathbb{R}^d$ |
| $r$ | Radius or length; used differently depending on context |
| $s$ | Cell |
| $t$ | Type or geometric dimension of a B-spline or subdivision element; in this work 1 for curves, 2 for surfaces, and 3 for volumes |
| $v$ | Node of a graph $\mathbf{G} = (\mathbf{V}, \mathbf{E})$ with $v \in \mathbf{V}$ |
| $x$ | Coordinate variable |
| $y$ | Coordinate variable |
| $z$ | Coordinate variable |
| $\alpha$ | Universal angle; used differently depending on context |
| $\beta$ | Universal angle; used differently depending on context |
| $\delta_{i,j}$ | Kronecker delta with $\delta_{i,j} = 1$ if $i = j$ and $\delta_{i,j} = 0$ otherwise |
| $\varphi$ | Universal parameter; used differently depending on context |
| $\vec{n}$ | Normal vector |

**Matrices**

| | |
|---|---|
| $0_n$ | Zero matrix in $\mathbb{R}^{n \times n}$ |
| $A$ | Adjacency matrix in $\{0, 1\}^{n \times n}$ |
| $B$ | General matrix |
| $C$ | Colin-de-Verdière-Matrix in $\mathbb{R}^{n \times n}$ |
| $D$ | Diagonal matrix |
| $E$ | List of edges or matrix of edge points |
| $E_n$ | Identity matrix in $\mathbb{R}^{n \times n}$ |
| $F$ | List of facets of a simple planar 3-connected graph; also used as matrix of facet points |
| $G$ | Rotation matrix |
| $H$ | Permutation matrix in $\mathbb{R}^{n \times n}$ |
| $I$ | Index matrix |
| $J$ | Matrix of Jordan blocks of a matrix in $\mathbb{C}^{n \times n}$ |
| $K$ | Index matrix |
| $L$ | Lower left triangular matrix in $\mathbb{R}^{n \times n}$ |
| $M$ | General matrix |
| $M^T$ | Transpose of a matrix |
| $M_{(:,j)}$ | $j$-th column of a matrix, with $j \in \mathbb{N}$ |
| $M_{(i,:)}$ | $i$-th row of a matrix, with $i \in \mathbb{N}$ |

$M_{(i,j)}$ $(i, j)$-th entry of a matrix, with $i, j \in \mathbb{N}$

$N$ Normalization matrix

$P$ Matrix of control points in $\mathbb{R}^{n\times d}$; each row represents a point in $\mathbb{R}^d$, thus $P$ consists of $n$ control points; also used as matrix of vertices of a (primal) polytope

$R$ Upper right triangular matrix in $\mathbb{R}^{n\times n}$

$S$ Subdivision matrix in $\mathbb{R}^{n\times n}$

$T$ Special adjacency matrix of an $n$-gon

$V$ Matrix of eigenvectors

**Mappings**

**c**(**G**) Colin-de-Verdière-number of the graph **G**

**d** **D**-equivalence relation

***f*** General mapping; used differently depending on context

**g** System of generating functions

**h** Permutation

**i** Index mapping

**q** Left-hand side of a system of equations

**r** Reparametrization mapping

**s** Subdivision evaluation function

**Special Vectors**

$\vec{1}$ Column vector $[1, \ldots, 1]^T$ with $n$ entries, adapted to the context

$\vec{0}$ Column vector $[0, \ldots, 0]^T$ with $n$ entries, adapted to the context

**Eigenvalues**

$\lambda$ Subdominant eigenvalue of a matrix

$\lambda_i$ Unspecified eigenvalue of a matrix

$\mu$ Subsubdominant eigenvalue of a matrix

**Graphs and Sets**

Aut(**G**) Automorphism group of the graph **G**

**C**(**G**) Set of all Colin-de-Verdière-matrices of the graph **G**

**D** Spline domain

**E** Set of edges of a graph

**F** Set of faces of a planar 3-connected graph **G**

**G** General graph

**G**$^*$ Graph dual to **G**

**GL**$_n(\mathbb{C})$ Set of all invertible matrices from $\mathbb{C}^{n\times n}$

**H**$_n$ Set of all permutations from $\{1,\dots,n\}$ to $\{1,\dots,n\}$

**I** Index set with **I** $= \{1,\dots,k\}$ for some $k \in \mathbb{N}$

**K** Set of primal and dual node pairs lying on a common quadrilateral in a restricted quadrilateral graph

**K**$(f)$ Set of primal points $p$ lying on a common quadrilateral with $f$ in a restricted quadrilateral graph

**K**$(p)$ Set of dual points $f$ lying on a common quadrilateral with $p$ in a restricted quadrilateral graph

**L** General set

**M** General set

**P** Polytope in $\mathbb{R}^3$ as realization of a planar 3-connected graph **G**

**P**$^*$ Dual polytope to **P**

**Q**$'$(**G**) Restricted quadrilateral graph of the planar 3-connected graph **G**

**Q**(**G**) Quadrilateral graph of the planar 3-connected graph **G**

**R** General set

**V** Set of vertices of a graph

**Special Functions**

alg$(\lambda)$ Algebraic multiplicity of an eigenvalue

bs Bobenko and Springborn functional

conv(**M**) Convex hull of the set **M** $\subseteq \mathbb{R}^d$

det$(M)$ Determinant of the matrix $M \in \mathbb{R}^{n\times n}$

exp Exponential function or mapping with exp$(a) = e^a$ or exp$(M) = e^M$

exp$_2$ Exponential function or mapping with base 2, i.e., exp$_2(a) = 2^a$ or exp$_2(M) = 2^M$

geo$(\lambda)$ Geometric multiplicity of an eigenvalue

ker$(M)$ Kernel of the matrix $M$

kron$(M_1, M_2)$ Kronecker product of two matrices $M_1$ and $M_2$

max Maximum value of a set or a vector

min Minimum value of a set or a vector

$\phi$ Stereographic projection from $\mathbb{S}^2 \setminus \{[0,0,1]\}$ to $\mathbb{R}^2$

$\phi^{-1}$ Inverse stereographic projection from $\mathbb{R}^2$ to $\mathbb{S}^2 \setminus \{[0,0,1]\}$

$\rho(M)$ Spectral radius of the square matrix $M$

sgn(**h**) Sign of the permutation **h**

sp$(M)$ Spectrum of the square matrix $M$